\documentclass[final]{elsarticle}
 \usepackage[colorlinks,citecolor=blue,linktoc=all,linkcolor=cyan]{hyperref}
\usepackage{graphicx}

\usepackage[T1]{fontenc}
\usepackage{mathrsfs}             
\usepackage{slashed}              
\usepackage{amsmath}
\usepackage{amssymb}
\usepackage{amsbsy}
\usepackage{amsfonts}
\usepackage{multirow}
\usepackage{multicol}
\usepackage{caption}
\usepackage{braket}
\usepackage{subcaption}

\numberwithin{equation}{section}
\numberwithin{table}{section}
\numberwithin{figure}{section}

\journal{Progress in Particle and Nuclear Physics}

\usepackage{titlesec}
\usepackage{sectsty}
\titleformat{\section}{\normalfont\Large\bfseries}{\thesection}{1em}{}
\titleformat{\subsection}{\normalfont\large\bfseries}{\thesubsection}{1em}{}
\titleformat{\subsubsection}{\normalfont\normalsize\bfseries}{\thesubsubsection}{1em}{}

\begin{document}
	
\begin{frontmatter}
		
\title{Neutrinos and their interactions with matter}

\author[mymainaddress]{M. Sajjad Athar\corref{mycorrespondingauthor}}
\cortext[mycorrespondingauthor]{Corresponding author}
\ead{sajathar@gmail.com}
	
\author[mymainaddress]{A. Fatima}
\author[mymainaddress]{S. K. Singh}

\address[mymainaddress]{Department of Physics, Aligarh Muslim University, Aligarh - 202002, India}

\begin{abstract}
We have presented a review of the properties of neutrinos and their interactions with matter. The different 
(anti)neutrino processes like the quasielastic scattering, inelastic production of mesons and hyperons, and the deep 
inelastic scattering from the free nucleons are discussed, and the results for the scattering cross sections are 
presented. The polarization observables for the leptons and hadrons produced in the final state, in the case of 
quasielastic scattering, are also studied. The importance of nuclear medium effects in the low, intermediate, and 
high energy regions, in the above processes along with the processes of the coherent neutrino-nucleus scattering, 
coherent meson production, and trident production, has been highlighted. In some cases, the results of the cross 
sections are also given and compared with the available experimental data as well as with the predictions in the 
different theoretical models. This study would be helpful in understanding the (anti)neutrino interaction cross 
section with matter in the few GeV energy region relevant to the next generation experiments like DUNE, 
Hyper-Kamiokande, and other experiments with accelerator and atmospheric neutrinos. We have emphasized the need of 
better theoretical models for some of these processes for studying the nuclear medium effects in nuclei. 

\end{abstract}		
\end{frontmatter}


	\thispagestyle{empty}
	\tableofcontents


\section{Introduction}\label{intro}
The idea of neutrino initially called ``neutron'', as neutral, weakly interacting, spin $\frac{1}{2}$ particle which 
obeys exclusion principle, having mass of the same order of magnitude as that of the electron mass was proposed by 
Pauli~\cite{Pauli:1930pc}, in 1930. This particle was hypothesized in order to explain the two outstanding problems in 
contemporary nuclear physics related with the observed continuous energy spectrum (violation of energy-momentum 
conservation) of electrons in $\beta$-decays of nuclei and the nuclear structure (anomalies in understanding the 
spin-statistics relation). Immediately after neutrinos were conjectured, the theoretical study of nuclear beta decay 
started with the works of Fermi~\cite{Fermi:1933jpa, Fermi:1934sk, Fermi:1934hr} and Perrin~\cite{Perrin:1933}, 
followed by the works of Henderson~\cite{Henderson:1934}. This may be considered to be the beginning of the study of 
neutrino interactions with electrons and nucleons. Fermi conceived the idea of ``four fermion current-current type 
point interaction'' with the strength of the interaction given by a coupling constant $G_F$ to describe the $\beta$ 
decay rates and the shape of the beta spectrum. He considered the interaction currents to be vector in nature 
following the analogy with the quantum electrodynamics~(QED). The experimental analyses of the shape of the 
$\beta$-spectrum from the various nuclear beta decays showed that the neutrino mass has to be very much smaller than 
the electron mass. Bethe and Peierls~\cite{Bethe:1934qn} were the first who performed the theoretical calculations for 
the total scattering cross sections for $\bar\nu + p \rightarrow e^+ + n$ process using $G_F$ as the strength of the 
interaction determined from nuclear beta decays. The calculated cross section was found to be too small~($10^{-44} 
\text { cm}^2$ for a 2.3~MeV neutrino beam) to be observed experimentally unless the neutrino flux and/or the mass of 
the detector material were increased by many orders of magnitude. This hindered any further progress in attempts to 
experimentally study the neutrino interactions with matter and supported Pauli's apprehension that ``I did a terrible 
thing, which no theorist should do, I postulated a particle that can not be detected''. After more than twenty five 
years of neutrino hypothesis and several experimental efforts, Reines and Cowan in 1956~\cite{Reines:1956rs, 
Cowan:1956rrn} were finally able to observe neutrinos at the Savannah River reactor, and sent a telegram to Pauli 
about their findings ``We are happy to inform you that we have definitely detected neutrinos....''. Pauli publicly 
announced this discovery in 1956 to the participants at the CERN Symposium, and replied to their message that 
``Everything comes to him who knows how to wait''. Since then the progress in understanding the physics of neutrinos 
and the development in neutrino physics has been amazing and full of surprises. The neutrinos continue to challenge 
our expectations even today regarding the validity of some symmetry principles and conservation laws in particle 
physics. A better understanding of these symmetry principles would be helpful in the fields of nuclear and particle 
physics as well as in the fields of cosmology and astrophysics~\cite{Athar:2020kqn}. 

The experimental study of $\beta$ decays of various nuclei made considerable progress and helped in the formulation of 
the theory of weak interactions by extending the Fermi theory of beta decay. During the next forty years following the 
idea of neutrinos, extensive work on the nuclear beta decays and many other weak decays of muons, nucleons, hyperons, 
and mesons led to the phenomenological theory of weak interactions known as the $V-A$~(Vector-Axial Vector) 
theory~\cite{Sudarshan:1958vf, Feynman:1958ty, Sakurai:1958zz}. This theory was formulated using the various properties 
of neutrinos determined experimentally like their mass, spin, helicity i.e. left~(right) handed 
neutrinos~(antineutrino) and the theoretical idea of the chiral~($\gamma_5$) invariance of neutrino interactions 
leading to the prediction of parity violation~\cite{Lee:1956vjd, Lee:1956qn}, which was confirmed subsequently by Wu 
et al.~\cite{Wu:1957my} and later by other experiments. With the discovery of the $\tau$ lepton in 1975 and various 
hadrons with heavy quark contents like the charm~($c$), bottom~($b$), and top~($t$) quarks and analyses of their weak 
decays, the $V-A$ theory of weak interactions was reformulated in terms of the leptons and quarks using the concept of 
quark mixing proposed by Cabibbo~\cite{Cabibbo:1963yz} and extended by Kobayashi and Maskawa~\cite{Kobayashi:1973fv} 
described in terms of the Cabibbo-Kobayashi-Maskawa~(CKM) matrix~\cite{ParticleDataGroup:2020ssz}. The experimental 
analyses of various weak interaction processes using the phenomenological $V-A$ theory were performed, which resulted 
in understanding the following properties of neutrinos and their interactions with matter:
\begin{itemize}
\item There are three flavors of (anti)neutrinos i.e. $\nu_e(\bar\nu_e)$, $\nu_\mu(\bar\nu_\mu)$, $\nu_\tau 
(\bar\nu_\tau)$ with limits on the masses so tiny that they can be considered to be massless. They are classified 
according to separate lepton flavor numbers for each flavor $i$ i.e. $L_i$($i=e,\mu,\tau$) and assigned $L_i=+1(-1)$ 
for the individual neutrino and antineutrino flavors.

\item The neutrinos and antineutrinos of each flavor are neutral, spin half fermions with helicity $-1~(+1)$ popularly 
known as left~(right)-handed fermions.

\item Neutrinos interact with the charged leptons and quarks through the exchange of massive charged vector fields 
${W_\mu}^{\pm}$ between the neutrino-charged lepton and quark-quark currents with the same strength for all the 
flavors. These currents transform as $V^\mu - A^\mu$ and are constructed as the charge carrying bilinear covariants 
from the lepton fields of the same flavor in the case of leptons and the quark fields in a CKM rotated flavor space in 
the case of quarks and carry linear momentum and energy. This is known as the phenomenological $V-A$ 
theory~\cite{Sudarshan:1958vf, Feynman:1958ty, Sakurai:1958zz}.

\item In this theory, the neutrino interactions are such that:
\begin{itemize}
 \item The lepton flavor number~(LFN) $L_i$~($i=e,\mu,\tau$) is conserved separately for each flavor and there are no 
 lepton flavor violating~(LFV) currents.
 
 \item The neutrinos of all flavors~($\nu_i;~i=e,\mu,\tau$) interact with the leptons of the same flavor and quarks 
 with the same strength for each flavor i.e. there exists lepton flavor universality~(LFU). 
 
 \item Most of the weak processes involving neutrinos and hadrons are charge changing with the hadronic currents 
 obeying $\Delta S=0$ or $|\Delta S|=1$ rule, where $S$ is the strangeness quantum number. The strength of the 
 couplings of $|\Delta S| = 1$ hadronic currents is suppressed as compared to the $\Delta S=0$ hadronic currents by a 
 factor of $\tan^{2} \theta_{C}$, where $\theta_{C} = 13.1^{0}$ is the Cabibbo angle. However, neutral currents~(NC) 
 are highly suppressed in $|\Delta S|=1$ sector leading to the principle of the absence of flavor changing neutral 
 current~(FCNC). There is no conclusive evidence of the existence of charge conserving NC in the $\Delta S=0$ sector. 
\end{itemize}
Therefore, the weak transitions like $\nu_l({\bar\nu}_l) \longrightarrow l^-(l^+);~~l=e,\mu,\tau$ occur with the same 
strength for each $l$. The weak transitions like  $\nu_l({\bar\nu}_l) \longrightarrow \nu_l({\bar\nu}_l)$ and 
$l^-(l^+) \longrightarrow l^-(l^+)$, without involving any change of charge are highly suppressed and the transitions 
like $\nu_l({\bar\nu}_l) \longrightarrow {l^\prime}^-({l^\prime}^+)$, $\nu_l({\bar\nu}_l) \longrightarrow \nu_{l^\prime}
({\bar\nu}_{l^\prime})$, where $l \neq l^\prime$ with $l,~ l^\prime= e ~\text{or}~ \mu ~\text{or}~ \tau$, and have not 
been observed are not allowed in the $V^{\mu} - A^{\mu}$ theory. 
\end{itemize}
The phenomenological $V-A$ theory of weak interaction successfully describes the neutrino interactions with matter 
specially at low energies. In the high energy region of neutrinos, the scattering cross section from the charged 
leptons and nucleons leads to divergences when calculated in higher orders of the perturbation theory and the theory 
is not renormalizable. Various attempts to find a renormalizable theory of weak interactions were not successful until 
a unified theory of weak and electromagnetic interactions of leptons was formulated by Weinberg~\cite{Weinberg:1967tq} 
and Salam~\cite{Salam:1968rm} and extended to the quark sector using GIM mechanism proposed by Glashow et 
al.~\cite{Glashow:1970gm}. This unified theory of electroweak interactions is known popularly as the standard 
model~(SM).

The SM was formulated using various experimental results on the neutrino properties and their weak interactions 
obtained from the phenomenological $V-A$ theory as described above and the theoretical ideas inspired from the gauge 
field theory of electromagnetic interactions based on the local $U(1)$ symmetry group and its extension to the higher 
nonabelian local symmetry groups by Yang and Mills~\cite{Yang:1954ek}. Such gauge field theories predict the existence 
of massless vector fields as the mediating field for generating the underlying interactions. The masses of the massless 
gauge fields are then generated using the idea of the spontaneously broken gauge theories by introducing the 
interacting scalar fields in the theory developed by Englert and Brout~\cite{Englert:1964et}, and 
Higgs~\cite{Higgs:1964pj}. In the SM, the group structure of the higher local gauge symmetry and the interacting 
scalar fields to break the symmetry spontaneously are chosen such that the four massless vector gauge fields appear by 
the requirement of the invariance under local gauge symmetry out of which masses are generated for the three vector 
fields and the fourth vector field remains massless. The three massive vector fields are identified as the fields 
mediating the weak interactions and the fourth massless field is identified as the vector field mediating the 
electromagnetic interactions thus providing a unified theory of electroweak interactions. The renormalizability of the 
theory was soon demonstrated by 't Hooft and Veltman~\cite{tHooft:1972tcz}, and Lee and Zinn-Justin~\cite{Lee:1972fj}.
The theory reproduces all the results obtained by the phenomenological $V-A$ theory and predicts various new physical 
processes which have been observed by the later experiments confirming the SM as a unified theory of electroweak 
interactions. For example, the prediction of:
\begin{itemize}
\item neutral weak currents$(\Delta Q = 0)$ in the neutrino interactions with $\Delta S=0$, which were first observed 
in neutrino experiments at CERN~\cite{GargamelleNeutrino:1974khg} and confirmed later in many other 
experiments~\cite{Cnops:1978gm, Heisterberg:1979aq, Faissner:1978qu}.

\item neutral weak currents in the electron sector leading to the parity violation in the polarized electron 
scattering, which were first observed in electron scattering experiments at SLAC~\cite{Prescott:1978tm} and confirmed 
later in many other experiments.

\item charged~($W^\pm$) and neutral~$(Z^0)$ gauge bosons which were observed at CERN in UA1 and UA2 
experiments~\cite{UA1:1983crd, UA2:1983tsx} with masses $M_{W^\pm}=80.38$~GeV and $M_{Z^0}=91.18$~GeV.

\item scalar Higgs boson~($H$) and its decays which were experimentally confirmed in CMS~\cite{CMS:2012qbp} and 
ATLAS~\cite{ATLAS:2012yve} experiments in 2012 with a mass of Higgs boson $M_{H} = 125.25$~GeV. 
\end{itemize}

However, there are some experimental results which are not explained by the SM and need physics beyond the standard 
model~(BSM). For example, the existence of:
\begin{itemize}
\item neutrino oscillations which imply 
\begin{itemize}
 \item [(i)] mixing of neutrino flavors, 
 \item [(ii)] the neutrinos to be massive, 
 \item [(iii)] additional flavor of nonstandard neutrino i.e. sterile neutrino which has no interaction with ordinary 
 matter. 
\end{itemize}

\item early indication of CP violation in neutrino interactions.
\item FCNC like $K_L^0 \longrightarrow \mu^+ \mu^-$, $K^{+} \longrightarrow \pi^+ \nu 
{\bar\nu}$, etc.~\cite{Cirigliano:2011ny, Lin:2022rxv}.
\end{itemize}

Furthermore, various experimental efforts are going on to observe rare processes that would require the existence of 
nonstandard interactions~(NSI). For example, the possible observation of~\cite{Bifani:2018zmi, 
Albrecht:2021tul, LHCb:2021trn, deSimone:2020kwi, Belle:2019rba, Celani:2021hni}: 
\begin{itemize}
\item  neutrinoless double beta decay~(NDBD), for which many experiments are being done implying neutrino to be its own 
antiparticle, known as the Majorana type of neutrino, requiring major changes in our understanding of neutrino 
interactions with matter.

\item decays like $K^+ \longrightarrow \pi^+ e^\mp \mu^\pm$, $K^- \longrightarrow \pi^- e^\mp \mu^\pm$, $B^+ 
\longrightarrow K^+ \mu^\pm \tau^\mp$, $B^+ \longrightarrow K^+ \mu^\pm e^\mp$, etc., which involve both FCNC and LFV.

\item LFV in purely leptonic processes with or without a photon like $\mu^- \longrightarrow  e^- \gamma$, $\mu^+ 
\longrightarrow e^- e^+ e^+$, or $\mu \leftrightarrow e$ conversion in nuclei. 

\item lepton flavor universality violation~(LFUV) in weak decays like $\pi^+ \longrightarrow \mu^- e^+ e^+ \nu_e$ as 
well as in the heavy quark sector like $b \longrightarrow s l\bar{l}$, $b\longrightarrow c l\bar{\nu}_{l}$, etc.
\end{itemize}

In the last 90 years since the neutrino was postulated and speculations were made about its interactions by Pauli, 
enormous progress has been made in understanding the neutrino interactions with matter but it is still far from being 
understood satisfactorily. Most of the observed electroweak processes are explained with the SM but the observation of 
certain phenomena like the neutrino oscillation, CP violation and FCNC requires BSM physics and there are many 
theoretical studies being made presently to study the BSM physics~\cite{Arguelles:2022xxa}. However, in view of the 
space limitations, we focus in this review only on the standard model interaction of neutrinos with matter. 
In literature, there are quite a few recent review articles~\cite{Katori:2016yel, Alvarez-Ruso:2014bla, Morfin:2012kn, Gallagher:2011zza, SajjadAthar:2021prg, Athar:2021xfr, NuSTEC:2017hzk} dealing with various aspects of the neutrino interactions with matter in the Standard Model. 
The present review deliberates at length on the interaction of neutrinos with nucleons and nuclei in the low, 
intermediate, and high energy regions focussing on the nuclear medium effects. 
The importance of various nuclear medium 
effects~(NME) like the Fermi motion, Pauli blocking, multinucleon correlation effects are discussed in various neutrino processes of quasielastic, inelastic, and deep inelastic scattering 
when the (anti)neutrino scattering takes place in nuclei like $^{12}$C, $^{16}$O, $^{40}$Ar, $^{208}$Pb, etc. relevant for the present and future $\nu(\bar{\nu})-$nucleus scattering experiments. 
The importance of understanding the role of quark-hadron duality in describing the neutrino scattering from nucleons and nuclei in the shallow inelastic region has been emphasized.

In Section~\ref{nu_properties}, we summarize the properties of neutrinos as we understand them today and describe 
various sources of neutrinos in the energy range of eV to EeV. In Section~\ref{nu:theory}, a brief discussion about 
the theoretical description of neutrinos and their interaction is presented. In Section~\ref{SM}, the basic theory of 
the neutrino interactions with leptons and quarks in the framework of the SM is obtained. In 
Sections~\ref{nu_interaction}, \ref{sec:inelastic:nucleon}, and \ref{dis:nucleon}, we describe the various processes 
of neutrino scattering from the nucleons viz. elastic, quasielastic, inelastic and deep inelastic scattering, 
respectively, and discuss NME in these processes in Section~\ref{nu:nuclei}. In 
Section~\ref{QH:duality}, we present in brief the concept of quark-hadron duality in the weak sector. The different 
neutrino event Monte Carlo generators are discussed in Section~\ref{MC}. Finally, we summarize the neutrino 
interaction physics presented in this review in Section~\ref{sec:sum}. 

\subsection{Experimental observations and properties of neutrinos}\label{nu_properties}
\subsubsection{Detection of neutrinos}\label{nu_detection}
The experimental attempts to make direct observation of neutrinos and antineutrinos started immediately after the 
formulation of the theory of beta decay, and the first attempt was made by Nahimas~\cite{Nahmias:1935yih} as early as 
1935 at the underground station Holborn in London, and later by Rodeback and Allen~\cite{Rodeback:1952gb}, 
Leipunski~\cite{Leipunski:1936}, Snell and Pleasonton~\cite{PhysRev.97.246}, Jaeobsen~\cite{Jaeobsen:1945}, 
Sherwin~\cite{PhysRev.73.216}, and Crane and Halpern~\cite{PhysRev.53.789}, which showed no conclusive evidence of the 
existence of neutrinos. The attempts took much longer time to succeed experimentally and the final success came when 
Reines and Cowan~\cite{Reines:1956rs, Cowan:1956rrn} in 1956 at the Savannah River reactor reported the observation of 
antineutrinos from the reactor in the reaction
\begin{eqnarray}\label{eq:nu_obser}
\bar{\nu}_{e} + p \rightarrow e^{+} + n
\end{eqnarray}
by making a coincidence measurement of the photons from particle annihilation $e^{+} + e^{-}\rightarrow \gamma + 
\gamma$ and a neutron capture $n + ^{108}\text{Cd}\rightarrow ^{109}\text{Cd} + \gamma$ reaction a few microseconds 
later~\cite{Reines:1956rs, Reines:1959nc} induced by $e^{+}$ and $n$ produced in reaction~(\ref{eq:nu_obser}).

The original proposal of Pontecorvo~\cite{Pontecorvo:1946} and Alvarez~\cite{Alvarez:1949} to use $^{37}$Cl as target 
to observe neutrinos from the reactors was followed by Davis~\cite{Davis:1964hf, Davis:1968cp} who looked for $\nu_e + 
^{37}$Cl $\rightarrow e^- + ^{37}$Ar reaction at the Brookhaven reactor using 4000 L of liquid CCl$_4$ and tried to 
observe the $^{37}$Ar produced in the reaction. No event was observed but a limit of $\bar{\sigma} (\bar{\nu} + ^{37} 
\text{Cl} \rightarrow e^- + ^{37}\text{Ar}) < 0.9 \times 10^{-45} \text{cm}^2$ was obtained while the theoretical 
prediction was $\approx 2.6 \times 10^{-45} \text{cm}^2$. This negative result was of importance as it showed that the 
neutrinos from the reactors do not produce electrons hinting that $\nu_e$ and ${\bar\nu}_e$ are different particles. 
In order to phenomenologically describe the situation, a new quantum number was proposed called the electron lepton 
number: $L_e$. The $\nu_e$ and $e^-$ were assigned $L_e=+1$, and  ${\bar\nu}_e$ and $e^+$ were assigned $L_e=-1$. 
 
It was suggested by  Markov~\cite{Markov}, Pontecorvo~\cite{Pontecorvo:1959sn}, and  Schwartz~\cite{Schwartz:1960hg} 
to use proton accelerators to produce high energy neutrino beam from pion decays to perform experiments like:
\begin{eqnarray}\label{ele-muo}
\nu+n &\longrightarrow& \mu^- + p \qquad \qquad \qquad \nu+n \longrightarrow e^- + p \\
\bar{\nu}+p &\longrightarrow& \mu^+ + n \qquad \qquad \qquad \bar{\nu}+p \longrightarrow e^+ + n 
\end{eqnarray}
to test whether the neutrinos from pion decays produce muons or electrons. Theoretical calculations for the above 
processes were done by Lee and Yang~\cite{Lee:1960qv}, Cabibbo and Gatto~\cite{Cabibbo:1961zz}, and 
Yamaguchi~\cite{Yamaguchi:1961sq} using the phenomenological $V-A$ theory. The experiments performed at the Brookhaven 
National Laboratory~(BNL) by Danby et al.~\cite{Danby:1962nd} and later at CERN by Bienlein et 
al.~\cite{Bienlein:1964zqi} observed that neutrinos from the pion decays, which were accompanied by muons, produce 
only muons in the above reactions~(Eq.~(\ref{ele-muo})) and never an electron/positron was observed. This confirmed 
that these neutrinos are different from the neutrinos produced in beta decay implying $\nu_{\mu} \ne \nu_{e}$. 
Consequently, for the muon family separate lepton number $L_{\mu}$ was defined. These lepton numbers were assumed to 
be conserved separately. The $L_e$ and $L_\mu$ were combined to define a new quantum number, i.e., LFN, $L_{f}~ (f=e,
\mu)$.
 
In 1975 when $\tau$-lepton was discovered~\cite{Perl:1975bf} and its weak decays were observed the existence of a new 
flavor of neutrinos $\nu_\tau$ was proposed, which was observed much later in the DONUT experiment~\cite{DONuT:2007bsg, 
DONUT:2000fbd} in 2000 at the Fermilab. More observations of $\nu_{\tau}$ induced events were later made in experiments 
with the accelerator and the atmospheric neutrinos by DONUT~\cite{DONUT:2000fbd}, OPERA~\cite{OPERA:2010pne, 
OPERA:2014fax, OPERA:2015wbl}, Super-Kamiokande~\cite{Super-Kamiokande:2017edb}, and IceCube~\cite{IceCube:2019dqi} 
collaborations. Future experiments like DsTau~\cite{DsTau:2019wjb}, SHiP~\cite{SHiP:2020sos, DiCrescenzo:2016irr} and 
DUNE~\cite{Machado:2020yxl, DUNE:2016evb, DUNE:2018mlo} are planning to observe significantly large number of events 
induced by the $\nu_{\tau}$ interactions.

\subsubsection{Sources of neutrinos and their fluxes}\label{nu_sources}
\begin{figure}
\centering
	\includegraphics[height=7cm, width=15 cm]{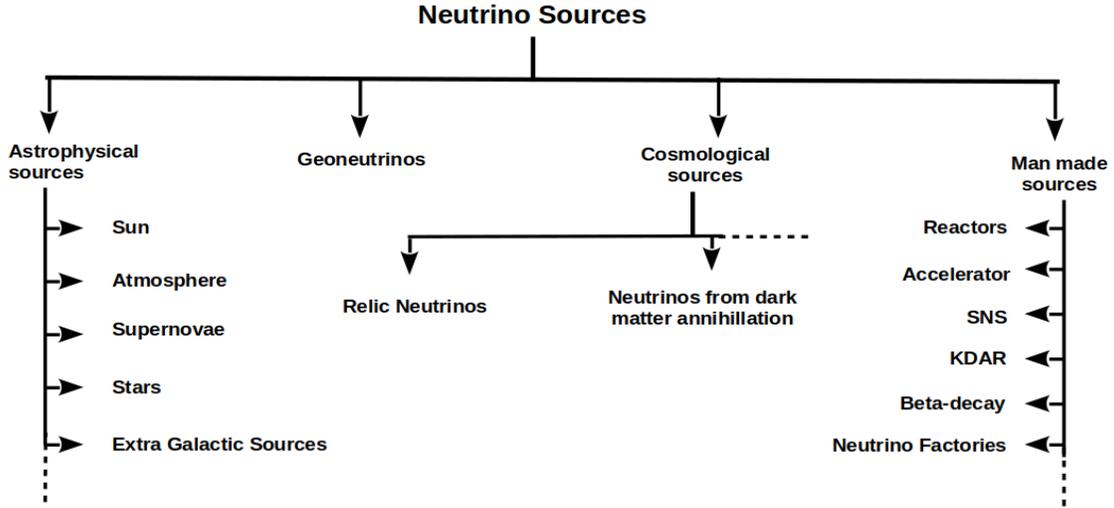}
	\caption{Different sources of neutrinos.}\label{Fig:sources1}
\end{figure}

The SM neutrinos are of three flavors viz. $\nu_e$, $\nu_\mu$ and $\nu_\tau$ and the corresponding antineutrinos. 
Initially the experiments were performed with the reactor antineutrinos and the solar neutrinos and later with the 
development of accelerators, $\nu_\mu$ and ${\bar\nu}_\mu$ beams were used. Today we know that there are various 
sources of neutrinos all around us and these sources may be broadly divided into two groups, one the natural sources 
and the other man made sources of (anti)neutrinos as shown in Fig.~\ref{Fig:sources1}. The neutrinos produced from the 
natural sources are the ones coming from the sun's core, earth's core and mantle, etc. Neutrinos are always produced 
during the birth, collision, and the death of stars. Particularly huge flux of neutrinos is emitted during a supernovae 
explosion. There are neutrinos around us which are relics of the Big Bang, and were produced almost 13.7 billion years 
ago, soon after the birth of the Universe. There are many other sources of astrophysical neutrinos like the cosmogenic 
neutrinos, neutrinos being produced in the violent collisions of high energy protons with active galactic nuclei, etc. 
Besides the various natural sources, there are man made sources of neutrinos and antineutrinos being produced at the 
particle accelerators, nuclear reactors, spallation neutron source~(SNS) facilities, etc. These neutrinos and 
antineutrinos from the various sources cover an energy span from $\mu$eV~($10^{-6}$~eV) to EeV~($10^{18}$~eV) as shown 
in Fig.~\ref{Fig:Data1}~\cite{Katz:2011ke}. Recently, Vitagliano et al.~\cite{Vitagliano:2019yzm} have also provided 
neutrino spectrum at earth obtained using different neutrino sources. The $\nu_\tau$ and ${\bar\nu}_\tau$ from the 
atmospheric source come with a very small flux which have been recently observed in the 
Super-Kamiokande~\cite{Super-Kamiokande:2012xtd, Super-Kamiokande:2017edb} and the IceCube~\cite{IceCube:2019dqi} 
experiments .

\begin{itemize}
 \item {\bf Natural neutrino sources}\\
All the stars including the sun create their energy through nuclear fusion reactions that take place in the star's 
core~\cite{Bahcall:2002ng}. The proton-proton chain reaction dominates in stars with mass of the order of the mass of the sun or smaller, 
while the Carbon-Nitrogen-Oxygen~(CNO) cycle reaction dominates in the stars with mass greater than 1.3 times mass of 
the sun. The process like hydrogen fusion to helium takes place via a sequence of chain reactions that begins with the 
fusion of two protons to form deuterium nucleus along with the emission of $e^+$ and $\nu_e$ and the complete process
may be written as
\begin{equation*}
 4p + 2e^- \longrightarrow ~~^4He + 2\nu_e + 26.7~\text{MeV}.
\end{equation*}
Corresponding to the luminosity of the sun as $3.9 \times 10^{26}$~Watt, almost $7 \times 10^{10}~ \nu_e/\text{cm}^2/
\text{sec}$ reach the earth's surface.

Atmospheric neutrinos~\cite{Gaisser:2002jj} are produced through the decay of secondary cosmic ray particles~($\pi, K,$ etc.) produced in the 
interaction of primary cosmic rays~(mainly protons) with the earth's atmosphere through the processes like:
\begin{eqnarray}
p+A_{air} \rightarrow n+\pi^++X; \qquad \qquad  n+A_{air}&\longrightarrow& p+\pi^-+X.\nonumber
\end{eqnarray}
The pions~(kaons) subsequently give rise to (anti)neutrinos
\begin{eqnarray}
\pi^\pm &\longrightarrow & \mu^\pm\;\nu_\mu(\bar\nu_\mu)\hspace{4 cm} (100\%)\nonumber\\
\mu^\pm &\longrightarrow & e^\pm\;\nu_e(\bar\nu_e)\bar\nu_\mu(\nu_\mu)\hspace{3.2 cm}(100\%)\nonumber\\
K^\pm&\longrightarrow &\mu^\pm\;\nu_\mu\;(\bar\nu_\mu)~~~(63.5\%); ~~~\pi^\pm\;\pi^0~~~(20.7\%);~~~
 \pi^\pm\;\pi^+\;\pi^-~~~(5.6\%) ...\nonumber
\end{eqnarray}
The spectrum of these secondaries peaks in the GeV range, extends to high energy region with approximately a power-law 
spectrum and therefore the neutrino flux decreases rapidly with the increasing energy. Up to the energies of about 
100~TeV, the neutrino flux is dominated by pion and kaon decays. 

Supernova neutrinos~\cite{Bethe:1985sox} are produced during the death phase of a massive star. When the core collapse-supernovae burst 
out, a colossal amount of energy is carried out mainly by all the flavors of neutrinos and antineutrinos. The energy 
released in a supernova explosion is the difference in the binding energy of the parent star and a neutron star and 
such explosions give rise to about $10^{58}$ $\nu$s and $\bar\nu$s  in a few tens of seconds of time, carrying out 
almost 99$\%$ of the gravitational binding energy of a dying star.

Active galactic nuclei~(AGN) are considered to be one of the sources of very high energy neutrinos~\cite{Halzen:2018fjs}. These AGN can 
accelerate protons up to about a maximum energy of $\sim 10^{20}$~eV and are surrounded by high intensity radiation 
fields, which act as the source of photo-hadron interactions and subsequently give rise to neutrinos.
\begin{figure}
\centering
\includegraphics[height=7.0 cm, width=15 cm]{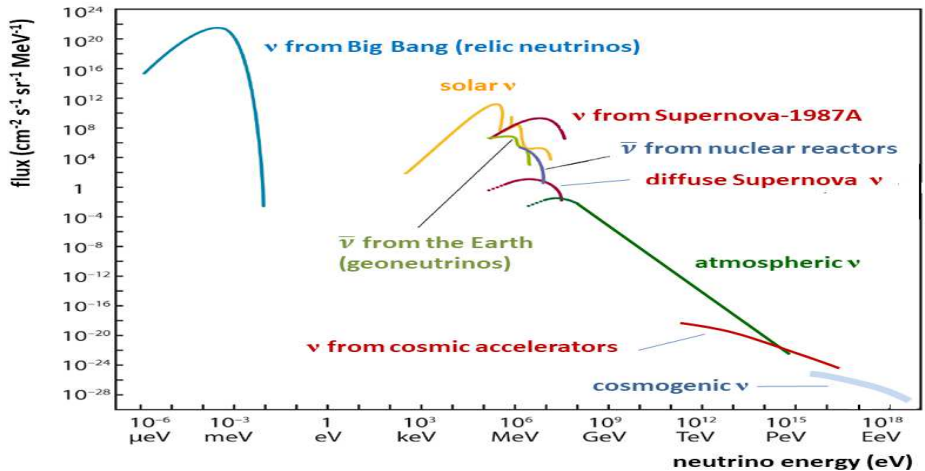}
\caption{Particle fluxes of neutrinos from different sources on earth. The flux is given in units of neutrinos per 
square centimeter, second, steradian and MeV. The neutrinos from the sun are indeed neutrinos, while those from the 
earth's interior and from nuclear reactors are antineutrinos. All other sources contain about as many neutrinos as 
antineutrinos. The relic neutrinos from the Big Bang, the diffuse supernova neutrinos and, at highest energies, 
cosmogenic neutrinos have not been detected yet~(courtesy C. Spiering)~\cite{Katz:2011ke}. }\label{Fig:Data1}
\end{figure}

Cosmogenic neutrinos are produced in the interaction of cosmic rays like the 
nucleons, whether they are free or bound in nuclei with the Lorentz boost factor $\Gamma \geq 10^{10}$ 
with the cosmic microwave background radiation  and gives rise to 
photo-pion production, where the pions decay to give rise to neutrinos:
\begin{eqnarray*}
 N + \gamma &\rightarrow& N^\prime + \pi^\pm;~~N, N^\prime=p ~\text{or}~ n. \nonumber
\end{eqnarray*}
The earth's interior radiates heat at the rate of about 47~TW. Some part of this heat loss is accounted for the heat 
generated upon the decay of radioactive isotopes like $^{40}$K, $^{232}$Th, $^{238}$U, etc. in the earth's interior 
which produce antineutrinos through a series of decays including beta decays like
\begin{eqnarray*}
 ^{238}U &\longrightarrow& ^{206}Pb + 8\alpha + 6e^- + 6{\bar\nu}_e + 51.7\text{~MeV},\nonumber\\
 ^{40}K &\longrightarrow& ^{40}Ca + e^- + {\bar\nu}_e + 1.311~\text{MeV},~~\text{etc.}
\end{eqnarray*}
which constitute geoneutrinos~\cite{Fiorentini:2007te}. It has been estimated that about $10^{6}~ \bar{\nu}_e/ \text{cm}^2$ reach the earth's 
surface from the decay of radioactive isotopes present in the earth's core. Recently, the information about spatial 
distribution of radionuclides has been studied and from this the size of the earth's core and mantle has been 
estimated. 
 
The cosmic-neutrino background~(C$\nu$B) or more commonly known as the relic neutrinos are the relics of the Big Bang 
and their origin is similar to the cosmic microwave background radiation observed by Penzias and Wilson in 1965. 
C$\nu$B are neutrinos which decoupled from matter when the universe was around one second old. It is estimated that 
these relic neutrinos have a temperature of about 1.95~K and an average density of around 330/cm$^3$.

\item {\bf Man made sources: accelerator and reactor (anti)neutrinos}\\
Accelerator and reactor based neutrino and antineutrino sources have been crucial to understand the neutrino properties. 
 Markov~\cite{Markov}, Pontecorvo~\cite{Pontecorvo:1959sn}, and  Schwartz~\cite{Schwartz:1960hg}, independently, proposed the 
idea of doing neutrino experiments with accelerators. They proposed the possibility of an experiment making use of a neutrino 
beam produced by pion decays at the proton accelerators. The more robust experiments with high energy neutrinos started with the 
development of synchrotron accelerators during 1960s, the AGS at Brookhaven and the PS at CERN operating at proton energies 
up to 30 GeV, and with this new window of studying neutrino interactions at the GeV scale opened. The first experiments with 
the accelerator neutrinos ran in 1962 at Brookhaven and CERN which showed $\nu_e$ and $\nu_\mu$ are different 
particles~\cite{Danby:1962nd}. The accelerator facilities are used to accelerate the protons to very high energies. These 
highly energetic protons are smashed into a target, the target can be any material, although it has to be able to withstand 
very high temperatures. When a proton traveling near the speed of light hits a target, it slows down and the proton's energy 
is used to produce a jet of hadrons. There are different kinds of particles in this jet, however, the most common are pions 
and kaons. The charged pions so produced are unstable and decay essentially into muons and neutrinos. A meson, carrying 
electric charge, can be collimated using electric and magnetic fields known as magnetic horns. Thus, to get a neutrino beam 
in a certain direction, one points the pions/kaons in the direction of the detector. A properly designed horn system can enhance the 
neutrino flux. To estimate the neutrino flux with better accuracy, it is important to precisely measure the momentum and the 
angular spectra of the mesons. In 1965, at BNL a new method to determine the flux of neutrinos as a function of the protons 
on target~(POT) was implemented which was later applied at CERN in 1967. Later accelerator neutrino experiments started at 
ANL. With the start of 1970s several accelerator neutrino experiments started to operate like the 350--400~GeV proton 
accelerator at Fermilab, the 70 GeV proton accelerator at Serpukhov, and in 1976, the 300 GeV super proton synchrotron~(SPS) 
at CERN and since then the tradition of using accelerator neutrinos have gradually strengthened~\cite{Dore:2018ldz, Mahn:2018mai}.

In the beginning of 21st century, several neutrino experiments started coming up around the globe like MiniBooNE, K2K, CNGS, 
MicroBooNE, NOvA, etc. and have used accelerators to produce pions and kaons which were collimated to produce neutrinos. The 
next generation experiment DUNE$@$Fermilab would be using imaging type of liquid argon time projection chamber~(LArTPC), and 
similarly T2HyperK in Japan would be important in addition to the current generation experiments to understand many of the 
neutrino properties. In Fig.~\ref{neu-flux}, we show the neutrino spectra of some the accelerator experiments. 
\begin{figure}[t]
\centering
	\includegraphics[height=7.0 cm, width=10 cm]{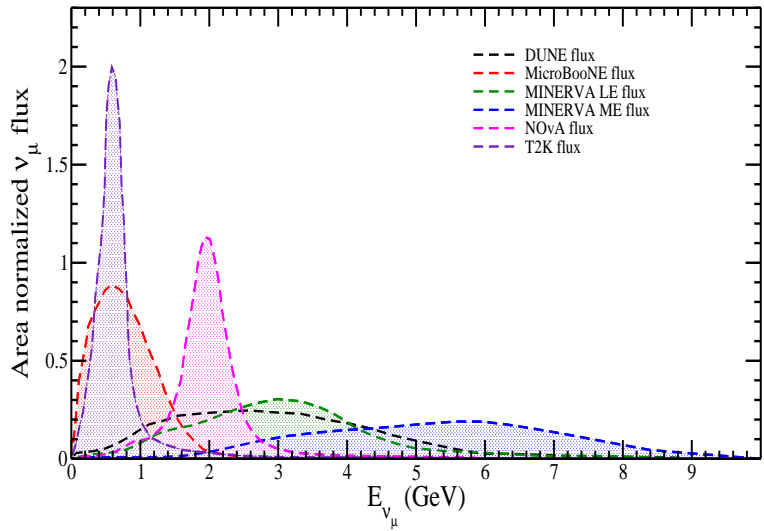}
	\caption{Neutrino flux as a function of neutrino energy for the accelerator neutrinos.}\label{neu-flux}
\end{figure}

The first antineutrino~(${\bar\nu}_e$) was observed at the nuclear reactor and since then many more studies using reactor 
antineutrinos have been performed. This is because the nuclear reactors are intense, pure and controllable sources of 
${\bar\nu}_e$. The recent experiments like Daya Bay, RENO, Chooz, Double Chooz, etc. have resulted precise information about 
neutrino properties. The next generation experiment JUNO is expected to shed more light on the neutrino properties. There 
are four main radioisotopes viz. $^{235}$U, $^{238}$U, $^{239}$Pu and $^{241}$Pu, which are responsible for the production 
of almost 99$\%$ of ${\bar\nu}_e$s, and in each fission reaction about six ${\bar\nu}_e$s are produced. Therefore, typically 
for each 1~GW of thermal energy power about $6 \times 10^{20}~{\bar\nu}_e$s are released. These antineutrinos have a 
maximum energy of about 8~MeV as shown in Fig.~\ref{Fig:Data1}.
\end{itemize}

\subsubsection{Masses, mixing and oscillation of neutrinos}\label{nu_oscillation}
\begin{itemize}
 \item {\bf Neutrino masses}\\
Experimentally, the neutrino masses are measured for the different flavors of neutrinos in different 
ways~\cite{ParticleDataGroup:2020ssz}:
\begin{itemize}
\item [(i)] direct determination of $m_{\nu_e}$ by studying the end point energy spectrum of electrons produced in the beta 
decay of nuclei.  
\item [(ii)] $m_{\nu_e}$ determination by the $e^-$-capture reaction on nuclei.
\item [(ii)] $m_{\nu_\mu}$ determination from the pion decay.
\item [(iii)] $m_{\nu_\tau}$ determination from the tau decay.
\item [(iv)] indirect determination of $m_{\nu_e}$ from astrophysics, cosmology and NDBD.
\end{itemize}
	
It was Fermi~\cite{Fermi:1934sk, Fermi:1934hr} and Perrin~\cite{Perrin:1933} who first discussed the determination of the 
neutrino mass from the study of the end-point spectrum of beta decay. These works were followed by the work of 
Henderson~\cite{Henderson:1934} who studied the thorium beta decay spectrum and concluded that the mass of neutrino must be 
much smaller than the electron mass. Hanna and Pontecorvo~\cite{Hanna:1949zab} in 1949, through the measurement of the 
beta-decay spectrum of tritium concluded that the mass of neutrino could not be larger than 500~eV. In 1972, 
Bergkvist~\cite{Bergkvist:1972xb} in his seminal work measured the energy spectrum of the electrons near threshold end point 
in tritium decay and concluded that $m_{\nu_e}<60$~eV, which was almost a factor of ten smaller than the limit given by Hanna 
and Pontecorvo~\cite{Hanna:1949zab}. Later many other attempts have been made to determine the ${\nu_e}$ mass.

The present upper limits on the neutrino masses are~\cite{ParticleDataGroup:2020ssz}:
\begin{itemize}
\item $\nu_e$, $\bar\nu_e$: $m_\nu \le$~ 1.1~eV, \qquad \qquad $\nu_\mu$, $\bar\nu_\mu$: $m_\nu \le$  
190~keV\;\;\; 90$\%$ CL \qquad \qquad 
$\nu_\tau$, $\bar\nu_\tau$: $m_\nu \le$ 18.2~MeV\;\;\; 95$\%$ CL
\end{itemize}

\item {\bf Neutrino mixing and oscillations}\label{nu:oscillation}\\
Pontecorvo~\cite{Pontecorvo:1957qd} in 1957 proposed the idea of neutrino oscillation by stating that the physical state of 
neutrinos produced in weak interaction processes is a superposition of neutrino and antineutrino states with definite masses. 
This was developed in analogy with the neutral kaon regeneration phenomenon which was proposed by Gell-Mann and 
Pais~\cite{Gell-Mann:1955ipe}, where $K^0$ and ${\bar K}^0$ could transform into each other via weak interaction with 
intermediate states of pions $K^0 \longleftrightarrow 2\pi \longleftrightarrow {\bar K}^0$, $K^0 \longleftrightarrow 3\pi
\longleftrightarrow {\bar K}^0$, which implies that a beam that initially consists of $\ket{K^0}$ pure state, would have 
some component of $\ket{\bar K^0}$ after some time and they propagate as the superposition of the states, $\ket{K_1}$ 
and $\ket{K_2}$, having definite masses and decay widths. Later Maki, Nakagawa and Sakata~\cite{Maki:1962mu} applied the idea 
of neutrino oscillation in flavor space in which neutrino oscillation between neutrinos of two flavor i.e. $\nu_e$ and 
$\nu_\mu$ was proposed and was later extended to three flavors of neutrinos. In the three flavor neutrino oscillation, a 
neutrino created in a specific flavor eigenstate is a specific quantum superposition of all three mass eigenstates. As a 
consequence the three flavor of neutrinos, viz. $\nu_e$, $\nu_\mu$, $\nu_\tau$, while propagating in space, travel as some 
admixture of three neutrino mass eigenstates viz. $\nu_i$~($i=1,2,3$) with masses $m_i$, where the strengths of the mixing of 
the mass eigenstates for the three neutrino flavors are different like $\nu_e$ has maximum contribution from $\nu_1$ or 
$\nu_\tau$ has maximum contribution from $\nu_3$ mass eigenstate. The idea of neutrino mixing leading to neutrino oscillations 
requires the neutrino mass states to be nondegenerate and in the case of $n$ flavor oscillation, $(n-1)$ neutrino mass states 
have nonzero masses.
	
The physics of neutrino mass, mixing and oscillations can be demonstrated by a simple example of two flavor mixing of 
$\nu_e$ and $\nu_\mu$ in analogy with the quark mixing~\cite{Kobayashi:1973fv}. A pure $\nu_e$ beam described by a wave 
function while traveling in space may develop a component of $\nu_\mu$ in this beam and the mixture of the $\nu_\mu$ wave 
function will describe the probability of finding $\nu_\mu$ component in the $\nu_e$ beam after a time $t$. We assume that 
the flavor states $\nu_e$ and $\nu_{\mu}$ participating in the weak interactions are mixture of the mass eigenstates 
$\nu_{1}$ and $\nu_{2}$ and the mixing is described by a unitary mixing matrix $U$ such that:
\begin{equation}\label{mmm}
\nu_{l={e,\mu}} = \sum_{i=1,2} U_{li} \nu_i.
\end{equation}
The unitarity of the $U$ matrix requires that in 2-dimensional space it is described by one parameter which is generally 
chosen to be $\theta$ such that:
\begin{eqnarray}
U= \begin{pmatrix}
    c_{12} & s_{12}\\
    -s_{12} & c_{12}
   \end{pmatrix}
\end{eqnarray}
where $c_{12}=\cos\theta$ and $s_{12}=\sin\theta$. As pure beam of $\nu_e$ at $t=0$ propagates, the mass eigenstates 
$|\nu_{1} \rangle$ and $|\nu_{2} \rangle$, occurring in Eq.~(\ref{mmm}), would evolve according to
\begin{eqnarray}\label{nu1}
|\nu_1 (t)\rangle = \nu_1 (0) e^{-i E_1 t}; \qquad \quad
|\nu_2 (t)\rangle = \nu_2 (0) e^{-i E_2 t},
\end{eqnarray}
where $E_1 = \sqrt{|\vec{p}|^2 + m_1^2}\approx |\vec{p}| + \frac{m_{1}^{2}}{2|\vec{p}|}$ and $E_2 = \sqrt{|\vec{p}|^2 + 
m_2^2} \approx |\vec{p}| + \frac{m_{2}^{2}}{2|\vec{p}|}$, with common momentum $\vec{p}$ and energies $E_{1}$ and 
$E_{2}$. $m_1$ and $m_2$ are the masses of $\nu_1$ and $\nu_2$ states, respectively. After a time $t$, the state $|\nu_e(t) 
\rangle$ will be a different admixture of $|\nu_1 \rangle$ and $|\nu_2 \rangle$. The probability of finding $\nu_\mu$ in the 
beam of $\nu_e$ at a later time $t$ is given by~\cite{Athar:2020kqn}:
\begin{equation}\label{111}
P(\nu_e \rightarrow \nu_\mu) = \sin^2 2\theta \sin^2\left( \frac{\Delta m^2}{4E}L\right) = \sin^2 2\theta \sin^2 \left(1.27 
\frac{\Delta 
m^2}{E} L\frac{[\text{eV}^2] [\text{km}]}{[\text{GeV}]}\right).
\end{equation}
Thus, we see that for $P(\nu_e \rightarrow \nu_\mu) \neq 0$ we need $\Delta m^2 \neq 0$ and $\theta \neq 0$ i.e. we need the 
mass difference between the neutrino mass eigenstates to be nonzero implying that at least one of them is massive and the 
mixing angle $\theta$ to be nonzero. Thus, if the explanation of the solar neutrino flux deficit and other deficits observed 
in the atmospheric, reactor and accelerator neutrinos are explained to be due to the neutrino oscillations, the neutrinos 
should have nonzero mass and the neutrino flavors should mix. In the case of three flavor neutrino mixing, these flavor 
and mass eigenstates are  related by a 
$3 \times 3$ unitary lepton mixing matrix~\cite{Athar:2020kqn}:
\begin{equation}\label{3flavormatrix}
|\nu_\alpha\rangle = \sum_{i=1}^3 U_{\alpha i} |\nu_i\rangle \quad (\alpha = e,\mu,\tau) \,,
\end{equation}
where $U$ is Pontecorvo-Maki-Nakagawa-Sakata~(PMNS) mixing matrix~\cite{Pontecorvo:1957qd, Maki:1962mu, Pontecorvo:1957cp}. The 
most popular parameterization of the PMNS matrix is given by~\cite{ParticleDataGroup:2020ssz}:
\begin{eqnarray}
U= 
\left( \begin{array}{ccc}
c_{13}c_{12} & c_{13}s_{12} & s_{13} e^{-i \delta_{CP}}\\
-c_{23}s_{12}-s_{13}s_{23}c_{12}e^{i\delta_{CP}} & 
c_{23}c_{12}-s_{13}s_{23}s_{12}e^{i\delta_{CP}} &
c_{13}s_{23} \\
s_{23}s_{12}-s_{13}c_{23}c_{12}e^{i\delta_{CP}} & 
-s_{23}c_{12}-s_{13}c_{23}s_{12}e^{i\delta_{CP}} &
c_{13}c_{23}
\end{array} \right) 
\end{eqnarray}
where $c_{ij} = \cos\theta_{ij}$ and $s_{ij} = \sin\theta_{ij}$, and $\delta$ is the CP violating phase. 

The general expression for the transition probability is given by~\cite{Athar:2020kqn}: 
\begin{equation}\label{3158}
P_{\nu_\alpha\rightarrow \nu_\beta} (L,E)=4\sum_{i>j}\Big(|{U_{\alpha i}}|^2|{U_{\alpha j}}|^2\Big)\sin^2\Big(
\frac{\Delta m{^2_{ij}}}{4E}L\Big).
\end{equation}
For the three flavors of neutrinos $i, j=1,2,3,$ with $i>j$, the mass squared difference terms are $\Delta m{^2_{32}}$, 
$\Delta m{^2_{31}}$, and $\Delta m{^2_{21}}$. Since
\begin{eqnarray}\label{3169}
\Delta m{^2_{32}} &=&m{^2_3}-m{^2_2}= (m{^2_3}-m{^2_1})+(m{^2_1}-m{^2_2})=\Delta m{^2_{31}}-\Delta m{^2_{21}},
\end{eqnarray}
therefore, only two of the three $\Delta m_{ij}$'s are independent. 

In deriving Eqs.~(\ref{111}) and (\ref{3158}) for the oscillation probability, a plane wave description of neutrino beam given in Eq.~(\ref{nu1}) has been assumed. 
However, to give a realistic description of the neutrino propagation, a wave packet description is used~\cite{Chan:2015mca, Jones:2022hme}. 
If one considers the neutrino described by the wave packet, then the transition probability of $\nu_{\alpha} \to \nu_\beta$ is obtained as~\cite{Chan:2015mca, Jones:2022hme}:
\begin{eqnarray}\label{P2}
 P_{\nu_\alpha\rightarrow \nu_\beta} (L,E) &\approx& \sum_{ij} \left\{U_{\alpha i}^{\ast} U_{\beta i} U_{\alpha j}^{\ast} U_{\beta j}^{\ast} \exp\left[-i \frac{2\pi L}{L_{ij}^{osc}} \right] \right\} \nonumber \\
 &\times& \left\{\left(\frac{1}{1+y_{ij}^2} \right)^{\frac{1}{4}} \exp \left(- \lambda_{ij} \right) \exp\left(- \frac{i}{2} \tan^{-1} \left(y_{ij} \right) \right) \exp \left(i \lambda_{ij} y_{ij} \right) \right\}
\end{eqnarray}
where
\begin{eqnarray}\label{P33}
 \lambda_{ij} &=& \frac{x_{ij}^2}{1 + y_{ij}^2}, \qquad\qquad y_{ij} ~=~ \frac{L}{L_{ij}^{dis}}, \qquad \qquad x_{ij} ~=~ \frac{L}{L_{ij}^{coh}}, \nonumber \\
  L_{ij}^{coh} &=& \frac{L_{ij}^{osc}}{\pi \sigma_{wp}}, \qquad\qquad L_{ij}^{dis} ~=~ \frac{L_{ij}^{osc}}{2\pi\sigma_{wp}^{2}}, \qquad \qquad L_{ij}^{osc} ~=~ \frac{4\pi E}{\Delta m_{ij}^{2}}, \nonumber \\ 
  \sigma_{wp} &=& \frac{\sigma_{\nu}}{E_{i} (p_{\nu})} .
\end{eqnarray}
 $E_{i}$ is the energy of the $\ket{\nu_{i}}$ eigenstate, and $p_{\nu}$ is the mean momentum.
The first term in Eq.~(\ref{P2}) is the plane wave neutrino oscillation probability, modified by a numerical factor depending upon 
$\sigma_{\nu}$, the width of the wave packet in the momentum space, which is independent of the neutrino energy. The term with quartic correction to $y_{ij}$ describes the dispersion effects and depends on the dispersion length $L_{ij}^{dis}$. The $\exp \left(-\lambda_{ij} \right)$ term corresponds to the decoherence effect arising due to the fact that the different neutrino mass eigenstates propagate at different speeds. 
The term  $x_{ij}$ is related to the coherence length $L_{ij}^{coh}$. It may also be observed from Eq.~(\ref{P33}) that $x_{ij} \propto \sigma_{wp}$ while $y_{ij} \propto \sigma_{wp}^2$, therefore, if the wave packet impact $\sigma_{wp} \ll 1$, the dispersion effect is expected to be more suppressed and negligible.

The various experimental efforts, with the the solar, reactor, atmospheric, and accelerator neutrinos made with the short 
and long baseline experiments are sensitive to the different parameters of the PMNS matrix which have been tabulated in 
Table-\ref{intro:table1}.
\end{itemize}

\begin{table}
\centering
\begin{tabular}{|c|cc|}
		\hline 
		$\nu(\bar\nu)$-Experiment & Dominant  & Important\\
		\hline 
		Solar & $\theta_{12}$  &$\Delta m_{21}^2$, $\theta_{13}$ \\
		Reactor LBL &$\Delta m_{21}^2$ & $\theta_{12}, \theta_{13}$  \\	
		Reactor MBL &$ \theta_{13}$, $|\Delta m_{31,32}^2| $ & \\
		Atmospheric &~~ &~~~~~ $\theta_{23}, |\Delta m_{31,32}^2|,  \theta_{13}, \delta_{CP}$\\
		Accelerator LBL $\nu_\mu(\bar\nu_\mu)$ disappearance&$|\Delta m_{31,32}^2|$, $\theta_{23}$  &\\
		Accelerator LBL $\nu_e(\bar\nu_e)$ appearance &  $\delta_{CP}$ &$\theta_{13}$ , $\theta_{23}$ \\
		\hline
\end{tabular}
\caption{Sensitivity of the (anti)neutrino sources to the oscillation parameters~\cite{ParticleDataGroup:2020ssz}.}
\label{intro:table1}
\end{table}

\subsubsection{Electromagnetic properties of neutrinos}\label{nu_EM}
Pauli in his neutrino proposal speculated that the magnetic moment of this particle should not be larger than $e \times 
10^{-13}$cm~\cite{Pauli:1930pc}. Very soon after the discovery of antineutrinos, in 1956, Reines and 
Cowan~\cite{Reines:1960pr} gave an upper limit on the neutrino magnetic moment $\mu_{{\bar \nu}_e}= 10^{-9} \mu_B$ ($\mu_B$ is 
the Bohr magneton), based on the extent of nonobservation of scintillator pulses along the path of reactor antineutrinos in 
their experiment. Their studies motivated Bernstein and Lee~\cite{Bernstein:1963jp} and many others to phenomenologically 
study neutrino magnetic moment. 
 
In general, the electroweak properties of a spin $\frac{1}{2}$ Dirac particle are described in terms of the two vector form 
factors called the electric and the magnetic form factors, which in the static limit define the charge and the magnetic moment, 
and the two axial-vector form factors called the axial-vector and the tensor form factors which in the static limit define 
the axial charge and the electric dipole moment, and that is 
related to the matrix element of the electromagnetic current between the initial and final neutrino mass 
states~\cite{Athar:2020kqn}:
\[\bra{\psi(p^\prime)}J_\mu^{EM}\ket{\psi(p)}= \bar{u}(p^\prime) \left[F_Q(Q^2) \gamma_\mu -  F_M(Q^2) i\sigma_{\mu\nu} 
q^\nu + F_E(Q^2) \sigma_{\mu\nu} q^\nu\gamma_5 + F_A(Q^2) \left(-Q^2 \gamma_\mu - q_\mu q \right)\gamma_5\right]u(p), \] 
where $q=p-p^{\prime}$, $F_Q (Q^2)$, $F_M (Q^2)$, $F_E (Q^2)$ and $F_A (Q^2)$ are, respectively, charge, magnetic dipole, 
electric dipole and axial charge neutrino electromagnetic form factors.
 
If the neutrino is considered to be the Dirac neutrino with nonzero mass, it could have these form factors to be 
nonvanishing and experimental attempts can be made to study them. In this case, they have magnetic dipole moment 
like neutrons and can have electric dipole moment if CP is violated in the lepton sector. Since neutrinos participate 
in weak interaction which violates CP invariance, they may have an electric dipole 
moment. If the neutrinos are Majorana fermions then from CPT invariance, regardless of whether CP invariance is violated or 
not, $F_Q(Q^2) = F_M(Q^2)=F_E(Q^2)=0$, and only the axial-vector form factor $F_A(Q^2)$ can be nonvanishing. Thus the 
electromagnetic properties of the (anti)neutrinos depend upon the type of (anti)neutrinos.
\begin{itemize}
\item [(i)] The SM calculations for the magnetic moment of a neutrino depends upon its mass $m_{\nu}$ and is therefore very 
small of the order 3$\times$10$^{-19}\frac{m_{\nu}}{eV}\mu_B$. There are models where the neutrino magnetic moment is not 
proportional to the neutrino mass and give larger magnetic moments~\cite{Babu:2020ivd, Canas:2015yoa}. Experimentally, the laboratory limits on the neutrino 
magnetic moments are obtained by performing the elastic $\nu_{e} - e, \bar{\nu}_{e} - e$  and $\nu_{\mu} - e$ scattering. 
The present upper limits on the neutrino magnetic moments are~\cite{ParticleDataGroup:2020ssz}:
\begin{itemize}
\item $\mu_{\nu_e} < 0.28 \times 10^{-10}   \mu_B ; \qquad \qquad \mu_{\nu_\mu} < 6.8 \times 10^{-10}   \mu_B; 
\qquad \qquad \mu_{\nu_\tau} < 3.9 \times 10^{-7}   \mu_B \qquad 90\%$ CL
\end{itemize}

\item [(ii)] The neutrinos are assumed to be electrically neutral, but there are attempts to measure the charge of the 
neutrino in $\beta$-decays by measuring the charge of the neutron $Q_{n}$ and the total charge of the proton and electron 
i.e. $|Q_{p}+Q_{e^{-}}|$ in the decay $ n \rightarrow p+e^{-}+ \bar{\nu}_{e}$~\cite{Dylla:1973zz, Zorn:1963zz}. This gives 
a limit on $Q_{\bar{\nu}}< (0.5\pm 2.9)\times 10^{-21}e$. The astrophysical limit derived from the SN1987A supernova 
observation is~\cite{Barbiellini:1987zz}: 
$$Q_{\bar{\nu}}< 2 \times 10^{-15}e.$$

\item [(iii)] The charge of neutrino is consistent with zero to a very high degree of precision but it may have a charge 
distribution like a neutron. Attempts to determine the charge radius have been made~\cite{Degrassi:1989ip} for $\nu_{e}$ 
and $\nu_{\mu}$ from $\nu_{e}e$~\cite{LSND:2001akn}, $\bar{\nu}_{e}e$~\cite{TEXONO:2009knm} and 
$\nu_{\mu}e$~\cite{Hirsch:2002uv} scattering. Like hadrons, the mean square charge radius of a neutrino is deduced from the 
measurement of the vector form factor in the $\nu_{e}e$ and $\nu_{\mu}e $ elastic scattering using the relation
\begin{equation}
\langle r^{2}\rangle = \left. 6\frac{d}{d Q^{2}} F(Q^{2})\right|_{Q^{2}=0},
\end{equation}
where $F(Q^{2})$ is the charge form factor corresponding to the matrix element of the vector current. { In the standard model, the value of $\langle r^{2}\rangle$ is estimated to be of the order of $10^{-32}$ cm$^2$~\cite{Bernstein:1963qh}.}
In the case of neutral 
particles, the value of $\langle r^{2}\rangle$ could be negative or positive and the following experimental 
limits~\cite{ParticleDataGroup:2020ssz, CHARM:1988tlj, Ahrens:1990fp} are obtained in the case of $\nu_e$ and $\nu_{\mu}$:
\begin{eqnarray*}
-5.3\times 10^{-32}< & \left[ \langle r^{2} \rangle_{\nu_{\mu}} \right] & < 1.3\times 10^{-32}\text{ cm}^{2} , \\
-0.77 \times 10^{-32} < & \left[ \langle r^{2} \rangle_{\nu_{\mu}}\right]& < 2.5\times10^{-32}\text{ cm}^{2},\\
-5.0\times 10^{-32}< & \left[ \langle r^{2} \rangle_{\nu_{e}} \right] & < 10.2\times 10^{-32} \text{ cm}^{2}. 
\end{eqnarray*}
\end{itemize} 

\subsection{Theoretical description of neutrinos and their interactions}\label{nu:theory}
\subsubsection{Dirac neutrinos}
The Dirac theory of electrons formulated in 1928~\cite{Dirac:1928hu} is conventionally used to describe the neutrinos. The 
Pauli's neutrinos proposed in 1930~\cite{Pauli:1930pc} were assumed to have a tiny mass but the later developments in the 
phenomenological study of neutrino interactions through the nuclear $\beta$ decays and the (anti)neutrino-nucleus scattering 
using the Fermi or the $V-A$ theory of weak interactions seem to be consistent with neutrinos being massless. This did not 
pose any problem in applying the Dirac theory of electrons to neutrinos as the theory can be extrapolated smoothly to the 
massless limit of the spin $\frac{1}{2}$ fermion of mass $m \rightarrow 0$. These neutrinos are called Dirac neutrinos, 
$\nu^{D}$, and the wave function $\Psi_{\nu^{D}} (x)$ describing these neutrinos satisfies the Dirac 
equation~\cite{Dirac:1928hu}:
\begin{equation}\label{Dirac1}
\left(i \gamma^{\mu} \partial_{\mu} - m \right)\Psi_{\nu^{D}} (x) = 0,
\end{equation}
where $\gamma^{\mu}$s~($\mu=0,1,2,3$) are four $4\times 4$ matrices and satisfy the algebra:
\begin{eqnarray}\label{eq:gamma}
 \left\{\gamma^{\mu},\gamma^{\nu} \right\} = 2g^{\mu\nu}, \qquad \quad g^{00}=1, \qquad \quad g^{ij}=-\delta_{ij} \qquad 
 \qquad(i,j=1,2,3), \qquad \quad {\gamma^{\mu}}^{\dagger} = \gamma^{0} \gamma^{\mu} \gamma^{0}.
\end{eqnarray}
These relations are independent of the representation used to parameterize the $\gamma^{\mu}$ matrices for which many 
representations exist. The most popular is the Pauli-Dirac representation in which
\begin{eqnarray*}
 \gamma^{0} = \begin{pmatrix}
               \mathbb{I}&0\\
               0&-\mathbb{I}
              \end{pmatrix},
\qquad \quad \gamma^{i} = \begin{pmatrix}
                                      0& \sigma^{i} \\
                                      -\sigma^{i} & 0
                                     \end{pmatrix},
\end{eqnarray*}
where $\sigma^{i}$ being Pauli matrices. But there are parameterizations like the Weyl, and Majorana representations, which are also used to describe the 
neutrinos~\cite{Athar:2020kqn}. The wave function $\Psi_{\nu^{D}}$ in Eq.~(\ref{Dirac1}) is a four component spinor and is 
generally written as
\begin{equation}
 \Psi_{\nu^{D}} (x) = \sum_{r,p} \frac{1}{\sqrt{2\omega_{\vec{p}}}} \left[a_{r}(p) u_{r} (\vec{p}\;) e^{-ip\cdot x} + 
 b_{r}^{\dagger} (p) v_{r} (\vec{p}\;) e^{ip\cdot x}\right],
\end{equation}
where $u_{r}(\vec{p})$ and $v_{r} (\vec{p})$ are the two component spinors which describe the two spin states of 
particles~(neutrinos) and antiparticles~(antineutrinos) corresponding to the spin states labeled by $\ket{s ~~s_{z}} = 
\ket{\frac{1}{2} ~~\pm \frac{1} {2}}$ and satisfy, in the momentum space, the equations 
\begin{equation}\label{Dirac:2}
 \left(\slashed{p} - m \right) u_{r}(\vec{p}) = 0; \qquad \qquad \left(\slashed{p} + m \right) v_{r} (\vec{p})=0.
\end{equation}
If the spin quantization axis is chosen in the direction of motion along the $Z$-axis, then the $\nu^{D}$ state $\ket{\frac{1}
{2} ~~+ \frac{1}{2}}$ with its spin along the $+Z$-axis is  denoted by $\nu_{+}^{D}$~(right handed), while the $\nu^{D}$ state 
$\ket{\frac{1}{2} ~~- \frac{1}{2}}$ has the spin opposite to $Z$-axis~(left handed) is denoted by $\nu^{D}_{-}$. Similarly, we 
have the two spin up and spin down states of the antineutrinos $\bar{\nu}_{+}^{D}$ and $\bar{\nu}_{-}^{D}$. It should be noted 
that under CPT transformation, a particle becomes an antiparticle with opposite spin, $\nu_{-}^{D} \rightarrow 
\bar{\nu}_{+}^{D}$ and $\nu_{+}^{D} \rightarrow \bar{\nu}_{-}^{D}$, with the same mass. Moreover, if the neutrinos have a mass 
then its speed is less than the speed of light and an observer can move faster than this speed. In this frame, an observer 
would see a right handed neutrino $\nu_{+}^{D}$ as the left handed $\nu_{-}^{D}$ but all other properties, if any, like the 
lepton number, etc., would be the same. In fact, $\nu_{+}^{D}$ and $\nu_{-}^{D}$ are the two spin states of the same 
particle neutrino. Similarly, $\bar{\nu}_{+}^{D}$ and $\bar{\nu}_{-}^{D}$ are the two spin states of the same antineutrino. 
There are, therefore, four states of a Dirac neutrino, described by $\Psi_{\nu^{D}}$. The phenomenological study of the weak 
interaction processes involving (anti)neutrinos establishes that for each flavor of neutrinos~\cite{Athar:2020kqn}:
\begin{itemize}
 \item [(i)] the neutrinos are left handed i.e. $\nu_{-}^{D}$ and the antineutrinos are right handed i.e. 
 $\bar{\nu}_{+}^{D}$, which take part in the weak interactions.
 
 \item [(ii)] $\nu_{-}^{D}$ always produces a charged lepton $l^{-}$ and $\bar{\nu}_{+}^{D}$ always produces a charged lepton 
 $l^{+}$ in charged current~(CC) interactions, which imply that $\nu_{-}^{D}$ and $\bar{\nu}_{+}^{D}$ are distinct particles.
 
 To ensure that $\nu_{-}^{D}$ and $\bar{\nu}_{+}^{D}$ are distinct particles like $l^{-}$ and $l^{+}$ and obey the selection 
 rules of weak processes, it was proposed that
 \begin{enumerate}
  \item there exists a new quantum number called lepton number $L_{l}$ for each flavor $l$ and $(\nu_{l-}^{D} ~~~l^{-})$ were 
  assigned $L_{l}=+1$ while $(\bar{\nu}_{l+}^{D} ~~~l^{+})$ were assigned $L_{l} = -1$.
  
  \item The lepton number $L_{l}$ is conserved for each flavor.
 \end{enumerate}
\item [(iii)] While the charged leptons and their antiparticles like $l^{-}$ and $l^{+}$ are different in their charge and 
 lepton number, the corresponding neutrinos and antineutrinos being neutral are different only in their lepton number $L_{l}$ and helicities. 
 It should be noted that $\nu_{l+}^{D}$ and $\nu_{l-}^{D}$ have the same lepton number $L_{l}=+1$. Similarly, $\bar{\nu}_{l+}^{D}$ 
 and $\bar{\nu}_{l-}^{D}$ also have the same lepton number $L_{l}=-1$.
\end{itemize}

\subsubsection{Weyl neutrinos}
In the limit of mass $m \rightarrow 0$, interesting features arise which become more intriguing in the case of neutrinos being 
neutral particles. In this limit, the Dirac equation becomes Weyl equation and the Weyl wave function $\Psi_{\nu^{W}}$ 
satisfies 
\begin{equation}
 i \gamma^{\mu} \partial_{\mu} \Psi_{\nu^{W}} (x) = 0.
\end{equation}
This equation of motion for a spin $\frac{1}{2}$ particle with $m=0$ was especially studied by Weyl in 
1929~\cite{Weyl:1929fm}, a year after the Dirac equation~\cite{Dirac:1928hu}, and is most easily solved using the Weyl 
representation for the $\gamma$ matrices~\cite{Weyl:1929fm}.

However, we discuss its solution using the chirality operator which is defined as $\gamma^{5} = i \gamma^{0} \gamma^{1} 
\gamma^{2} \gamma^{3} = \begin{pmatrix}
                                                                               0 & \mathbb{I} \\
                                                                               \mathbb{I} & 0
                                                                              \end{pmatrix}
$ for the following reason. Using the 4-dimensional representation of spin $\vec{\Sigma} = 
\begin{pmatrix}
 \vec{\sigma} & 0 \\
 0 & \vec{\sigma}
\end{pmatrix} = \gamma^{5} \gamma^{0} \vec{\gamma}$, the helicity operator $\vec{\Sigma} \cdot \hat{p}$ is written as 
$\vec{\Sigma} \cdot \hat{p} = \gamma^{5} \gamma_{0} \vec{\gamma} \cdot \hat{p}$. In the case of $m \rightarrow 0$, the Weyl 
equation is written, in momentum space, as
\begin{equation}\label{Weyl:eq}
 \slashed{p} \;\Psi_{\nu^{W}} (p) = 0.
\end{equation}
Now, consider the equation
\begin{equation}\label{Weyl:1}
 \vec{\Sigma} \cdot \vec{p}\; \Psi_{\nu^{W}} (p) = \gamma^{5} \gamma_{0} \vec{\gamma} \cdot \vec{p} \;\Psi_{\nu^{W}} (p).
\end{equation}
Using $p_{0} = |\vec{p}\;|$ and Eq.~(\ref{Weyl:eq}) in the case of $m=0$, we get
\begin{equation}
  \vec{\Sigma} \cdot \hat{p} \;\Psi_{\nu^{W}} (p) = \gamma^{5} \;\Psi_{\nu^{W}} (p).
\end{equation}
Thus, in the case of $m=0$, $\gamma^{5}$ is the helicity operator $\vec{\Sigma} \cdot \hat{p}$, which is also called the 
chirality operator. Since $\vec{\Sigma} \cdot \hat{p} \; \vec{\Sigma} \cdot \hat{p} \equiv ({\gamma^{5}})^{2} = 1$, 
$\gamma^{5}$ has two eigenvalues $\pm 1$ corresponding to helicity~$+1$ and $-1$, also called the right handed~($R$) and left 
handed~($L$) helicity states of the massless neutrinos. The eigen functions corresponding to the eigenvalues $+1$ and $-1$ are, 
respectively, $\Psi_{R}^{W}$ and $\Psi_{L}^{W}$, which satisfy
\begin{eqnarray}\label{eq:Weyl:1}
 \vec{\Sigma} \cdot \hat{p} \;\Psi_{R}^{W} (p) = \gamma_{5} \;\Psi_{R}^{W} (p) = (+1) \;\Psi_{R}^{W} (p), \qquad \qquad
 \vec{\Sigma} \cdot \hat{p} \;\Psi_{L}^{W} (p) = \gamma_{5} \;\Psi_{L}^{W} (p) = (-1)\;\Psi_{L}^{W} (p).
\end{eqnarray}
It should be noted that in $m\rightarrow 0$, $\nu_{R}^{W}$ and $\nu_{L}^{W}$ are two distinct particles and not the two spin 
states of one particle as in the case of the Dirac neutrinos $\nu_{+}^{D}$ and $\nu_{-}^{D}$~(in the case of $m \neq 0$) 
because there exists no frame in which $\nu_{R}^{W}$ would appear as $\nu_{L}^{W}$ due to the Weyl neutrinos moving with the 
speed of light. In principle, while $\nu_{+}^{D}$ and $\nu_{-}^{D}$ have the same lepton number, $\nu_{R}^{W}$ and 
$\nu_{L}^{W}$ could have different lepton numbers. If neutrinos exist in $\nu_{L}^{W}$ state, then they cannot exist in 
$\nu_{R}^{W}$ state. Consequently, the antineutrinos will exist in $\bar{\nu}_{R}^{W}$ state and not in $\bar{\nu}_{L}^{W}$ 
state. Thus, the Weyl (anti)neutrinos have only two states unlike the Dirac (anti)neutrinos which have four states. If 
physical neutrinos observed in nuclear $\beta$ decays or other weak processes are $\nu_{L}^{W}$~(or $\nu_{R}^{W}$), the 
massless Weyl neutrinos imply maximal violation of the left-right symmetry i.e., parity violation. This is the reason that 
the Weyl equation was disfavored during 1930--1957. After the parity violation was proposed and observed 
experimentally~\cite{Wu:1957my}, the two component theory of neutrinos with chiral invariance was proposed by Lee and 
Yang~\cite{Lee:1957qr}, Landau~\cite{Landau:1957tp}, and Salam~\cite{Salam:1957st}. If the two states $\nu_{L}^{W}$ and 
$\nu_{R}^{W}$ are independent, in the case of $m=0$, then we can write a neutrino state $\nu^{W}$ as
\begin{equation}\label{eq:Weyl:3}
 \Psi^{W} = \Psi_{L}^{W} + \Psi_{R}^{W}.
\end{equation}
Using Eqs.~(\ref{eq:Weyl:1}) and (\ref{eq:Weyl:3}), we obtain 
\begin{eqnarray}
  \Psi_{L}^W = \frac{\mathbb{I}-\gamma_{5}}{2} \Psi^{W}, \qquad \qquad && \qquad \qquad \Psi_{R}^W = \frac{\mathbb{I} +
  \gamma_{5}}{2} \Psi^{W},
\end{eqnarray}
as the left-handed and right-handed Weyl neutrinos. Conversely, if $\nu^{W}$ exists either in $\nu_{L}^{W}$ or in 
$\nu_{R}^{W}$ state, it has to be massless as the mass term in the Lagrangian given by
\begin{equation}
 {\cal L}^{W}_{mass} = - m\bar{\Psi}^{W} \Psi^{W} = -m \left(\bar{\Psi}_{L}^{W} \Psi_{R}^{W} + \bar{\Psi}_{R}^{W} \Psi_{L}^{W} 
 \right)
\end{equation}
vanishes.

The $V-A$ theory of weak interaction was formulated using the two component neutrinos by Sudarshan and 
Marshak~\cite{Sudarshan:1958vf}, and Feynman and Gell-Mann~\cite{Feynman:1958ty} using left handed neutrinos $\nu_{L}^{W}$. 
The antineutrino in the Weyl theory are obtained in a similar manner by performing a CPT transformation such that 
\begin{equation}
 \nu_{R}^{W} \xrightarrow{CPT} \bar{\nu}_{L}^{W}, \qquad \qquad \quad \nu_{L}^{W} \xrightarrow{CPT} \bar{\nu}_{R}^{W}.
\end{equation}
The relation between the Dirac and Weyl neutrinos can be expressed as
\begin{itemize}
 \item [(i)] Four component Dirac spinor is equivalent to two two-component Weyl spinors.
 
 \item [(ii)] While Dirac neutrinos could have nonvanishing mass~($m$) and can be extrapolated to $m \rightarrow 0$, Weyl 
 neutrinos are necessarily massless.
 
 \item [(iii)] $\nu_{+(-)}^{D} \xrightarrow{m \rightarrow 0} \nu_{R(L)}^{W}$; $\bar{\nu}_{+(-)}^{D} \xrightarrow{m \rightarrow 
 0} \bar{\nu}_{R(L)}^{W}$.
\end{itemize}

\subsubsection{Majorana neutrinos}
While the phenomenology of the weak interaction processes was consistent with the massless neutrinos, the experimental 
attempts to measure their masses were continuing relentlessly. Theoretically also the mass of $\nu_{e}(\bar{\nu}_{e})$ was 
being inferred from the experimental observations made in astrophysics and cosmology. The improvements in the experimental 
limits of the neutrino masses of various flavors are reported periodically and a nonzero mass for neutrino is not ruled out. 
However, the observation of neutrino oscillations involving all the three flavors of neutrinos $\nu_{e}$, $\nu_{\mu}$ and 
$\nu_{\tau}$ in the experiments with solar, reactor, atmospheric, and accelerator neutrinos, confirmed that the neutrinos~(at 
least two flavors) have masses even though very small. This rules out the neutrinos being Weyl neutrinos. However, the 
neutrinos being neutral particles could be still described by a two component neutrino, if they are their own antiparticles. 
Such a possibility was studied by Majorana in his celebrated paper on ``The symmetry of the theory of electrons and 
positrons''~\cite{Majorana:1937vz}. These neutrinos are called Majorana neutrinos $\nu^{M}$. If the Majorana neutrino is its 
own antiparticle, then its wave function described by $\Psi_{\nu^{M}} (x)$ satisfies the equation
\begin{equation}\label{Majorana:1}
 \Psi_{\nu^{M}} (x) = \Psi_{\nu^{M}}^{\star} (x)
\end{equation}
implying that $\Psi_{\nu^{M}} (x)$ is real. But the wave function of the neutrinos written in Eq.~(\ref{Dirac1}) or 
Eq.~(\ref{Majorana:1}) is complex due to some of the coefficients $\gamma^\mu$ being complex. If a representation could be 
found in which all the $\gamma^{\mu}$'s are imaginary such that the coefficients $(i\gamma_{\mu}\partial^{\mu} - m)$ are real, 
then the solutions $\Psi_{\nu}(x)$ and Eq.~(\ref{Majorana:1}) would be satisfied. This was done by Majorana by using Majorana 
representation of the gamma matrices, in which $\tilde{\gamma}_{\mu}$'s are defined as:
\begin{equation*}
 \tilde{\gamma}^{0} = \begin{pmatrix}
                       0 & \sigma_{2} \\
                       \sigma_{2} & 0
                      \end{pmatrix}, \qquad \quad 
 \tilde{\gamma}^{1} = \begin{pmatrix}
                       i\sigma_{3} & 0 \\
                       0 & i\sigma_{3}
                      \end{pmatrix}, \qquad \quad 
 \tilde{\gamma}^{2} = \begin{pmatrix}
                       0 & -\sigma_{2} \\
                       \sigma_{2} & 0
                      \end{pmatrix}, \qquad \quad 
 \tilde{\gamma}^{3} = \begin{pmatrix}
                       -i\sigma_{1} & 0 \\
                       0 & -i\sigma_{1}
                      \end{pmatrix}, \qquad \quad                       
\end{equation*}
$\tilde{\gamma}^{5} = i  \tilde{\gamma}^{0} \tilde{\gamma}^{1} \tilde{\gamma}^{2} \tilde{\gamma}^{3} = 
\begin{pmatrix}
 \sigma_{2} & 0 \\
 0 & -\sigma_{2}
\end{pmatrix}$, and all of them are purely imaginary. This Majorana representation $\tilde{\gamma}^{\mu}$ of gamma matrices 
satisfy the algebra given in Eq.~(\ref{eq:gamma}). However, Eq.~(\ref{Majorana:1}) is not covariant i.e. 
if this equation is satisfied in Majorana representation in one Lorentz frame, it will not be 
satisfied in another Lorentz frame as the Lorentz transformation of spinors depend on $\tilde{\gamma}^\mu$ matrices which change in 
another frame. For making this equation valid in other frames a conjugate field $\Psi_\nu^c(x)$ is defined as 
\begin{equation}\label{eq:C:Majorana2}
 \Psi_\nu^c(x)= C\gamma^0 \Psi_\nu^*(x),
\end{equation}
where $C$ is chosen such that $\Psi_{\nu} (x)$ and $\Psi_{\nu}^{c}$ satisfy
\begin{equation}\label{eq:C:Majorana}
 \Psi_\nu^c(x)= \Psi_\nu^{c*}(x).
\end{equation}
The matrix $C$ is a unitary matrix, which satisfies $CC^{\dagger} = C^{\dagger} C = \mathbb{I}_{4 \times 4}$, $C^{T} = 
C^{\dagger} = -C$, and $C^{2} = - \mathbb{I}_{4 \times 4}$ and depends upon the representation used for defining the 
$\gamma^\mu$ matrices. Obviously in Majorana representation $C=C^{M}=-i\tilde{\gamma}^0$ such that Eq.~(\ref{eq:C:Majorana}) is 
recovered. In Pauli-Dirac representation, $C=C^D=i\gamma^2\gamma^0$ with $\gamma^2$ and $\gamma^0$ being the Pauli-Dirac gamma 
matrices.

Eq.~(\ref{eq:C:Majorana}) implies that in Eq.~(\ref{Dirac1}) $u_{r} (\vec{p})$ and $v_{r} (\vec{p})$ satisfy~\cite{Mohapatra:1991ng}
\begin{eqnarray}\label{Majorana:2}
 C\gamma^0 v_r^*({\vec p})=v_r({\vec p})\qquad \qquad
 C\gamma^0 u_r^*({\vec p})=u_r({\vec p}).
\end{eqnarray}
The Majorana neutrino is, therefore, described by a wave function $\Psi_{\nu^M}(x)$ given by
\begin{equation}
 \Psi_{\nu^M}(x)= \Psi_{\nu}^c(x)
\end{equation}
and $ \Psi_{\nu^M}(x)$ can be defined in any representation provided $C$ and $\Psi_{\nu}^*(x)$ are chosen in the same 
representation. {The field theory of Majorana particles is obtained by treating $\Psi_{\nu^M} (x)$ as fields and formulating its quantization as discussed in Refs.~\cite{Majorana:1937vz, Blasone:2003hh, Fujikawa:2020ijt}.}
 
Eq.~(\ref{eq:C:Majorana2}) relating the neutrino wave functions in various representation ensures that the wave function of 
neutrino has the required covariance properties under the Lorentz transformation in any given representation. The Majorana 
neutrinos could have mass like the Dirac neutrinos or could be massless like the Weyl neutrinos. In the case of Majorana 
neutrinos with mass, the Lagrangian would contain a term like $m\bar\Psi_{\nu}^{C}(x)\Psi_{\nu}(x) $ or $m 
\bar{\Psi}_{\nu}^{C}(x) \Psi_{\nu}^{C}(x)$ or $m\bar\Psi_{\nu}(x)\Psi_{\nu}^C(x) $. The mechanism for generation of mass or a 
field theoretic description of Majorana neutrinos and its properties under $C$, $CP$ and $CPT$ transformations is beyond the 
scope of this article. However, some interesting features appear due to the neutrinos being neutral Majorana particles which 
we mention in the following:
\begin{itemize}
 \item [(i)] If neutrinos are their own antiparticles i.e. $\nu^M={\bar\nu}^M$ then there are only two neutrino states with 
 spin states $\ket{\frac{1}{2}, +\frac{1}{2}}$ and $\ket{\frac{1}{2}, -\frac{1}{2}}$, even in the case of massive Majorana 
 neutrinos. In the case of massless Majorana neutrinos, the two spin states become helicity states and describe the two 
 independent particles.
 
 \item [(ii)] The familiar picture of neutrino and antineutrino interactions conceived in the Dirac's neutrino picture is 
 replaced by the spin dependent interaction of neutrinos in which a left-handed Majorana neutrinos $\nu_L^M$ produces a $l^-$ 
 and a right-handed Majorana neutrinos $\nu_R^M$ produces a $l^+$ through the weak CC interactions.
 
 \item [(iii)] The concept of lepton number~($L_{l}$) and its conservation is irrelevant in the case of Majorana neutrinos.
 \begin{figure} 
 \begin{center}
  \includegraphics[height=5cm,width=12cm]{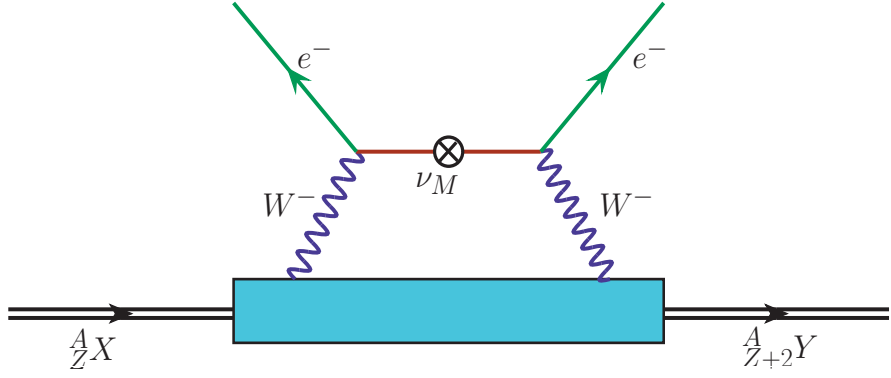}
 \end{center}
\caption{Feynman diagram of the transition $dd \longrightarrow uue^{-} e^{-}$, which induced NDBD.}
\label{NDBD}
\end{figure}
 \item [(iv)] Under the $CPT$ transformations  $\nu^M(s)$ and $\nu^M(-s)$ are related 
 \begin{equation}
  CPT \ket{\nu^M(s)}= \eta^s \ket{\nu^M(-s)}.
 \end{equation}
 \item [(v)] The CPT properties of the Majorana neutrino ensure that they do not have vector current interaction implying that 
 the charge and magnetic moment of Majorana neutrinos vanish~\cite{Mohapatra:1991ng}.
 
 \item [(vi)] It is a challenging task to discriminate between the Dirac and Majorana neutrinos specially if the neutrinos are 
 completely relativistic or ultrarelativistic. This is because, in this case, all the three types of neutrinos~(antineutrinos) 
 are left~(right) handed particles distinguished by their helicities $-1(+1)$, notwithstanding the fact that in the case of 
 Dirac and Weyl neutrinos~(antineutrinos) they are also distinguished by an additional quantum number, i.e., lepton number. 
 \end{itemize}
  There is extensive discussion of various processes, in which there is a possibility to distinguish between the 
 Dirac and the Majorana neutrinos~\cite{Cirigliano:2022oqy, Balantekin:2018ukw}. However, the most distinct process which 
 establishes the existence of Majorana neutrinos is the process of NDBD of nuclei in which the 
 $\bar{\nu}_{e}$ produced in the process $n \longrightarrow p + e^{-} + \bar{\nu}_{e}$ is absorbed by another neutron i.e. 
 $n + \bar{\nu}_{e}(=\nu_{e}) \longrightarrow e^{-} + p$ such that $n + n \longrightarrow p + p + e^{-} + e^{-}$ in the 
 nucleus leading to $^{A}_{Z} X \longrightarrow ^{A}_{Z+2} Y + e^{-} + e^{-}$ as shown in Fig.~\ref{NDBD}. These processes 
 were discussed by Racah~\cite{Racah:1937qq} and Furry~\cite{Furry:1938zz, Furry:1939qr} soon after Majorana's theory. In 
 Fig.~\ref{NDBD}, $\otimes$ denotes the neutrino interaction in the Majorana mass term, which changes the helicity of the 
 neutrino. Such an interaction requires the Majorana neutrino to have mass or the presence of right handed currents. Various 
 theoretical models have been used to calculate NDBD using BSM physics. Experimentally, 
 there are enormous efforts being made to observe such nuclear decays in various experiments being done 
 around the world, for example, EXO-200, KamLAND-Zen, NEMO-3, CUORE, ELEGANT-IV, GERDA, etc. For a review, see 
 Ref.~\cite{Dolinski:2019nrj}.
 
In this work, we focus on the neutrino interactions with matter using the SM. The SM is presented 
briefly in the following Section.

\subsection{Standard model of electroweak interactions}\label{SM}
\subsubsection{Introduction}
The SM was formulated by Weinberg~\cite{Weinberg:1967tq} and Salam~\cite{Salam:1968rm} as the theory of the 
electroweak interaction of leptons. It was extended to the quark sector using the Glashow, Illiopolis and 
Maiani~\cite{Glashow:1970gm} scheme of quark mixing proposed earlier by Cabibbo~\cite{Cabibbo:1963yz}. The formulation 
of SM makes use of the 
experimental results on the properties and interactions of neutrinos obtained from the phenomenological $V-A$ theory of weak 
interactions and the theoretical ideas from the local gauge field theories based on the invariance under continuous symmetry, 
to generate the interactions. Such gauge field theories require the existence of massless vector bosons known as 
Nambu-Goldstone bosons, which mediate the interaction between the matter fields describing the physical particles in field 
theories. This mechanism of generating interactions works in the case of electromagnetic interactions where the invariance of 
the Lagrangian describing the charged leptons $l(=e,~\mu, ~\tau)$ under the local gauge $U(1)$ symmetry, generates a massless 
vector field $A^\mu(x)$, which is identified as the electromagnetic field and mediates the electromagnetic interaction between the 
charged particles. However, this mechanism is not sufficient to generate CC weak interactions, which are 
mediated by the two massive vector fields $W^{\mu +} (x)$ and $W^{\mu -}(x)$. Consequently, a symmetry group higher than 
$U(1)$, which can generate more than one vector field and includes a mechanism to generate masses of the vector fields is 
needed. In the SM proposed by Weinberg~\cite{Weinberg:1967tq} and Salam~\cite{Salam:1968rm}, a higher group $SU(2)_{I_{W}} 
\times U(1)_{Y_{W}}$ (where 
$I_W$ and $Y_W$ are the isospin and hypercharge operators in weak interactions defined in analogy with the strong 
interactions), 
is considered, which requires the existence of four massless vector fields, when the invariance under this symmetry is imposed 
on the Lagrangian. The masses of three of these vector fields leaving one field massless are generated using the mechanism 
of spontaneous breaking of symmetry proposed by Englert and Brout~\cite{Englert:1964et}, and Higgs~\cite{Higgs:1964pj} by 
introducing a doublet of interacting scalar fields $\phi^+(x)$ and $\phi^0(x)$ in the theory. The two out of the three massive fields are 
identified as $W^{\mu +}(x)$ and $W^{\mu -}(x)$ fields, mediating the CC weak interactions and the third massive field is the 
neutral vector field $Z^\mu$, which is new and is predicted to mediate NC interactions in the weak 
sector. The massless field $A^{\mu}(x)$ is identified as the electromagnetic field. The SM was shown later, to be 
renormalizable by 't Hooft and Veltman~\cite{tHooft:1972tcz}, and Lee and Zinn-Justin~\cite{Lee:1972fj}.

For a review of the local gauge field theories based on the continuous symmetries, implying the existence of massless 
Nambu-Goldstone bosons and the phenomenon of Higgs mechanism to generate the masses of the Nambu-Goldstone bosons and the 
renormalizability of the SM, the reader is referred to a general text on quantum field theory~\cite{Quigg:1983gw}.

\subsubsection{SM of electroweak interaction of leptons}\label{section_electron_nue}
The essential results about the neutrino properties and their interactions obtained from the phenomenological $V-A$ theory 
used in formulating the SM are summarized as:
\begin{itemize}
\item [(i)] the (anti)neutrinos are considered to be neutral, massless, left-handed spin $\frac{1}{2}$ particles with helicity 
$(+1)-1$, which exist in three flavors i.e. $\nu_l = \nu_e$, $\nu_\mu$, $\nu_\tau$.

\item [(ii)] the (anti)neutrino of each flavor $l$ are assigned a lepton number $L_{l}=(-1)+1$, which is conserved in weak 
interactions.

\item [(iii)] the neutrinos of flavor $l(=e,\mu,\tau)$ interact with other leptons through the interaction of leptonic 
currents $l^\mu(x)$, which has $V-A$ structure defined as
\begin{equation}
l^\mu(x)=\sum_{{l=e,\mu,\tau}}\bar{\Psi}_l(x)\gamma_{\mu}(1-\gamma_5)\Psi_{\nu_l}(x)
\end{equation}
and interact with $W^{+}_\mu (x)$ to produce  charged leptons of the same flavor. In the lowest order, the interaction 
Lagrangian for describing the $\nu_l lW$ vertex is given by:
 \begin{equation}
   {\it L}_{\text{WI}}^{\text{int}}=\frac{g}{2\sqrt{2}}\left(l^\mu(x)W_\mu^+(x) + h.c.\right),
  \end{equation}
where $\frac{g}{2\sqrt{2}}$ is the strength of the $\nu_l lW$ interaction. As a consequence of the $V-A$ structure of the 
leptonic currents, the left handed neutrino~($\nu_L=\frac{1}{2}(1-\gamma_5)\Psi_\nu$) interacts only with the left handed 
component of the electron ($e_L=\frac{1}{2}(1-\gamma_5)\Psi_e$)  and $\Bar{\Psi}_{e_R}\gamma_\mu(1-\gamma_5)\Psi_{\nu_{e_L}} 
=0$. Therefore, only $\nu_L$ and $e_L$ participate in the weak interaction. Moreover, $\nu_L$ and $e_L$ always interact in 
pairs of $(\nu_L,e_L)$. The Feynman diagrams describing the various vertices $\nu_l l^{-} W^{+}$ and $\bar{\nu}_l l^{+} W^{-} 
(l=e,~\mu,~\tau)$ are represented in Fig.~\ref{SM:1}.
\begin{figure} 
 \begin{center}
  \includegraphics[height=2.5cm,width=4cm]{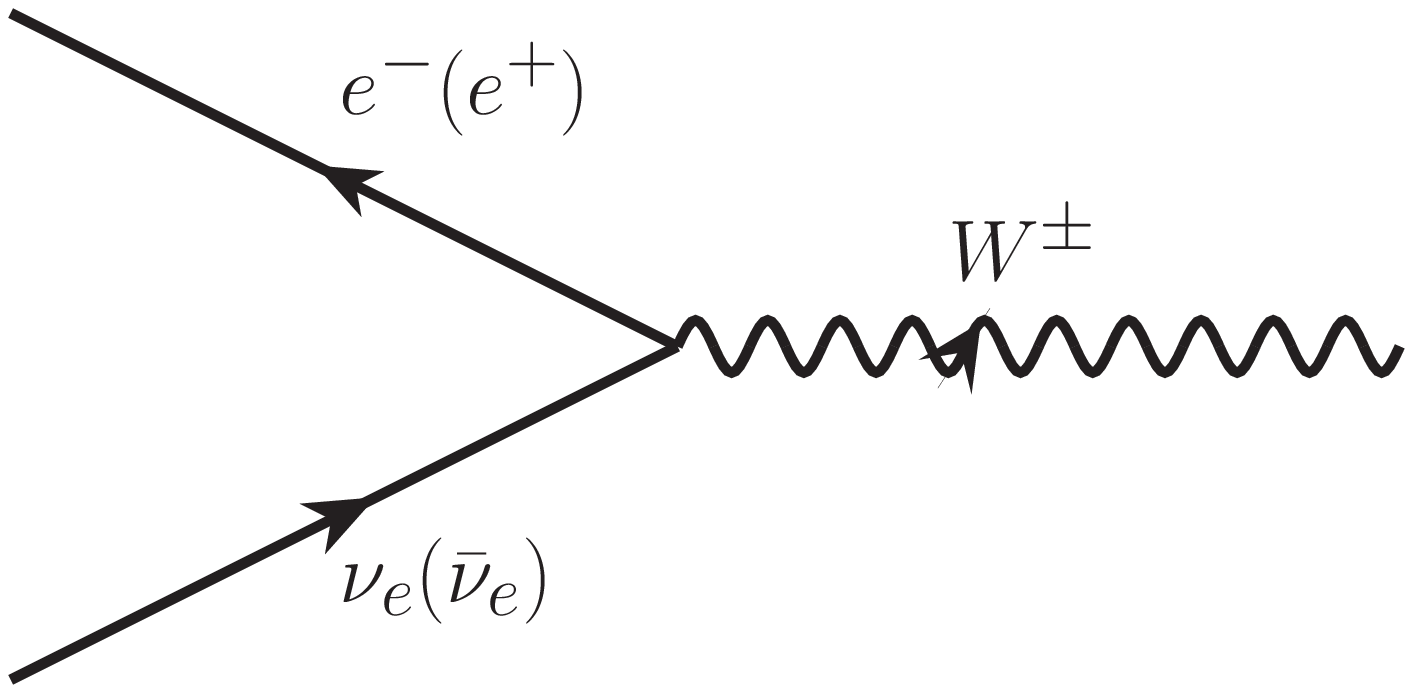}
  \includegraphics[height=2.5cm,width=4cm]{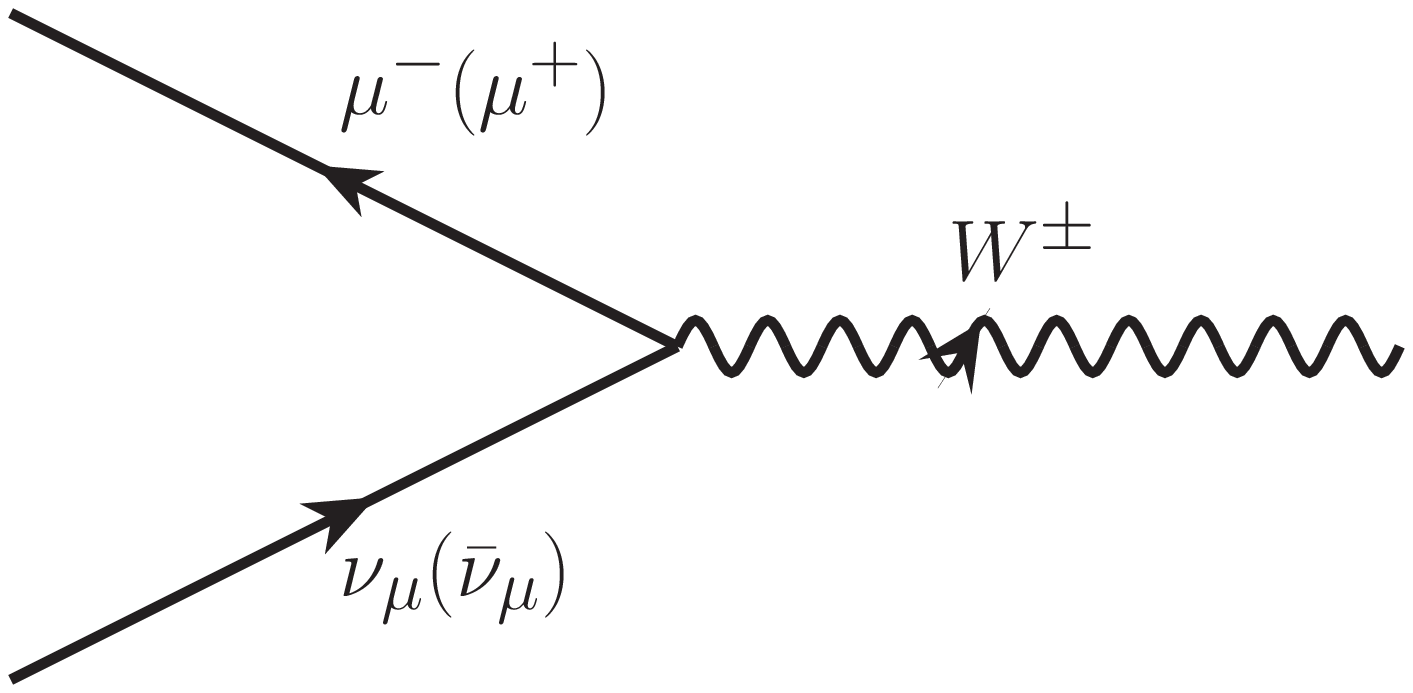}
  \includegraphics[height=2.5cm,width=4cm]{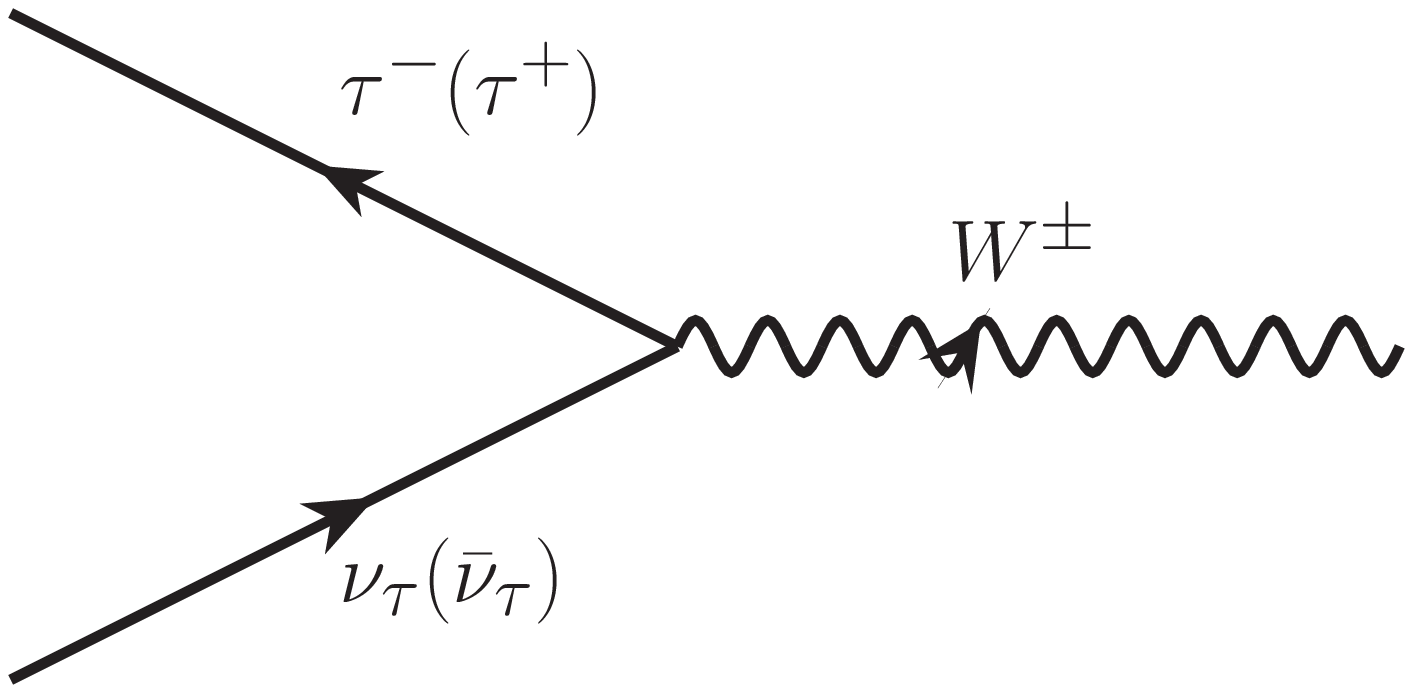}
 \end{center}
\caption{Feynman diagrams for $\nu_l l^{-} W^{+} (l=e,~\mu,~\tau)$ and $\bar{\nu}_l l^{+} W^{-} (l=e,~\mu,~\tau)$ vertices.}
\label{SM:1}
\end{figure}
                                   
\item [(iv)] The physical processes like $\mu^-\rightarrow e^-\bar{\nu}_e\nu_\mu$ and $\nu_\mu e^-\rightarrow \nu_e 
\mu^-$~(shown in Fig.~\ref{SM:2}), etc., take place in the second order such that at low energies, the effective interaction 
is given by the phenomenological $V-A$ interaction Lagrangian with the strength $G_F$ by 
\begin{figure} 
 \begin{center}
 \includegraphics[height=3.5cm,width=6cm]{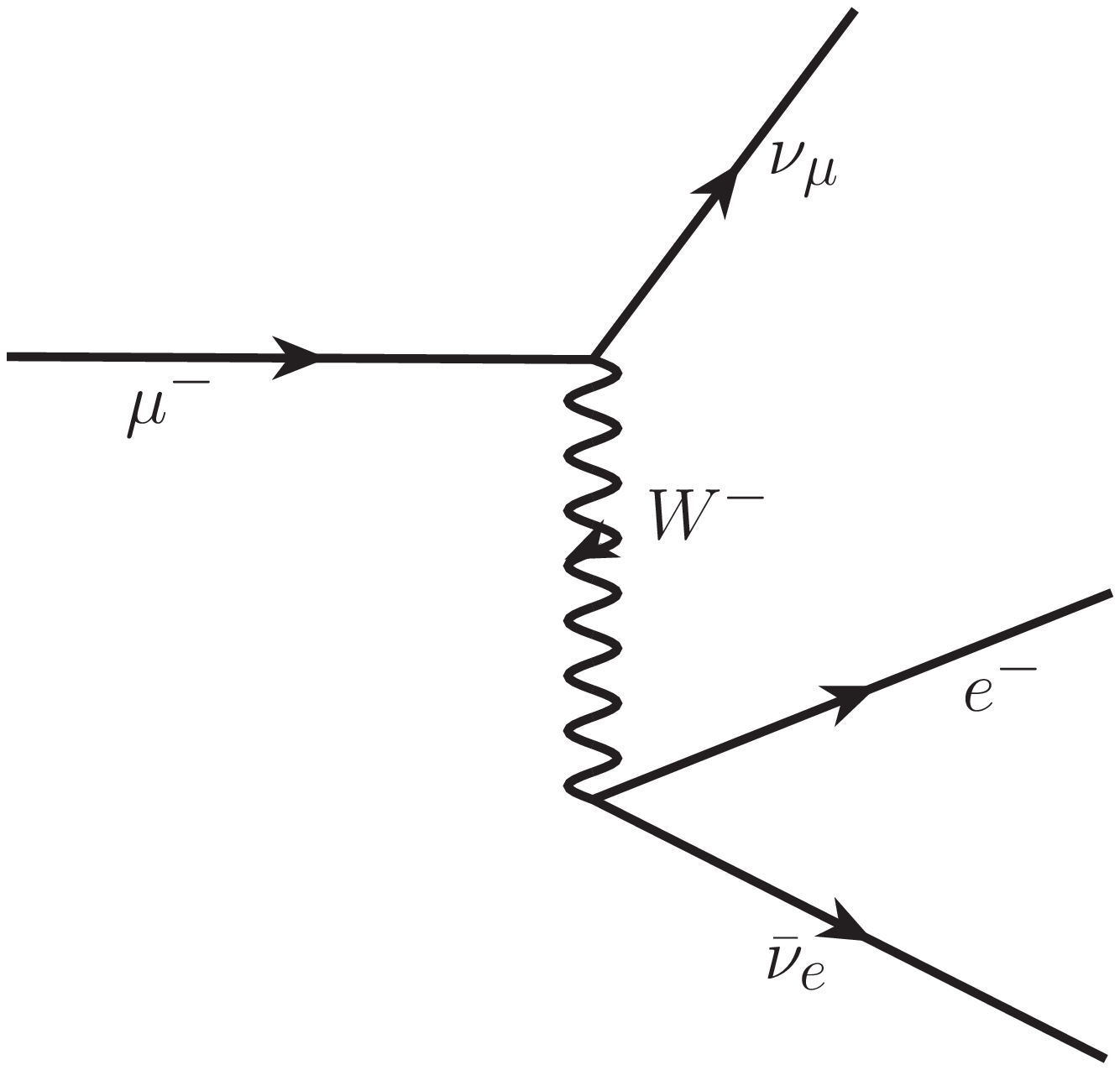} 
 \hspace{1.5cm}
 \includegraphics[height=2.5cm,width=6cm]{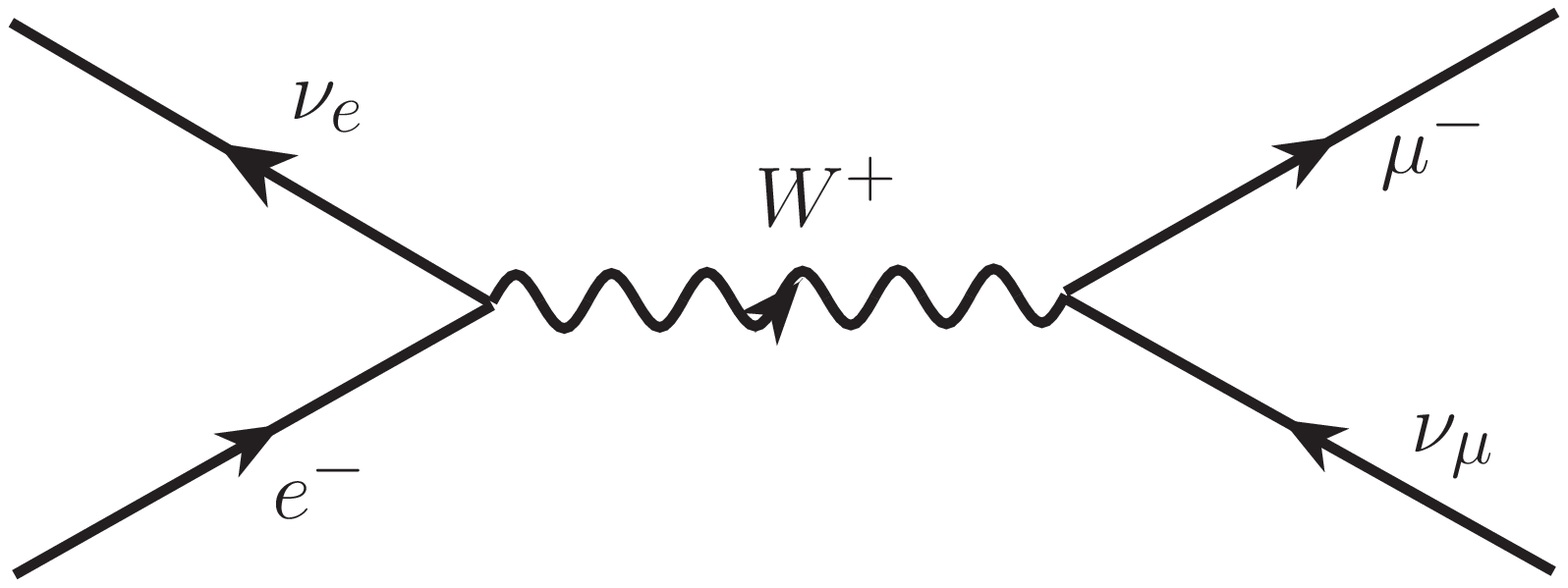} 
 \end{center}
 \caption{Second order Feynman diagram for the processes $\mu^-\rightarrow e^-\bar{\nu}_e\nu_\mu$~(left) and $\nu_\mu 
 e^-\rightarrow \nu_e \mu^-$~(right).}\label{SM:2}
\end{figure}
\begin{equation}
\frac{G_F}{\sqrt{2}}=\frac{g^2}{8M_W^2}.
\end{equation}
\item [(v)] On the other hand, in the theory of electromagnetic interaction described by QED, the 
interaction Lagrangian for the interaction of the charged leptons $l$ with the electromagnetic field $A_\mu(x)$ is given 
by:
\begin{equation}
\mathcal{L}_{int}= -eQ_{|l|}A_\mu(x)\bar{\Psi}_l(x)\gamma^\mu\Psi_l(x),
\end{equation}
where $Q_{|l|}$ is the electronic charge of the lepton in units of $|e|$. It may be noticed that the interaction Lagrangian 
for the electromagnetic interactions of the charged leptons $l$ involve both the left~($l_L$) as well as the right~($l_R$) 
handed components of the lepton, as:
\begin{equation}
\Psi_l(x)=\Psi_{lL}(x)+\Psi_{lR}(x)
\end{equation}
\end{itemize}
Therefore, while the weak interaction Lagrangian involves only the left handed component of leptons i.e. $\nu_{lL}$ and $l_L$, 
the electromagnetic interaction Lagrangian involves both the left handed as well as the right handed components of the charged 
lepton fields $\Psi_{lL}(x)$ and $\Psi_{lR}(x)$. 

In the SM of Weinberg and Salam, the local gauge symmetry group is chosen to be $SU(2)_{I_W}\times U(1)_{Y_{W}}$. Since the 
left handed component of the neutrinos and the corresponding leptons i.e. $\nu_{L}$ and $l_L$ $(l=e,
~\mu,\tau)$ interact in pairs, they are assigned to a doublet under $SU(2)_{I_{W}}$ corresponding to the $\Ket{\frac{1}{2}~~~+
\frac{1}{2}}$ and $\Ket{\frac{1}{2}~~~-\frac{1}{2}}$ states of $I_W$ and $I_{3W}$.
Accordingly, the right handed components $\nu_{lR},~l_{R}$ are assigned to singlet $\ket{0~~~ 0}$ under $SU(2)_{I_{W}}$.
The weak hypercharge $Y_W$ is assigned so that the charge of the leptons $\nu_l$ and $l$ are reproduced using the weak
interaction analogue of the Gell-Mann Nishijima relation in strong interactions and the relation $Y_W=2(Q-I_{3_W})$ is used in 
this case. In Table-\ref{intro:table2}, we tabulate the weak isospin and weak hypercharge of all the left and right handed 
leptons in the upper panel where we also show these assignments for the scalar particles and quarks in the middle and 
lower panels for further use in Sections~\ref{SM:Higgs} and \ref{SM:quarks}.

In the following, we summarize the main steps in formulating the SM for leptons and for simplicity consider the case 
of $\nu_e$ and $e^-$ which can be generalized to other flavors of leptons. We introduce the notation $\Psi_L(x)$ and 
$\Psi_R(x)$ to represent the doublet state of the left handed component of leptons $(\nu_{L}, e_{L})$ and the singlet state 
of the right handed component of the leptons $\nu_{R}$ and $e_{R}$ as:
\begin{eqnarray}
\Psi_L=\begin{pmatrix}
          \Psi_{\nu_e}\\ \Psi_{e}\end{pmatrix}_L= \begin{pmatrix}
         \nu_L \\ e_L\end{pmatrix}, ~\Psi_{e_{R}}=e_R, ~\Psi_{\nu_{R}}=\nu_{R}
\end{eqnarray}
where $\Psi_L=\frac{1}{2}(1-\gamma_5)\Psi$, with $\Psi = \begin{pmatrix}
                                                          \Psi_{\nu_{e}}\\
                                                          \Psi_{e}
                                                         \end{pmatrix}
$, $\Psi_{e_R} = \frac{1}{2}(1+\gamma_5)\Psi_e$ and $\Psi_{\nu_R}=\frac{1}{2}(1+\gamma_5)\Psi_{\nu_e}$.

A Lagrangian for the free massless leptons $\nu_L$, $e_L$ and $e_R$ is written as
 \begin{equation}\label{lg_fr_lp}
  \mathcal{L}=\sum_{j=L,e_R,\nu_R}{{\bar\Psi}_j}\slashed{\partial} \Psi_j(x)
 \end{equation}
with $\slashed{\partial} =\gamma^\mu \frac{\partial}{\partial x_\mu}$. The Lagrangian is invariant under the transformations 
of the global symmetry group $SU(2)_{I_W} \times U(1)_{Y_W}$ generated by the gauge transformations $U= U_1U_2$, where 
$U_1=e^{i \vec{\alpha}\cdot\frac{\vec\tau}{2}}$, $U_2=e^{i\beta I}$, and $\vec{\alpha}(\alpha_1,\alpha_2,\alpha_3)$ and 
$\beta$ are the parameters describing the transformation of $U_1$ and $U_2$, respectively, and $\tau_1 =\begin{pmatrix} 
0&1\\1&0	\end{pmatrix}$, $\tau_2=\begin{pmatrix} 0&-i\\i&0	\end{pmatrix}$ and $\tau_3= \begin{pmatrix} 1&0\\0&-1 
\end{pmatrix}$ are the Pauli matrices, $I$ is the unit matrix. A mass term like $m\bar{\Psi}_{j} \Psi_{j} ~(= \bar{\Psi}_{jL} 
\Psi_{jR} + \bar{\Psi}_{jR} \Psi_{jL})$ is not included as it is not invariant under global $SU(2)_{I_W} \times U(1)_{Y_W}$. 
However, when the transformations are made local by replacing $\vec{\alpha}\to \vec{\alpha}(x)$ and $\beta \to \beta(x)$ then 
the Lagrangian given in Eq.~(\ref{lg_fr_lp}) is not invariant under the local gauge group generated by the local gauge 
transformations $U_1 (x)U_2(x)$ due to the presence of the derivation term $\frac{\partial}{\partial x_\mu}$ in the 
Lagrangian. In order to restore the invariance of the Lagrangian under local transformation, the Lagrangian is rewritten in 
terms of the covariant derivative $\frac{D}{Dx^\mu}$ instead of the ordinary derivative $\frac{\partial}{\partial x^\mu}$ by 
introducing the matrix valued gauge fields $W^\mu=\sum_i \frac{\tau^i}{2}\cdot W^{\mu i}$ corresponding to $U_1(x)$ and the 
field $B^\mu$ corresponding to $U_2(x)$ transformation of $SU(2)_{I_{W}}$ and $U(1)_{Y_{W}}$ and defining the covariant 
derivative $\frac{D}{Dx^\mu}$ as
\begin{equation}\label{eq:4.8}
 \frac{D}{Dx^\mu} = \frac{\partial}{\partial x^\mu} +ig\frac{\vec{\tau}\cdot \vec{W}^\mu}{2} + i\frac{g'}{2}Y_W B^\mu(x),
\end{equation}
$g$ and $g'$ being the coupling constant corresponding to $SU(2)_{I_W}$ and $U(1)_{Y_W}$ gauge fields. A factor of 
$\frac{1}{2}$ is introduced with the $B^\mu(x)$ field in analogy with the $\vec{W}^\mu(x)$. Requiring that the new gauge 
vector field $W^\mu$  and $B^\mu$ transform under $U_1(x)$ and $U_2(x)$ as:
\begin{eqnarray}\label{eq:4.1011}
 \vec{W}^\mu(x) \to \vec{W'}^\mu(x) =\vec{W}^\mu(x)  - \vec{\alpha}\times \vec{W}^\mu -\frac{i}{g}\partial_\mu \vec{\alpha},\qquad \qquad
 B^\mu(x) \to B'^\mu(x) = B^\mu(x) -\frac{i}{g'}\partial_\mu  \beta(x)
\end{eqnarray}
ensures that under the local gauge transformation
\begin{eqnarray}\label{eq:4.12}
 \Psi(x) \to \Psi^\prime(x) = U \Psi(x), \qquad \qquad
  D\Psi(x) \to (D\Psi)^\prime(x) = U(D \Psi(x))
\end{eqnarray}
making the redefined Lagrangian invariant under the local gauge transformations $U(x)$.
It can be shown using 
Eq.~(\ref{eq:4.8}), that
\begin{eqnarray}
[D^\mu, D^\nu]&=& \frac{g}{2}\vec{\tau}\cdot\vec{G}^{\mu\nu}(x)+\frac{g'}{2}Y_W B^{\mu\nu}(x),
\end{eqnarray}
where $B^{\mu\nu}$ and $\vec{G}^{\mu\nu}$ being the field tensors for $B^{\mu}(x)$ and $\vec{W}^{\mu}(x)$ fields given by:
\begin{eqnarray}
B^{\mu\nu}=\partial^\mu B^\nu(x)-\partial^\nu B^\mu(x),\qquad \qquad
\vec{G}^{\mu\nu}(x)=\partial^\mu\vec{W}^\nu (x)-\partial^\nu \vec{W}^{\mu}(x)+g \vec{W}^{\mu}(x)\times\vec{W}^{\nu}(x),
\end{eqnarray}
and are used to define the kinetic energy of the vector $ B^\mu$ and $\vec{W}^{\mu}$ fields.

Consequently, the free particle Lagrangian is redefined as
\begin{equation}\label{eq:4.16}
 \mathcal{L}= \sum_{j=L,e_R,\nu_R} \bar{\Psi}_{j}(x) \slashed{D}\Psi_j.
\end{equation}
Writing the expressions for $D^\mu \Psi_L$, $D^\mu \Psi_{e_R}$ and $D^\mu \Psi_{\nu_R}$, using the values of $Y_W$ for 
$\Psi_{L}$, $\Psi_{e_R}$ and $\Psi_{\nu_R}$ given in Table.\ref{intro:table2}, we obtain
 \begin{eqnarray}
   D^\mu\Psi_L(x)=\left(\partial^\mu +\frac{ig}{2}\vec{\tau}\cdot \vec{W}^{\mu}(x)-\frac{ig'}{2}B^\mu(x)\right)\Psi_L,~~~
   D^\mu \Psi_{e_R}(x)=(\partial^\mu-ig'B^\mu(x))\Psi_R,~~~
   D^\mu \Psi_{\nu_R}=\partial^\mu \Psi_{\nu_R}
  \end{eqnarray}
The Lagrangian in Eq.~(\ref{eq:4.16}) is expanded over $j$ and is written as
\begin{eqnarray}
 \mathcal{L} &=& \mathcal{L}_0 + \mathcal{L}_{int}, \qquad \qquad \text{with} \nonumber\\
 \mathcal{L}_0 &=& i\bar{\Psi}_L \slashed{\partial} \Psi_L + i \bar{\Psi}_{e_R} \slashed{\partial} \Psi_{e_R} +i
 \bar{\Psi}_{\nu_R} \slashed{\partial} \Psi_{\nu_R},\qquad \qquad \text{and} \nonumber \\
 \label{WS:lint1}
\mathcal{L}_{int}&=& -\frac{g}{2\sqrt{2}}\Big( \overline{\nu}_{e} \gamma^\mu (1-\gamma_5)e W^+_\mu +\overline{e} \gamma^\mu
(1-\gamma_5)\nu_e W^-_\mu \Big) -\frac{\sqrt{g^2 + g^{\prime 2}}}{2} \overline{\nu}_L\gamma^\mu \nu_L Z_\mu\nonumber\\
&& + \frac{g g^\prime }{\sqrt{g^2+g^{\prime 2}}} \overline{e}\gamma^\mu eA_\mu + \frac{1}{\sqrt{g^2+g^{\prime 2}}}
\left[-g^{\prime 2} \overline{e}_R \gamma^\mu e_R +\frac{g^2-g^{\prime 2}}{2}  \overline{e}_L \gamma^\mu e_L \right] Z_\mu,
\end{eqnarray}
where
\begin{eqnarray}
W_{\mu}^{\pm}=\frac{W_{\mu} ^{1} \mp i W_{\mu} ^{2}}{\sqrt{2}}, \qquad 
 Z_{\mu}=\frac{g W^{3}_{\mu}-g^{\prime}B_{\mu}}{\sqrt{g^2 + g^{\prime 2}}} \qquad \text{and} \qquad
 A_{\mu}= \frac{g^{\prime} W^{3}_{\mu}+g B_{\mu}}{\sqrt{g^2 + g^{\prime 2}}}.
\end{eqnarray}
We can observe from $\mathcal{L}$ that
\begin{itemize}
\item[(i)] no terms like $W_i^\mu W_{i\mu}$ and $B^\mu B_\mu$~(or equivalently like $A^\mu A_\mu$, $Z^\mu Z_\mu$ or $W^{\pm 
\mu} W^\mp_\mu$) appear in ${\cal L}$, implying that all the fields $W^{+ \mu}$, $W^{- \mu}$, $Z^\mu$ and $A^\mu$ are massless.

\item[(ii)] ${\cal L}_{int}$ correctly reproduces 
\begin{itemize}
 \item [1.] CC weak interaction of $\nu_e$ and $e$ with strength $\frac{g}{2\sqrt{2}}$ given by
\begin{equation}\label{eq:20}
 \mathcal{L}_{int}^{CC}= -\frac{g}{2\sqrt{2}}\bar{\Psi}_{{\nu}_{e}} \gamma^\mu (1-\gamma_5)\Psi_e W_\mu^+ + h.c.
\end{equation}

\item [2.] the electromagnetic interaction of electrons with the electromagnetic coupling given by
\begin{equation}\label{eq:21}
 \mathcal{L}_{int}^{EM}= \frac{gg'}{\sqrt{g^2+g^{\prime 2}}}\bar{\Psi}_{{e}}\gamma^\mu\Psi_e A_\mu.
\end{equation}
\end{itemize}

\item[(iii)] ${\cal L}_{int}$ predicts
\begin{itemize}
 \item [1.] NC interaction of neutrinos is given by:
\begin{equation}
\mathcal{L}^\nu_{NC}=-\frac{\sqrt{g^2+g^{\prime 2}}}{4}\bar{\Psi}_\nu \gamma_\mu(1-\gamma_5)\Psi_\nu Z^\mu, 
\qquad \text{with strength } \frac{\sqrt{g^2+g^{\prime 2}}}{2}.
\end{equation}
\item [2.] NC interaction of electrons is given by:
\begin{equation}
{\cal{L}}^{e}_{NC} = -\frac{Z^\mu}{\sqrt{g^2+g^{\prime 2}}}\left[
  \frac{g^2-g^{\prime 2}}{4}  \overline{\Psi}_{e} \gamma_\mu(1-\gamma_5) \Psi_e -\frac{g^{\prime 2}}{2} \overline{\Psi}_{e} 
  \gamma_\mu(1+\gamma_5) \Psi_e\right].
\end{equation}
\end{itemize}
\begin{table}
	\centering
\begin{tabular}{|c|c|c|c|c|}
\hline 
	Quantum numbers$\rightarrow$ & $I_W$  & $I_{W3}$& $Y_W$ &$Q$\\
	Particles$\downarrow$ &   &&  &\\
	\hline 
	$\nu_{eL}$,$\nu_{{\mu}L}$,$\nu_{{\tau}L}$ & $\frac{1}{2}$ &+$\frac{1}{2}$ &-1&0 \\
	$e_L$,$\mu_L$,$\tau_L$ & $\frac{1}{2}$ &-$\frac{1}{2}$ &-1&$-1$ \\
	$e_R$,$\mu_R$,$\tau_R$ & 0 &0 &$-$2&$-1$ \\ 
	\hline
	$\phi^+$ & $\frac{1}{2}$ &+$\frac{1}{2}$ &1&+1 \\
	$\phi^0$ & $\frac{1}{2}$ &-$\frac{1}{2}$ &1&0 \\ 
	\hline
	$u_L$,$c_L$,$t_L$ & $\frac{1}{2}$ &+$\frac{1}{2}$ &$\frac{1}{3}$&+$\frac{2}{3}$ \\
	${d_L}^\prime$,${s_L}^\prime$,${b_L}^\prime$ & $\frac{1}{2}$ &-$\frac{1}{2}$ &$\frac{1}{3}$&-$\frac{1}{3}$ \\
	$u_R$,$c_R$,$t_R$ & 0 &0&$\frac{4}{3}$&+$\frac{2}{3}$ \\
	${d_R}^\prime$,${s_R}^\prime$,${b_R}^\prime$ & 0 &0 &-$\frac{2}{3}$&-$\frac{1}{3}$ \\ \hline
\end{tabular}
\caption{Weak isospin($I_W$), its third component($I_{W3}$), weak-hypercharge($Y=2(Q-I_3)$), charge(Q$(|e|)$) of the leptons, 
scalar mesons and quarks in the SM model.}\label{intro:table2}
\end{table}

\item[(iv)] The SM therefore describes the electroweak interaction of leptons in terms of the two parameters $g$ and $g'$. 
Comparing ${\cal L}_{int}^{CC}$ and ${\cal L}_{int}^{EM}$ given in Eqs.~(\ref{eq:20}) and (\ref{eq:21}), respectively, with 
the $V-A$ theory of weak interactions and QED of the charged leptons, we see that $g$ and $g^{\prime}$ are related with the 
strength of Fermi interaction $G_F$ and the electromagnetic coupling $e$ through the relations
\begin{eqnarray}
\frac{G_F}{\sqrt{2}}  = \frac{g^2}{8M_W^2}, \qquad \qquad
\frac{1}{e^2} = \frac{1}{g^2} + \frac{1}{g^{\prime 2}},
\end{eqnarray}
where $M_W$ is the mass of $W^{\mu(\pm)}$ vector fields.

\item [(v)] The Lagrangian obtained using the local gauge field theory, thus, predicts the electromagnetic and weak 
interactions mediated by four vector gauge fields, $W^{\mu +}(x),~W^{\mu -}(x),~Z^{\mu}(x)$ and $A^{\mu}(x)$, all being 
massless as there are no mass terms like $M_V^2 V_i^\mu V_{i\mu}(V_i=W^+, W^-,Z,A)$ for any of the fields. While the model can 
describe the electromagnetic interaction, it can not describe the weak interaction which is mediated by massive vector fields 
$W^{\mu\pm}$. Therefore, the model in this form is inconsistent with the phenomenological $V-A$ theory of weak interactions 
unless a mechanism is devised to generate the masses of these fields. This is done using the Higgs mechanism.
\end{itemize}

Since all the fields $W^{\mu}_{+}$, $W^{\mu}_{-}$, $Z^{\mu}$, and $A^{\mu}$ are massless, the kinetic energy terms are 
added by hand to redefine the free Lagrangian $\mathcal{L}_0$ as
\begin{equation*}
 \mathcal{L}_0 \to \mathcal{L}_0 - \frac{1}{4}B^\mu B_\mu - \frac{1}{4}G^{\mu\nu}G_{\mu\nu}
\end{equation*}
in analogy with the kinetic energy term for massless electromagnetic field $A^{\mu}$ in QED.

\subsubsection{Higgs mechanism and generation of mass}\label{SM:Higgs}
The phenomenon of spontaneous breaking of continuous symmetry in field theory proposed by  Englert and 
Brout~\cite{Englert:1964et} and Higgs~\cite{Higgs:1964pj}, generally called the Higgs mechanism was used by 
Weinberg~\cite{Weinberg:1967tq} and Salam~\cite{Salam:1968rm} to generate the mass of the gauge vector bosons. In this 
phenomenon, strongly interacting doublet of scalar fields $\phi (x)$ are introduced in the Lagrangian. The vacuum state 
of this Lagrangian breaks the symmetry while the Lagrangian respects the symmetry. Hence, the name spontaneous breaking of 
symmetry instead of the explicit breaking of symmetry in field theory is given to this phenomenon. In local gauge field 
theories the invariance of the Lagrangian generates massless vector gauge fields corresponding to each generator of the 
symmetry. In the SM,  the spontaneous breaking of symmetry based on $SU(2)_{I_W} \times U(1)_{Y_W}$, is realized by 
introducing interacting scalar fields $\phi(x)$ which transform as doublet under $SU(2)_{I_{W}}$ i.e.
\begin{eqnarray}
\phi(x)=\begin{pmatrix} \phi^+(x)\\ \phi^0(x)\end{pmatrix}=\frac{1}{\sqrt{2}}\begin{pmatrix} \phi_1(x)+i\phi_2(x)\\ 
\phi_3(x)+i\phi_4(x)\end{pmatrix}
\end{eqnarray}
with fields $\phi^+(x)$ and $\phi^0(x)$ having $I=\frac{1}{2}$ and $I_3=\pm\frac{1}{2}$ and are assigned $Y_W=+1$ to 
reproduce their charges as shown in Table-\ref{intro:table2}~(middle panel). The interaction Lagrangian for the scalar fields 
$\phi$ is written in a locally gauge invariant way under the $SU(2)_{I_{W}}\times U(1)_{Y_{W}}$ transformation using the covariant 
derivative $D^\mu$ given by:
\begin{equation}\label{Lag:D}
D^\mu\phi =(\partial_\mu +ig\vec{\tau}\cdot \vec{W}^\mu +i\frac{g'}{2}Y_{W}B^\mu)\phi
\end{equation}
and is written as:
\begin{equation}\label{Lag:Higgs}
\mathcal{L}_\phi =D^\mu \phi^\dagger D_\mu\phi-V(\phi^* \phi)
\end{equation}
where the potential $V(\phi^*\phi)$ is given by:
\begin{equation}
V(\phi^*\phi)=-\mu^2\phi^*(x)\phi(x)+\lambda(\phi^*\phi)^2
\end{equation}
and has minimum value given by the condition:
\begin{equation}
\frac{\partial V}{\partial \phi^*}=\phi(x)(-\mu^2+2\lambda \phi^* \phi)=0
\end{equation}
which implies that for $\mu^2<0$, the minimum occurs at $\phi(x)=0$, but for $\mu^2>0$, there is minima at $\phi^\dagger 
(x)\phi(x)=\frac{\mu^2}{2\lambda}$. While $\phi(x)=0$ is a trivial ground state, $\phi^\dagger (x)\phi(x)=\frac{\mu^2}
{2\lambda}$ implies an infinitely degenerate value to $\phi(x)$ since it is a complex field given by $\phi(x) = e^{i\theta} 
\phi(x)$, $\theta$ being arbitrary.

The $SU(2)_{I_{W}}\times U(1)_{Y_{W}}$ symmetry is spontaneously broken in the SM by choosing:
$$\phi_1(x)=\phi_2(x)=\phi_4(x)=0, ~~~~~~~~~~~~~~~~~~~~\phi_3(x)\neq 0$$
such that vacuum expectation value~(VEV) of $\phi(x)$ is given by:
\begin{eqnarray}\label{VEV2}
\bra{0}\phi(x)\ket{0}=\begin{pmatrix}0\\ \frac{v}{\sqrt{2}}\end{pmatrix}, \qquad v=\frac{1}{\sqrt{2}}\bra{0}\phi_3(x)\ket{0}
\end{eqnarray} 
This choice of ground state ensures that the ground state $\phi_0(x)$ remains invariant under the symmetry group 
transformations of $U(1)_Q$, where $Q=\frac{1}{2}\tau_3 +\frac{Y}{2}$ is the generator of the group because $Q\ket{\phi_0} 
=0$. This means that $SU(2)_{I_{W}}\times U(1)_{Y_{W}}$ is spontaneously broken to $U(1)_Q$, keeping the gauge fields corresponding to 
$U(1)_Q$ symmetry i.e. the electromagnetic field massless, while generating the mass corresponding to other three generators 
$\tau^+,~ \tau^-$ and $\frac{1}{2}(\tau_3 -Y)$. The Lagrangian for the Higgs field $\phi(x)$, invariant under the local 
gauge group $SU(2)_{I_W}\times U(1)_{I_W}$, written in terms of the covariant derivative $D^\mu \phi$ given in Eq.~(\ref{Lag:D}) is written 
explicitly using the value of $Y_{W}$ for $\phi$ from Table-\ref{intro:table2}~(middle panel) as
\begin{eqnarray}
\mathcal{L}_\phi &=& \left(\partial^\mu +i\frac{g}{2}\vec{\tau}\cdot\vec{W^\mu}+i\frac{g'}{2}B^\mu\right)\phi^*(x)
\left(\partial_\mu+i\frac{g}{2}\vec{\tau}\cdot\vec{W}_\mu +i\frac{g'}{2}B_\mu\right)\phi (x) -V(\phi^* (x)\phi (x))
\end{eqnarray} 
Expanding the field $\phi(x)$ around its VEV $\langle\phi(x)\rangle_0$ given in Eq.~(\ref{VEV2}), and writing 
\begin{equation}\label{VEV}
\phi(x)=\frac{1}{\sqrt{2}}\begin{pmatrix}
0 \\ v+H(x)
\end{pmatrix},
\end{equation}
we write
\begin{eqnarray}\label{UT_mgw5}
  {\cal L}_{\phi}&=&\frac{1}{2}\partial^{\mu} H(x)\partial_{\mu} H(x) 
  +\frac{v^2 g^2}{8}(|W_{\mu}^{+}|^2 + |W_{\mu}^{-}|^2)+\frac{g^2}{8}(H^{2}+2Hv)(|W_{\mu}^{+}|^2 + |W_{\mu}^{-}|^2) 
  +\left(\frac{g^2 +g^{\prime 2}}{4}\right)\left(\frac{H^2 +2Hv +v^2}{2}\right) \nonumber\\
  &\times&Z_\mu Z^\mu 
  +\left[ \frac{ g^2 g'^2}{4(g^2 +g^{\prime 2})}(H^2 +2Hv)-\frac{g^2 g'^2}{4(g^2 +g^{\prime 2})}(H^2 +2Hv)
 \right] A_\mu A^\mu - V(\phi^* (x) \phi(x)).
\end{eqnarray}  
The above Lagrangian predicts the masses of the vector gauge bosons $W^{+\mu},~W^{-\mu}, ~Z^{\mu}$, where $Z_{\mu}=~\frac{g 
W^{3}_{\mu}-g^{\prime}B_{\mu}}{\sqrt{g^2 + g^{\prime 2}}}$ and $ A_{\mu}= \frac{g^{\prime} W^{3}_{\mu}+g B_{\mu}}{\sqrt{g^2 + 
g^{\prime 2}}}$  to obtain
\begin{eqnarray}
M_{W^+}= M_{W^-}=\frac{vg}{2}; \qquad \qquad
M_Z = \frac{v\sqrt{g^2+g^{\prime 2}}}{2};\qquad \qquad
M_A = 0,
\end{eqnarray}
and mass of the Higgs scalar $H(x)$ is predicted by expanding $V(\phi^* (x) \phi(x))$ using 
Eq.~(\ref{VEV}) to obtain:
\begin{eqnarray}
 M_{H} &=& \sqrt{2\lambda} v.
\end{eqnarray}
We see that the mass of $Z^\mu$ and $W^\mu$ vector fields are related by 
\begin{equation}\label{Mass:W:Z}
\frac{M_Z}{M_W}=\sqrt{\left(1+\frac{g^{\prime 2}}{g^2} \right)}\geqslant 1
\end{equation}
and the absolute values of $M_W$ and $M_Z$ are determined by $g~,g'$, and $v$.

\subsubsection{Neutral current interactions and the weak mixing angle}
It has been shown by Eqs.~(\ref{eq:20}) and (\ref{eq:21}) that the SM reproduces CC weak and 
electromagnetic interactions of leptons mediated by $W^{\mu\pm}$ vector bosons with mass $M_W=\frac{vg}{2}$ and the massless 
electromagnetic vector field $\vec{A}^\mu$, as well as predicts NC weak interactions for the neutrinos 
and electrons which are mediated by the neutral vector boson $Z^\mu$ with mass $M_Z=M_W\sqrt{1+\frac{g^{\prime 2}}{g^2}}$, as 
shown by Eq.~(\ref{Mass:W:Z}). The strength of the NC weak interaction is alternatively defined in terms of a weak mixing angle $\theta_W$ 
defined as
\begin{eqnarray}
\tan\theta_W=\frac{g'}{g}&\qquad \qquad& \text{
such that}\nonumber \\ 
M_W = M_Z\cos\theta_W; &\qquad \qquad&
e = g\sin\theta_W=g'\cos\theta_W \\
\label{Zmu}
 Z^\mu =\cos\theta_W W_3^\mu-\sin\theta_WB^\mu; &\qquad\qquad&
A^\mu = \sin\theta_W W_3^\mu + \cos\theta_W B^\mu.
\end{eqnarray}
The weak mixing angle $\theta_{W}$ mixes the neutral gauge vector bosons $W_3^\mu$ and $B^\mu$ corresponding to the 
$SU(2)_{I_W}$ and $U(1)_{Y_{W}}$ gauge bosons to produce the physical gauge vector fields $Z^\mu$ and $A^\mu$ responsible for the 
weak NC and the electromagnetic current carrying vector fields. In terms of the weak mixing angle, the 
Lagrangians for the weak CC and NC as well as the electromagnetic interactions written in Eq.~(\ref{WS:lint1}) are 
rewritten as:
\begin{eqnarray}
\label{weak_cc_lag}
\mathcal{L}_{CC}&=&-\frac{g}{2\sqrt{2}}\left[ \overline{\nu}_{e}\gamma^{\mu}(1-\gamma_5)e W_{\mu}^{+}+h.c.\right], \\ 
\mathcal{L}_{NC}&=&-\frac{g}{2\cos\theta_W}\left[\overline{\nu}_{e} \gamma^{\mu} (g_{V}^{\nu_{e}} - g_{A}^{\nu_{e}} 
\gamma_{5}) \nu_{e} 
  +\bar{e}\gamma^\mu (g_V^e -g_A^e \gamma_5)e \right] Z_\mu , \label{weak_nc_lag}\\
  \mathcal{L}_{EM} &=& - |e| \bar{e} \gamma^{\mu} e A_{\mu},
  \end{eqnarray}
where
\begin{eqnarray}\label{gve_ref}
g_V^e = 2 \sin^2\theta_W -\frac{1}{2} , \qquad\qquad g_V^{\nu_{e}} = \frac{1}{2}, \qquad \qquad
g_A^e = -\frac{1}{2}, \qquad \qquad g_A^{\nu_{e}} = \frac{1}{2}. 
\end{eqnarray}  

\subsubsection{Extension of the SM to the leptons, quarks and nucleons}\label{SM:quarks}
The extension of the SM to the leptons of other flavors is straightforward. The left and right handed components of 
$(\nu_\mu, ~\mu^-)$ and $(\nu_\tau ,~\tau^-)$ are assigned to the $SU(2)_{I_{W}}\times U(1)_{Y_{W}}$ representations in same way as done 
for the $(\nu_e,e^-)$ leptons and shown in Table-\ref{intro:table2} and interaction can be generated following the procedure 
in Section~\ref{section_electron_nue} with implicit assumption of LFU and the interactions for all the 
flavors of leptons can be written as:
\begin{eqnarray}
\label{Lag:CC:all}
\mathcal{L}_{CC}&=&-\frac{g}{2\sqrt{2}}\sum_{l=e,\mu,\tau}\left[ \overline{\nu}_{l}\gamma^{\mu}(1-\gamma_5)l W_{\mu}^{+}+h.c.
\right], \\
\label{Lag:NC:all}
\mathcal{L}_{NC}&=&-\frac{g}{2\cos\theta_W}\sum_{l=e,\mu,\tau}\left[\overline{\nu}_{l}\gamma^{\mu} (g_{V}^{\nu_{l}} - 
g_{A}^{\nu_{l}} \gamma_{5})\nu_{l} 
  +\bar{l}\gamma^\mu (g_V^l -g_A^l \gamma_5)l \right] Z_\mu , \label{NC_lag_allf}
\end{eqnarray}
with $g_V^e$, $g_A^e$, $g_V^{\nu_e}$, and $g_A^{\nu_e}$ given in Eq.~(\ref{gve_ref}), and have the same values for all the lepton flavors $l$. 

However, the formalism presented in Section~\ref{section_electron_nue} can be reformulated in terms of the weak 
isospin~($\vec{\tau}_f$) and the charge operators $Q_f$ for the fermions $f$ instead of the weak
hypercharge. The weak CC and NC currents are written using $SU(2)_{I_{Y}}$ doublets,
$\Psi_{f_L}=\frac{1-\gamma_5}{2}\Psi_f$ and $\Psi_{f_R}=\frac{1+\gamma_5}{2}\Psi_f$ for a fermion $f$, so that they can be 
applied to a lepton and quark in a unified way. 

It is straightforward to see that the weak CC interaction Lagrangian is given by:
\begin{equation}
\mathcal{L}_{CC}=-\frac{g}{2\sqrt{2}}\sum_{f=e,\mu,\tau}\bar{\Psi}_{\nu f} \gamma^\mu (1-\gamma_5)\Psi_f W_\mu^+ + h.c.
\end{equation}
 The weak NC Lagrangian using Eq.~(\ref{NC_lag_allf}) is written as: 
\begin{equation}\label{SM:Lag:NC}
\mathcal{L}_{NC}=-\sum_{f=e,\mu,\tau}\left[\bar{\Psi}_{f_L}\left(g\frac{\tau_3^f}{2}\slashed{W}^3+g'\frac{Y_L^f}{2}
\slashed{B}\right)\Psi_{f_L}+\bar{\Psi}_{f_R}g'\frac{Y_R^f}{2}\slashed{B}\Psi_{f_R} \right]
\end{equation}
where $\tau_3^f$ is isospin operator and  $Y^f_L$ and $Y^f_R$ are the hypercharges of the left- and right- handed fermions in 
Table-\ref{intro:table2}. Since, $\tau_3 \Psi_{f_R}=0$ as $\Psi_{f_R}$ is isosinglet, the two terms in Eq.~(\ref{SM:Lag:NC}) 
are combined to write:
\begin{eqnarray}
\mathcal{L}_{NC}=-\sum_{f=e.\mu,\tau}\sum_{i=L,R}\bar{\Psi}_{f_i}\left(g\frac{\tau_3^{f_i}}{2}\slashed{W}^3 +g'
\frac{Y^{f_i}}{2}\slashed{B}\right)\Psi_{f_i}.
\end{eqnarray}
Using Eq.~(\ref{Zmu}) to express $W^{3\mu}$ and $B^\mu$ in terms of $Z^\mu$ and $A^\mu$ and $\Psi^{f_i} = 
2Q^{f_i}- \tau_3^{f_i}$, $\mathcal{L}_{NC}$ can be expressed as:
\begin{equation}\label{SM:Lag:NC2}
\mathcal{L}_{NC}=-\frac{g}{2\cos\theta_W}\sum_{f=e,\mu,\tau}\sum_{i=L,R}\bar{\Psi}_{f_i}\gamma^\mu (\tau_3^{f_i}-2Q^{f_i}
\sin^2\theta_W)\Psi_{f_i}Z_\mu .
\end{equation}
After further expanding over $i=L,R$ and using $\tau_3^{f_R} \Psi_{f_R}=0$, for $f=e,\mu,\tau$ the following expression is 
obtained
\begin{eqnarray}
\mathcal{L}_{NC}&=&-\frac{g}{2\cos\theta_W}\sum_{f=e,\mu,\tau}\bar{\Psi}_f \gamma^\mu (g_V^f-g_A^f\gamma_5)\Psi_f, \qquad \qquad \text
{with}\nonumber \\
\label{gve_tau_theta}
g_V^f &=& \frac{1}{2}\tau_3^f -2Q^f \sin^2\theta_W, \qquad \qquad\qquad 
g_A^f = \frac{1}{2}\tau_3^f 
\end{eqnarray}
After operating $\tau_3^f$ and $Q^f$ on leptons, in Table-\ref{intro:table2}, $\mathcal{L}_{NC}^{\text{lepton}}$ is
obtained as stated in Eq.~(\ref{SM:Lag:NC2}).

In this form, it can be used to generate the weak CC and NC interactions of quarks which are classified under 
$SU(2)_{I_{W}}\times U(1)_{Y_{W}}$ as shown in Table-\ref{intro:table2}~(lower panel) for their left handed and right handed components 
and arranged in three flavors of doublets as:
\begin{equation}
q_L =\begin{pmatrix}
u \\ d'
\end{pmatrix}_L ,\begin{pmatrix}
c \\ s'
\end{pmatrix}_L,\begin{pmatrix}
t \\ b'
\end{pmatrix}_L 
\end{equation} 
and singlet as $q_R=u_R,d^\prime_R,c_R,s^\prime_R,t_R,b^\prime_R$, where 
\begin{equation}
\begin{pmatrix}
d_{R}^{\prime}\\
s_{R}^{\prime}\\
b_{R}^{\prime}
\end{pmatrix} = U\begin{pmatrix}
d_{R}\\
s_{R}\\
b_{R}
\end{pmatrix}
\end{equation} 
and $U$ being the CKM matrix~\cite{ParticleDataGroup:2020ssz}.
The $\mathcal{L}_{\text{CC,NC}}^{\text{quarks}}$ are then written as:
\begin{eqnarray}
\mathcal{L}_{CC} &=& -\frac{g}{2\sqrt{2}}\sum_{q}\bar{\Psi}_q \gamma^\mu (1-\gamma_5)\tau^+ \Psi_q W^+_\mu + h.c. \\
\mathcal{L}_{NC} &=& -\frac{g}{2\cos\theta_W}\sum_q \bar{\Psi}_q\gamma^\mu (g_V^q-g_A^q\gamma_5)\Psi_q
\end{eqnarray}
where $g_V^q$ and $g_A^q$ are given by Eq.~(\ref{gve_tau_theta}), and the explicit values of $g_V^q$ 
and $g_A^q$ for each quark are shown in the Table-\ref{WS:gvga}.
\begin{table}
\begin{center}
 \begin{tabular}{|c|c|c|c|c|}\hline
States $\to$ &$\nu_l$&$l$&u,c,t&$d^\prime,s^\prime,b^\prime$\\
Couplings$\downarrow$&&&&\\\hline
2$g_V$&$1$&$-1+4\sin^2\theta_W$&$1-\frac{8}{3}\sin^2\theta_W$&$-1+\frac{4}{3}\sin^2\theta_W$\\\hline
2$g_A$&$1$&$-1$&$1$&$-1$\\\hline
 \end{tabular}
 \caption{Couplings of the leptons and quarks to $Z_{\mu}$ field.}
 \label{WS:gvga}
 \end{center}
\end{table}

The weak interaction Lagrangian for the nucleons is evaluated in a straightforward manner assuming the quark structure of the  
protons and neutrons as composed of antisymmetrized $uud$ and $udd$ quarks and using the isospin structure of CC and NC 
currents. Since the weak CC currents are charge raising and charge lowering components, they can be written in a 
straightforward way for the nucleons as: 
{
\begin{eqnarray}
\mathcal{L}_{CC}^N = -\frac{g}{2\sqrt{2}} J^\mu_{CC} W_\mu^{+}  + h.c., \qquad \qquad \text{with} \qquad
J^\mu_{CC} =  \bar{\Psi}_N\gamma^\mu (1-\gamma_5)\tau^+ \Psi_N,
\end{eqnarray}
where $\Psi_{N} = \begin{pmatrix}
                   u\\
                   d^\prime
                  \end{pmatrix}
$ is the quark isodoublet after implementing GIM mechanism, with $d^\prime = V_{ud} ~d + V_{us} ~s$. 

In the case of NC
\begin{eqnarray}
\mathcal{L}_{NC}^N = -\frac{g}{2\cos\theta_W} J^\mu_{NC} Z_\mu\;, \qquad \qquad
\text{where}~~~~~~~~J^\mu_{NC} = V^\mu_{NC}- A^\mu_{NC}.
\end{eqnarray}
The expressions for $V_{NC}^{\mu}$ and $A_{NC}^{\mu}$ can be written in a straightforward manner using the isospin structure 
of these currents given in Eq.~(\ref{gve_tau_theta}) as~\cite{Alberico:2001sd}:
\begin{eqnarray}
V^{\mu}_{NC} &=& V^{\mu 3} - \left( 2\sin^2\theta_W J^\mu_{EM} + \frac{1}{2}V^{\mu}_S \right); ~~~~~~~A^{\mu}_{NC} = A^{\mu 3} + \frac{1}{2} A_S^\mu,\\
\qquad \text{with} \qquad V^{\mu 3}&=&  \bar{\Psi}_N \gamma^\mu \frac{\tau_3}{2} \Psi_N, \qquad 
\qquad ~~~~~~ A^{\mu 3} =  \bar{\Psi}_N \gamma^\mu \gamma_5 
\frac{\tau_3}{2} \Psi_N
\qquad 
\qquad ~~~~~~ J_{EM}^{\mu} = e_N \bar{\Psi}_N \gamma^\mu \Psi_N~~~~~~~~~~~~ \nonumber \\
\qquad V^{\mu}_S&=&  \bar{s} \gamma^\mu  s \qquad \qquad A^{\mu}_S=  \bar{s} \gamma^\mu \gamma_5  s, 
\end{eqnarray}
 where $\Psi_N = \begin{pmatrix}
                      u\\
                      d
                     \end{pmatrix}
$ is the isodoublet of nonstrange quarks with charge $e_N$, and $s$ is an isoscalar, which represents the strange quark contribution.}

\begin{table}
\begin{center}
 \begin{tabular}{|c|c|c|c|}\hline
 S. No. & Quantity & SM prediction & Experimental value \\ \hline
 1 & Mass of the $W^{\pm}$ boson & 80.361 $\pm 0.006$ GeV & $80.376 \pm 0.033$ GeV \\
 2& Mass of the $Z^{0}$ boson& 91.1882 $\pm 0.002$ GeV& $91.1876 \pm 0.0021$ GeV\\
 3& $W^{\pm}$ total decay width, $\Gamma_{W}$& 2.090 $\pm 0.001$ GeV&$2.046 \pm 0.049$ GeV\\
 4& $Z^{0}$ total decay width, $\Gamma_{Z}$& 2.4942 $\pm 0.0009$ GeV&$2.4955 \pm 0.0023$ GeV\\
 5& Mass of Higgs boson& $125.30 \pm 0.13$ GeV&$125.30 \pm 0.13$ GeV\\
 6& Vector coupling $g_{V}^{\nu_{e}}$& $-0.0398 \pm 0.0001$& $-0.040 \pm 0.015$\\
 7& Axial-vector coupling $g_{A}^{\nu_{e}}$& $-0.5064$&$-0.507 \pm 0.014$\\
 8& Weak charge of electron & $-0.0476 \pm 0.0002$&$-0.0403 \pm 0.0053$\\
 9 & $\sin^{2} \theta_{W}$ & 0.23121 $\pm 0.00004$ & $0.2299 \pm 0.0043$\\ \hline
 \end{tabular}
 \end{center}
  \caption{Predictions of the SM and the experimentally observed values~\cite{ParticleDataGroup:2020ssz}.}
 \label{tab:SM2}
\end{table}
The triumphs of the SM are many, like the predictions of the existence of $W$, $Z$ bosons, and Higgs boson as well 
as the prediction of NC in the neutrino and electron sectors. The model also predicts various relations between the weak 
decays of the charm, bottom and top quarks. The agreement between the SM values and the experimentally observed results for 
many observables are unprecedented,  
and in Table~\ref{tab:SM2}, some of the experimentally observed values and their SM predictions that will be used later in 
this article, have been tabulated. 

\subsubsection{Higher order effects in electroweak interactions}
The theoretical calculations of the various electroweak observables from which the parameters shown in Table~\ref{tab:SM2} are extracted, are done including the corrections due to the higher order loop effects beyond the lowest order nonvanishing contributions in the Born approximation, 
using the standard model of the electroweak interactions in order to compare them with the very high precision experimental results for these parameters obtained in various experiments~\cite{ParticleDataGroup:2020ssz}. 
The physical processes studied in these experiments involve the interactions of leptons and quarks, and  therefore, are also subject to the corrections due to the higher order loop effects in QCD, in addition to the higher order loop effects in the electroweak interactions. 
The corrections due to the higher order effects generally fall in the following categories and have been discussed extensively in the literature~\cite{ParticleDataGroup:2020ssz, Erler:2019hds, Dubovyk:2019szj, Freitas:2014hra, Pich:2012sx}:
\begin{itemize}
 \item [(i)] {\bf QED and QCD corrections} \\
 There are two types of these corrections. The first type of corrections arise due to the vacuum polarization effects of the QED and QCD vacuum. In the case of QED, the one loop, two loops, and higher loop corrections in the photon propagator due to the fermion-antifermion pairs in the intermediate state lead to the renormalization of the electromagnetic coupling $\alpha$, and make it energy dependent. 
 Similarly, in the case of QCD, the renormalization of the gluon propagator due to the higher order loop effects arising due to the quark-antiquark pairs and higher order self interactions of the gluons in the intermediate states make the strong coupling $\alpha_s$ energy dependent 
 making it smaller at higher energies leading to the asymptotic freedom of QCD. 
 Since most of the electroweak measurements are made at higher energies corresponding to the weak gauge boson~($W/Z$) mass, except the low energy process of muon decay, a value of $\alpha_{s} (M_Z) = 0.1185 \pm 0.0016$ is used in fitting the electroweak observables. 
 Using this value of $\alpha_s (M_Z)$, the QED and QCD vacuum polarization effects change the value of electromagnetic coupling $\alpha$ from its value of $\alpha^{-1} (m_e) = 133.472 \pm 0.007$ to $\alpha^{-1} (M_Z) = 127.952 \pm 0.009$. 
 These changes due to the vacuum polarization effects in the value of $\alpha (M_Z)$ and $\alpha_s (M_Z)$ affect the extraction of various parameters~(shown in Table~\ref{tab:SM2}) from the electroweak observables like the weak decay widths of heavy leptons~($\mu$ and $\tau$), and gauge bosons $W$ and $Z$, as well as the asymmetries observed in the electron scattering experiments. 
 
 The second type of the higher order corrections arise due to the photons~(gluons) appearing in the one, two, and higher loop diagrams. These corrections are generally gauge invariant and finite but energy dependent. 
 Therefore, depending upon the individual physical processes and the energy involved in these processes, these effects are, in general, different for every process and need to be calculated accordingly.
 
 \item [(ii)] {\bf Electroweak corrections}\\
 The corrections due to the higher order loop effects calculated in the standard model of the electroweak interactions arise due to the vacuum polarization effects of $W$ and $Z$ propagator due to the virtual $\gamma Z$, $ZZ$, and $WW$ pairs as well as the fermion-antifermion pairs~($q\bar{q}$) in the one loop, two loop, and higher loop diagrams. 
 In addition to the vacuum polarization effects on the gauge boson propagators, the contribution due to the vertex corrections, box diagrams involving virtual $W$ and $Z$ bosons and the contribution due to the virtual $q\bar{q}$ loops in the intermediate state are important. 
 These corrections at the one loop level have been calculated for most of the weak processes of the weak decays and parity violating asymmetries. 
 The two loop and higher order loop contributions have also been calculated for some weak processes. 
 
 \item [(iii)] {\bf Mixed QCD-electroweak corrections}\\
 The higher order loop corrections due to the mixed QCD-electroweak interactions, where $q\bar{q}$ pairs are not involved, are calculated upto order $\alpha\alpha_s$ and $\alpha\alpha_s^2$. 
 The corrections due to the higher order loop diagrams involving $q\bar{q}$ pairs of the order of $\alpha\alpha_s m_t^2$, $\alpha\alpha_s^2 m_t^2$, $\alpha\alpha_s^3 m_t^2$, $\alpha^2 \alpha_s m_t^4$, etc., where $m_t$ is the top quark mass, are calculated in some processes. 
 \end{itemize}
 
  The combined corrections because of the higher loop diagrams arising due to the QED, QCD, electroweak, and mixed QCD-electroweak interactions outlined above have been calculated for all the electroweak processes and their influence on extracting various parameters like $G_F$, $M_W$, $M_Z$, $\sin^2 \theta_W$, etc. have been discussed in the literature. 
 For some recent reviews, see Refs.~\cite{ParticleDataGroup:2020ssz, Erler:2019hds, Dubovyk:2019szj, Freitas:2014hra, Pich:2012sx}.
 
 In the following sections, we illustrate some simple examples of the neutrino scattering from the electrons and quarks in the nonvanishing lowest order perturbation theory using the standard model.

\subsubsection{$\nu_{l}-e$ and $\bar{\nu}_{l}-e$ scattering}
First let us consider the process
\begin{equation}
 \nu_{e} (\vec{k}, E_{\nu_{e}}) + e^{-} (\vec{p}, E_{e}) \longrightarrow \nu_{e} (\vec{k}^{\prime}, E_{\nu_{e}}^{\prime}) + 
 e^{-} (\vec{p}^{\prime},  E_{e}^{\prime}) 
\end{equation}
which is mediated by the neutral~($Z^{0}$) as well as the charged~($W^{+}$) current interactions and using the Lagrangian given in 
Eqs.~(\ref{weak_cc_lag}) and (\ref{weak_nc_lag}) one may write the invariant matrix element for CC interaction as:
    \begin{eqnarray}
 {\cal M}^{CC}&=& \frac{G_F}{\sqrt2}\left[ {\bar u}(\vec{p}^{\prime})\gamma_\mu(1- \gamma_5)u(\vec{k})
 \right]\cdot\left[{\bar u}(\vec{k}^{\prime}) \gamma^{\mu}(1- \gamma_5)u(\vec{p})\right],
 \end{eqnarray}
 and for NC interaction as:
    \begin{eqnarray}
{\cal M}_{NC}&=& \frac{G_F}{\sqrt2}\left[ \bar{u}(\vec{k}^{\prime})\gamma_\mu (1-\gamma_5)u(\vec{k})\right]
\cdot \left[ \bar{u}(\vec{p}^{\prime})\gamma^\mu (g_V^e-g_A^e \gamma_5)u(\vec{p})\right],
\end{eqnarray}
where the value of $g_{V}^{e}$ and $g_{A}^{e}$ are given in Table-\ref{WS:gvga}.

Using the Fierz transformation, the total contribution for CC and NC induced reactions may be written as:
 \begin{eqnarray}
 {\cal M}_{CC}+{\cal M}_{NC}= \frac{G_F}{\sqrt2}\left[ \bar{u}(\vec{k}^{\prime})\gamma_\mu (1-\gamma_5)u(\vec{k})\right]\cdot
 \left[ \bar{u}(\vec{p}^{\prime})\gamma^\mu (g^{\prime}_{V}-g^{\prime}_{A} \gamma_5)u(\vec{p})\right],
 \end{eqnarray}
where $g^{\prime}_{V} = g_{V}^{e} + 1$, $g^{\prime}_{A} = g_{A}^{e} +1$.

The matrix element square $|{\cal M}|^{2}$ averaged over the initial spin state and summed over the final spin state is given by
\begin{equation}
   \overline{\sum_i}\sum_f |{\cal M}|^2 = 16~G^2_F \left[\alpha~ (k'\cdot p')(k\cdot p) + \beta~ (k'\cdot p)(k\cdot p') 
   - \gamma~m_{e}^{2}(k\cdot k') \right],
\end{equation}
where the values of $\alpha$, $\beta$ and $\gamma$ are given in Table~\ref{process}.

 \begin{table}
 \begin{center}
\begin{tabular}{ c c c c }
\hline\hline
 Process & $\alpha$ & $\beta$ & $\gamma$\\ 
 \hline
 $\nu_{l^{\prime}}e^{-}\rightarrow\nu_{l^{\prime}}e^{-}$ & $(g_{V}^e + g_{A}^e)^{2}$ & $(g_{V}^e - g_{A}^e)^{2}$ & 
 $ \left(g_{A}^e\right)^{2}- \left(g_{V}^e\right)^{2}$ \\
$\overline{\nu}_{l^{\prime}}e^{-}\rightarrow \overline{\nu}_{l^{\prime}}e^{-}$&  $(g_{V}^e - g_{A}^e)^{2}$ &  
$(g_{V}^e + g_{A}^e)^{2}$ & 
$ \left(g_{A}^e\right)^{2}- \left(g_{V}^e\right)^{2}$ \\ 
 $\nu_{e}e^{-}\rightarrow\nu_{e}e^{-}$ & $(g_{V}^{\prime} + g_{A}^{\prime})^{2}$ & $(g_{V}^{\prime} - 
 g_{A}^{\prime})^{2}$ & $g_{A}^{\prime^2} - g_{V}^{\prime^2}$\\
 $\overline{\nu}_{e}e^{-}\rightarrow \overline{\nu}_{e}e^{-}$& $(g_{V}^{\prime} - g_{A}^{\prime})^{2}$ & 
 $(g_{V}^{\prime} + g_{A}^{\prime})^{2}$ & $ g_{A}^{\prime^2} - g_{V}^{\prime^2}$\\ \hline\hline
\end{tabular}
 \end{center}
 \caption{Values of $\alpha$, $\beta$ and $\gamma$ for $\nu_{l^{\prime}}e^{-}$, $\overline{\nu}_{l^{\prime}}e^{-}$, $\nu_{e}e^{-}$ 
 and $\overline{\nu}_{e}e^{-}$ scattering, where $l^{\prime}=\mu,\tau$.}\label{process}
\end{table}
The expression for the differential cross section in 
CM frame is obtained as~\cite{Athar:2020kqn}:
\begin{eqnarray}\label{nue_domega:eq}
\left. \frac{d\sigma}{d\Omega} \right|_{CM} &&= \frac{1}{4 \pi^{2} s}G_{F}^{2}\Big[\alpha
\left(\frac{s-m_{e}^2}{2}\right)^2 +\beta \left(\frac{u- m_{e}^{2}}{2}\right)^2+\gamma\frac{m_e^2}{2}t \Big],~~~~~
\end{eqnarray}
where the values of $\alpha$, $\beta$ and $\gamma$ are given in Table~\ref{process}, and $ s,t,u$ are the Mandelstam variables. 
The $\nu_{l^{\prime}} 
e^{-}$ and $\bar{\nu}_{l^{\prime}} e^{-}$~($l^{\prime} = \mu,\tau$) scattering take place via. NC only, and the corresponding 
values of $\alpha$, $\beta$ and $\gamma$  are tabulated in Table~\ref{process}.  

In the massless limit of electron, the differential and total scattering cross sections for CC induced $\nu_{e} e^{-}$ 
scattering process are obtained as
\begin{eqnarray}
 \left.\frac{d\sigma}{d\Omega} \right|_{CC}(\nu_{e} e^{-}) = \frac{G_{F}^{2}s}{4\pi^{2}} \qquad \text{and }\qquad \left.\sigma
 \right|_{CC}(\nu_{e} e^{-}) = \frac{G_{F}^{2}s}{3\pi}.
\end{eqnarray}
Similarly, for $\bar{\nu}_{e} e^{-}$ scattering, the differential and total scattering cross sections for CC 
induced process are obtained as
\begin{eqnarray}\label{nue_domega}
 \left.\frac{d\sigma}{d\Omega} \right|_{CC}(\bar{\nu}_{e} e^{-}) = \frac{G_{F}^{2}s}{16\pi^{2}} (1-\cos\theta_{CM})^{2} 
 \qquad \text{and }
 \qquad \left.\sigma\right|_{CC}(\bar{\nu}_{e} e^{-}) = \frac{G_{F}^{2}s}{3\pi},
\end{eqnarray}
where $\theta_{CM}$ is the angle between the incoming $\bar{\nu}_{e}$ and the outgoing electron.

\subsubsection{(Anti)neutrino-quark scattering}
For (anti)neutrino-quark scattering like the processes
\begin{equation}
\nu_{l} + d \longrightarrow l^{-} + u, \qquad \qquad \bar{\nu}_{l} + u \longrightarrow l^{+} + d,
\end{equation}
which can take place only through CC channel, the general expression for the differential scattering cross section is 
similarly obtained with the values of $g_{V}$ and $g_{A}$ for the quarks defined in Table~\ref{WS:gvga}.

In the massless lepton limit, the differential scattering cross sections are given by
\begin{eqnarray}
 \frac{d\sigma}{d\Omega} (\nu_{l} + d \rightarrow l^{-} + u)= \frac{G_{F}^{2} s}{4\pi^{2}}, \qquad \text{and} 
 \qquad \frac{d\sigma}{d\Omega} (\bar{\nu}_{l} + u \rightarrow l^{+} + d)= \frac{G_{F}^{2}s}{16\pi^{2}} 
 (1+\cos\theta_{CM})^{2}.
\end{eqnarray}
For $\bar{\nu}_{l} \bar{d} \rightarrow l^{+} \bar{u}$ and $\nu_{l} \bar{u} \rightarrow l^{-} \bar{d}$ processes, 
the differential scattering cross sections are given by
\begin{eqnarray}
 \frac{d\sigma}{d\Omega} (\bar{\nu}_{l} + \bar{d} \rightarrow l^{+} + \bar{u})= \frac{G_{F}^{2} s}{4\pi^{2}}, 
 \qquad \text{and} \qquad \frac{d\sigma}{d\Omega} ({\nu}_{l} + \bar{u} \rightarrow l^{-} + \bar{d})= 
 \frac{G_{F}^{2}s}{16\pi^{2}} (1+\cos\theta_{CM})^{2}.
\end{eqnarray}
 
\subsection{Resonance scattering of neutrinos: Glashow resonance}
In the early days of the development of the theory of weak processes mediated by the intermediate vector bosons~($W$), 
Glashow~\cite{Glashow:1960zz} considered the reaction
\begin{equation}\label{glashow-reac}
 \bar\nu + e^- \rightarrow \bar\nu + \mu^-
\end{equation}
and speculated about the resonance scattering of $\bar\nu$ through the process $\bar\nu + e^- \rightarrow W^- \rightarrow 
\bar\nu + \mu^-$ which would radically enhance the cross section for the reaction shown in Eq.~(\ref{glashow-reac}). This 
would happen for an antineutrino~($\bar\nu$) scattering from the electron at rest. This resonance is commonly known as the Glashow 
resonance~\cite{Glashow:1960zz}. In the SM of the electroweak interactions, the weak processes are mediated by the 
charged $W^{\pm}$ and neutral $Z^0$ bosons, respectively. Consequently, the resonance scattering is predicted to occur at an 
antineutrino energy $E_{{\bar\nu}_e} \approx \frac{M_W^2}{2m_e}=6.3$~PeV,  where $M_W$ is the mass of the vector boson. 
Such antineutrino energies are too high to be 
produced in the terrestrial accelerators but can be produced in the case of astrophysical sources of neutrinos. The 
astrophysical neutrinos are produced as decay products of the unstable mesons and baryons created in various cosmic reactions 
involving very high energy $pp$ and $\gamma p$ collisions in space. The flavor composition of the very high energy 
astrophysical neutrinos and antineutrinos and their energy distribution has been recently studied by many 
authors like Barger et al.~\cite{Barger:2014iua}, Biehl et al.~\cite{Biehl:2016psj}, Loewy et al.~\cite{Loewy:2014zva}, 
Bhattacharya et al.~\cite{Bhattacharya:2011qu}, in the context of the observation of these (anti)neutrinos in the PeV energy 
region recently by the IceCube Collaboration~\cite{IceCube:2021rpz}.

It has been shown that the resonant cross section for ${\bar\nu}_e + e^- \rightarrow W^-$ production assuming a Breit-Wigner 
form for the $W^-$ resonance is given by~\cite{Barger:2014iua}:
\begin{equation}
 \sigma_{\text{res}}(s)= \frac{\left(s\; \Gamma_W^2\right)}{\left(s - M_W^2 \right)^2 + \left(M_W \Gamma_W\right)^2} 
 \sigma_{\text{res}}^{\text{peak}}
\end{equation}
where $s=\left(k+p\right)^2$ and $\Gamma_W$ is the $W$'s full width at half maximum (2.1~GeV).

\begin{figure} 
\centering
 \includegraphics[height=5.5 cm, width=0.55\textwidth]{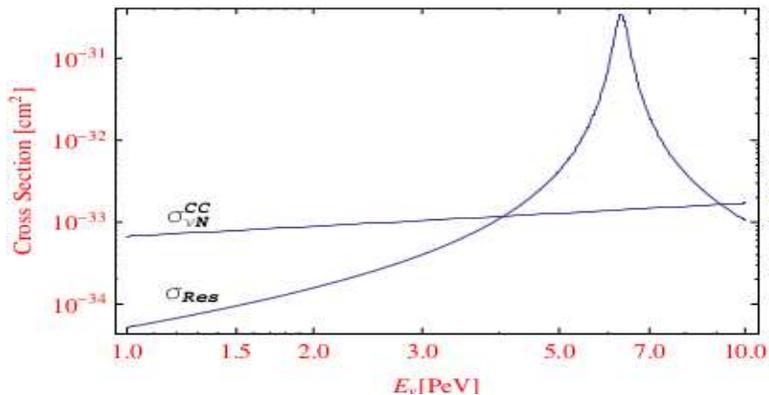}
 \caption{Cross sections for the resonant process ${\bar\nu}_e + e^- \rightarrow W^- \rightarrow {\text{hadrons}}$, and 
the nonresonant~(NR) process $\nu_e + N \rightarrow e^- + {\text{hadrons}}$ in the 1 - 10 PeV region. The figure has been 
taken from Ref.~\cite{Barger:2014iua}.}\label{Fig-Glashow}
\end{figure}

$ \sigma_{\text{res}}^{\text{peak}}$ is the cross section of $W$-resonance given by
\begin{equation}
  \sigma_{\text{res}}^{\text{peak}}=\frac{24\pi}{M_W^2} B\left({W^- \rightarrow {\bar\nu}_e + e^-}\right)=5.02 \times 
  10^{-31}{\text{cm}}^2,
\end{equation}
where $ B\left({W^- \rightarrow {\bar\nu}_e + e^-}\right)$ is the branching ratio for the $W^- \rightarrow {\bar\nu}_e +
e^-$ mode. Since the process ${\bar\nu}_e + e^- \rightarrow W^- \rightarrow $ hadrons is more frequent because the branching 
ratio for the $ W^- \rightarrow {\text{hadrons}}$ is 67\%, it is the more likely mode for the detection of $W^-$, the Glashow 
resonance. Moreover, the hadron production through the ${\bar\nu}_e + e^- \rightarrow W^- \rightarrow {\text{hadrons}}$ is 
considerably larger than the hadron production in the neutrino-nucleon scattering through the $\nu_e + N \rightarrow e^- + 
{\text{hadrons}}$ process in the energy region of Glashow resonance as shown in Fig.~\ref{Fig-Glashow}.

The resonance production peak is affected by the Doppler effect of the moving electrons in the case of ${\bar\nu}_e$ 
scattering from atomic electrons in atoms leading to the  broadening of the peak shown in Fig.~\ref{Fig-Glashow} but the effect 
is shown to be small~\cite{Loewy:2014zva}. While the cross section at the peak energy of $E_{{\bar\nu}_e}=6.3$~PeV for the 
resonance production of hadrons is more than 300 times larger than CC neutrino nucleon cross section, 
the production of $e^-$ events, i.e. the rate of resonance events also depends upon 
the $\bar{\nu}_e$ content in the neutrino flux arriving at the detector which is affected by the flavor oscillations of the 
antineutrinos. In general the ${\bar\nu}_e$ content in the neutrino flux is smaller than the ${\bar\nu}_\mu$ content, but it 
is enhanced by the ${\bar\nu}_\mu \rightarrow {\bar\nu}_e$ oscillations during their propagation from the source to the 
detector. An experimental observation of resonant $e^-$ events by the $W^-$ resonance production and a theoretical study of 
the flavor decomposition of the antineutrino flux generated in the various neutrino producing reactions from the high energy 
$pp$ and $\gamma p$ reactions in space including the effect of flavor oscillations of neutrinos, will provide important 
information about the source and production mechanism of very high energy neutrinos in the PeV energy region.

A recent report of the IceCube neutrino observatory~\cite{IceCube:2021rpz} has claimed to observe one event of a cascade 
of high energy particle shower with a visible energy of $6.05 \pm 0.72$~PeV detected from the Cherenkov radiation of the 
shower particles, which is claimed to be due to the Glashow resonance. After correcting the visible energy for shower
particles which do not radiate, the neutrino energy is 
inferred to be 6.3~PeV consistent with the prediction of the SM. The IceCube Generation-2 
experiment~\cite{IceCube-Gen2:2020qha} planned for future would improve the statistics and enable to measure the high energy 
antineutrino flux which would give information about the different mechanism for producing high energy astrophysical neutrinos 
in the PeV region and enrich our knowledge of the neutrino astronomy. 

\section{Neutrino scattering from nucleons}\label{nu_interaction}
Neutrino experiments are done in the wide range of energy starting from a few MeV to TeV region  using solar, reactor, 
atmospheric, and accelerator (anti)neutrinos. The present goal of the experimenters is to measure with better precision the 
various neutrino oscillation parameters, like the mixing angles, the mass-squared-difference of the neutrino mass 
eigenstates, CP violating phase $\delta$ in the lepton sector as well as to determine the mass hierarchy of the neutrino 
mass eigenstates. These parameters are sensitive to the neutrinos of different energy range which are obtained from 
accelerator, atmospheric, reactor and solar neutrino and/or (anti)neutrinos sources as mentioned in 
Section-\ref{nu:oscillation} and summarized in Table-\ref{intro:table1}. Almost all the current generation (anti)neutrino 
experiments are using moderate to heavy nuclear targets. These experiments are measuring (anti)neutrino events which are 
convolution of energy dependent (anti)neutrino flux and the energy dependent neutrino-nucleus cross section where NME 
play very important role. In the precision era of neutrino physics, to achieve an accuracy of a few 
percent~(2--3\%) in the systematics, a good understanding of the neutrino-nucleon and neutrino-nucleus cross sections is 
highly desirable both experimentally as well as theoretically, which has been highlighted by various review 
articles~\cite{SajjadAthar:2021prg, Athar:2021xfr, NuSTEC:2017hzk, Alvarez-Ruso:2014bla}. Apart from being significant to 
the determination of neutrino oscillation parameters, the neutrino-nucleon and neutrino-nucleus cross sections are important 
in their own right as they provide information about the axial-vector response of the nucleons bound inside the nucleus, 
which is not accessible via. photon or electron induced reactions, and recently it has been 
suggested~\cite{Alvarez-Ruso:2022ctb} to perform neutrino experiments using hydrogen and deuterium targets. In this 
section, we focus on the neutrino-nucleon reactions and take up the 
neutrino-nucleus reactions in Section~\ref{nu:nuclei}.
 
The (anti)neutrino interaction with a nucleon target starts with the elastic and quasielastic~(QE) scattering processes.  
With the increase in available neutrino energy, the inelastic~(IE) reactions in which new particles like 1$\pi$, multiple 
pions, 1$\eta$, 1$K$, $Y\pi$, and 
$YK$~($Y = \Lambda, \Sigma, \Xi$), etc. are created as well as the deep inelastic scattering~(DIS) become possible which are 
diagrammatically shown in Fig.~\ref{feynman-intro} and described below~\cite{Athar:2020kqn}:
\begin{itemize}
 \item {\bf Elastic and quasielastic scattering:} Neutrinos and antineutrinos of all flavors interact with a nucleon 
 through CC as well as NC interactions, 
\begin{eqnarray}\label{Ch-4:process1_nu}
 {\nu}_\ell/\bar{\nu}_\ell  + N  &\longrightarrow& \ell^-/\ell^{+} + N^{\prime}, \quad \quad \qquad ~~~~~~ \;\; 
 ({\rm CC})\\
 \label{Ch-4:process5_nu1}
 \text{and}~~\qquad~{\nu}_\ell/\bar{\nu}_\ell  + N  &\longrightarrow& {\nu}_\ell/\bar{\nu}_\ell + N,   \qquad \qquad 
 \qquad ~~~~ ({\rm NC})
\end{eqnarray}
in the $\Delta S=0$ sector, and 
\begin{eqnarray}\label{Ch-4:process1_nuy}
 \bar{\nu}_\ell  + N  &\longrightarrow& \ell^{+} + Y, \quad \quad \;\; \qquad \qquad \qquad \qquad\qquad ~~~~
 ~~({\rm CC}),
\end{eqnarray}
in the $\Delta S=1$ sector; where $N,N^{\prime}=n,p$ and $Y=\Lambda, \Sigma^{-,0}$.
\begin{figure}  
\begin{center}
\includegraphics[height=4.0 cm, width=4.2 cm]{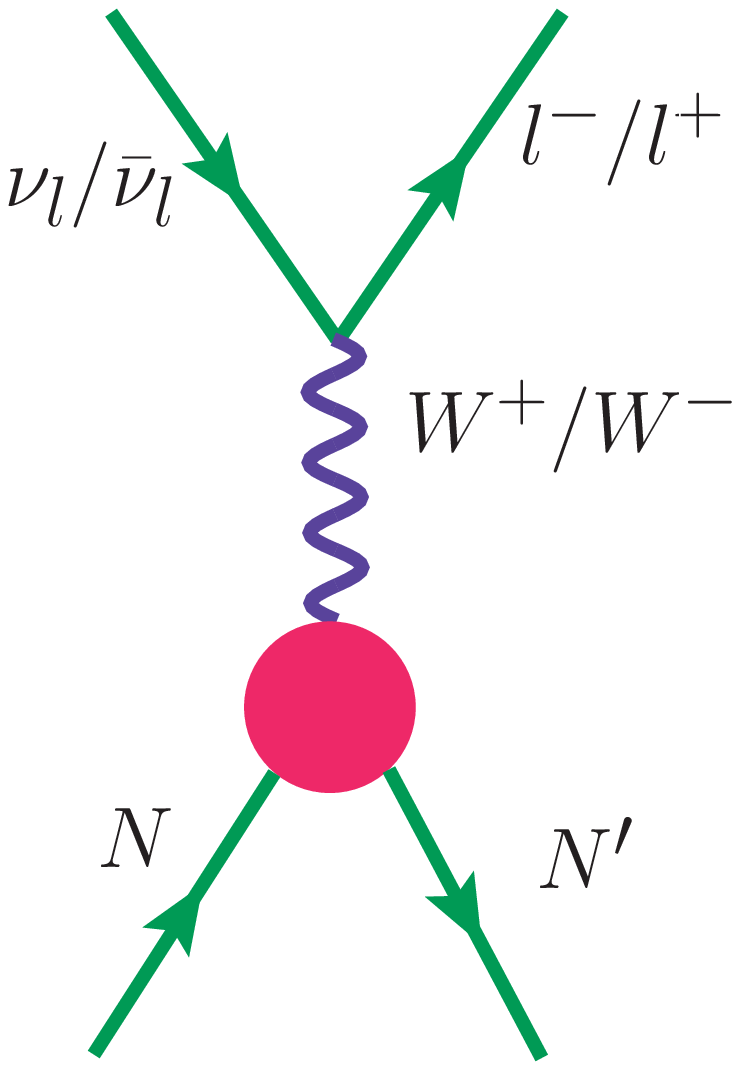}
\includegraphics[height=4.0 cm, width=4.2 cm]{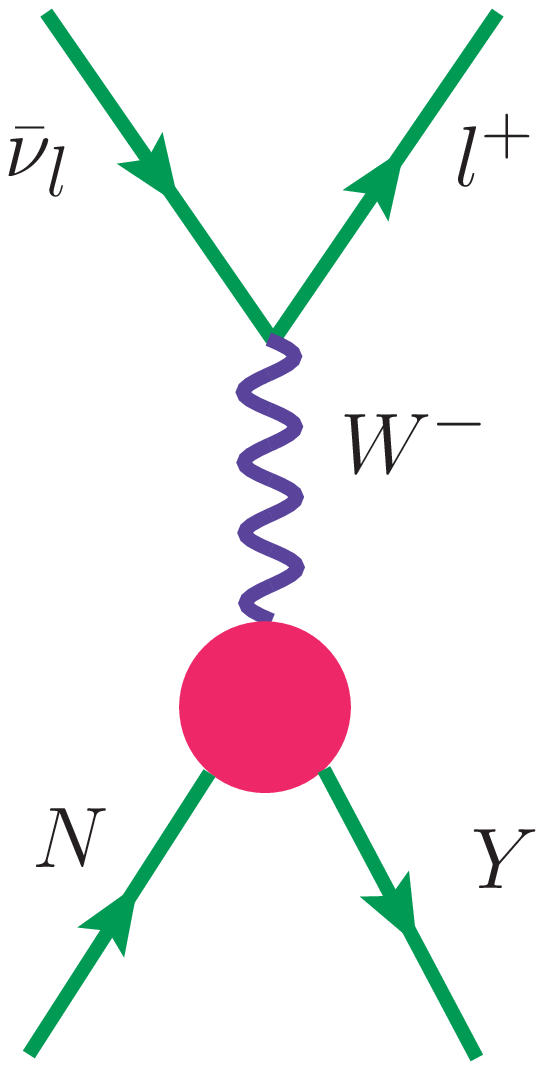}
\includegraphics[height=4.0 cm, width=4.2 cm]{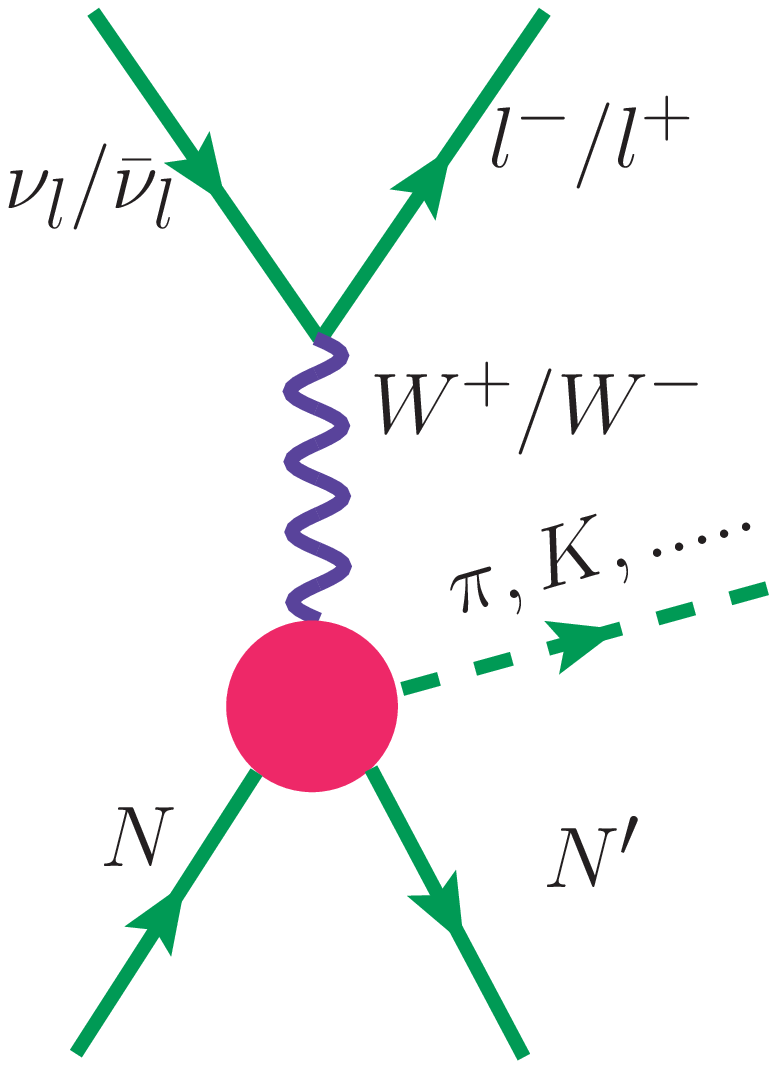}
\includegraphics[height=4.0 cm, width=4.2 cm]{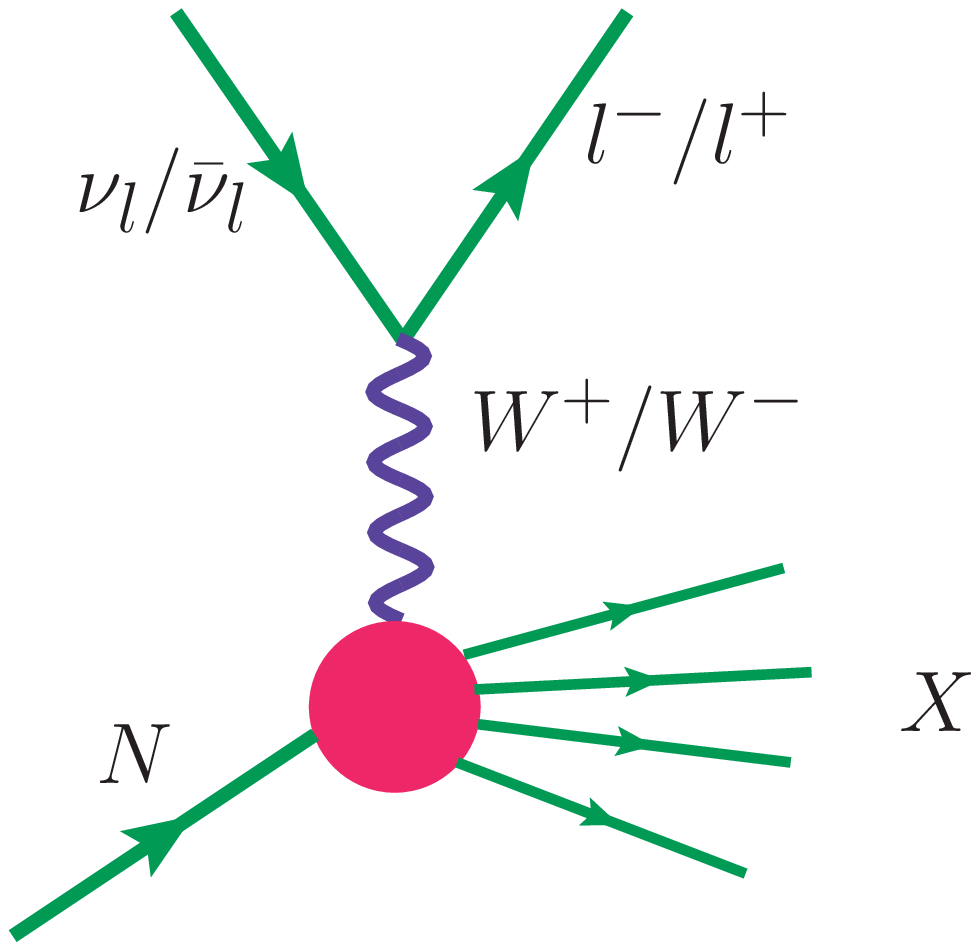}
\end{center}
\caption{(Left to right) Feynman diagram depicting the $\Delta S=0$ QE process, $\Delta S=1$ QE process, 
IE process, and the DIS process in CC induced reactions. Similar processes for 
${\nu}_{l}(\bar{\nu}_{l}) \rightarrow {\nu}_{l}(\bar{\nu}_{l})$ reactions are also induced by NC through the $Z$ exchange.}
\label{feynman-intro}
\end{figure}

\begin{figure}[h]
\begin{center}
\includegraphics[height=7.0 cm, width=15 cm]{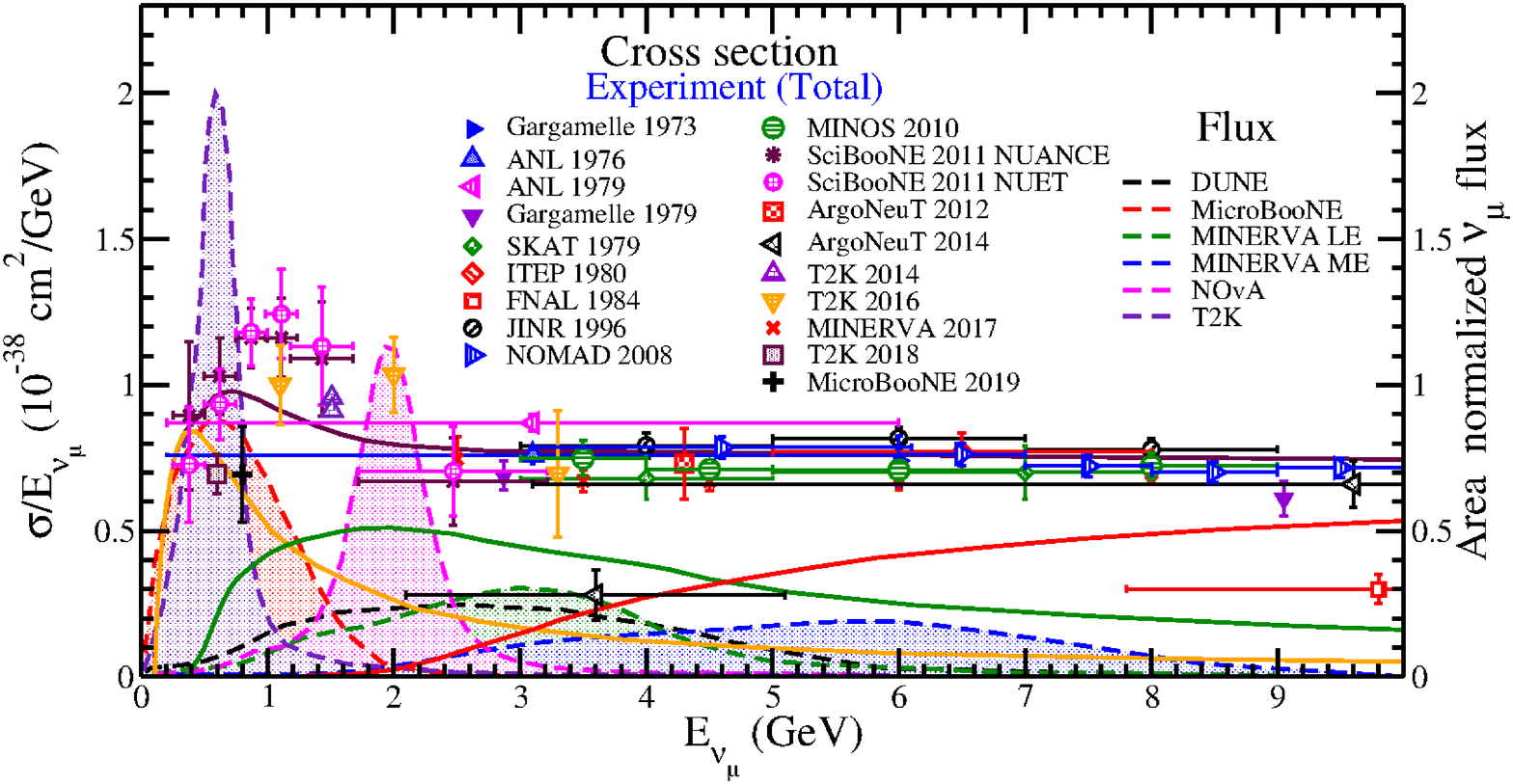}
\includegraphics[height=7.0 cm, width=15 cm]{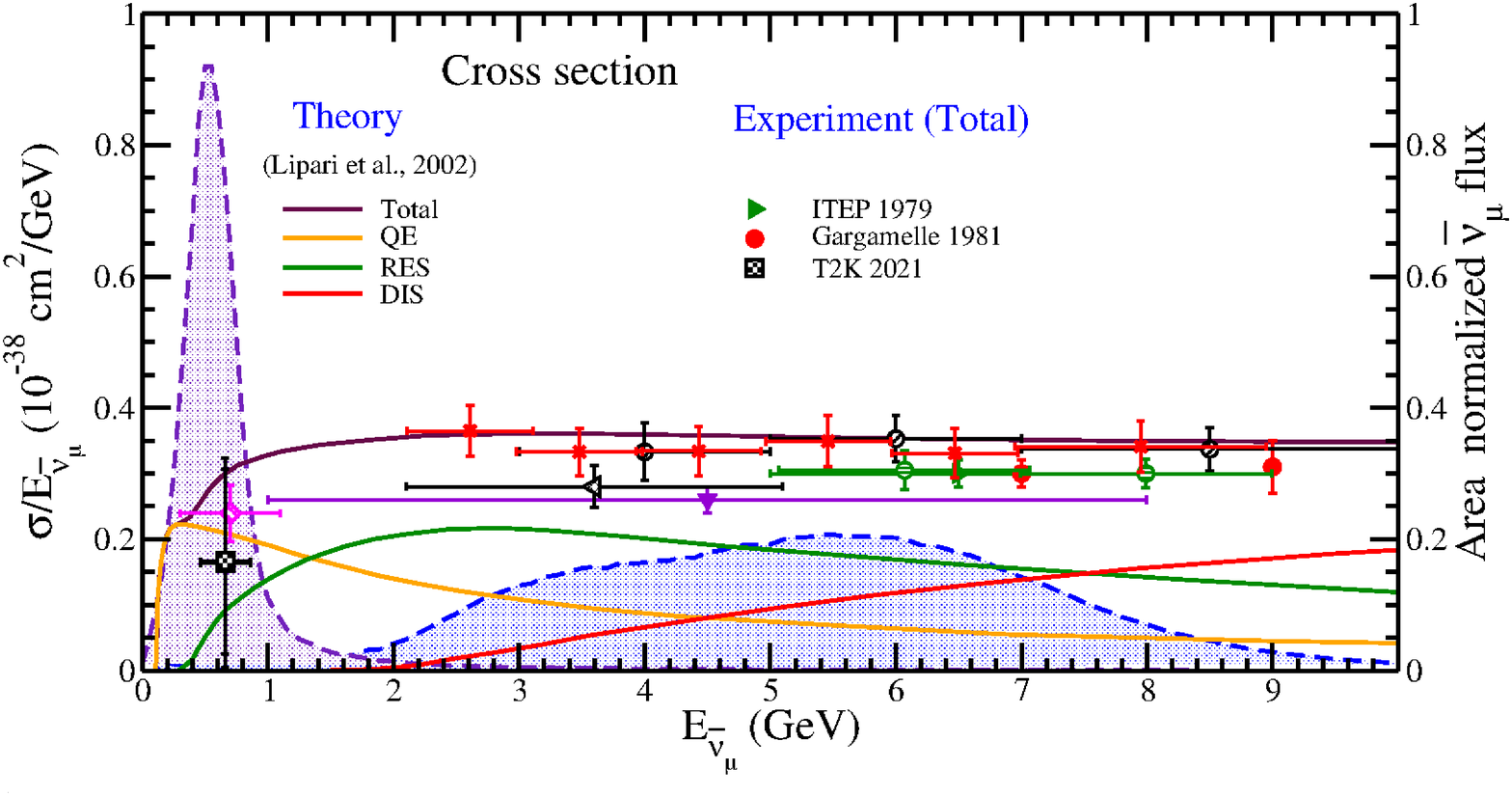}
\end{center}
\caption{$\frac{\sigma}{E_{\nu_\mu}}$ vs $E_{\nu_\mu}$~(top panel) and $\frac{\sigma}{E_{\bar{\nu}_\mu}}$ vs 
$E_{\bar{\nu}_\mu}$~(bottom panel) for an isoscalar target. The data are the experimental points for the inclusive cross 
section~($\sigma$) in various nuclear targets.  The theoretical result for $\frac{\sigma}{E_{\nu_{\mu} 
(\bar{\nu}_{\mu})}}$~(solid line) has the contribution from total cross section~(maroon line), QE 
scattering~(orange line), resonance production~(green line), and DIS~(red line) provided by the NUANCE 
generator (Casper, 2002) and compiled by Lipari et al. The various neutrino fluxes which are being used in the T2K, MINERvA 
low energy~($<E_{\nu_\mu}>= 3$~GeV), medium energy~($<E_{\nu_\mu}>= 6$~GeV), NOvA, MicroBooNE experiments along with the proposed DUNE 
experiment at the Fermilab are shown to highlight the importance of the understanding of the cross section in the few GeV 
energy region. These neutrino fluxes are normalized to unit area.}
\label{flux}
\end{figure}
It may be noticed that in the strangeness sector, single hyperon~(like $\Lambda$,$\Sigma^{-,0}$, etc.) produced in the final 
states, are possible only in the antineutrino induced reactions while it is prohibited in neutrino channel due to the $\Delta
S = \Delta Q$ and FCNC rules.	

\begin{table}[h]
\begin{center}
\begin{tabular}{c c c }
	\hline \hline
	~~~S.\ No.\               & CC induced $\nu(\bar{\nu})$ reactions & NC induced $\nu(\bar{\nu})$ reactions~~~\\ \hline
1. & $\nu_{\ell} (\bar{\nu}_{\ell}) + N \longrightarrow \ell^{-} (\ell^{+})+ N^{\prime} + \pi$ & $\nu_{\ell} 
(\bar{\nu}_{\ell}) + N \longrightarrow \nu_{\ell}(\bar{\nu_{\ell}})+ N^{\prime} + \pi$ \\ 
			
2. & $\nu_{\ell} (\bar{\nu}_{\ell}) + N \longrightarrow \ell^{-} (\ell^{+})+ N^{\prime} + n \pi$ & $\nu_{\ell} 
(\bar{\nu}_{\ell}) + N \longrightarrow \nu_{\ell} (\bar{\nu}_{\ell})+ N^{\prime} + n\pi$ \\ 
			
3. & $\nu_{\ell} (\bar{\nu}_{\ell}) + N \longrightarrow \ell^{-} (\ell^{+})+ N^{\prime} + \eta$ &  $\nu_{\ell} 
(\bar{\nu}_{\ell}) + N \longrightarrow \nu_{\ell}(\bar{\nu}_{\ell}) + N^{\prime} + \eta$\\ 
			
4. & $\nu_{\ell} (\bar{\nu}_{\ell}) + N \longrightarrow \ell^{-} (\ell^{+})+ Y + K$ & $\nu_{l} (\bar{\nu}_{\ell}) + N 
\longrightarrow \nu_{\ell} (\bar{\nu}_{\ell})+ Y + K$ \\ 
			
5. & $\nu_{\ell} (\bar{\nu}_{\ell}) + N \longrightarrow \ell^{-} (\ell^{+})+ N^{\prime} + K(\bar{K})$ &  \\ 

6. & $\bar\nu_l+N 
\longrightarrow l^{+}+Y+\pi$ & \\ \hline \hline
\end{tabular}
\end{center}
\caption{CC and NC induced IE processes. Here $N,N^{\prime}$ represent proton and neutron, $Y = 
\Lambda, \Sigma$ represents the hyperons, $K=K^{+}, K^{0}$ represents the kaons, $\bar{K}=K^-, \bar{K}^{0}$ represents the 
antikaons and $\ell=e,\mu,\tau$ represents the leptons.}\label{sec2:Table1}
\end{table}

\item {\bf Inelastic scattering:} The IE processes like the single  and multiple mesons are produced in the CC 
and NC reactions subject to the absence of FCNC. A list of such reactions is given in 
Table-\ref{sec2:Table1}. 
		
\item {\bf Deep inelastic scattering:} The DIS processes induced by the CC and NC interactions are represented by the 
reactions
\begin{eqnarray}\label{Ch-4:process1_nu}
{\nu}_\ell/\bar{\nu}_\ell  + N  \longrightarrow \ell^-/\ell^{+} + X, \qquad \qquad 
{\nu}_\ell/\bar{\nu}_\ell  + N  \longrightarrow 	{\nu}_\ell/\bar{\nu}_\ell + X.
\end{eqnarray}
where $X$ is jet of hadrons in the final state.
\end{itemize}
In the region of intermediate and high energies relevant to the atmospheric and accelerator neutrinos, the inclusive 
reactions discussed above become important in various regions of energy as shown in Fig.~\ref{flux}, where the total 
scattering cross section per nucleon per unit energy of the incoming (anti)neutrino is presented as a function of the 
(anti)neutrino energy. 
The individual contributions to the QE, IE, and DIS cross 
sections as well as the sum of all the processes are shown and compared with the available experimental data starting from 
the Gargamelle collaboration in 1973 to MicroBooNE collaboration in 2019, extracted from the interaction of accelerator and 
atmospheric (anti)neutrinos with free nucleons as well as with nuclear targets. Also, in the same plot, we have shown the 
area normalized flux for present and future neutrino experiments like MicroBooNE, T2K, MINERvA, NOvA, and DUNE. It is 
evident from the figure that in the few GeV energy region all the three processes, viz., QE, IE, and DIS, have 
contributions to the neutrino and antineutrino induced processes. The different neutrino 
experiments have their flux peaked at different average energies for the corresponding experiment.

In the high energy region of neutrinos of the order of TeV and PeV, 
Bustamante and Connolly~\cite{Bustamante:2017xuy} have studied the energy dependence of the neutrino-nucleon cross section 
measured in the IceCube experiment and concluded that the results are compatible with 
predictions based on nucleon structure extracted from scattering experiments at lower energies and disfavor extreme 
deviations that could result from new physics in the TeV--PeV range. This has been further discussed in Ref.~\cite{Valera:2022ylt}.
Experimentally, the IceCube collaboration~\cite{IceCube:2020rnc} 
has measured the neutrino-nucleon cross section between 60TeV and 10PeV energy range and found the results 
to be compatible with the SM predictions.

While the QE and elastic scattering processes are kinematically well defined, the kinematic region defining the IE 
scattering and the onset of DIS is not free from ambiguities. We discuss $\Delta S=0$ QE scattering of (anti)neutrinos with 
the free nucleon in Section~\ref{nu_QE} and the antineutrino induced $|\Delta S|=1$ QE scattering in 
Section~\ref{qe_hyperon}. The IE scattering processes start with the single pion production which is dominated by the 
$\Delta$ resonance. But the NR production of single pion starts earlier at the threshold of pion production corresponding 
to $W=1.08$~GeV, where $W$ is the CM energy of the final pion-nucleon system. 
In recent years, some authors have advocated to 
consider the onset of IE processes much earlier in energy with the production of single photon at $M < W < M + 
m_\pi$~\cite{Ruso:2022qes}, where $M~(m_{\pi})$ is the nucleon~(pion) mass. Traditionally, the kinematic region of the IE 
scattering is considered to be from $W=1.08$~GeV to the onset of DIS for which $W=2$~GeV is generally taken 
but a precise value is not defined. The kinematic region of the IE scattering above $W=1.08$~GeV with moderate $Q^2 \le 
1$~GeV$^2$ is quite intriguing and is called the shallow inelastic scattering~(SIS) region. Recently the need to understand the 
IE processes has been highlighted in many workshops and conferences like NuINT, NUSTEC, etc. A recent compilation of articles 
by several experimenters, theorists and phenomenologists have highlighted  the development in the area of neutrino interactions 
in the intermediate and high energy regions~\cite{Athar:2021xfr}. The 
present accelerator experiments like NOvA, and MINERvA~(low energy beam) and the future experiment like DUNE have average 
energies of about 3~GeV. For example at DUNE, it is expected that more than 50$\%$ events would come from the SIS plus 
DIS regions. Moreover, the atmospheric neutrino studies in the next generation Hyper-Kamiokande experiment will also have 
significant contributions from the SIS and DIS regions. With the increase in $Q^2$, one approaches the onset of DIS. 
Therefore, it becomes essential to understand the dynamics of this kinematic region which is presently neither well understood 
theoretically nor experimentally~\cite{NuSTEC:2017hzk, SajjadAthar:2020nvy, Andreopoulos:2019gvw}. Most of the present neutrino 
event generators use the prescription of Bodek {  et 
al.}~\cite{Bodek:2002ps, Bodek:2004pc, Bodek:2010km} to take care of the transition region using parton distribution functions 
empirically extrapolated from the DIS region to lower $W$ and $Q^2$. 
\begin{figure}
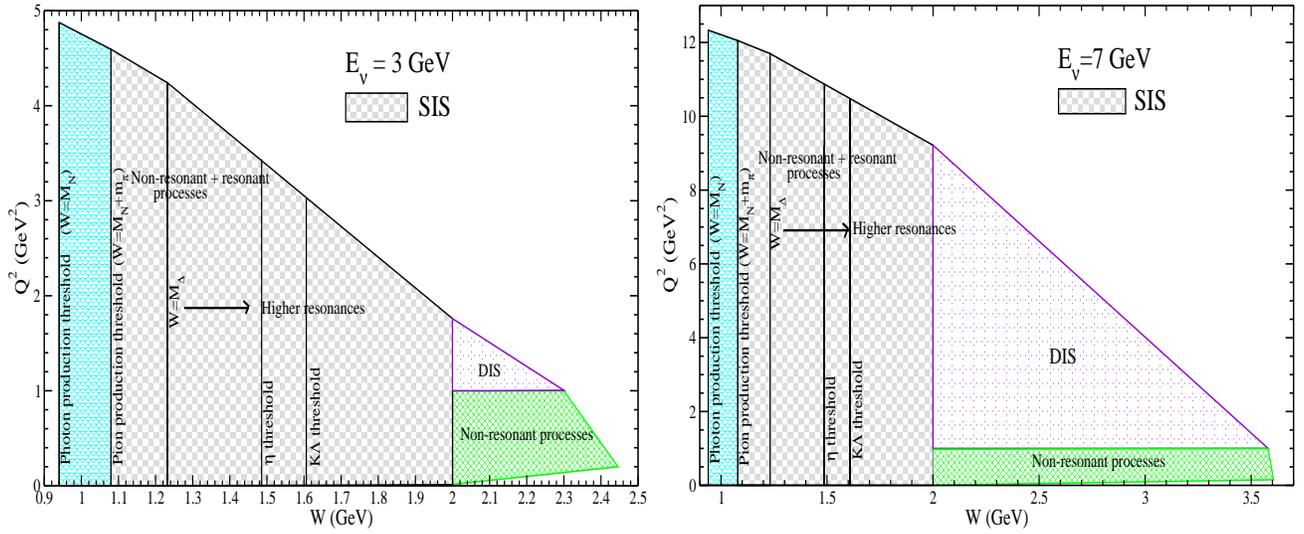
  
\begin{center}
	\includegraphics[height=7.0 cm, width=8.5 cm]{wq2_3gev_new.eps}
	\includegraphics[height=7.0 cm, width=8.5 cm]{wq2_7gev_new.eps}
\end{center}
\caption{$Q^2,W$ plane depicting neutrino-nucleon scattering at two representative laboratory neutrino energies, where $Q^2 
\ge 0$ is the negative of the four momentum transfer squared $q^2 (\le 0)$ and $W$ is the center of mass~(CM) energy.}
\label{q2W}
\end{figure}
To understand the significance of the IE region in Fig.~\ref{q2W}, we have shown different $(Q^2,W)$ regions at the 
two representative incident neutrino energies of 3~GeV and 7~GeV. It may be observed that above the pion production 
threshold $W \approx 1.08$~GeV the excitation of the $\Delta(1232)$ resonance dominates, but at higher $W$ the hadron 
dynamics results from a nontrivial interplay of overlapping baryon resonances, NR amplitudes and their 
interference. 
 
In most of the neutrino event generators, the Rein-Sehgal approach~\cite{Rein:1980wg} has been used to describe the pion production, 
associated particle production, etc., which considers nucleon-resonance transition form factors obtained using a 
constituent quark model. Some modifications in this approach have been recently done by updating the resonance properties like 
resonance masses, decay widths and branching ratios but interferences are neglected. Recently this model has been updated by 
Kabirnezhad~\cite{Kabirnezhad:2017jmf, Kabirnezhad:2020wtp} by considering NR part of the amplitude and the empirical 
inputs for the vector part of the transition current. To understand weak pion production off the nucleon, several 
authors~\cite{Hernandez:2007qq, Hernandez:2007ej, RafiAlam:2010kf, Alam:2011vwg, Wang:2013wva} have used approximate chiral 
symmetry of quantum chromodynamics~(QCD) to construct the transition amplitude in the region of small energy and 
momentum transfers. Though the single pion 
production is dominated by the $\Delta(1232)$ resonance, there are other resonances like $N^{\star}(1440)$, $N^{\star}(1520)$, 
$N^{\star}(1535)$, $N^{\star}(1650)$, etc. which contribute in $n\pi^+$~($p\pi^-$) and $p\pi^0$~($n\pi^0$) channels for 
(anti)neutrino induced interaction. Furthermore, there are  additional contributions from the NR amplitudes as well 
as their interferences with the resonance counterpart~\cite{Leitner:2008ue, RafiAlam:2015fcw} and these will be discussed in 
Section~\ref{sec:1pion}. Similarly in the case of single kaon production due to the absence of $S=1$ baryonic resonances it 
is possible to obtain model independent predictions for the scattering cross sections using chiral perturbation theory which 
has been done by our group~\cite{RafiAlam:2010kf}, while for the antikaon production there is additional contribution from 
the $\Sigma^* (1385)$ resonance, which has been studied by us~\cite{RafiAlam:2012bro}. Some of the results are  
presented for the scattering cross section for the (anti)neutrino scattering off nucleon leading to (anti)kaon production in 
Sections~\ref{kaon} and \ref{sec:1antiKaon}. In the case of eta production, it is well known from $\pi N$ scattering, that 
it is dominated by $N^{\star}(1535)$ resonance besides very small contributions from the higher resonances like 
$N^{\star}(1650)$ and $N^{\star}(1710)$, and the NR terms. We have discussed first the eta production 
induced by the real photons off the nucleon targets in Section~\ref{sec:eta:photo} and compared the results with the MAMI 
data~\cite{CrystalBallatMAMI:2010slt, A2:2014pie} and then extended the formalism to the weak sector in Section~\ref{sec:eta}. 
Similarly for the associated particle production, our group~\cite{Fatima:2020tyh} has studied associated particle production 
induced by photons off proton target and compared the results with the CLAS data~\cite{CLAS:2005lui} and extended this study 
to include the (anti)neutrino induced associated particle production off nucleon target. These are discussed in 
Sections~\ref{sec:associated:photo} and \ref{sec:associated}. We also discuss in brief $Y\pi$ production following the 
works of Benitez Galan et al.~\cite{BenitezGalan:2021jdm}, $\Xi K$ production following the works of Alam et 
al.~\cite{RafiAlam:2019rft} and two pion production~\cite{Hernandez:2007ej, Valverde:2008jj} in Sections~\ref{sec:Ypi}, 
\ref{XK} and \ref{sec:2pion}, respectively. Finally, in Section~\ref{dis:nucleon}, we discuss DIS of (anti)neutrinos with 
free nucleon, where a jet of hadron is produced in the final state.

\subsection{Quasielastic and elastic $\nu-$scattering processes on nucleons}\label{nu_QE}
\subsubsection{Introduction} 
\begin{figure}  
\begin{center}
	\includegraphics[height=5cm,width=4cm]{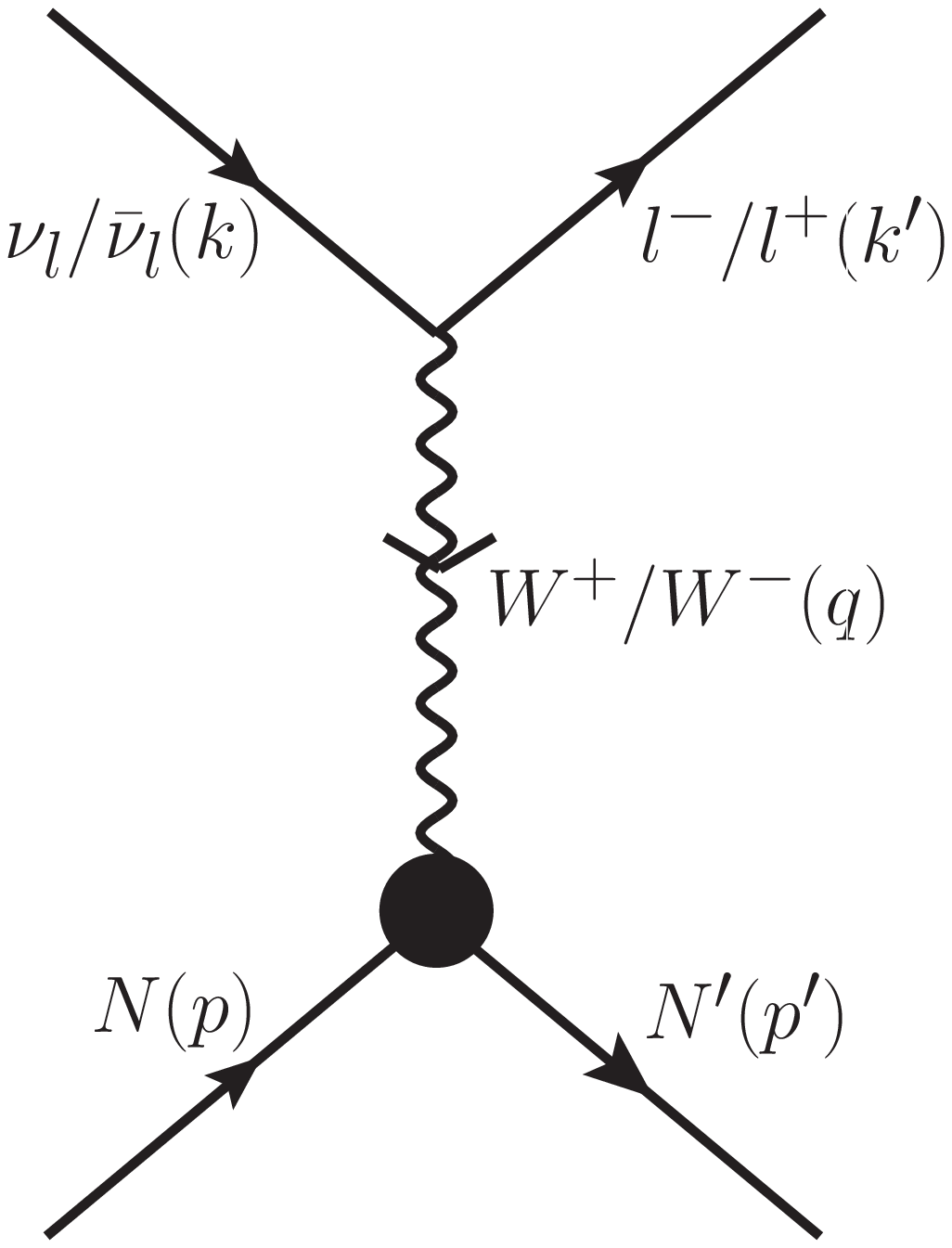}
	\hspace{1cm}
	\includegraphics[height=5cm,width=4cm]{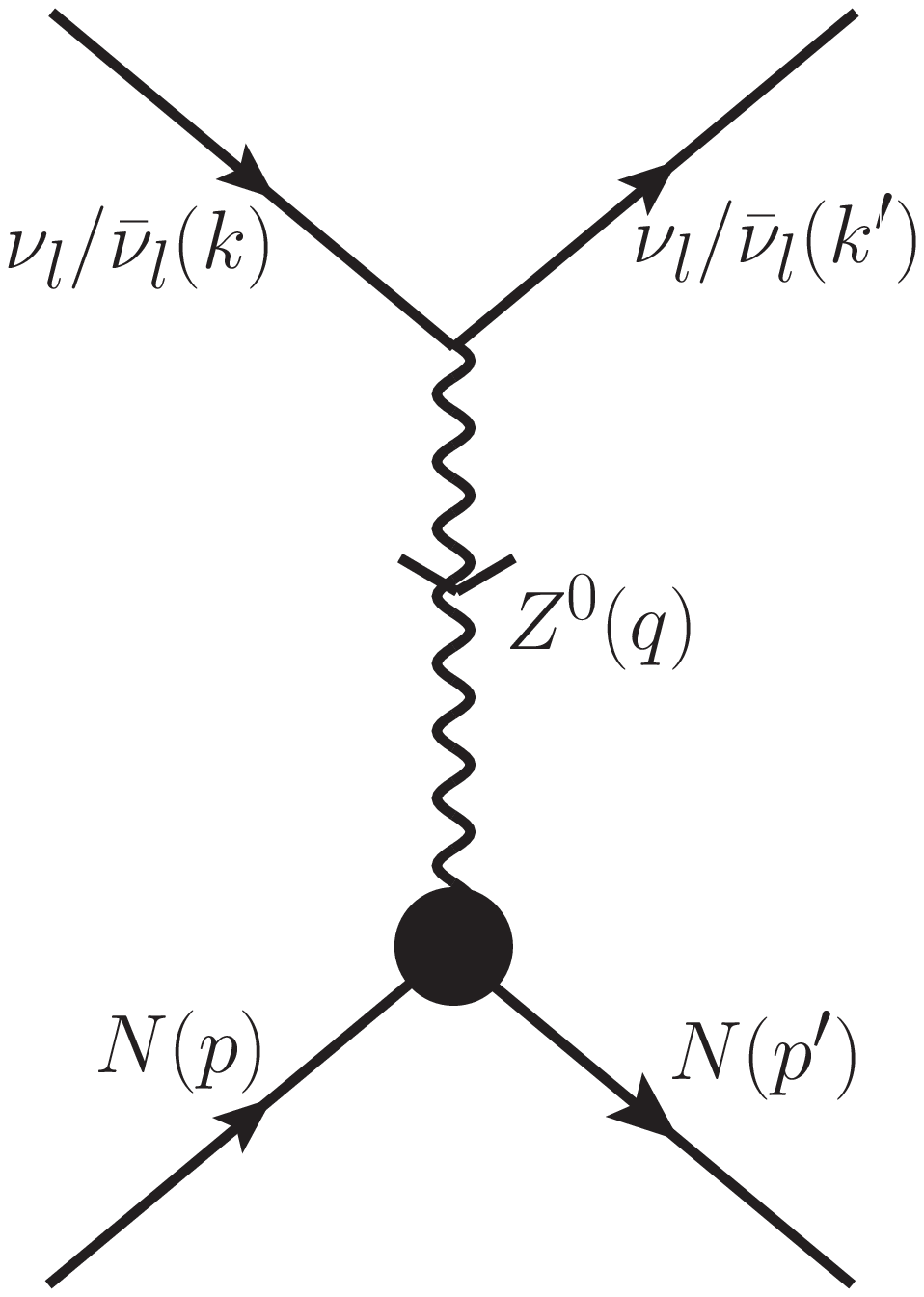}
\caption{QE~(left panel) and elastic~(right panel) $\nu-$scattering processes on the nucleons ($N=n,p$ and $N^\prime =p,n$) target.}
\label{Feynman_diagram}
\end{center}
\end{figure}

Neutrinos and antineutrinos interact with the free nucleons via the CC as well as NC induced weak 
processes like:
\begin{equation}\label{cc_quasi_reaction}
\begin{array}{l}
\nu_l(k)~+~n(p)~\rightarrow~l^-(k^\prime)~+~p(p^\prime), \quad \qquad 
{\bar{\nu}}_l(k)~+~p(p)~\rightarrow~l^+(k^\prime)~+~n(p^\prime)
\end{array}~~~\mbox{(CC)}
\end{equation}
\begin{equation}\label{nc_elas_reaction}
\left.
\begin{array}{l}
\nu_l(k)~+~n(p)~\rightarrow~\nu_l(k^\prime)~+~n(p^\prime), \qquad \quad
\nu_l(k)~+~p(p)~\rightarrow~\nu_l(k^\prime)~+~p(p^\prime)\\
{\bar{\nu}}_l(k)~+~n(p)~\rightarrow~{\bar{\nu}}_l(k^\prime)~+~n(p^\prime), \quad \qquad
{\bar{\nu}}_l(k)~+~p(p)~\rightarrow~{\bar{\nu}}_l(k^\prime)~+~p(p^\prime)\\
\end{array}\right\}~~~\mbox{(NC)}
\end{equation}             
In the above processes, $k$ and $k^\prime$ are, respectively, the four momenta of the (anti)neutrino and the corresponding 
charged/neutral lepton and $p$ and $p^\prime$ are four momenta of the incoming and outgoing nucleons. Feynman diagrams 
corresponding to reactions given in Eqs.~(\ref{cc_quasi_reaction}) and (\ref{nc_elas_reaction}) are shown in 
Fig.~\ref{Feynman_diagram}.

\subsubsection{Charged current quasielastic reactions and weak nucleon form factors}\label{CCQE}
The general expression for the differential scattering cross section for reactions given in Eqs.~(\ref{cc_quasi_reaction}) 
and (\ref{nc_elas_reaction}) is written as,
\begin{eqnarray}\label{diff_xsect_quasi}
d\sigma=\frac{(2\pi)^{4}\delta^{4}(k+p-p^\prime-k^\prime)}{4(k\cdot p)}\frac {d{\vec{k}^{\;\prime}}}{(2\pi)^{3}2
E_{l}^{\prime}}\frac {d{\vec{p}^{\;\prime}}}{(2\pi)^{3}2E^{\prime}} {\bar\Sigma}\Sigma| \mathcal{M} |^2,
\end{eqnarray}
which results in the expression of the double differential cross section $\sigma_{free}(E_l^{\prime},\Omega_l^{\prime})$ on 
the free nucleon target in the laboratory frame as
\begin{equation}\label{sig_zero}
\sigma_{free}(E_l^{\prime},\Omega_l^{\prime})\equiv \left(\frac{d^2 \sigma}{ d E_l^{\prime} \; d \Omega_l^{\prime}}
\right)_{\nu/\bar{\nu}-N}=\frac{{|\vec k^\prime|}}{64\pi^2 E_\nu E E^{\prime}}\overline{\sum}\sum{|{\cal M}|^2}\delta[q_0 + 
E - E^{\prime}],
\end{equation}
where $q_0(=E_\nu - E_l^{\prime})$ is the energy transferred to the hadronic system; $E_{\nu}$, $E_{l}^{\prime}$ are the 
energies of the incoming neutrino and outgoing lepton, $E~(=M)$ and $E^{\prime}~(=M+q_{0})$, respectively, are the energies 
of the incoming and outgoing nucleons in the laboratory frame, ${\cal M}$ is the invariant matrix element and for 
Eq.~(\ref{cc_quasi_reaction}) is given by
\begin{eqnarray}\label{qe_lep_matrix}
{\cal M}=\frac{G_F}{\sqrt{2}}~l^\mu~J_\mu.
\end{eqnarray}
In the above expression $G_F$ is the Fermi coupling constant,  and the leptonic weak current 
is given by
\begin{equation}\label{lep_curr}
l^\mu=\bar{u}(\vec{k}^{\prime})\gamma^\mu(1 \mp \gamma_5)u(\vec{k}),
\end{equation}
and $-(+)$ represents the neutrino~(antineutrino) induced QE scattering processes. The hadronic current~($J_\mu$) 
for CC induced interaction is given by
\begin{equation}\label{had_curr}
J_\mu^{CC}=\bar{u}(\vec{p}^{\,\prime}) \;{\cal O}_{\mu}^{CC}\; u(\vec{p}),
\end{equation}
where ${\cal O}_{\mu}^{CC} = V_{\mu}^{CC}-A_{\mu}^{CC}$ is CC weak hadronic vertex, and the matrix elements 
of the vector~($V_\mu^{CC}$) and the axial-vector~($A_\mu^{CC}$) currents are given by~\cite{LlewellynSmith:1971uhs, 
Fatima:2018tzs}:
\begin{eqnarray}\label{vx}
\langle N^\prime(\vec{p}^{\,\prime}) | V_\mu^{CC}| N(\vec{p}) \rangle &=& \cos \theta_C\bar{u}(\vec{p}^{\,\prime}) \left[ \gamma_\mu 
f_1(Q^2)+i\sigma_{\mu \nu} \frac{q^\nu}{(M+M^\prime)} f_2(Q^2) + \frac{2 q_\mu}{(M + M^\prime)} f_3(Q^2) \right] u(\vec{p}),
\\
\label{vy}
\langle N^\prime(\vec{p}^{\,\prime}) | A_\mu^{CC}| N(\vec{p}) \rangle &=& \cos \theta_C\bar{u} (\vec{p}^{\,\prime}) \left[ \gamma_\mu 
\gamma_5 g_1(Q^2) +  i\sigma_{\mu \nu} \frac{q^\nu}{(M+M^\prime)} \gamma_5 g_2(Q^2) +  \frac{2 q_\mu}{(M+M^\prime)} \gamma_5 
g_3(Q^2) \right] u(\vec{p}).
\end{eqnarray}
In the above expression, $N,N^\prime=n,p$; $\theta_C$ is the Cabibbo angle, $Q^2 = -q^2=-(k-k^{\prime})^2$ is the four momentum transfer squared. $M$ and 
$M^\prime$ are the masses of the initial and the final nucleons, respectively. $f_1(Q^2)$, $f_2(Q^2)$ and $f_3 (Q^2)$ are 
the vector, weak magnetic and induced scalar form factors and $g_1(Q^2)$, $g_2(Q^2)$ and $g_3(Q^2)$ are the axial-vector, 
induced tensor~(or weak electric) and induced pseudoscalar form factors, respectively. 

Using the leptonic and hadronic currents given in Eqs.~(\ref{lep_curr}) and (\ref{had_curr}) in Eq.~(\ref{qe_lep_matrix}), 
the matrix element squared is obtained as: 
\begin{equation}\label{mat_quasi}
{|{\cal M}|^2}=\frac{G_F^2}{2}~{ L}^{\mu\nu} {J}_{\mu\nu},
\end{equation} 
where the leptonic tensor ${ L}^{\mu\nu}$ is calculated to be
\begin{eqnarray}\label{lep_tens}
{ L}^{\mu\nu}&=&8~\left[k^\mu {k^\prime}^{\nu}+{k^\prime}^{\mu} k^\nu-g^{\mu\nu}~k\cdot k^{\prime} \pm i \epsilon^{\mu
\nu\alpha\beta}~k^{\prime}_{\alpha} k_{\beta} \right],
\end{eqnarray}
$+(-)$ is for the neutrino~(antineutrino) induced processes.

The hadronic tensor ${J}_{\mu\nu}$ given in Eq.~(\ref{mat_quasi}), is obtained using Eq.~(\ref{had_curr}) averaged over the 
initial spin state of the nucleon and summed over the final spin state as:
\begin{eqnarray}\label{had_tens}
J_{\mu\nu}&=&\overline{\sum}\sum {J_{\mu}^{CC}}^{\dagger} J_\nu^{CC}=\frac{1}{2}\mbox{Tr}\left[(\slashed{p}^{\;\prime}+M)
{\cal O}_\mu (\slashed{p}+M)\tilde{\cal O}_\nu\right],
\end{eqnarray}
where $\tilde{\cal O}_{\nu}=\gamma^0~{{\cal O}_{\nu}}^\dagger ~\gamma^0$. The expression for $J_{\mu\nu}$ is given in 
Appendix~\ref{appendix}.

The differential scattering cross section $\frac{d\sigma}{dQ^2}$ for CC and NC induced processes, in 
the laboratory frame is then obtained as
\begin{equation}\label{dsig:QE}
\frac{d\sigma}{dQ^2}=\frac{G_F^2}{8 \pi {M}^2 {E^2_{\nu_{l}}}} N(Q^2),
\end{equation}
where $N(Q^2) = J_{\mu\nu}L^{\mu\nu}$ and the expression for $N(Q^2)$ is given in Appendix~\ref{appendix1}.

\subsubsection{Neutral current elastic reactions and weak nucleon form factors}\label{QE:NC}
We define the hadronic current for the weak NC induced processes on the proton and neutron targets, given 
in Eq.~(\ref{nc_elas_reaction}), in terms of NC form factors $\tilde{f}_i^{p,n}(Q^2) \text{ and } 
\tilde{g}_i^{p,n}(Q^2) (i=1,2,3)$ for the protons and neutrons, respectively, as
\begin{eqnarray}\label{jmu:NC}
 J_\mu^{NC}|_{i}&=& \bar u(\vec{p}^{\,\prime}) {\cal O}_{\mu}^{NC}  u(\vec{p})= \bar u(\vec{p}^{\,\prime})\left[\gamma_\mu 
 \tilde{f}^{i}_1 (Q^2) + \frac{i\sigma_{\mu\nu}q^\nu \tilde{f}^{i}_2 (Q^2)}{2M} + \frac{q_\mu}{M} \gamma_5 \tilde{f}_3^{i}
 (Q^2) \right.\nonumber \\
&+&\left. \gamma_\mu \gamma_5 \tilde{g}^{i}_1 (Q^2) + \frac{(p_\mu+ p^{\prime}_{\mu})}{M}\gamma_5 \tilde{g}_2^{i}(Q^2) + 
\frac{q_\mu \gamma_5 \tilde{g}^{i}_3 (Q^2)}{M}   \right]u(\vec{p}), 
\end{eqnarray}
with $i=p,n$. Here the nomenclature of the form factors is the same as in the case of CC QE 
process~(Eqs.~(\ref{vx}) and (\ref{vy})) and the parameterizations for $\tilde{f}_{i}(Q^{2})$ and $\tilde{g}_{i}(Q^{2})$ are 
given in Section~\ref{para_FF}. 

\subsubsection{Symmetry properties of the weak hadronic current}\label{had_symmetry}
The weak hadronic current $J_\mu$ has the vector $V_{\mu}$ and the axial-vector $A_\mu$ terms constructed using the bilinear 
covariants associated with the nucleon fields as well as the four momenta of the incoming and outgoing nucleons. These 
bilinear covariants have certain definite properties under discrete transformation like C, P and T as well as the internal 
symmetries like the isospin and unitary symmetry~\cite{Athar:2020kqn, Fatima:2018tzs}. These symmetry properties are exploited 
in writing the matrix elements of these currents between the initial and final states of spin 0, $\frac{1}{2}$, and 
$\frac{3}{2}$ particles. We discuss below these symmetry properties and their role in writing the general structure of the 
matrix elements.
\begin{itemize}
\item [(i)] {\bf Isospin properties of the weak hadronic current}

The weak hadronic currents between the neutron and proton states involve a change of charge $\Delta Q=\pm 1$ in the case of 
$n \rightarrow p$ and $p\rightarrow n$ transitions. Since $Q=I_3+\frac{B}{2}$ for the nonstrange baryons, therefore 
$\Delta Q=\pm1$ implies~$\Delta I_3=\pm1$ using baryon number conservation. Since protons and neutrons are assigned to a 
doublet, therefore, they can be written as a two component isospinor under the group of isospin transformation, i.e.
\begin{eqnarray}
u=
\begin{pmatrix}
u_{p}\\ u_{n}
\end{pmatrix}.
\end{eqnarray}
The isospin group of transformations is generated by the three $2\times 2$ Pauli matrices $\tau_{i}~(i=1-3)$, which along 
with the vector currents constitute the isovector part of the hadronic current. By defining the isospin raising and lowering 
operators $\tau^\pm=\frac{\tau_1\pm i\tau_2}{2}$, we can write 
\begin{eqnarray}
\bar{u}_p {V}_\mu^{CC} u_{n}=\bar{u} {V}_\mu^{CC} \tau^+ u = \bar{u} {V}_\mu^{CC+} u, \qquad \quad
\bar{u}_n {V}_\mu^{CC} u_{p}=\bar{u} {V}_\mu^{CC} \tau^- u= \bar{u} {V}_\mu^{CC-} u.
\end{eqnarray}
It may be observed from the above relations that the charged weak vector currents are purely isovector in nature.

Similarly, for the electromagnetic vector current, the hadronic current is given by
\begin{equation}\label{em:QE}
 J_{\mu (p,n)}^{em}(p,p^{\prime}) = \bar u(\vec{p}^{\,\prime}_{p,n}) {V}_{\mu}^{em} u(\vec{p}_{p,n}),
\end{equation}
with
\begin{equation}\label{O:EM}
 {V}_{\mu}^{em}(p,n) = \left[ \gamma_\mu F_1^{p,n}(Q^2)+ i\sigma_{\mu \nu} \frac{q^\nu}{(2M)} F_2^{p,n}(Q^2) \right] ,
\end{equation}
where $q=p^{\prime} - p$ with $Q^{2} = -q^{2}$. $F_{1}^{p,n}(Q^2)$ and $F_{2}^{p,n}(Q^2)$ are, respectively, the Dirac and 
Pauli form factors of the nucleon. 
In terms of the Pauli matrices, the hadronic currents for the electromagnetic induced interactions are written as
\begin{eqnarray}
\bar{u}_p {V}_\mu^{em} u_{p}=\bar{u}{V}_\mu^{em} \frac{\mathbb{I}+\tau_3}{2} u , \qquad \quad
\bar{u}_n {V}_\mu^{em} u_{n}=\bar{u}{V}_\mu^{em} \frac{\mathbb{I}-\tau_3}{2} u ,
\end{eqnarray}
implying the isoscalar and isovector current matrix elements as
\begin{eqnarray}
\bar{u} \mathbb{I} {V}_\mu^{em} u = \bar{u}_{p}{V}_\mu^{em} u_{p}+\bar{u}_{n} {V}^{em}_\mu 
u_{n},\qquad \quad
\bar{u}\tau_3 {V}_\mu^{em} u =\bar{u}_p {V}_\mu^{em}u_p -\bar{u}_n {V}_\mu^{em}u_n.
\end{eqnarray}   
If we parameterize the matrix element of the isoscalar~(with form factors $F_{1,2}^{S} (Q^2)$) and isovector~(with form 
factors $F_{1,2}^{V} (Q^2)$) components as
\begin{eqnarray}
\bar{u}\mathbb{I} {V}_\mu^{em} u = \bar{u}\left[F_{1}^S(Q^2)\gamma_{\mu}+iF_2^S(Q^2)
\frac{\sigma_{\mu\nu}q^\nu}{2M}\right]u ,&&
\bar{u}\tau_{3} {V}_\mu^{em} u =\bar{u}\left[F_{1}^V(q^2)\gamma_{\mu}+i\frac{\sigma_{\mu\nu}q^\nu}{2M}F_{2}^V(q^2)
\right] \tau_3 u, 
\end{eqnarray}
and the electromagnetic matrix element of protons and neutrons given in Eq.~(\ref{em:QE}) with ${ V}_{\mu}^{em}$ given 
in Eq.~(\ref{O:EM}), then we can write:
\begin{eqnarray}
F_{1,2}^S (Q^2) = F_{1,2}^p(Q^2) + F_{1,2}^n(Q^2), \qquad \quad
F_{1,2}^V (Q^2) = F_{1,2}^p(Q^2) - F_{1,2}^n(Q^2).
\end{eqnarray}
The above expression shows that the electromagnetic current transforms as the sum of the isoscalar and isovector currents.

\item [(ii)] {\bf T invariance}

Time reversal invariance holds if
\begin{equation}
{\cal{M}^{\prime}} = {\cal{M}}^{*},
\end{equation}
where ${\cal{M}^{\prime}}$ represents the time reversed matrix element and ${\cal{M}}^{*}$ represents the Hermitian 
conjugate of the unreversed matrix element. Under time-reversal invariance~(T invariance), the initial and final state particles are 
interchanged as well as their spin and angular momenta are reversed.

Taking all the bilinear covariants used with the form factors in the vector and the axial-vector currents individually, we 
obtain the transformation of the vector and axial-vector form factors under T invariance as~\cite{Athar:2020kqn}:
\begin{eqnarray}
\bar{u}_{p} u_{n} &\xrightarrow{\text{~~T~~}}& \bar{u}_{n}u_{p}, \qquad~~~~~~~ \qquad \bar{u}_{p}\gamma_{5} u_{n} 
\xrightarrow{\text{~~T~~}} \bar{u}_{n}\gamma_{5} u_{p},  \nonumber \\
\bar{u}_{p}\gamma^{\mu}u_{n} &\xrightarrow{\text{~~T~~}}& \bar{u}_{n}\gamma_{\mu}u_{p},\qquad ~~~~\qquad \bar{u}_{p}
\gamma^{\mu}\gamma_{5} u_{n} \xrightarrow{\text{~~T~~}} \bar{u}_{n}\gamma_{\mu}\gamma_{5} u_{p},\nonumber \\
\bar{u}_{p}\sigma^{\mu\nu} u_{n} &\xrightarrow{\text{~~T~~}}& -\bar{u}_{n}\sigma_{\mu\nu} u_{p}, \qquad \qquad \bar{u}_{p}
\sigma^{\mu\nu}\gamma_{5} u_{n} \xrightarrow{\text{~~T~~}} \bar{u}_{n}\sigma_{\mu\nu}\gamma_{5} u_{p}.\nonumber 
\end{eqnarray}
The hadronic current $J_{\mu}$ is defined in Eq.~(\ref{had_curr}) with $V_{\mu}$ and $A_{\mu}$ defined in Eqs.~(\ref{vx}) 
and (\ref{vy}), respectively. The time reversed current $J_{\mu}^{\prime}$ is obtained as~\cite{Athar:2020kqn}:
\begin{eqnarray}\label{tr}
J_\mu^{\prime}&=&\cos\theta_C~\bar{u}_{n} \left[f_1(Q^2)\gamma_\mu+ i\sigma_{\mu\nu}\frac{q^\nu}{M+M^{\prime}} f_2(Q^2) +
\frac{2 q_\mu}{M+M^{\prime}} f_3(Q^2) \right. \nonumber \\
&-& \left. g_1(Q^2) \gamma_\mu\gamma_5 - i\sigma_{\mu \nu} \frac{q^\nu}{M+M^{\prime}} \gamma_5 g_2(Q^2)- \frac{2q_\mu}
{M+M^{\prime}}\gamma_5  g_3(Q^2)\right] u_{p}.
\end{eqnarray}
Hermitian conjugate of Eq.~(\ref{had_curr}) is written as
\begin{eqnarray}\label{hc}
J_\mu^{*}&=&\cos\theta_C~\bar{u}_{p} \left[f_1^{*}(Q^2)\gamma_\mu+i\sigma_{\mu\nu}\frac{q^\nu}{M+M^{\prime}} f_2^{*}(Q^2) +
\frac{2 q_\mu}{M+M^{\prime}} f_3^{*}(Q^2)  - g_1^{*}(Q^2) \gamma_\mu\gamma_5\right. 
\nonumber \\ 
&-& \left. i\sigma_{\mu \nu} \frac{q^\nu}{M+M^{\prime}} \gamma_5 g_2^{*}(Q^2)- \frac{2q_\mu}{M+M^{\prime}}\gamma_5  
g_3^{*}(Q^2)\right] u_{n}.
\end{eqnarray}
Comparing Eqs.~(\ref{tr}) and (\ref{hc}), we find that $f_{i} (Q^2) = f^{*}_{i} (Q^2)$ and  $g_{i}(Q^2) = g_{i}^{*} (Q^2)$ 
which implies that if time reversal invariance holds, the form factors must be real.

\item [(iii)] {\bf Conserved vector current hypothesis}

The hypothesis of the conserved vector current~(CVC) was proposed by Gershtein and Zeldovich~\cite{Gershtein:1955fb} and 
Feynman and Gell-Mann~\cite{Feynman:1958ty}. They made an important observation in the study of the nuclear $\beta$ decays 
in Fermi transition driven by the vector currents, with no change in parity. They observed that the strength of the weak 
vector coupling~(weak charge) for the muon and neutron decays are the same, just like in the case of the electromagnetic 
interactions where the strength of the electromagnetic coupling i.e. $e$, remains the same for the electrons and protons. 
Since the equality of the charge coupling, also known as the universality of the electromagnetic interactions follows from 
the conservation of the electromagnetic current, therefore, it was suggested that the weak vector current is also conserved 
i.e. $\partial_{\mu} V^{\mu} (x)=0$, which leads to the equality of the weak coupling for the leptons and hadrons.

In fact, they proposed a stronger hypothesis of the isotriplet of the vector currents which goes beyond the hypothesis of 
CVC and predicts the form factors $f_{1,2}(Q^2)$ describing the matrix elements of the weak vector current in terms of the 
electromagnetic form factors of hadrons. According to the isotriplet hypothesis, the weak vector currents $V_{\mu}^+$,~$V_{\mu}^-$ 
and the isovector part of the electromagnetic current $V_{\mu}^{em}$ are assumed to form an isotriplet under the isospin 
symmetry such that $f_{1}$ and $f_{2}$ are given in terms of the isovector electromagnetic form factors i.e.
\begin{eqnarray}
f_{1}(Q^2)= F_{1}^{V} (Q^2) = F_{1}^p(Q^2)-F_{1}^n(Q^2), \qquad \quad
f_{2}(Q^2)= F_{2}^{V} (Q^2) = F_{2}^p(Q^2)-F_{2}^n(Q^2).
\end{eqnarray}

The CVC hypothesis, i.e. $\partial_{\mu} V^{\mu}(x) =0$ implies $f_{3}(Q^2)=0$. It should be 
noted that while the isotriplet current hypothesis implies CVC due to the isospin symmetry, the vice versa is not true. In 
the literature, the term CVC is mostly used meaning both the isotriplet hypothesis of weak vector currents $V_{\mu}^{+}$ 
and $V_{\mu}^{-}$ and the CVC hypothesis.

\item [(iv)] {\bf Partial conservation of axial-vector current}

In contrast to the vector current which is conserved, the axial-vector current is not conserved. To see this explicitly, 
consider the matrix element of the axial-vector current between one pion state and vacuum which enters in the $\pi l_{2}$ 
decay of pion i.e. $\bra{0} A^{\mu}(x) \ket{{\pi}^{-}} = i f_{\pi} q^{\mu} e^{-iq \cdot x}$, where $q$ is the four momentum 
of the pion. Taking its divergence leads to
\begin{eqnarray}\label{pheno2:pcac1}
<0|{\partial}_{\mu}A^{\mu}(x)|{\pi}^{-}(q)> & = &(-i)i f_{\pi}q_{\mu}q^{\mu}e^{-iq.x} ~=~ f_{\pi}m_{\pi}^{2}e^{-iq.x},
\end{eqnarray}
as $q^{2}=m_{\pi}^{2}$. If the axial-vector current $A^{\mu}$ is divergenceless then either $m_{\pi}=0$ or $f_{\pi}=0$, 
implying the pion to be massless or it does not decay. Since $m_{\pi}\neq 0$, conservation of axial-vector current implies 
$f_{\pi}=0$, which is also not true. Therefore, the axial-vector current is not conserved. However, since the pion is the 
lightest hadron, we can work in the limit of $m_{\pi}\rightarrow 0$, and say that the axial-vector current is conserved in 
the limit
\begin{equation}
 \underset{m_{\pi} \longrightarrow 0}{\lim}{\partial}_{\mu}{A}^{\mu}(x)=0,
\end{equation}
which is termed as the partial conservation of axial-vector current~(PCAC). The hypothesis of PCAC has been very useful in 
calculating many processes in the weak interaction physics and deriving relations between various processes in the limit of 
$m_{\pi} \rightarrow 0$. However, the real predictive power of PCAC lies in making further assumptions about the divergence 
of the axial-vector field $\partial_{\mu}A^{\mu} (x)$ and identifying with the pion field $\phi_{\pi}(x)$ that establishes 
a connection between the weak and 
strong interaction physics and assuming that the transition amplitudes derived in the $m_{\pi} \rightarrow 0$ limit can be 
smoothly extrapolated to the physical mass of the pion. The success of PCAC in various applications of calculating the 
physical processes is based on the following assumptions:
\begin{enumerate}
\item [(i)] The divergence of the axial-vector field is a pseudoscalar field and the pion is also described by a 
pseudoscalar field. If it is assumed that both are related then the physical pion field is described by the divergence 
of the axial-vector current, i.e. $\partial_{\mu} A^{\mu}(x)\propto \phi_{\pi}(x)$, such that $\partial_{\mu} A^{\mu}(x) = 
e_{\pi} \phi_{\pi}(x)$. This assumption makes it possible to relate the weak interaction processes induced by $A^{\mu}$ to 
the pion physics in the 
strong interaction processes through the matrix element of its derivative i.e. $\partial_{\mu} A^{\mu}$.

\item [(ii)] Taking the limit $m_\pi \rightarrow 0$~(corresponding to the conserved axial-vector current) in the processes 
involving pions and nucleons, makes it easier to evaluate the transition amplitude in many weak processes. If further 
assumption is made that these amplitudes vary smoothly with $q^2$ and do not change much over the range of $q^{2}$ involved 
in the processes, then the amplitudes evaluated at $q^{2}=0$ can be extrapolated to the physical limit of $q^{2}=
m_{\pi}^{2}$, i.e. $f(0) \rightarrow f(m_{\pi}^{2})$, where $f(q^{2})$ is the pion form factor. This is called the soft pion 
limit widely used in the weak interaction physics. However, there remains an ambiguity whether to take the limit as 
$m_{\pi}^{2}\rightarrow 0$ or $m_{\pi}\rightarrow 0$~\cite{Treiman:1967cua, Jackiw:1968xt}. For more discussion see 
Ref.~\cite{Athar:2020kqn}.
\end{enumerate}

\item [(v)] {\bf G-parity and second class currents}

G-parity is a multiplicative quantum number, first used to classify the multipion states in $pp$ and $\pi p$ 
collisions~\cite{10.1143/ptp/4.3.389, PhysRev.80.487.3, PhysRev.87.871, PhysRev.103.258} and later used by 
Weinberg~\cite{Weinberg:1958ut} to classify the weak hadronic currents. It is defined as the product of C, the charge 
conjugation operation and a rotation by $180^{0}$ about the Y-axis in the isospin space i.e.
\begin{equation}
G=Ce^{i\pi I_{Y}}.
\end{equation}
Since strong interactions are invariant under C and isospin, they are also invariant under G-parity. The G-parity is a very 
useful concept in the study of pion production in N$ \overline{\text{N}}$ collisions. Since the weak currents involve 
bilinear covariants formed out of the nucleon fields $\overline{\psi}(p^\prime)$ and $\psi(p)$, their transformations can be 
well defined under G-parity. The weak vector and axial-vector currents between a neutron and a proton are defined in 
Eqs.~(\ref{vx}) and (\ref{vy}). Since the currents belong to the triplet representation of the isospin, therefore, all the 
terms have similar transformation under the rotation $e^{i \pi I_Y}$. It is their transformation under 
C-parity which defines their relative transformation under G-parity. Under C-parity, the bilinear terms in 
Eqs.~(\ref{vx}) and (\ref{vy}) transforms as: 
\begin{eqnarray}\label{pheno2_f3}
\bar{u}_{p}u_{n} &\xrightarrow{\text{~C~}}& -\bar{u}_{n}u_{p}~~~~~~~~~~~~~~~(\text{assumed with $f_3$})\\
\label{pheno2_fp}
\bar{u}_{p}\gamma_{5}u_{n} &\xrightarrow{\text{~C~}}& -\bar{u}_{p}\gamma_{5}u_{n}~~~~~~~~~~~~(\text{assumed with $g_3$})\\
\bar{u}_{p}\gamma_{\mu}\gamma_{5}u_{n} &\xrightarrow{\text{~C~}}& -\bar{u}_{p}\gamma_{\mu}\gamma_{5}u_{p}~~~~~~~
~~(\text{assumed with $g_1$})\label{pheno2_fa}
\end{eqnarray}
while
\begin{eqnarray}
\bar{u}_{p}\gamma^\mu u_{n} &\xrightarrow{\text{~C~}}& \bar{u}_{n}\gamma^\mu u_{p}~~~~~~~~~~~~~~~~(\text{assumed with 
$f_1$})
\label{pheno2_f1}\\
\bar{u}_{p}\sigma^{\mu\nu} u_{n} &\xrightarrow{\text{~C~}}& \bar{u}_{n}\sigma^{\mu\nu} u_{p}~~~~~~~~~~~~~~~(\text{assumed 
with $f_2$})\label{pheno2_f2}\\
\bar{u}_{p}\sigma^{\mu\nu}\gamma_5 u_{n} &\xrightarrow{\text{~C~}}& \bar{u}_{n}\sigma^{\mu\nu}\gamma_5 u_{p}~~~~~~~~~~~~
(\text{assumed with $g_2$})\label{pheno2_fe}
\end{eqnarray}
What is observed from Eqs.~(\ref {pheno2_f3})--(\ref{pheno2_fe}) is that the bilinear terms associated with $f_2$ 
transforms the same way as $f_{1}$ does, while $f_3$ transforms in opposite way. Similarly, $g_3$ transforms the same way 
as $g_{1}$ does while $g_2$ transforms in a different way. It was Weinberg~\cite{Weinberg:1958ut} who first used the 
G-parity to classify the weak currents. He called the currents associated with $f_1,~f_2,~g_1$ and $g_3$ which are invariant 
under G-parity as the first class currents, and the currents associated with $f_3$ and $g_2$ which violate G-parity as the 
second class currents~(SCC). Consequently, if G invariance is valid in the weak interactions then the currents with form factors 
$f_{1}(Q^{2})$, $f_{2}(Q^{2})$, $g_{1}(Q^{2})$ and $g_{3}(Q^{2})$ should exist and $f_{3}(Q^{2}) = g_{2}(Q^{2})=0$. It 
should be noted that $f_{3}(Q^{2}) =0$ is also predicted as a consequence of CVC hypothesis.
\end{itemize}

\subsubsection{Parameterization of the weak form factors}\label{para_FF}
\begin{itemize}
 \item [(i)] {\bf Vector form factors}\\
In the case of CC interactions, the hadronic current contains two isovector form factors $f_{1,2}(Q^2)$ of the 
nucleons, which can be related to the isovector combination of the Dirac~($F_{1,2}^p(Q^2)$) and Pauli~($F_{1,2}^n(Q^2)$) 
form factors of the proton and the neutron using the relation
\begin{equation}\label{f1v_f2v}
f_{1,2}(Q^2)=F_{1,2}^p(Q^2)- F_{1,2}^n(Q^2) .
\end{equation}
The Dirac~($F_{1}(Q^2)$) and Pauli~($F_{2}(Q^2)$) form factors are, in turn, expressed in terms of the experimentally 
determined Sachs' electric~($G_E^{p,n} (Q^2)$) and magnetic~($G_M^{p,n} (Q^2)$) form factors of the nucleon as: 
\begin{eqnarray}\label{f1pn}
F_1^{p,n}(Q^2)&=&\left(1+\frac{Q^2}{4M^2}\right)^{-1}~\left[G_E^{p,n}(Q^2)+\frac{Q^2}{4M^2}~G_M^{p,n}(Q^2)\right],\\
\label{f2pn}
F_2^{p,n}(Q^2)&=&\left(1+\frac{Q^2}{4M^2}\right)^{-1}~\left[G_M^{p,n}(Q^2)-G_E^{p,n}(Q^2)\right].
\end{eqnarray}
Initially, it was observed from the experimental data of the electromagnetic scattering that the Sachs' form factors may have 
a dipole form. However, with the development of electron beam accelerator experiments, it was observed that the Sachs' form 
factors deviate from the dipole form. Galster et al.~\cite{Galster:1971kv} parameterized the deviated Sachs' form 
factors as
\begin{eqnarray}
G_E^p(Q^2)&=&G_D(Q^2) \qquad \qquad \qquad G_M^p(Q^2)=(1+\mu_p)G_{D}(Q^2) \nonumber\\
G_M^n(Q^2)&=&\mu_nG_{D}(Q^2) \qquad \qquad ~~~G_E^n(Q^2)=-(\frac{Q^2}{4M^2})\mu_nG_{D}(Q^2)\xi_n \nonumber \\
\xi_n&=&\frac{1}{\left(1-\lambda_n\frac{q^2}{4M^2}\right)},\qquad \qquad ~~~ G_{D} (Q^2) = \frac{1}{\left(1+ \frac{Q^2}
{M_{V}^2}\right)^{2}},\nonumber
\end{eqnarray}
with $ \mu_p=1.7927\mu_N,~      \mu_n=-1.913\mu_N,~        M_V=0.84 \text{ GeV}$ and $\lambda_n=5.6$.

Recently for $G_E^{p,n}(Q^2)$ and $G_M^{p,n}(Q^2)$, various parameterizations are available in the literature like BBBA05 
parameterized by Bradford et al.~\cite{Bradford:2006yz}, BBA03 parameterized by Budd et al.~\cite{Budd:2004bp}, Alberico et 
al.~\cite{Alberico:2008sz}, Bosted~\cite{Bosted:1994tm}, modified Galster parameterization given by Platchkov et 
al.~\cite{Platchkov:1989ch}, Kelly~\cite{Kelly:2004hm}, and modified Kelly parameterization given by Punjabi et 
al.~\cite{Punjabi:2015bba}. We have used, in our numerical calculations, the parameterization given by Bradford et 
al.~\cite{Bradford:2006yz}.

The vector form factors for NC induced processes are obtained as 
\begin{eqnarray}
\tilde{f}^p_{1,2} (Q^2) &=&\left(\frac{1}{2}-2\sin^2\theta_W \right) F^p_{1,2} (Q^2)-\frac{1}{2}F^n_{1,2} (Q^2)-
\frac{1}{2}F^s_{1,2} (Q^2),  \\
\tilde{f}^n_{1,2} (Q^2)&=&\left(\frac{1}{2}-2\sin^2\theta_W \right)F^n_{1,2} (Q^2)-\frac{1}{2}F^p_{1,2}(Q^2)-
\frac{1}{2}F^s_{1,2} (Q^2),
\end{eqnarray}
where $\theta_{W}$ is the Weinberg angle, and $F^s_{1} (Q^2)$ and $F^s_{2} (Q^2)$ are the strangeness vector form factors, 
which are discussed later in this section.

\item [(ii)] {\bf Axial-vector form factor}\\
The isovector axial-vector form factor is parameterized as
\begin{equation}\label{fa}
g_1(Q^2)=g_A(0)~\left(1+\frac{Q^2}{M_A^2}\right)^{-2},
\end{equation}
where $g_{A}(0)$ is determined experimentally from the $\beta$ decay of neutron. $M_A$ is known as the axial-dipole mass and 
is obtained from the QE neutrino and antineutrino scattering as well as from the pion electroproduction 
data~(Fig.~\ref{m10}). The dipole parameterization is extensively used in the analysis of various experiments in 
the QE (anti)neutrino scattering. However, a new parameterization based on $Z$-expansion has been recently proposed in 
literature~\cite{Gupta:2017dwj, Meyer:2016oeg}. Theoretically $g_{1}(Q^{2})$ is also calculated in various models of lattice 
gauge theory~\cite{Gupta:2017dwj, Green:2017keo, Alexandrou:2017hac, Yao:2017fym, Capitani:2017qpc}. 

\begin{figure}  
\begin{center}
	\includegraphics[height=7cm, width=7.5cm]{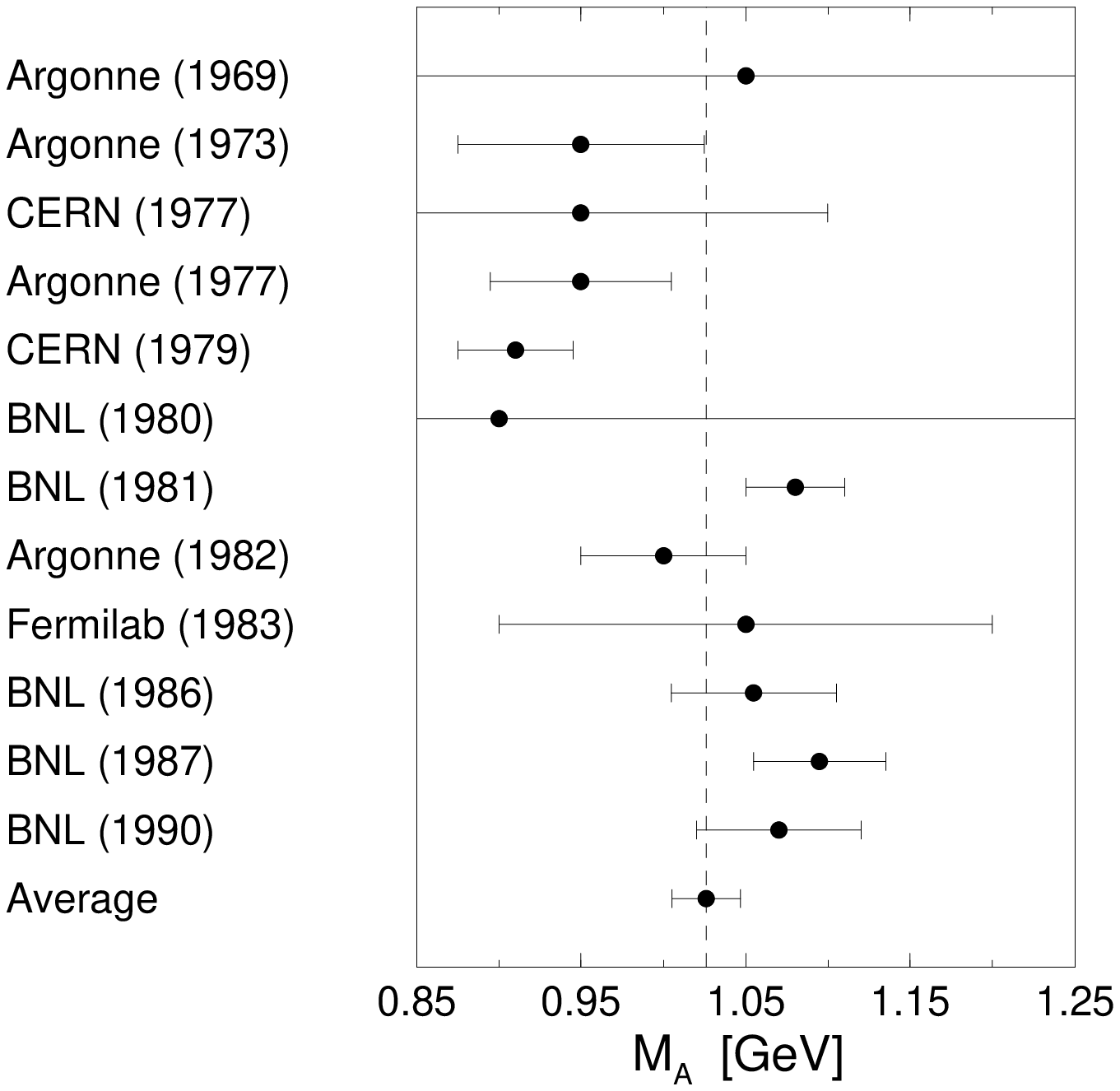}
	\includegraphics[height=7cm, width=7.5cm]{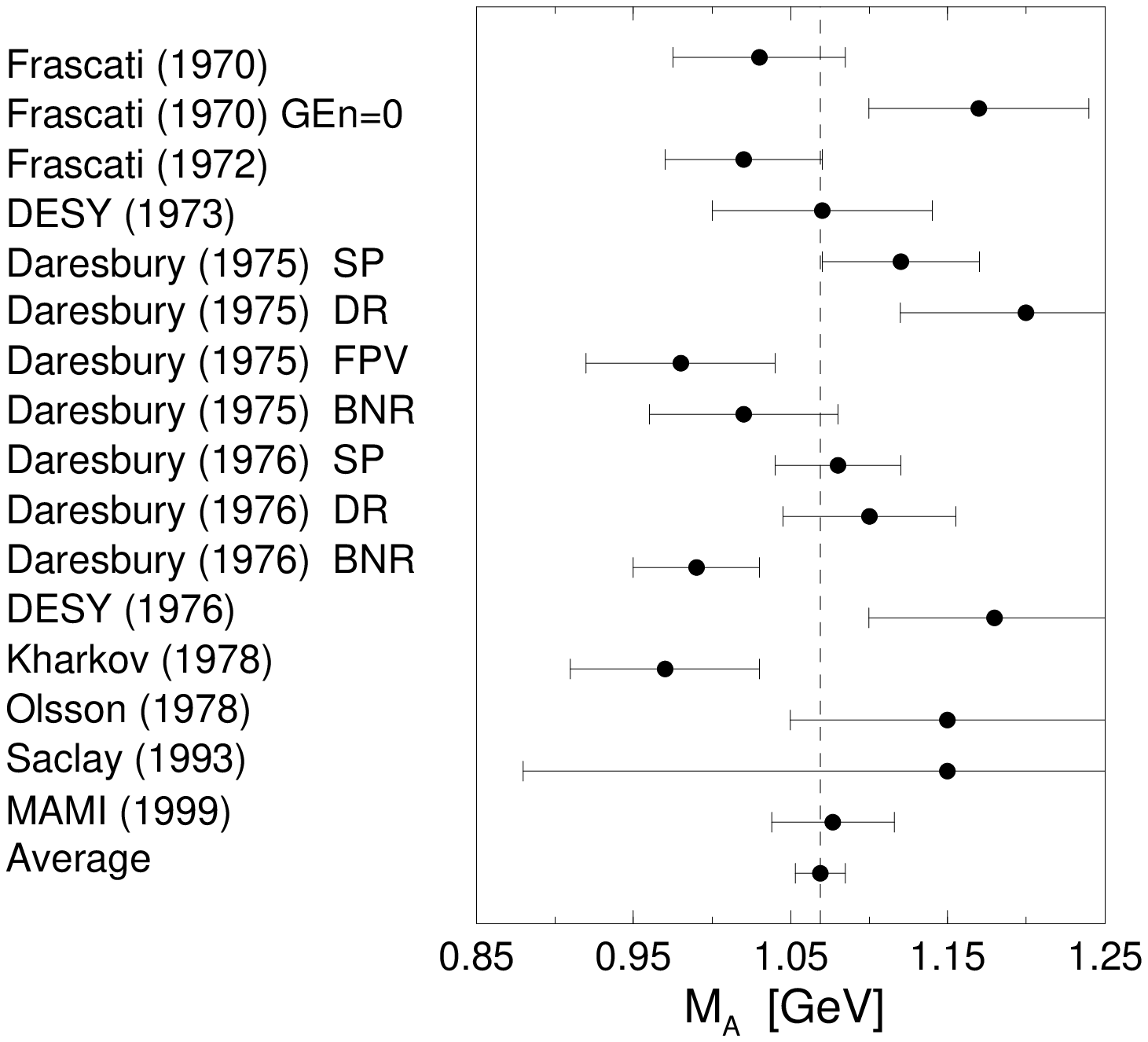}
\end{center}
\caption{Axial mass $M_A$ extractions from (quasi)elastic neutrino and antineutrino scattering experiments on hydrogen and 
deuterium targets~(left) and from the charged pion electroproduction experiments~(right). The weighted average from the left 
panel is $M_A= 1.026 \pm 0.021$ GeV and from the right panel is $M_A = 1.069 \pm 0.016$ GeV.}\label{m10}
\end{figure}

The numerical value of $M_A$ to be used in the calculations of neutrino-nucleon cross section has been a subject of intense 
discussion in the neutrino physics community in recent years and a wide range of the values of $M_A$ has been discussed in 
the literature~\cite{Alvarez-Ruso:2014bla, Morfin:2012kn, Gallagher:2011zza}. The old data available on (anti)neutrino 
scattering on hydrogen and deuterium targets~\cite{Miller:1982qi, Baker:1981su, Kitagaki:1983px} reanalyzed by Bodek et 
al.~\cite{Bodek:2007ym} gives a value of $M_A=1.014\pm0.014$ GeV, while the analysis of the same data by Meyer et 
al.~\cite{Meyer:2016oeg} gives a value in the range of 1.02-1.17 GeV depending upon which, data of ANL~\cite{Miller:1982qi}, 
BNL~\cite{Baker:1981su} and FNAL~\cite{Kitagaki:1983px} experiments are considered. Bernard {  et al.}~\cite{Bernard:2001rs} 
had earlier reanalyzed the data of the neutrino and antineutrino scattering on the hydrogen and deuterium targets as well as 
the electroproduction data and got the best $\chi^{2}$ fit for $M_A$ as:
\begin{equation*}
M_{A} = 1.026 \pm 0.021 \text{ GeV}.
\end{equation*}
Since then high statistics data on QE neutrino-nucleus scattering have been obtained and analyzed from 
neutrino and antineutrino scattering on the nuclear targets both at the low and intermediate energies. The data from 
NOMAD~\cite{Lyubushkin:2008pe} and MINERvA~\cite{MINERvA:2013bcy, MINERvA:2013kdn} favor a lower value of $M_A$ around 1.03 GeV, 
while the data from MiniBooNE~\cite{MiniBooNE:2007iti, MiniBooNE:2010xqw, MiniBooNE:2010bsu}, MINOS~\cite{Dorman:2009zz, 
MINOS:2014axb}, K2K~\cite{K2K:2006odf}, T2K~\cite{T2K:2015gcs} and SciBooNE~\cite{SciBooNE:2010slc, MiniBooNE:2012meu} favor 
a higher value of $M_A$ which lies in the range of 1.2--1.35 GeV. The suggested values of $M_{A}$ from these experiments have 
been tabulated in Table-\ref{tab:axial_mass:MA}.  Since the data from NOMAD~\cite{Lyubushkin:2008pe} and 
MINERvA~\cite{MINERvA:2013bcy} collaborations are at relatively higher energies than the data from the other experiments, the higher value of $M_{A}$ could be the manifestation of NME in the region of low energies of few hundred MeV. This has been discussed in recent literature. Alternatively it 
could be an indication of an energy dependent $M_{A}$. Such a possibility and the energy dependence of $M_{A}$ has recently 
been discussed by Kuzmin et al.~\cite{Kuzmin:2007kr, Kakorin:2021axo}.

In the case of NC induced reactions, the axial-vector form factor for the nucleon is given by:
\begin{equation}
\tilde{g}_1^{p,n} (Q^2)=\pm\frac{1}{2}g_1 (Q^2)-\frac{1}{2}F_A^s (Q^2), 
\end{equation}
where $g_{1} (Q^2)$ is given in Eq.~(\ref{fa}) with $M_{A}=1.026$ GeV and $F_A^s (Q^2)$ is the strangeness axial-vector form 
factor.
\begin{table}
\begin{center}
\begin{tabular}{|c|c|c|c|}  \hline 
	Experiment & $M_A~(GeV)$ & Experiment & $M_A~(GeV)$\\ \hline \hline 
	MINERvA~\cite{MINERvA:2013bcy, MINERvA:2013kdn} & 0.99 & SciBooNE~\cite{SciBooNE:2010slc} & 1.21$\pm$0.22 \\ 
	\hline
	NOMAD~\cite{Lyubushkin:2008pe} & 1.05$\pm$0.02$\pm$0.06 & K2K-SciBar~\cite{K2K:2006odf} & 1.144$\pm$0.077 \\ 
	\hline
	MiniBooNE~\cite{MiniBooNE:2007iti, MiniBooNE:2010xqw, MiniBooNE:2010bsu} & 1.23$\pm$0.20 & 
	K2K-SciFi~\cite{K2K:2006odf} & 1.20$\pm$0.12\\ \hline
        MINOS~\cite{Dorman:2009zz, MINOS:2014axb} & 1.19$(Q^2>0)$ &  World Average & 1.026$\pm$ 
        0.021~\cite{Bernard:2001rs} \\ 
	& 1.26$(Q^2>0.3GeV^2)$ & & 1.014$\pm 0.014$~\cite{Bodek:2007ym} \\  \hline
\end{tabular}
\end{center}
\caption{Recent measurements of the axial dipole mass($M_A$).}\label{tab:axial_mass:MA}
\end{table}

\item [(iii)] {\bf Pseudoscalar form factor}\\
In CC sector where PCAC is assumed, the pseudoscalar form factor $g_3(Q^2)$ is dominated by the pion pole dominance 
of the divergence of the axial-vector current~(PDDAC)  
and is given using the Goldberger-Treiman relation~\cite{Goldberger:1958vp}
as
\begin{equation}\label{fp_NN}
g_3(Q^2)=\frac{2M^2 g_1(Q^2)}{m_\pi^2 + Q^2}.
\end{equation}
However, in the literature, there are various other versions of the pseudoscalar form factor like the one in 
Ref.~\cite{Schindler:2006it}:
\begin{eqnarray}\label{fp1}
g_3(Q^2)&=&\frac{M}{Q^2}\left[\left(\frac{2 m_\pi^2 f_\pi}{m_\pi^2 + Q^2}\right)\left(\frac{M g_A(0)}{f_\pi} +
\frac{g_{\pi NN}(0)\Delta Q^2}{m_\pi^2} \right)+2 M g_1(Q^2) \right],
\end{eqnarray}
where $g_{\pi NN}(0)=13.21$, $f_\pi=92.42~\text{MeV}$ and $\Delta=1+\frac{M g_A(0)}{f_\pi g_{\pi NN}(0)}$.

Pseudoscalar form factor is also calculated in the chiral perturbation theory and is given by~\cite{Schindler:2006it, Tsapalis:2007qqc}
\begin{equation}\label{fp2}
g_{3}(0)=\frac{2M g_{\pi NN}(0) f_\pi}{m_\pi^2 + Q^2}+\frac{g_A(0) M^2 r_A^2}{3},
\end{equation}
where axial radius $r_A=\frac{2\sqrt{3}}{M_A}$.

The contribution of the pseudoscalar form factor to 
the cross section in the QE reactions is proportional to the square of the lepton mass and hence, it vanishes in the case of NC
interactions. 

\item [(iv)] {\bf Second class current form factors}\\
In Section~\ref{had_symmetry}(v), we have discussed G-parity and the classification of the first and the SCC. Here, we discuss 
the parameterization of the form factor associated with the SCC. In 
the $\Delta S = 0$ sector, the violation of G-parity due to the difference in the masses of $u$ and $d$ quarks or the 
intrinsic charge symmetry violation of the strong interaction, is very small, and the form factors $f_3 (Q^2)$ and $g_2 (Q^2)$ 
are expected to be very small. Moreover, in the vector sector, the hypothesis 
of CVC predicts $f_3 (Q^2) = 0$. However, in the axial-vector sector there is no such constraint on the form factor $g_2 
(Q^2)$ and it could be nonvanishing albeit small. It is because of this reason that most of the experiments in $\Delta S = 
0$ sector are analyzed for the search of SCC assuming $f_3 (Q^2)=0$ with a nonvanishing $g_2 
(Q^2)$ which is found to be quite small. Generally, the form factor $g_{2} (Q^{2})$, in analogy with $g_{1}(Q^{2})$, is parameterized 
as
\begin{equation}\label{g2}
g_2(Q^2)=g_2(0)~\left[1+\frac{Q^2}{M_2^2}\right]^{-2},
\end{equation}
where for simplicity $M_{2}=M_{A}$. 

This form factor $g_{2}(Q^{2})$ may also give information about the time reversal invariance~(TRI). If TRI is assumed, then 
$g_{2}(Q^2)$ must be real while in the absence of TRI the form factor $g_{2}(Q^{2})$ can be taken as imaginary. We have 
explored the possibility of both real and imaginary $g_{2}(Q^{2})$, and discussed the effect of TRI. { In the numerical calculations, we have taken the real and imaginary values of $g_2(0)$ in the range $-3$ to 3, which have been guided 
 phenomenologically by the works of Fearing et al.~\cite{Fearing:1969nr}, and Berman and 
 Veltman~\cite{BERMAN1964275}, and discussed recently in Refs.~\cite{Fatima:2018tzs, Fatima:2018wsy}. 
 However, the experimental limits of $g_{2}(0)$ obtained from the muon capture and high precision $\beta$ decays are too stringent and lie between $10^{-3} - 10^{-2}$~\cite{Cirigliano:2013xha}. }

\item [(v)] {\bf Strangeness form factors}
\begin{itemize}
 \item [(a)] {\bf Strangeness vector form factors}\\
 The strangeness vector form factors $F_{1}^{s} (Q^2)$ and $F_{2}^{s} (Q^2)$ may be redefined in terms of the strangeness 
Sachs' electric and magnetic form factors as:
\begin{eqnarray}
G_{E}^{s} (Q^2) = F_{1}^{s} (Q^2) - \tau F_{2}^{s}  (Q^2), \qquad \qquad 
G_{M}^{s} (Q^2) = F_{1}^{s} (Q^2) + F_{2}^{s} (Q^2),
\end{eqnarray}
where $\tau=\frac{Q^2}{4M^{2}}$. At $Q^2 = 0$, the Sachs' electric form factor gives the net strangeness of the 
nucleon, i.e. $ G_{E}^{s} (0) =0$. At low 
momentum transfer, the electric form factor is expressed in terms of $\rho^{s}$, i.e.
\begin{eqnarray}
\rho^{s} = \frac{d G_{E}^{s} (Q^2)}{d Q^{2}} = - \frac{1}{6} \langle r_{s}^2\rangle,
\end{eqnarray}
where $\langle r_{s}^2\rangle$ is the strangeness radius. Similarly, at $Q^2 = 0$, $G_{M}^{s} (Q^2) = \mu^{s}$, the 
strangeness magnetic moment. Therefore, these two parameters $\rho^{s}$ and $\mu^{s}$ determine NC form 
factors $F^s_{1} (Q^2)$ and $F^s_{2} (Q^2)$ in the low $Q^{2}$ region. The $Q^2$ dependence of $G_{E}^{s} (Q^2)$ and 
$G_{M}^{s} (Q^2)$ is parameterized as:
\begin{eqnarray}
G_{E}^{s} = \frac{\rho^{s} \tau}{\left(1+\frac{Q^2}{\Lambda_{E}^2} \right)}, \qquad G_{M}^{s} = \frac{\mu^{s}}{\left(1+
	\frac{Q^2}{\Lambda_{M}^2} \right)}, 
\end{eqnarray}
where the best fits for $\rho^{s}$ and $\mu^{s}$ assuming $\Lambda_{E,M}^{s}$ to be very large, are given 
as~\cite{Pate:2010kz}:
\begin{equation*}
\rho^{s} = 0.13 \pm 0.21 \qquad \text{and}\qquad \mu^{s} = 0.035 \pm 0.053.
\end{equation*}
 \item [(a)] {\bf Strangeness axial-vector form factor}\\
 The strangeness axial-vector form factor $g_{1}^{s} (Q^2)$ is taken to be of the dipole form:
\begin{eqnarray}
g_1^s (Q^2) = \frac{\Delta s}{\left(1 + \frac{Q^2}{M_A^2}\right)^2} ,
\end{eqnarray}
where $\Delta s = 0.08$ is the strange quark contribution to the nucleon spin~\cite{MiniBooNE:2010xqw, Formaggio:2012cpf}.
\end{itemize}
\end{itemize}

\subsubsection{Cross sections for charged current processes}
\begin{figure}
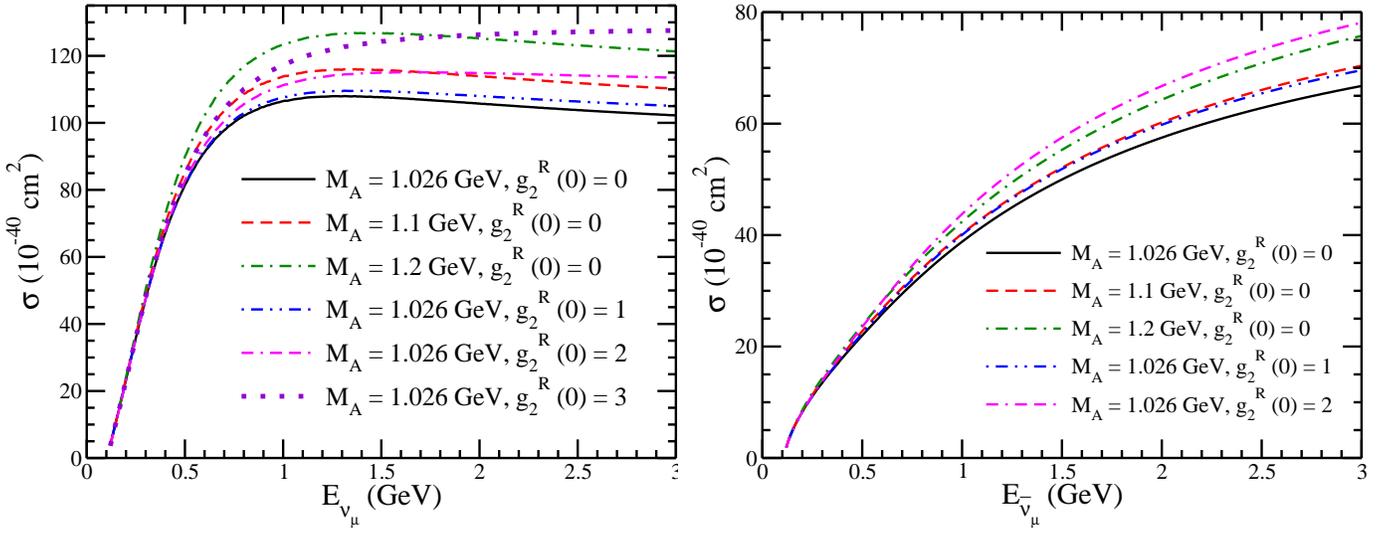
  
\centering
	\includegraphics[height=7cm,width=9cm]{total-s-g2-Ma-proton-2.eps}
	\includegraphics[height=7cm,width=9cm]{total-s-g2-Ma-neutron-1.eps} 
\caption{$\sigma~ \text{vs}~E_{{\nu}_{\mu}(\bar{\nu}_\mu)}$ for the process ${{\nu}_\mu + n \rightarrow \mu^- + p}$~(left 
panel) and ${\bar{\nu}_\mu + p \rightarrow \mu^+ + n}$~(right panel) for the different combinations of $M_A$, and ${{g_2^R 
(0)}}$ viz. $M_A = 1.026$ GeV and ${{g_2^{R} (0)}} = 0$~(solid line), $M_A = 1.1$ GeV and ${{g_2^{R} (0)}} = 0$~(dashed 
line), $M_A = 1.2$ GeV and ${{g_2^{R} (0)}} = 0$~(dashed-dotted line), $M_A = 1.026$ GeV and ${{g_2^{R} (0)}} = 
1$~(double-dotted-dashed line), $M_A = 1.026$ GeV and ${{g_2^{R} (0)}} = 2$~(double-dashed-dotted line) and $M_A = 
1.026$~GeV and ${{g_2^R (0)}} = 3$~(dotted line)~\cite{Fatima:2018tzs}.} \label{figb}
\end{figure}
The differential scattering cross section for the QE (anti)neutrino scattering from the free nucleons is then 
calculated using Eq.~(\ref{dsig:QE}) and the total cross section is obtained by integrating $\frac{d\sigma}{dQ^2}$ over $Q^2$. 
These results have been discussed in detail in Refs.~\cite{Fatima:2018tzs, Fatima:2020pvv}. It has been observed that
\begin{itemize}
 \item [(i)] in the case (anti)neutrino induced $\Delta S=0$ QE scattering processes, the total as well as the 
 differential cross sections are almost insensitive to the change in the parameterization of the vector form factors, i.e., 
 by taking into account the various parameterizations for the Sachs' electric and magnetic form factors~\cite{Fatima:2018tzs}.
 
 \item [(ii)] in the case of (anti)neutrino scattering from the free nucleon the effect of pseudoscalar form factor is almost 
 negligible in the case of $\nu_{l}, \bar{\nu}_{l};~(l=e,\mu)$ induced reactions~\cite{Fatima:2018tzs}. However, in the case 
 of $\nu_{\tau}, \bar{\nu}_{\tau}$ scattering~\cite{Fatima:2020pvv}, it has been observed that there is some dependence of the 
 pseudoscalar form factor on the differential cross section for neutrino induced reactions. Moreover, in the case of 
 $\bar{\nu}_{\tau}$ induced reactions, the different choice of the pseudoscalar form factors leads to a variation of about 50\% in 
 $\frac{d\sigma}{dQ^2}$ in the threshold region which increases with the increase in antineutrino energy.
 
 \item [(iii)] the cross section for CCQE reactions increases with increase in the value of $M_{A}$, for example, an 
 increase~(decrease) in the value of $M_{A}$ by 10\% increases~(decreases) $\sigma$ by about 15\% at $E_{\nu}=1$~GeV, which 
 becomes 10\% at $E_{\nu}=2$~GeV in the case of $\nu_{\mu}$ induced CCQE reactions while in the case of $\bar{\nu}_{\mu}$, 
 the variation with change in $M_{A}$ is about 10\% at $E_{\nu}=1$~GeV, which becomes 6\% at $E_{\nu}=2$~GeV, as may be observed 
 from Fig.~\ref{figb}. Similar observations have been made in the case of $\nu_{\tau}, \bar{\nu}_{\tau}$ induced reactions~\cite{Fatima:2020pvv}. 
 
 \item [(iv)] the presence of SCC form factor $g_{2}(Q^2)$ increases the cross sections for both neutrino and antineutrino induced processes.
\end{itemize}
Since the cross section for CCQE processes is sensitive to both $M_{A}$ and $g_{2}(Q^2)$, therefore, the dependence of the 
total scattering cross section on $M_A$ is shown in Fig.~\ref{figb} by varying $M_{A}$ in the range 1.026--1.2 GeV, with or 
without the presence of SCC by varying $g_{2}^{R}(0)$ in the range $0-3$, using the 
BBBA05~\cite{Bradford:2006yz} parameterization of the Sachs' electric and magnetic form factors. It may be observed from the 
figure that:
\begin{itemize}
 \item [(a)] in the absence of SCC~($g_{2}^{R}(0) = 0$)~\cite{Fatima:2018tzs}, the cross 
 section for both neutrino and antineutrino induced reactions increases with increase in the value of $M_{A}$ as discussed 
 above.
 
 \item [(b)] for the  process ${{\nu}_\mu + n \longrightarrow \mu^- + p}$, the results obtained 
 by taking $M_A = 1.1$ GeV and $g_2^R(0) = 0$ are comparable to the results obtained with $M_A = 1.026$ GeV and $g_2^R(0) = 
 2$, whereas the results obtained by taking $M_A = 1.2$ GeV and $g_2^R(0) = 0$ are comparable to the results obtained using 
 $M_A = 1.026$ GeV and $g_2^R(0) = 3$.
 
 \item [(c)] for the process ${\bar{\nu}_\mu + p \longrightarrow \mu^+ + n}$, the results 
 obtained by taking $M_A = 1.1$ GeV and $g_2^R(0) = 0$ are comparable to the results obtained with $M_A = 1.026$ GeV and 
 $g_2^R(0) = 1$, whereas the results obtained by taking $M_A = 1.2$ GeV and $g_2^R(0) = 0$ are slightly lower than the results 
 obtained using $M_A =1.026$ GeV and $g_2^R(0) = 2$. Thus, a higher value of $\sigma (E_{\bar{\nu}_\mu})$ may be obtained by 
 either taking a nonzero value of $g_2 (0)$ or increasing the value of $M_A$. 
 
 \item [(d)] The cross section measurements may give information only about the nonzero value of $g_2 (0)$ irrespective of the 
 nature of the SCC current i.e. with or without time reversal invariance. 
However, one may obtain the nature of the SCC by measuring the polarization observables which gives different results with 
the real and imaginary values of $g_2 (0)$, corresponding to the SCC with or without time reversal invariance, and this has 
been discussed by us in brief in Section~\ref{sec:polarization}. For more discussion, readers are referred to Fatima et 
al.~\cite{Fatima:2018tzs, Fatima:2020pvv}.
\end{itemize}

\subsection{Quasielastic hyperon production}\label{qe_hyperon}
\begin{figure}  
\begin{center}
	\includegraphics[height=3cm,width=6cm]{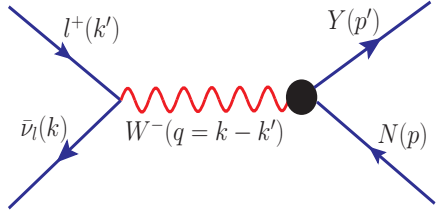}
\caption{Feynman diagram for the process $\bar{\nu}_l (k) + N (p) \rightarrow l^+ (k^\prime) + Y (p^\prime)$, where 
$N(=p,n)$ and $Y(=\Lambda, \Sigma^0, \Sigma^-)$ represent the initial nucleon and the final hyperon, respectively.}
\label{fyn_hyp}
\end{center}
\end{figure}

The following processes are induced when an antineutrino interacts with a nucleon to produce a hyperon and an
antilepton~(Fig.~\ref{fyn_hyp}):
\begin{eqnarray}\label{process1}
\bar{\nu}_l (k) + p (p) &\rightarrow& l^+ (k^\prime) + \Lambda/\Sigma^{0} (p^\prime), \\
\label{process3}
\bar{\nu}_l (k) + n (p) &\rightarrow& l^+ (k^\prime) + \Sigma^- (p^\prime);\qquad l=e,\mu,\tau, 
\end{eqnarray}
where the quantities in the brackets represent the four momenta of the particles. 

\subsubsection{Matrix element and form factors}
The transition matrix element for the processes presented in Eqs.~(\ref{process1}) and (\ref{process3}) is written as
\begin{eqnarray}
\label{matrixelement}
{\cal{M}} = \frac{G_F}{\sqrt{2}} ~ l^\mu {{J}}_\mu.
\end{eqnarray}
The leptonic current~($l^\mu$) is given in Eq.~(\ref{lep_curr}). The hadronic current~($J_\mu$) for the QE hyperon 
production can be written in analogy with the antineutrino-nucleon scattering except that the mass of the final nucleon is 
replaced by the mass of the hyperon and the electroweak form factors of the nucleons are replaced by the $N-Y$ transition 
form factors. The general expression for $J_\mu$ is given in Eq.~(\ref{had_curr}) and the matrix elements of the 
vector~($V_\mu$) and the axial-vector~($A_\mu$) currents between a hyperon $Y(=\Lambda, \Sigma^0 \text{ and } \Sigma^-)$ and a 
nucleon $N=n,p$ are written as:
\begin{eqnarray}\label{vmu}
\langle Y(\vec{p}^{\,\prime}) | V_\mu| N(\vec{p}) \rangle &=& \sin \theta_c\bar{u}(\vec{p}^{\,\prime}) \left[ \gamma_\mu f_1^{NY}(Q^2) +
i\sigma_{\mu \nu} \frac{q^\nu}{M+M^\prime} f_2^{NY}(Q^2)+ \frac{2 ~q_\mu}{M+M^\prime} f_3^{NY}(Q^2) \right] u(\vec{p}), \\
\label{amu}
\langle Y(\vec{p}^{\,\prime}) | A_\mu| N(\vec{p}) \rangle &=& \sin \theta_c\bar{u} (\vec{p}^{\,\prime}) \left[ \gamma_\mu \gamma_5 
g_1^{NY}(Q^2) + i\sigma_{\mu \nu} \frac{q^\nu}{M+M^\prime} \gamma_5 g_2^{NY}(Q^2) +\frac{2 ~q_\mu} {M+M^\prime} \gamma_5 
g_3^{NY}(Q^2) \right] u(\vec{p}), 
\end{eqnarray}
where $M$ and $M^\prime$ are the masses of the nucleon and hyperon, respectively. $f_1^{NY}(Q^2)$, $f_2^{NY}(Q^2)$ and 
$f_3^{NY}(Q^2)$ are the vector, weak magnetic and induced scalar $N-Y$ transition form factors and $g_1^{NY}(Q^2)$, 
$g_2^{NY}(Q^2)$ and $g_3^{NY}(Q^2)$ are the axial-vector, induced tensor (or weak electric) and induced pseudoscalar form 
factors, respectively.

The transition matrix element squared is obtained as
\begin{equation}\label{matrix}
\overline{\sum} \sum |{\cal{M}}|^2 = \frac{G_F^2 }{2} {J}_{\mu \nu} {L}^{\mu \nu},
\end{equation} 
where $J_{\mu \nu}$ and $L^{\mu \nu}$ are obtained in a similar way, as mentioned in Section~\ref{CCQE}.

The weak transition form factors $f_{i} (Q^2)$ and $g_{i} (Q^2);~ i=1-3$ are determined using Cabibbo theory of $V-A$ 
interaction extended to the strange sector with the application of $SU(3)$ symmetry. The details are given in Appendix-\ref{Cabibbo}.
The expressions for the vector form factors in terms of the electromagnetic form factors $F_{1,2}^p(Q^2)$ and $F_{1,2}^n 
(Q^2)$ for the various processes given in Eqs.~(\ref{process1}) and (\ref{process3}), are given as~\cite{Athar:2020kqn, Fatima:2018tzs}:
\begin{eqnarray}
f_{1,2}^{p \Lambda}(Q^2)= -\sqrt{\frac{3}{2}}~F_{1,2}^p(Q^2),~~
f_{1,2}^{n \Sigma^-}(Q^2)= -\left[F_{1,2}^p(Q^2) + 2 F_{1,2}^n(Q^2) \right],  ~~ 
f_{1,2}^{p \Sigma^0}(Q^2)= -\frac{1}{\sqrt2}\left[F_{1,2}^p(Q^2) + 2 F_{1,2}^n(Q^2) \right].
\end{eqnarray}
The axial-vector form factors $g_i^{NY} (Q^2) (i=1,2,3)$ are expressed in terms of the two functions $F_i^{A} (Q^2)$ and 
$D_i^{A} (Q^2)$ corresponding to the antisymmetric and symmetric couplings of the two octets. But we express the form 
factors $g_i^{NY} (Q^2)$ in terms of $g_i (Q^2)$ and $x_i (Q^2)$, which are defined as~\cite{Athar:2020kqn, Fatima:2018tzs}: 
\begin{eqnarray}\label{gnp}
g_i (Q^2) = F_i^A (Q^2) + D_i^A (Q^2) = g_i^{np} (Q^2), \qquad \qquad
x_i (Q^2) = \frac{F_i^A (Q^2)}{F_i^A (Q^2) + D_i^A (Q^2)}; \qquad i=1-3
\end{eqnarray}
and the expressions for the axial-vector transition form factors for the various processes given in 
Eqs.~(\ref{process1}) and (\ref{process3}) are given as~\cite{Athar:2020kqn, Fatima:2018tzs}:
\begin{eqnarray}
\label{gplam}
g_{1,2}^{p \Lambda}(Q^2)= -\frac{1}{\sqrt{6}}(1+2x_{1,2}) g_{A,2}^{np} (Q^2), ~~
g_{1,2}^{n \Sigma^-}(Q^2)= (1-2x_{1,2})g_{A,2}^{np}(Q^2),  ~~
g_{1,2}^{p \Sigma^0}(Q^2)= \frac{1}{\sqrt2}(1-2x_{1,2})g_{A,2}^{np}(Q^2).
\end{eqnarray}
In the following we describe the explicit forms of the axial-vector form factors used for calculating the numerical results.
\begin{itemize}
\item [(a)] {Axial vector form factor $g_1^{NY} (Q^2)$:} \\
We note from Eq.~(\ref{gnp}), that $g_1 (Q^2)$ is the axial-vector form factor for $n \rightarrow p$ transition and is 
defined in Eq.~(\ref{fa}). The parameter $x_1 (Q^2)$ occurring in Eq.~(\ref{gplam}) for $g_1^{NY}(Q^2)$ 
($Y = \Lambda, \Sigma^0, \Sigma^-$) is determined at low $Q^2$ from the analysis of semileptonic hyperon decay and is 
found to be $x_1 (Q^2 \approx 0) = 0.364$. There is no experimental information about the $Q^2$ dependence of $x_1 (Q^2)$, 
therefore, we assume it to be constant i.e. $x_1 (Q^2) \approx x_1(0) = 0.364$ for convenience.

\item [(b)] {Second class current form factor $g_2^{NY} (Q^2)$:}\\
The expression for $g_2 (Q^2)$ for the hyperons $\Lambda, \Sigma^-, \Sigma^0$ are given in Eq.~(\ref{gplam}) 
in terms of $g_2^{np} (Q^2)$ and $x_2 (Q^2)$, where $g_2^{np} (Q^2)$ is parameterized in Eq.~(\ref{g2}). There is some 
information on $g_2^{np} (Q^2)$ from neutrino and antineutrino scattering off the nucleons. It is shown that the value of 
$g_2^{np} (0)$ is correlated with the value of $M_2$ used in the analysis. There exists theoretical calculations for the 
$g_2^{R(np)} (0)$ and $g_2^{R(NY)} (0)$ for $Y=\Lambda, \Sigma^-, \Sigma^0$. In the literature, various values of 
$g_{2}^{I}(0)$ for the nucleons and hyperons have been used, which are in the range 1--10~\cite{Fearing:1969nr, BERMAN1964275, 
10.1143/PTP.33.552}. However, there is no information about $x_2(Q^2)$. To see the dependence of $g_2^{R}(0)$ and $g_2^{I}(0)$ 
on the differential and the total scattering cross sections, we have varied $g_2^{R}(0)$ and $g_2^{I}(0)$ in the range of 
$0-3$ and use $M_2 = M_A$~\cite{Fatima:2018tzs}. For the $Q^2$ dependence of the form factor $g_2^{NY}(Q^2)$, we use 
the $SU(3)$ symmetric expressions for $g_2^{np}(Q^2)$ taken to be of the dipole form given in Eq.~(\ref{g2}) for the various 
transitions given in Eq.~(\ref{gplam}), treating $x_2 (Q^2)$ to be constant and assuming $x_2 = 
x_1$~\cite{Fatima:2018tzs}.
	
\item [(c)] {The induced pseudoscalar form factor $g_3^{NY} (Q^2)$:}\\
In general, the contribution of $g_3^{NY} (Q^2)$ to the (anti)neutrino scattering cross sections is proportional to $m_l^2$, 
where $m_l$ is the mass of the corresponding charged lepton, and is small in $e^{\pm}$ and $\mu^{\pm}$ productions but is 
significant in the processes involving $\tau^{\pm}$ leptons. For $g_3^{NY} (Q^2)$, Nambu~\cite{Nambu:1960xd} has given a 
generalized parameterization using PCAC and Goldberger-Treiman~(GT) relation for the $\Delta S = 1$ currents
\begin{equation}\label{g3_Nambu}
g_3^{NY}(Q^2)=\frac{(M + M^{\prime})^2}{2(m_K^2 + Q^2)}g_1^{NY}(Q^2),
\end{equation}
where $m_K$ is the mass of kaon and $g_1^{NY} (Q^2)$ is given in Eq.~(\ref{gplam}), for $Y = \Lambda, 
\Sigma^-, \Sigma^0$.

Another parameterization for the pseudoscalar form factor in the case of $\Delta S=1$ processes is given by Marshak et 
al.~\cite{Marshak:1969}:
\begin{equation}\label{g3_Marshak}
g_3^{NY}(Q^2)=\frac{(M+M^{\prime})^2}{2Q^2} \frac{g_{1}^{NY} (Q^2) (m_{K}^{2} + Q^2) - m_{K}^{2} g_{1}^{NY} (0)}{m_{K}^{2} + 
Q^2}.
\end{equation}
\end{itemize}

\subsubsection{Cross sections: Experimental results}
\begin{figure}
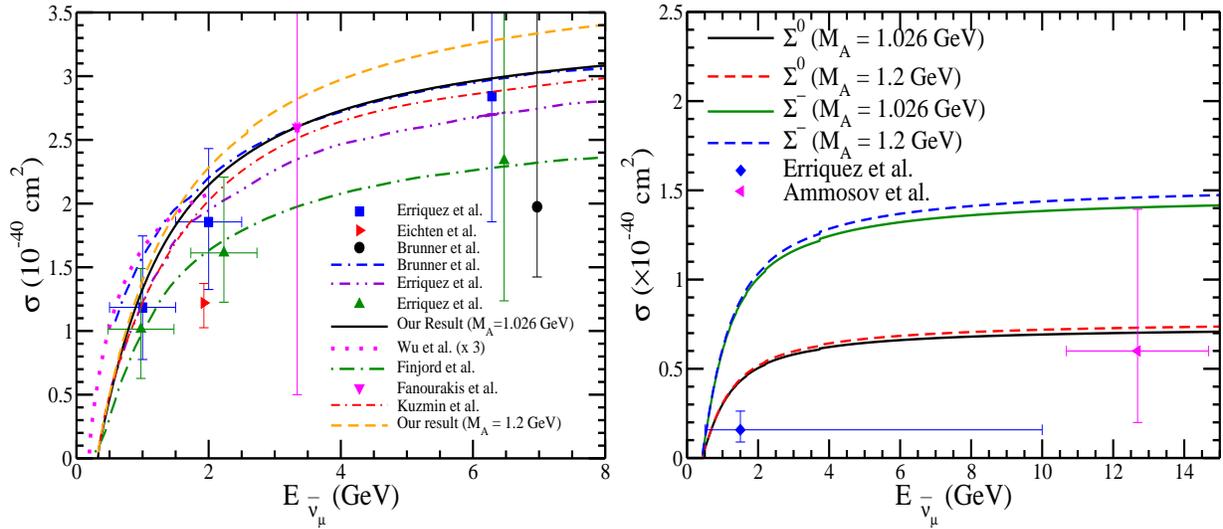
  
 \begin{center}
    \includegraphics[height=7cm,width=8cm]{xsec_comp2.eps} 
    \includegraphics[height=7cm,width=8cm]{xsctn_sig0_vs_enu.eps} 
    \end{center}
\caption{$\sigma$ vs. $E_{\bar{\nu}_\mu}$ for the $\Lambda$ production~(left panel), $\Sigma^{0}$ and $\Sigma^{-}$
production~(right panel) cross sections~\cite{Fatima:2018wsy}. Solid~(dashed) line represents the result using $M_A = 
1.026~(1.2)$ GeV. Experimental results for the process $\bar \nu_\mu p \to \mu^+ \Lambda$ (triangle 
right~\cite{Eichten:1972bb}, triangle up~\cite{Erriquez:1977tr}, square~\cite{Erriquez:1978pg}, triangle down~($\sigma = 
2.6^{+5.9}_{-2.1} \times 10^{-40} cm^2$)~\cite{Fanourakis:1980si}, circle~\cite{SKAT:1989nel}) and for the process 
$\bar{\nu}_\mu p \to \mu^+ \Sigma^0$~(diamond~\cite{Erriquez:1977tr}) are shown with error bars. Theoretical curves are of 
Kuzmin and Naumov~\cite{Kuzmin:2008zz}~(double dashed-dotted line), Brunner et al.~\cite{SKAT:1989nel}~(dashed line), Erriquez 
et al.~\cite{Erriquez:1978pg}~(dashed-double dotted line) obtained using Cabibbo theory with axial-vector dipole mass as 
1~GeV, 1.1 GeV and 1 GeV, respectively, while the results of Wu et al.~\cite{Wu:2013kla}~(dotted line) and Finjord and 
Ravndal~\cite{Finjord:1975zy}~(dashed dotted line) are obtained using quark model.}\label{free_hyperon}
 \end{figure}

The results for the hyperon production cross sections from the free nucleons given in Eqs.~(\ref{process1}) and (\ref{process3}) 
as a function of antineutrino energies are presented in Fig.~\ref{free_hyperon}. These results are presented for $\Lambda$, 
$\Sigma^-$ and $\Sigma^0$ production cross sections at the two values of $M_A$ viz. $M_A = 1.026$ GeV and 1.2 GeV. In this 
region, there is very little dependence of $M_A$ on the cross section for $\Sigma^-$ and $\Sigma^0$ productions, while in the 
case of $\Lambda$ production, the cross section increases with energy and the increase is about 5$\%$ at $E_{\bar{\nu}_\mu} = 
1$ GeV. In the case of free nucleon, the cross sections for $\bar \nu_\mu + n \rightarrow \mu^+ + \Sigma^- $ and 
$\bar{\nu}_{\mu} + p \rightarrow \mu^+ +  \Sigma^0$ are related by an isospin relation i.e. $\sigma(\bar \nu_\mu p \rightarrow 
\mu^+ \Sigma^0) = \frac12 \sigma(\bar \nu_\mu n \rightarrow \mu^+ \Sigma^-)$, while no $\Sigma^+$ is produced off the free 
nucleon target due to $\Delta S \neq \Delta Q$ rule. A comparison is made with available experimental results from 
CERN~\cite{Eichten:1972bb, Erriquez:1977tr, Erriquez:1978pg}, BNL~\cite{Fanourakis:1980si}, FNAL~\cite{Ammosov:1986jn, 
Ammosov:1986xv} and SKAT~\cite{SKAT:1989nel} experiments as well as with the theoretical calculations performed by Wu et 
al.~\cite{Wu:2013kla} and Finjord and Ravndal~\cite{Finjord:1975zy} using quark model and the calculations performed by 
Erriquez et al.~\cite{Erriquez:1978pg}, Brunner et al.~\cite{SKAT:1989nel} and Kuzmin and Naumov~\cite{Kuzmin:2008zz} based 
on the prediction using Cabibbo theory. A reasonable agreement with the experimental results can be seen.

\subsection{Polarization of final hadrons and leptons}\label{sec:polarization}
In the case of elastic $e^- p$ scattering, the polarized electron beam and the polarized proton target have played an 
important role in determining the vector form factors. In the weak sector, the vector form factors are expressed in terms 
of the electromagnetic form factors of the nucleons. In the axial-vector sector, the information on the form factors is 
obtained form the semileptonic decays of nucleons and hyperons at low $Q^2$, one may also obtain information about these 
form factors by measuring the polarization of the final hadron. In the case of the QE scattering, experimentally, 
it is difficult to study the polarization of the final nucleon as one requires the double polarization measurement. However, 
in the case of the QE hyperon production, it is easier to study the polarization observables as the produced 
hyperon decays into pions which gives information about the polarization of the final hyperon. The calculations for the 
polarization observables of the final hadrons and leptons produced in the $\Delta S = 0$ and $|\Delta S| = 1$ QE 
scattering of (anti)neutrinos with free nucleons have been done earlier~\cite{Block:1965, Block:1968} and summarized by 
Llewellyn Smith~\cite{LlewellynSmith:1971uhs} but recently these calculations have been done by Bilenky et 
al.~\cite{Bilenky:2013fra, 
Bilenky:2013iua}, Graczyk and Kowal~\cite{Graczyk:2021oyl, Graczyk:2019xwg, Graczyk:2019blt}, Tomalak~\cite{Tomalak:2020zlv}, 
Thorpe et al.~\cite{Thorpe:2020tym} and our group~\cite{Fatima:2018tzs, Fatima:2018wsy, Fatima:2020pvv, Fatima:2018gjy, 
Akbar:2016awk, Fatima:2021ctt, Fatima:2022tlf, Akbar:2017qsf} in the SM. In Refs.~\cite{Fatima:2018tzs, Fatima:2018wsy, 
Akbar:2016awk, Fatima:2021ctt}, we have calculated the polarization observables of the proton, neutron, $\Lambda$, and 
$\Sigma^{-}$ produced in the $\bar{\nu}_{\mu}$ induced QE processes. In Refs.~\cite{Fatima:2020pvv, Fatima:2022tlf}, 
we have studied the $\tau^{\pm}$ polarization in the processes $\nu_{\tau} (\bar{\nu}_{\tau}) + N \longrightarrow \tau^{\pm} 
+ N$ and $\bar{\nu}_{\tau} + N \longrightarrow \tau^{+} + \Lambda(\Sigma)$.

\subsubsection{Polarization of the final hadron}
The polarization 4-vector $\xi^\tau$ of the hadron produced in the final state in reactions (\ref{cc_quasi_reaction}), 
(\ref{process1}), and (\ref{process3}) is written as~\cite{Athar:2020kqn, Bilenky}:
\begin{equation}\label{polar}
\xi^{\tau}=\frac{\mathrm{Tr}[\gamma^{\tau}\gamma_{5}~\rho_{f}(p^\prime)]}
{\mathrm{Tr}[\rho_{f}(p^\prime)]},
\end{equation}
where the spin density matrix $\rho_f(p^\prime)$ corresponding to the final hadron of momentum $p^\prime$ is given by 
\begin{equation}\label{polar1}
\rho_{f}(p^\prime)= { L}^{\alpha \beta} ~\mathrm{Tr}[\Lambda(p'){\cal O}_{\alpha} \Lambda(p)\tilde{\cal O}_{\beta} 
\Lambda(p')],
\end{equation}
where $\Lambda (p^{\prime}) = \slashed{p}^{\prime} + M^{\prime}$ is the projection operator for spin $\frac{1}{2}$ 
fermions with momentum $p^{\prime}$.

Using the following relations:
\begin{eqnarray}\label{polar3}
\Lambda(p')\gamma^{\tau}\gamma_{5}\Lambda(p')=2M^\prime \left(g^{\tau\sigma}-\frac{p'^{\tau}p'^{\sigma}}{M^{\prime 2}}
\right)\Lambda(p')\gamma_{\sigma}\gamma_{5}, \qquad \quad
\Lambda(p^\prime)\Lambda(p^\prime) = 2M^\prime \Lambda(p^\prime),
\end{eqnarray}
where $M^{\prime}$ corresponds to the mass of the final hadron. $\xi^\tau$ defined in Eq.~(\ref{polar}) may be rewritten 
as~\cite{Athar:2020kqn, Bilenky}:
\begin{equation}\label{polar4}
\xi^{\tau}=\left( g^{\tau\sigma}-\frac{p'^{\tau}p'^{\sigma}}{M^{\prime 2}}\right) 
\frac{  { L}^{\alpha \beta}  \mathrm{Tr}
	\left[\gamma_{\sigma}\gamma_{5}\Lambda(p'){\cal O}_{\alpha} \Lambda(p)\tilde{\cal O}_{\beta} \right]}
{ { L}^{\alpha \beta} \mathrm{Tr}\left[\Lambda(p'){\cal O}_{\alpha} \Lambda(p)\tilde{\cal O}_{\beta} \right]}.
\end{equation}
Note that in Eq.~(\ref{polar4}), $\xi^\tau$ is manifestly orthogonal to $p^{\prime \tau}$, \textit{i.e.} $p^\prime \cdot 
\xi=0$. Moreover, the denominator is directly related to the differential scattering cross section given in 
Eq.~(\ref{dsig:QE}).
With ${ J}^{\alpha \beta}$ and ${ L}_{\alpha \beta}$ given in Eqs.~(\ref{had_tens}) and (\ref{lep_tens}), respectively, 
an expression for $\xi^\tau$ is obtained in terms of the 4-momenta of the particles. Here, we have considered two cases:

{\bf Case I: When time reversal invariance is assumed.}

\begin{figure}  
\begin{center}  
	\includegraphics[height=7cm,width=13cm]{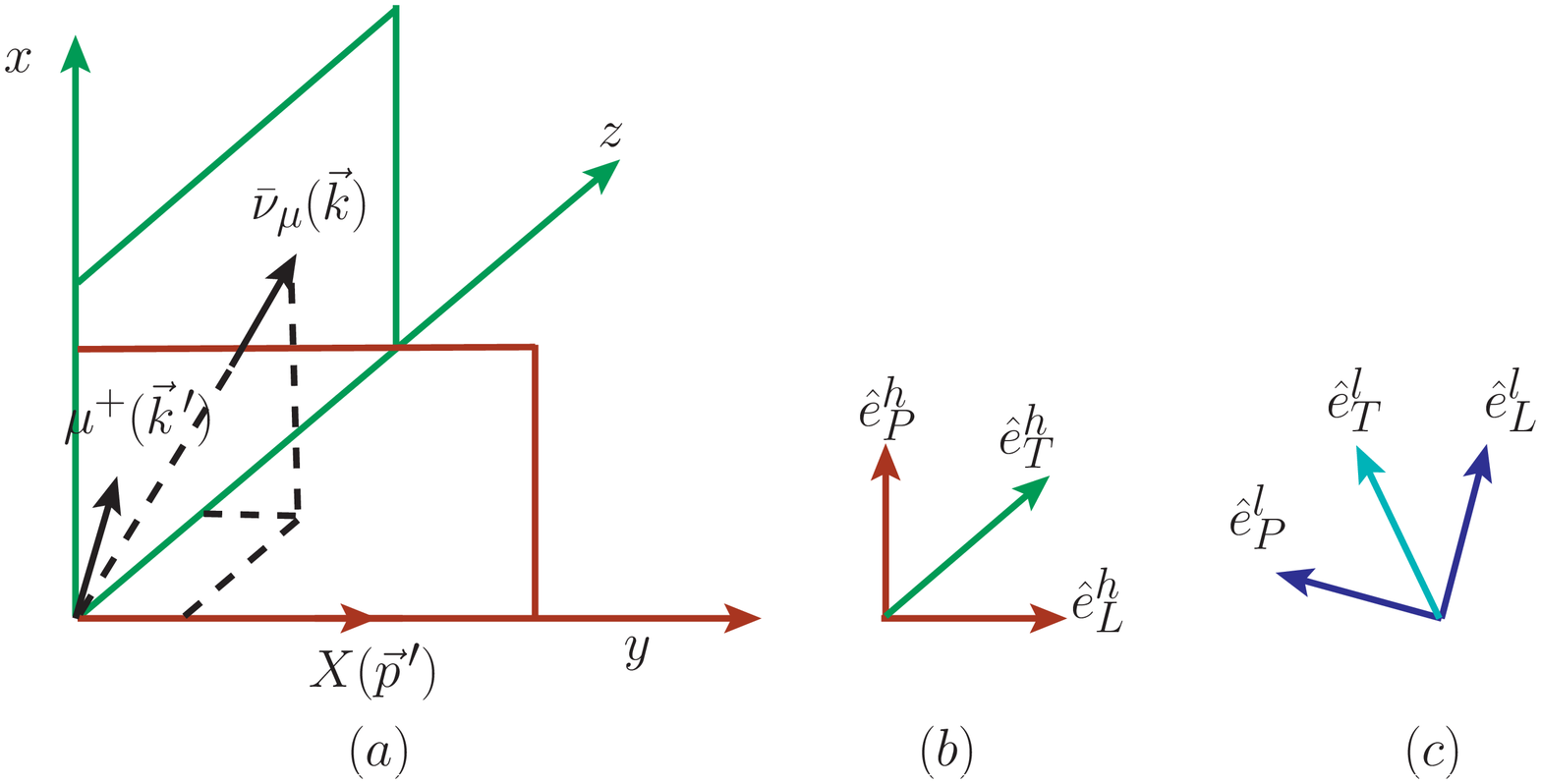}
\caption{(a) Momentum and polarization directions of the final baryon and the lepton. $\hat{e}_{L}^{h,l}$, $\hat{e}_{P}^{h,
l}$ and $\hat{e}_{T}^{h,l}$ represent the orthogonal unit vectors corresponding to the longitudinal, perpendicular and 
transverse directions with respect to the momentum of the final hadron in (b) and the final lepton in (c).}\label{TRI}
\end{center}
\end{figure}

The polarization vector $\xi^{\tau}$ defined in Eq.~(\ref{polar4}) is evaluated in the laboratory frame, i.e. when the 
initial nucleon is at rest, $\vec{p} = 0$, and the momentum directions are depicted in Fig.~\ref{TRI}(a). 
If the time reversal invariance is assumed then all the form factors defined in 
Eqs.~(\ref{vx}), (\ref{vy}), (\ref{vmu}), and (\ref{amu}) are real and $\xi^{\tau}$ is expressed as
\begin{equation}\label{3pol}
\vec{\xi} =\frac{\left[{A^h(Q^2)\vec{ k}} + B^h(Q^2){\vec{p}}\,^{\prime}  \right]}{N(Q^2)}, 
\end{equation}
where the expressions of $A^h(Q^2)$, $B^h(Q^2)$ and $N(Q^2)$ are given in Appendix-\ref{appendix1}, and 
are taken in the limit $f_{3} (Q^2) = 0$ and $g_{2}(0) 
= g_{2}^{R}(0)$ to ensure the time reversal invariance.

From Eq.~(\ref{3pol}), it follows that the polarization vector lies in the plane of reaction and there is no component of 
polarization in a direction perpendicular to the reaction plane. This is a consequence of time reversal invariance which 
makes the transverse component of polarization, perpendicular to the reaction plane, to vanish. We now expand $\vec{\xi}$ 
along the orthogonal directions, ${\hat{e}}_L^h$, ${\hat{e}}_P^h$ and ${\hat{e}_T^h}$ in 
the reaction plane corresponding to the longitudinal, perpendicular and transverse directions, defined as
\begin{equation}\label{vectors}
\hat{e}_{L}^h=\frac{\vec{ p}^{\, \prime}}{|\vec{ p}^{\,\prime}|},\qquad
\hat{e}_{P}^h=\hat{e}_{L}^h \times \hat{e}_T^h,\qquad   {\rm where}~~~~~ 
\hat{e}_T^h=\frac{\vec{ p}^{\,\prime}\times \vec{ k}}{|\vec{ p}^{\,\prime}\times \vec{ k}|},
\end{equation}
and have depicted in Fig.~\ref{TRI}(b). We then write $\vec{\xi}$ as:
\begin{equation}\label{polarLab}
\vec{\xi}=\xi_{L} \hat{e}_{L}^h+\xi_{P} \hat{e}_{P}^h,
\end{equation}
such that the longitudinal and perpendicular components of $\vec{\xi}$ in the laboratory frame are 
given by
\begin{equation}\label{PL}
\xi_L(Q^2)=\vec{\xi} \cdot \hat{e}_L^h,\qquad \xi_P(Q^2)= \vec{\xi} \cdot \hat{e}_P^h.
\end{equation}
From Eq.~(\ref{PL}), the longitudinal $P_L^h(Q^2)$ and perpendicular $P_P^h(Q^2)$ components of the $\vec{\xi}$
defined in the rest frame of the final hadron are given by 
\begin{equation}\label{PL1}
P_L^h(Q^2)=\frac{M^\prime}{E^{\prime}} \xi_L(Q^2), \qquad P_P^h(q^2)=\xi_P(Q^2),
\end{equation}
where $\frac{M^\prime}{E^{\prime}}$ is the Lorentz boost factor along ${\vec p}\, ^\prime$. With the help of 
Eqs.~(\ref{3pol}), (\ref{vectors}), (\ref{PL}) and (\ref{PL1}), the longitudinal $P_L^h(Q^2)$ and perpendicular $P_P^h 
(Q^2)$ components of polarization are calculated to be~\cite{Athar:2020kqn, Fatima:2018tzs}:
\begin{eqnarray}
P_L^h (Q^2) &=& \frac{M^\prime}{E^{{\prime}}} \frac{A^h(Q^2) \vec{k} \cdot \vec{p}^{\,\prime} + B^h (Q^2) 
	|\vec{p}^{\,\prime}|^2}{N(Q^2)~|\vec{p}^{\,\prime}|},\label{Pl} \\
P_P^h (Q^2) &=& \frac{A^h(Q^2) [(\vec{k} \cdot \vec{p}^{\,\prime})^2 - |\vec{k}|^2 |\vec{p}^{\,\prime}|^2]}{N(Q^2)~
	|\vec{p}^{\,\prime}| ~ |\vec{p}^{\,\prime}\times \vec{k}|}.\label{Pp}
\end{eqnarray}

{\bf Case II: When time reversal violation is assumed.}

In the absence of time reversal invariance, $\vec{\xi}$ is calculated as~\cite{Athar:2020kqn, Fatima:2018tzs}
\begin{eqnarray}
\vec{\xi} &=& \frac{A^h(Q^2)\vec{ k} + B^h(Q^2){\vec{p}}\,^{\prime} + C^h (Q^2) M (\vec{k} \times {\vec{p}}\,^{\prime})}
{N(Q^2)},
\end{eqnarray}
where the expressions of $C^h(Q^2)$ is given in Appendix-\ref{appendix1}.

The $\vec{\xi}$ may be written in terms of the longitudinal, perpendicular and transverse components as
\begin{equation}\label{polarLab_1}
\vec{\xi}=\xi_{L} \hat{e}_{L}^h+\xi_{P} \hat{e}_{P}^h +\xi_{T} \hat{e}_{T}^h,
\end{equation}
where the unit vectors are defined in Eq.~(\ref{vectors}). The longitudinal and perpendicular components are given in 
Eqs.~(\ref{Pl}) and (\ref{Pp}), respectively. The transverse component of polarization in the rest frame of the final hadron 
is given as
\begin{equation}\label{xi_t}
P_T(Q^2) = \xi_T(Q^2) = \vec{\xi}.\hat{e}_T .
\end{equation}
Using Eqs.~(\ref{vectors}) and (\ref{polarLab_1}) in Eq.~(\ref{xi_t}), we obtain~\cite{Athar:2020kqn, Fatima:2018tzs}
\begin{equation}
P_T^h (Q^2) = \frac{C^h(Q^2) M [(\vec{k} \cdot \vec{p}^{\,\prime})^2 - |\vec{k}|^2 |\vec{p}^{\,\prime}|^2]}{N(Q^2)~
	|\vec{p}^{\,\prime} \times \vec{k}|}. \label{Pt}
\end{equation}
If the T invariance is assumed then all the vector and the axial-vector form factors are real and the expression for 
$C^h(Q^2)$ vanishes which implies that the transverse component of the polarization perpendicular to the production plane, 
$ P_T^h (Q^2)$ vanishes.

\begin{figure}
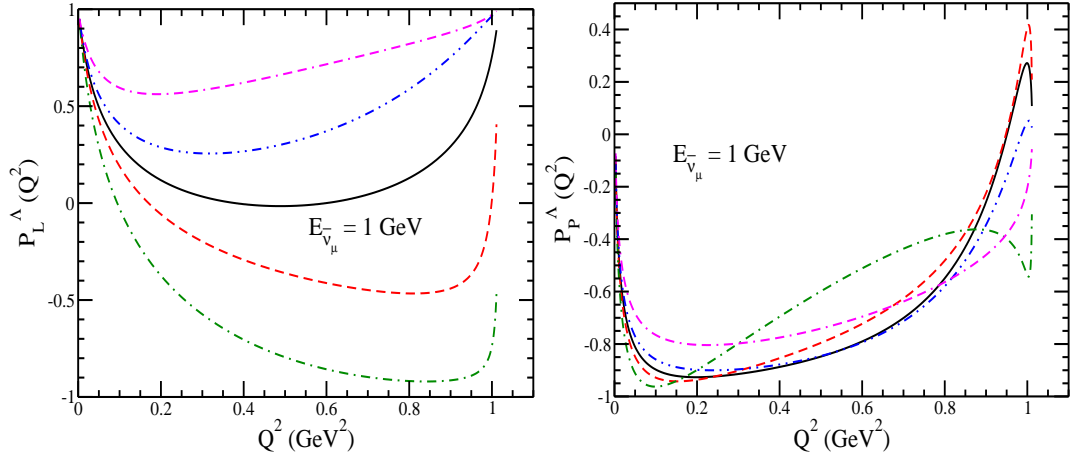
  
\begin{center}  
\includegraphics[height=6cm,width=7cm]{Pl-g2-Fp-1.eps}
\includegraphics[height=6cm,width=7cm]{Pp-g2-Nambu-1GeV-1.eps}
\caption{$P_L (Q^2)~vs.~ Q^2$ (left panel) and $P_P (Q^2)~vs.~ Q^2$ (right panel) for the process ${\bar{\nu}_\mu + p 
\rightarrow \mu^+ + {\Lambda}}$ at the incoming antineutrino energy, $E_{\bar{\nu}_{\mu}} =$ 1~GeV for the polarized 
$\Lambda$ in the final state, at the different values of $g_2^R (0)$ viz. $g_2^R (0) = $ 0~(solid line), 1~(dashed line), 
3~(dashed-dotted line), $-1$~(double-dotted-dashed line) and $-3$~(double-dashed-dotted line)~\cite{Fatima:2018tzs}.} \label{fig1_pol}
\end{center}
\end{figure} 
Using Eqs.~(\ref{Pl}), (\ref{Pp}) and (\ref{Pt}), the polarization components of the $\Lambda$ produced in the reaction 
$\bar{\nu}_{\mu} + p \longrightarrow \mu^{+} + \Lambda$ are calculated, where the expressions of $A^{h}(Q^2)$, $B^{h} 
(Q^2)$, and $C^h(Q^2)$ are given in Appendix~\ref{appendix1}. To see the dependence of $g_2^{R} (0)$ on the polarization 
observables, in Fig.~\ref{fig1_pol}, the results of $P_{L} (Q^2)$ and $P_{P} (Q^2)$ are presented as a function of $Q^2$ 
using $g_2^{R} (0) =$ 0, $\pm$1 and $\pm$3 at $E_{\bar{\nu}_{\mu}} =$ 1~GeV. It may be observed that $P_L (Q^2)$ shows 
large variation as we change $|g_2^{R} (0)|$ from 0 to 3. For example, in the peak region of $Q^2$, the difference is 
about 50$\%$ as $|g_2^R (0)|$ is changed from 0 to 3. In the case of $P_P (Q^2)$, $Q^2$ dependence is quite strong and 
similar to $P_L (Q^2)$.

\begin{figure}
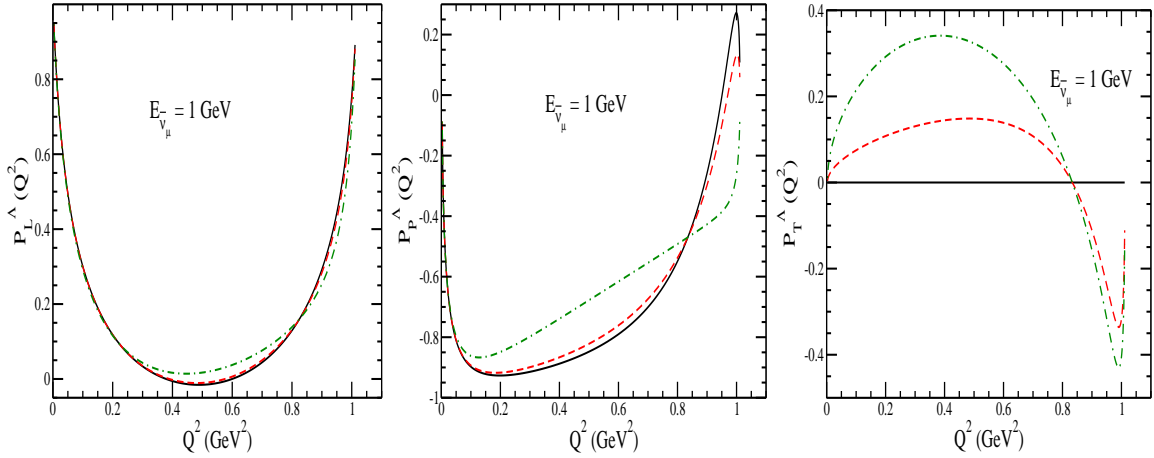
  
\begin{center}  
	\includegraphics[height=6cm,width=5cm]{Pl-im-g2-Nambu-1GeV.eps}
	\includegraphics[height=6cm,width=5cm]{Pp-im-g2-Nambu-1GeV.eps}
	\includegraphics[height=6cm,width=5cm]{Pt-im-g2-Nambu-1GeV.eps}
\caption{$P_L (Q^2)~vs.~ Q^2$ (left panel), $P_P (Q^2)~vs.~ Q^2$ (middle panel) and $P_T (Q^2)~vs.~ Q^2$ (right panel) for 
the process ${\bar{\nu}_\mu + p \rightarrow \mu^+ + {\Lambda}}$ at the incoming antineutrino energy $E_{\bar{\nu}_{\mu}} 
=$ 1~GeV for the polarized $\Lambda$ in the final state, at the different values of $g_2^I (0)$ viz. $g_2^I (0) = $ 
0~(solid line), 1~(dashed line) and 3~(dashed-dotted line)~\cite{Fatima:2018tzs}.}\label{fig2_pol}
\end{center}
\end{figure}

To see the dependence of $g_2^{I} (0)$ on the polarization observables, in Fig.~\ref{fig2_pol}, the results are presented 
for $P_L (Q^2)$, $P_P (Q^2)$ and $P_T (Q^2)$ as a function of $Q^2$ using $g_2^{I} (0) =$ 0, 1 and 3 at 
$E_{\bar{\nu}_{\mu}} =$ 1~GeV. It may be deduced that while $P_L(Q^2)$ is less sensitive to $g_2^I (0)$ at low antineutrino 
energies, $P_P(Q^2)$ is sensitive to $g_2^{I}(0)$ at $E_{\bar{\nu}_{\mu}} =$ 1~GeV. Moreover, $P_T(Q^2)$ shows 40$\%$ 
variations at $Q^2 = $ 0.4 GeV$^2$, $E_{\bar{\nu}_{\mu}} =$ 1 GeV, when $g_2^{I} (0)$ is varied from 0 to 3.

\subsubsection{Polarization of the final lepton}
In the case of final lepton polarization in CC reactions,  the polarization 
4-vector~($\zeta^\tau$) in reactions~(\ref{cc_quasi_reaction}), (\ref{process1}), and (\ref{process3}) is written 
as~\cite{Athar:2020kqn, Fatima:2018tzs}:
\begin{equation}\label{polarl}
\zeta^{\tau}=\frac{\mathrm{Tr}[\gamma^{\tau}\gamma_{5}~\rho_{f}(k^\prime)]}
{\mathrm{Tr}[\rho_{f}(k^\prime)]},
\end{equation}
and the spin density matrix for the final lepton $\rho_f(k^\prime)$ is given by 
\begin{equation}\label{polar1l}
\rho_{f}(k^\prime)= { J}^{\alpha \beta}  \text{ Tr}[\Lambda(k') \gamma_\alpha (1 + \gamma_5) \Lambda(k) \tilde\gamma_{\beta} 
(1 + \tilde\gamma_5)\Lambda(k')], 
\end{equation} 
with $\tilde{\gamma}_{\alpha} =\gamma^0 \gamma^{\dagger}_{\alpha} \gamma^0$ and $\tilde{\gamma}_{5} =\gamma^0 
\gamma^{\dagger}_{5} \gamma^0$.

Using Eq.~(\ref{polar3}), $\zeta^\tau$ defined in Eq.~(\ref{polarl}) may also be rewritten as~\cite{Athar:2020kqn, Fatima:2018tzs}
\begin{equation}\label{polar4l}
\zeta^{\tau}=\left( g^{\tau\sigma}-\frac{k'^{\tau}k'^{\sigma}}{m_{l}^2}\right) \frac{  { J}^{\alpha \beta} 
	\mathrm{Tr} \left[\gamma_{\sigma}\gamma_{5}\Lambda(k') \gamma_\alpha (1 + \gamma_5) \Lambda(k) \tilde\gamma_ {\beta} 
	(1 + \tilde\gamma_5) \right]} { { J}^{\alpha \beta} \mathrm{Tr}\left[\Lambda(k') \gamma_\alpha (1 + \gamma_5) 
	\Lambda(k) \tilde\gamma_ {\beta} (1 + \tilde\gamma_5) \right]},
\end{equation}
where $m_{l}$ is the charged lepton mass. 

With ${ J}^{\alpha \beta}$ and ${ L}_{\alpha \beta}$ given in Eqs.~(\ref{had_tens}) and (\ref{lep_tens}), respectively, 
an expression for $\zeta^\tau$ is obtained. In the laboratory frame where the initial nucleon is at rest, the polarization 
vector $\vec{\zeta}$ is calculated to be a function of 3-momenta of incoming antineutrino $({\vec{k}})$ and outgoing lepton 
$({\vec{k}}\,^{\prime})$, and is given as~\cite{Athar:2020kqn, Fatima:2018tzs}  
\begin{equation}\label{3poll}
\vec{\zeta} =\frac{\left[{A^l(Q^2)\vec{ k}} + B^l(Q^2){\vec{k}}\,^{\prime} + C^l(Q^2) M (\vec{k} \times 
	{\vec{k}}\,^{\prime}) \right]}{N(Q^2)}, 
\end{equation}
where the expressions of $A^l(Q^2)$, $B^l(Q^2)$ and $C^l(Q^2)$ are given in Appendix-\ref{appendix2}.

One may expand $\vec{\zeta}$ along the orthogonal directions, ${\hat{e}}_L^l$, ${\hat{e}}_P^l$ and 
${\hat{e}_T^l}$ in the reaction plane corresponding to the longitudinal, perpendicular and transverse directions, defined as
\begin{equation}\label{vectorsl}
\hat{e}_{L}^l=\frac{\vec{ k}^{\, \prime}}{|\vec{ k}^{\,\prime}|} ,\qquad
\hat{e}_{P}^l=\hat{e}_{L}^l \times \hat{e}_T^l ,\qquad   {\rm where}~~~~~ 
\hat{e}_T^l=\frac{\vec{ k}\times \vec{ k}^{\,\prime}}{|\vec{ k}\times \vec{ k}^{\,\prime}|} ,
\end{equation}
and depicted in Fig.~\ref{TRI}(c). We then write $\vec{\zeta}$ as:
\begin{equation}\label{polarLabl}
\vec{\zeta}=\zeta_{L} \hat{e}_{L}^l+\zeta_{P} \hat{e}_{P}^l + \zeta_{T} \hat{e}_{T}^l ,
\end{equation}
such that the longitudinal, perpendicular and transverse components of the $\vec{\zeta}$ in the laboratory frame are given 
by
\begin{equation}\label{PLl}
\zeta_L(Q^2)=\vec{\zeta} \cdot \hat{e}_L^l,\qquad \zeta_P(Q^2)= \vec{\zeta} \cdot \hat{e}_P^l, 
\qquad \zeta_T(Q^2) = \vec{\zeta} \cdot \hat{e}_T^l.
\end{equation}
From Eq.~(\ref{PLl}), $P_L^l(Q^2)$, $P_P^l(Q^2)$ and $P_T^l(Q^2)$ defined in the rest frame of the final lepton are given by 
\begin{equation}\label{PL1l}
P_L^l(Q^2)=\frac{m_{l}}{E_{k^\prime}} \zeta_L(Q^2), \qquad P_P^l(Q^2)=\zeta_P(Q^2), \qquad P_T^l(Q^2)=\zeta_T(Q^2),
\end{equation}
where $\frac{m_{l}}{E_{k^\prime}}$ is the Lorentz boost factor along ${\vec k}\, ^\prime$. Using Eqs.~(\ref{3poll}), 
(\ref{vectorsl}) and (\ref{PLl}) in Eq.~(\ref{PL1l}), $P_L^l(Q^2)$, $P_P^l (Q^2)$ and 
 $P_T^l (Q^2)$ are calculated to be~\cite{Athar:2020kqn, Fatima:2018tzs}
\begin{eqnarray}
P_L^l (Q^2) &=& \frac{m_{l}}{E_{k^{\prime}}} \frac{A^l(Q^2) \vec{k} \cdot\vec{k}^{\,\prime} + B^l (Q^2) 
	|\vec{k}^{\,\prime}|^2}{N(Q^2)~|\vec{k}^{\,\prime}|},\label{Pll} \\
P_P^l (Q^2) &=& \frac{A^l(Q^2) [|\vec{k}|^2 |\vec{k}^{\,\prime}|^2 - (\vec{k} \cdot \vec{k}^{\,\prime})^2]}{N(Q^2)~
	|\vec{k}^{\,\prime}| ~ |\vec{k}\times \vec{k}^{\,\prime}|},\label{Ppl} \\
P_T^l (Q^2) &=& \frac{C^l(Q^2) M [(\vec{k}\cdot \vec{k}^{\,\prime})^2 - |\vec{k}|^2 |\vec{k}^{\,\prime}|^2]}{N(Q^2)~
	|\vec{k} \times \vec{k}^{\,\prime} |}. \label{Ptl}
\end{eqnarray}

\begin{figure}[h]
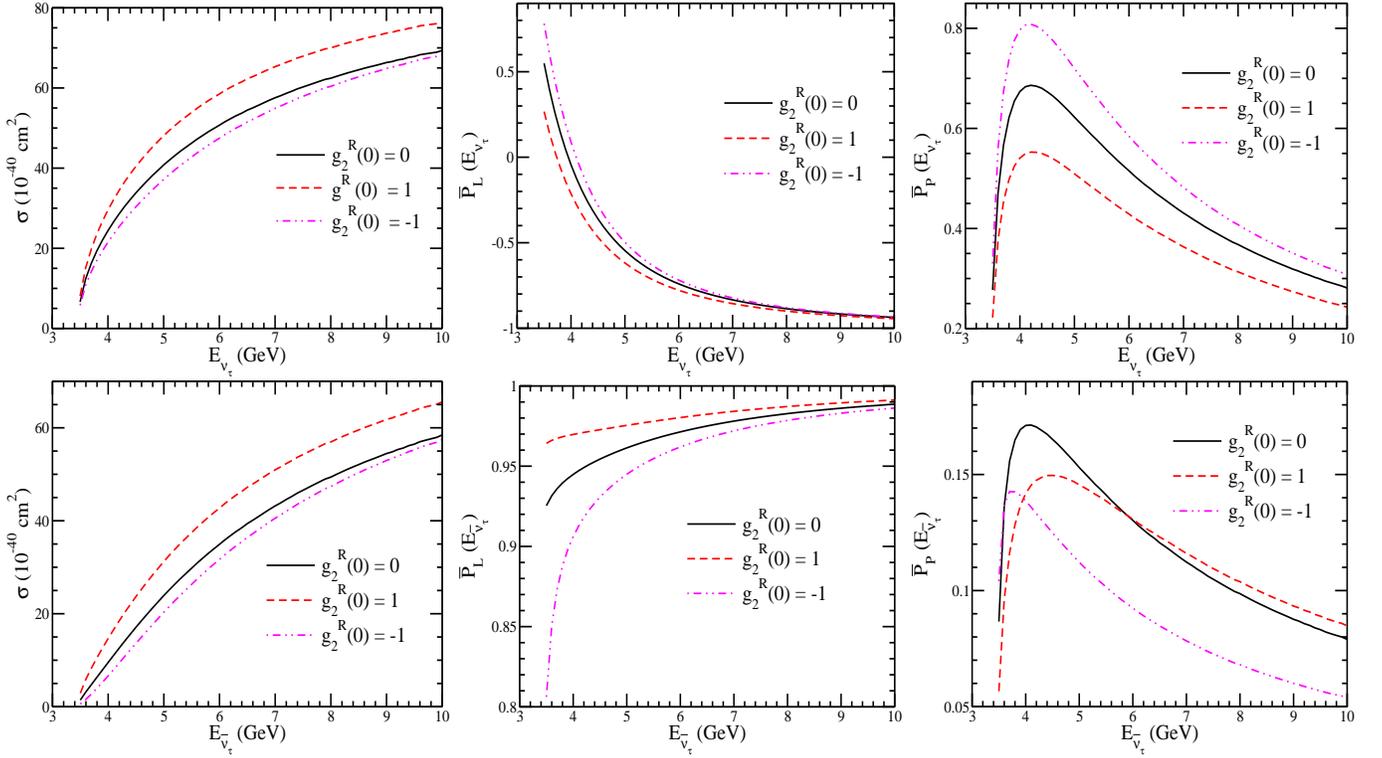

\centering
\includegraphics[height=5cm,width=5.9cm]{total_sigma_proton_g2r.eps}
\includegraphics[height=5cm,width=5.9cm]{total_Pl_proton_g2r.eps}
\includegraphics[height=5cm,width=5.9cm]{total_Pp_proton_g2r.eps}
\includegraphics[height=5cm,width=5.9cm]{total_sigma_neutron_g2r.eps}
\includegraphics[height=5cm,width=5.9cm]{total_Pl_neutron_g2r.eps}
\includegraphics[height=5cm,width=5.9cm]{total_Pp_neutron_g2r.eps}
\caption{(Top panel) left to right:  $\sigma$ vs $E_{\nu_{\tau}}$, $\overline{P}_{L} (E_{\nu_{\tau}})$ vs $E_{\nu_{\tau}}$,
 and $\overline{P}_{P} (E_{\nu_{\tau}})$ vs $E_{\nu_{\tau}}$ for the $\nu_{\tau} + n \rightarrow \tau^{-} + p$ process.
 (Bottom panel) left to right:  $\sigma$ vs $E_{\bar{\nu}_{\tau}}$, $\overline{P}_{L} (E_{\bar{\nu}_{\tau}})$ vs 
 $E_{\bar{\nu}_{\tau}}$, and $\overline{P}_{P} (E_{\bar{\nu}_{\tau}})$ vs $E_{\bar{\nu}_{\tau}}$ for the $\bar{\nu}_{\tau} + 
 p \rightarrow \tau^{+} + n$ process. The calculations have been performed using electric and magnetic Sachs' form 
 factors parameterized by Bradford { et al.}~\cite{Bradford:2006yz} with $M_{A} = 1.026$~GeV, and with the different values of $g_{2}^{R} (0)$ {  viz.} 
 $g_{2}^{R} (0) = 0$~(solid line), 1~(dashed line) and $-1$~(double-dotted-dashed line) used in Eq.~(\ref{g2})~\cite{Fatima:2020pvv}.}
 \label{sigma_g2R_nu}
\end{figure}
\begin{figure}[tbp]
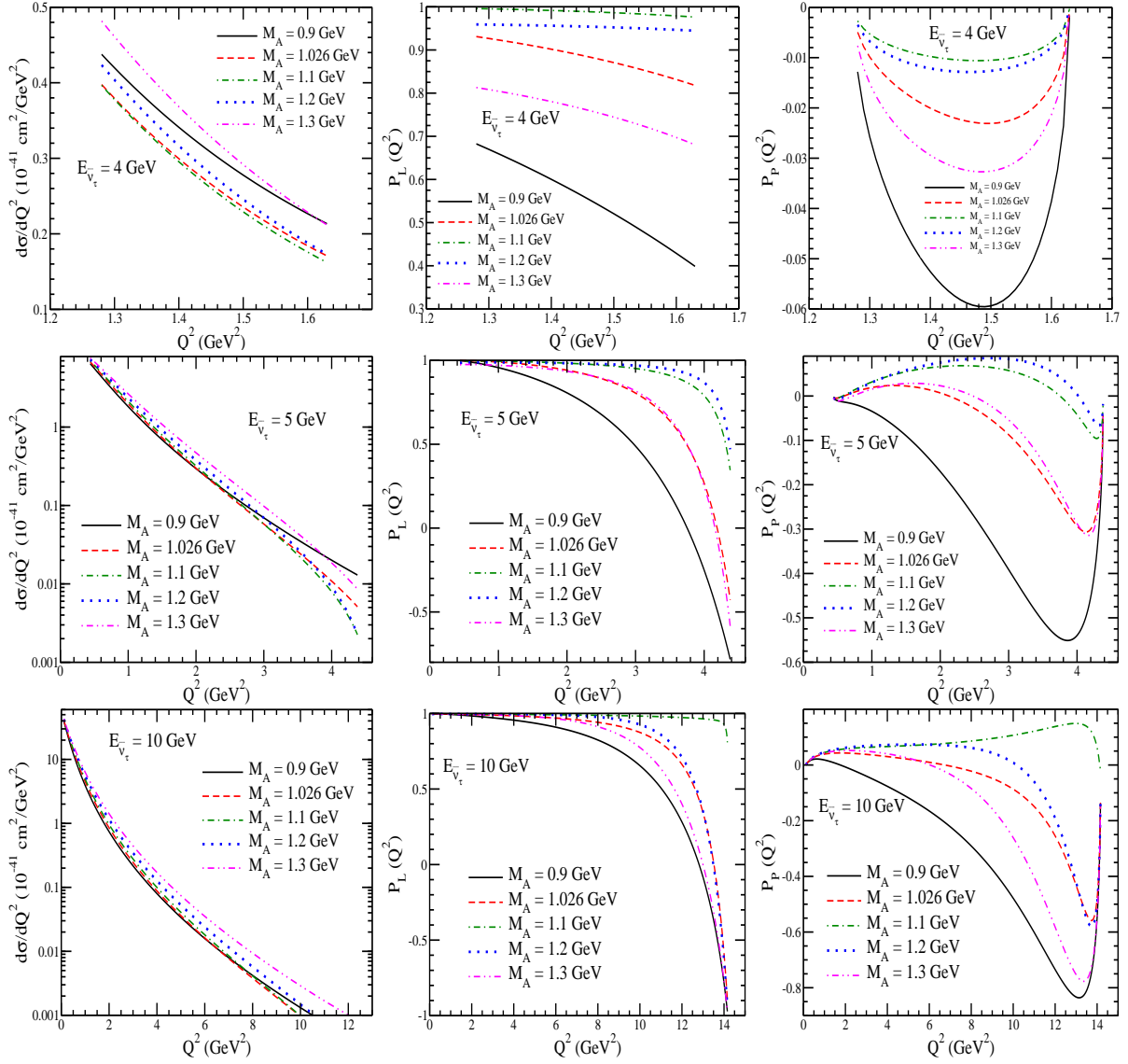

\begin{center}
\includegraphics[height=5cm,width=5.2cm]{dsig_dq2_Ma_4GeV.eps}
\includegraphics[height=5cm,width=5.2cm]{Pl-q2-Ma-4GeV-1.eps}
\includegraphics[height=5cm,width=5.2cm]{Pp-q2-Ma-4GeV-A.eps}
\includegraphics[height=5cm,width=5.2cm]{dsig_dq2_Ma_5GeV.eps}
\includegraphics[height=5cm,width=5.2cm]{Pl-q2-Ma-5GeV-2.eps}
\includegraphics[height=5cm,width=5.2cm]{Pp-q2-Ma-5GeV-B.eps}
\includegraphics[height=5cm,width=5.2cm]{dsig_dq2_Ma_10GeV.eps} 
\includegraphics[height=5cm,width=5.2cm]{Pl-q2-Ma-10GeV-3.eps}
\includegraphics[height=5cm,width=5.2cm]{Pp-q2-Ma-10GeV-C.eps}
\end{center}
\caption{$\frac{d\sigma}{dQ^2}$~(left panel), $P_L (Q^2)$~(middle panel) and $P_P (Q^2)$~(right panel) versus $Q^2$ for the 
process $\bar{\nu}_{\tau} + p \longrightarrow \tau^+ + \Lambda$ at $E_{\bar{\nu}_{\tau}}$ = 4 GeV~(upper panel), 5 
GeV~(middle panel) and 10 GeV~(lower panel). The calculations have been performed using the electric 
and magnetic Sachs' form factors parameterized by Bradford {  et al.}~\cite{Bradford:2006yz} and for the axial form 
factor~(Eq.~(\ref{gplam})), 
the different values of $M_{A}$ have been used {  viz.} $M_{A} =$ 0.9 GeV~(solid line), 1.026 GeV~(dashed line), 1.1 
GeV~(dashed-dotted line), 1.2 GeV~(dotted line) and 1.3 GeV~(double-dotted-dashed line)~\cite{Fatima:2022tlf}.}\label{dsigma_pol_MA_lambda}
\end{figure}

Using Eqs.~(\ref{dsig:QE}), (\ref{Pll}), and (\ref{Ppl}) the differential scattering cross section as well as the 
polarization observables of the final lepton produced in the (anti)neutrino induced processes are calculated. It has been 
observed that 
\begin{itemize}
 \item [(i)] in the case of $\bar{\nu}_{\mu}$ induced processes, the outgoing $\mu^{+}$ is almost longitudinally 
 polarized because of the small mass of $\mu^{+}$ while the perpendicular and transverse polarizations show some effect at 
 lower $\bar{\nu}_{\mu}$ energy but become almost negligible at $E_{\bar{\nu}_{\mu}}=1$~GeV~\cite{Fatima:2018tzs}. However for 
 the $\nu_{\tau}(\bar{\nu}_{\tau})$ induced processes, the effect of polarization observables of the $\tau^{\pm}$ is 
 significant~\cite{Fatima:2020pvv, Fatima:2022tlf} and is shown in Figs.~\ref{sigma_g2R_nu} and \ref{dsigma_pol_MA_lambda} for 
 nucleon and $\Lambda$ productions, respectively.

 \item [(ii)] The effect of the second class current form factor $g_{2}^{R} (0)$ on the total cross section and average 
 polarizations is studied by integrating the expressions of $\frac{d\sigma}{dQ^2}$, $P_{L}^{l}(Q^2)$, and $P_{P}^{l}(Q^2)$ 
 over $Q^2$. In Fig.~\ref{sigma_g2R_nu}, the results for $\sigma$, ${P}_{L} (E_{\nu_{\tau}(\bar{\nu}_{\tau})})$ and ${P}_{P} 
 (E_{\nu_{\tau}(\bar{\nu}_{\tau})})$ are presented as a function of (anti)neutrino energies by taking $g_{2}^{R} (0)=0$ and 
 $\pm 1$. It may be observed from the figure that in the case of $\sigma$, for both the processes $\nu_{\tau} + n \rightarrow 
 \tau^{-} + p$ and $\bar{\nu}_{\tau} + p \rightarrow \tau^{+} + n$, the results obtained with $g_{2}^{R} (0) =-1$ are 
 slightly lower~(1 -- 2$\%$) than the results obtained with $g_{2}^{R} (0) =0$ in the range of $E_{\nu_{\tau}, 
 \bar{\nu}_{\tau}}$ from threshold up to 10 GeV, while the results obtained with $g_{2}^{R} (0) =+1$, are higher from the 
 results obtained with $g_{2}^{R} (0) =0$  and the difference decreases with the increase in energy. For example, 
 at $E_{\nu_{\tau} (\bar{\nu}_{\tau})}=5$~GeV, the results obtained with $g_{2}^{R} (0) = +1$ are 
  higher by about 18~(30)$\%$ from the results of $g_{2}^{R} (0)=0$, while at 10 GeV, this difference becomes 10~(12)$\%$ for 
  the (anti)neutrino induced processes. 
  
  \item [(iii)] In the case of ${P}_{L} (E_{\nu_{\tau}, \bar{\nu}_{\tau}})$, there is a slight 
  variation due to the change in the value of $g_{2}^{R} (0)$ for neutrino induced process, while for the antineutrino 
  induced process, this difference is large at lower antineutrino energies which gradually becomes smaller with the increase 
  in energy.
  
  \item [(iv)] For ${P}_{P} (E_{\nu_{\tau} (\bar{\nu}_{\tau})})$, the results for both the neutrino as well as antineutrino 
  induced processes show dependence on the choice of $g_{2}^{R} (0)$, while the nature of dependence is different. 
  In the case of $\nu_\tau$ induced reaction, in the peak region, the results are $\sim$ 20$\%$ smaller for 
  $g_{2}^{R} (0)=+1$ from the results obtained with $g_{2}^{R} (0)=0$, while using $g_{2}^{R} (0)=-1$ the results are 18$\%$ 
  higher than the results obtained using $g_{2}^{R} (0) =0$. However, in the case of $\bar{\nu}_{\tau}$ induced processes, 
  the results obtained with $g_{2}^{R} (0) = \pm 1$ are lower than the results obtained with $g_{2}^{R} (0)=0$ in the region 
  of threshold up to $E_{\bar{\nu}_{\tau}}=6$ GeV.
\end{itemize}
To study the effect of $M_{A}$ variation in the range 0.9--1.3~GeV on the differential cross section and polarization 
observables, in Fig.~\ref{dsigma_pol_MA_lambda}, the results for $\frac{d\sigma}{dQ^2}$, $P_L (Q^2)$ and $P_P (Q^2)$ as a 
function of $Q^2$ for $\bar{\nu}_{\tau} + p \longrightarrow \tau^{+} + \Lambda$ at $E_{\bar{\nu}_{\tau}}$ = 4 GeV, 5 GeV and 
10 GeV are presented. It has been found that at low ${\bar\nu}_\tau$ energies, there is some dependence of the differential 
cross section as well as the polarization observables on the choice of $M_A$. With the increase in ${\bar\nu}_\tau$ energy, 
this dependence on the variation in $M_A$ decreases, especially for $\frac{d\sigma}{dQ^2}$ and to some extent for $P_L (Q^2)$ 
but not for $P_P (Q^2)$ distribution. Moreover, it is important to point out that in the case of $\bar{\nu}_{\tau} + p 
\longrightarrow \tau^{+} + \Lambda$ reaction, with the increase in $M_{A}$, $\frac{d\sigma}{dQ^2}$ decreases~(0.9 GeV to 1.1 
GeV), but with the further increase in $M_{A}$~(1.1 GeV to 1.3 GeV), $\frac{d\sigma}{dQ^2}$ increases, which is not generally 
the case in ${\nu}_{l} + n \longrightarrow l^{-} + p;~ (l=e,\mu,\tau)$ scattering. Furthermore, in the case of $\bar{\nu}_{l} + 
p \longrightarrow l^{+} + n$, it has been shown that with the increase in $M_{A}$, $\frac{d\sigma}{dQ^2}$ decreases~(from 0.9 
GeV to 1.1 GeV) and with further increase in $M_{A}=1.2$~GeV, $\frac{d\sigma}{dQ^2}$ increases~\cite{Fatima:2020pvv}. A 
similar trend is observed in the case of $\Lambda$ production induced by $\bar{\nu}_{\mu}$~\cite{Akbar:2016awk}, as in 
the case of $\bar{\nu}_{\tau}$ induced CCQE reaction~\cite{Fatima:2020pvv} with $\Delta S = 0$ currents. It may be pointed out 
that with the 
increase in antineutrino energy, the polarization observables show a significant dependence on the axial dipole mass.

\section{Inelastic $\nu-$scattering processes from nucleons}\label{sec:inelastic:nucleon}
\subsection{Introduction}
With the increase in energy of the neutrinos, the IE processes start to appear in which new particles are produced. The 
production of a single pion is the simplest IE process which starts at a threshold energy of $E_\nu \sim 135$~MeV in 
the reactions induced by the weak NC interactions of $\nu_e, \nu_\mu, \nu_\tau$ and their antiparticles 
${\bar\nu}_e, {\bar\nu}_\mu, {\bar\nu}_\tau$. In case of the IE processes induced by the weak CC in which 
pions are accompanied by the corresponding charged leptons $e^\mp, \mu^\mp, \tau^\mp$, the threshold energies are $E_\nu \sim 
150$~MeV, 280~MeV and 3.8~GeV, respectively. With the further increase in energy, various particles with masses higher than 
the pion mass like $\eta$, $K$, $\rho$, $\omega$, $\Lambda$, etc. are produced, subject to the selection rules satisfied by 
the weak charge and NC. Specifically, we focus in this section, on the reactions given in 
Table~\ref{sec2:Table1}, which are induced by the CC and NC weak interaction processes. 

Some of the IE processes listed in Table~\ref{sec2:Table1}, specially in the $\Delta S=0$ sector have been studied for 
many years in the reactions induced by photons and electrons where the contribution comes from the electromagnetic vector 
current only. The contribution of the weak vector current in the neutrino scattering processes are determined in the $\Delta 
S=0$ sector using isospin symmetry which are extended to the $\Delta S=1$ sector assuming $SU(3)$ symmetry. Therefore, the experimental and theoretical 
studies of IE production of various mesons like $\pi$, $K$, $\eta$, etc. induced by photons and electrons play very 
important role in the study of the weak IE production of various mesons induced by (anti)neutrinos as listed in 
Table~\ref{sec2:Table1}. The contribution of the axial-vector current to the weak IE processes induced by 
(anti)neutrinos is determined in terms of the axial-vector transition form factors calculated using the generalized form of PCAC and 
the GT relation. However, all the transition form factors in the axial-vector sector in the case of 
nucleon-resonance transitions are not determined in this way and a phenomenological approach is used following the seminal 
work of Adler~\cite{Adler:1968tw}.
\begin{figure}  
 \begin{center}
  \includegraphics[height=4cm, width=15cm]{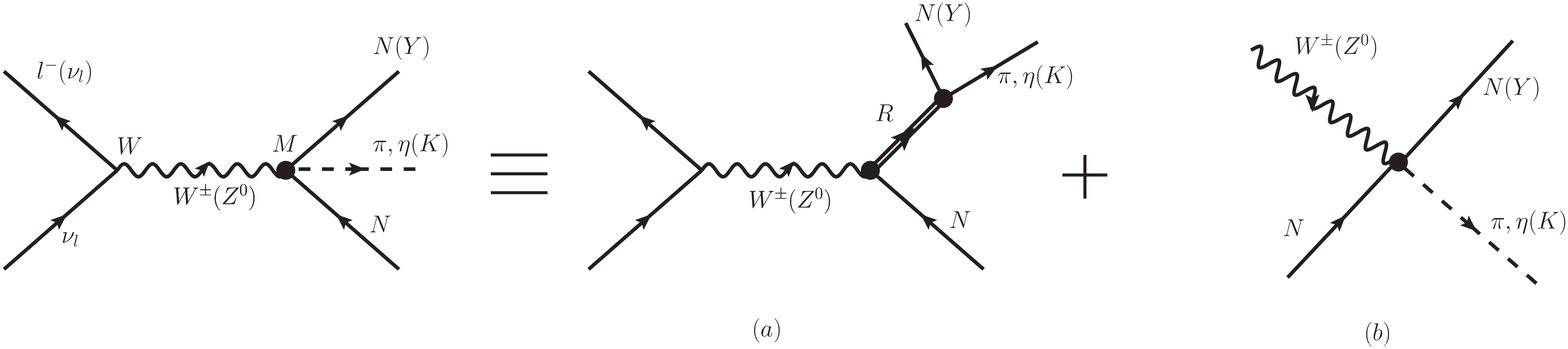}
 \end{center}
\caption{Generic Feynman diagrams representing CC and NC induced IE processes given in 
Table-\ref{sec2:Table1}. In Fig.~(a), $R$ is the resonance excited by the (anti)neutrino interactions induced by 
$W^\pm(Z^0)$, which subsequently decays to a baryon and a meson and Fig.~(b) shows the 
NR terms. }\label{Ch11_Ch12-fig1}
\end{figure}

\begin{figure}  
 \begin{center}
  \includegraphics[height=4cm, width=15cm]{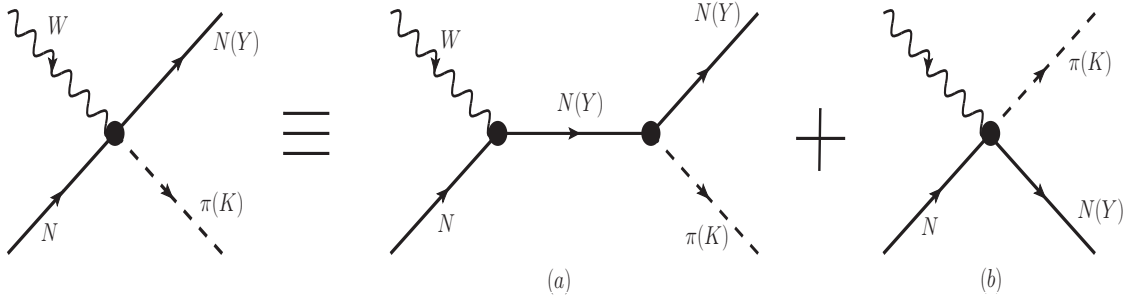}
 \end{center}
\caption{Generic Feynman diagrams representing the NR background terms contributing to the IE 
processes, where Fig.~(a) shows the meson and baryon pole terms and Fig.~(b) shows the contact diagram.}
\label{Ch11_Ch12-fig2}
\end{figure}
The study of the IE processes in the EM interactions induced by photons and electrons { shows  that the 
reactions receives contribution from the resonance excitations as well as from the nonresonant Born diagrams. These contributions are diagrammatically shown in Figs.~\ref{Ch11_Ch12-fig1} and \ref{Ch11_Ch12-fig2}. While the dominant contribution comes from the resonant diagrams specially from the $\Delta$ resonance in the case of pion production, and $S_{11}(1535)$ resonance in the case of $\eta$ production, the contribution from the nonresonant diagrams is quite important in almost the entire range of energy and not only in the threshold region.} This perception of the dynamics of these 
IE reactions is also expected to be valid in the weak IE reactions induced by (anti)neutrinos which are shown 
in Figs.~\ref{Ch11_Ch12-fig1} and \ref{Ch11_Ch12-fig2}. In Fig.~\ref{Ch11_Ch12-fig1}, $R$ is the resonance 
excited by the (anti)neutrino interactions induced by $W^\pm(Z^0)$ intermediate vector bosons and decays into  nucleons and 
mesons. In Fig.~\ref{Ch11_Ch12-fig2}, the interaction vertex includes the contribution of all the NR diagrams to 
the IE processes in $s$, $t$ and $u$ channels and the contact diagrams.
 
The weak IE processes induced by (anti)neutrinos play very important role in modeling the (anti)neutrino-nucleon 
cross sections to be used in formulating the neutrino event generators in simulating the neutrino oscillation experiments in 
the few GeV energy region. Moreover, the weak IE reactions on the nucleon targets also help to probe some aspects of 
hadronic structure in the axial-vector sector in conjunction with the hadronic structure being probed by the electromagnetic 
current in the vector sector using photons and electrons.
 
In the following sections, we first describe the general kinematics of the IE reactions with the single meson 
production in Section~\ref{sec:kinematics}. In Sections~\ref{sec:1pion}, \ref{sec:eta}, \ref{kaon}, and \ref{sec:associated} 
we discuss the single pion, single eta, single kaon, and associated production of kaons, respectively. In 
Sections~\ref{sec:Ypi}, \ref{XK}, and \ref{sec:2pion}, we discuss briefly $Y\pi$, $\Xi K$ and $2\pi$ productions, respectively.
 
 \subsection{Kinematics}\label{sec:kinematics}
\begin{figure}  
\centering
 \includegraphics[height=6cm,width=15cm]{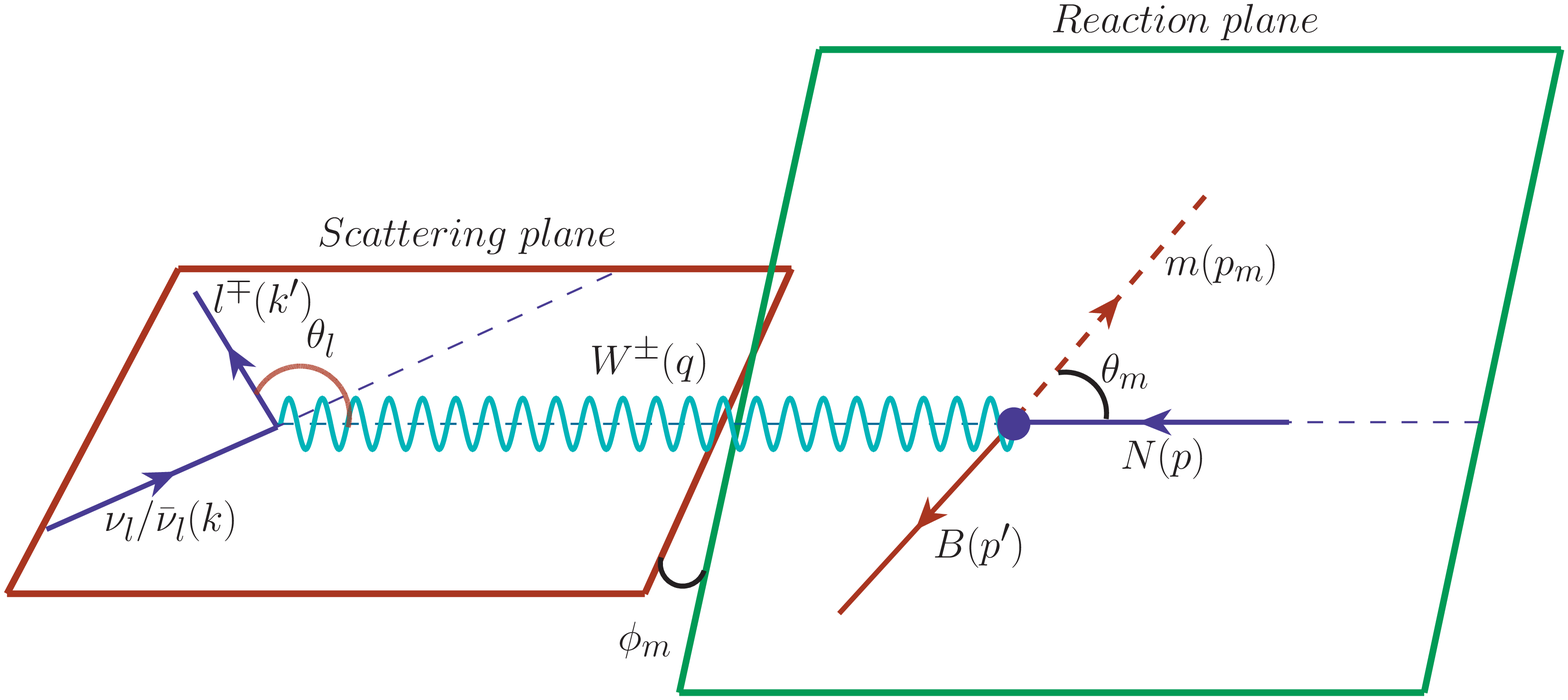}
 \caption{(Anti)neutrino scattering and reaction planes, depicting the hadronic plane in CM frame 
 and scattering plane in the laboratory frame. The kinematical variables used in the calculation of the different IE 
 scattering processes are defined in the figure.}
 \label{pion:kinematics}
\end{figure}

The general expression for the differential scattering cross section of the IE processes discussed in 
Table-\ref{sec2:Table1} and written in general as
\begin{equation}\label{eq:inelastic:reaction}
\nu_{l}/\bar{\nu}_{l} (k) + N(p) \longrightarrow l^{\mp} (k^{\prime}) + B(p^{\prime}) + m(p_{m})
\end{equation}
in the laboratory frame is given by
\begin{eqnarray}\label{eq:sigma_inelas}
d\sigma &=& \frac{1}{4 ME_\nu(2\pi)^{5}} \frac{d{\vec k}^{\prime}}{ (2 E_{l})} 
\frac{d{\vec p\,}^{\prime}}{(2 E_{B})} \frac{d{\vec p}_{m}}{ (2 E_m)}
 \delta^{4}(k+p-k^{\prime}-p^{\prime}-p_{m})\overline{\sum}\sum | \mathcal M |^2,\;\;\;\;\;
\end{eqnarray}
where in Eq.~(\ref{eq:inelastic:reaction}), $m(=\pi, \eta, K$, etc.) is a meson produced with a baryon~($B=N,Y$, etc.) in 
the final state. $ k( k^\prime) $ is the four momentum of the incoming~(outgoing) lepton having energy $E_\nu( E_l)$; $p$ is 
the four momentum of the incoming nucleon which is at rest, $E_B$ and $p^\prime$ are respectively the energy and four 
momentum of the outgoing baryon, and the meson four momentum is $p_m$ with energy $ E_m$, and $M$ is the nucleon mass. The 
different kinematical variables used in the numerical calculations of the scattering cross section are depicted in 
Fig.~\ref{pion:kinematics}, where the scattering plane is in the laboratory frame while the reaction plane is in the center 
of mass frame. $\overline{\sum}\sum | \mathcal M |^2  $ is the square of the transition amplitude averaged~(summed) over the 
spins of the initial~(final) states and the transition matrix element is written in terms of the leptonic and the hadronic 
currents as 
\begin{equation}
\label{eq:Gg}
 \mathcal M = \frac{G_F}{\sqrt{2}}\, {l_\mu} j^{\mu},
\end{equation}
where the leptonic current $l_\mu$, and the constant
 $G_F$ are defined after Eq.~(\ref{lep_curr}). $  j^{\mu}_{CC(NC)}$  is the hadronic current for $W^i + N 
\longrightarrow B + \text{meson}$  interaction for CC~$(W^i \equiv W^\pm \; ; i=\pm)$ and NC~$(W^{0} = Z^0)$ 
induced processes.

Integrating over the three momentum of the outgoing baryon, the expression for the differential scattering cross section 
given in Eq.~(\ref{eq:inelastic:reaction}) becomes
\begin{equation}\label{dsigma:pion}
\frac{d \sigma}{dE_{m} ~d \Omega_{m} } = \frac{1}{32(2\pi)^{5}} \int d\Omega_{l} dE_{l} ~\delta(E_{\nu} + M - E_{l} - 
E_{B} - E_{\pi}) \frac{|\vec{k}^{\prime}| |\vec{p}_{m}|}{E_{\nu}M E_{B}} \overline{\sum}\sum | \mathcal M |^2,
\end{equation}
which after integration over $dE_{m}$ becomes
\begin{equation}\label{eq:sigma_inelastic}
\frac{d^4 \sigma}{dE_{l} ~d\cos \theta_{l} ~d\cos \theta_{m} ~d\phi_{m}} = \frac{|\vec{k}^{\prime}| |\vec{p}_{m}|^{2}}
{32(2\pi)^{4}E_{\nu}M} \frac{G_{F}^2  L_{\mu\nu} J^{\mu\nu}}{4} \frac{1}{(E_{\nu} + M - E_{l})|\vec{p}_{m}|^{2} -
E_{m} (\vec{p}_{m} \cdot \vec{q}\;)},
\end{equation}
$L_{\mu\nu}$ is given in Eq.~(\ref{lep_tens}) and $J^{\mu\nu} = \sum j^{\mu}{j^{\nu}}^{\dagger}$, where the hadronic 
current~($j^{\mu}$) receives contribution from the nonresonant background~(NRB) terms as well as from the resonance excitations 
and their decay into a particular meson-baryon final state. The different IE channels receive contribution from the 
different background terms as well as from the different resonance excitations. In the next section, we present the 
structure of the hadronic currents for the background and the resonance terms in general. Moreover, for the different IE 
channels, the specific couplings and the contribution from the different terms are discussed in the respective sections. 
Specifically, the hadronic current for an IE scattering process is written as
\begin{equation}\label{allterms}
 j^{\mu}=j^{\mu}_{\text NR} ~+~ j^{\mu}_{R_{\frac{1}{2}}}~+~j^{\mu}_{R_{\frac{3}{2}}}, 
\end{equation}
where $j^{\mu}_{\text NR}$, $j^{\mu}_{R_{\frac{1}{2}}}$, and $j^{\mu}_{R_{\frac{3}{2}}}$, respectively, represent the 
contribution of the hadronic current from the NRB terms, spin $\frac{1}{2}$ resonance, and spin 
$\frac{3}{2}$ resonance excitations. 

\subsection{Structure of matrix elements}\label{inelastic:mat}
As already discussed in the previous section, in the following we discuss the NRB contribution in 
Section~\ref{NRB} and the resonance excitation and their subsequent decay to different meson-baryon 
final state are discussed in Section~\ref{res:inelastic}.

\subsubsection{Nonresonant contribution}\label{NRB}
The nonresonant~(NR) contributions for the pion production have been calculated using a microscopic model based on the $SU(2)$ 
chiral Lagrangians. We have used $SU(2)$ nonlinear $\sigma$ model involving pions and nucleons and the corresponding vector and 
axial-vector currents generated by the chiral symmetry transformations to determine the structure of the chiral NR 
terms~\cite{Athar:2020kqn}. It has been observed that the NR contributions are particularly important in the meson production 
threshold region, for values of $W \simeq M + m_{m}$, with $m_{m}$ being the mass of the produced meson. To include the 
strange meson production, this model is extended to the $SU(3)$ chiral Lagrangians. The basic parameters of the model are the 
meson decay constant $f_{m}$, the Cabibbo's angle, the proton and neutron magnetic moments, and the asymmetric and symmetric 
axial-vector coupling constants for the two baryon octets, $D$ and $F$, respectively, that are obtained from the analysis of 
the semileptonic decays of neutron and hyperons.

The Lagrangian for QCD is written as
\begin{equation}\label{Ch11_eq2:qcdLag}
 {\cal L}_{\textrm QCD} = \overline{q} ( i \slashed{D} - m_{q}) q - {\frac14}G^\alpha_{\mu\nu}G^{\alpha\mu\nu} ,
\end{equation}
where $q = \begin{pmatrix}
            u\\
            d\\
            s
           \end{pmatrix}
$ denotes the quark field, $G^\alpha_{\mu\nu}$ is the gluon field-strength tensor with $\alpha$ as a color index and 
$D_{\mu}$ is defined as
\begin{equation}
 D_{\mu} =  \partial_{\mu} + ig \frac{\lambda^{\alpha}}{2} G_{\mu \alpha},
\end{equation}
where $g$ is the quark-gluon coupling strength and $G_{\mu \alpha}$ is the vector gluon field. The Lagrangian written in 
Eq.~(\ref{Ch11_eq2:qcdLag}) does not preserve chiral symmetry in its present form, however, in the limit when quark masses are 
assumed to be zero, the QCD Lagrangian preserves chiral symmetry. Today it is well established that all the quarks have 
nonzero mass although the current quark masses for $u,d,s$ are small as compared to the nucleon mass. Thus, in the case of 
strong interactions, chiral symmetry is a symmetry of the Lagrangian in the limit of $m_{u}, m_{d}, m_{s} \rightarrow 0$. The consequence of 
the symmetries of the Lagrangian leads to the conserved currents. The vector current is conserved in nature due to the 
isospin symmetry. Similarly, the axial-vector current is conserved in the presence of the chiral symmetry. In case, the chiral 
symmetry based on $SU(2) \times SU(2)$ is broken spontaneously, it leads to the existence of massless Goldstone bosons, 
which are identified as the pions in 
the limit $m_{u}, m_{d}\rightarrow 0$ and as the octet of pseudoscalar mesons in the case of chiral symmetry based on  $SU(3) 
\times SU(3)$ symmetry in the limit $m_{u}, m_{d}, m_{s} \rightarrow 
0$. The local gauge symmetry of QCD leads to the construction of the chiral effective theory of the Goldstone bosons as well as 
their interaction with the baryons.
 
In order to get the Lagrangian, which describes the dynamics of these pseudoscalar mesons, we need continuous fields which 
are described in terms of these Goldstone modes. The elements of $SU(3)$ pseudoscalar meson fields are written in terms of a 
unitary matrix~\cite{Athar:2020kqn} 
\begin{equation}
 U(\Theta) = \exp\left( -i \Theta_k \frac{\lambda_k}{2} \right)\;,
\end{equation}
where $\Theta_k;~(k=1-8)$ are the real set of parameters and $\lambda_k$ are the traceless, Hermitian $3 \times 3$ Gell-Mann 
matrices. 

Each Goldstone boson corresponds to the $x$-dependent Cartesian component of the fields $\phi_k (x) $, which in turn, are 
expressed in terms of the physical fields as~\cite{Athar:2020kqn}: 
\begin{eqnarray}\label{eq2:ps_matrix_final}
 \Phi(x) =\sum_{k=1}^{8} \phi_k(x) \lambda_k =
\left(\begin{array}{ccc}
\pi^0+\frac{1}{\sqrt{3}}\eta &\sqrt{2}\pi^+&\sqrt{2}K^+\\
\sqrt{2}\pi^-&-\pi^0+\frac{1}{\sqrt{3}}\eta&\sqrt{2}K^0\\
\sqrt{2}K^- &\sqrt{2}\bar{K}^0&-\frac{2}{\sqrt{3}}\eta
\end{array}\right).
\end{eqnarray}
For the baryons, we follow the same procedure as we do for the mesons. However, unlike the pseudoscalar mesons where the 
fields are real, in the case of baryon fields, represented by a $B$ matrix, each entry is a complex-field and the general 
representation is given by~\cite{Athar:2020kqn}:
\begin{eqnarray}\label{eq2:pmatrix2}
B(x)=\sum_{k=1}^{8} \frac{1}{\sqrt2} b_k(x) \lambda_k = \left(\begin{array}{ccc}
\frac{1}{\sqrt{2}}\Sigma^0+\frac{1}{\sqrt{6}}\Lambda&\Sigma^+&p\\
\Sigma^-&-\frac{1}{\sqrt{2}}\Sigma^0+\frac{1}{\sqrt{6}}\Lambda&n\\
\Xi^-&\Xi^0&-\frac{2}{\sqrt{6}}\Lambda
\end{array}\right).
\end{eqnarray}
After getting the representation of the pseudoscalar meson fields octet $\Phi(x)$ in Eq.~(\ref{eq2:ps_matrix_final}) and 
baryon fields octet $B(x)$ in Eq.~(\ref{eq2:pmatrix2}), we now discuss the construction of the Lagrangian for meson-meson, 
baryon-meson interactions and their interaction with the external fields. 

\subsubsection{Meson - meson interaction}\label{Sec2:MMinter}
The lowest-order $SU(3)$ chiral Lagrangian describing the pseudoscalar mesons in the presence of an external current is 
obtained as~\cite{Scherer:2002tk, Scherer:2012xha}:
\begin{equation}\label{eq2:lagM}
{\cal L}_M=\frac{f_\pi^2}{4}\mbox{Tr}[D_\mu U (D^\mu U)^\dagger].
\end{equation}
 The covariant derivatives $D^{\mu} U$ and $D^{\mu} U^{\dagger}$ appearing 
in Eq.~(\ref{eq2:lagM}) are expressed in terms of the partial derivatives as
\begin{eqnarray}\label{eq2:coDer}
 D^\mu U \equiv \partial^\mu U - i r^\mu U + i U l^\mu, \qquad \qquad
  D^\mu U^\dagger \equiv \partial^\mu U^\dagger + i U^\dagger r^\mu - i l^\mu U^\dagger,
\end{eqnarray}
where $U$ is the $SU(3)$ unitary matrix given as
\begin{equation}
 U(x) = \exp\left(i\frac{\Phi(x)}{ f_m } \right), 
\end{equation}
where $\Phi(x)$ is given in Eq.~(\ref{eq2:ps_matrix_final}). $r_{\mu}$ and $l_{\mu}$, respectively, represent the right and 
left handed currents, defined in terms of the vector~($v_{\mu}$) and axial-vector~($a_{\mu}$) fields as
\begin{equation}\label{Ch11_NLSM_VA}
 l_{\mu} = \frac{1}{2}(v_{\mu} - a_{\mu}), \qquad \qquad r_{\mu} = \frac{1}{2}(v_{\mu} + a_{\mu}).
\end{equation}
The vector and axial-vector fields are different for the interaction of the different gauge bosons with the meson fields. 

In the case of electromagnetic interactions, the left and right handed currents are identical and are expressed as
\begin{equation}\label{Ch11_lr_EM}
 l_{\mu} = r_{\mu} = -e \hat{Q} A_{\mu} ,
\end{equation}
where $e$ is the strength of the electromagnetic interaction, $A_{\mu}$ represents the photon field and $\hat{Q}=\begin{pmatrix}
                                                                                      2/3 & 0 & 0 \\
                                                                                      0 & -1/3 & 0 \\
                                                                                      0 & 0 & -1/3
                                                                                      \end{pmatrix}$ represents the charge 
                                                                                      of the
                                                                                      $u,~ 
d$, and $s$ quarks. In the case of CC induced processes, the left and right handed currents are expressed as
\begin{equation}\label{Ch11_lr_CC}
 l_{\mu} = - \frac{g}{2} (W_{\mu}^{+} T_{+} + W_{\mu}^{-}T_{-}), \qquad \qquad r_{\mu} = 0,
\end{equation}
where $g = \frac{e}{\sin \theta_{W}}$, $\theta_{W}$ is the Weinberg angle, $W_{\mu}^{\pm}$ represents the W-boson field 
and $T_{\pm}$ is defined as
\begin{equation}
T_{+} = \begin{pmatrix}
        0 & V_{ud} & V_{us} \\
        0 & 0 & 0 \\
        0 & 0 & 0
        \end{pmatrix}, \qquad \text{and} \qquad T_{-} = \begin{pmatrix}
                                                        0 & 0 & 0 \\
                                                        V_{ud} & 0 & 0 \\
                                                        V_{us} & 0 & 0
        \end{pmatrix},
\end{equation}
with $V_{ud} = \cos \theta_{C}$ and $V_{us} = \sin \theta_{C}$ being the elements of the Cabibbo-Kobayashi-Maskawa matrix 
and $\theta_{C}$ being the Cabibbo angle.

The left and right handed currents for NC induced processes are expressed as
\begin{equation}\label{Ch11_lr_NC}
 l_{\mu} = \left(- \frac{g}{\cos \theta_{W}} + e \tan \theta_{W} \right) Z_{\mu} \frac{\lambda_{3}}{2} , \qquad \qquad 
 r_{\mu} =  g \tan \theta_{W} \sin \theta_{W} Z_{\mu} \frac{\lambda_{3}}{2} ,
\end{equation}
where $Z_{\mu}$ represents the Z-boson field and $\lambda_{3}$ is the third component of the Gell-Mann matrices.

\subsubsection{Baryon - meson interaction}
To incorporate baryons in the theory, we have to take care of their masses, which do not vanish in the chiral 
limit~\cite{Kubis:2007iy}. However, if we take nucleons as massive matter fields which couples to external currents and the 
pseudoscalar mesons, we have to then expand the Lagrangian according to their increasing number of momenta. Here, we shall 
present in brief the extension of the formalism to incorporate the heavy matter fields. 

The lowest-order chiral Lagrangian for the baryon octet in the presence of an external current, may be written in terms of 
the $SU(3)$ matrix $B$ as~\cite{Scherer:2002tk, Scherer:2012xha},
\begin{equation}\label{eq2:lagB}
{\cal L}_{MB}=\mbox{Tr}\left[\bar{B}\left(i\slashed{D}
-M\right)B\right]
-\frac{D}{2}\mbox{Tr}\left(\bar{B}\gamma^\mu\gamma_5\{u_\mu,B\}\right)
-\frac{F}{2}\mbox{Tr}\left(\bar{B}\gamma^\mu\gamma_5[u_\mu,B]\right),
\end{equation}
where $M$ denotes the mass of the baryon octet, $D=0.804$ and $F=0.463$ are the symmetric and antisymmetric axial-vector 
coupling constants for the baryon octet, 
the matrix $B$ is given in Eq.~(\ref{eq2:pmatrix2}) and the Lorentz vector $ u^\mu$ is given by~\cite{Scherer:2012xha}:
\begin{equation}\label{eq2:vielbein}
u^\mu = i \left[ u^\dagger ( \partial^\mu - i r^\mu) u - u ( \partial^\mu - i l^\mu) u^\dagger \right].
\end{equation} 
In the case of meson-baryon interactions, the unitary matrix for the pseudoscalar field is expressed as 
\begin{equation}
 u = \sqrt U \equiv \exp \left( i \frac{\Phi(x)}{ 2 f_m } \right),
\end{equation}
and the covariant derivative $D_{\mu}$ on the baryon fields $B$ is given by
\begin{equation}\label{dmuB}
D_\mu B=\partial_\mu B +[\Gamma_\mu,B], \qquad \text{with} \qquad \Gamma^\mu=\frac{1}{2}\left[u^\dagger(\partial^\mu-
ir^\mu)u
+u(\partial^\mu-il^\mu)u^\dagger\right],
\end{equation}
which is known as the chiral connection.

\subsubsection{Decuplet baryon - octet baryon - meson interaction}\label{DEC}
A systematic way of obtaining the relationships~($SU(3)$ factors) between the weak vertices for all the allowed transitions 
and that for the $n\longrightarrow\Delta^+$ is to use the lowest order Lagrangian that couples the decuplet baryons with the 
octet baryons and mesons in the presence of an external current~\cite{Butler:1992pn, Doring:2006ub} and that has been used
in Refs.~\cite{Alam:2011vwg, RafiAlam:2012bro, RafiAlam:2019rft}. Its form is
\begin{equation}\label{eq:lag_dec}
\mathcal{L}_{\rm DBM}= \mathcal{C}
\left( \epsilon^{abc}\,
\overline{T}^\mu_{ade} (u_{\mu})^{d}_{b}\, B^{e}_{c} 
+\epsilon^{abc} \bar{B}^{e}_{c} (u_\mu)^{d}_{b}\, T^{\mu}_{aed}\right),
\end{equation}
where $B$ is given by Eq.~(\ref{eq2:pmatrix2}), $u_\mu$ is given in Eq.~(\ref{eq2:vielbein}), and $T^\mu_{aed}$ is the $SU(3)$
representation of the Rarita-Schwinger fields for the decuplet baryons. This representation is completely symmetric,
which in the present notation is given by a $3\times3 \times3$ array of matrices 
\begin{eqnarray}\label{m3}
T_{abc} &=& 
    \left( 
    \hspace{-0.5em}
    \begin{array}{ccc}
\Delta^{++} & \frac{1}{\sqrt{3}} \Delta^{+} & \frac{1}{\sqrt{3}} \Sigma^{*+} \\
\frac{1}{\sqrt{3}} \Delta^{+} & \frac{1}{\sqrt{3}} \Delta^{0} &
\frac{1}{\sqrt{6}} \Sigma^{*0} \\
\frac{1}{\sqrt{3}} \Sigma^{*+} & \frac{1}{\sqrt{6}} \Sigma^{*0} &
\frac{1}{\sqrt{3}} \Xi^{*0}
    \end{array}
    \hspace{-0.5em}
    \right)
    \left(
    \hspace{-0.5em}
    \begin{array}{ccc}
\frac{1}{\sqrt{3}} \Delta^{+} & \frac{1}{\sqrt{3}} \Delta^{0} &
\frac{1}{\sqrt{6}} \Sigma^{*0} \\
\frac{1}{\sqrt{3}} \Delta^{0} & \Delta^{-} & \frac{1}{\sqrt{3}} \Sigma^{*-} \\
\frac{1}{\sqrt{6}} \Sigma^{*0} & \frac{1}{\sqrt{3}} \Sigma^{*-} &
\frac{1}{\sqrt{3}} \Xi^{*-}
    \end{array}
    \hspace{-0.5em}
    \right)  \left(
    \begin{array}{ccc}
\frac{1}{\sqrt{3}} \Sigma^{*+} & \frac{1}{\sqrt{6}} \Sigma^{*0} &
\frac{1}{\sqrt{3}} \Xi^{*0} \\
\frac{1}{\sqrt{6}} \Sigma^{*0} & \frac{1}{\sqrt{3}} \Sigma^{*-} &
\frac{1}{\sqrt{3}} \Xi^{*-} \\
\frac{1}{\sqrt{3}} \Xi^{*0} & \frac{1}{\sqrt{3}} \Xi^{*-} & \Omega^{-}
    \end{array}
    \hspace{-0.5em}
    \right) . 
\end{eqnarray}
in the three flavor indices~($u,d,s$), and an implicit sum over flavor indices ($a,b,...=1,2,3$) is understood in 
Eq.~(\ref{eq:lag_dec}). It is worth relating the $T_{abc}$ representation to the physical states as:
\begin{eqnarray}
&&T_{111}= \Delta^{++}; \quad T_{112}=\frac{\Delta^{+}}{\sqrt{3}};
\quad T_{122}=\frac{\Delta^{0}}{\sqrt{3}}; \quad
T_{222}=\Delta^{-}; \quad T_{113}=\frac{\Sigma^{*+}}{\sqrt{3}};
\quad T_{123}=\frac{\Sigma^{*0}}{\sqrt{6}}\nonumber\\
&& T_{223}=\frac{\Sigma^{*-}}{\sqrt{3}}; \quad 
T_{133}=\frac{\Xi^{*0}}{\sqrt{3}}; \quad 
T_{233}=\frac{\Xi^{*-}}{\sqrt{3}}; \quad T_{333}=\Omega^{-}.
\end{eqnarray}
The hadronic current for the NRB terms contributing to the different IE scattering processes 
are obtained using the chiral Lagrangians of the nonlinear sigma model discussed in this section. 

\subsubsection{Resonance~($R_{J}$; $J=\frac{1}{2}, \frac{3}{2}$) contribution}\label{res:inelastic}
Besides the NRB contribution to the IE scattering processes, there are several resonances with 
spin $\frac{1}{2}$, $\frac{3}{2}$, $\frac{5}{2}$, etc., which contribute to these processes. In Table-\ref{Tab:Resonance}, 
we have tabulated the properties of those resonances which have been considered in this work and these will be separately 
discussed for each process of present interest. It may be noticed from the table that considered resonances are spin 
$\frac12$  and spin $\frac32$ resonant states with positive or negative parity. We discuss in brief the structure of 
the transition current for these resonant states. The nucleon and delta resonances which are excited in the IE reactions are 
characterized by their mass, parity, spin and isospin and are represented by the symbol $R_{IJ} 
(M_{R})$~(Table-\ref{Tab:Resonance}), where $R$ is the name of the resonance given on the basis of its orbital angular 
momentum i.e. $L=0,1,2$ and named $S,~ P,~ D$, etc., showing its parity, $M_{R}$ is the mass while $I$ and $J$ specify 
their isospin and spin quantum numbers. 

\subsubsection{Charged current induced resonance excitation}\label{CC}
The basic (anti)neutrino induced CC reactions on the nucleon target for the IE processes through the 
resonance excitations are
\begin{eqnarray}\label{eq:nucc_res}
\nu_{_l}(k) + N(p) &\longrightarrow &l^{-}(k^{\prime}) + { R}(p_R) ~\longrightarrow~ l^{-}(k^{\prime}) + m(p_{m}) + 
B(p^{\prime}), \\ 
\label{eq:anucc_res}
\bar\nu_{_l}(k) + N(p) &\longrightarrow& l^{+}(k^{\prime}) + {R}(p_R)  ~\longrightarrow~ l^{+}(k^{\prime}) + m(p_{m}) + 
B(p^{\prime}).
\end{eqnarray}
    \begin{table}
  \begin{center}
    \begin{tabular*}{170mm}{@{\extracolsep{\fill}}c c c c  c c cc c}
      \hline \hline
  Resonance           & $M_R$ & $\Gamma$ &  $I(J^P) $    &  \multicolumn{5}{c}{Branching Ratios(in $\%$)}  \\
   & (GeV) & (GeV)&  & $N\pi$& $N\eta$ & $K\Lambda$ & $K\Sigma$&$\pi\pi N$ \\ \hline
      
  $P_{11} (1440)$  & $1.370 \pm 0.01$ & $0.175 \pm 0.015$ & $1/2(1/2^+)$ & $65$ & $<1$ & - & -&34 \\ \hline
      
  $S_{11} (1535)$  & $1.510 \pm 0.01$ & $0.130 \pm 0.020$ & $1/2(1/2^-)$ & $42$ & $42$ & - & -& 8\\ \hline
      
  $S_{31} (1620)$  & $1.600 \pm 0.01$ & $0.120 \pm 0.020$ & $3/2(1/2^-)$ & $30$ & - & - & -& 67\\ \hline
   
  $S_{11} (1650)$  & $1.655 \pm 0.015$ & $0.135 \pm 0.035$ & $1/2(1/2^+)$ & $60$ & $25$ & $10$ & -&22 \\ \hline
     
  $P_{11} (1710)$  & $1.700 \pm 0.02$ & $0.120 \pm 0.040$ & $1/2(1/2^+)$ & $10$ & $30$ & $15$ & $<1$&- \\ \hline
      
  $P_{11} (1880)$  & $1.860 \pm 0.04$ & $0.230 \pm 0.050$ & $1/2(1/2^+)$ & $6$ & $30$ & $20$ & $17$&55 \\ \hline
      
  $S_{11} (1895)$  & $1.910 \pm 0.02$ & $0.110 \pm 0.030$ & $1/2(1/2^-)$ & $10$ & $25$ & $18$ & $13$&- \\ \hline
%
      
      $P_{33} (1232)$  & $1.210 \pm 0.001$ & $0.100 \pm 0.002$ & $3/2(3/2^+)$  & $99.4$ & - & - & -&- \\ \hline
      
 $D_{13} (1520)$  & $1.510 \pm 0.005$ & $0.110 \pm^{0.010}_{0.005}$ & $1/2(3/2^-)$  & $60$ & - & - & - & 30
 \\ \hline
%
      
      
 $D_{33} (1700)$  & $1.665 \pm 0.025$ & $0.250 \pm 0.05$ & $3/2(3/2^-)$ & $15$ & - & - & - &32\\ \hline
      
 $P_{13} (1720)$  & $1.675 \pm 0.015$ & $0.250 \pm_{0.150}^{0.100}$ & $1/2(3/2^+)$& $11$ & $3$ & $4.5$ & - &70
 \\ \hline
      
      
 $P_{13} (1900)$  & $1.920 \pm 0.02$ & $0.150 \pm 0.05$ & $1/2(3/2^+)$ & $10$ & $8$ & $11$ & $5$&60 \\ \hline
%
  \end{tabular*}
  \end{center}
\caption{Properties of the spin $1/2$ and $3/2$ resonances available in the PDG~\cite{ParticleDataGroup:2020ssz}, with 
Breit-Wigner mass $M_R$, the total decay width $\Gamma$, isospin $I$, spin $J$, parity $P$, and the central value of the 
branching ratio into different meson-baryon like $N\pi$, $N\eta$, $K \Lambda$, $K \Sigma$ and $\pi \pi N$.}
\label{Tab:Resonance}
\end{table}
In the following, we will first discuss the excitation of spin $\frac12$ resonances and their subsequent decay to 
meson-baryon final state, followed by the discussion of spin $\frac32$ resonances.
\begin{itemize}
 \item [A.] {\bf Spin $\frac12$ resonances}\\
 The hadronic current for nucleon to spin $\frac12$ resonance state is given by  
\begin{eqnarray}\label{had_curr_1/2}
j^{\mu}_{\frac{1}{2}}=\bar{u}(p') \Gamma^\mu_{\frac12} u(p), 
\end{eqnarray}
where $u(p)$ and $\bar u(p^\prime)$ are respectively the Dirac spinor and adjoint Dirac spinor for spin $\frac{1}{2}$ 
particles and $\Gamma^\mu_\frac12$ is the vertex function, given by 
\begin{align}\label{eq:vec_half_pos}
  \Gamma^{\mu}_{\frac{1}{2}^{\pm}} &= [{V}^{\mu}_\frac{1}{2} -  {A}^{\mu}_\frac{1}{2} ]\cdot \begin{pmatrix}
\mathbb{I}_{4}             \\                                                                       
 \gamma_{5}                                                                                   \end{pmatrix}
  \end{align}
where upper~(lower) sign stands for a positive~(negative) parity resonance, $V^{\mu}_{\frac{1}{2}}$ and 
$A^{\mu}_{\frac{1}{2}}$, respectively, represent the vector and axial-vector currents, 
which are parameterized in terms of the vector~$(f_{1,2}(Q^2))$ and the axial-vector~(${g_{1,3}}(Q^2)$) form factors, assuming 
the absence of SCC, and are written as, 
\begin{align}  \label{eq:vectorspinhalfcurrent}
  V^{\mu}_{\frac{1}{2}} & =\frac{{f_{1}^{CC}}(Q^2)}{(2 M)^2}
  \left( Q^2 \gamma^\mu + {q\hspace{-.5em}/} q^\mu \right) + \frac{f_2^{CC}(Q^2)}{2 M} 
  i \sigma^{\mu\alpha} q_\alpha ,  \\
    \label{eq:axialspinhalfcurrent}
  A^{\mu}_\frac{1}{2} &=  \left[{g_1^{CC}}(Q^2) \gamma^\mu  +  \frac{g_3^{CC}(Q^2)}{M} q^\mu\right] \gamma_5 ,
\end{align}
\begin{table}
\begin{center}
\begin{tabular*}{170mm}{@{\extracolsep{\fill}}ccccc}
\hline
 $N^*\;$ & Amplitude & ${\mathcal A}_{\alpha}(0)$ &
$a_1$ & $b_1$ \\
\hline
$S_{11}(1535)$ & $A_{\frac{1}{2}}$ & $ 95.0$ & $ 0.5$ & $0.51$\\
                & $S_{\frac{1}{2}}$ & $ -2.0$ & $  23.9$& $0.81$\\
$S_{11}(1650)$ & $A_{\frac{1}{2}}$ & $ 33.3$ & $ 1.45 $ & $0.62$\\
                & $S_{\frac{1}{2}}$ & $ -3.5$ & $ 2.88 $ & $0.76$\\ 
$P_{11}(1710)$ & $A_\frac{1}{2}$ & $50.0$ & $ 1.4$  & $0.95$\\
                & $S_\frac{1}{2}$ & $  27.4$ & $  0.18 $  & $0.88$\\                
$P_{13}(1720)$ & $A_{\frac{1}{2}}$ & $ 100.0$ & $ 1.89$ &  $1.55$\\
                & $A_{\frac{3}{2}}$ & $ 30.0$ & $ 1.83$ &  $1.0$\\
                & $S_{\frac{1}{2}}$ & $ -53.0$ & $  2.46$&  $1.55$\\             
                \hline
\end{tabular*}
\end{center}
\caption{Parameterization of the transition form factors for the spin $\frac{1}{2}$ and $\frac{3}{2}$ resonances on proton 
target. ${\mathcal A}_{\alpha}(0)$ is given in units of $10^{-3}\,{\textrm {GeV}}^{-\frac{1}{2}}$ and the coefficients $a_1$, 
and $b_1$ in units of ${\textrm {GeV}}^{-2}$, and ${\textrm {GeV}}^{-2}$, respectively.}
\label{tab:param-p2}
\end{table}
where $f_i^{CC}(Q^2)$~($i=1,2$) are the isovector transition form factors which in turn are expressed in terms of the 
charged~($f_{i}^{R+} (Q^2)$) and neutral~($f_{i}^{R0} (Q^2)$) electromagnetic transition form factors as:
\begin{equation}\label{eq:f12vec_res_12}
f_i^{CC}(Q^2) = f_i^{R+}(Q^2) - f_i^{R0}(Q^2), \quad \quad i=1,2
\end{equation}
for isospin $\frac{1}{2}$ resonant states like $P_{11} (1440)$, $S_{11} (1535)$, etc., and as
\begin{equation}\label{eq:f12vec_res_12_neg}
 f_i^{CC}(Q^2) = - f_i^{R}(Q^2), \quad \quad i=1,2
\end{equation}
where $R=R^+$ for the proton target and $R=R^0$ for the neutron target, for  isospin $\frac{3}{2}$ resonant states like 
$S_{31} (1620)$. The electromagnetic form factors are extracted from the meson electroproduction data, especially from the 
pion electroproduction data.

The electromagnetic transition form factors $f^{R+,R0}_{i}(Q^2)$ are derived from the helicity amplitudes $A_{\frac{1}{2}}$ 
and $S_{\frac{1}{2}}$ extracted from the real and/or virtual photon scattering experiments. In order to determine the 
helicity amplitudes $A_{1/2}$ and $S_{1/2}$, one assumes the interaction of a nucleon with a virtual/real photon to produce a 
spin $1/2$ resonance. The helicity amplitudes for the process $\gamma N \longrightarrow R_{1/2}$ are expressed in terms of 
the polarization of the photon and the spins of the incoming nucleon and the outgoing spin $1/2$ resonance, where the spin of 
the resonance is fixed in the positive Z-direction, i.e. $J_{z}^{R} = +1/2$. The expressions for $A_{1/2}$ and $S_{1/2}$ are 
defined as~\cite{Leitner:2008ue}:
\begin{eqnarray}\label{Ch11_a12}
 A_{1/2}^{N} &=& \sqrt{\frac{2 \pi \alpha}{K_{R}}} <{R, J_{z}^{R} = +1/2} |\epsilon_{\mu}^{+} V^{\mu}| {N, 
 J_{z}^{N} = -1/2}> e^{i\phi}, \\
 \label{Ch11_s12}
 S_{1/2}^{N} &=& - \sqrt{\frac{2 \pi \alpha}{K_{R}}} \frac{|\vec{q}|}{\sqrt{Q^{2}}} <{R, J_{z}^{R} = +1/2} |
 \epsilon_{\mu}^{0} V^{\mu} |{N, J_{z}^{N} = +1/2}> e^{i\phi},
\end{eqnarray}
where $\phi$ is the phase factor, which relates the amplitude for the production of the resonances and the nucleons in the final state, $K_{R} = 
(M_{R}^{2} - M^{2})/2M_{R}$ is the momentum of the real photon measured in the resonance rest frame and $|\vec{q}|$ is the 
momentum of the virtual photon measured in the laboratory frame given as
\begin{equation}
 |\vec{q}| = \sqrt{\frac{(M_{R}^{2} - M^{2} - Q^{2})^{2}}{(2M_{R})^{2}} + Q^{2}}.
\end{equation}
The expressions for $V^{\mu}$ is given in Eq.~(\ref{eq:vectorspinhalfcurrent}) and $\epsilon_{\mu}$ represents the photon 
polarization vector. The transverse polarized photon vector $\epsilon_{\mu}^{\pm}$ is defined as
\begin{equation}\label{Ch11_transverse}
 \epsilon_{\mu}^{\pm} = \mp \frac{1}{\sqrt{2}} (0,1,\pm i,0),
\end{equation}
and for the longitudinal polarization of the photon $\epsilon_{\mu}^{0}$ is defined as
\begin{equation}\label{Ch11_longitudinal}
 \epsilon_{\mu}^{0} = \frac{1}{\sqrt{Q^{2}}} (|\vec{q}|, 0, 0, q^{0}).
\end{equation}
From Eqs.~(\ref{Ch11_a12}), (\ref{Ch11_s12}), (\ref{Ch11_transverse}) and (\ref{Ch11_longitudinal}), one may observe that 
for the spin $1/2$ resonances, $A_{1/2}$ represents the interaction of the transverse polarized photons with the $NR_{1/2}$ 
vertex whereas $S_{1/2}$ represents the interaction of the longitudinally polarized photons with the $NR_{1/2}$ vertex. 

\begin{table}
\begin{center}
\begin{tabular*}{170mm}{@{\extracolsep{\fill}}ccccc}
\hline
 $N^*\;$ & Amplitude & ${\mathcal A}_{\alpha}(0)$ &
$a_1$ & $b_1$ \\
\hline
$S_{11}(1535)$ & $A_{\frac{1}{2}}$ & $ -78.0$ & $ 1.75$ & $1.75$\\
                & $S_{\frac{1}{2}}$ & $ 32.5$ & $  0.4$& $1.0$\\
$S_{11}(1650)$ & $A_{\frac{1}{2}}$ & $ 26.0$ & $ 0.1 $ & $2.5$\\
                & $S_{\frac{1}{2}}$ & $ 3.8$ & $ 0.4 $ & $0.71$\\ 
$P_{11}(1710)$ & $A_\frac{1}{2}$ & $-45.0$ & $ -0.02$  & $0.95$\\
                & $S_\frac{1}{2}$ & $  -31.5$ & $  0.35 $  & $0.85$\\                
$P_{13}(1720)$ & $A_{\frac{1}{2}}$ & $ -2.9$ & $ 12.7$ &  $1.55$\\
                & $A_{\frac{3}{2}}$ & $ -31.0$ & $ 3.0$ &  $1.55$\\
                & $S_{\frac{1}{2}}$ & $ 0.0$ & $  0.0$&  $0.0$\\ \hline
\end{tabular*}
\end{center}
\caption{Parameterization of the transition form factors for the spin $\frac{1}{2}$ and $\frac{3}{2}$ resonances on neutron 
target. ${\mathcal A}_{\alpha}(0)$ is given in units of $10^{-3}\,{\textrm {GeV}}^{-\frac{1}{2}}$ and the coefficients $a_1$, 
and $b_1$ in units of ${\textrm {GeV}}^{-2}$, and ${\textrm {GeV}}^{-2}$, respectively.}
\label{tab:param-p2n}
\end{table}

The explicit relations between the form factors $f_i^{R+,R0}(Q^2)$ and the helicity amplitudes $A_{\frac{1}{2}}^{p,n}(Q^2)$ 
and $S_{\frac{1}{2}}^{p,n}(Q^2)$, for $\phi=0$, are given by~\cite{Leitner:2008ue}:
\begin{eqnarray}\label{eq:hel_spin_12_x}
A_\frac{1}{2}^{p,n}&=& \sqrt{\frac{2 \pi \alpha}{M} \frac{(M_R \mp M)^2+Q^2}{M_R^2 - M^2}} \left[  \frac{Q^2}{4 M^2}
f_1^{R+,R0} + \frac{M_R \pm M}{2 M} f_2^{R+,R0} \right] , \nonumber \\
S_\frac{1}{2}^{p,n}&=&\mp~\sqrt{\frac{ \pi \alpha}{M} \frac{(M \pm M_R)^2+Q^2}{M_R^2 - M^2}}
 \frac{(M_R \mp M)^2 +Q^2}{4 M_R M} \left[
\frac{M_R \pm M}{2 M} f_1^{R+,R0} - f_2^{R+,R0}\right],
\end{eqnarray}
where the upper sign represents the positive parity state and the lower sign denotes the negative parity state. $M_R$ is the 
mass of corresponding resonance and $f^{R+,R0}_{1,2}(Q^2)$ are the electromagnetic transition form factors. The vector form 
factors $f^{R+,R0}_i(Q^2)$ are related with the helicity amplitudes~(Eq.~(\ref{eq:hel_spin_12_x})) for which the $Q^2$ 
dependence is parameterized as~\cite{Tiator:2011pw}:
\begin{equation}\label{eq:ffpar}
{\mathcal A}_{\alpha}(Q^2) = {\mathcal A}_{\alpha}(0) (1+a_1 Q^2)\, e^{-b_1 Q^2} ,
\end{equation}
where $ {\mathcal A}_{\alpha}(Q^2)$ are the helicity amplitudes; $A_{\frac12}(Q^2)$ and $S_{\frac12}(Q^2)$ and parameters 
${\mathcal A}_{\alpha}(0)$ are generally determined by a fit to the photoproduction data of the corresponding resonance. 
While the parameters $a_1$ and $b_1$ { in the case of proton target} for each amplitude are obtained from the electroproduction data available at different 
 $Q^2$. {While for the neutron target, these parameters are determined using the data available for the inverse pion photoproduction~($\pi^- p \rightarrow \gamma n$) process~\cite{CrystalBall:2004tnd}.} Not all the resonances quoted in Table~\ref{Tab:Resonance} are well understood by the photo- and electro- production 
data. The MAID group~\cite{Tiator:2011pw} has parameterized the values of these parameters for the resonances which have been 
experimentally studied in the photo- and electro- production processes and the values of these parameters for the proton and 
neutron targets are taken from Ref.~\cite{Tiator:2011pw} for $P_{11} (1440)$, $D_{13} (1520)$, $S_{31} (1620)$, and $D_{33} 
(1700)$ resonances. However, for some resonances, like $S_{11}(1535)$, $S_{11}(1650)$, and $P_{13}(1720)$, there are latest experimental 
data for the photo- and electro- production processes as well as for the helicity amplitudes, therefore, we have refitted the 
values of ${\mathcal A}_{\alpha}(0)$, $a_1$ and $b_1$, to explain the latest data, and the refitted values for these 
resonances are given in Tables~\ref{tab:param-p2} and \ref{tab:param-p2n}, respectively, for proton and neutron targets. 
Moreover, for the resonances which are not parameterized by the MAID group, we have taken the value of ${\mathcal A}_{\alpha} 
(0)$ from PDG~\cite{ParticleDataGroup:2020ssz} and fitted the values of $a_{1}$ and $b_{1}$ to the available data. For 
example, in the case of $P_{11}(1710)$ resonance, we have fitted the $Q^2$ dependence to explain the pion electroproduction 
data from the CLAS collaboration given in Ref.~\cite{CLAS:2014fml}. 

    \begin{table}
  \begin{center}
    \begin{tabular*}{120mm}{@{\extracolsep{\fill}}c   c c c c}
      \hline 
  Resonance           &    $g_{RN\pi}$& $g_{RN\eta}$ & $g_{RK\Lambda}$ & $g_{RK\Sigma}$ \\ \hline
      
  $P_{11} (1440)$   & 0.38 & - & - & - \\ \hline
      
  $S_{11} (1535)$ & $0.10195$ & $-0.3696$ & - & -\\ \hline
      
  $S_{31} (1620)$   & 0.18 & - & - & -\\ \hline
   
  $S_{11} (1650)$   & $0.0915$ & $0.1481$ & $0.09766$ & - \\ \hline
     
  $P_{11} (1710)$  & $0.04182$ & $0.15675$ & $-0.2386$ \\ \hline
      
  $P_{11} (1880)$   & $0.0277$ & $0.137$ & $-0.2218$ & $0.1276$ \\ \hline
      
  $S_{11} (1895)$  & $0.0261$ & $0.0961$ & $0.0758$ & $0.05587$ \\ \hline

      $P_{33} (1232)$   & 2.14 & - & - & - \\ \hline
      
 $D_{13} (1520)$   & 1.6 & - & - & - 
 \\ \hline
      
 $D_{33} (1700)$   & 1.288 & - & - & - \\ \hline
      
 $P_{13} (1720)$  & $0.1165$ & $0.2248$ & $0.35$ & - 
 \\ \hline
      
 $P_{13} (1900)$   & $0.068$ & $0.149$ & $-0.091$ & $0.1023$ \\ \hline
  \end{tabular*}
  \end{center}
\caption{Strong coupling constants $g_{RMB}$ for the different resonances considered in the present work.}
\label{Tab:Resonance:para}
\end{table}

The axial-vector current consists of two form factors viz. $g_1^{CC}(Q^2)$ and $g_3^{CC}(Q^2)$, which are determined 
assuming the PCAC hypothesis and PDDAC through the off diagonal GT relation for $N 
\longrightarrow R$ transition. This assumption allows us to relate the axial-vector form factor at $Q^2=0$ to the pion-nucleon 
scattering~(see Ref.~\cite{Athar:2020kqn}), which is also well understood experimentally, and leads to the following relation
\begin{equation}\label{eq:g1_pos}
g_1^{CC}(0)= 2 g_{RN\pi},
\end{equation}
for the isospin $\frac{1}{2}$ resonances, and 
\begin{equation}\label{eq:g1_neg}
g_1^{CC}(0)= -\sqrt{\frac{2}{3}} g_{RN\pi},
\end{equation}
for the isospin $\frac{3}{2}$ resonances, with $g_{RN\pi}$ being the coupling strength for $R_{\frac12} \to N\pi$ decay, 
which has been determined by the partial decay width of the resonance. Since no information about the $Q^2$ dependence of 
the axial-vector form factor is known experimentally, therefore, a dipole form is assumed as in the case of $N \rightarrow 
N^{\prime}$ or $N \rightarrow Y$ transitions:
\begin{equation}
 g_1^{CC}(Q^2) = \frac{g_1^{CC}(0)}{\left(1+\frac{Q^2}{M_{A}^2}\right)^2},
\end{equation}
with $M_{A}=1.026$~GeV, and the pseudoscalar form factor $g_3^{CC}(Q^2)$  is given by
\begin{equation}\label{eq:fp_res_spinhalf}
g_{3}^{CC}(Q^2) = \frac{(MM_{R}\pm M^{2})}{m_{\pi}^{2}+Q^{2}} g_1^{CC}(Q^2) ,
\end{equation}
where $+(-)$ sign is for positive~(negative) parity resonances.

The most general form of the hadronic currents for the s-channel~(direct resonance pole diagram) and u-channel~(cross 
resonance pole diagram) processes where a positive~(negative) parity resonance state $R^{\frac{1}{2}\pm}$ is produced and decays to a 
meson and baryon in the final state, are written as 
\begin{eqnarray}\label{eq:res1/2_had_current_pos}
j^\mu\big|_{sR}^{\frac{1}{2}\pm}&=& 
i \; a\;  \mathcal{C^{R}} 
  \bar u({p}\,') 
{{p\hspace{-.5em}/}_m} \Gamma_{s}\frac{{p\hspace{-.5em}/}+{q\hspace{-.5em}/}+M_{R}}{(p+q)^2-M_{R}^2+ iM_{R}\Gamma_{R}}
\Gamma^\mu_{\frac{1}{2}\pm}  
u({p}\,),\\
j^\mu\big|_{uR}^{\frac{1}{2}\pm} &=& 
i \; a\; \mathcal{C^{R}} 
  \bar u({p}\,') \Gamma^\mu_{\frac{1}{2}\pm}
\frac{{p\hspace{-.5em}/}'-{q\hspace{-.5em}/}+M_{R}}{(p'-q)^2-M_{R}^2+ iM_{R}\Gamma_{R}} {{p\hspace{-.5em}/}_m} \Gamma_{s}
u({p}\,),
\label{eq:res1/2_had_current_pos_u}
\end{eqnarray}
where $\Gamma_{s} = \gamma_{5}~(\mathbb{I}_{4})$ stands for the positive~(negative) parity resonances, 
$a=\cos \theta_c~(\sin \theta_{c})$ for CC $\Delta S=0~(\Delta S =1)$ process and $a= 1$ for 
NC process. $M_{R}$ and $\Gamma_{R}$ are, respectively, the masses and total decay width of these resonances 
and are given in Table~\ref{Tab:Resonance}. $\mathcal{C^{R}}$ is a constant which includes the coupling strength
tabulated in Table~\ref{Tab:Resonance:para}, and the isospin factor involve in ${\cal R} \longrightarrow MB$ 
transition. 

\item [B.] {\bf Spin $\frac{3}{2}$ resonances}

The general structure for the hadronic current for spin three-half resonance excitation is determined by the following 
equation~\cite{LlewellynSmith:1971uhs} 
\begin{eqnarray}\label{eq:had_current_3/2}
J_{\mu}^{\frac{3}{2}}=\bar{\psi}^{\nu}(p') \Gamma_{\nu \mu}^{\frac32} u(p), 
\end{eqnarray}
where u(p) is the Dirac spinor for nucleon, ${\psi}^{\mu}(p)$ is the Rarita-Schwinger spinor for spin three-half particle 
and $\Gamma_{\nu \mu}^{\frac32}$ is the weak $WNR_{\frac{3}{2}}$ vertex, given as 
\begin{eqnarray}\label{eq:gamma_3half_pos}
  \Gamma_{\nu \mu }^{\frac{3}{2}^{\pm}} &=& \left[{V}_{\nu \mu }^\frac{3}{2} - {A}_{\nu \mu }^\frac{3}{2}\right] 
  \cdot \begin{pmatrix}
           \gamma_{5}\\                                                                                                             
                       \mathbb{I}_{4}                                  \end{pmatrix} ,
\end{eqnarray}
where upper~(lower) sign stands for a positive~(negative) parity resonance, and $V_{\frac32}~(A_{\frac32})$ is the 
vector~(axial-vector) current for spin three-half resonances. The vector and the 
axial-vector part of the currents are given by
\begin{eqnarray}\label{eq:vec_3half_pos}
  V_{\nu \mu }^{\frac{3}{2}} &=& \left[ \frac{{ C}_3^V}{M} (g_{\mu \nu} {q\hspace{-.5em}/} \, - q_{\nu} \gamma_{\mu})+
  \frac{{ C}_4^V}{M^2} (g_{\mu \nu} q\cdot p' - q_{\nu} p'_{\mu}) 
  + \frac{{ C}_5^V}{M^2} (g_{\mu \nu} q\cdot p - q_{\nu} p_{\mu}) + 
  g_{\mu \nu} { C}_6^V\right] ,  \\
  \label{eq:axial_3half_pos}
  A_{\nu \mu }^{\frac{3}{2}} &=&- \left[ \frac{{ C}_3^A}{M} (g_{\mu \nu} {q\hspace{-.5em}/} \, - q_{\nu} \gamma_{\mu})+
  \frac{{ C}_4^A}{M^2} (g_{\mu \nu} q\cdot p' - q_{\nu} p'_{\mu})+
 {{ C}_5^A} g_{\mu \nu}
  + \frac{{ C}_6^A}{M^2} q_{\nu} q_{\mu}\right] \gamma_5 ,
\end{eqnarray}
where ${ C}^V_i$ and ${ C}^A_i$ are the vector and axial-vector CC transition form factors which are functions 
of $Q^2$. The CVC hypothesis leads to ${ C}_6^V(Q^2)=0$. 

The isovector ${C}_i^V; (i=3,4,5)$ form factors for the resonance which have $J=\frac32 , \, I=\frac12$, like $D_{13} 
(1520)$, $P_{13}(1720)$, etc., are written in terms of the electromagnetic charged~($C^{R+}_i(Q^2)$) and 
neutral~($C^{R0}_i(Q^2)$) transition form factors through a simple relation~\cite{Leitner:2006ww} as
\begin{equation}\label{eq:civ_NC}
{ C}_i^V = C^{R+}_i  - C^{R0}_i ; \;\; \qquad i = 3,4,5\, ,
\end{equation}
while for the resonance with $J=\frac32$ and $I=\frac32$ like $P_{33}(1232)$, $D_{33} (1700)$, etc., the isovector form 
factors ${C}_i^V; (i=3,4,5)$ are expressed as
\begin{equation}\label{eq:civ_NC_32}
{C}_i^V = - C^{N}_i ; \;\; \qquad i = 3,4,5 \, ,
\end{equation}
with $N=R0~(R+)$ stands for the neutral~(charged) electromagnetic form factor.

In the case of spin $3/2$ resonances, along with i.e. $J_{z}^{R} = +1/2$, $J_{z}^{R} = +3/2$ also contributes in the 
positive Z-direction. Again it is our choice to take $J_{z}^{R}$ in the positive Z-direction, one may obtain the expressions 
for the helicity amplitudes by fixing $J_{z}^{R}$ in the negative Z-direction. The expressions for $A_{1/2}$ and $S_{1/2}$ 
in terms of the matrix element of $V^{\mu}$ are given in Eqs.~(\ref{Ch11_a12}) and (\ref{Ch11_s12}) with $V^{\mu}$ defined in 
Eq.~(\ref{eq:vec_3half_pos}). The expression for $A_{3/2}$ is given below:
\begin{eqnarray}\label{Ch11_a32}
 A_{3/2}^{N} &=& \sqrt{\frac{2 \pi \alpha}{K_{R}}} <{R, J_{z}^{R} = +3/2} |\epsilon_{\mu}^{+} V^{\mu}| 
 {N, J_{z}^{N} = +1/2} >e^{i\phi}.
\end{eqnarray}
The relations between the vector form factors $C^{R+,R0}_i(Q^2)$ and helicity amplitudes are given as~\cite{Drechsel:2007if}:
\begin{eqnarray}\label{helicity1}
A_{\frac{3}{2}}^{p,n} &=& \sqrt{\frac{\pi \alpha}{M}  \frac{(M_R \mp M)^2+Q^2}{M_R^2-M^2}}  
\left[ \frac{C^{R+,R0}_3}{M} (M \pm  M_R) \pm  \frac{C^{R+,R0}_4}{M^2} \frac{M_R^2-M^2 - Q^2}{2} \right. \nonumber\\
&\pm& \left. \frac{C^{R+,R0}_5}{M^2} \frac{M_R^2-M^2 + Q^2}{2} \right] , \\
\label{helicity2}
 A_{\frac{1}{2}}^{p,n} &=&\sqrt{\frac{\pi \alpha}{3 M}  \frac{(M_R \mp M)^2+Q^2}{M_R^2-M^2}}  
 \left[ \frac{C^{R+,R0}_3}{M} \frac{M^2+M M_R +Q^2}{M_R} -  \frac{C^{R+,R0}_4}{M^2} \frac{M_R^2-M^2 -
Q^2}{2} \right. \nonumber \\
&-& \left.  \frac{C^{R+,R0}_5}{M^2} \frac{M_R^2-M^2 + Q^2}{2} \right] ,\\ 
\label{helicity3}
S_{\frac{1}{2}}^{p,n} &=& \pm \sqrt{\frac{\pi \alpha}{6 M}  \frac{(M_R \mp  M)^2+Q^2}{M_R^2-M^2}} 
\frac{\sqrt{ Q^4 + 2 Q^2 (M_R^2 + M^2) + (M_R^2-M^2)^2    }}{M_R^2} \nonumber \\
 &\times&  \left[ \frac{C^{R+,R0}_3}{M} M_R +  \frac{C^{R+,R0}_4}{M^2} M_R^2 +  \frac{C^{R+,R0}_5}{M^2}
\frac{M_R^2+M^2 + Q^2}{2}  \right],
\end{eqnarray}
where upper~(lower) signs stand for the positive~(negative) parity resonances, $A_{\frac32}(Q^2)$, $A_{\frac12}(Q^2)$, and $S_{\frac12} 
(Q^2)$ are the amplitudes corresponding to the transverse and longitudinal polarizations, respectively, and are parameterized as a function of  $Q^2$ 
using Eq.~(\ref{eq:ffpar}). Once the parameters $a_1$ and $b_1$ are fixed for 
$A_{\frac32}(Q^2)$, $A_{\frac12}(Q^2)$, and $S_{\frac12}(Q^2)$ amplitudes, one gets the form factors $C^{R+ , R0}_i(Q^2)$.
 
For the 
$\Delta(1232)$ resonance, the three vector form factors $C_i^V,i=3,4,5$ are given in terms of the isovector 
electromagnetic form factors for $p \longrightarrow \Delta^+$  transition and the parameterization of which are taken from 
the Ref.~\cite{Lalakulich:2006sw}, 
\begin{eqnarray}\label{vecff}
C_3^V(Q^2) &=& \frac{2.13}{(1+Q^2/M_V^2)^2}\times
\frac{1}{1+\frac{Q^2}{4M_V^2}}, \qquad 
\qquad C_4^V(Q^2) = \frac{-1.51}{(1+Q^2/M_V^2)^2}\times \frac{1}{1+\frac{Q^2}{4M_V^2}},\nonumber\\
\qquad C_5^V(Q^2) &=& \frac{0.48}{(1+Q^2/M_V^2)^2}\times
\frac{1}{1+\frac{Q^2}{0.776M_V^2}} .
\end{eqnarray}
with the vector dipole mass taken as $M_V= 0.84$ GeV. 
 
The axial-vector form factors are determined from the early analysis of weak pion production data at 
ANL~\cite{Radecky:1981fn} and BNL~\cite{Kitagaki:1986ct} by Schreiner and von Hippel~\cite{Schreiner:1973mj} using Adler's 
model, which are consistent with the hypothesis of PCAC and generalized GT relation. These considerations 
give $C_6^A(Q^2)$ in terms of $C_5^A(Q^2)$: 
\begin{align}\label{c6-c5}
 C_6^A(Q^2) =& C_5^A(Q^2) \frac{M^2}{Q^2 + m_\pi^2}.
 \end{align}
The $Q^2$ dependence of $C_5^A$ is parameterized by Schreiner and von Hippel~\cite{Schreiner:1973mj} and is given by 
 \begin{eqnarray}\label{c5aq}
  C_5^A(Q^2) &=&  \frac{C_5^A(0) \left( 1+ \frac{a\, Q^2}{b~+~Q^2} \right)}{\left( 1 + Q^2 /M_{A\Delta}^2 \right)^2} ,
\end{eqnarray}
with $a$ and $b$ determined from the experiments and found to be $a=-1.21$ and $b=2$~GeV$^2$~\cite{Radecky:1981fn, 
Kitagaki:1990vs}. $M_{A\Delta}$ is the axial dipole mass, and  $C_5^A(0)$ is given in terms of $g_{\Delta N \pi}$ as
 \begin{align}
 C_5^A(0) =& f_\pi \frac{ g_{\Delta N \pi}  }{2 \sqrt3 M },
\end{align}
with $g_{\Delta N \pi}$ being the $\Delta N \pi$ coupling strength  for $\Delta \longrightarrow N \pi$ decay.

The $Q^2$ dependence of $C_3^A(Q^2)$ and $C_4^A(Q^2)$ are obtained in the Adler's model as~\cite{Adler:1968tw, Schreiner:1973mj} 
\begin{equation}\label{c4-c3}
 C_4^A(Q^2) = -\frac{1}{4}C_5^A(Q^2);  \qquad            C_3^A(Q^2) =0.
\end{equation}
The form factors ${C}_i^A(Q^2), \; (i=3,4,5,6)$ corresponding to the axial current have not been studied in the case of 
higher resonances. The earlier calculations have used PCAC to determine ${C}_5^A(Q^2)$ and ${C}_6^A(Q^2)$ and taken other 
form factors to be zero. In view of this, we have also taken a simple model for the determination of the axial form factors 
based on PCAC and GT relation and use the relation between ${C}_5^A(Q^2)$ and ${C}_6^A(Q^2)$ given in 
Eq.~(\ref{c6-c5}) to write ${C}_6^A(Q^2)$ in terms of ${C}_5^A(Q^2)$ as
\begin{equation}\label{c6_CC}
 C_6^A(Q^2) = C_5^A(Q^2) \frac{M^2}{Q^2 + m_\pi^2} .
\end{equation}
For ${C}_5^A(Q^2)$, a dipole form has been assumed 
 \begin{equation}\label{c5a-r}
{C}_5^A(Q^2) = \frac{{C}_5^A(0)}{ \left( 1 + Q^2 /{M_A^{\it R}}^2 \right)^2 } ,
\end{equation}
with ${C}_5^A(0)= -2 g_{R N \pi}~(\sqrt{\frac{2}{3}} g_{R N \pi})$ for isospin $\frac{1}{2}~(\frac{3}{2})$ resonances, 
$g_{R N \pi}$ is the coupling for $R \longrightarrow N \pi$ decay for each resonance $R$. $M_{A}^{\it R}$ is taken as 
$1.026~{\rm GeV}$. ${C}_3^A(Q^2)$ as well as ${C}_4^A(Q^2)$ are taken as zero. 
  
One may write the most general form of the hadronic current for the s-channel~(direct resonance pole diagram) and the 
u-channel~(cross resonance pole diagram) processes where a positive~(negative) parity resonance state $R^{\frac{3}{2}\pm}$ is produced 
and decays to a meson and a baryon in the final state as
\begin{eqnarray}\label{eq:res_had_current_pos}
j^\mu\big|_{R}^{\frac{3}{2}\pm} &=& i \; a\; \mathcal{C^{R}} 
   \frac{p_m^\alpha \Gamma_{s}}{p_R^2-M_R^2+ i M_R \Gamma_R}
   \bar u({p}\,') P_{\alpha\beta}^{3/2}(p_R) \Gamma^{\beta\mu}_{\frac{3}{2}\pm}(p,q)
   u({p}\,),\quad \qquad ~~~p_R=p+q,  \\ 
   \label{eq:res_had_current_pos_u}
 j^\mu\big|_{C R}^{\frac{3}{2}\pm} &=& i \; a\; \mathcal{C^{R}} 
   \frac{p_m^\beta }{p_R^2-M_R^2+ i M_R \Gamma_R}
   \bar u(\vec{p}\,')  {\hat \Gamma}^{\mu\alpha}_{\frac{3}{2}\pm}(p',-q) P_{\alpha\beta}^{3/2}(p_R) \Gamma_{s}
   u({p}\,),\qquad p_R=p'-q,
\end{eqnarray}
where $\Gamma_{s} = \mathbb{I}_{4}~(\gamma_{5})$ stands for positive~(negative) parity resonances, 
with $\mathbb{I}_{4}$ being the $4 \times 4$ identity matrix.
${\hat \Gamma}^{\mu\alpha}_{\frac32}(p',-q) = \gamma_{0} {{ \Gamma}^{\mu\alpha}}^{\dagger}_{\frac32}(p',-q) 
\gamma_{0}$, $a=\cos \theta_c~(\sin \theta_{c})$ for CC $\Delta S=0~(\Delta S =1)$ process and $a= 1$ for 
 NC process. $M_{R}$ and $\Gamma_{R}$ are, respectively, the masses and total decay width of these 
resonances and are given in Table~\ref{Tab:Resonance}. The constant $\mathcal{C^{R}}$ includes the coupling strength, isospin 
factor involve in ${\cal R} \longrightarrow MB$ transition, etc., and has been tabulated in different sections for the 
corresponding IE processes. These resonances are generally off-shell and their off-shell effects are also taken into 
account. $P_{\alpha\beta}^{3/2}$ is spin three-half projection operator and is given by
\begin{equation}\label{propagator:32}
P_{\alpha\beta}^{3/2} (p^{\prime})=- \left({p\hspace{-.5em}/}^\prime + M_R \right) \left( g_{\alpha \beta}
- \frac{2}{3} \frac{p'_{\alpha } p'_{\beta}}{M_R^2} 
+ \frac{1}{3} \frac{p'_{\alpha } \gamma_{\beta} - p'_{\beta } \gamma_{\alpha}}{M_R} 
- \frac{1}{3} \gamma_{\alpha} \gamma_{\beta} \right).
\end{equation}
The structure of the matrix element for the hadronic current is given in 
Eqs.~(\ref{eq:res_had_current_pos})--(\ref{eq:res_had_current_pos_u}) for positive and negative parity resonances, 
respectively, and the weak vertex for positive and negative parity states are given in Eq.~(\ref{eq:gamma_3half_pos}). The vector and axial-vector pieces are written in Eqs.~(\ref{eq:vec_3half_pos}) and 
(\ref{eq:axial_3half_pos}), respectively, with corresponding form factors, ${C}^V_i$ and ${C}^A_i$, defined for each 
resonances. 
\end{itemize}

\subsubsection{Neutral current induced resonance excitation}\label{nc1pi}
In this section, we present in brief the structure of resonance terms that may contribute to the hadronic current of 
(anti)neutrino induced NC processes. The basic NC (anti)neutrino induced reactions for meson 
production through resonance excitations are the following:
\begin{eqnarray}\label{eq:nc_res}
\nu_{_l}(k) + N(p) &\longrightarrow& \nu_{l}(k^{\prime}) + {\cal R}(p_R) ~\longrightarrow ~ \nu_{l}(k^{\prime}) +  
B(p^\prime) + m (p_m),    \\
\label{eq:anc_res}
\bar\nu_{_l}(k) + N(p) &\longrightarrow& \bar\nu_{l}(k^{\prime}) + {\cal R}(p_R) ~\longrightarrow ~ \bar{\nu}_{l}(k^{\prime}) + 
B (p^\prime) + m (p_m),    
\end{eqnarray}
where ${\cal R}$ stands for the resonances~(${\cal R}$) which contribute to the meson production. We will discuss separately 
the contribution of spin $\frac{1}{2}$ and $\frac{3}{2}$ resonances to NC induced single meson production.

\begin{itemize}
 \item [A.] {\bf  Spin $\frac{1}{2}$ resonances}\\
For NC process producing a  spin $\frac{1}{2}$ resonance in the intermediate state, the hadronic current is 
given by Eq.~(\ref{had_curr_1/2}). $\Gamma^\mu_{\frac12}$ is the vertex function which for positive and negative parity states is given 
in Eq.~(\ref{eq:vec_half_pos}). The vector and 
axial-vector parts of the current are written in terms of vector and axial-vector form factors and have the same form as given in 
Eqs.~(\ref{eq:vectorspinhalfcurrent}) and (\ref{eq:axialspinhalfcurrent}), but with a modified form factor and a different 
expression for charged~(${f}_i^{R+}$) and neutral~(${f}_i^{R0}$) resonance states with the replacement of $f_{1,2}^{CC}$ by 
$\tilde{f}_{1,2}^{R+,R0}$, corresponding to isospin $\frac{1}{2}$ resonance.

In the case of isospin $\frac{1}{2}$ resonances, the explicit expressions for the vector and axial-vector form factors are 
written as  
\begin{eqnarray}
 \tilde{f}_i^p (Q^2) &=& \left(\frac{1}{2} -2 \sin^{2}\theta_W \right) f_i^{R+} (Q^2) - \frac{1}{2} f_i^{R0} (Q^2),~~~ 
 \qquad\tilde{g}_1^p (Q^2) = \frac{1}{2} g_1^{CC} (Q^2),
\end{eqnarray}
for the positive charged state and 
\begin{eqnarray}
  \tilde{f}_i^n (Q^2)&=&\left(\frac{1}{2} -2 \sin^{2}\theta_W \right) f_i^{R0} (Q^2) - \frac{1}{2} f_i^{R+} (Q^2) ,~~~ 
  \qquad\tilde{g}_1^n (Q^2) =-\frac{1}{2} g_{1}^{CC} (Q^2),
\end{eqnarray}
for the neutral state. While for the case of isospin $\frac{3}{2}$ resonances, these form factors $\tilde{f}_i^p$ and 
$\tilde{f}_i^n$ are given as:
   \begin{eqnarray}
 \tilde{f}_i^p (Q^2) &=& (1-2 \sin^{2}\theta_W) f_i^{R+} (Q^2), ~~~~~~~ \qquad
 \tilde{g}_1^p  (Q^2)=- g_{1}^{CC} (Q^2),
 \end{eqnarray}
for the positive charged state and 
\begin{eqnarray}
  \tilde{f}_i^n (Q^2)&=&(1 -2 \sin^{2}\theta_W) F_i^{R0} (Q^2), ~~~~~~~~~ \qquad
  \tilde{g}_1^n (Q^2)=- g_{1}^{CC} (Q^2) ,
 \end{eqnarray}
for the neutral state.

\item [B.] {\bf  Spin $\frac{3}{2}$ resonances}\\ 
The general structure for the hadronic current $J_{\mu}^{\frac{3}{2}}$ for NC induced spin $\frac{3}{2}$ 
resonance in the intermediate state is given by Eq.~(\ref{eq:had_current_3/2}), for which $\Gamma_{\nu \mu}^{\frac32 +,-}$ 
is given by Eq.~(\ref{eq:gamma_3half_pos}) for positive and negative parity 
states. The vector and axial-vector parts of the current are given by Eqs.~(\ref{eq:vec_3half_pos}) and 
(\ref{eq:axial_3half_pos}) with the corresponding NC form factors $({\tilde C}^V_i)~(i=3,4,5)$ and 
$({\tilde C}^A_i)~(i=4,5,6)$ which in the SM are given 
in terms of ${C}^{V}_i$ and ${C}^{A}_i$.  

The NC form factors $\tilde{C}^V_i$ and $\tilde{C}^A_i$~($i=3,4,5$) for the case of isospin $\frac{1}{2}$ resonances, are 
given by: 
\begin{align} 
({\tilde C}^V_i) \mathrel{\mathop{\longrightarrow}^{\mathrm{for \; p}}}  (\frac{1}{2}-2 \sin^2 \theta_W)  C^{R+}_i  - 
\frac12   C^{R0}_i     ,~~~~~
({\tilde C}^V_i) \mathrel{\mathop{\longrightarrow}^{\mathrm{for \; n}}}  (\frac{1}{2}-2 \sin^2 \theta_W)  C^{R0}_i  - 
\frac12   C^{R+}_i     ,~~~~~
({\tilde C}^A_i) \mathrel{\mathop{\longrightarrow}^{\mathrm{for \; p,n}}}  \pm  \frac12 \tilde{C}^A_i   ,
\end{align}
while for the isospin $\frac{3}{2}$ resonances, NC form factors for the proton and neutron targets, are given as:
\begin{align} 
({\tilde C}^V_i) \mathrel{\mathop{\longrightarrow}^{\mathrm{for \; p}}}  (1-2 \sin^2 \theta_W)  C^{R+}_i ,~~~~~
({\tilde C}^V_i) \mathrel{\mathop{\longrightarrow}^{\mathrm{for \; n}}}  (1-2 \sin^2 \theta_W)  C^{R0}_i  ,~~~~~
({\tilde C}^A_i) \mathrel{\mathop{\longrightarrow}^{\mathrm{for \; p,n}}} - \tilde{C}^A_i    .
\end{align}
\end{itemize}

\subsubsection{Strong couplings of the resonances}\label{coupling}
Due to the lack of experimental data there is large uncertainty associated with ${\cal R}MB$ coupling at the ${\cal R}\to 
MB$ vertex. We have fixed ${\cal R}MB$ coupling using the data of branching ratio and decay width of these resonances 
from PDG~\cite{ParticleDataGroup:2020ssz} and use the expression for the decay rate which is obtained by writing the most 
general form of $RMB$ Lagrangian given by~\cite{Athar:2020kqn}:
\begin{align}\label{eq:spin12_lag}
 \mathcal{L}_{R_{\frac{1}{2}} MB} &= \frac{g_{R\frac12 MB} }{f_m}\bar{\Psi}_{R_{\frac{1}{2}}} \; 
 \Gamma^{\mu}_{\frac{1}{2}}  \;
  \partial_\mu  \phi^i T_i \,\Psi \\
 \label{eq:spin32_lag} 
 \mathcal{L}_{R_{\frac{3}{2}} MB} &= \frac{g_{R\frac32 MB}}{f_m} \bar{\Psi}_{R_{\frac{3}{2}}} \; \Gamma^{\mu}_{\frac{3}{2}} 
 \;
\partial^\mu \phi^i T_i \,\, \Psi  
\end{align}
where $f_{m}$ is the meson decay constant, which in the case of pion production becomes $f_{m}=f_{\pi} = 
92.4$~MeV~\cite{ParticleDataGroup:2020ssz} and for eta and kaon production becomes $f_{m} = f_{\eta} = f_{K} = 
105$~MeV~\cite{Faessler:2008ix}. $g_{R\frac12 MB}$ and $g_{R\frac32 
MB}$ are, respectively, the ${\cal R}MB$ coupling strength for spin $\frac{1}{2}$ and $\frac{3}{2}$ resonances. $\Psi$ is 
the nucleon field and ${\Psi}_{R_{\frac{1}{2}}}$ and ${\Psi}_{R_{\frac{3}{2}}}$ are the fields associated with the 
resonances of spin $\frac12$ and spin $\frac32$, respectively. $\phi^i$ are the mesonic field and $T_i$ are the isospin 
operator which is $T = \vec{\tau}$ for isospin  $\frac12$ states and $T = T^\dagger$ for isospin $\frac32$ 
states~($\vec{\tau}$ and $T^\dagger$ are the isospin operator for  doublet and quartet, respectively). The interaction vertex 
$\Gamma^{\mu}_{\frac{1}{2}}$ is $\gamma^\mu \gamma^5$~($\gamma^\mu$) for spin $\frac12$ resonances with positive~(negative) 
parity. Similarly, the interaction vertex $\Gamma^{\mu}_{\frac{3}{2}}$, for spin $\frac32$ resonances for 
positive~(negative) parity state, are ${\mathbb I}_4 $~($\gamma_5$). Using the above Lagrangian one may obtain the 
expression for the decay width in the resonance rest frame as
\begin{align}\label{eq:12_width}
 \Gamma_{R_{\frac{1}{2}} \longrightarrow MB} &= \frac{\mathcal{C}}{4\pi} \left(\frac{g_{R\frac12 MB}}{f_m}\right)^2 
 \left(M_R \pm M_{B}\right)^2 
\frac{E_N \mp M_{B}}{M_R} |\vec{q}_{\mathrm{cm}}|, \\ 
\label{eq:32_width}
\Gamma_{R_{\frac{3}{2}} \longrightarrow MB} &= \frac{\mathcal{C}}{12\pi} \left(\frac{g_{R\frac32 MB}}{f_m}\right)^2 
\frac{E_N \pm M_{B}}{M_R} |\vec q_{cm}|^3,
\end{align}
where the upper(lower) sign represents the positive(negative) parity resonance state. 
The parameter $\mathcal{C}$ is obtained from the isospin analysis and found out to be  $3$ for isospin $\frac12$ state and 
$1$ for isospin $\frac32$ states. $|\vec q_{cm}|$ is the outgoing pion momentum measured from resonance rest frame and $E_N$ is the nucleon energy, which are 
given by, 
\begin{equation}\label{eq:pi_mom}
|\vec{q}_{\mathrm{cm}}| = \frac{\sqrt{(W^2-m_{m}^2-M_{B}^2)^2 - 4 m_{M}^2 M_{B}^2}}{2 M_R}, \qquad \qquad   E_N=\frac{W^2+M_{B}^2-m_{m}^2}{2 M_R}, 
\end{equation}
where $W$ is the CM energy carried by the resonance. 

In view of the above, we fix $N \Delta \pi$ coupling$(g_{\pi N \Delta})$ by comparing $\Delta \to N \pi$ decay width 
evaluated in the rest frame of $\Delta$ resonance, 
\begin{equation}
\Gamma_\Delta(s) = \frac{1}{6\pi} \left ( \frac{g_{\pi N \Delta}}{m_\pi}\right )^2
 \frac{M}{\sqrt s} \left [ \frac{\lambda^\frac12
  (s,m_\pi^2,M^2)}{2\sqrt s} \right ]^3 \Theta(\sqrt s
-M-m_\pi),\qquad s= p_\Delta^2 ,
\end{equation}
where  $\lambda(x,y,z) = x^2+y^2+z^2-2xy-2xz-2yz$ is K\"{a}llen function. To get the offshell effect of $\Delta(1232)$ 
resonance we have taken momentum dependent width. 

\subsection{Single pion production}\label{sec:1pion}
Historically, the weak pion production induced by (anti)neutrinos has been studied for a long time starting from 
1962~\cite{Dennery:1962zz, Dombey:1962mi, Bell:1962fyy, Bell:1970mc} in the energy region of (anti)neutrinos relevant for the 
early experiments done at CERN, ANL and BNL. These early calculations used various approaches based on the 
\begin{itemize}
\item[(i)] dynamical models with dispersion theory,
\item[(ii)] quark models with higher symmetry like $SU(6)$, and 
\item[(iii)] phenomenological Lagrangians for describing the interaction of mesons with nucleons and excitation of higher 
resonances.
\end{itemize}
These calculations have been comprehensively summarized by Adler~\cite{Adler:1968tw}, Llewellyn 
Smith~\cite{LlewellynSmith:1971uhs} and Schreiner and von Hippel~\cite{Schreiner:1973mj}. 

In the low energy region corresponding to the threshold production of pions various theoretical models motivated by the 
chiral symmetry were used to study these processes. For example, the low energy theorems~(LET) based on 
PCAC~\cite{Gershtein:1980vd, Komachenko:1983jv} and/or current 
algebra~(CA) as well as the effective Lagrangians incorporating the chiral symmetry which were formulated to study the photo 
and electroproduction of pions were extended to study the weak production of pions induced by (anti)neutrinos. The early work 
using this approach has been summarized by Adler and Dashen~\cite{Adler:1968} and Treiman et al.~\cite{Treiman:1972}. In 
recent years, the advances made in 
the field of chiral perturbation theory have been used to study the (anti)neutrino induced pion production in the threshold 
region~\cite{Hernandez:2007qq, RafiAlam:2015fcw, Nikolakopoulos:2018gtf, Gonzalez-Jimenez:2017fea, RafiAlam:2015zqz}.
 
After the experimental results from the hydrogen and deuterium bubble chamber experiments from ANL~\cite{Radecky:1981fn} and 
BNL~\cite{Kitagaki:1990vs} and later experiments from CERN~\cite{Lee:1976wr, Bell:1978qu, 
Aachen-Bonn-CERN-Munich-Oxford:1980iaz} and other laboratories on the nuclear targets, many new calculations were made 
using the phenomenological Lagrangian~\cite{Schreiner:1973mj, Fogli:1979cz, Sato:2003rq, Fogli:1979qj, Kamano:2012id, 
Nakamura:2013zaa, Paschos:2003qr, Lalakulich:2005cs, Barbero:2008zza, Barbero:2013eqa}, the Lagrangian based on the chiral 
symmetry~\cite{Hernandez:2007qq, RafiAlam:2010kf, Leitner:2008ue, RafiAlam:2015fcw, Leitner:2006ww, Lalakulich:2006sw, RafiAlam:2015zqz, 
Lalakulich:2010ss, Gonzalez-Jimenez:2016qqq, RafiAlam:2013jcs, Alam:2011vwg} and the quark 
model~\cite{Rein:1980wg}. In this article it is not possible to describe all the approaches mentioned above and we 
choose to focus on the effective Lagrangian approach to describe the single pion production induced by (anti)neutrinos from 
the nucleon targets. We use an effective Lagrangian obtained using the nonlinear realization of chiral symmetry to calculate 
the NR contribution and a phenomenological Lagrangian to calculate the resonance excitations and its decay to pions, 
as discussed earlier in Section~\ref{res:inelastic} and for details, readers are referred to Ref.~\cite{Athar:2020kqn}.
 
In the following, we first discuss the pion production induced by CC in Section~\ref{CC:pion} and then
the pion production from NC induced processes are discussed in Section~\ref{NC:pion}.

\subsubsection{Charged current (anti)neutrino induced processes}\label{CC:pion}
\begin{figure}  
\centering
\includegraphics[height=7cm]{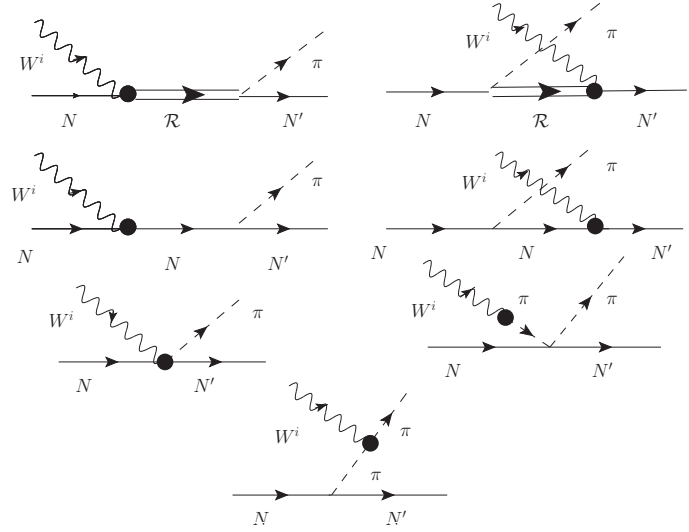}
\caption{Feynman diagrams contributing to the hadronic current corresponding 
to $W^{i} N \to N^{\prime} \pi^{\pm,0}$, where $(W^i \equiv W^\pm \; ; i=\pm)$ for CC processes and $(W^i 
\equiv Z^0 \; ; i=0)$ for NC processes with $N,N^{\prime}=p \;{\rm or}\; n$. First row: direct and cross 
diagrams for resonance production where intermediate term ${\cal R}$ stands for different resonances. Second row: nucleon 
pole~(NP and CNP) terms. The contact term~(CT) and pion pole~(PP) term (third row left to right) and pion in 
flight~(PF)~(fourth row).}
\label{fig:feynmann}
\end{figure}

The various possible reactions which may contribute to the single pion production through CC 
(anti)neutrino induced reaction on a nucleon target are the following:
\begin{eqnarray}
 \nu_{_l} p &\to l^- p \pi^+,  \qquad \nu_{_l} n \to l^- n \pi^+, \qquad  \nu_{_l} n \to l^- p \pi^0& \nonumber\\
\bar \nu_{_l} n &\to l^+ n \pi^-, \qquad  \bar \nu_{_l} p \to l^+ p \pi^-, \qquad 
    \bar \nu_{_l} p \to l^+ n \pi^0 &\quad \qquad ; \; l = e,\mu
\label{eq:CC_all}
\end{eqnarray}
The Feynman diagrams which may contribute to the matrix element of the hadronic current are shown in 
Fig.~\ref{fig:feynmann}. The NRB terms include five diagrams viz. direct~(NP) and cross nucleon 
pole~(CP), contact term~(CT), pion pole~(PP) and pion in flight~(PF) terms. For the $\Delta(1232)$ resonance we have included 
both direct~(s-channel) and cross~(u-channel) diagrams. Apart from the $\Delta(1232)$ resonance, which 
mainly~(Table~\ref{tab:born_para_1}) decays to $N\pi$, we have also taken contributions from $P_{11}$(1440), $S_{11}$(1535), 
$S_{31}(1620)$, and $S_{11}$(1650) spin half resonances and $D_{13}$(1520), $D_{33}(1700)$, and $P_{13}$(1720) spin three-half 
resonances and considered both s-channel and u-channel contributions. In the following, we present the formalism in 
brief which has been used for the NRB terms and the resonant spin half and spin three-half contributions 
to the one pion production processes. 

The contribution from the NRB terms in the case of CC~$(W^i \equiv W^\pm \; ; i=\pm)$ and 
NC~$(W^i \equiv Z^0 \; ; i=0)$ reactions $W^i N \to N^{\prime}\pi$ has been obtained using nonlinear sigma 
model~\cite{Athar:2020kqn, Hernandez:2007qq, RafiAlam:2015fcw} described in Section~\ref{NRB}. In the lowest order, the 
contributions to the hadronic current are written as~\cite{Hernandez:2007qq}:
\begin{eqnarray} \label{eq:background}
j^\mu\big|_{NP} &=& 
a~\mathcal{A}^{NP}
  \bar u(\vec{p}\,') 
 {p\hspace{-.5em}/}_\pi \gamma_5\frac{{{p\hspace{-.5em}/}}+{q\hspace{-.5em}/}+M}{(p+q)^2-M^2}\left [V^\mu_N(Q^2)-A^\mu_N(Q^2) 
 \right]  u(\vec{p}\,), \\
j^\mu\big|_{CP} &=& 
a~\mathcal{A}^{CP}
  \bar u(\vec{p}\,') \left [V^\mu_N(Q^2)-A^\mu_N(Q^2) \right]
\frac{{p\hspace{-.5em}/}^\prime-{q\hspace{-.5em}/}+M}{(p'-q)^2-M^2} {p\hspace{-.5em}/}_\pi \gamma_5  u(\vec{p}\,),
\\
j^\mu\big|_{CT} &=&
a~\mathcal{A}^{CT}
  \bar u(\vec{p}\,') \gamma^\mu\left (
  g_A  f_{CT}^V(Q^2)\gamma_5 - f_\rho\left((q-p_\pi)^2\right) \right ) u(\vec{p}\,),  \\
j^\mu\big|_{PP} &=& 
a~\mathcal{A}^{PP}f_\rho\left((q-p_\pi)^2\right)
  \frac{q^\mu}{m_\pi^2+Q^2}
  \bar u(\vec{p}\,')\ {q\hspace{-.5em}/} \ u(\vec{p}\,),  \\
j^\mu\big|_{PF} &=& 
a~\mathcal{A}^{PF}f_{PF}(Q^2)
  \frac{(2p_\pi-q)^\mu}{(p_\pi-q)^2-m_\pi^2}
  2M\bar u(\vec{p}\,')  \gamma_5 u(\vec{p}\,),\label{eq:eqscc}
\end{eqnarray}
with $ a = \cos \theta_C$ for CC induced process. $q$ is the four momentum transfer(=$k-k^\prime$), $q^2(=-Q^2) \le 
0$ and $p_\pi$ is the pion momentum and $m_\pi$ is the mass of pion. The constant factor $\mathcal{A}^i, \; i=NP,CP, CT, PP 
\; {\rm and} \; PF$, and are tabulated in  Table--\ref{tab:born_para_1}.

The  vector$(V^\mu_N(Q^{2}))$ and axial-vector$(A^\mu_N(Q^{2}))$ currents for nucleon pole diagrams in the case of CC and 
NC interactions are calculated neglecting SCC and are given by,
\begin{eqnarray}\label{eq:vec_curr}
V^\mu_N(Q^2)=f_1(Q^2)\gamma^\mu+ f_2(Q^2)i\sigma^{\mu\nu}\frac{q_\nu}{2M} , \qquad \quad
A^\mu_N(Q^2)= \left( g_1(Q^2)\gamma^\mu 
+ g_3(Q^2) \frac{q^\mu}{M}  \right)\gamma^5,
\end{eqnarray}
where  $f_{1,2}(Q^2)$ and $g_{1,3}(Q^2)$ are the vector and axial-vector form factors for the nucleons. The form factors 
$f_{1,2}(Q^2)$ are expressed in terms of the electromagnetic nucleon form factors~($F_{1,2}^{p,n}(Q^2)$) as:
\begin{equation}\label{f1v_f2v}
f_{1,2}(Q^2) = F_{1,2}^p(Q^2)- F_{1,2}^n(Q^2), 
\end{equation}
where $F_{i}^{p,n}(Q^2);  i=1,2$ are the Dirac and Pauli form factors of the nucleons. These form factors 
are in turn expressed in terms of the experimentally determined Sachs' electric $G_E^{p,n}(Q^2)$ and magnetic $G_M^{p,n}
(Q^2)$ form factors~\cite{Galster:1971kv}. 

On the other hand, the axial-vector form factor~($g_{1}(Q^2)$) is generally taken to be of the dipole form and is given by
\begin{equation}\label{fa:in}
g_{1}(Q^2)=g_A(Q^2)=g_A(0)~\left(1+\frac{Q^2}{M_A^2}\right)^{-2},
\end{equation}
where $g_A(0)$ is the axial charge and is  obtained from the QE neutrino and antineutrino scattering as well as 
from the pion electro-production data. We have used $g_A(0)$=1.267 and $M_A$=1.026GeV~\cite{Bernard:2001rs}, in the 
numerical calculations.

The next contribution from the axial-vector part comes from the pseudoscalar form factor $ g_3(Q^2)$, the determination of 
which is based on PCAC and PDDAC and is related to $g_1(Q^2)$ through the relation 
\begin{equation}\label{fp}
g_3 (Q^2)=\frac{2M^2 \; g_1 (Q^2)}{m_\pi^2+Q^2}.
\end{equation}
In order to conserve vector current for CC processes at the weak vertex, the two form factors viz. $f_{PF}(Q^2)$ and 
$f_{CT}^{V}(Q^2)$ are expressed in terms of the isovector nucleon form factor as~\cite{Hernandez:2007qq}
\begin{equation}
 f_{PF}(Q^2) = f_{CT}^{V}(Q^2) = 2 f_{1}(Q^2).
\end{equation}
The $\pi \pi NN$ vertex has the dominant $\rho$--meson cloud contribution and following Ref.~\cite{Hernandez:2007qq}, we 
have introduced $\rho-$form factor $(f_{\rho}(Q^2))$ at $\pi \pi NN$ vertex and taken it to be of monopole form:
\begin{equation}
 f_{\rho}(Q^2) = \frac{1}{1+Q^2/m_{\rho}^2}; \qquad \qquad {\rm with } \; m_\rho = 0.776~\text{ GeV}.
\end{equation}
$f_{\rho}(Q^2)$ also  has been used with axial part of the CT diagram in order to be consistent with the assumption of PCAC. 

\begin{table*}
  \begin{center}
    \renewcommand{\arraystretch}{1.2}
    \begin{tabular*}{170mm}{@{\extracolsep{\fill}}|c|ccc|ccc|cccc|}
      \noalign{\vspace{-1pt}}
     \hline
      Constant term $\longrightarrow$         & \multicolumn{3}{c|}{$\mathcal{A}$(CC $\nu$)}&\multicolumn{3}{c|}
      {$\mathcal{A}$(CC $\bar \nu$)}
      &\multicolumn{4}{c|}{$\mathcal{A}$(NC $\nu(\bar\nu)$)}     \\ \hline
      Final states  $\longrightarrow$  &   $p\pi^{+}$ & $n\pi^{+}$ & $p\pi^{0}$&
      $n\pi^{-}$ & $n\pi^{0}$ & $p\pi^{-}$&
      $n\pi^{+}$ & $p\pi^{0}$ & $p\pi^{-}$&$n\pi^{0}$ \\ \hline
      NP  &   0   & $\frac{-ig_A}{\sqrt{2}f_\pi}$&{$\frac{-ig_A}{2 f_\pi}$ }&
              0   &{ $\frac{ig_A}{2f_\pi}$}&{ $\frac{-ig_A}
              {\sqrt{2} f_\pi}$} &
              $\frac{-ig_A}{\sqrt{2}f_\pi}$& { $\frac{-ig_A}{2 f_\pi}$} &{ $\frac{ig_A}{\sqrt{2}f_\pi}$}&{ $\frac{-ig_A}{2 
              f_\pi}$}\\
      CP &  {$\frac{-ig_A}{\sqrt{2}f_\pi}$}  &0 &{$\frac{ig_A}
              {2f_\pi}$}  &
             {$\frac{-ig_A}{\sqrt{2}f_\pi}$} & {$\frac{-ig_A}{2
              f_\pi}$} &  0 &
            { $\frac{ig_A}{\sqrt{2}f_\pi}$}& { $\frac{-ig_A}{2 
              f_\pi}$} & $\frac{-ig_A}{\sqrt{2}f_\pi}$& { $\frac{-ig_A}{2f_\pi}$} \\
      CT  &   $\frac{-i}{\sqrt{2}f_\pi}$ &$\frac{i}{\sqrt{2}f_\pi}$ & {$\frac{i}{ f_\pi}$} &
              $\frac{-i}{\sqrt{2}f_\pi}$ & {$\frac{-i}{f_\pi}$} & $\frac{i}{\sqrt{2}f_\pi}$&
           $\frac{\sqrt{2}i}{f_{\pi}}$& 0 & {$\frac{-\sqrt{2}
              i}{f_{\pi}}$} & 0\\
      PP  &   $\frac{i}{\sqrt{2}f_\pi}$ & $\frac{-i}{\sqrt{2}f_\pi}$& {$\frac{-i}{ f_\pi}$} &
              $\frac{i}{\sqrt{2}f_\pi}$&$\frac{i}{f_\pi}$& $\frac{-i}{\sqrt{2}f_\pi}$  &
             {$\frac{\sqrt{2}i}{f_{\pi}}$} & 0 &{$\frac{-\sqrt{2}
              i}{f_{\pi}}$} & 0\\
      PF  &  { $\frac{-i g_A}{\sqrt{2}f_\pi}$} &{ $\frac{i g_A}
              {\sqrt{2}f_\pi}$} & {$\frac{i g_A}{ f_\pi}$}   &
           {$\frac{-i g_A}{\sqrt{2}f_\pi}$} & { $\frac{-i g_A}
              {f_\pi}$} &{$\frac{i g_A}{\sqrt{2}f_\pi}$} &
             {$\frac{\sqrt{2}i g_A}{f_{\pi}}$} & 0 & { $\frac{-
              \sqrt{2}i g_A}{f_{\pi}}$}& 0\\
      \hline
    \end{tabular*}
  \end{center}
\caption{The values of constant term($\mathcal{A}^i$) appearing in Eq.~(\ref{eq:eqscc}), where $i$ corresponds to the 
nucleon pole~(NP), cross nucleon pole~(CP), contact term~(CT), pion pole~(PP) and pion in flight~(PF) terms. $f_\pi$ is 
pion weak decay constant and $g_A$ is axial nucleon coupling.}\label{tab:born_para_1}
\end{table*}
\begin{table*}
  \begin{center}
    \renewcommand{\arraystretch}{1.2}
    {\small
        \begin{tabular*}{150mm}{@{\extracolsep{\fill}}|c|ccc|ccc|}
      \noalign{\vspace{-8pt}}
      \hline
      I~(J) & \multicolumn{3}{c|}{$\mathcal{C^R}$(CC $\nu$)}&\multicolumn{3}{c|}{$\mathcal{C^R}$(CC $\bar \nu$)} 
           \\ \hline
      &    $p \longrightarrow p\pi^{+}$ & $n \longrightarrow n\pi^{+}$ & $n \longrightarrow p\pi^{0}$&
      $n \longrightarrow n\pi^{-}$ & $p \longrightarrow n\pi^{0}$ & $p \longrightarrow p \pi^{-}$ 
      \\ \hline
     $\frac{3}{2}$  ($\frac{3}{2}$) &   $\frac{\sqrt{3}f^\star}{m_\pi}$   
      & $\sqrt{\frac{1}{3}}\frac{f^\star}{m_\pi}$& $-\sqrt{\frac{2}{3}}\frac{f^\star}{m_\pi}$ &
              $\frac{\sqrt{3}f^\star}{m_\pi}$  & $\sqrt{\frac{2}{3}}\frac{f^\star}{m_\pi}$& $\sqrt{\frac{1}{3}}
              \frac{f^\star}{m_\pi}$   \\ 
              
          $\frac{3}{2}$  ($\frac{1}{2}$) &   $\frac{\sqrt{3}f^\star}{m_\pi}$   
      & $\sqrt{\frac{1}{3}}\frac{f^\star}{m_\pi}$& $-\sqrt{\frac{2}{3}}\frac{f^\star}{m_\pi}$ &
              $\frac{\sqrt{3}f^\star}{m_\pi}$  & $\sqrt{\frac{2}{3}}\frac{f^\star}{m_\pi}$& $\sqrt{\frac{1}{3}}
              \frac{f^\star}{m_\pi}$  \\      
              
$\frac{1}{2}$ ($\frac{3}{2}$)&   0  & {  $\sqrt{2}\frac{f^\star}{m_\pi}$} & 
       {  $\frac{f^\star}{m_\pi}$}  &
              0 &  {  $-\frac{f^\star}{m_\pi}$} &  {  $\sqrt{2}
              \frac{f^\star}{m_\pi}$}\\
              
              $\frac{1}{2}$  ($\frac{1}{2}$)&   0  & {  $\sqrt{2}\frac{f^\star}{m_\pi}$} & 
       {  $\frac{f^\star}{m_\pi}$}  &
              0 &  {  $-\frac{f^\star}{m_\pi}$} &  {  $\sqrt{2}
              \frac{f^\star}{m_\pi}$}  \\ \hline
    \end{tabular*}
    }
    \caption{ 
  Coupling constant($\mathcal{C^R}$) for spin and isospin $\frac12$ and spin $\frac32$ resonances for the charge current 
  (anti)neutrino induced pion production.
Here $f^\star$ stands for ${\cal R} \to N \pi$ coupling which for $\Delta(1232)$ resonance is $g_{\Delta N \pi}$
and $g_{R\frac12 N\pi}(g_{R\frac32 N\pi})$ for spin $\frac12(\frac32)$ resonances. }
  \label{tab:Coupling constant_s}
  \end{center}
\end{table*}

We have already discussed in Section~\ref{CC}, the excitation and decay of spin $\frac{1}{2}$ and $\frac{3}{2}$ resonances 
into a meson and a baryon in the final state. In the case of single pion production, we have taken the contribution from 
spin $\frac{1}{2}$ resonances like $P_{11} (1440)$, $S_{11} (1535)$, $S_{31} (1620)$, $S_{11} (1650)$, and spin $\frac{3}{2}$ 
resonances like $P_{33} (1232)$, $D_{13} (1520)$, $D_{33} (1700)$, $P_{13} (1720)$. It should 
be noted that in the vector sector, the helicity amplitudes for all these resonance excitations are given by the MAID 
parameterization~\cite{Tiator:2011pw}. In the case of spin $\frac{1}{2}$ resonances, the s-channel and u-channel hadronic 
currents for the positive and negative parity resonances are given in 
Eqs.~(\ref{eq:res1/2_had_current_pos}) and (\ref{eq:res1/2_had_current_pos_u}), with the explicit form of the vector and 
axial-vector form factors given in Eqs.~(\ref{eq:f12vec_res_12}) and (\ref{eq:f12vec_res_12_neg}) for the isospin 
$\frac{1}{2}$ resonances and in Eqs.~(\ref{eq:g1_pos}) and (\ref{eq:g1_neg}) for the isospin $\frac{3}{2}$ resonances. The 
coefficient ${\cal C}$ for CC and NC induced processes is given in Tables~\ref{tab:Coupling constant_s} 
and \ref{tab:Coupling constant_u}, respectively. Similarly in the case of positive and negative parity spin $\frac{3}{2}$ 
resonances, the general expression of the hadronic current for the s- and u-channels are given in 
Eqs.~(\ref{eq:res_had_current_pos}) and (\ref{eq:res_had_current_pos_u}). The vector and axial-vector form factors used in the 
case of isospin $\frac{1}{2}$ resonances are given in Eqs.~(\ref{eq:civ_NC}) and (\ref{c5a-r}), respectively while for the 
isospin $\frac{3}{2}$ resonances, these form factors are given in Eqs.~(\ref{eq:civ_NC_32}) and (\ref{c5a-r}). 

The axial-vector form factors as discussed in Eqs.~(\ref{c6-c5}) and (\ref{c5aq}) along with the vector form factors given 
in Eq.~(\ref{vecff}), have been used to analyze the present experimental cross sections for the weak pion 
production. Most of the recent theoretical calculations~\cite{Lalakulich:2006sw, Hernandez:2007qq, Leitner:2006ww, 
Lalakulich:2010ss} use a simpler modification to the dipole form viz.
\begin{equation}\label{delta}
C_5^A(Q^2) = \frac{C_5^A(0)}{ \left( 1 + Q^2 /M_{A\Delta}^2 \right)^2 } \; \frac{1}{ 1 + Q^2/(3 M_{A\Delta}^2)}.
\end{equation}
With the nonvanishing axial-vector form factors determined in terms of $C_5^A(Q^2)$ and the vector form factors determined 
from the electron scattering experiments, the weak pion production cross section is described in terms of $C_5^A(Q^2)$ with 
the parameters $C_5^A(0)$ and axial mass $M_{A\Delta}$. Keeping $M_{A\Delta}=1.026 {\rm GeV}$ and then varying $C_{5}^{A}(0)$, 
we obtain the best possible value to be $C_{5}^{A}(0)=1$ to obtain a good description of reanalyzed 
data~\cite{Wilkinson:2014yfa} of ANL and BNL experiments for $\nu_\mu p \longrightarrow \mu^- p \pi^+$ 
reaction~\cite{RafiAlam:2015fcw}. We find that while fitting the reanalyzed data for the reaction $\nu_\mu  p \longrightarrow 
\mu^- p \pi^+$, the contributions to the cross section is predominantly obtained from $\Delta(1232)$ resonant terms and the 
background terms give small contribution. This has been further discussed in Section~\ref{results:pion}.

\subsubsection{Neutral current (anti)neutrino induced processes}\label{NC:pion}
In this section, we briefly discuss the single pion production induced by NC~(NC1$\pi$). The older data on 
NC1$\pi$ production are available from ANL~\cite{Derrick:1980nr} and Gargamelle~\cite{Pohl:1978iy} experiments. Recently, 
NC1$\pi$ production measurements have been performed by the MiniBooNE~\cite{Anderson:2009zzc}, K2K~\cite{K2K:2004qpv}, 
SciBooNE~\cite{SciBooNE:2009nlf}, MicroBooNE~\cite{MicroBooNE:2022lvx}, and other collaborations. The NC $\pi^0$ 
production in neutrino interactions plays an important role in the background studies of $\nu_\mu \leftrightarrow \nu_e$ or 
$\bar\nu_\mu \leftrightarrow \bar\nu_e$ oscillations in the appearance mode as well as in discriminating between $\nu_\mu 
\longrightarrow \nu_\tau$ and $\nu_\mu \longrightarrow \nu_s$ modes. Furthermore, in the reconstruction of neutrino energy 
using QE events like $\nu_e + n \longrightarrow e^- + p$ or $\bar\nu_e + p \longrightarrow e^+ + n$, a missing 
electron or positron produced by the photon from the $\pi^0$ decay may be mistaken as QE event. Moreover, NC induced pion 
production may also help to distinguish between the production of $\nu_\tau$ and $\bar \nu_\tau$ in some oscillation 
scenarios at neutrino energies much below the $\tau$ production threshold but above the pion threshold~\cite{Hernandez:2006yg}.

The NC induced (anti)neutrino scattering processes form the free nucleon target are given by:
 \begin{align}
 \nu_{_l} (\bar{\nu}_{l}) p \to \nu_{_l} (\bar{\nu}_{l})n \pi^+ ,\qquad \qquad \nu_{_l}(\bar{\nu}_{l}) p \to \nu_{_l}(\bar{\nu}_{l}) p \pi^0 , \qquad \qquad
  \nu_{_l}(\bar{\nu}_{l}) n \to \nu_{_l}(\bar{\nu}_{l}) n \pi^0 ,  \qquad \qquad \nu_{_l}  (\bar{\nu}_{l})n \to \nu_{_l}  (\bar{\nu}_{l})p \pi^-.
  \label{eq:NC_all}
\end{align}

\begin{table*}
  \begin{center}
    \renewcommand{\arraystretch}{1.2}
    {\small
        \begin{tabular*}{120mm}{@{\extracolsep{\fill}}|c|cccc|}
      \noalign{\vspace{-8pt}}
      \hline
     I(J)         &  
     \multicolumn{4}{c|}{$\mathcal{C^R}$(NC $\bar \nu (\bar{\nu})$)}
           \\ \hline
      &     
       $p \longrightarrow n\pi^+$ &
        $p\longrightarrow p \pi^0$ & $n \longrightarrow p \pi^-$ & $n  \longrightarrow n \pi^0$
      \\ \hline
     $\frac{3}{2}$  ($\frac{3}{2}$) &  $\frac{1}{\sqrt{3}} \frac{f^\star}
     {m_\pi}$ 
     & $\sqrt{\frac{2}{3}}\frac{f^\star}{m_\pi}$ & $-\frac{1}{\sqrt{3}}\frac{f^\star}{m_\pi}$ & 
     $\sqrt{\frac{2}{3}}\frac{f^\star}{m_\pi}$  \\ 
              
     $\frac{3}{2}$  ($\frac{1}{2}$)  & 
     $\frac{1}{\sqrt{3}}\frac{f^\star}
     {m_\pi}$ 
     & $\sqrt{\frac{2}{3}}\frac{f^\star}{m_\pi}$& $-\frac{1}{\sqrt{3}}\frac{f^\star}{m_\pi}$ & 
     $\sqrt{\frac{2}{3}}\frac{f^\star}{m_\pi}$  \\          
              
     $\frac{1}{2}$ ($\frac{3}{2}$) & $-\frac{1}{\sqrt{2}}\frac{f^\star}{m_\pi}$ & 
     $\frac{1}{2}\frac{f^\star}{m_\pi}$ & $ \frac{1}{\sqrt{2}}\frac{f^\star}{m_\pi}$ & $\frac{1}{2}\frac{f^\star}{m_\pi}$ \\
              
     $\frac{1}{2}$  ($\frac{1}{2}$)  & $-\frac{1}{\sqrt{2}}\frac{f^\star}{m_\pi}$ & 
     $\frac{1}{2}\frac{f^\star}{m_\pi}$ & $ \frac{1}{\sqrt{2}}\frac{f^\star}{m_\pi}$ & $\frac{1}{2}\frac{f^\star}{m_\pi}$ \\ 
     \hline
    \end{tabular*}
    }
    \caption{ 
  Coupling constant($\mathcal{C^R}$) for spin and isospin $\frac12$ and spin $\frac32$ resonances for the charge current 
  (anti)neutrino induced pion production. Here $f^\star$ stands for ${\cal R} \to N \pi$ coupling which for $\Delta(1232)$ 
  resonance is $g_{\Delta N \pi}$ and $g_{R\frac12 N\pi}(g_{R\frac32 N\pi})$ for spin $\frac12(\frac32)$ resonances. }
  \label{tab:Coupling constant_u}
  \end{center}
\end{table*}

In the case of NC induced processes, the expressions for the different terms contributing to the NRB are given in 
Eqs.~(\ref{eq:background})--(\ref{eq:eqscc}), with $a=1$ and the values of ${\cal C}$ for the 
different pion production channels given in Table~\ref{tab:born_para_1}. The neural current vector form factors are 
expressed as:
\begin{align}
f_{1,2}(Q^2) &  \mathrel{\mathop{\longrightarrow}^{\mathrm{for \; p}}}  \tilde f_{1,2}^{p}(Q^2)=\left(\frac12 - 2 
\sin^2 \theta_W\right)F_{1,2}^{p}(Q^2)- \frac12 F_{1,2}^{n}(Q^2),\\
 f_{1,2}(Q^2) &\mathrel{\mathop{\longrightarrow}^{\mathrm{for \; n}}} \tilde f_{1,2}^{n}(Q^2)=\left(\frac12 - 2 \sin^2 
 \theta_W\right)F_{1,2}^{n}(Q^2)- \frac12 F_{1,2}^{p}(Q^2),
\end{align}
where $\theta_W$ is the weak mixing angle, and the axial-vector form factor is given by 
\begin{equation}\label{fa_nc}
\tilde g^{p,n}_{1}(Q^2)= \pm \frac12 g_{1}(Q^2),
\end{equation}
where the plus~(minus) sign stands for proton~(neutron) target. 
 The contribution of the pseudoscalar form factor being proportional to the lepton mass vanishes for NC 
 induced processes.
 \begin{figure}[h]
\includegraphics[width=18cm,height=14cm]{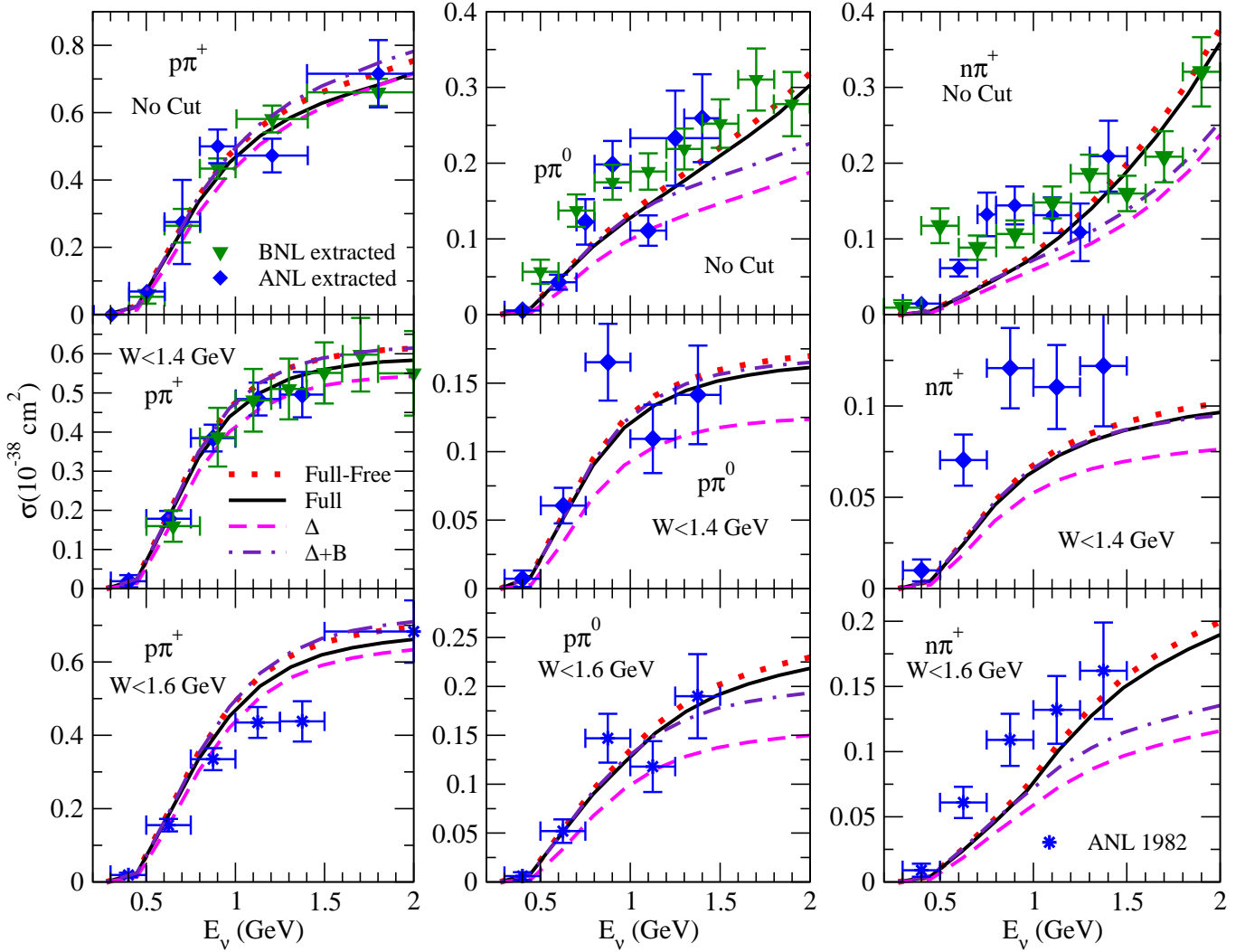}
\caption{Total scattering cross section for CC neutrino induced pion production processes: $\nu_{\mu}   p 
\longrightarrow \mu^{-}   p   \pi^{+}$~(left panel), $\nu_{\mu}   n \longrightarrow \mu^{-}   p  \pi^{0}$~(central panel), 
$\nu_{\mu}   n \longrightarrow \mu^{-}   n   \pi^{+}$~(right panel). The dashed line is the result calculated in the 
$\Delta(1232)$ dominance model, dashed-dotted line is the result obtained when we include NRB terms in 
our calculations. The solid line is the result of our full calculation when other resonances like $P_{11}(1440)$, $D_{13}
(1520)$, $S_{11}(1535)$, $S_{31}(1620)$, $S_{11}(1650)$, $D_{33}(1700)$ and $P_{13}(1720)$ are also included. All the above 
three cases are with deuteron effect. The dotted line is the result of the full calculation without deuteron effect. The 
results in the top panels are obtained when we have not included any cut on the invariant mass. The middle panel shows the 
results with a cut of 1.4~GeV on the $W$, while in the bottom panel a cut of $W<1.6$~GeV is 
introduced while calculating total scattering cross section. Data points quoted as ANL extracted and BNL extracted are the 
reanalyzed data by Wilkinson et al.~\cite{Wilkinson:2014yfa} and Rodrigues et al.~\cite{Rodrigues:2016xjj}. Other data 
points in figures are the results from ANL~\cite{Radecky:1981fn} experiment.}\label{fig:sigma_Nu_CC1pion}
\end{figure} 
The structure of the hadronic current for NC induced s-channel and u-channel spin as well as isospin $\frac{1}{2}$ 
and $\frac{3}{2}$ resonances is similar to that of CC induced reactions. In the case of NC 
induced single pion production, only the couplings ${\cal C^R}$ and the form factors corresponding to $N\longrightarrow {\cal 
R}$ transitions are different as compared to CC induced processes. In Section~\ref{nc1pi}, we have already discussed 
in detail NC vector and axial-vector form factors for spin $\frac{1}{2}$ and $\frac{3}{2}$ resonances and 
 NC couplings ${\cal C^R}$ are given in Table~\ref{tab:Coupling constant_u}. For the numerical calculations, 
we have used NC vector and axial-vector form factors described in Section~\ref{nc1pi} with the hadronic 
current given in Eqs.~(\ref{eq:res1/2_had_current_pos}) and (\ref{eq:res1/2_had_current_pos_u}) for spin $\frac{1}{2}$ 
resonances and Eqs.~(\ref{eq:res_had_current_pos}) and (\ref{eq:res_had_current_pos_u}) for spin $\frac{3}{2}$ resonances.

\subsubsection{Results and discussion}\label{results:pion}
In this section, we present the results of the numerical calculations and discuss the findings. Due to the limitations on the 
validity of the nonlinear sigma model at higher energies~\cite{Hernandez:2007qq}, we have put a constraint on $W$ as $W_{min} 
= M + m_{\pi}$ and $W_{max} = 1.2 ~GeV$ while evaluating the NRB terms. This 
constraint on $W$~(i.e. $M + m_{\pi} \le W \le 1.2GeV$) has been put in all numerical evaluations while considering 
NRB contribution. 

Since earlier experiments to measure CC neutrino induced single pion production were mainly performed using 
hydrogen/deuteron target like the experiments at  ANL~\cite{Radecky:1981fn} and BNL~\cite{Kitagaki:1986ct}, therefore, 
deuteron correction factor must be taken into account. In recent analyses by Wilkinson et al.~\cite{Wilkinson:2014yfa} and 
Rodrigues et al.~\cite{Rodrigues:2016xjj}, experimental results of ANL~\cite{Radecky:1981fn} and BNL~\cite{Kitagaki:1986ct} 
have been normalized to the deuteron data. Therefore, we have taken deuteron effect by writing~\cite{RafiAlam:2015fcw}:
\begin{equation} \label{de}
  \left(\frac{d\sigma}{dQ^2dW}\right)_{\nu d}=\int d{\vec{p}}_p^d |\Psi_d({\vec{p}}_p^d)|^2 \frac{M}{E_p^d} \left(\frac{d\sigma}
  {dQ^2dW}
  \right)_{\rm{ off~ shell}},
\end{equation}
where the four momentum of the proton inside the deuteron is described by $p^\mu=(E_p^d, \vec{p}_p^d)$ with $E_p^d 
(=M_{\text Deuteron}-\sqrt{M^2+|{\vec{p}}_p^d|^2})$ as the energy of the off shell proton inside the deuteron and $M_{\text{ 
Deuteron}}$ is the deuteron mass. $\left(\frac{d\sigma}{dQ^2dW}\right)_{\text{off~shell}}$ is obtained by using 
Eq.~(\ref{eq:sigma_inelas}). In the above expression $\Psi_d(\vec{p}_{d})$ is the deuteron wave function taken from the works of Lacombe et 
al.~\cite{Lacombe:1981eg}.
\begin{figure}  
\centering
\includegraphics[width=18cm,height=7cm]{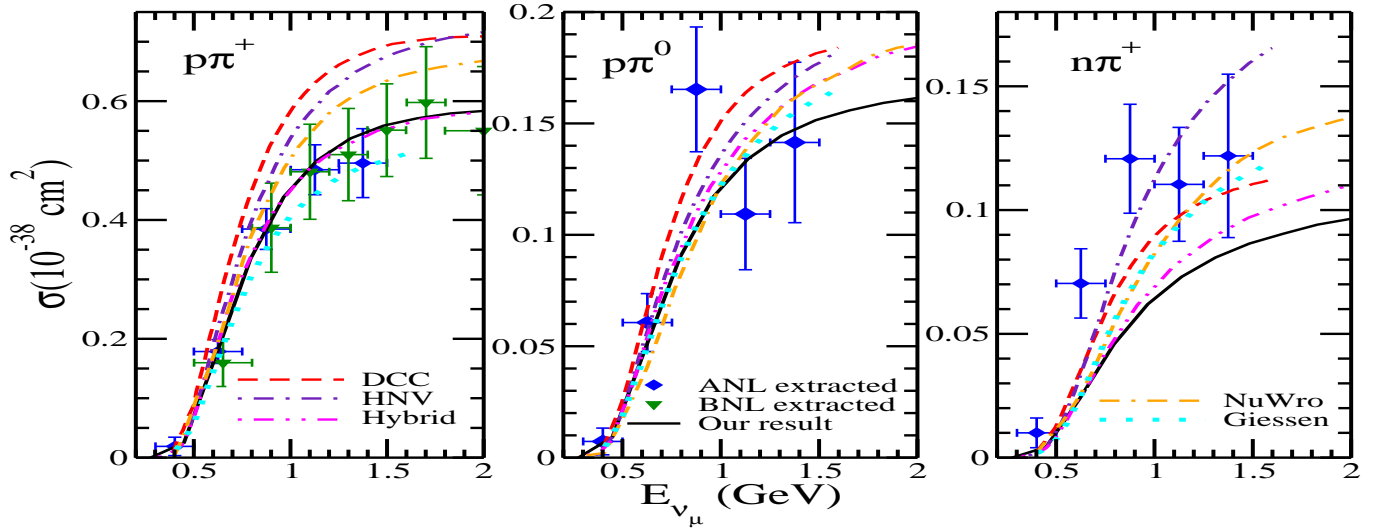}
\caption{$\sigma$ vs. $E_{\nu_{\mu}}$ for the processes $\nu_{\mu} p \longrightarrow \mu^{-}   p   \pi^{+}$~(left panel), 
$\nu_{\mu}   n \longrightarrow \mu^{-}   p   \pi^{0}$~(central panel), $\nu_{\mu}   n \longrightarrow \mu^{-}   n  
\pi^{+}$~(right panel) with $W<1.4$~GeV. Solid line is the result of the present model with deuteron effect; compared with 
other theoretical models like DCC~\cite{Nakamura:2015rta}~(dashed line), HNV~\cite{Hernandez:2007qq}~(dashed-dotted line), 
Hybrid~\cite{Gonzalez-Jimenez:2016qqq}~(double-dotted-dashed line), 
NuWro~\cite{Gonzalez-Jimenez:2016qqq}~(double-dashed-dotted line) and Giessen~\cite{Lalakulich:2010ss}~(dotted line). Data 
points have the same meaning as in Fig.~\ref{fig:sigma_Nu_CC1pion}.}\label{fig:cc_compare}
\end{figure}

We have calculated the total scattering cross section for CC neutrino induced pion production processes 
and the results are presented in Fig.~\ref{fig:sigma_Nu_CC1pion}. The experimental data for $\pi^+ p$ channel where no cut 
on $W$ is applied is the reanalyzed data by Wilkinson et al.~\cite{Wilkinson:2014yfa} of the ANL~\cite{Radecky:1981fn} and 
BNL~\cite{Kitagaki:1986ct} experiments. The experimental data for $\pi^0 p$ and $\pi^+ n$ channels for no cut on $W$ are the 
reanalyzed data by Rodrigues et al.~\cite{Rodrigues:2016xjj} of the ANL~\cite{Radecky:1981fn}, BNL~\cite{Kitagaki:1986ct}, 
and other experiments. While for all the three channels with a cut of 1.4~GeV on $W$, we have used the reanalyzed data by 
Rodrigues et al.~\cite{Rodrigues:2016xjj} and for all the pion production channels with $W<1.6$~GeV, the experimental data are 
of ANL~\cite{Radecky:1981fn} experiment. 
\begin{itemize}
 \item [(i)] In the case of $\nu_\mu  p \longrightarrow \mu^-  p  
\pi^+$ induced reaction, the main contribution to the total scattering cross section comes from the $\Delta(1232)$ resonance 
when no cut on $W$ is applied while when this cut is applied there is some contribution from the higher resonances and the 
background terms which are considered in this work. It should be noticed that in the case when no cut on $W$ is applied or 
when $W<1.4$~GeV is considered, our theoretical results are in very good agreement with the reanalyzed experimental data 
by Wilkinson et al.~\cite{Wilkinson:2014yfa} and Rodrigues et al.~\cite{Rodrigues:2016xjj}. While in the case when $W<1.6$~GeV 
cut is applied, we are consistent with the experimental data obtained by the ANL~\cite{Radecky:1981fn}. Quantitatively, we 
find that due to the presence of the NRB terms there is an increase in the cross section which is about 
$14\%$ at $E_{\nu_\mu}=1$~GeV and becomes $\sim 9\%$ at $E_{\nu_\mu}$=2GeV, when no cut on $W$ is applied. However, when 
the cuts on $W$ are applied, then due to the presence of background contributions, this increase in the cross section further 
increases and become $\sim 13\%$ at 2~GeV for $W<1.4$~GeV and $12\%$ for $W<1.6$~GeV.

\item [(ii)] For $\nu_\mu  n \longrightarrow \mu^-  p  \pi^0$ as well as $\nu_\mu  n \longrightarrow \mu^-  n  \pi^+$ processes, there 
are significant contributions from the NRB terms as well as other higher resonant terms besides the 
$\Delta(1232)$ dominance. In the case of $\nu_\mu  n \longrightarrow \mu^- p  \pi^0$ our results with deuteron effects are in 
a good agreement with the reanalyzed data as well as with the original data from ANL experiment at $W<1.6$~GeV. Due to the 
presence of the background terms, without any constraint on $W$, the total increase is about $32\%$ at $E_{\nu_\mu}=1$~GeV which becomes $20\%$ at 
$E_{\nu_\mu}=2$~GeV. With the inclusion of higher resonances, there is a further increase of about $3\%$ at $E_{\nu_\mu} = 
1$~GeV and $40\%$ at $E_{\nu_\mu}=2$~GeV. It may be observed from the figure the our results for without cut on $W$ and with 
a cut of $W<1.6$~GeV are quite consistent with the experimental data for the $n\pi^+$ channel. The net contribution to the 
total pion production due to 
the presence of the NRB terms in $\nu_\mu  n \longrightarrow \mu^- n  \pi^+$ reaction results in an 
increase in the cross section of about $22\%$ at $E_{\nu_\mu}=1$~GeV which becomes $8\%$ at $E_{\nu_\mu}=2$~GeV with no cut 
on $W$. When other higher resonances are also taken into account there is further increase in the cross section by about $3\%$ 
at $E_{\nu_\mu}=1$~GeV which becomes $40\%$ at $E_{\nu_\mu}=2$~GeV. Thus, we find that the inclusion of higher resonant terms 
lead to a significant increase in the cross section for $\nu_\mu  n \longrightarrow \mu^- n  \pi^+$ and $\nu_\mu  n 
\longrightarrow \mu^-  p  \pi^0$ processes. Furthermore, it may also be concluded from the above observations that 
contribution from NRB terms decreases with the increase in neutrino energy, while the total scattering 
cross section increases when we include other higher resonances in our calculations. When a cut of 1.4~GeV or 
1.6~GeV on the CM energy is applied, then due to the presence of background terms, the increase in the cross 
section is about $24\%$ at $E_{\nu_\mu}$=1~GeV which becomes about $20\%$ at $E_{\nu_\mu}=2$~GeV for $\nu_\mu  n 
\longrightarrow \mu^-  n  \pi^+$ reaction. When higher resonances are also taken into account there is a further increase in 
the cross section which is about $8\%$ at $E_{\nu_\mu}=1$~GeV and $2\%$ at $E_{\nu_\mu}=2$~GeV for $W < 1.6$~GeV. Similarly, in the case 
of $\nu_\mu  n \longrightarrow \mu^-  p  \pi^0$ there is a significant increase in the total cross section due to the presence 
of the NRB terms and higher resonances.
\end{itemize}

\begin{figure}  
\centering
\includegraphics[width=18cm,height=10cm]{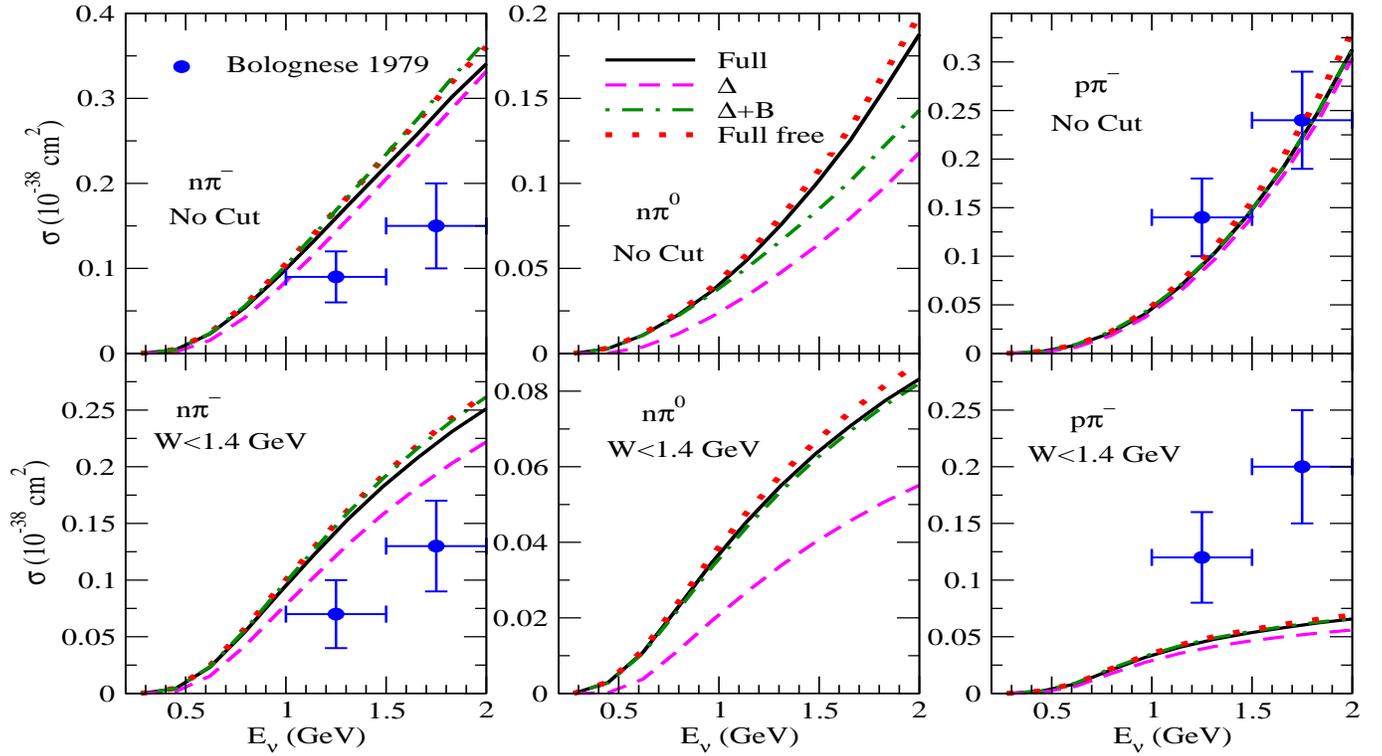}
\caption{$\sigma$ vs. $E_{\bar{\nu}_{\mu}}$ for CC induced $\bar\nu_{\mu}n \longrightarrow \mu^{+}n\pi^{-}$~(left panel), 
$\bar\nu_{\mu}  p \longrightarrow \mu^{+} n  \pi^{0}$~(central panel), and $\bar\nu_{\mu}  p \longrightarrow \mu^{+}  p  
\pi^{-}$~(right panel) processes, with deuteron effect. Data points are the experimental results from 
Ref.~\cite{Bolognese:1979gf}. Lines have the same meaning as in Fig.~\ref{fig:sigma_Nu_CC1pion}.}
\label{fig:sigma_Nubar_CC1pion}
\end{figure}
\begin{figure}
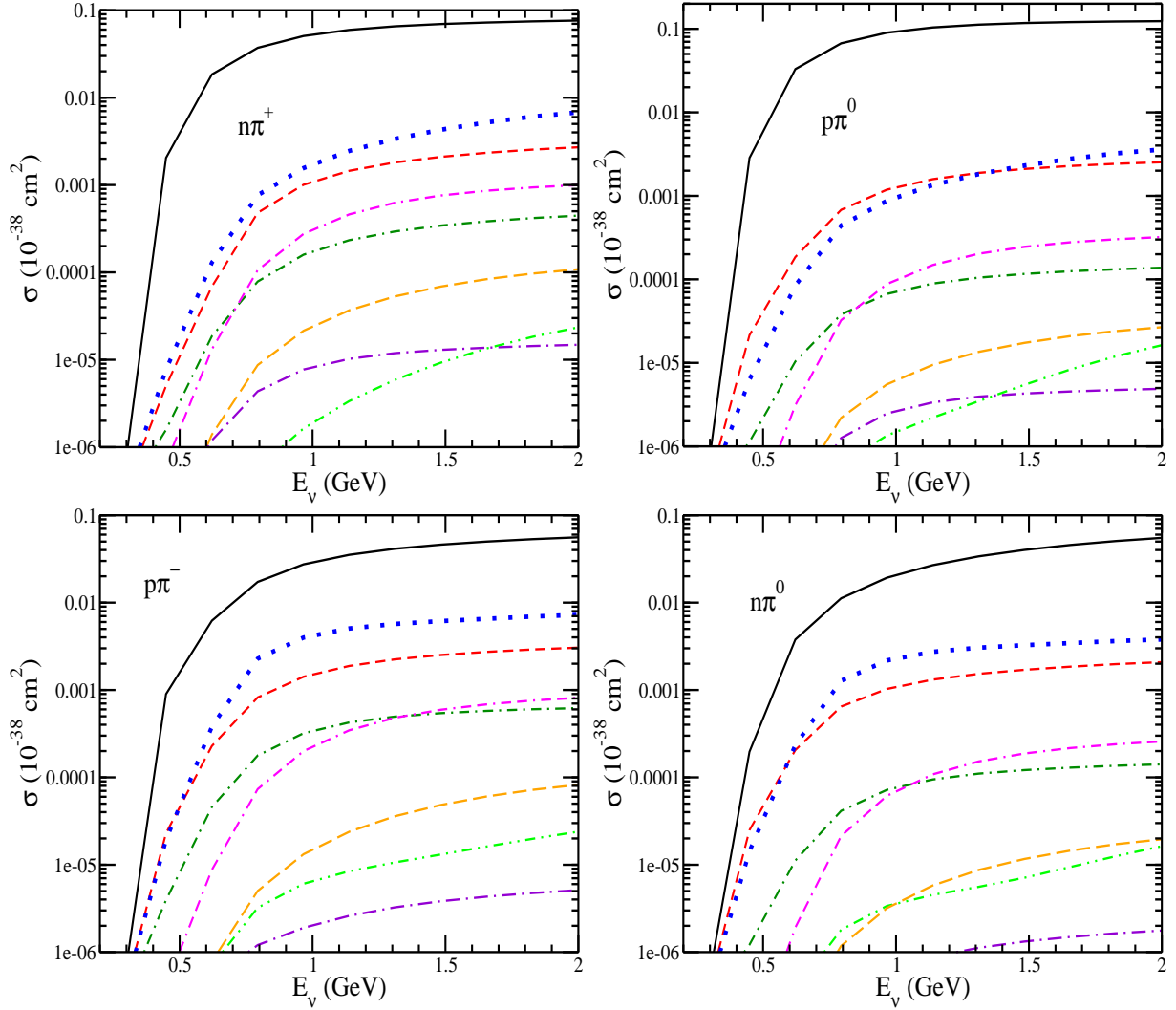
 
 \centering
\includegraphics[width=8cm,height=7cm]{nu_ch2_resonance.eps}
\includegraphics[width=8cm,height=7cm]{nu_ch3_resonance.eps}
\includegraphics[width=8cm,height=7cm]{anu_ch2_resonance.eps}
\includegraphics[width=8cm,height=7cm]{anu_ch3_resonance.eps}
\caption{The results are presented for the total scattering cross section for $\nu_{\mu}  n \longrightarrow \mu^{-}  n  
\pi^{+}$~(upper left panel), $\nu_{\mu}  n \longrightarrow \mu^{-}  p  \pi^{0}$~(upper right panel), $\bar{\nu}_{\mu} p 
\longrightarrow \mu^{+} p \pi^{-}$~(lower left panel), and $\bar{\nu}_{\mu}  p \longrightarrow \mu^{+}  n  \pi^{0}$~(lower 
right panel) processes where the individual contribution of various resonances like $P_{33}(1232)$~(solid line), $P_{11} 
(1440)$~(dotted line), $D_{13}(1520)$~(short dashed line), $S_{11} (1535)$~(double-dashed-dotted line), $S_{31} 
(1620)$~(long dashed-dotted line), $S_{11}(1650)$~(long dashed line), $D_{33} (1700)$~(short dashed-dotted line) and $
P_{13} (1720)$~(double-dotted-dashed line) have been shown.}\label{fig:sigma_CC_res}
\end{figure} 
 \begin{figure}  
\centering
\includegraphics[width=18cm,height=9cm]{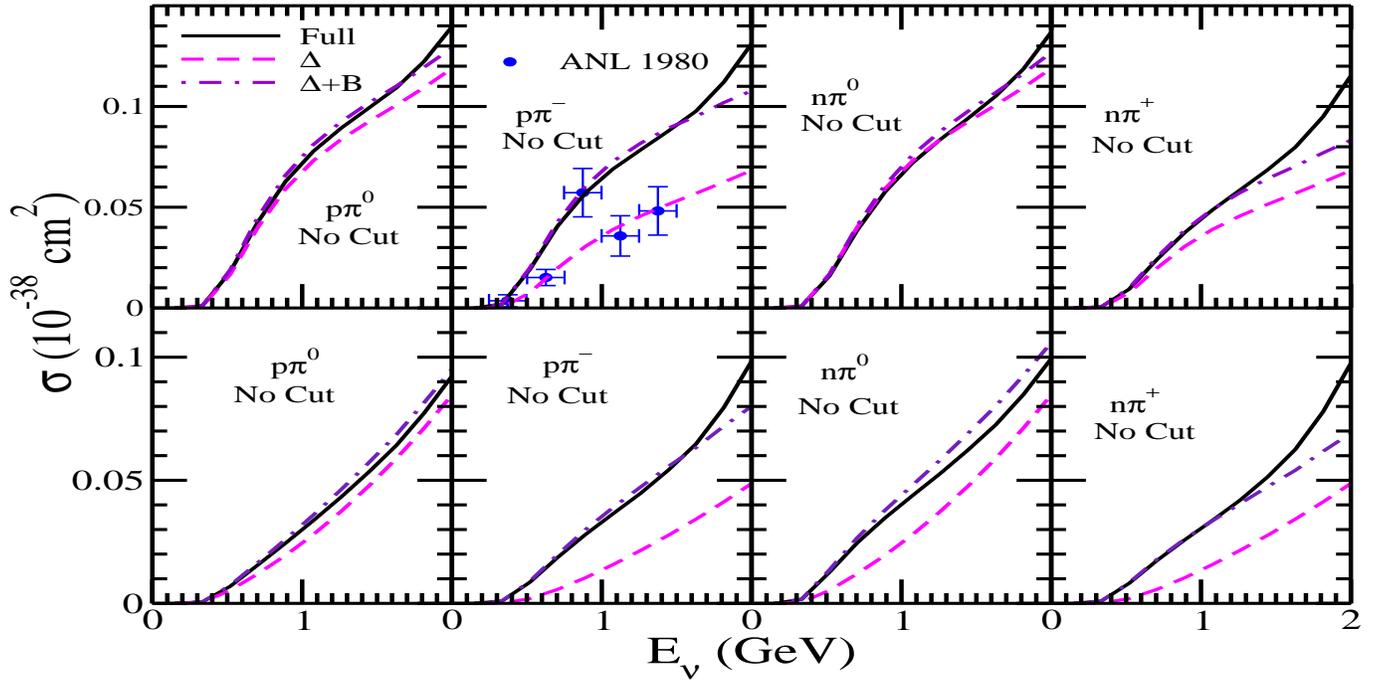}
\caption{Total scattering cross section for NC neutrino~(upper panel) and antineutrino~(lower panel) 
induced pion production processes with deuteron effect. The results are presented from the left for $\nu p \longrightarrow  
\nu p \pi^{0}$, $\nu  n \longrightarrow \nu p  \pi^{-}$, $\nu n \longrightarrow  \nu n \pi^{0}$ and $\nu p \longrightarrow 
\nu n \pi^{+}$ processes. Data points are the experimental results from Ref.~\cite{Derrick:1980nr}. The theoretical 
results presented here should be corrected for NME before making any comparison with the experimental 
data. Lines have the same meaning as in Fig.~\ref{fig:sigma_Nu_CC1pion}.}\label{fig:sigma_Nu_Nubar_NC1pion}
\end{figure}
In Fig.~\ref{fig:cc_compare}, we have compared the theoretical results for the single pion production induced by neutrinos 
when a cut of 1.4~GeV is applied on $W$, obtained in the different models like the present 
model~(Fig~\ref{fig:sigma_Nu_CC1pion}), the dynamical coupled channel~(DCC) model by Nakamura et al.~\cite{Nakamura:2015rta}, 
the HNV model by Hernandez et al.~\cite{Hernandez:2007qq}, the extension of HNV model by incorporating Regge model at high 
energies~(Hybrid) by Gonzalez-Jimenez et al.~\cite{Gonzalez-Jimenez:2016qqq}, the results from 
NuWro~\cite{Gonzalez-Jimenez:2016qqq} Monte Carlo generator, and the Giessen model by Lalakulich et 
al.~\cite{Lalakulich:2010ss}. It may be observed from the figure that in the case of $p\pi^+$ channel, the results obtained in 
our model are quite consistent with the results obtained by the hybrid model and are in a very good agreement with the 
reanalyzed data of ANL~\cite{Radecky:1981fn} and BNL~\cite{Kitagaki:1986ct} by Wilkinson et al.~\cite{Wilkinson:2014yfa} and 
Rodrigues et al.~\cite{Rodrigues:2016xjj}. However, the results obtained in the other models like DCC, HNV, etc., are higher 
than the results obtained by us as well as the experimental data, but are consistent with one another. Moreover, the 
results obtained by the Giessen group~\cite{Lalakulich:2010ss} are lower than our results. In the case of $p\pi^0$ channel, 
our results are in a quite good agreement with the experimental data, while the results obtained in the other  theoretical 
models are higher than our results. At energies $E_{\nu_{\mu}}<0.8$~GeV, the results obtained in the various models are 
consistent with each other. Furthermore, in the case of $n\pi^+$ production, our results are smaller than the experimental data, while the 
other theoretical models give higher values of the cross section than obtained in the present model. In this case, the results 
obtained by HNV model as well as by the NuWro generator show a good agreement with the experimental data. It may be noticed 
from the figure that there is a large difference among the various theoretical models and Monte Carlo generators available in 
the literature. In order to understand the dynamics of the single pion production, which is the simplest IE process, 
further theoretical and experimental work is required.

In Fig.~\ref{fig:sigma_Nubar_CC1pion}, we have shown the results for CC antineutrino induced pion 
production processes. These results are presented in the $\Delta(1232)$ dominance model, including NRB 
terms as well as with our full model. 
\begin{itemize}
 \item [(i)] In the case of $\bar\nu_\mu  n \longrightarrow \mu^+  n \pi^-$ reaction, there is 
very small contribution from the higher resonances other than $\Delta(1232)$ resonance for both cases {  i.e.}, when the 
results are obtained with no cut on $W$ as well as when a cut of $W<1.4$~GeV is applied. The inclusion of NRB terms increases the cross section by around $24\%~(26\%)$ at $E_{\bar{\nu}_\mu}$=1~(2)~GeV when no cut on $W$ is 
applied, which becomes around $27\%~(17\%)$ at $E_{\bar{\nu}_\mu} = 1~(2)$~GeV when a cut of $W<1.4$~GeV is applied.

\item [(ii)]  In  
$\bar\nu_\mu  p \longrightarrow \mu^+  n  \pi^0$ reaction, inclusion of NRB terms increases the cross 
section by around $62\%~(21\%)$ at $E_{\bar{\nu}_\mu} = 1~(2)$~GeV when no cut on $W$ is applied, and becomes $76\%~(50\%)$ 
at $E_{\bar{\nu}_\mu}=1~(2)$~GeV when a cut of $W<1.4$~GeV is applied. When other higher resonances are included, the cross 
section further increases by $\sim 10\%~(40\%)$ at $E_{\bar{\nu}_\mu}=1~(2)$~GeV at no cut and becomes almost $10\%~(6\%)$ 
at $E_{\bar{\nu}_\mu}=1~(2)$~GeV.

\item [(iii)]  In the case of $\bar\nu_\mu  p \longrightarrow \mu^+  p  \pi^-$ reaction, when no cut on 
$W$ is applied the effect of NRB terms as well as contribution from higher resonances is very small even 
at $E_{\bar{\nu}_{\mu}} = 2$~GeV. We find the theoretical results to be consistent with the experimental data. However, 
when $W<1.4$~GeV cut is applied, due to the inclusion of NRB terms, the cross section increases by about 
$20\%$ at $E_{\bar{\nu}_{\mu}}=1$~GeV, which becomes $18\%$ at $E_{\bar{\nu}_{\mu}}=2$~GeV. While in the case of $p\pi^-$ 
production, there is almost negligible contribution from the higher resonances.
\end{itemize}
 We have compared the present results with 
the experimental data of Gargamelle experiment~\cite{Bolognese:1979gf} performed at CERN PS where propane was used as the nuclear target. Since 
propane is a composite target with more than one nuclear target, therefore, the cross sections would get modulated due to 
NME. Thus, the theoretical results presented in Fig.~\ref{fig:sigma_Nubar_CC1pion} should be corrected for 
NME before making any comparison with the experimental data. We would like to point out that, in our 
earlier works~\cite{Ahmad:2006cy, SajjadAthar:2009rc, SajjadAthar:2009rd} on CC and NC pion production 
in the $\Delta(1232)$ dominance model, we have observed that NME reduces the cross section significantly 
when the calculations are performed for nuclear targets, which will be discussed later in Section~\ref{SPP:nucleus}.
      
To explicitly show the contribution of individual resonances to the total scattering cross section, in 
Fig.~\ref{fig:sigma_CC_res}, we have presented the results for $\nu_{\mu}  n \longrightarrow \mu^{-}  n  \pi^{+}$, 
$\nu_{\mu}  n \longrightarrow \mu^{-}  p  \pi^{0}$, $\bar{\nu}_{\mu} p \longrightarrow \mu^{+} p \pi^{-}$, and 
$\bar{\nu}_{\mu}  p \longrightarrow \mu^{+}  n  \pi^{0}$ processes as a function of incoming (anti)neutrino energy. It may 
be observed that the dominant contribution comes from $\Delta(1232)$ resonance, followed by $P_{11}(1440)$ and 
$D_{13}(1520)$ resonances. However, the contribution for the neutrino and the antineutrino induced processes are not similar, for 
example, larger $\Delta(1232)$ dominance may be observed in the neutrino case than in the case of antineutrino induced 
processes. For the case of neutrino induced $n\pi^+$ process, at $E_{\nu} = 1 GeV$, the contribution to the total scattering 
cross section from $P_{11} (1440)$~($D_{13}(1520)$) resonances is around $3 \%$($2\%$) as that of the contribution from 
$\Delta(1232)$ resonance, which increases and becomes around $9 \%$($4 \%$) at $E_{\nu} = 2$~GeV. However, for the case of 
neutrino induced $p\pi^0$ production, the contribution from $P_{11} (1440)$ and $D_{13}(1520)$ resonances are almost similar 
and is around $1\%$ at $E_{\nu} = 1$~GeV, which becomes $3\%$ at $E_{\nu} = 2$~GeV. For antineutrino induced $p\pi^-$ 
production, at $E_{\nu} = 1$~GeV, the contribution to the total scattering cross section from $P_{11}(1440)$($D_{13} 
(1520)$) resonance is around $14 \%$($5 \%$) at $E_{\nu} = 1$ GeV which becomes around $13 \%$($5 \%$) at $E_{\nu} = 2$ GeV as that of the 
contribution from $\Delta(1232)$ resonance. Similar results are obtained in the case of $n\pi^0$ production induced by 
antineutrinos.
      
In Fig.~\ref{fig:sigma_Nu_Nubar_NC1pion}, we have plotted the total scattering cross section for NC 
(anti)neutrino induced pion production processes on proton and neutron targets. The experimental points are the data from 
ANL experiment~\cite{Derrick:1980nr}. It may be observed from the figure that in the case of $\nu(\bar{\nu}) n 
\longrightarrow \nu(\bar{\nu}) p \pi^-$ and $\nu(\bar{\nu}) p \longrightarrow \nu(\bar{\nu}) n \pi^+$ processes, besides $\Delta 
(1232)$ resonant term, there is significant contribution from the NRB terms which results in an increase in 
the total scattering cross section in these channels. However, in the case of $ p \pi^0$ and $n \pi^0$ production reactions, 
the effect of NRB is small as compared to the other processes. We also observe that when higher 
resonances are included, there is no appreciable change in the cross sections in the case of (anti)neutrino induced 
$p\pi^0$ and $n\pi^0$ production cross sections, while in the case of $p\pi^-$ and $n\pi^+$ productions, we observe a 
significant contribution from the higher resonances, especially at $E_{\nu}>1.5$~GeV in the case of $p\pi^-$ production and 
at $E_{\nu}>1.2$~GeV in the case of $n\pi^+$ production. 

\begin{figure}  
\centering
\includegraphics[width=18cm,height=7cm]{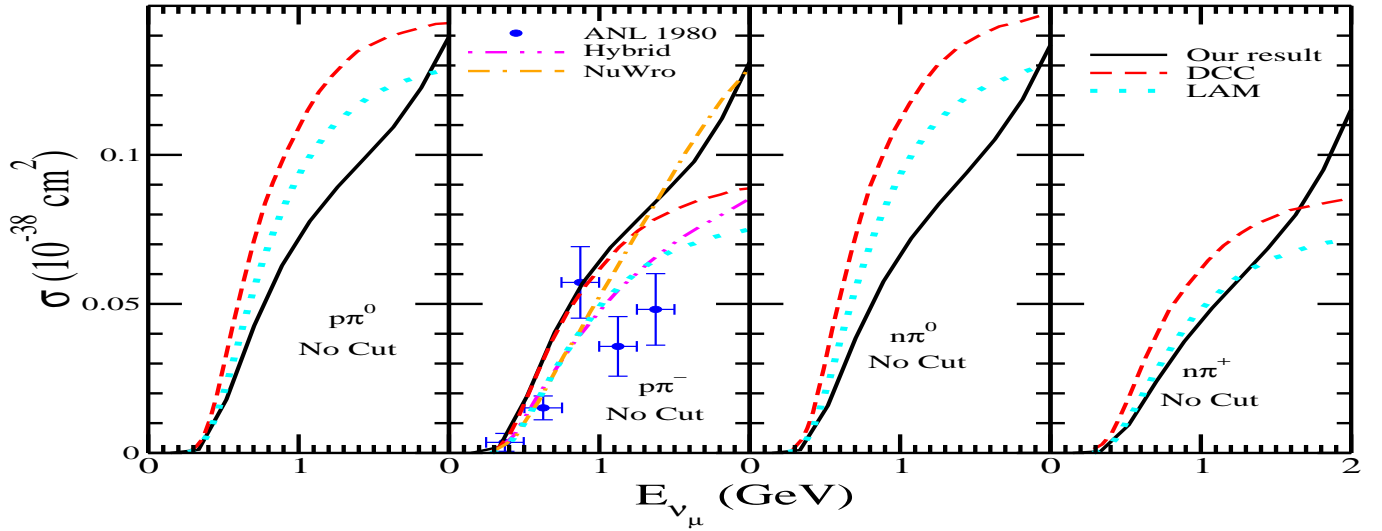}
\caption{Comparison of our results for the total scattering cross section for NC neutrino induced pion 
production processes with deuteron effect. The results are presented from the left for $\nu_{\mu} p \longrightarrow  
\nu_{\mu} p \pi^{0}$, $\nu_{\mu}  n \longrightarrow \nu_{\mu} p  \pi^{-}$, $\nu_{\mu} n \longrightarrow  \nu_{\mu} n 
\pi^{0}$ and $\nu_{\mu} p \longrightarrow \nu_{\mu} n \pi^{+}$ processes. Data points are the experimental results from 
Ref.~\cite{Derrick:1980nr}. Lines have the same meaning as in Fig.~\ref{fig:cc_compare}.}
\label{fig:sigma_Nu_Nubar_NC1pion_compare}
\end{figure}
In Fig.~\ref{fig:sigma_Nu_Nubar_NC1pion_compare}, we have compared our results for neutrino induced NC 
processes {  viz.}, $\nu_{\mu} p \longrightarrow  \nu_{\mu} p \pi^{0}$, $\nu_{\mu}  n \longrightarrow \nu_{\mu} p  
\pi^{-}$, $\nu_{\mu} n \longrightarrow  \nu_{\mu} n \pi^{0}$ and $\nu_{\mu} p \longrightarrow \nu_{\mu} n \pi^{+}$, 
with other theoretical models like the DCC~\cite{Nakamura:2015rta}, the Hybrid~\cite{Gonzalez-Jimenez:2016qqq}, 
NuWro~\cite{Gonzalez-Jimenez:2016qqq}, and Giessen~\cite{Leitner:2006sp} models. 
It may be observed from the figure that in the case of $p\pi^-$ production, our results are in good agreement with the results 
of NuWro Monte Carlo generator, while the results of the other models are quite lower. However, in the case of other channels, 
our results are consistent with the results of DCC and Giessen models.

\subsection{Eta production}\label{sec:eta}
$\eta$-meson is an isoscalar pseudoscalar particle~($I=0,~J^P=0^-$) with mass 547.86~MeV. As the (anti)neutrino energy 
increases, these particles are produced at $E_{\nu_l({\bar\nu}_l)} \ge 0.71(.88)$~GeV for $\nu_e ({\nu}_\mu)$ induced 
CC reactions. The (anti)neutrino induced eta production is interesting because of the several reasons. Being an isoscalar 
particle, the $\eta$ meson is one of the important probes to search for the strange quark~($s\bar{s}$) content of the 
nucleons~\cite{Dover:1990ic}. A precise determination of the $\eta$ production cross section would also help in understanding 
the background in the proton decay searches through the $p \longrightarrow \eta e^{+}$ decays. Therefore, the 
background contribution of $\eta$ production due to the atmospheric neutrino interactions in search of proton decays should 
be well estimated. Furthermore, since the $\eta$ production is expected to be dominated by $S_{11}(1535)$ resonance excitation 
as this state appears near the threshold of the $N \eta$ system and has large branching ratio to $N\eta$ decay modes, a 
precise measurement of the cross section for $\eta$ production will also allow to determine the axial-vector properties of 
this resonance. The production of $\eta$ particle in electromagnetic reactions induced by photons and electrons have been 
studied theoretically and experimentally and the contribution of the vector currents to these processes is fairly 
known.

\begin{figure}  
\begin{center}
\includegraphics[width=0.65\textwidth,height=.3\textwidth]{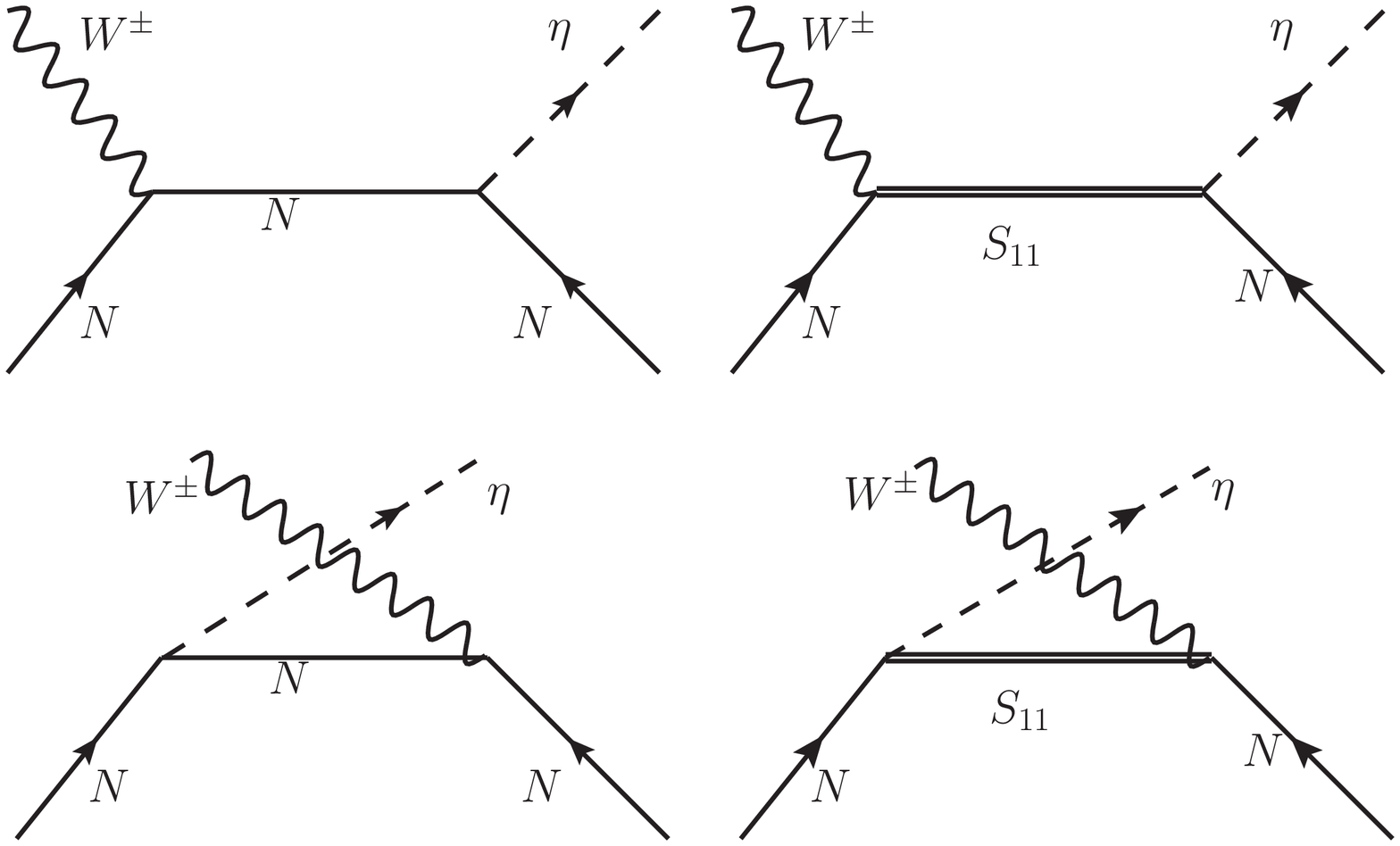}
\caption{Feynman diagrams for the processes $\nu/\bar{\nu} (k) + N (p)  \longrightarrow \mu^\mp (k^\prime) + \eta (p_\eta) + 
N^\prime (p^\prime)$. First row from left to right: s-channel nucleon pole~(SC) and $S_{11}$ resonance~(SC $N^*$); second 
row: u-channel nucleon pole~(UC) and $S_{11}$ resonance~(UC $N^*$).}\label{Ch12_fg_eta:cc_weak_feynman}
\end{center}
\end{figure}

The weak production of $\eta$ mesons via CC interactions which are produced by $\nu_l({\bar\nu}_l)$ from 
the nucleon targets (Fig.~\ref{Ch12_fg_eta:cc_weak_feynman}) are given by
\begin{eqnarray}\label{Ch12_eq:eta_weak_process_cc}
\nu_\ell + n \longrightarrow l^- + \eta + p,  \qquad \qquad
\bar \nu_\ell  + p \longrightarrow l^+ + \eta  + n .
\end{eqnarray}
In the case of electromagnetic production of 
$\eta$ meson and associated particle production, the 
experimental data are available from the MAMI and CLAS collaborations, respectively, for the total as well as differential 
scattering cross sections. Also several theoretical models are available in the literature to study these processes induced 
by photons and electrons. While in the case of weak interactions, these processes are almost unexplored both theoretically 
as well as experimentally. Due to this reason, in order to study the eta production in the 
weak sector, a model that would explain the experimentally available data from photon and electron induced processes has to 
be developed. In the case of pion production, the weak vector part of the hadronic current, in general, is related to the 
electromagnetic current {  via} the CVC hypothesis. Since the electromagnetic production of pions is very well studied, 
therefore, in the case of weak production of pions, we have directly used those information as inputs. Keeping this in mind, 
we have developed a model to study eta production induced by photons to fix the strong and electromagnetic couplings by 
fitting our results with the experimental data available in the literature. Then we have applied the same model to study 
electron induced reactions where the $Q^2$ dependence in the form factors comes into play. With these inputs from the 
electromagnetic sector, we study the single $\eta$ production induced by (anti)neutrinos in Section~\ref{sec:eta}.
For completeness, in Section~\ref{sec:eta:photo}, we discuss the single $\eta$ production induced by photons.

\subsubsection{$\eta$ production induced by photons}\label{sec:eta:photo}
The differential cross section for the photoproduction of $\eta$ mesons off the free nucleon, {  i.e.},
\begin{equation}\label{eq:eta}
 \gamma(q) + N(p ) \longrightarrow N (p^{\prime}) + \eta(p_{\eta}),
\end{equation}
is written as
\begin{eqnarray}\label{eq:sigma_gen}
d\sigma &=& \frac{1}{4 (q\cdot p)} (2 \pi)^{4} \delta^{4}(q+p-p_{\eta}-p^{\prime}) \frac{d{\vec{p}_{\eta}}}{(2 \pi)^{3} (2 
E_{\eta})} \frac{d{\vec p\,}^{\prime}}{(2 \pi)^{3} (2 E^{\prime})} \overline{\sum_{r}} \sum | \mathcal M^{r} |^2,
\end{eqnarray}
where $N = p$ or $n$, the quantities in the parentheses of Eq.~(\ref{eq:eta}) represent the four momenta of the 
corresponding particles, $E_{\eta}$ and $E^{\prime}$, respectively, are the energies of the outgoing eta and nucleon. 
$ \overline{\sum} \sum | \mathcal M^{r} |^2$ is the square of the transition matrix element $\mathcal{M}^{r}$, for photon 
polarization state $r$, averaged and summed over the initial and final spin states. $\mathcal{M}^{r}$ is written in terms of 
the real photon polarization vector $\epsilon_{\mu}^{r}$ and the matrix element of the electromagnetic current taken between 
the hadronic states of $\ket{N}$ and $\ket{N\eta}$, {  i.e.}
\begin{equation}
\mathcal{M}^{r} = e \epsilon_{\mu}^{r} (q) \bra{N(p^{\prime}) \eta(p_{\eta})} {J}^{\mu} \ket{N},
\end{equation}
where $e = \sqrt{4\pi \alpha}$ is the strength of the electromagnetic interaction, with $\alpha = \frac{1}{137}$ being the 
fine-structure constant. In the case when the photon polarization remains undetected, the summation over all the 
polarization states is performed which gives
\begin{equation}\label{lep:photo}
 \sum_{r=\pm1} \epsilon^{*(r)}_{\mu}\epsilon^{(r)}_{\nu} \longrightarrow - g_{\mu \nu}. 
\end{equation}
The hadronic tensor 
${\cal J}^{\mu \nu}$ is written in terms of the hadronic current $J^\mu$ as
\begin{equation}\label{had}
  {\cal J}^{\mu \nu} = \overline{\sum} \sum_{spins} {J^{\mu}}^{\dagger} J^{\nu} = \text{Tr} \left[(\slashed{p}+M) \tilde 
  J^{\mu}(\slashed{p}^{\prime}+M^{\prime})J^\nu\right], \qquad \tilde J^\mu=\gamma_0(J^\mu)^{\dagger}\gamma_0,
\end{equation}
where $M$ and $M^{\prime}$ are the masses of the incoming and outgoing nucleons, respectively. The hadronic matrix element 
of the electromagnetic current $J^{\mu}$ receives the contribution from the background terms and the terms contributing to 
the resonance excitations. 

Using Eqs.~(\ref{lep:photo}) and (\ref{had}), the transition matrix element squared is obtained as
\begin{equation}\label{mat}
\overline{\sum_{r}} \sum_{spin} |\mathcal{M}^{r}|^2 = -\frac{e^2}{4}g_{\mu \nu} {\cal J}^{\mu \nu}.
\end{equation}

\begin{figure} 
\begin{center}
\includegraphics[height=4cm,width=8cm]{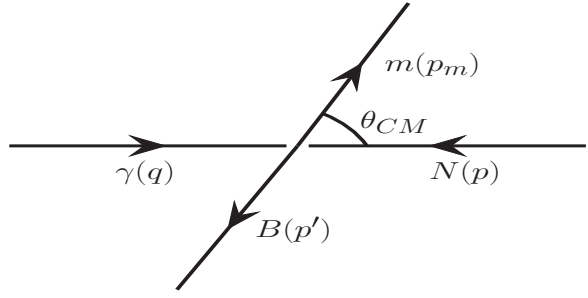}
\caption{Diagrammatic representation of the process $ \gamma (q) + N(p) \longrightarrow m(p_{m}) + B(p^{\prime})$ in CM frame. 
The quantities in the parentheses represent the four momenta of the corresponding particles. 
 $\theta^{CM}$ is the angle between photon and meson in the CM frame.}\label{CM}
 \end{center}
 \end{figure}
Following the above expressions, the differential cross section $\frac{d\sigma}{d\Omega}$ in the CM frame is written as
\begin{equation}\label{dsig}
\left. \frac{d \sigma}{d \Omega}\right|_{CM} = \frac{1}{64 \pi^{2} s} \frac{|\vec{p}\;^{\prime}|}{|\vec{p}|} 
\overline{\sum_{r}} \sum_{spin} |\mathcal{M}^{r}|^2,
\end{equation}
where $s$ is the CM energy squared obtained as
\begin{equation}\label{s}
 s = W^2 = (q + p)^2 = M^{2} + 2M E_{\gamma} ,
\end{equation}
with $E_{\gamma}$ being the energy of the incoming photon in the laboratory frame. 

The hadronic currents for the various NR terms shown in Fig~\ref{Ch12_fg_eta:cc_weak_feynman} are 
obtained using the nonlinear sigma model described in Section~\ref{NRB}. The expressions of the hadronic currents for $s$-, 
$u$- channels are obtained as~\cite{Hernandez:2007qq, RafiAlam:2015fcw}:
\begin{eqnarray}\label{j:s}
J^\mu \arrowvert_{sN} &=&- A_{s}~F_{s}(s) \bar u(p^\prime) \slashed{p}_{\eta} \gamma_5 \frac{\slashed{p} + \slashed{q} + M}
  {s -M^2} \left(\gamma^\mu e_{N} +i \frac{\kappa_{N}}{2 M} \sigma^{\mu \nu} q_\nu \right) u(p), \\
  \label{j:ulam}
J^\mu \arrowvert_{uN} &=&- A_{u} ~F_{u} (u) \bar u(p^\prime) \left(\gamma^\mu e_{N} + i \frac{\kappa_{N}}{2 M} \sigma^{\mu 
\nu} q_\nu \right) \frac{ \slashed{p}^{\prime} -\slashed{q} + M}{u - M^2} \slashed{p}_\eta \gamma_5 u(p),
\end{eqnarray}
where $N$ stands for a proton or a neutron in the initial and final states, $s$ is defined in Eq.~(\ref{s}) and $u = 
(p^{\prime} - q)^{2}$,
$A_{i}$'s; $i=s,u$ are the coupling strengths of $s$, and $u$ channels, respectively, and are obtained as
\begin{eqnarray}\label{eq:coupling}
 A_{s} = A_{u} &=& \left(\frac{D - 3F}{2 \sqrt{3} f_{\eta}}\right) ,
 \end{eqnarray}
$D$ and $F$ are the axial-vector couplings of the baryon octet and $f_{\eta}=105$~MeV is the $\eta$ decay constant. 
The value of $\kappa$ for proton, and neutron are $\kappa_{p} = 1.793$, and $\kappa_{n} = -1.91$ in units of $\mu_{N}$. 

In order to take into account the hadronic structure of the nucleons, the form factors $F_{s} (s)$, and $F_{u} (u)$, are 
introduced at the strong vertex. Various parameterizations of these form factors are available in the 
literature~\cite{Skoupil:2016ast}. We use the most general form of the hadronic form factor which is 
taken to be of the dipole form~\cite{Fatima:2020tyh}:
\begin{equation}\label{FF_Born}
F_{x} (x) = \frac{\Lambda_{B}^{4}}{\Lambda_{B}^{4} + (x - M_{x}^{2})^{2}}, \qquad \qquad \quad x=s,u
\end{equation}
where $\Lambda_{B}$ is the cut-off parameter for the s- and u-channel NRB terms. The value of $\Lambda_{B}$ 
is fitted to the experimental data for both the proton and neutron targets simultaneously and the best fitted value is 
$\Lambda_{B}=0.78$~GeV for $s$- and $u$-channel diagrams. $x$ represents the Mandelstam variables $s,~u$, and $M_{x} = M$ 
corresponds to the mass of the exchanged nucleons in the $s$ and $u$ channels. 
One of the most important property of the electromagnetic current is gauge invariance which corresponds to the current 
conservation and is implemented in the case of $\eta$ production.

In Section~\ref{res:inelastic}, we have already discussed the structure of hadronic current contributing to spin $\frac{1}{2}$ 
and $\frac{3}{2}$ resonance excitations and their subsequent decays to meson-baryon final state. In the case of photon induced 
resonance excitations, the hadronic current is purely vector in nature. Since $\eta$ is an isoscalar meson, therefore, it couples 
to spin $\frac{1}{2}$ resonances only. The vertex function for positive and negative parity spin $\frac{1}{2}$ resonances are 
given in Eq.~(\ref{eq:vec_half_pos}). In the case of real photons, which are 
purely transverse in nature {  i.e.}, the amplitude  $S_{\frac{1}{2}}=0$, the vector form factors are expressed only in terms of 
$A_{\frac{1}{2}}$ helicity amplitude. 

The explicit relation between the coupling $F_2^{R^{+}, R^{0}}$ and 
the helicity amplitude $A_{\frac{1}{2}}^{p,n}$, in the limit $Q^2=0$, is given by:
\begin{eqnarray}\label{eq:hel_spin_12}
A_\frac{1}{2}^{p,n}&=& \sqrt{\frac{2 \pi \alpha}{M} \frac{(M_R \mp M)^2}{M_R^2 - M^2}} \left[ \frac{M_R \pm M}{2 M} 
F_2^{R^{+},R^0} \right] ,
\end{eqnarray}
where the upper~(lower) sign stands for the positive~(negative) parity resonance. $M_R$ is the mass of corresponding 
resonance. In the case of $\eta$ production, we have considered three spin $\frac{1}{2}$ resonances {  viz.} $S_{11} 
(1535)$, $S_{11} (1650)$, and $P_{11} (1710)$, where the dominant contribution to the total scattering cross section comes 
from $S_{11}(1535)$ resonance. In the present work, we have used the value of $A_{\frac{1}{2}}$ for these resonances given 
in PDG~\cite{ParticleDataGroup:2020ssz}.

The most general form of the hadronic currents for the $s-$ and $u-$ channel processes where a resonance state $R_{\frac12}$ 
is produced and decays to a $\eta$ and a nucleon in the final state, are written as
\begin{eqnarray}
j^\mu\big|_{s}&=&  \frac{g_{RN\eta}}{f_{\eta}} \bar u({p}\,') 
 \slashed{p}_{\eta} \Gamma_{s} \left( \frac{\slashed{p}+\slashed{q}+M_{R}}{s-M_{R}^2+ iM_{R} \Gamma_{R}}\right) 
 \Gamma^\mu_{\frac12 
 \pm} u({p}\,), \nonumber\\
 \label{eq:res1/2_had_current}
 j^\mu\big|_{u}&=&  \frac{g_{RN\eta}}{f_{\eta}} \bar u({p}\,') 
 \Gamma^\mu_{\frac12 \pm}\left(\frac{\slashed{p}^{\prime}-\slashed{q}+M_{R}}{u-M_{R}^2+ iM_{R} \Gamma_{R}}\right) 
 \slashed{p}_{\eta} \Gamma_{s}  u({p}\,),
\end{eqnarray}
where $\Gamma_{R}$ is the decay width of the resonance, $\Gamma_{s} = 1(\gamma_{5})$ stands for the positive~(negative) 
parity resonances. $\Gamma_{\frac{1}{2}^{+}}$ and $\Gamma_{\frac{1}{2}^{-}}$ are, respectively, the vertex function for the 
positive and negative parity resonances, defined in Eq.~(\ref{eq:vec_half_pos}). 
$g_{RN\eta}$ is the coupling strength for the process $ R \to N\eta$, given in Table~\ref{Tab:Resonance:para}.

In analogy with the NR terms,  we have considered the following form factors at the strong 
vertex, in order to take into account the hadronic structure:
\begin{equation}
F^{*}_{x} (x) = \frac{\Lambda_{R}^{4}}{\Lambda_{R}^{4} + (x - M_{x}^{2})^{2}},
\end{equation}
where $\Lambda_{R}$ is the cut-off parameter whose value is fitted to the experimental data, $x$ represents the Mandelstam 
variables $s,~u$, and $M_{x} = M_{R}$ corresponding to the mass of the nucleon resonances exchanged in the $s,$ and $u$ 
channels. In general, $\Lambda_{R}$ would be different from $\Lambda_{B}$, however, in the case of $\eta$ production by 
photons, it happens that the same value of $\Lambda_{R}$ as that of $\Lambda_{B}$ i.e. $\Lambda_{R} = \Lambda_{B} =
0.78$~GeV gives the best results. The same values of $\Lambda_{R}$ and $\Lambda_{B}$ help us to minimize the number of free parameters used to fit 
the experimental data.

\begin{figure}  
\begin{center}
\includegraphics[width=0.75\textwidth,height=6.5cm]{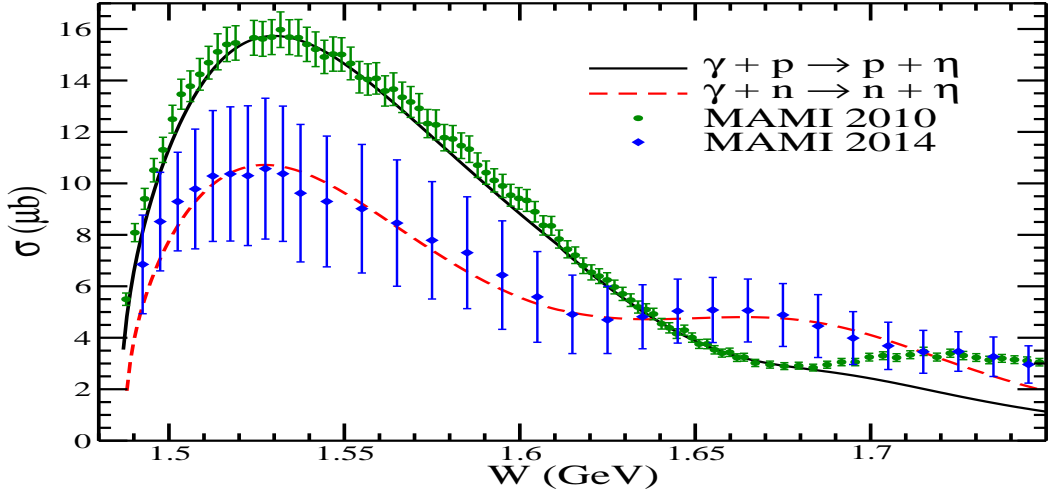}
\caption{Cross section for $\gamma  N \longrightarrow \eta N,\; N=n,p$ process. The experimental points for proton target are 
obtained from MAMI crystal ball~\cite{CrystalBallatMAMI:2010slt} and for neutron target, we have used the quasifree neutron data from 
Werthmuller et. al. \cite{A2:2014pie}, and the results are shown up to the $\Lambda K$ threshold.}
\label{fg_eta:photo_xsec_mami}
\end{center}
\end{figure}

In Fig.~\ref{fg_eta:photo_xsec_mami}, we have presented the results for the total scattering cross section $\sigma$ as a 
function of $W$ for $\gamma + p \longrightarrow p + \eta$ and $\gamma + n \longrightarrow n + \eta$ 
processes in the region of $W$ from $\eta$ production threshold to $K\Lambda$ production threshold. We have compared our 
theoretical results with the experimental data obtained by the MAMI crystal ball~\cite{CrystalBallatMAMI:2010slt} 
collaboration for the proton target and the quasifree neutron data from Werthmuller et 
al.~\cite{A2:2014pie}. It may be observed from the figure that in the case of $\eta$ production from the proton and neutron 
targets, our results are in very good agreement with the available experimental data with a very few free parameters. We 
have fitted the value of strong coupling constant $g_{RN\eta}$ from the photoproduction channels that would be used as an 
input in the weak production of $\eta$ mesons, discussed in the next section.

However, 
in the case of electron induced eta production one has to include the structure of the hadronic current by taking into 
account the nucleon-nucleon and nucleon-resonance transition form factors at the electromagnetic vertex. The $Q^2$ dependence 
of the nucleon-resonance vector form factors is obtained by fitting the experimental data available for electroproduction of 
the eta mesons, where electromagnetic coupling fixed from the photoproduction are used as the values of these form factors at 
$Q^2=0$. The results for the electroproduction of $\eta$ mesons from the nucleons will be reported 
elsewhere~\cite{Fatima:2022}.

\subsubsection{$\eta$ production induced by (anti)neutrinos}\label{sec:eta}
(Anti)neutrino induced single $\eta$ production off the nucleon target~(Fig.~\ref{Ch12_fg_eta:cc_weak_feynman}) are given by the 
following reactions
\begin{eqnarray}\label{Ch12_eq:eta_weak_process_cc}
\nu_\mu (k)  + n (p) &\longrightarrow& \mu^- (k^\prime) + \eta ( p_\eta) + p (p^\prime),   \\
\label{Ch12_eq:eta_weak_process_cc1}
\bar \nu_\mu (k)  + p (p) &\longrightarrow& \mu^+ (k^\prime) + \eta( p_\eta)  + n(p^\prime) ,
\end{eqnarray}
where the quantities in the parenthesis are the four momenta of the particles.

The general expression of the differential scattering cross section for the reactions shown in 
Eqs.~(\ref{Ch12_eq:eta_weak_process_cc}) and (\ref{Ch12_eq:eta_weak_process_cc1}) in the laboratory frame is given in 
Eq.~(\ref{eq:sigma_inelastic}), with $\vec{p}_{m}=\vec{p}_{\eta}$ as the three-momentum of the outgoing eta-meson and 
$E_{m} = E_{\eta}$, the energy of the eta-meson. The transition matrix element, in terms of the leptonic and the hadronic 
currents, is given in Eq.~(\ref{eq:Gg}). The leptonic current is given in Eq.~(\ref{lep_curr}) and the hadronic current 
receives contribution from the NRB terms as well as from the resonance excitations and their subsequent 
decay to $N\eta$ final state.  

The hadronic currents for the NRB terms, \textit{i.e.}, Born diagrams~(s- and u-channels) with nucleon 
poles, using the nonlinear sigma model discussed in Section~\ref{NRB}, are obtained as~\cite{RafiAlam:2015zqz, 
RafiAlam:2013jcs}:
\begin{eqnarray}\label{Ch12_Eq_eta:cc_weak_amp_nucl_pole}
J_{N(s)}^\mu &=& a 
\frac{D-3F}{2\sqrt3 f_\pi} \bar u_N (p^\prime) {p\hspace{-.5em}/}_\eta \gamma^5  
\frac{{p}\hspace{-.5em}/+{q}\hspace{-.5em}/+M}{(p+q)^2-M^2} 
\left[V_{N}^{\mu} - A_{N}^{\mu} \right] u_N (p), \\ 
J_{N(u)}^\mu &=& a \frac{D-3F}{2\sqrt3 f_\pi} 
\bar u_N (p^\prime) \left[V_{N}^{\mu} - A_{N}^{\mu} \right]
  \frac{{p}\hspace{-.5em}/-{p\hspace{-.5em}/}_\eta+M}{(p - p_\eta)^2-M^2} 
{p\hspace{-.5em}/}_\eta \gamma^5 u_N (p), 
\end{eqnarray}
where $a=\cos\theta_{C}$, and $V_{N}^{\mu}$, $A_{N}^{\mu}$, are defined in Eq.~(\ref{eq:vec_curr}), respectively, 
in terms of the vector and axial-vector form factors discussed in Section~\ref{CC}. 

In analogy with the photoproduction of $\eta$ mesons, in the case of (anti)neutrino interactions we have considered only 
$S_{11}(1535)$, $S_{11}(1650)$, and $P_{11}(1710)$ resonances, which decay to $N\eta$ in the final state. The hadronic 
states for these resonance excitations and their subsequent decays in the $s$- and $u$- channels are given in 
Eqs.~(\ref{eq:res1/2_had_current_pos}) and (\ref{eq:res1/2_had_current_pos_u}). The determination of the vector and 
axial-vector $N-R$ transition form factors are discussed in detail in Section~\ref{CC}, and the strong coupling $g_{RN\eta}$ 
is fixed by the photoproduction processes, obtained using the method discussed in Section~\ref{coupling} and are tabulated 
in Table.~\ref{Tab:Resonance:para}.

\begin{figure}[h]
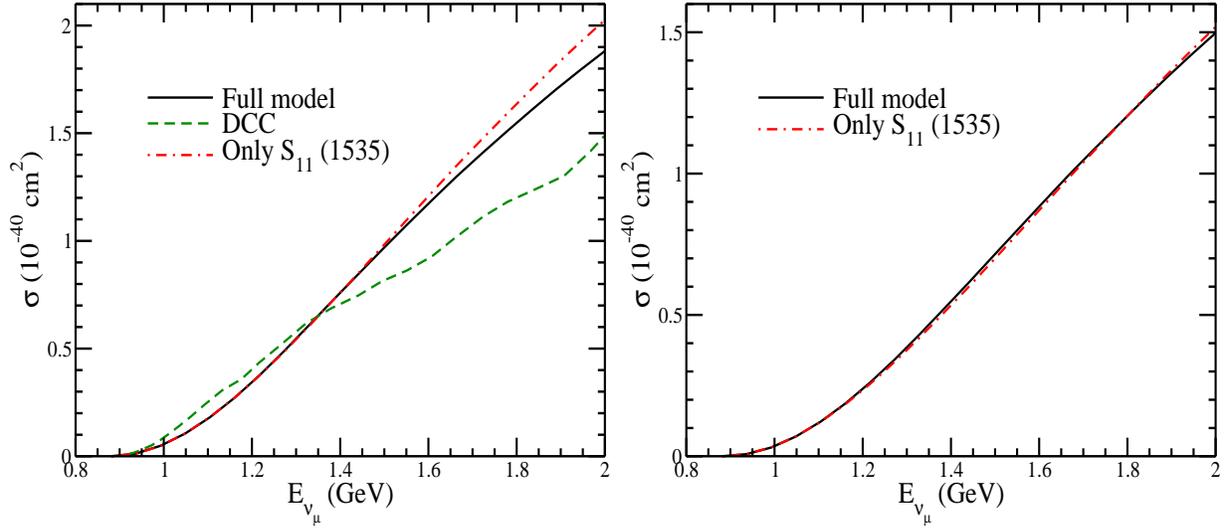

\begin{center}
\includegraphics[width=8cm,height=7cm]{neutrino_eta.eps}
\includegraphics[width=8cm,height=7cm]{antineutrino_eta.eps}
\caption{Total scattering cross section for CC induced $\eta$  production i.e. $\nu_{\mu} + n 
\longrightarrow \mu^{-} + \eta + p$~(left panel) and $\bar{\nu}_{\mu} + p \longrightarrow \mu^{+} + \eta + n$~(right panel). 
Full model consists of the contributions from all the diagrams including $S_{11}(1535)$, $S_{11}(1650)$, and $P_{11}(1710)$. 
In the case of neutrino induced $\eta$ production, we have also compared our results of the full model with the results 
obtained in the DCC model by Nakamura et al.~\cite{Nakamura:2015rta}.}\label{Ch12_fg_eta:cc_xsec_weak}
\end{center}
\end{figure}
Fig.~\ref{Ch12_fg_eta:cc_xsec_weak} shows the results for the total scattering cross sections for the processes $\nu_{\mu} + 
n \longrightarrow \mu^{-} + \eta + p$ and $\bar{\nu}_{\mu} + p \longrightarrow \mu^{+} + \eta + n$. The individual 
contributions from $S_{11}(1535)$ resonance excitations, where both the direct and crossed diagrams are considered, as well as 
the full model~(sum of all the diagrams) are shown. It may be observed from the figure that in the case of both neutrino and 
antineutrino induced reactions, $S_{11}(1535)$ has the dominant contribution. We have also compared the results for the neutrino induced $\eta$ 
with the results of DCC model~\cite{Nakamura:2015rta} and found that from threshold up to $E_{\nu_{\mu}} \sim 1.3$~GeV our 
results are consistent with the results of DCC model. While at $E_{\nu_{\mu}} > 1.3$~GeV, our results are higher than the 
results obtained using DCC model. The $Q^2$-distribution, momentum-distribution, etc. will be reported 
elsewhere~\cite{Fatima:2022}.

\subsection{Strange particle production}\label{sec:Kaon:eta}
With the increase in (anti)neutrino energy single kaon is produced for $E_{\nu_l({\bar\nu}_l)} \ge 0.62~(0.79)$~GeV for 
$\nu_e({\nu}_\mu)$ induced reactions off the nucleon target, by the strangeness changing $|\Delta S|=1$ CC 
interaction like
\begin{eqnarray}\label{Ch12_reaction}
\nu_{l} + n &\longrightarrow& l^{-} + n + K^{+},~~~ \qquad \bar{\nu}_{l} + p \longrightarrow l^{+} + p + K^{-}, \nonumber \\
\nu_{l} + p &\longrightarrow& l^{-} + p + K^{+},~~~ \qquad \bar{\nu}_{l} + n \longrightarrow l^{+} + n + K^{-}, \nonumber \\
\nu_{l} + n &\longrightarrow& l^{-} + p + K^{0},~~~ \qquad \bar{\nu}_{l} + p \longrightarrow l^{+} + n + \bar{K}^{0}, 
\end{eqnarray}
where $l=e,\mu$. Due to lowest threshold among the processes which give rise to a strange particle, the single kaon 
production becomes an important source of kaons for a wide range of energies, and thus their study is important for the 
lower energy accelerator experiments as well as for the atmospheric neutrino experiments.

For the antineutrinos, single hyperon~(like $\Lambda$, $\Sigma$, etc.) is produced in ${\bar\nu}_e({\bar\nu}_\mu)$ induced 
reactions off the nucleon target for $E_{{\bar\nu}_l} \ge 0.19~(0.32)$~GeV (for $\Lambda$ production), by the strangeness 
changing $|\Delta S|=1$ CC interaction like
 \begin{eqnarray}\label{eq10}
\bar{\nu}_\ell  + p  &\longrightarrow& l^{+} + \Lambda,~~~ \bar{\nu}_\ell  + p  \longrightarrow l^{+} + \Sigma^0,~~~ 
\bar{\nu}_\ell  + n  \longrightarrow l^{+} + \Sigma^-, \nonumber
	\end{eqnarray}
which are prohibited for neutrino induced process due to the $\Delta S= \Delta Q$ selection rule. This process is an 
additional source of pion production through hyperon decays, which is significant in the energy region of antineutrinos up to 1~GeV, especially in 
the presence of nuclear medium and final state interaction effects, which are discussed later in 
Section~\ref{hyperon:nucleus}.

With the further increase in antineutrino energies, besides a hyperon, a pion may also be produced in the final state with a 
threshold of $E_{{\bar\nu}_l} \ge 0.37~(0.52)$~GeV (for $\Lambda$ production) in ${\bar\nu}_e({\bar\nu}_\mu)$ induced 
processes, like
 \begin{eqnarray}\label{eq10}
\bar{\nu}_\ell  + p  &\longrightarrow& l^{+} + \Lambda + \pi^0,~~~\bar{\nu}_\ell + n \longrightarrow l^{+} + \Lambda + 
\pi^-, \nonumber \\
\bar{\nu}_\ell  + p  &\longrightarrow& l^{+} + \Sigma^0 +\pi^0,~~~\bar{\nu}_\ell  + n \longrightarrow l^{+} + \Sigma^0 + 
\pi^-, \nonumber \\
\bar{\nu}_\ell  + p  &\longrightarrow& l^{+} + \Sigma^+ +\pi^-,~~~\bar{\nu}_\ell  + n \longrightarrow l^{+} + \Sigma^- + 
\pi^0 .  
 \end{eqnarray}
Then we have associated particle production accompanied by a kaon and a hyperon where strangeness quantum number is 
conserved while all the above processes of strange particle production~($|\Delta S|=1$) are Cabibbo 
suppressed. The threshold for $\Lambda~(\Sigma)$ production for $\nu_{\mu}(\bar{\nu}_{\mu})$ induced reactions is about 
$E_{\nu_\mu({\bar\nu}_\mu)} \ge 1.05~(1.25)$~GeV:
\begin{eqnarray}\label{Ch12_Eq_app:ccAssproduction}
 \nu_{l} + p &\longrightarrow& l^- + \Sigma^{+} + K^{+},~~~\bar{\nu}_{l} + p \longrightarrow l^+ + \Lambda + K^{0},\nonumber 
 \\ 
 \nu_{l} + n &\longrightarrow& l^- + \Lambda + K^{+},~~~\bar{\nu}_{l} + p \longrightarrow l^+ + \Sigma^{0} + K^{0},  
 \nonumber \\ 
 \nu_{l} + n &\longrightarrow& l^- + \Sigma^{0} + K^{+},~~~\bar{\nu}_{l} + p \longrightarrow l^+ + \Sigma^{-} + K^{+},
 \nonumber \\ 
 \nu_{l} + n &\longrightarrow& l^- + \Sigma^{+} + K^{0},~~~\bar{\nu}_{l} + n \longrightarrow l^+ + \Sigma^{-} + K^{0}.
\end{eqnarray}
Similarly for the antineutrinos, $\Xi~(S=-2)$ is produced along with a kaon for $E_{{\bar\nu}_l} \ge 1.28 (1.5)$GeV in 
${\bar\nu}_e({\bar\nu}_\mu)$ induced reactions off the nucleon target, by the strangeness changing $|\Delta S|=1$ CC 
interaction, like  
\begin{align}\label{eq:all_weak_ch}
    \bar{\nu}_\ell + p & \longrightarrow l^+ + K^+ + \Xi^-,~~~ \bar{\nu}_\ell + p  \longrightarrow l^+ + K^0 + \Xi^0,~~~    
    \bar{\nu}_\ell + n  \longrightarrow l^+ + K^0 + \Xi^- . \nonumber    
\end{align}
In the following we briefly describe all the above processes of single kaon production taking up CC 
production induced by (anti)neutrino in the $|\Delta S|=1$ sector in Sections~\ref{kaon} and \ref{sec:1antiKaon}, and the 
associated particle production~($\Delta S=0$) in Sections~\ref{sec:associated:photo} and \ref{sec:associated}.

\subsubsection{Charged current $\nu_l$ induced $K^{+}/K^{0}$ production}\label{kaon}
\begin{figure}  
\begin{center}
	\includegraphics[height=6cm, width=12cm]{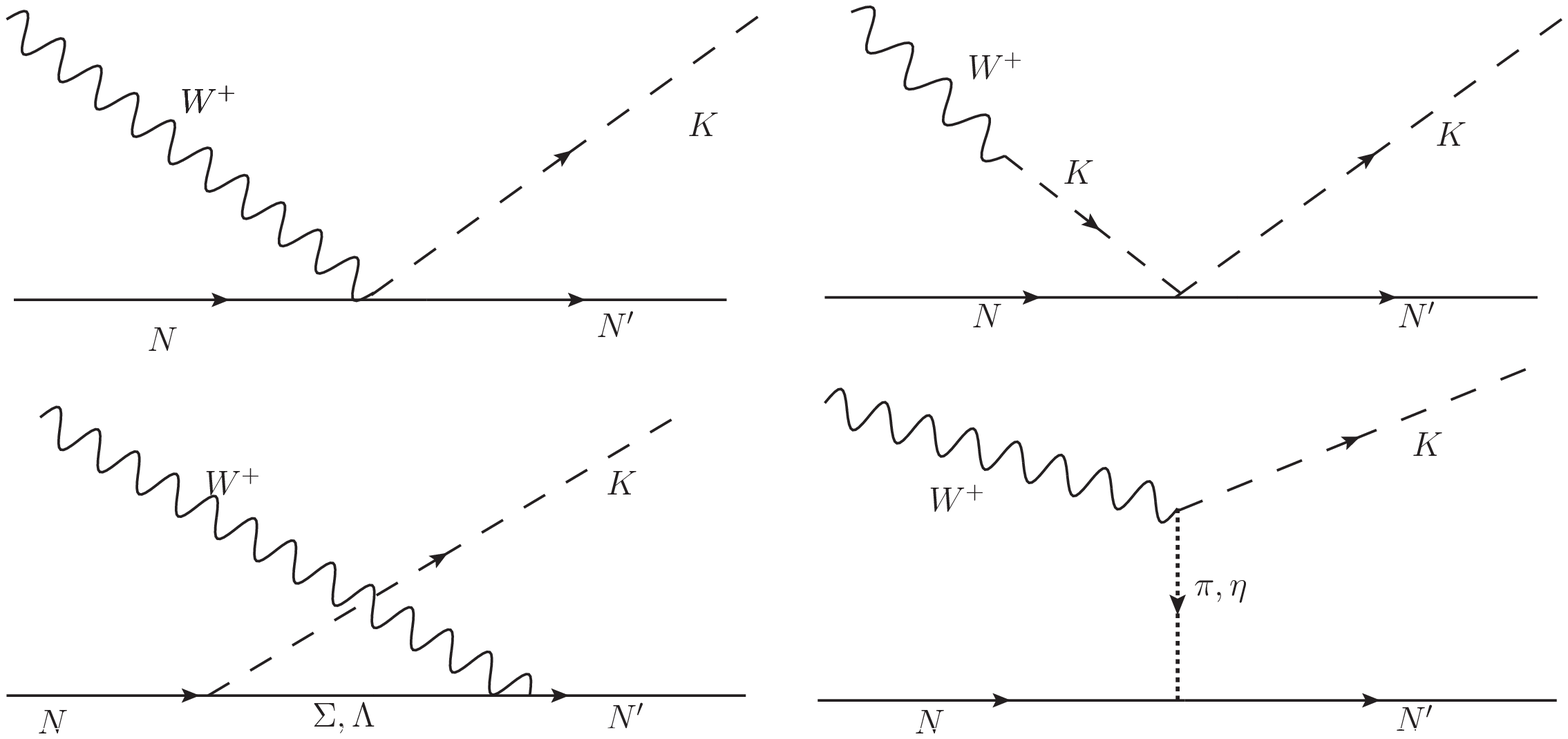}
\caption{Feynman diagrams for the process $ \nu N\longrightarrow l N^\prime K $.  First row from left to right: contact 
term~(CT), kaon pole term~(KP); second row: u-channel diagram~($ C\Sigma$, $C\Lambda $) and pion~(eta) 
in flight~($ \pi P$, $ (\eta P) $. }\label{Ch12_fg:terms}
\end{center}
\end{figure}

The basic reaction for the neutrino induced CC kaon production is
\begin{eqnarray}\label{Ch12_reaction:K}
\nu_{l}(k) + N(p) &\longrightarrow& l^{-}(k^{\prime}) + N^\prime(p^{\prime}) + K^j(p_{K}),  
\end{eqnarray}
where $N, N^\prime =p~ {\text{or}}~ n;~l=e,\mu;~j=K^+ ~\text{or}~ K^0$. 

The reaction shown in the above equation produces a $K^+$ or a $K^0$ meson on proton/neutron target, using neutrino beam, 
which has $S=+1$ in the final hadronic state. Since there are no resonance with $S=+1$, therefore, there is no contribution 
from the resonance excitation to these reactions, and only NR Born diagrams contribute.

The first calculation for these processes were done by Shrock~\cite{Shrock:1975an}, Amer~\cite{Amer:1977fy}, and 
Dewan~\cite{Dewan:1981ab} and the first experimental results were reported by the ANL/BNL experiments with very low 
statistics~\cite{Bell:1978qu, Barish:1978pj}. It is expected that in future the experiments like the DUNE~\cite{DUNE:2022aul} 
and Hyper-Kamiokande~\cite{Hyper-Kamiokande:2022smq} will be able to observe more events of kaon production.

We describe here the latest calculation by our group~\cite{RafiAlam:2010kf} using the effective Lagrangian based on the 
nonlinear sigma model described in Section~\ref{NRB}, and its extension to include the description of strange particles 
using $SU(3)$ symmetry.
\begin{table*}
\begin{center}
\begin{tabular}{|c|c c c c c c c| }\hline\hline
Process & $A_{CT} $&$ B_{CT}$ & $A_{C\Sigma}$ & $A_{C\Lambda}$ & $A_{KP}$ &$ A_{\pi P}$ & $A_{\eta P}$ \\ \hline\hline
$ \nu_{l} + n \longrightarrow l^{-} + K^{+} + n$ & 1 & $D-F$ & $-(D-F)$ & 0 & 1 & 1 & 1\\\hline
$ \nu_{l} + p \longrightarrow l^{-} + K^{+} + p$ & 2 & $-F$  & $-(D-F)/2$ & $(D+3F)$ & 2 & $-1$ & 1\\\hline
$ \nu_{l} + n \longrightarrow l^{-} + K^{0} + p$ & 1& $-(D+F)$ & $(D-F)/2$ & $(D+3F)$ & 1 & $-2$ & 0\\\hline\hline
\end{tabular}
\caption{Values of the constant parameters appearing in Eq.~(\ref{Ch12_NRB:kaon}) for the hadronic currents.}
\label{Ch12_tab:1}
\end{center}
\end{table*}
The expression for the differential scattering cross section is given in Eq.~(\ref{eq:sigma_inelastic}), where $E_{m} = 
E_{K}$, is the energy of the outgoing kaon and $\vec{p}_{m} = \vec{p}_{K}$ represents the three-momentum of the kaon. 
The transition matrix element for these processes~(Eq.~(\ref{Ch12_reaction:K})) is given in Eq.~(\ref{eq:Gg}) with $a=\sin 
\theta_{C}$, where the expression for the leptonic current $l_\mu$ is given in Eq.~(\ref{lep_curr}) and the hadronic current 
matrix element for the different diagrams shown in Fig.~\ref{Ch12_fg:terms}, using effective Lagrangian approach discussed 
in Section~\ref{NRB} and obtained in  Ref.~\cite{RafiAlam:2010kf}, is given by:
\begin{eqnarray}
j^\mu \big|_{CT} &=& -i A_{CT} \frac{\sqrt{2}}{2 f_\pi} \bar{u}(p^\prime) (\gamma^\mu+ \gamma^\mu \gamma^5 B_{CT}) u(p),
\nonumber\\
j^{\mu}\big|_{Cr\Sigma} &=& i A_{Cr\Sigma} \frac{\sqrt{2}}{2 f_\pi} \bar{u}(p^\prime) \left( \gamma^\mu +i
\frac{\kappa_p+2\kappa_n}{2M} \sigma^{\mu \nu} q_\nu + (D-F)\left(\gamma^\mu+\frac{q^\mu}{Q^2+m_K^2}{q\hspace{-.5em}/} \right)
\gamma^5 \right)\frac{{p\hspace{-.5em}/} - {p\hspace{-.5em}/}_K + M_\Sigma}{( p -  p_K)^2 -M_\Sigma^2} {p\hspace{-.5em}/}_K 
\gamma^5  u(p),\nonumber\\
j^{\mu}\big|_{Cr\Lambda}&=& i A_{Cr\Lambda} \frac{\sqrt{2}}{4 f_\pi} \bar{u}(p^\prime)\left( \gamma^\mu +i\frac{\kappa_p}{2M} 
\sigma^{\mu\nu}q_\nu -\frac{D+3F}{3} \left(\gamma^\mu +\frac{q^\mu}{Q^2+m_K^2}{q\hspace{-.5em}/} \right)\gamma^5  \right)
\frac{{p\hspace{-.5em}/} - {p\hspace{-.5em}/}_K +M_\Lambda}{( p -  p_K)^2 -M_\Lambda^2}   {p\hspace{-.5em}/}_K 
\gamma^5 u(p),\nonumber \\
j^{\mu}\big|_{KP}&=&- i A_{KP} \frac{\sqrt{2}}{4 f_\pi} \bar{u}(p^\prime) ({q\hspace{-.5em}/}+{p\hspace{-.5em}/}_K) u(p) 
\frac{1}{Q^2+m_K^2} q^\mu,\nonumber \\
j^{\mu}\big|_{\pi}&=& i A_{\pi P} (D+F) \frac{\sqrt{2}}{2 f_\pi} \frac{M}{(q-p_K)^2 - {M_{\pi}^2}} \bar{u}(p^\prime) 
\gamma^5 (q^\mu - 2 {p_K}^\mu) u(p),\nonumber \\
j^{\mu}\big|_{\eta }&=& i A_{\eta P} (D-3F) \frac{\sqrt{2}}{2 f_\pi} \frac{M}{(q-p_K)^2 - {M_{\eta}^2}} \bar{u}(p^\prime)
\gamma^5 (q^\mu - 2 {p_K}^\mu) u(p), \label{Ch12_NRB:kaon}
\end{eqnarray}
where, $q=k-k^\prime$ is the four momentum transfer, $\kappa_p$ and  $\kappa_n$ are, respectively, the 
proton and neutron anomalous magnetic moments. The value of the various parameters appearing in the expressions of the 
hadronic currents of the different channels are shown in Table-\ref{Ch12_tab:1}. 

\begin{figure}
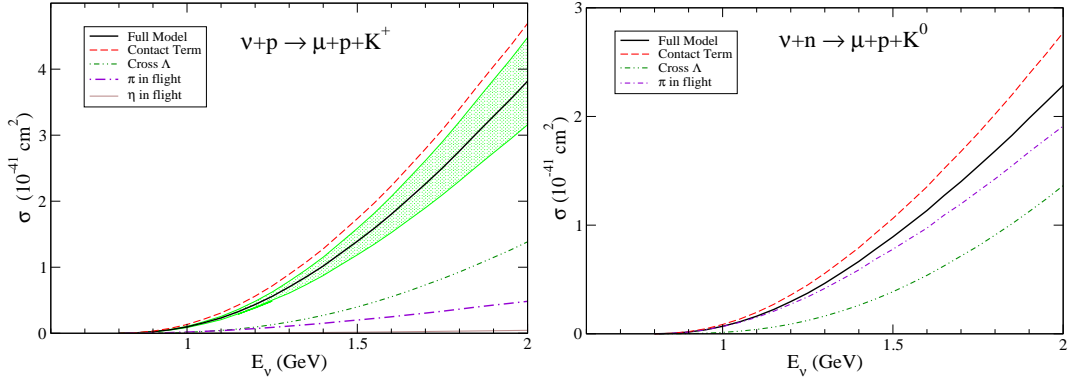
  
\begin{center}
	\includegraphics[height=5cm,width=7cm]{PP_Kaon.eps}
	\includegraphics[height=5cm,width=7cm]{NP_Kaon.eps}
\caption{Contribution of the different terms to the total scattering cross section for the $ \nu_{\mu} + p \longrightarrow 
\mu^{-} + K^+ + p $~(left panel) and $ \nu_{\mu} + n \longrightarrow \mu^{-} + K^0 + p $~(right panel) processes.}
\label{Ch12_fg:full_pp}
\end{center}
\end{figure}
To incorporate the hadronic structure in the matrix element, a dipole form factor 
\begin{equation}\label{Ch12_global_ff}
F(Q^2)=\frac{1}{\left(1+\frac{Q^2}{M_F^2}\right)^2}, 
\end{equation} 
is used with mass $M_F\simeq 1$ GeV. 

Fig.~\ref{Ch12_fg:full_pp} shows the results of the contributions of the different diagrams to the total scattering cross 
sections for the processes $\nu_\mu p \longrightarrow \mu^{-} K^+ p $ and $ \nu_\mu n \longrightarrow \mu^{-} K^0 p$. It may 
be observed that the contact term has a dominant contribution to the total scattering cross section in both the processes 
discussed above. The curve labeled as the full model is calculated with a dipole form factor with $M_{F}=$ 1 GeV. The band 
corresponds to variation of $M_{F}$ by 10$\%$. The process $\nu_\mu n \longrightarrow \mu^- K^0 p$ has a cross section 
of a similar size and the contact term is the largest followed by the $\pi$ exchange diagram and the u-channel~($\Lambda$) 
term. A destructive interference between the different terms has been observed and this resulted in the total 
cross section obtained with the full model to be smaller than that produced by the contact term. For more details and results, 
see Ref.~\cite{RafiAlam:2010kf}.

\subsubsection{Charged current $\bar{\nu}_{l}$ induced ${K}^{-}/\bar{K}^{0}$ production}\label{sec:1antiKaon}
\begin{figure}  
\begin{center}
\includegraphics[width=0.8\textwidth,height=.45\textwidth]{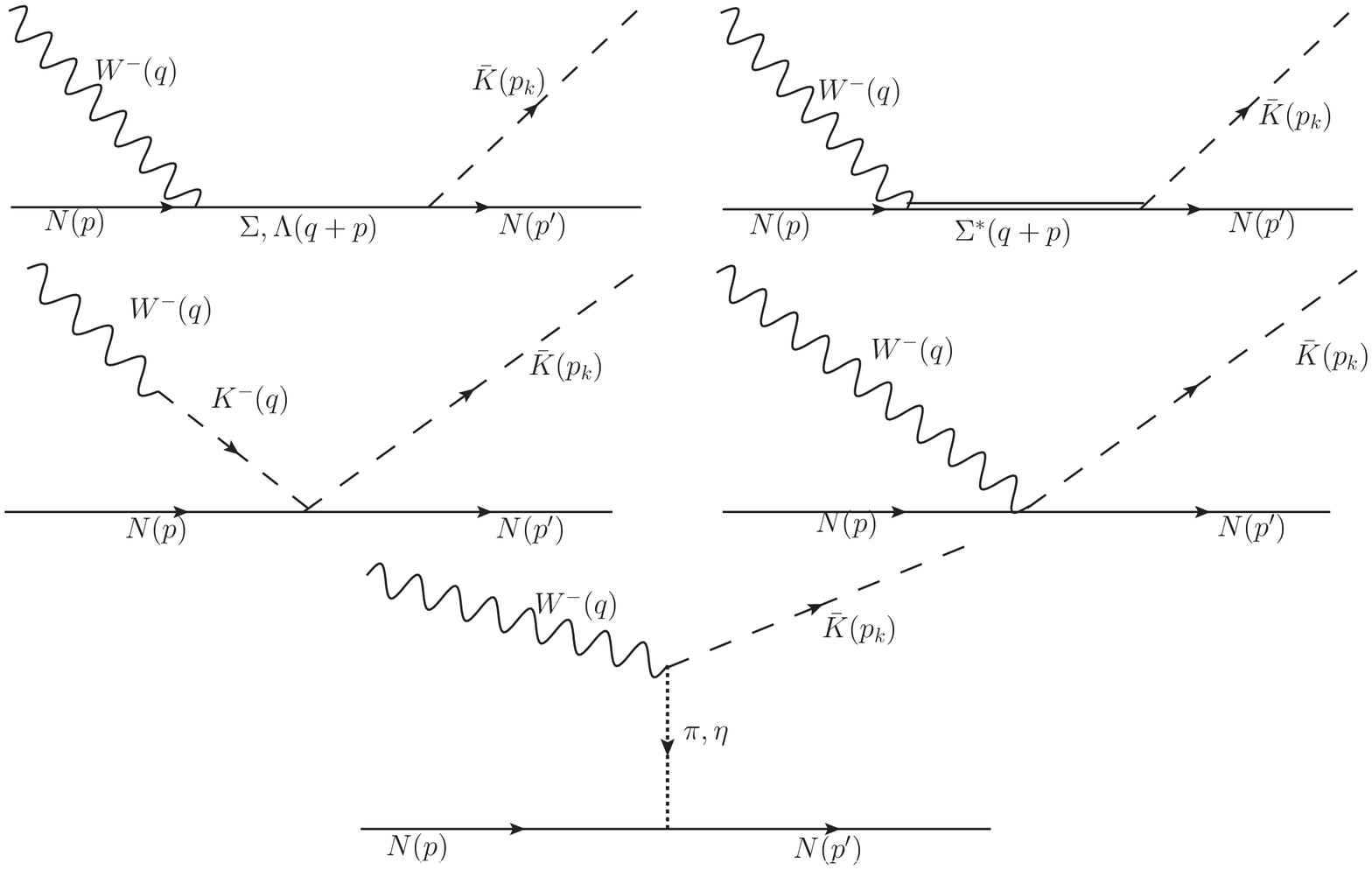}
\caption{Feynman diagrams for the process $\bar \nu N\longrightarrow l N^\prime \bar K$.  First row from left to right: 
s-channel $\Sigma,\Lambda $ propagator (SC), s-channel $\Sigma^*$ Resonance (SCR), second row: kaon 
pole term (KP); contact term (CT) and last row: pion~(eta) in flight ($ \pi P/ \eta P $). }
\label{Ch12_fg:terms_antiKaon}
\end{center}
\end{figure}

In the case of antineutrino induced reactions $K^-$ or ${\bar K}^0$ particle are produced off a nucleon target with $S=-1$ 
in the final state. Consequently there would be resonance excitation with $S=-1$ in the final state which will decay to 
produce $K^-$ or ${\bar K}^0$ particles as shown in Fig.~\ref{Ch12_fg:terms_antiKaon} where $\Sigma^*(1385)$ is a resonance with 
spin $J=\frac{3}{2}$ and isospin $I=1$ along with the NR Born diagrams. In this section, we are briefly discussing 
antikaon production, for details see Ref.~\cite{Alam:2011vwg}.

The basic reaction for the antineutrino induced CC antikaon production is
\begin{eqnarray}\label{Ch12_reaction}
\bar{\nu}_{l}(k) + N(p) &\longrightarrow& l^{+}(k^{\prime}) + N^\prime (p^{\prime}) + K^i(p_{K}) ,\nonumber 
\end{eqnarray}
where $N, N^\prime =p~{\text{or}}~ n;~l=e,\mu;~i=K^- ~\text{or}~ {\bar K}^0$.  

The expression for the differential scattering cross section is given in Eq.~(\ref{eq:sigma_inelastic}), where $E_{m} = 
E_{K}$ is the energy of the outgoing antikaon and $\vec{p}_{m} = \vec{p}_{K}$ represents the three-momentum of the antikaon. 
The transition matrix element is defined in Eq.~(\ref{eq:Gg}) with $a=\sin\theta_C$ and the leptonic current, given in 
Eq.~(\ref{lep_curr}). The different channels which contribute to the hadronic currents are the s-channel with $\Sigma,
~\Lambda$~(SC) and $\Sigma^*$~(SCR) as the intermediate states, the kaon pole~(KP) term, the contact term~(CT), and the 
meson~($\pi$P, $\eta$P) exchange terms~\cite{Alam:2011vwg}. 
\begin{table}
\begin{center}
\begin{tabular}{|l|c|c|c|c|c|c|c|c|} \hline \hline
	Process & $B_{CT}$  & $A_{CT}$ & $A_{\Sigma}$ & $A_{\Lambda}$ & $A_{KP}$ & $A_{\pi }$ & $A_{\eta }$ & $ 
	A_{\Sigma^*}$
	\\ \hline
	$ \bar{\nu}_{l} + n \longrightarrow l^+ + K^- + n $ & $D-F$ &  1 & $-1$ & 0 & $-1$ & 1 & 1 &  2 \\ 
	$ \bar{\nu}_{l} + p \longrightarrow l^+ + K^- + p $ & $-F$  & 2 & $-\frac{1}{2}$  &1 & $-2$ & $-1$ & 1 & 1 \\ 
	$ \bar{\nu}_{l} + p \longrightarrow l^+ + \bar{K}^0 + n $ & $-(D+F)$ & 1 & $\frac{1}{2}$ & 1 & $-1$ & $-2$ & 0 & 
	$-1$ 
	\\ 
	\hline \hline
\end{tabular}
\caption{Values of the constant parameters appearing in Eq.~(\ref{Ch12_NRB:antikaon}) for the hadronic currents.}
\label{Ch12_tb:currents}
\end{center}
\end{table}

The hadronic currents for the background terms discussed in Section~\ref{NRB} and obtained in Ref.~\cite{Alam:2011vwg}, are 
written as
\begin{eqnarray}
	J^\mu \arrowvert_{CT} &=&i A_{CT}  \frac{ \sqrt{2}}{2 f_\pi}  \bar{u}(p^\prime) \; (\gamma^\mu + B_{CT} \; 
	\gamma^\mu \gamma_5 ) \; u(p) \nonumber\\
	J^\mu \arrowvert_{\Sigma} &=&i A_{\Sigma} (D-F) \frac{\sqrt{2}}{2 f_\pi} \bar{u}(p^\prime) \slashed{p}_K
	\gamma_5  \frac{ p\hspace{-.5em}/ +
		q\hspace{-.5em}/ + M_\Sigma}
	{( p +  q)^2 -M_\Sigma^2} \left(\gamma^\mu +i  \frac{(\kappa_p + 2\kappa_n)}{2 M} \sigma^{\mu \nu} q_\nu + (D-F) \left\{ \gamma^\mu 
	+ \frac{q^\mu}{ Q^2+{m_K}^2 } q\hspace{-.5em}/ \right\} \gamma^5 \right) u(p) \nonumber\\
	J^\mu \arrowvert_{\Lambda} &=&  i A_{\Lambda}  (D+3F)  \frac{1} {2 \sqrt{2} f_\pi} \bar{u}(p^\prime) \slashed{p}_K
	\gamma^5 \frac{ p\hspace{-.5em}/ +
		q\hspace{-.5em}/ +M_\Lambda}
	{( p +  q)^2 -M_\Lambda^2} \left(\gamma^\mu +i \frac{\kappa_p}{2 M}  \sigma^{\mu \nu} q_\nu \right. \nonumber\\
	&-& \left. \frac{(D + 3 F)}{3} \left\lbrace \gamma^\mu  + \frac{q^\mu }{ Q^2+{m_K}^2 } 
	q\hspace{-.5em}/ \right\rbrace \gamma^5 \right) u(p)\nonumber \\
	J^\mu \arrowvert_{KP}&=& -i A_{KP}  \frac{\sqrt{2}}{2 f_\pi}  \bar{u}(p^\prime)  q\hspace{-.5em}/ \; u(p) 
	\frac{q^\mu}{Q^2+m_K^2}  \nonumber  \\
	J^\mu \arrowvert_{\pi} &=& iA_{\pi } \frac{M\sqrt{2}}{2 f_\pi}    (D + F)\frac{ 2 {p_K}^\mu -q^\mu}{(q-p_K)^2 - 
		{m_\pi}^2} \bar{u}(p^\prime)  \gamma_5  u(p) \nonumber \\
		 	\label{Ch12_NRB:antikaon}
	J^\mu \arrowvert_{\eta} &=&i A_{\eta } \frac{M\sqrt{2}}{2 f_\pi}    (D - 3 F)\frac{2 {p_K}^\mu - q^\mu}{(q-p_K)^2 - 
		{m_\eta}^2} \bar{u}(p^\prime) 
	\gamma_5 u(p) 
\end{eqnarray}
The hadronic current for $\Sigma^{*}(1385)~(J=\frac{3}{2},~I=1)$ is written in analogy with the hadronic current of the $\Delta$ 
resonance 
as discussed in Section~\ref{res:inelastic}. The factors $A_{i}$ for each diagram contributing to the hadronic current are 
tabulated in Table-\ref{Ch12_tb:currents}. In analogy with the single kaon production, a global dipole form factor given in 
Eq.~(\ref{Ch12_global_ff}) with $M_F\simeq 1$ GeV is used in the hadronic currents, except for the resonance excitation, for 
which the form factors are related to the $\Delta^+$ excitation discussed in Section~\ref{res:inelastic}. 

\begin{figure}
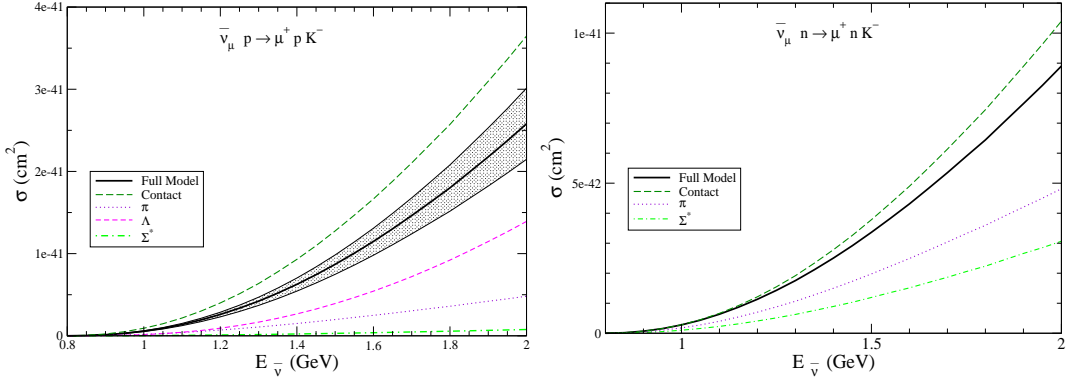
  
\begin{center}
\includegraphics[height=5cm,width=7cm]{pp_antikaon.eps}
\includegraphics[height=5cm,width=7cm]{nn_antikaon.eps}
\caption{Total scattering cross section for the processes $\bar \nu_\mu p \longrightarrow \mu^+ p K^- $ and $\bar \nu_\mu n 
\longrightarrow \mu^+ n K^- $.}
\label{Ch12_fg:xsec_mu_pp}
\end{center}
\end{figure}

In Fig.~\ref{Ch12_fg:xsec_mu_pp}, the different contributions of the hadronic current to the $\bar \nu_\mu p \longrightarrow 
\mu^+ p K^- $ and $\bar \nu_\mu n \longrightarrow \mu^+ n K^- $ reactions are presented. 
It may be observed that the cross section is dominated by the NR terms, where the contact term gives the largest 
contribution among all the NR terms. The destructive interference leads to a total scattering cross section smaller 
than that obtained by the contact term only. It should be noted that in the case of $\bar \nu_\mu p \longrightarrow \mu^+ p 
K^- $ process, $\Sigma^*(1385)$ has negligible contribution. This can be understood because the mass of $\Sigma^*$ is below 
the kaon production threshold. The curve labeled as full model is calculated with a dipole form factor with a mass of 1 GeV. 
The band corresponds to a 10$\%$ variation in $M_{F}$. For the $\bar \nu_\mu n \longrightarrow \mu^+ n K^- $ case, the 
contribution of $\Sigma^*$ resonance is substantial due to the larger value of the couplings (see 
Table-\ref{Ch12_tb:currents}). 

\subsubsection{Associated particle production induced by photons}\label{sec:associated:photo}
\begin{figure}
 \begin{center}
    \includegraphics[height=3cm,width=4.9cm]{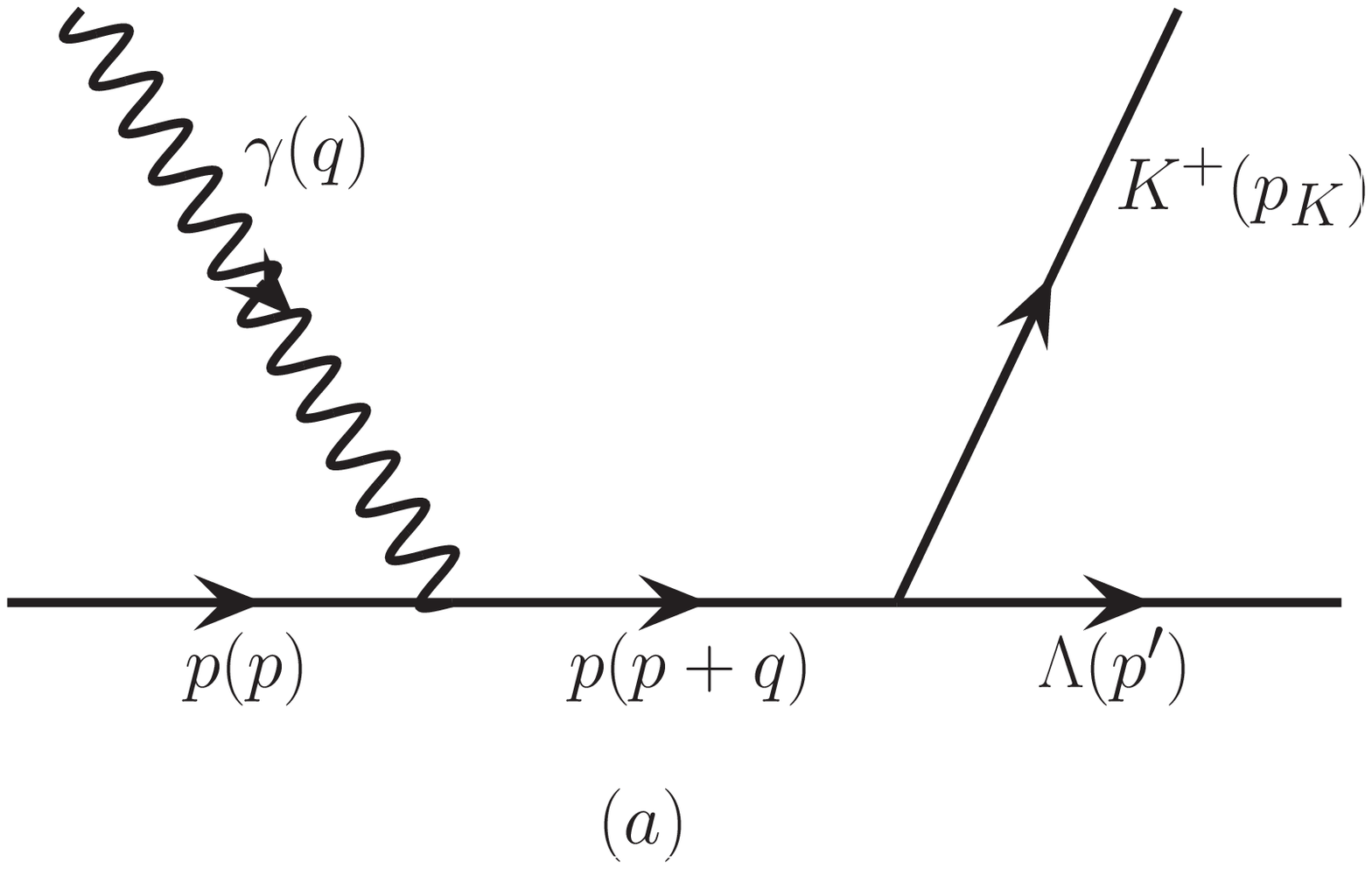}
    \hspace{5mm}
    \includegraphics[height=3.5cm,width=4.3cm]{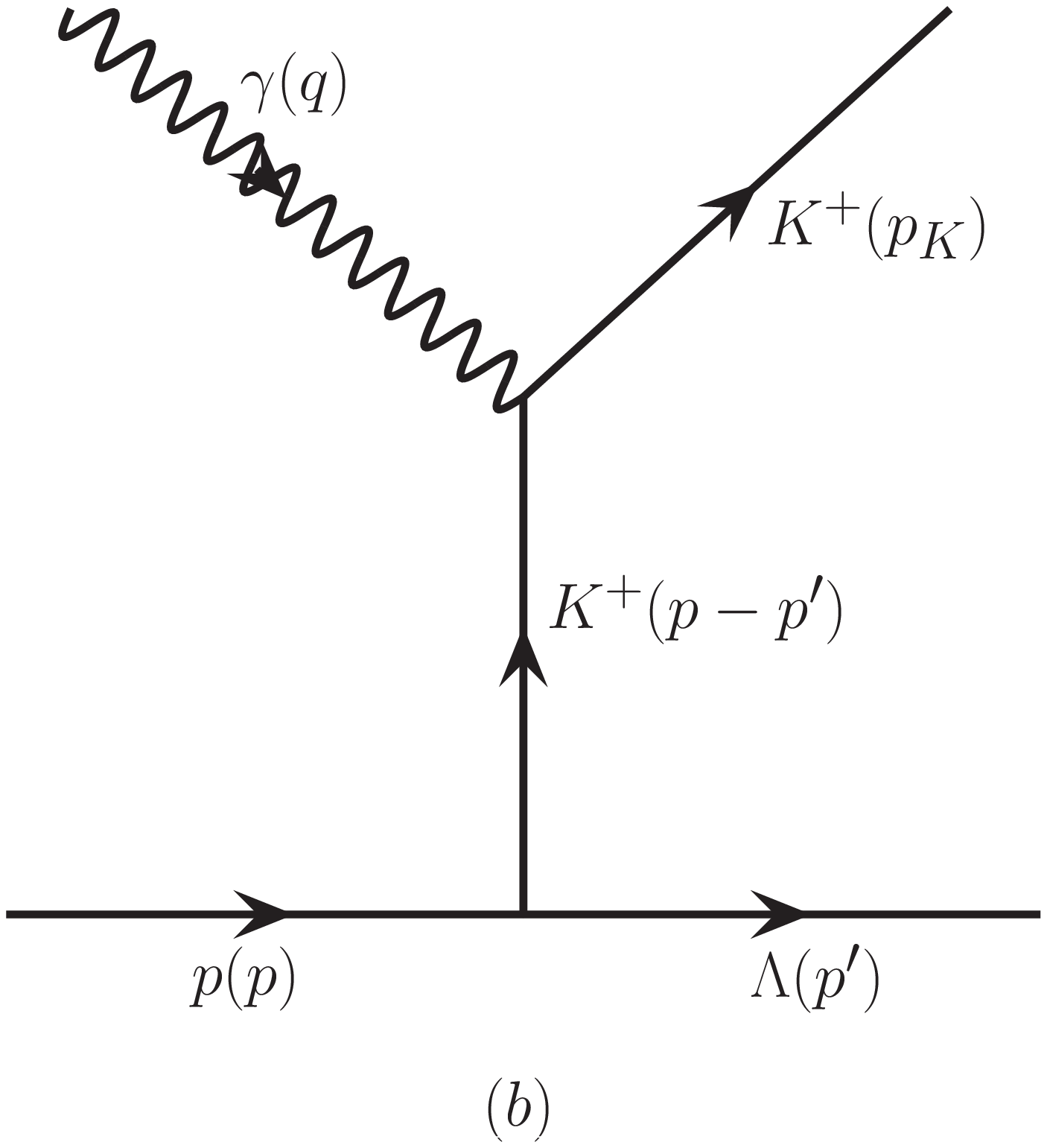}
    \hspace{5mm}
    \includegraphics[height=3cm,width=4.9cm]{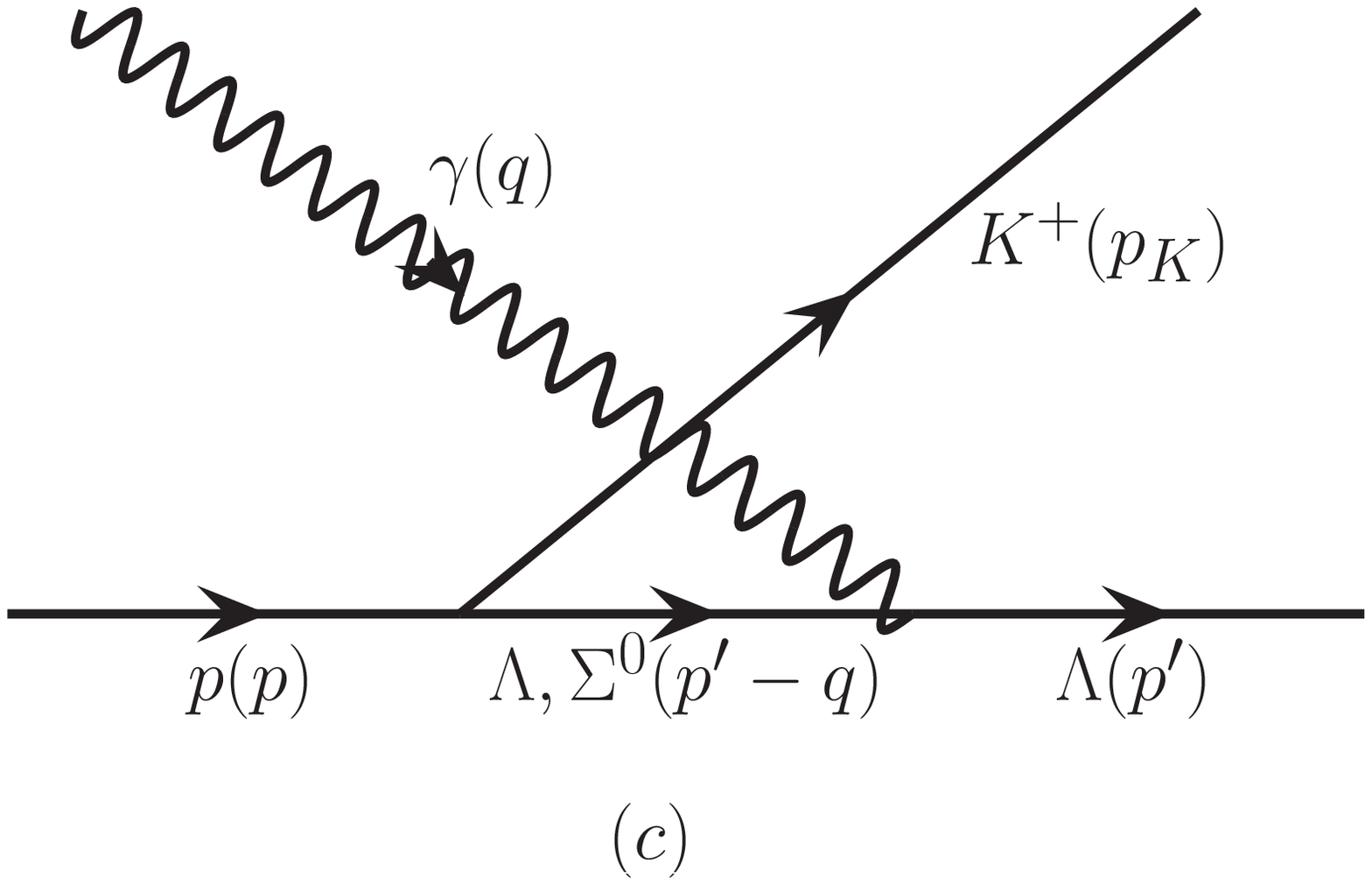}
    \vspace{5mm}
    
    \includegraphics[height=3cm,width=5cm]{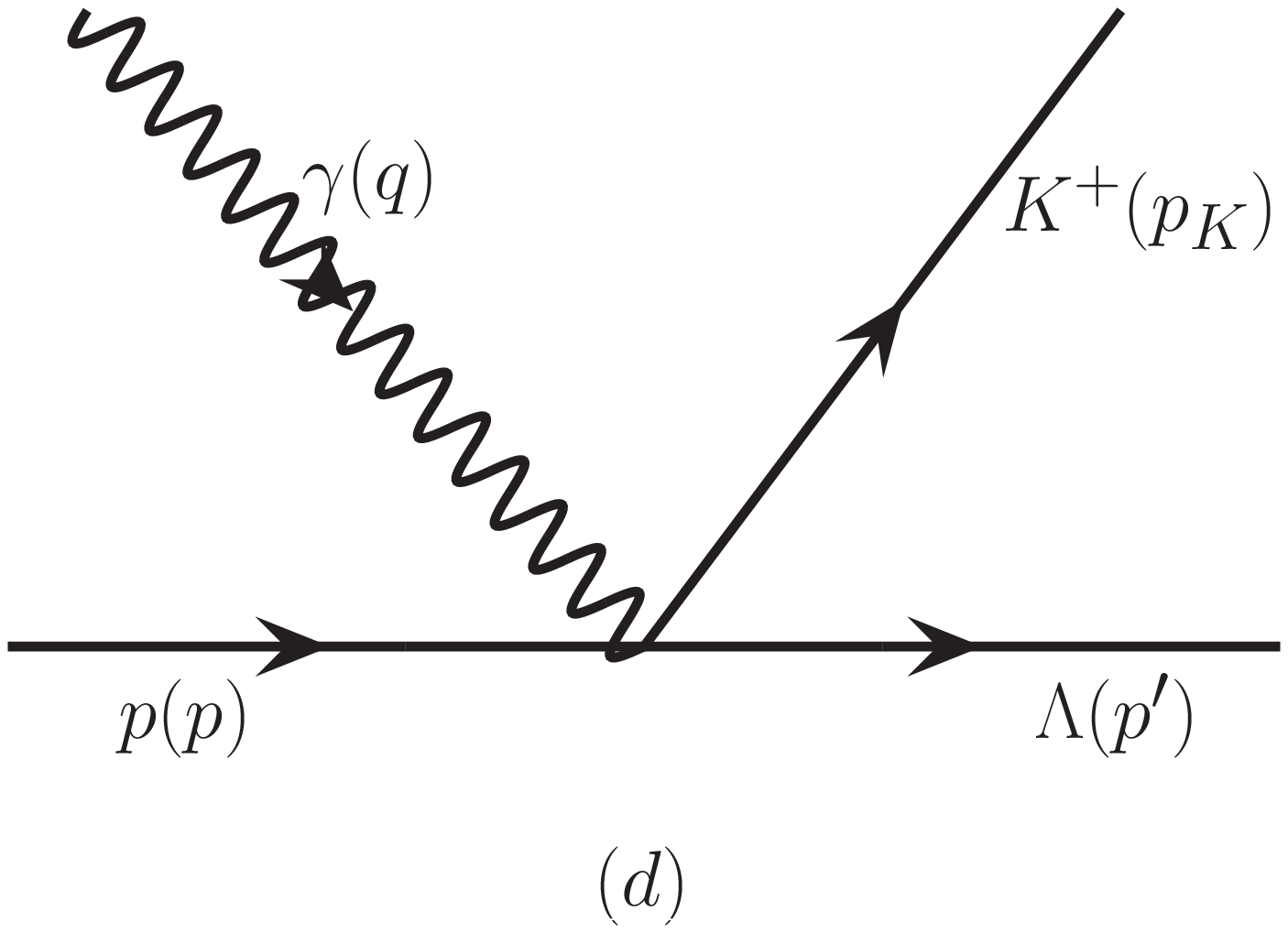}
    \hspace{5mm}
    \includegraphics[height=3.5cm,width=5.5cm]{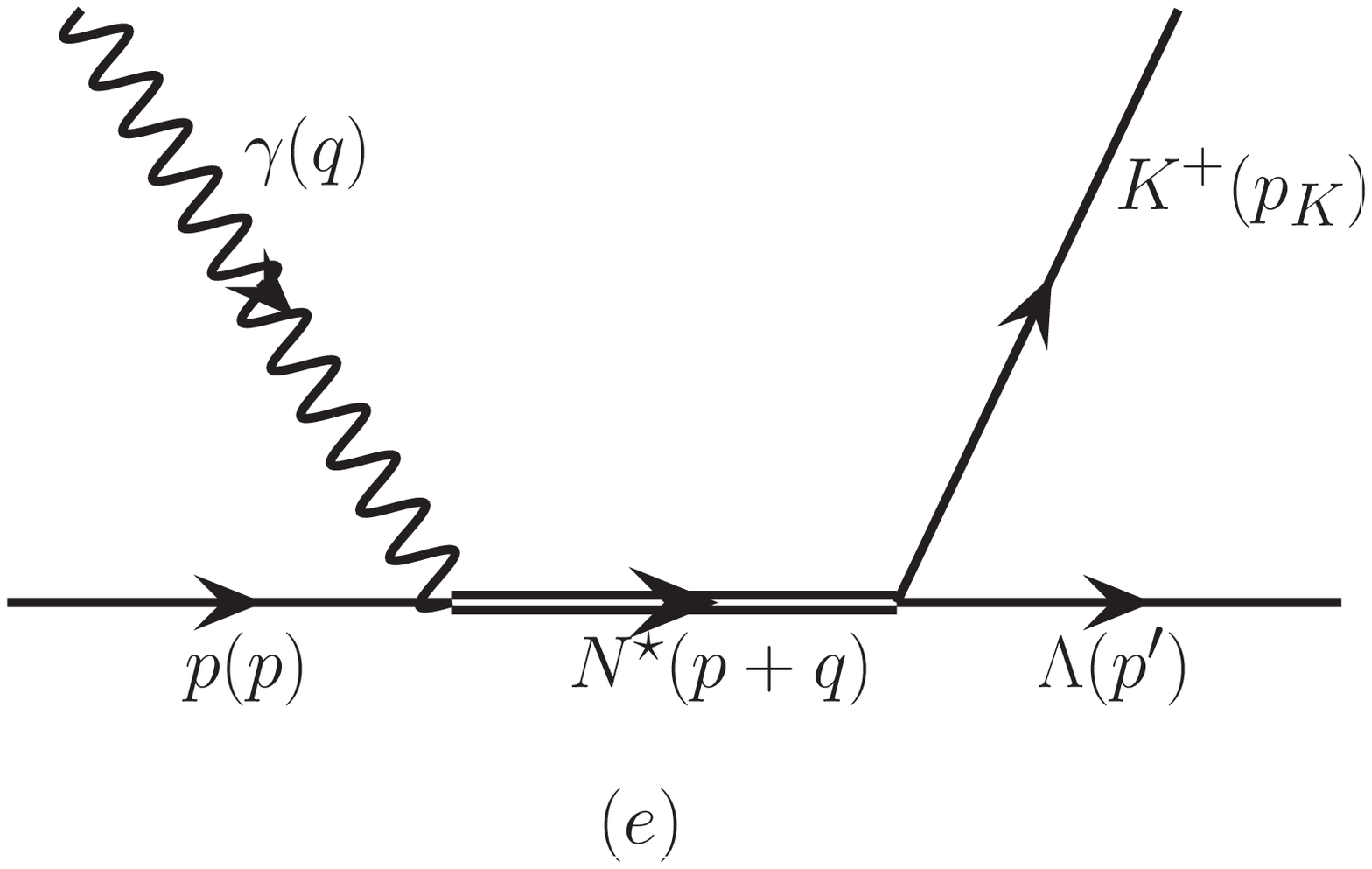}
\caption{Feynman diagram for the various channels possible for the process $ \gamma (q) + p(p) \rightarrow K^{+}(p_{k}) + 
\Lambda(p^{\prime})$. (a)~$s$ channel, (b)~$t$ channel, (c)~$u$ channel and (d)~contact term constitute the NR 
terms. (e)~nucleon resonances in the $s$ channel. The quantities in the bracket represent four momenta of the corresponding 
particles.}\label{fyn_dia}
\end{center}
 \end{figure}

As discussed in Section~\ref{sec:eta:photo}, before calculating the scattering cross section for (anti)neutrino induced 
associated particle production, the strong and electromagnetic couplings are fixed by calculating the total cross section of 
associated production induced by photons. Here, we focus only on the production of $K\Lambda$ in the final state induced by 
photon, where the general reaction may be written as:
\begin{equation}\label{eq:ass:photo}
 \gamma(q) + p (p) \longrightarrow \Lambda (p^{\prime}) + K^{+} (p_{K}).
\end{equation}
The differential scattering cross section for the above reaction is calculated in the same way as presented in 
Section~\ref{sec:eta:photo}, with the expression of $\frac{d\sigma}{d\Omega}$ given in Eq.~(\ref{dsig}). The hadronic 
current receives contribution from both the NR and resonance excitations, as shown in Fig.~\ref{fyn_dia}. Unlike 
the case of $\eta$ production, in this case, spin $\frac{3}{2}$ resonances with $I=\frac{1}{2}$ also contribute 
significantly.

The hadronic currents for the various NR terms shown in Fig.~\ref{fyn_dia}(a)--(d) are obtained using the nonlinear 
sigma model described in Section~\ref{NRB} and are obtained as~\cite{Fatima:2020tyh}:
\begin{eqnarray}\label{j:s}
J^\mu \arrowvert_{s} &=&- A_{s}~F_{s}(s) \bar u(p^\prime) \slashed{p}_K \gamma_5 \frac{ \slashed{p} + \slashed{q} + M}
  {s -M^2} \left(\gamma^\mu e_{p} +i \frac{\kappa_{p}}{2 M} \sigma^{\mu \nu} q_\nu \right) u(p), \\
  \label{j:t}
  J^\mu \arrowvert_{t}&=& - A_{t}~ F_{t}(t) \bar u(p^\prime)\left[(\slashed{p} - \slashed{p}^{\prime}) \cdot \gamma_{5} 
  \right] u(p) \frac{(2 p_{K}^{\mu} - q^{\mu})}{t - m_{K}^{2}} , \\
  \label{j:ulam}
J^\mu \arrowvert_{u \Lambda} &=&- A_{u}^{\Lambda} ~F_{u}^{\Lambda} (u) \bar u(p^\prime) \left(\gamma^\mu e_{\Lambda} + i 
\frac{\kappa_{\Lambda}}{2 M_{\Lambda}} \sigma^{\mu \nu} q_\nu \right) \frac{ \slashed{p}^{\prime} -\slashed{q} + M_{\Lambda}}
{u - M_{\Lambda}^2} \slashed{p}_K \gamma_5 u(p), \\
\label{j:usig}
J^\mu \arrowvert_{u \Sigma^{0}} &=&- A_{u}^{\Sigma^{0}} ~F_{u}^{\Sigma^{0}} (u) \bar u(p^\prime) \left(\gamma^\mu 
e_{\Sigma^{0}} + i \frac{\kappa_{\Sigma^{0}}}{2 M_{\Sigma^{0}}} \sigma^{\mu \nu} q_\nu \right) 
\frac{\slashed{p}^{\prime} -\slashed{q} + M_{\Sigma^{0}}} {u -M_{\Sigma^{0}}^2} \slashed{p}_K \gamma_5 u(p), \\
\label{j:CT}
J^\mu \arrowvert_{CT} &=& A_{CT} ~F_{CT} \bar u(p^\prime) \; \gamma^\mu \gamma_5 \; u(p),
\end{eqnarray}
where $s$ and $u$ are already defined and $t=(p - p^{\prime})^{2}$. 
The couplings $A_{i}$'s for the different terms, obtained in the nonlinear sigma model are:
\begin{eqnarray}\label{eq:coupling:associated}
 A_{s} = A_{t} = A_{u}^{\Lambda} = A_{CT} = - \left(\frac{D + 3F}{2 \sqrt{3} f_{K}}\right),  \qquad \quad
 A_{u}^{\Sigma^{0}} =  \left(\frac{D-F}{2 f_{K}} \right).
\end{eqnarray} 
The value of $\kappa$ for lambda i.e. $\kappa_{\Lambda} = -0.613$ and for sigma i.e. $\kappa_{\Sigma^{0}} = 1.61$, are in 
units of $\mu_{N}$.

One of the most important property of the electromagnetic current is the gauge invariance that corresponds to the current 
conservation, which is implemented for the full current. The total hadronic current for the NR terms is given by
\begin{eqnarray}
 J^{\mu} = J^\mu \arrowvert_{s} +J^\mu \arrowvert_{t} +J^\mu \arrowvert_{u \Lambda} +J^\mu \arrowvert_{u \Sigma^{0}} + 
 J^\mu \arrowvert_{CT} .
\end{eqnarray}
The condition to fulfill gauge invariance is $q_{\mu} J^{\mu} = 0$, which gives
\begin{equation}\label{GI}
 q_{\mu} J^{\mu} = - \frac{D+F}{2\sqrt{3} f_{K}} \bar{u} (p^{\prime}) \left[(\slashed{p}_{k} F_{s} + (\slashed{p}^{\prime} - 
 \slashed{p})F_{t} - \slashed{q} F_{CT}) \gamma_{5} \right] u(p) .
\end{equation}
From the above equation, it is evident that the hadronic current is not gauge 
invariant. Therefore, in order to restore gauge invariance, the following term is added to Eq.~(\ref{GI})
\begin{equation}\label{GI1}
 q_{\mu} J^{\mu}_{add} = - \frac{D+F}{2\sqrt{3} f_{K}} \bar{u} (p^{\prime}) \left[ \slashed{p}_{k} \left(F_{CT} - F_{s} 
 \right) + (\slashed{p}^{\prime} - \slashed{p})(F_{CT} - F_{t})\right] \gamma_{5} u(p).
\end{equation}
Thus, the presence of the additional terms given in Eq.~(\ref{GI1}) implies that the gauge invariance can be achieved if 
the hadronic current $J^\mu$ is supplemented by adding an additional term $J^\mu_{add}$ given by 
\begin{equation}\label{GI2}
 J^{\mu}_{add} = - \frac{D+F}{2\sqrt{3} f_{K}} \bar{u} (p^{\prime}) \left[\frac{2 \slashed{p}_{k} p^{\mu}}{s - M^{2}} 
 (F_{CT} - F_{s}) + \frac{2p_{k}^{\mu}}{t - M_{k}^{2}}(\slashed{p} - \slashed{p}^{\prime}) (F_{CT} - F_{t}) \right] u(p).
\end{equation}
In order to take into account the effect of the form factor for the contact term, there are different prescriptions 
available in the literature, for example that of Ohta~\cite{Ohta:1989ji}, Haberzettl et al.~\cite{Haberzettl:1998aqi}, 
Davidson and Workman\cite{Davidson:2001rk}, etc. In the present work, we have followed the prescription of Davidson and 
Workman~\cite{Davidson:2001rk}, where $F_{CT}$ is given by:  
\begin{equation}\label{FF_CT}
 F_{CT} = F_{s}(s) + F_{t}(t) - F_{s}(s) \times F_{t}(t).
\end{equation}
In the case of associated particle production, we have considered six nucleon resonances exchanged in the $s$ channel, out 
of which four are spin $\frac{1}{2}$ {viz.} $S_{11}(1650)$, $P_{11}(1710)$, $P_{11} (1880)$, and $S_{11} (1895)$, and two are 
spin $\frac{3}{2}$ resonances {viz.} $P_{13} (1720)$, and $P_{13}(1900)$. We have already discussed the case of spin 
$\frac{1}{2}$ resonances in Section~\ref{sec:eta:photo}. However, for completeness, in this section we write the general 
form of the hadronic currents for the $s$ channel processes where a resonance state with spin $\frac{1}{2}$ is produced and 
decays to a kaon and a lambda in the final state as~\cite{Fatima:2020tyh}: 
\begin{eqnarray}\label{eq:res1/2_had_current}
j^\mu\big|_{R}^{\frac{1}{2}\pm}&=& 
-~ \bar u({p}\,') \frac{g_{RK \Lambda}}{f_{K}} 
 \slashed{p}_{K} \Gamma_{s} \frac{\slashed{p}+\slashed{q}+M_{R}}{s-M_{R}^2+ iM_{R} \Gamma_{R}} \Gamma^\mu_{\frac12 
 \pm} u({p}\,),
\end{eqnarray}
where $\Gamma_{R}$ is the decay width of the resonance, $\Gamma_{s} = 1(\gamma_{5})$ stands for the positive~(negative) 
parity resonances. $\Gamma_{\frac{1}{2}^{+}}$ and $\Gamma_{\frac{1}{2}^{-}}$ are, respectively, the vertex function for the 
positive and negative parity resonances, defined in Eq.~(\ref{eq:vec_half_pos}). $g_{R K 
\Lambda}$ is the coupling strength for the process $ R \to K \Lambda$, given in Table~\ref{Tab:Resonance:para}. For 
a more detailed discussion, see Ref.~\cite{Fatima:2020tyh}.

In the following, we briefly discuss spin $\frac{3}{2}$ resonance excitations. The general structure of the hadronic current 
for $N-R$ transition has already been discussed in Section~\ref{res:inelastic}. In the case of real photon scattering, the 
electromagnetic couplings $C_{i}^{p,n}$ are related to the helicity amplitudes $A_{\frac{1}{2}}, ~A_{\frac{3}{2}}$ and 
$S_{\frac{1}{2}}$, which are obtained using Eqs.~(\ref{helicity1})--(\ref{helicity3}) in the limit $Q^2=0$. In the numerical 
calculations, we have taken $S_{\frac{1}{2}} =0$ as we are dealing with the real photons. The fitted values of 
$A_{\frac{1}{2}}$ and $A_{\frac{3}{2}}$ have been taken from PDG~\cite{ParticleDataGroup:2020ssz} for spin $\frac{1}{2}$ and 
$\frac{3}{2}$ resonances, and are quoted in Tables~\ref{tab:param-p2} and \ref{tab:param-p20}. 

\begin{table}
\begin{center}
\begin{tabular*}{170mm}{@{\extracolsep{\fill}}ccccc}
\hline \hline
 Resonance & \multicolumn{4}{c}{Helicity amplitude} \\
   & $A_{\frac{1}{2}}^{p}$ ($10^{-3}$ GeV$^{-1/2}$)  & $A_{\frac{3}{2}}^{p}$  ($10^{-3}$ GeV$^{-1/2}$)  
   & $A_{\frac{1}{2}}^{n}$ ($10^{-3}$ GeV$^{-1/2}$)  & $A_{\frac{3}{2}}^{n}$  ($10^{-3}$ GeV$^{-1/2}$)\\
\hline
$P_{11}(1880)$ &  $ 21$ & -  &$-60$ &-\\
$S_{11}(1895)$ &  $ -16$ & - & 13 &- \\
$P_{13}(1900)$ &  $ 24$ & $-67$ & $0.7$ &$0.7$ \\
\hline \hline
\end{tabular*}
\end{center}
\caption{Values of the helicity amplitudes $A_{\frac{1}{2}}^{p,n}$ and $A_{\frac{3}{2}}^{p,n}$ for $P_{11}(1880)$, 
$S_{11} (1895)$, and $P_{13}(100)$ resonances taken from PDG~\cite{ParticleDataGroup:2020ssz}.}
\label{tab:param-p20}
\end{table}

The most general expression of the hadronic current for the $s$ channel spin $\frac{3}{2}$ resonance exchange may be 
written as~\cite{Fatima:2020tyh}:
\begin{eqnarray}\label{eq:res_had_current_pos:associated}
j^\mu\big|_{R}^{\frac32 \pm} &=& ie~ \frac{g_{RK \Lambda}}{f_{K}} 
   \frac{p_{K}^{\alpha}\Gamma_{s}}{s - M_R^2+ i M_R \Gamma_R}
   \bar u({p}\,') P_{\alpha\beta}^{3/2}(p_R) \Gamma^{\beta\mu}_{\frac32 \pm}
   u({p}\,),\quad p_R=p+q,
\end{eqnarray}
where $\Gamma_{s} = 1 (\gamma_{5})$ for positive~(negative) parity resonances, $f_{K}$ is defined in Section~\ref{coupling}, 
$g_{RK \Lambda}$ is the coupling strength for $R \to K \Lambda$ transition, the values of which are given in 
Table~\ref{Tab:Resonance:para}. $M_R$ is the mass of the resonance, $\Gamma_R$ is its decay width and $P_{\alpha 
\beta}^{3/2}(p_R)$ is given in Eq.~(\ref{propagator:32}). 

\begin{figure}  
\begin{center}
 \includegraphics[height=7cm,width=15cm]{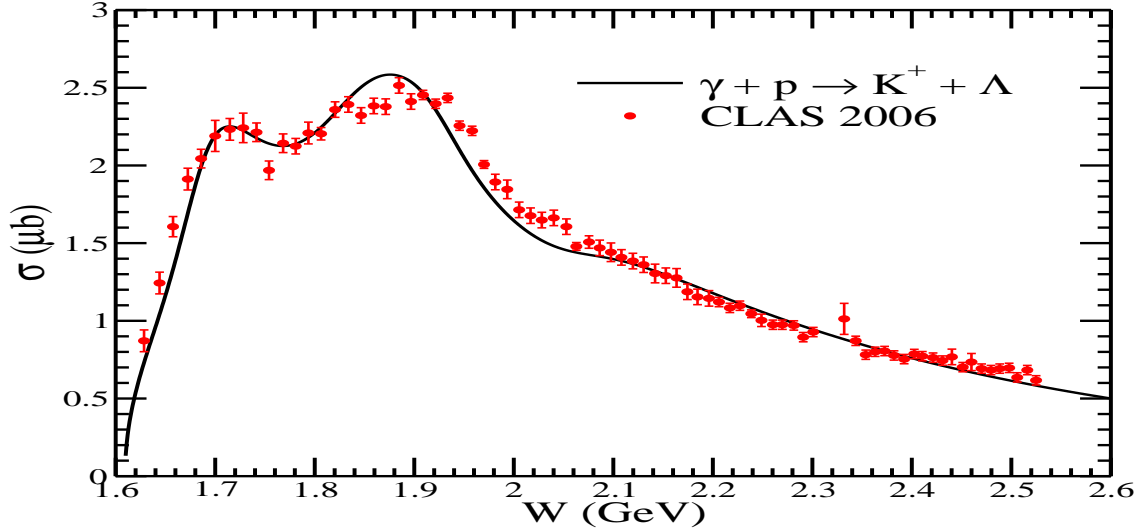}
\caption{Comparison of $\sigma$ {  vs.} $W$ for the process $\gamma + p \longrightarrow K^{+} + \Lambda$~(black solid 
line) calculated in our model with the experimental data taken from the CLAS 2006~\cite{CLAS:2005lui}~(solid circle).}
\label{fig_data}
\end{center}
\end{figure}
In Fig.~\ref{fig_data}, we have presented the results for the total scattering cross section as a function of $W$ for the 
photon induced $K\Lambda$ production. The theoretical calculations are compared with the 
experimental data from the CLAS experiment~\cite{CLAS:2005lui}. It may be observed from the figure that there is  good 
agreement of our results with the experimental data.

\subsubsection{Associated particle production induced by (anti)neutrinos}\label{sec:associated}
The study of the neutrino induced $\Delta S=0$ associated particle production processes provide an improved understanding 
of the basic symmetries of the SM, structure of the weak hadronic form factors, strange-quark content of the 
nucleon, coupling constants, etc. Moreover the kaon production through the associated production also constitutes a 
background in the proton decay searches i.e. $p \longrightarrow K \bar \nu$. Therefore, an understanding and reliable 
estimate of the cross sections for the neutrino induced kaon production contributing as the background event is important 
and has been emphasized~\cite{Solomey:2005rs, Mann:1986ht}. The experimental observations of the neutrino induced associated 
particle production processes were performed earlier at BNL~\cite{Baker:1981tx}, ANL~\cite{Barish:1974ye} and 
CERN~\cite{Erriquez:1977tr, Erriquez:1978pg, Deden:1975pa}. However, these experiments have very low statistics and 
large systematic errors. Attempts are being made to study them in the context of the present day neutrino experiments with 
high intensity $\nu(\bar{\nu})$ beams.

Theoretically, the early attempts were made by Shrock~\cite{Shrock:1975an}, Amer~\cite{Amer:1977fy}, Dewan~\cite{Dewan:1981ab}
and Mecklenburg~\cite{Mecklenburg:1976pk}. The associated particle 
production cross sections used for example in the NUANCE Monte Carlo generator~\cite{Casper:2002sd} consider only the 
resonant kaon production based on the Rein and Sehgal model for the pion production~\cite{Rein:1980wg}. Moreover, these 
cross sections miss the experimental data points by almost a factor of four~\cite{Datchev:2002}. Therefore, a better 
estimation of the weak interaction induced associated particle production cross section is needed.

\begin{figure}  
\centering
\includegraphics[height=6cm, width=18cm]{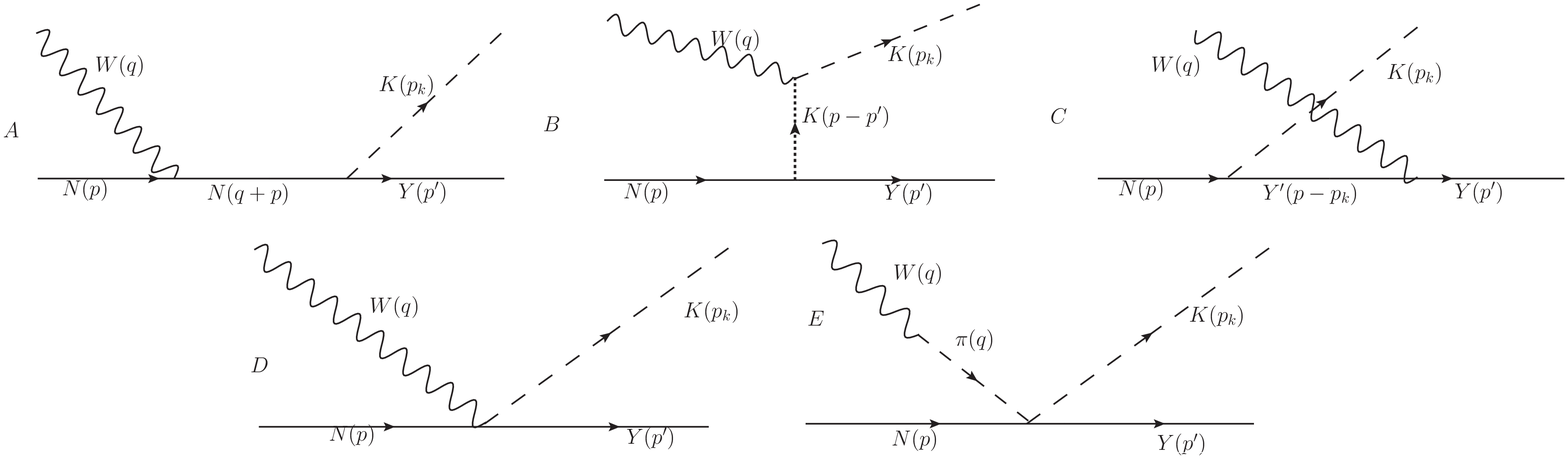}
\caption{Feynman diagrams corresponding to the (anti)neutrino induced $\Delta S=0$ associated particle production processes.}
\label{Ch12_fig:feyn_app}
\end{figure}

Here, the formalism for writing the hadronic current is the same as adopted in the case of pion and eta meson production 
processes discussed in Section~\ref{sec:1pion} and \ref{sec:eta}, respectively. The CC induced $\Delta S = 0$ 
processes are the following 
\begin{eqnarray}\label{Ch12_Eq_app:ccAssproduction}
 \nu_{l}(k) + N(p) \longrightarrow l^-(k^{\prime}) + Y(p^{\prime}) + K(p_{K}), \quad \quad
 \bar{\nu}_{l}(k) + N(p) \longrightarrow l^+(k^{\prime}) + Y(p^{\prime}) + K(p_{K}), \quad \text{where } l=e,\mu. 
\end{eqnarray}
For demonstrating the results, in this work we have focused only on the production of $K\Lambda$ induced by (anti)neutrinos, 
and the results for the other channels will be reported elsewhere~\cite{Fatima:2022associated}. We have considered the 
contribution of the NRB terms shown in Fig.~\ref{Ch12_fig:feyn_app} as well as from the isospin 
$\frac{1}{2}$ resonances exchanged in $s$-channel, as $\Lambda$ being an isoscalar particle, does not couple to the isospin 
$\frac{3}{2}$ resonances in order to conserve isospin at the strong vertex. We have taken only those resonances in the 
numerical calculations, which make significant contribution to the cross section for $W<2$~GeV.

The differential scattering cross section for the processes given in Eq.~(\ref{Ch12_Eq_app:ccAssproduction}) is given in 
Eq.~(\ref{eq:sigma_inelastic}) with $E_{m} =E_{K}$ and $\vec{p}_{m} = \vec{p}_{K}$, the outgoing kaon's energy and 
three-momentum, respectively, and $E_{p}^{\prime}$ is replaced with $E_{Y}$, the energy of the outgoing hyperon. The 
transition matrix element for the associated particle production process is given in Eq.~(\ref{eq:Gg}) with the leptonic 
current defined in Eq.~(\ref{lep_curr}). The contribution to the hadronic current $J^\mu$ comes from the different pieces of 
the Lagrangian corresponding to the Feynman diagrams shown in Fig.~\ref{Ch12_fig:feyn_app}. 

\begin{table}
\begin{center}
\centering
 \renewcommand{\arraystretch}{1.3}
\begin{tabular}{ccccccc} \hline \hline
 $A_{CT}$	            &$B_{CT}$		      &     $A_{SY}$
& \multicolumn{2}{c}{$A_{UY}$}                        & $A_{TY}$                     &  $A_{\pi }$ \\ 
                                                                       &                         &    
                                               & $Y^\prime=\Sigma$             & $Y^\prime=\Lambda$  &       
                                               &             \\     \hline
$-\frac{\sqrt{3}}{2f_{K}}$   & $\frac{-1}{3}(D+3F)$ & $\frac{D+3F}{2\sqrt{3}f_{K}}$ &  $\frac{\sqrt{2}}{3f_{K}}(D-F)$ & 0 & 
$\frac{D+3F}{2\sqrt{3}f_{K}}$ & -$\frac{\sqrt{3}}{4f_{K}}$ \\ \hline\hline
\end{tabular}
\caption{Constant factors  appearing in the hadronic current in Eq.~(\ref{Ch12_Eqapp:amplitude}). }
\label{Ch12_tb_app:currents}
\end{center}
\end{table}
In analogy with the weak pion production discussed in Section~\ref{sec:1pion}, the hadronic currents corresponding to the 
diagrams shown in Fig.~\ref{Ch12_fig:feyn_app} are obtained as:
\begin{eqnarray}\label{Ch12_Eqapp:amplitude}
j^\mu \arrowvert_{s} &=& ia~ A_{SY}   \; \bar u (p^\prime) \slashed{p}_K \gamma_5 
	    \frac{{p\hspace{-.5em}/} + {q\hspace{-.5em}/} + M}{(p+q)^2-M^2} 
	    \left[V^{\mu} - A^{\mu} \right] u(p) \nonumber \\
j^\mu \arrowvert_{u} &=& ia~ A_{UY} \; \bar u (p^\prime) \left[V^{\mu} - A^{\mu} \right]
	      \frac{{p\hspace{-.5em}/} - \slashed{p}_K + M_{\Sigma}}{(p - p_K)^2-M_{\Sigma}^2}  
	      \slashed{p}_K \gamma_5 u (p)\nonumber\\
j^\mu \arrowvert_{PF} &=& ia~ A_{TY} \; f_{PF}(Q^2)\; (M+M_\Lambda) \; \bar u (p^\prime) \gamma_5 \; u (p) \;\;
	    \frac{2 p_K^\mu - q^\mu }{(p-p^\prime)^2-m_K^2}\nonumber \\
j^\mu \arrowvert_{CT} &=& ia~ A_{CT} \; \bar u (p^\prime) 
		    \left[ \gamma^\mu f_{\rho}((q-p_{K})^2) + B_{CT} \; f_{CT}(Q^2)\;\gamma^\mu  \gamma_5 \right] u (p) \nonumber\\
j^\mu \arrowvert_{PP} &=& ia~ A_{\pi} \; f_{\rho}((q-p_{K})^2) \; \bar u (p^\prime) 
	      \left[{q\hspace{-.5em}/} + \slashed{p}_K\right] u(p) \frac{q^\mu}{q^2-m_\pi^2}
\end{eqnarray}
where,
\begin{eqnarray}
 V^\mu &=& f_1^{YY^{\prime}} (Q^{2}) \gamma^\mu + i \frac{f_2^{YY^{\prime}}(Q^{2})}{M+M^{\prime}} \sigma^{\mu \nu} q_\nu \\
 A^{\mu} &=& g_1^{YY^{\prime}}(Q^{2}) \gamma^\mu \gamma_{5} + g_3^{YY^{\prime}}(Q^{2})\frac{2 q^\mu}{M+M^{\prime}}\gamma_5
\end{eqnarray}
are the vector~($V^{\mu}$) and axial-vector~($A^{\mu}$) transition currents for $Y \leftrightarrows Y^\prime $ with $Y= 
Y^\prime \equiv$ nucleon and/or hyperon. The vector and axial-vector form factors $f_{1,2}^{YY^{\prime}} (Q^2)$ and $g_{1,
3}^{YY^{\prime}}(Q^2)$ are determined assuming the Cabibbo theory and the various symmetry properties of the weak hadronic 
current discussed in Section~\ref{qe_hyperon} and Appendix~\ref{Cabibbo}. The form factors $f_{CT}(Q^2)$, $f_{PF}(Q^2)$ and 
$f_{\rho}((q-p_{K})^2)$ are introduced in the contact, pion pole and pion in flight terms to taken into account the hadronic 
structure. It may be observed from the Feynman diagrams~(Fig.~\ref{Ch12_fig:feyn_app}) that the pion in flight term is purely 
vector in nature while the pion pole diagram is possible only with axial-vector current. In the case of contact term, the 
term associated with $B_{CT}$ represents the vector part of the weak hadronic current while the term with $\gamma^{\mu}$ is 
associated with the axial-vector part. CVC hypothesis imposes the following condition on the form factors $f_{CT}(Q^2)$ and 
$f_{PF}(Q^2)$, {  i.e.},
\begin{figure}[h]
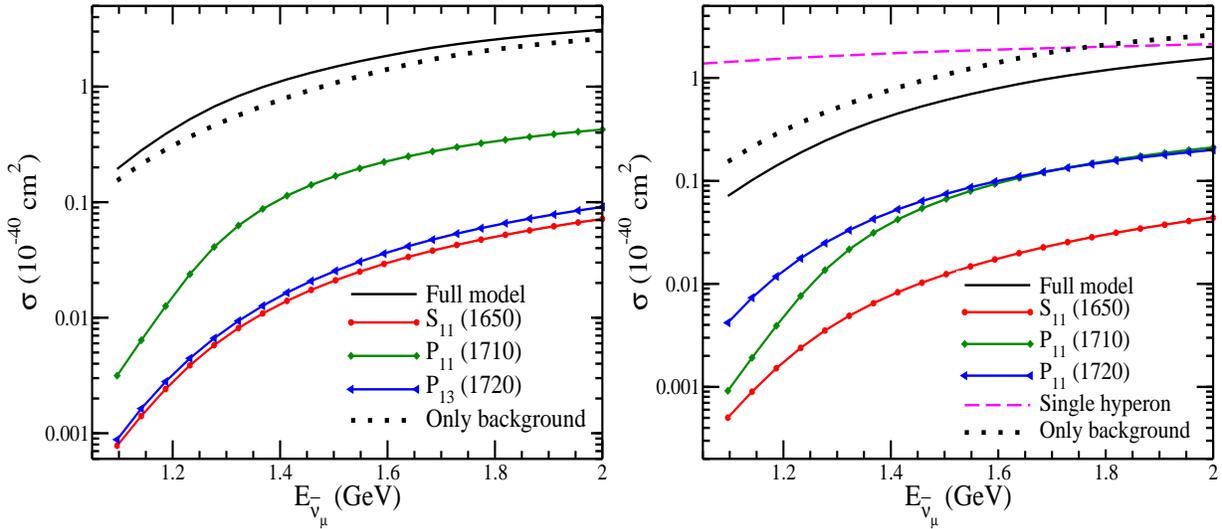
  
\centering
\includegraphics[height=7cm,width=8cm]{neutrino_associated_total.eps}
\includegraphics[height=7cm,width=8cm]{antineutrino_associated_total.eps}
\caption{Cross section for $\nu_{\mu} + n \longrightarrow \mu^{-} + \Lambda + K^{+}$~(left panel) and $\bar{\nu}_{\mu} + p 
\longrightarrow \mu^{+} + \Lambda + K^0$~(right panel) processes. Solid~(dotted) line represents the results of the full 
model~(only NRB terms), solid line with circle, diamond and left triangle, respectively, represents the 
individual contribution from $S_{11}(1650)$, $P_{11} (1710)$, and $P_{13} (1720)$ resonances. For comparison, in the right 
panel we have also shown the result of single hyperon production~(dashed line) induced by the $|\Delta S|=1$ reaction.}
\label{Ch12_fig:xsecap_nu}
\end{figure}
\begin{eqnarray}
 f_{CT} (Q^2) &=& f_{1} (Q^2) - 2F_{1}^{n} (Q^2) \; \left(\frac{D-F}{D+3F}\right) \; \left(\frac{u - M_{\Sigma} 
 M_{\Lambda} + MM_{\Sigma} - MM_{\Lambda}}{M_{\Sigma}^{2} - u} \right),\\
 f_{PF} (Q^2) &=& 2F_{1}^{n} (Q^2) \left(\frac{D-F}{D+3F}\right) \; \left(\frac{(M+M_{\Sigma})(u-M_{\Lambda}^2)}{(M + 
 M_{\Lambda})(M_{\Sigma}^2 - u)}\right) - f_{1} (Q^2),
\end{eqnarray}
where $u=(p-p_{K})^2$, $f_{1} (Q^2) = F_{1}^{p} (Q^2) - F_{1}^{n} (Q^2)$ is the vector form factor with $F_{1}^{p} 
(Q^2)$ and $F_{1}^{n} (Q^2)$ being the nucleon electromagnetic form factors, discussed in Section~\ref{nu_QE}. In analogy with 
the single pion production, the form factor $f_{\rho} (Q^2)$ corresponding to the axial-vector current is given by:
\begin{equation}
 f_{\rho}(Q^2) = \frac{1}{1+Q^2/m_{\rho}^2}; \qquad \qquad {\rm with } \; m_\rho = 0.776~\text{ GeV}.
\end{equation}
We have already discussed in Section~\ref{CC}, the excitation and decay of spin $\frac{1}{2}$ and $\frac{3}{2}$ resonances 
into a meson and a baryon in the final state. In this case of associated particle production, we have taken the contribution 
from both spin $\frac{1}{2}$ resonances like $S_{11} (1650)$, $P_{11} (1710)$, and spin $\frac{3}{2}$ resonances like 
$P_{13} (1720)$, in the numerical calculations. It should be noted that the helicity amplitudes for some of these 
resonances~($S_{11}(1650)$ and $P_{13}(1720)$) are given by the MAID parameterization~\cite{Tiator:2011pw}. While for $P_{11}
(1710)$, we have fitted the $Q^2$ dependence of the helicity amplitudes to the experimental data of Ref.~\cite{CLAS:2014fml}. 
In the case of spin $\frac{1}{2}$ resonances, the s-channel hadronic currents for the positive~($P_{11} (1710)$) and 
negative~($S_{11} (1650)$) parity resonances are given in Eq.~(\ref{eq:res1/2_had_current_pos}), with the explicit form of 
the vector and axial-vector form factors given 
in Eqs.~(\ref{eq:f12vec_res_12}) and (\ref{eq:f12vec_res_12_neg}) for the isospin $\frac{1}{2}$ resonances. Similarly in the case 
of positive parity spin $\frac{3}{2}$ resonances~($P_{13}(1720)$), the general expression of the hadronic current for the 
s-channel is given in Eq.~(\ref{eq:res_had_current_pos}). In this expression, the vector 
and axial-vector form factors used in the case of isospin $\frac{1}{2}$ resonances are given in Eqs.~(\ref{eq:civ_NC}) and 
(\ref{c5a-r}), respectively. ${\cal C^{R}}=\frac{g_{K\Lambda R}}{f_{K}}$ for both spin $\frac{1}{2}$ and $\frac{3}{2}$ 
resonances. Using the expression for the hadronic current given in Eq.~(\ref{Ch12_Eqapp:amplitude}) for the background terms 
and Eqs.~(\ref{eq:res1/2_had_current_pos}) and (\ref{eq:res_had_current_pos}) for the resonance excitations, the hadronic 
tensor $J^{\mu\nu}$ is obtained, which contracts 
with the leptonic tensor $L^{\mu\nu}$ to get the expression for the matrix element squared.

In Fig.~\ref{Ch12_fig:xsecap_nu}, we have presented the results for $\nu_{\mu} n \longrightarrow \mu^{-} \Lambda  
K^{+}$ and $\bar{\nu}_{\mu} p \longrightarrow \mu^{+} \Lambda K^0$ processes as a function of incoming (anti)neutrino 
energy. The results are presented for the full model as well as for the individual contribution from $S_{11}(1650)$, $P_{11} 
(1710)$ and $P_{13} (1720)$ resonances. For comparison, the results for the background contribution are also 
presented. In the case of neutrino induced associated particle production, there is constructive interference between the background 
and resonance terms in the entire range of neutrino energies considered in this work. The most dominant contribution among the resonances is of $P_{11} 
(1710)$, which is an order of magnitude smaller than the results obtained for the full model at $E_{\nu_{\mu}}=2$~GeV. While 
the contributions from $S_{11} (1650)$ and $P_{13} (1720)$ are almost comparable to each other and both are about 15 times 
smaller than the results of the full model.

However, in the case of antineutrino induced associated production, there is a destructive interference between the 
background and the resonance terms and the results obtained with the background terms only are almost two times the results of the full 
model in the entire antineutrino energy range. Among the resonances, in the low energy region~($E_{\bar{\nu}_{\mu}}<1.5$~GeV), the most 
dominant contribution is from $P_{13} (1720)$ resonance followed by $P_{11} (1710)$ and $S_{11}(1650)$ resonances. With the 
increase in antineutrino energy~($E_{\bar{\nu}_{\mu}}>1.6$~GeV), the results obtained from the individual contribution of 
$P_{13} (1720)$ and $P_{11} (1710)$ resonances overlap and are found to be an order of magnitude smaller than the results 
obtained using full model. As we have already discussed in Section~\ref{qe_hyperon}, the hyperons are also produced in the 
$\Delta S=1$ QE scattering of antineutrinos with the nucleon target, which are Cabibbo suppressed, while in the present 
case hyperons are produced in association with a kaon and are not Cabibbo suppressed. In view of this, we have 
compared the results of $\Lambda$ production in the $\Delta S=0$ associated particle production with the results of $\Lambda$ 
production in the $\Delta S=1$ QE process, and found that the $\Lambda$ production induced by the $\Delta S=1$ 
QE process dominates the associated particle production in the region of antineutrino energy $E_{\bar{\nu}_{\mu}} 
\lesssim 2$~GeV~(Fig.~\ref{Ch12_fig:xsecap_nu}).
\begin{figure*}
\begin{subfigure}[h]{0.5\textwidth}
\includegraphics[width=0.95\textwidth,height=.65\textwidth]{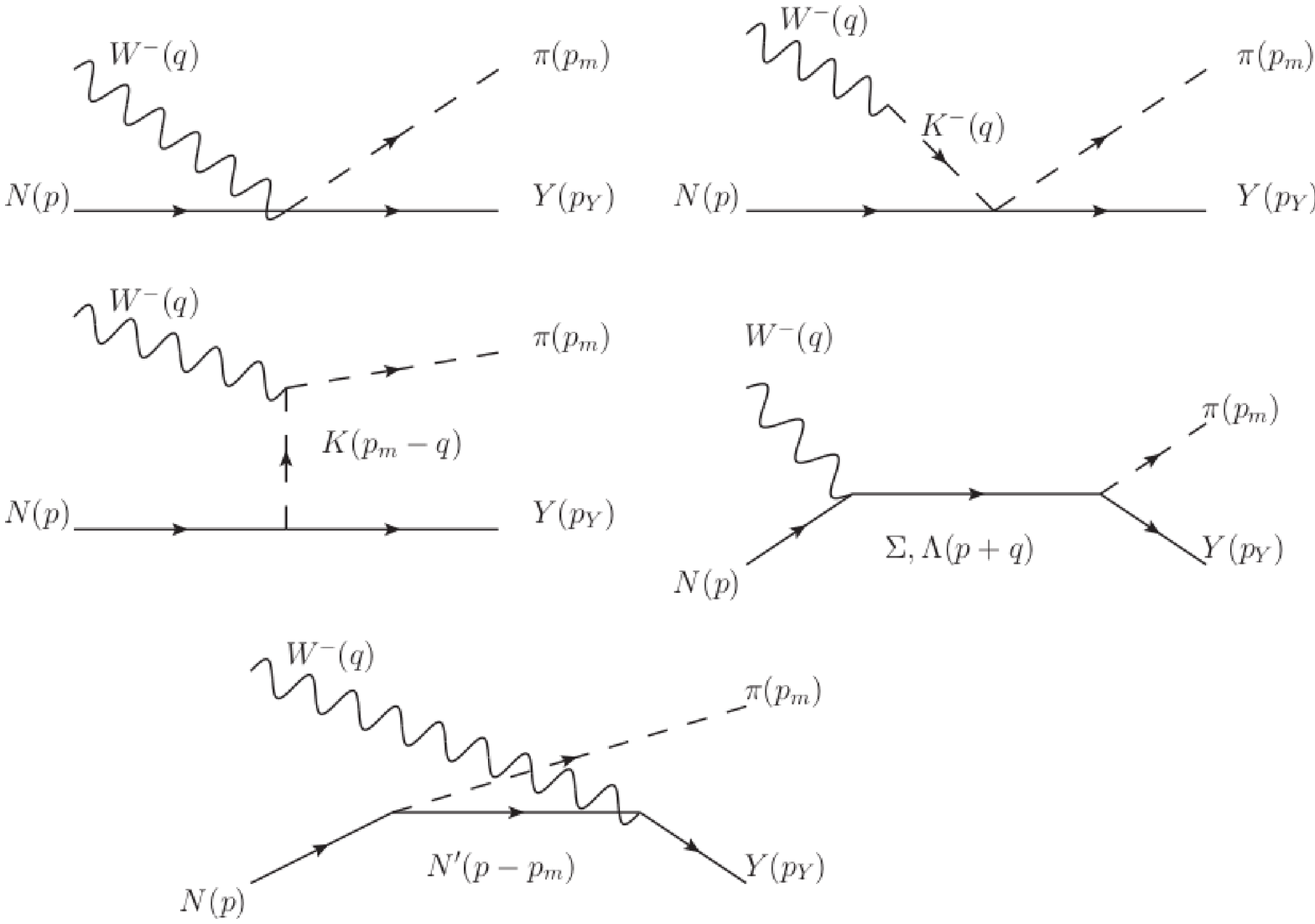}
\caption{From top to bottom and from left to right: the contact term (CT), 
the kaon pole (KP), the kaon-in-flight (KF), the s-channel $\Sigma$ and $\Lambda$ (s-$\Sigma$ and s-$\Lambda$) and the 
u-channel $N$ (u-$N$) diagrams.}\label{fig:background}
\end{subfigure}
\hspace{1cm}
\begin{subfigure}[h]{0.44\textwidth}
\includegraphics[width=0.95\textwidth,height=.65\textwidth]{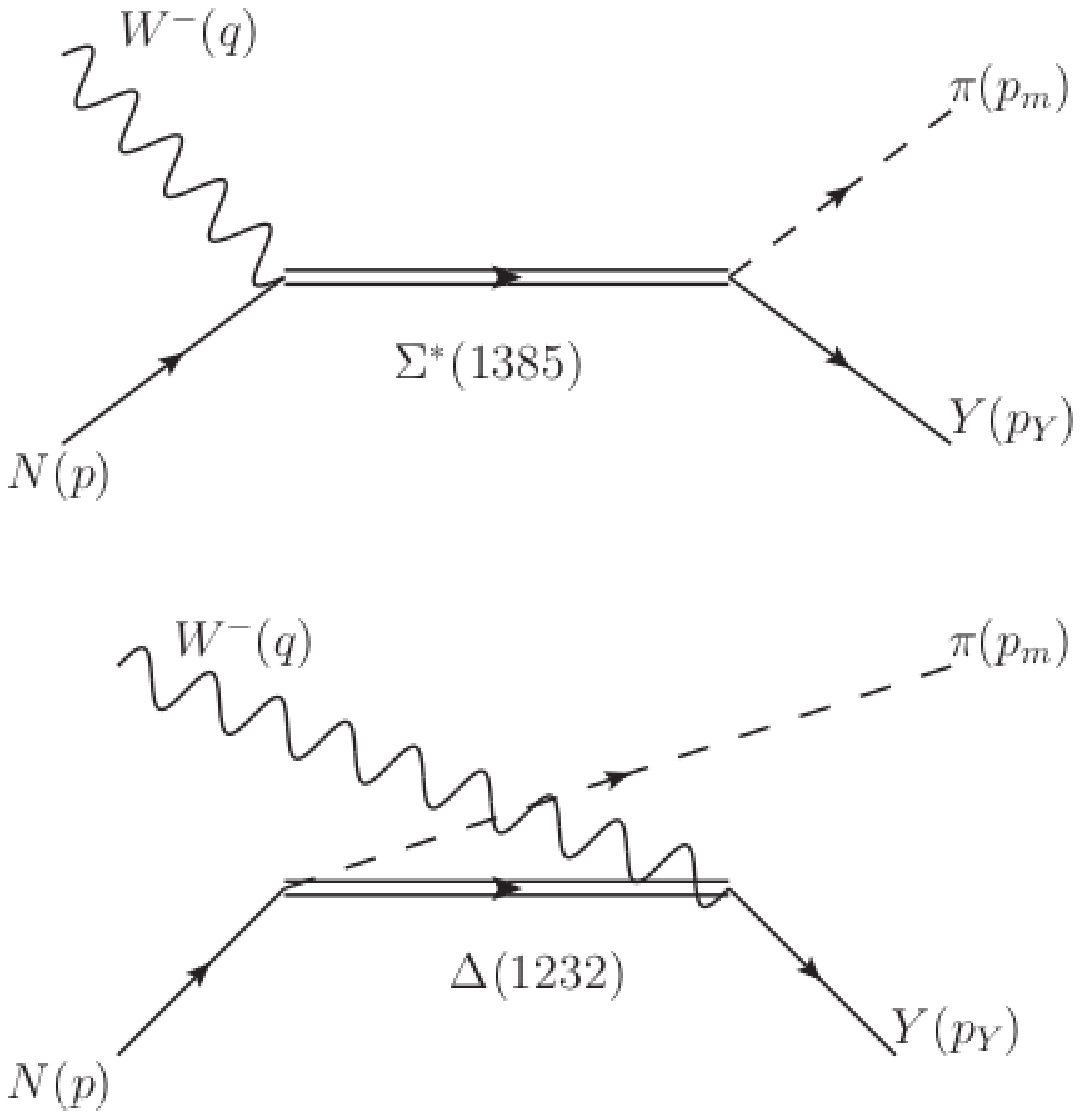}
\caption{Resonance diagrams included in the s-channel: $\Sigma^*(1385)$ and 
the u-channel: $\Delta(1232)$.}\label{fig:resonances}
\end{subfigure}
\caption{Feynman diagrams for the Cabibbo suppressed $\pi Y$ production process off nucleons induced by
antineutrinos~\cite{BenitezGalan:2021jdm}.}\label{fig: model}
\end{figure*}

\subsubsection{Single pion production with hyperon~($Y\pi$)}\label{sec:Ypi}
In antineutrino induced reactions, single pion can be produced along with a hyperon i.e.
\begin{eqnarray} \label{eq:reaction-Ypi}
 \bar{\nu}_l(k) + N(p) &\longrightarrow&  l^{+}(k^\prime) + \pi(p_\pi) + Y(p_Y),
\end{eqnarray}
where $N$ stands for a nucleon and $Y$ can be a $\Sigma$ or $\Lambda$ hyperon. The four-momenta of particles are given in
parentheses.  

These processes get contribution from the NR as well as resonance channels~(Fig.~\ref{fig: model}) specially from
$\Delta(1232)$ and $\Sigma^*(1385)$ resonances~(Fig.~\ref{fig:resonances}). Recently, Benitez Galan et 
al.~\cite{BenitezGalan:2021jdm} have studied such processes~(Eq.~(\ref{eq:reaction-Ypi})) where the hadronic matrix element 
are calculated using effective $V-A$ strangeness-changing weak CC with vector and axial-vector form factors for 
the $N-Y$ and $N-Y^\prime$ transitions. The vector and axial-vector form factors are determined in the same way as discussed in 
Section~\ref{sec:associated}, and the values of the coefficients $a$ and $b$ are tabulated in Table~\ref{APB:table3}.
\begin{table*} 
\centering
\begin{tabular}{|c|c|c|c|c|c|c|c|}
\hline
Reaction  & $\mathcal{A}^{N\longrightarrow Y\pi}_{\rm CT}$ & 
$a^{N\longrightarrow Y\pi}$ & $\mathcal{A}^{N\longrightarrow Y\pi}_{\rm KP}$ & 
$\mathcal{A}^{N\longrightarrow Y\pi}_{\rm KF}$ & 
$\mathcal{A}^{N\longrightarrow Y\pi}_{\rm s-\Sigma}$ & 
$\mathcal{A}^{N\longrightarrow Y\pi}_{\rm u-N^\prime}$ & 
$\mathcal{A}^{N\longrightarrow Y\pi}_{\rm s-\Lambda}$\\
\hline
$\bar{\nu}_l + p \longrightarrow l^{+} + \pi^0 + \Lambda$ & 
$\frac{\sqrt{3}}{2\sqrt{2}f_\pi}$ & $F + \frac{D}{3}$ & 
$-\frac{\sqrt{3}}{2\sqrt{2}f_\pi}$ &$-\frac{(D+3F)}{2\sqrt{6}f_\pi}$ & 
$\frac{D}{\sqrt{3}f_\pi}$ & $\frac{D+F}{2f_\pi}$ & 0\\
\hline
$\bar{\nu}_l + n \longrightarrow l^{+} + \pi^{-} + \Lambda$ & 
$\frac{\sqrt{3}}{2f_\pi}$ & $F + \frac{D}{3}$ & 
$-\frac{\sqrt{3}}{2 f_\pi}$ &$-\frac{(D+3F)}{2\sqrt{3}f_\pi}$ & 
$\frac{D}{\sqrt{3}f_\pi}$ & $\frac{D+F}{\sqrt{2}f_\pi}$ & 0\\
\hline
$\bar{\nu}_l + p \longrightarrow l^{+} + \pi^{0} + \Sigma^0$ & 
$\frac{1}{2\sqrt{2}f_\pi}$ & $F - D$ & 
$-\frac{1}{2\sqrt{2}f_\pi}$ &$\frac{(D-F)}{2\sqrt{2}f_\pi}$ & 
$0$ & $\frac{D+F}{2f_\pi}$ &$\frac{D}{\sqrt{3}f_\pi}$ \\
\hline
$\bar{\nu}_l + p \longrightarrow l^{+} + \pi^{-} + \Sigma^{+}$ & 
$\frac{1}{\sqrt{2}f_\pi}$ & $F - D$ & 
$-\frac{1}{\sqrt{2}f_\pi}$ &$\frac{(D-F)}{\sqrt{2}f_\pi}$ & 
$-\frac{F}{f_\pi}$ & $0$ &$\frac{D}{\sqrt{3}f_\pi}$ \\
\hline
$\bar{\nu}_l + p \longrightarrow l^{+} + \pi^{+} + \Sigma^{-}$ & 
$0$ & $0$ & $0$ &$0$ & $\frac{F}{f_\pi}$ & 
$\frac{D+F}{\sqrt{2}f_\pi}$ &$\frac{D}{\sqrt{3}f_\pi}$ \\
\hline
$\bar{\nu}_l + n \longrightarrow l^{+} + \pi^{-} + \Sigma^{0}$ & 
$-\frac{1}{2f_\pi}$ & $F-D$ & $\frac{1}{2f_\pi}$ &
$\frac{(F-D)}{2f_\pi}$ & $\frac{F}{f_\pi}$ & 
$\frac{D+F}{\sqrt{2}f_\pi}$ &$0$ \\
\hline
$\bar{\nu}_l + n \longrightarrow l^{+} + \pi^{0} + \Sigma^{-}$ & 
$\frac{1}{2f_\pi}$ & $F-D$ & $-\frac{1}{2f_\pi}$ &
$\frac{(D-F)}{2f_\pi}$ & $-\frac{F}{f_\pi}$ & 
$-\frac{D+F}{2f_\pi}$ &$0$ \\
\hline
\end{tabular}
\caption{Constants $\mathcal{A}^{N\longrightarrow Y\pi}_i$ and $a^{N\longrightarrow Y\pi}$ (for the axial-vector piece of 
the CT diagram) for each reaction and diagram shown in Fig.~\ref{fig:background}.}\label{tab:constants_diagrams}
\end{table*}

The matrix element of the hadronic currents for the NR Born diagrams shown in Fig~\ref{fig:background} are obtained 
as~\cite{BenitezGalan:2021jdm}:
\begin{eqnarray}
J^\mu_{\rm CT}&=&i\,a\,\mathcal{A}^{N\longrightarrow Y\pi}_{\rm CT}\; 
F_D(Q^2)\;\bar{u}(\vec{p}_Y) \left[ \gamma^\mu - a^{N\longrightarrow Y\pi}
\gamma^\mu \gamma_5 \right] u(\vec{p})\label{eq:currentCT},\\
J^\mu_{\rm KP}&=&-i\,a\,\mathcal{A}^{N\longrightarrow Y\pi}_{\rm KP}\; 
F_D(Q^2)\; \frac{q^\mu}{Q^2+m^2_K}\; 
\bar{u}(\vec{p}_Y) \left[\slashed{q} - 
\frac{\left(M_Y-M\right)}{2} \right]u(\vec{p}),\label{eq:currentKP}\\
J^\mu_{\rm KF}&=&i\,a\,\mathcal{A}^{N\longrightarrow Y\pi}_{\rm KF}\; 
F_D(Q^2)\; \frac{2p^\mu_{\pi} - q^\mu}{(p_{\pi}-q)^2-m^2_K}\; \left(M_Y+M\right)
\bar{u}(\vec{p}_Y) \gamma_5 u(\vec{p}),\label{eq:currentKF}\\
J^\mu_{\rm s-Y^\prime}&=&i\,a\,
\mathcal{A}^{N\longrightarrow Y\pi}_{\rm s-Y^{\prime}}\; 
\bar{u}(\vec{p}_Y)\slashed{p}_{\pi}\gamma_5 \;
\frac{\slashed{p}+\slashed{q} + M_{Y^\prime}}{(p+q)^2-M^2_{Y^\prime}}\;
\left[ V^\mu_{NY^\prime}(Q^2) - A^\mu_{NY^\prime}(Q^2) \right]u(\vec{p}),
\label{eq:current_sY}\\
J^\mu_{\rm u-N^\prime}&=&i\,a\,
\mathcal{A}^{N\longrightarrow Y\pi}_{\rm u-N^\prime}\; 
\bar{u}(\vec{p}_Y)\left[ V^\mu_{N^\prime Y}(Q^2) - 
A^\mu_{N^\prime Y}(Q^2) \right]
\frac{\slashed{p} - \slashed{p}_{\pi} + M}{(p-p_{\pi})^2-M^2}\;
\slashed{p}_{\pi}\gamma_5 u(\vec{p}),\label{eq:current_uN}
\end{eqnarray}
where $a=\sin\theta_{C}$, $Y,=\Sigma,\Lambda$; $Y^{\prime}=\Sigma^{\star}$; $N,N^\prime=p,n$; $F_D(q^2)$ is a global dipole form factor, taken as
\begin{equation}
F_D(Q^2)=\frac{1}{\left(1 + \frac{Q^2}{M^2_D} \right)^2}, \quad \qquad
M_D\simeq 1\; {\rm GeV}.
\end{equation}
for the CT, KP and KF diagrams.  In Eqs.~(\ref{eq:currentCT})--(\ref{eq:current_uN}), the $\mathcal{A}^{N\longrightarrow Y 
\pi}_i$ are global constants that depend on the particular reaction and are  given in Table~\ref{tab:constants_diagrams}.

The vector and axial-vector weak vertices of Eqs.~(\ref{eq:current_sY}) and (\ref{eq:current_uN}) are given by 
\begin{eqnarray*}
V^\mu_{NY^\prime}(Q^2)=f^{NY^\prime}_1(Q^2) \gamma^\mu + 
\frac{if^{NY^\prime}_2(Q^2)}{M+M_{Y^\prime}} 
\sigma^{\mu\nu} q_{\nu},\qquad \qquad
A^\mu_{NY^\prime}(Q^2)=g^{NY^\prime}_1(Q^2)\left( 
\gamma^\mu + \frac{q^\mu \slashed{q}}{Q^2+m^2_K}\right)\gamma_5,
\end{eqnarray*}
with the vector $f^{NY^\prime}_{1,2}(Q^2)$ and axial-vector $g^{NY^\prime}_1(Q^2)$ form factors, discussed in 
Section~\ref{sec:associated}.

\begin{table*} 
\centering
\begin{tabular}{|c|c|c|}
\hline
Reaction  & $\mathcal{A}^{N\longrightarrow Y\pi}_{\rm s-\Sigma^*}$ & 
$\mathcal{A}^{N\longrightarrow Y\pi}_{\rm u-\Delta}$\\
\hline
$\bar{\nu}_l + p \longrightarrow l^{+} + \pi^0 + \Lambda$ & 
$\frac{\mathcal{C}}{\sqrt{2}f_\pi}$ & $0$\\
\hline
$\bar{\nu}_l + n \longrightarrow l^{+} + \pi^{-} + \Lambda$ & 
$\frac{\mathcal{C}}{f_\pi}$ & $0$\\
\hline
$\bar{\nu}_l + p \longrightarrow l^{+} + \pi^{0} + \Sigma^0$ & 
$0$ & $2\sqrt{\frac23}\frac{\mathcal{C}}{f_\pi}$\\
\hline
$\bar{\nu}_l + p \longrightarrow l^{+} + \pi^{-} + \Sigma^{+}$ & 
$\frac{\mathcal{C}}{\sqrt{6}f_\pi}$ & $\frac{\mathcal{C}\sqrt{6}}{f_\pi}$\\
\hline
$\bar{\nu}_l + p \longrightarrow l^{+} + \pi^{+} + \Sigma^{-}$ & 
$-\frac{\mathcal{C}}{\sqrt{6}f_\pi}$ & $\sqrt{\frac23}\frac{\mathcal{C}}{f_\pi}$\\
\hline
$\bar{\nu}_l + n \longrightarrow l^{+} + \pi^{-} + \Sigma^{0}$ & 
$-\frac{\mathcal{C}}{\sqrt{3}f_\pi}$ & $-\frac{2\mathcal{C}}{\sqrt{3}f_\pi}$\\
\hline
$\bar{\nu}_l + n \longrightarrow l^{+} + \pi^{0} + \Sigma^{-}$ & 
$\frac{\mathcal{C}}{\sqrt{3}f_\pi}$ & $\frac{2\mathcal{C}}{\sqrt{3}f_\pi}$\\
\hline
\end{tabular}
\caption{Constants $\mathcal{A}^{N\longrightarrow Y\pi}_i$ for each reaction and the resonance~(s-$\Sigma^*$ and 
u-$\Delta$) diagrams shown in Fig.\ref{fig:resonances}.}\label{tab:constants_diagrams_resonances}
\end{table*}

\begin{figure}  
\centering
 \includegraphics[width=0.6\textwidth,height=6cm]{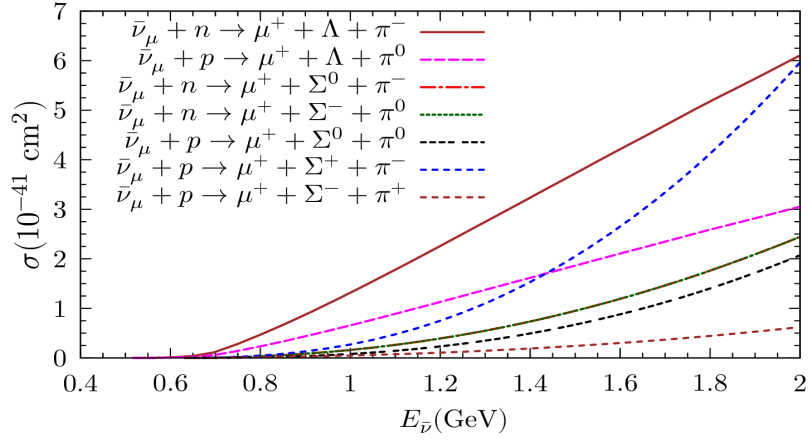}
\caption{Plot of the total cross sections for $Y\pi$ production off nucleons induced by muon antineutrinos as a function of 
the antineutrino energy in the laboratory frame. The figure is taken from Ref.~\cite{BenitezGalan:2021jdm}.}
\label{fig:numubar_total_xsect}
 \end{figure}

The $Y\pi$ states in the reaction induced by the antineutrinos can also be produced by exciting the $\Sigma^*$ and 
$\Delta$-resonances in the s and u channels. Since $\Sigma^*(1385)$ and $\Delta(1232)$ are members of the same decuplet, 
therefore, under the assumption of exact $SU(3)$ flavor symmetry for the couplings, the weak transition form factors 
connecting an octet state to a decuplet state can be obtained. We have already discussed the coupling of baryon decuplet and 
octet with mesons in Section~\ref{DEC}. The general structure of the hadronic current, the $N-R$ transition form factors, and 
the propagator for an intermediate baryon decuplet exchange is presented in Section~\ref{res:inelastic}.

The results for the total cross sections in case of the full model corresponding to all the possible $Y\pi$ channels induced 
by muon antineutrinos off nucleons as a function of the antineutrino energy in the laboratory frame are presented in 
Fig.~\ref{fig:numubar_total_xsect}.  It may be observed that the total cross sections have the same order of magnitude as
those of the single $K$ and $\bar{K}$ production ($1K/\bar K$) cross sections off nucleons studied in 
Refs.~\cite{RafiAlam:2010kf,Alam:2011vwg}.  While the $1K/\bar K$ cross sections are smaller than the single pion cross 
sections because of the smallness of the Cabibbo angle; the $Y\pi$ cross section misses the strong $\Delta(1232)$-like 
mechanism, apart from the threshold effect.

\subsubsection{Kaon production with $\Xi$ hyperon}\label{XK}
\begin{figure}  
\begin{center}
\includegraphics[width=0.55\textwidth,height=5cm]{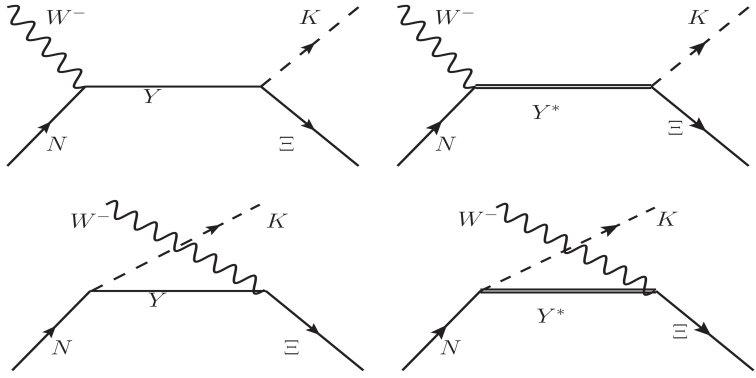}
\caption{Feynman diagrams for the $\Xi$ production. The intermediate states $Y$ are the ($S=-1$) $\Lambda,\Sigma$ hyperons, 
and $Y^*$ is $\Sigma^\ast(1385)$ resonance.}\label{fg:feynman}
\end{center}
\end{figure}

The K meson can also be produced in the antineutrino reactions accompanied by a $\Xi$ baryon through the reactions like
\begin{align}\label{eq:all_weak_ch}
    \bar \nu_\mu + N & \longrightarrow \mu^+ + K + \Xi 
\end{align}
In Fig.~\ref{fg:feynman} the Feynman diagrams that contribute to the matrix element of the hadronic current have been shown. 
Recently Alam et al.~\cite{RafiAlam:2019rft} have studied such processes, by considering NRB terms and 
$\Sigma^\ast(1385)$ resonance following the formalism discussed by us in Sections~\ref{NRB} and \ref{res:inelastic} for 
NRB and resonance contribution, respectively. The intermediate states contributing to this process are 
$Y=\Lambda, \Sigma$ baryons in the s and u channels as shown in Fig.~\ref{fg:feynman}.

The NRB terms have direct and cross diagrams, the corresponding matrix elements are calculated using the effective 
Lagrangian based on $SU(3)$ symmetry(Section~\ref{NRB}) and are given by
\begin{align}\label{eq:born_amp}
    j^\mu_{cc} \big \arrowvert_{sY} &=   
    \frac{i A_{s} a}{f_\pi}\;
    \bar u(\vec{p}^{\;\prime}_\Xi)\,  
    \slashed{p}_{K} \gamma^5  \frac{ 
    \slashed{p} + \slashed{q} +M_Y}{(p+q)^2 - M_Y^2}
    \Gamma^\mu_{N Y}\,  
    u(\vec{p}) , \nonumber \\          
    j^\mu_{cc} \big \arrowvert_{uY} &=  
    \frac{i A_{u} a}{f_\pi}\;
    \bar u(\vec{p}^{\;\prime}_\Xi)\, 
    \Gamma^\mu_{Y \Xi} \frac{ \slashed{p} - \slashed{p}_K + M_Y}{(p - 
    p_K)^2 - M_Y^2} \slashed{p}_{K} 
    \gamma^5\, u(\vec{p}) , \nonumber \\ 
\Gamma^\mu_{B_i B_j}(Q^2) 
&= f^{B_i B_j}_1(Q^2) \gamma^\mu 
+ i f^{B_i B_j}_2(Q^2) 
\frac{ \sigma^{\mu \nu}}{M_{B_i} + 
        M_{B_j}} q_\nu    - g^{B_i B_j}_1(Q^2) 
    \gamma^\mu  \gamma^5  - 
    g^{B_i B_j}_3(Q^2) \frac{2q^\mu}{M_{B_i} + 
        M_{B_j}}  \gamma^5 ,
\end{align} 
where $a= \sin \theta_C$. The 
weak vertex function $\Gamma^\mu_{B_i B_j}(Q^2)$ denotes the weak transition from baryon $B_i$ to $B_j$ and it is written in 
terms of transition vector $(f^{B_i B_j}_{1,2}(Q^2))$ and axial-vector $(g^{B_i B_j}_{1,3}(Q^2))$ form factors. The 
determination of these form factors has been discussed in detail in Section~\ref{qe_hyperon} and Appendix~\ref{Cabibbo}.

\begin{table} 
\begin{center}
\caption{Constant factors ($A_s$, $A_u$) in Eq.~(\ref{eq:born_amp}).}
\renewcommand{\arraystretch}{1.2}
\begin{tabular}{| l |c | c |c |  c |c |  c |c |   }\hline\hline
Process &     
\multicolumn{3}{|c|}{Direct 
term $(A_s)$ } &    
\multicolumn{3}{|c|}{Cross 
term $(A_u)$}  \\ 
        &  $Y=\Lambda$ & $Y=\Sigma$ & $Y=\Sigma^\ast$ & $Y=\Lambda$ & $Y=\Sigma$ & $Y=\Sigma^\ast$ \\ \hline
$\bar \nu_l + p \longrightarrow l^+ + K^+ + \Xi^-$ &  $ - \frac{D - 3F}{2 \sqrt3 }$ &  $ \frac{D +F}{2}$ & $\frac{1}
{\sqrt{6}}$ &$- \frac{D + 3F}{2 \sqrt3 }$ &   $ \frac{D-F}{2}$ & $\frac{1}{\sqrt{6}}$ \\        
$\bar \nu_l + p \longrightarrow l^+ + K^0 + \Xi^0$ &  $ - \frac{D - 3F}{2 \sqrt3 }$ & $-$ $\frac{D+F}{2}$ & $-\frac{1}
{\sqrt6}$ & 0 & $\frac{D-F}{\sqrt2}$ & 
$\sqrt{\frac{2}{3}}$ \\
$\bar \nu_l + n \longrightarrow l^+ + K^0 + \Xi^-$ &  
0 & $  \frac{D+F}{\sqrt2}$ & $\sqrt{\frac{2}{3}}$
& $ - \frac{D+3F}{2 \sqrt3 }$ & $ - \frac{D-F}{2}$ &  $-\frac{1}{\sqrt{6}}$  \\\hline 
\end{tabular}
\label{tb:coupling}
\end{center}
\end{table}

The couplings $A_s$ and $A_u$ in Eqs.~(\ref{eq:born_amp}) are obtained from the $SU(3)$ rotations at strong vertices of the
diagrams given in Fig.~\ref{fg:feynman} and are given in Table~\ref{tb:coupling}. The axial-vector couplings are used at 
the strong $BB^\prime K$ vertices. As in the case of $Y\pi$ production, the contribution of $\Sigma^{*} (1385)$ both in the 
s- and u-channels is also taken into account in the case of $K\Xi$ production. The details of the Lagrangian for strong 
vertex of the decuplet baryons with mesons and octet baryons are given in Section~\ref{DEC}, while for the weak vertex the 
details are given in Section~\ref{res:inelastic}.

\begin{figure}
\centering
\includegraphics[width=0.5\textwidth,height=6cm]{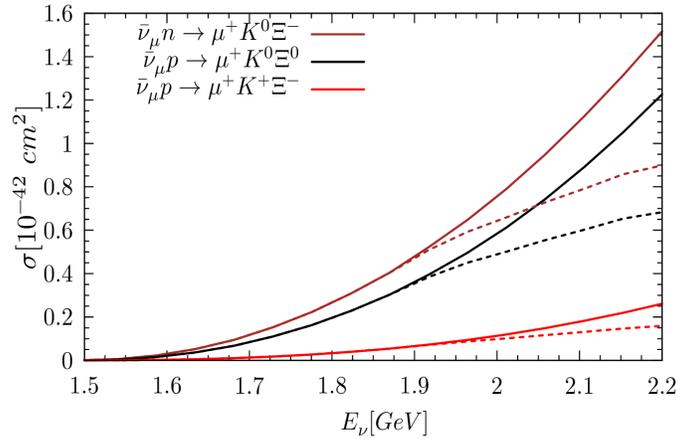}
\caption{Total cross section $\sigma$ vs. $E_\nu$ for the different channels of Eq.~(\ref{eq:all_weak_ch}). Dashed lines 
show the results with $W_{cut}=2.0$ GeV for each process (same color). The figure has been taken from 
Ref.~\cite{RafiAlam:2019rft}.}\label{fg:xsec_all}
\end{figure}

The results are presented for ${\bar\nu}_\mu$ induced total cross section in Fig.~\ref{fg:xsec_all}~\cite{RafiAlam:2019rft}. 
The full model results are shown by solid curves, while dashed lines show the results by applying a cut in the $K\Xi$ 
invariant mass of $W_{cut}=2$ GeV for the corresponding processes (identified by the same color). It is found that among the 
three channels, $ n \to K^0 \Xi^-$ is the most dominant one followed by $p \to K^0 \Xi^0$ and $p \to K^+ \Xi^-$, and these 
results are compared with the results for the inclusive kaon production~(Section~\ref{sec:associated}) with $\Delta S=0$ 
mechanisms and it has been observed that the cross section for $K^0$ and $K^+$  are about 3 and 6 percent of the 
corresponding $\Delta S=0$ processes, respectively~\cite{RafiAlam:2019rft}. This is in agreement with Cabibbo suppression 
for $|\Delta S|=1$ processes with respect to their $\Delta S=0$ counterparts.

\subsection{Two pion production}\label{sec:2pion}
There exists very few attempts to measure the two pion production induced by neutrinos and antineutrinos. Experiments done at
ANL~\cite{Barish:1978pj, Day:1983itd} and BNL~\cite{Kitagaki:1986ct} investigated the two pion production processes in the 
threshold region, in order to test the predictions of chiral symmetry.  Biswas et al.~\cite{Biswas:1978ey} used PCAC and 
current algebra methods to calculate the threshold production of two pions. Adjei et al.~\cite{Adjei:1980nj} made specific 
predictions using an effective Lagrangian incorporating the chiral symmetry. However, these models did not include any 
resonance production, and kept only terms up to ${\cal O}(1/f_\pi^2)$. 

In general the reaction for neutrino induced two pion production off the nucleon target can be written as
\begin{equation}
\nu_l(k) + N(p) \to l^-(k^\prime) + N(p^\prime) + \pi(k_{\pi_1})+\pi(k_{\pi_2})
\label{eq:reac} \, .
\end{equation}
In recent years, Hernandez et al.~\cite{Hernandez:2007ej} were the first one to study the two pion production using the 
effective Lagrangian given by the $SU(2)$ nonlinear $\sigma$ model discussed in Section~\ref{NRB}. This model provides 
expressions for the NR hadronic currents that couple with the lepton current, in terms of the first sixteen Feynman 
diagrams depicted in Ref.~\cite{Hernandez:2007ej} and includes the contribution from the Roper resonance~($P_{11}(1440)$) to 
the two pion production, which has significant coupling to the $2\pi$ channel.
 
 \begin{figure}
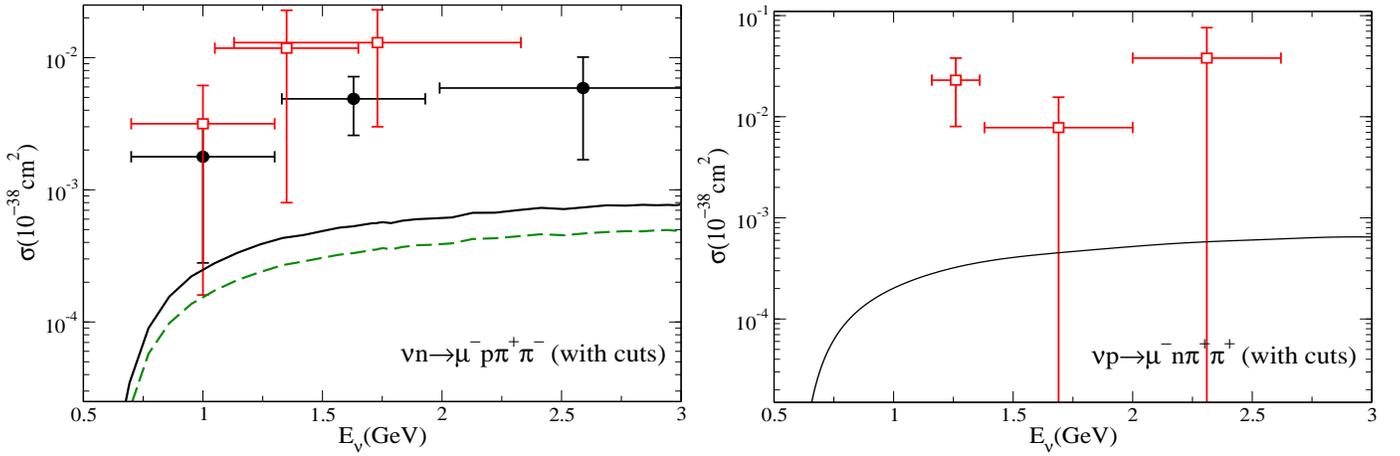
 
\centering
\includegraphics[width=0.49\textwidth,height=6cm]{Fig5_FF1.eps}
\includegraphics[width=0.49\textwidth, height=6cm]{Fig6.eps}
\caption{Cross section for the $\nu n\to \mu^- p \pi^+\pi^-$ (left panel) and $\nu p\to \mu^- n \pi^+\pi^+$ (right panel) 
processes with cuts as 
explained in Ref.~\cite{Hernandez:2007ej}. Dashed line: background terms. Solid line: full model with set FF1 of 
nucleon-Roper transition form factors. Data from Ref.~\cite{Kitagaki:1986ct}~(solid circles) and 
Ref.~\cite{Day:1983itd}~(open squares).\label{fig:5}}
\end{figure}

The results for the cross section for the $\nu n\to \mu^- p \pi^+\pi^-$ channel in the left panel of Fig.~\ref{fig:5} and for 
the channel $\nu p\to \mu^- n \pi^+\pi^+$ in the right panel are presented. 
Recently, Nakamura et al.~\cite{Nakamura:2015rta} have also studied two pion production in the DCC model in the CC 
induced reactions on proton and neutron targets and compared their results with ANL~\cite{Barish:1978pj, Day:1983itd} 
and BNL~\cite{Kitagaki:1986ct} data. These results show that more work both theoretically as well as experimentally are needed 
to understand the $2\pi$ production.

\subsection{NC $\nu(\bar{\nu})$ induced single photon production off the nucleon target}
The photon production can take place through the basic reactions on neutron and proton targets initited by (anti)neutrinos 
through the CC and NC processes i.e.
\begin{eqnarray}\label{singlephoton}
\nu_l + n \rightarrow l^- + p + \gamma, (CC)\nonumber\\
{\bar\nu}_l + p \rightarrow l^+ + n + \gamma, (CC) \nonumber\\
\nu_l({\bar\nu}_l) + N \rightarrow \nu_l({\bar\nu}_l) + N + \gamma. (NC)
\end{eqnarray}
 
In the case of nuclear targets the incoherent and coherent production of photons induced by the CC and NC interactions can take place.
 The one photon production induced by the NC processes has been recently discussed in considerable detail due to its possible relevance in 
 explaining the low energy excess of electron events observed in some neutrino oscillations experiments
 ~\cite{Efrosinin:2009zz, Hill:2009ek, Zhang:2012aka, Zhang:2012xi, Zhang:2012xn, Barbero:2012sb, Wang:2013wva}.
  In the oscillation experiments for the $\nu_e({\bar\nu}_e)$  appearance mode in $\nu_\mu({\bar\nu}_\mu)$ beam i.e. 
$\nu_\mu({\bar\nu}_\mu) \rightarrow \nu_e({\bar\nu}_e)$, for $E_{\nu({\bar\nu})}$ of $\sim$ 1GeV, detecting $e^-(e^+)$ via Cerenkov radiation in
CCQE scattering of  $\nu_e({\bar\nu}_e)$  has an important background arising due to the neutral current induced 1$\pi^0$ production, 
where a $\pi^0$ decays into two photons($\pi^0 \rightarrow \gamma\gamma$), and may give rise to overlapping rings or with one photon missing in the detection,
 resembles with an  $e^-(e^+)$ event. 

 The process of photon production in weak processes has been discussed quite early in case of radiative muon capture, for a
 review see Refs.~\cite{Bergbusch:1999ms, Cheoun:2003ah} and neutrino reactions~\cite{Gershtein:1980wu} using phenomenological effective Lagrangians.
 Gershtein et al.~\cite{Gershtein:1980wu} estimated the cross 
section for the NC reaction given in Eq.~(\ref{singlephoton}) by including the production through the exchange of virtual mesons like
$\pi^0$, $\omega^0$, $\rho^0$ and $f^0$. 
However, in recent calculations of the one photon production in neutrino reactions
 the approaches based on the phenomenological effective Lagrangians as well as the effective Lagrangians based on the chiral symmetry have been 
 used~\cite{Hill:2009ek, Zhang:2012aka, Barbero:2012sb, Wang:2013wva}, in 
order to understand the electron excess events in the MiniBooNE experiment~\cite{MiniBooNE:2008yuf} 
and also for the experiments which are being performed in the few GeV energy region like the T2K experiment. 
Hill et al.~\cite{Hill:2009ek}, and Zhang and Serot~\cite{Zhang:2012aka}, have calculated the one photon production cross section using
effective field theory, have
taken into account the 
contribution to the transition amplitude from nucleon pole~(NP and CNP), $\Delta(1232)$ pole~($\Delta$P and C$\Delta$P), 
and pion, $\sigma$, $\omega$, $\rho$ meson exchange in the t-channel. In addition, Zhang and Serot~\cite{Zhang:2012aka} 
have also 
considered contact terms arising due to symmetry. Wang et al.~\cite{Wang:2013wva, Wang:2015ivq} have used the model described here in Section-\ref{CC:pion}, 
where they considered NP and CNP, $\Delta$P and C$\Delta$P, pion exchange term in the t-channel, as well as the contributions from the higher 
resonances viz. $P_{11}(1440)$, $D_{13}(1520)$ and  $S_{11}(1535)$. They observed that the cross section is dominated by $\Delta$
production and its subsequent decay to $N\gamma$, similar to the observations made by earlier groups. They also find that
there is significant contributions from the NP, CNP nonresonant terms in the $\sim$ 1GeV 
energy region, and for $E_{\nu(\bar\nu)} > 1.5$GeV, the contribution from the $D_{13}(1520)$ resonance has been found to be significant.
Furthermore, with the same values of $C_5^A(0)$ and $M_{A}^\Delta$ (Eq.\ref{c5a-r}), all these 
results~\cite{Hill:2009ek, Zhang:2012aka,Zhang:2012xi, Zhang:2012xn, Barbero:2012sb, Wang:2013wva},  are consistent for $E_{\nu(\bar\nu)} \le $ 1.2GeV.
In the case of incoherent single photon production off nuclear targets, it has been found that when nuclear medium effects on the $\Delta$ properties
(Section~\ref{SPP:nucleus}) are taken into account, there is about 30$\%$ reduction in the cross section from the free nucleon case for
$E_{\nu(\bar\nu)} \sim 1$GeV~\cite{Wang:2013wva}. These calculations have also been performed for the coherent NC 1$\gamma$ production by 
these authors ~\cite{Wang:2013wva}, and observed that $\Delta$ contribution alone gives about 90$\%$ contribution to the total production cross section. 
These results are qualitatively in agreement with the results of Zhang and Serot~\cite{Zhang:2012xn} for $E_{\nu(\bar\nu)} < 1.5$GeV. 
Moreover, it was found that (anti)neutrino induced coherent NC$\gamma$ cross sections are (10)15$\%$ of the incoherent NC$\gamma$ cross sections.

%

\section{Deep inelastic scattering}\label{dis:nucleon}
\subsection{Introduction}
It is well known that with electrons of energy in the region of few hundreds of MeV, which corresponds to the de Broglie 
wavelength of { the virtual photons} being of the order of nuclear radius, the QE electron-nucleus scattering is used to study 
the structure of the nucleus specially its charge and magnetic moment distributions. With the increase in energy in the region 
of a few GeV, when the de Broglie wavelength becomes smaller, the electron scattering takes place from the nuclear 
constituents like the protons and neutrons and determines the charge and magnetic moment distributions of the nucleon~(nucleus) 
which are discussed in some detail in Section~\ref{para_FF}. These distributions are obtained in terms of the electromagnetic 
charge ($G_{E}(Q^2)$) and magnetic moment ($G_{M}(Q^2)$) form factors which are defined in terms of the deviation of the 
electron-nucleon~(nucleus) scattering cross sections from the Mott scattering cross sections corresponding to the point 
particles. In the elastic~(QE) scattering of electrons from nucleons~(nuclei), the nucleon~(nuclear) electromagnetic
($G_E^{N(A)}(Q^2)$ and $G_M^{N(A)}(Q^2)$) form factors depend upon only one independent kinematic variable chosen to be 
$Q^2(=-q^2\ge 0)$ due to the condition of the scattering being elastic~(QE) i.e. $(q+p)^2=p'^2$ which in the laboratory  
frame reduces to $Q^2=2M(E-E')$, where $M$ is the mass of the target nucleon(nucleus). $G_E^{N(A)}(Q^2)$ and $G_M^{N(A)} 
(Q^2)$ are generally characterized  by a steep fall with increase in $Q^2$ discussed in Section~\ref{para_FF}, and the radius 
of the nucleon~(nucleus) charge and magnetic moment distributions are obtained using the relation
 \begin{equation}
  \langle r^{2,charge}_{N(A)}\rangle=-6 \frac{d G_E^{N(A)}}{d Q^2}|_{Q^2=0},\;\;\;\; \langle r^{2,mag.~ mom.}_{N(A)}\rangle = 
  -\frac{6}{\mu} \frac{d G_M^{N(A)}}{d Q^2}|_{Q^2=0} .
 \end{equation}
    \begin{figure}[h]
  \begin{center}
  \includegraphics[height=4cm, width=0.35\textwidth]{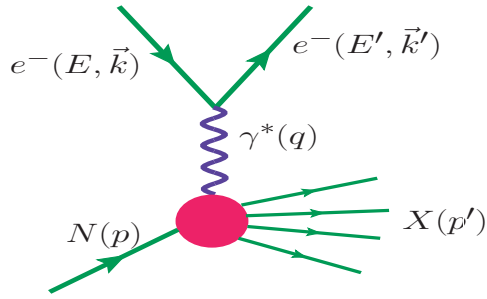}
      \end{center}
  \caption{Feynman diagram representing the electron-proton DIS process. }
  \label{fig:feynep}
  \end{figure}
With further increase in energy of the electrons, the de Broglie wavelength becomes very small which enables the electrons 
to probe deep into the composite structure of the nucleons. When the energy of the electrons is large enough to break the 
nucleons into the jet of hadrons the inelastic scattering takes place, as shown in Fig.~\ref{fig:feynep}, the process is known
as DIS. In these processes the energy-momentum conservation implies $(q+p)^2=p'^2=W^2$, and 
$q^2=(k-k')^2 \simeq -2 E E'(1-cos\theta)$, $\theta$ being the laboratory scattering angle. In case of the inclusive DIS no 
measurement is made on the final state hadrons $X$, while in the case of the exclusive IE scattering the excitation of 
nucleons to the definite resonance states $X$ as discussed in Section~\ref{nu_interaction} is studied. In these cases, the 
cross section is described in terms of the two kinematic variables i.e. energy ($E'$) and scattering angle ($\theta$) of the 
final state leptons or equivalently $Q^2$ and the energy transferred $\nu$ to the target, defined as 
\begin{equation*}
 \nu=E-E'=\frac{M_X^2+Q^2-M^2}{2M},~~M_X^2=W^2,
\end{equation*}
where $M_X=W$ is the mass of the hadronic system $X$. Out of these variables $\nu$, $Q^2$ and $W$ defining the kinematics of 
the DIS reactions, only two are linearly independent which are generally chosen to be $\nu$ and $Q^2$. The first set of DIS 
experiments with electron beams of different energies were done at SLAC in 1968 with the 20 GeV electron accelerator with 
$Q^2$ in the range of $1<Q^2<10$ GeV$^2$~\cite{Friedman:1972sy}. The first results on the cross sections were analyzed in 
terms of two functions $\nu W_2(\nu,Q^2)$ and $M W_1(\nu,Q^2)$ in analogy with the form factors in the case of elastic 
scattering and are called structure functions. These results were very surprising and led to a new understanding of the 
nucleon structure and its dynamic properties which complemented our knowledge of the nuclear structure obtained through the 
study of its static properties in terms of the quark model proposed by Gell-Mann~\cite{Gell-Mann:1964ewy} and 
Zweig~\cite{Zweig:1964jf}. Specifically the DIS experiments by Taylor, Friedman and Kendall~\cite{Friedman:1972sy, 
Friedman:1990ur, Taylor:1991ew, Kendall:1991np, Friedman:1991nq} showed that:
\begin{itemize}
 \item The DIS cross sections are an order of magnitude larger than the elastic cross sections from the proton target with a 
 very weak $Q^2$ dependence as shown in Fig.~\ref{fig:ep}~\cite{Friedman:1990}. This is indicative of the electron scattering 
 taking place not from the proton as a composite object but from its constituents which seem to be point particles without any 
 structure.
 
 \item The structure functions $\nu W_2(\nu,Q^2)$ and $M W_1(\nu,Q^2)$ do not depend upon the two variables $Q^2$ and $\nu$ 
 as expected but when studied as a function of $Q^2$ and $x=\frac{Q^2}{2 M \nu}$, the cross sections are almost independent 
 of $Q^2$ in the region of high $Q^2$ and depend only upon the variable $x$. This shows that $Q^2$ and $\nu$ dependence of the 
 cross sections scale, and there is dependence on only one variable $x$. Such behavior of the cross sections was theoretically 
 predicted by Bjorken assuming that the scattering takes place from the point like constituents of nucleons and the phenomenon 
 is called Bjorken scaling with $x$ as the Bjorken variable.
 
The physical interpretation of Bjorken scaling and the variable $x$ was given by Feynman who explained the DIS results in 
terms of a parton model in which electrons are assumed to scatter incoherently from point like constituents of nucleon called 
partons. The partons were later identified with quarks as proposed by Gell-Mann~\cite{Gell-Mann:1964ewy} and 
Zweig~\cite{Zweig:1964jf}. Therefore the model is popularly known as the quark-parton model. In the following sections we 
describe briefly the formalism of DIS in the quark-parton model in the region of high $Q^2$ and $\nu$, and various corrections 
needed to extend it to the region of lower $Q^2$.
\end{itemize}

\subsection{DIS of electrons from nucleons}
   \begin{figure}[h]
  \begin{center}
  \includegraphics[height=6 cm, width=0.75\textwidth]{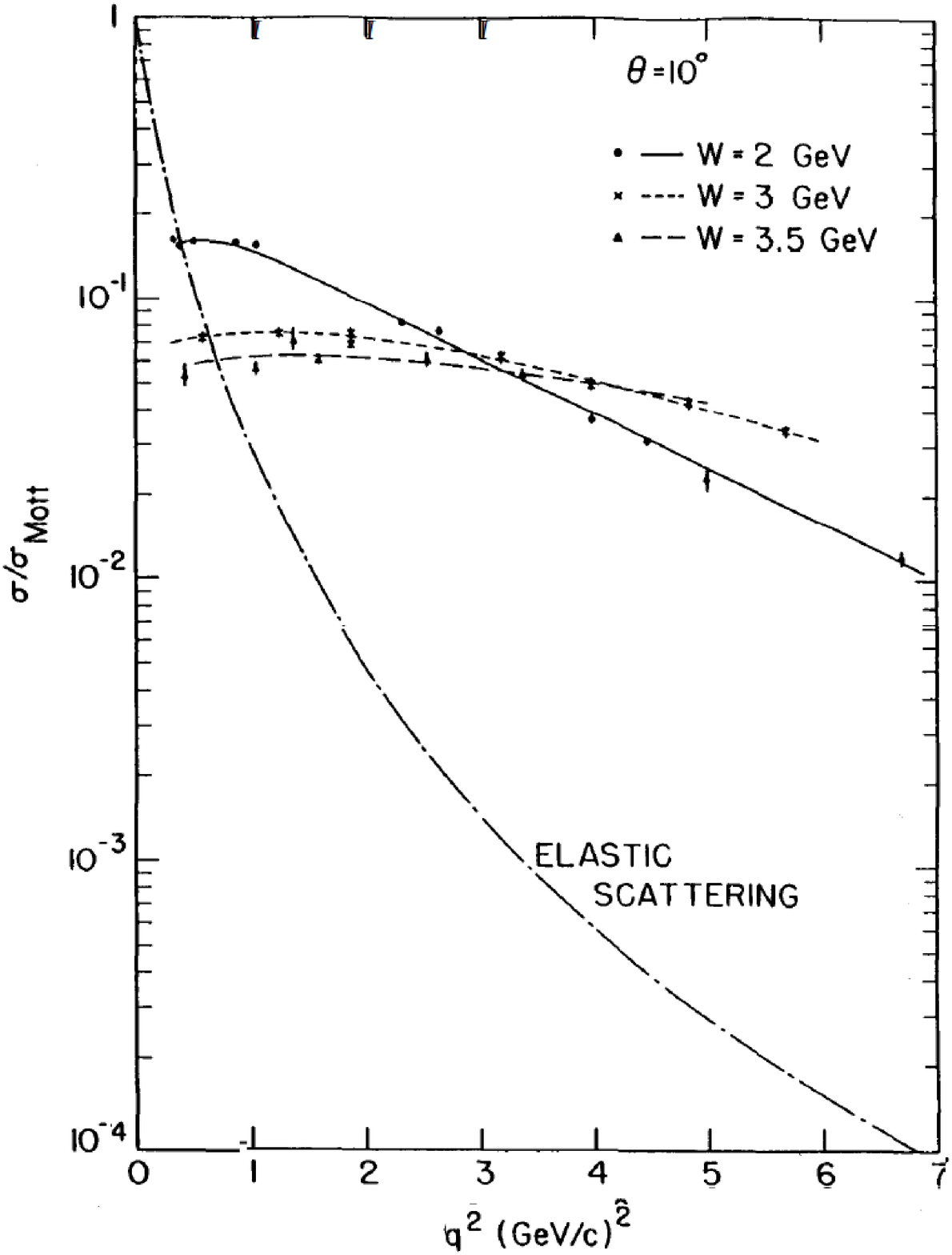}
      \end{center}
  \caption{Ratio of the double differential scattering cross section~($\sigma/\sigma_{Mott}$) vs $Q^2$ for the process 
  $e+p\rightarrow e+X$ at the 
  different values of CM energy $W$ and scattering angle of $10^0$. Figure
  has been taken from Ref.~\cite{Friedman:1990}. }
  \label{fig:ep}
  \end{figure}
 The matrix element for the DIS of electrons on nucleons corresponding to the Fig.~\ref{fig:feynep}, is written as
 \begin{equation}
  {\cal M}=-\left(\frac{e^2}{q^2}\right)\;l_\mu J^\mu,
 \end{equation}
where $l_\mu=\bar u(k')\gamma_\mu u(k)$, and $J^\mu(p,p')=\langle p|J_{em}^\mu|X\rangle$. The differential cross section 
$d\sigma$ for production of $X$ particles, summed over all $X$ in the laboratory frame is given by
\begin{eqnarray}\label{dxsig}
 d\sigma&=&\frac{1}{4 M E}\;\frac{d^3 k'}{(2\pi)^3 2 E'} \;L_{\mu\nu}\;J^{\mu\nu}, \qquad \qquad  \text{where} \\
 L_{\mu\nu}&=&\frac{1}{2}\sum_{spins}l_\mu^\dagger \;l_{\nu}=2(k_\mu k'_\nu+k_\nu k'_\mu-(k\cdot k'-m_l^2)g_{\mu\nu}), \\
 \label{hada}
 J^{\mu\nu}&=&\sum_N\;\frac{1}{2}\sum_s\int \prod_{n=1}^{N}\frac{d^3\;p'_n}{(2\pi)^3 (2E'_n)}\sum_{s_n}\langle p s|
 \tilde{J_\mu}^\dagger|X\rangle\;\langle X|J_\nu|ps\rangle\;(2\pi)^4 \;\delta^4(p+q-\sum_{n} p'_n)
\end{eqnarray}
where $N$ number of $X$ particles are produced. Since all the final hadronic momenta $p'_n$ are integrated and a sum over all 
the final hadrons are performed in Eq.~(\ref{hada}), the hadronic tensor $J^{\mu\nu}$ will depend only upon the momenta 
$q^\mu$ and $p^\nu$. However, for convenience of interpretation, we redefine the second rank tensor $J^{\mu\nu}$ in terms of 
$W^{\mu\nu}$ as $J^{\mu\nu}=4\pi MW^{\mu\nu}$ and construct the most general form for $W^{\mu\nu}$ using $p^\mu$, $q^\nu$ and 
$g^{\mu\nu}$ as
\begin{eqnarray}\label{hadtensor}
W^{\mu\nu}_N&=&-g^{\mu\nu}~W_{1N}^{EM}(\nu,Q^2)+\frac{p^\mu p^\nu}{M^2}~W_{2N}^{EM}(\nu,Q^2)-i\epsilon^{\mu\nu\lambda\sigma}
\frac{p_\lambda q_\sigma}{2 M^2}~W_{3N}^{EM}(\nu,Q^2)+\frac{q^\mu q^\nu}{M^2}~W_{4N}^{EM}(\nu,Q^2)\nonumber\\
&& +\frac{(p^\mu q^\nu+p^\nu q^\mu)}{M^2}~W_{5N}^{EM}(\nu,Q^2) +i \frac{(p^\mu q^\nu-p^\nu q^\mu)}{M^2}~W_{6N}^{EM}(\nu,
Q^2)\,,
\end{eqnarray}
$W_{iN}^{EM}(\nu,Q^2),~(i=1-6)$ are the nucleon structure functions which are functions of $\nu$ and $Q^2$. Since $L_{\mu\nu}$ 
is symmetric tensor, the terms involving $W_{3N}^{EM}(\nu,Q^2)$ and $W_{6N}^{EM}(\nu,Q^2)$ would not contribute in the 
electromagnetic interaction processes. The CVC at the hadronic vertex implies $q_\nu W^{\mu\nu}_N=q_\mu  W^{\mu\nu}_N=0$, 
which leads to the following relations 
 \begin{equation}
 \label{cvc}
 \left.
\begin{array}{l}
  W_{4N}^{EM}(\nu,Q^2)=\frac{M^2}{q^2}W_{1N}^{EM}(\nu,Q^2)+\left(\frac{p \cdot q}{q^2}\right)^2 W_{2N}^{EM}(\nu,Q^2),\;\;
  \textrm{and}\\ 
  W_{5N}^{EM}(\nu,Q^2)= \frac{- p \cdot q}{q^2}W_{2N}^{EM}(\nu,Q^2).
 \end{array}
 \right\}
 \end{equation}
Thus there are only two independent structure functions, which are generally chosen to be $W_{1N}^{EM}(\nu,Q^2)$ and 
$W_{2N}^{EM}(\nu,Q^2)$ and the expression of $W^{\mu\nu}_N$ is written in terms of these two structure functions as:
\begin{eqnarray}\label{nucleonht}
W^{\mu \nu}_N = 
\left( \frac{q^{\mu} q^{\nu}}{q^2} - g^{\mu \nu} \right) \;
W_{1N}^{EM}(\nu,Q^2) + \left( p^{\mu} - \frac{p . q}{q^2} \; q^{\mu} \right)
\left( p^{\nu} - \frac{p . q}{q^2} \; q^{\nu} \right)
\frac{W_{2N}^{EM}(\nu,Q^2)}{M^2}.
\end{eqnarray}
Contraction of $L_{\mu\nu}$ with $W^{\mu\nu}_N$ in the limit of massless lepton results
\begin{eqnarray}
L_{\mu\nu} W^{\mu \nu}_N &=&4W_{1N}^{EM}(\nu,Q^2)\left[-q^2\right] + 4 \frac{W_{2N}^{EM}(\nu,Q^2)}{M^2} \left[2 p \cdot k p 
\cdot k^\prime - M^2 k \cdot k^\prime\right].
\end{eqnarray}
Using the above equation in Eq.~(\ref{dxsig}), the expression for the differential scattering cross section is obtained 
as~\cite{Athar:2020kqn}:
\begin{eqnarray}\label{sigma_2}
\frac{d^2\sigma}{d\Omega' dE'}&=&
\frac{4 \alpha^2 E'^2}{Q^4} \left\{2\sin^2\left(\frac{\theta}{2}\right) 
W_{1N}^{EM}(\nu,Q^2) + \cos^2\left(\frac{\theta}{2}\right) {W_{2N}^{EM}(\nu,Q^2)} \right\}.
\end{eqnarray}
This is analogous to the expression for the differential scattering cross section $\frac{d^2\sigma}{d\Omega' dE'}$ for elastic 
scattering from a point particle like the $e \mu \rightarrow e\mu$ scattering~\cite{Athar:2020kqn}:
\begin{equation}\label{eqa}
 \left( \frac{d^2\sigma}{dE' d\Omega'}\right)_{e\mu \rightarrow e\mu}=\left(\frac{4\alpha^2 E'^2}{Q^4}\right)\left[\cos^2 
 \left(\frac{\theta}{2}\right)+\frac{Q^2}{2 m^2}\;\sin^2\left(\frac{\theta}{2}\right) \right]\;\delta\left(\nu-\frac{Q^2} 
 {2 m}\right),
\end{equation}
where $m$ is the muon mass.
Rewriting the expression given in Eq.~(\ref{eqa}) in terms of $W_{1N}^{EM}(\nu,Q^2) $ and $W_{2N}^{EM}(\nu,Q^2)$, we identify that
\begin{equation}\label{eqb}
 W_{2N}^{EM}(\nu,Q^2)=\delta\left(\nu-\frac{Q^2}{2 m}\right) \;\;\textrm{and}\;\; W_{1N}^{EM}(\nu,Q^2)=\frac{Q^2}{4 m^2}\;
 \delta\left(\nu-\frac{Q^2}{2 m}\right).
\end{equation}
Therefore, in case of the elastic electron scattering from spin 1/2 point particles, Eq.~(\ref{eqb}) implies that
\begin{equation}\label{eqc}
 \nu W_{2N}^{EM}(\nu,Q^2)=\delta\left(1-\frac{Q^2}{2 m \nu}\right) \;\;\textrm{and}\;\; m W_{1N}^{EM}(\nu,Q^2)=\frac{1}{2}\;
 \frac{Q^2}{2 m \nu}\;\delta\left(1-\frac{Q^2}{2 m \nu}\right) .
\end{equation}
It may be noticed that $\nu W_2(\nu,Q^2)$ and $m W_1(\nu,Q^2)$ which represent now point structure functions depend only upon 
the variable $\frac{Q^2}{2 m \nu} $ and not separately on $\nu$ and $Q^2$. Similar behavior of $\nu W_{2N}^{EM}(\nu,Q^2)$ and 
$m W_{1N}^{EM}(\nu,Q^2)$ in the case of DIS shows that the DIS of electrons from proton takes place from the point like 
constituents of the proton and not from the proton as a composite particle. It should be noted that this behavior of $\nu 
W_{2N}^{EM}(\nu,Q^2)$ and $m W_{1N}^{EM}(\nu,Q^2)$ in the case of electron muon scattering is due to the kinematics which in the case 
of DIS of electrons in $ep\rightarrow eX$ scattering is due to the dynamics of DIS as a result of electrons scattering from 
the point like constituents of the proton which leads to the phenomenon of scaling proposed by Bjorken and elaborated by 
Feynman as the parton model of DIS.

\subsection{Parton model of DIS}\label{forep}
Feynman proposed that it is convenient to visualize the DIS in an infinite momentum frame in which the electron scattering 
takes place from its constituents called partons. In this frame, parton motion is time dilated and hadron is Lorentz contracted 
as shown in Fig.~\ref{reacrate}. Moreover, the basic assumptions of the parton model are:
\begin{itemize}
 \item[i)] In an infinite momentum frame, a rapidly moving nucleon appears as a jet of partons, all of which travel more or 
 less in the same direction as that of the parent hadron such that the transverse momentum of the parton $p_T=0$.
 
 \item[ii)] The basic process of electron scattering takes place from free partons 
 to which all the energy $\nu=E-E'$ is transferred. The cross section is then summed incoherently over the contributions of 
 partons in the nucleon~(represented by the LHS of Fig.~\ref{reacrate}).
 
\begin{figure} 
\begin{center}
  \includegraphics[height=3.2 cm, width=15 cm]{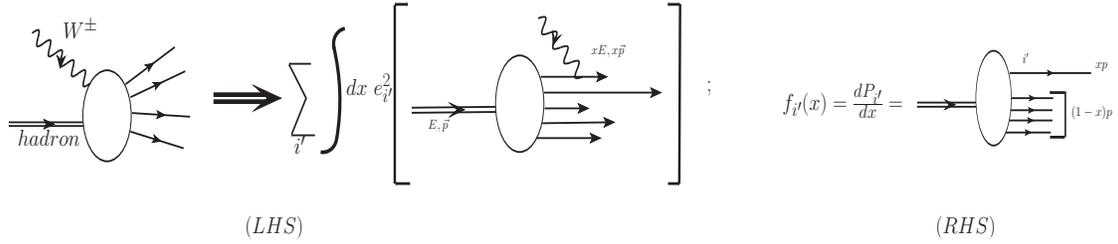}
  \end{center}
  \caption{Figure on the LHS depicts incoherent sum of the contributions and on the RHS represents momentum shared by the 
  charged partons. $E$, $p$ and $M$ are the energy, momentum and mass of the parent hadron. }\label{reacrate}
\end{figure}

 \item[iii)] The momentum and energy of the nucleon is shared among the partons such that for partons transverse momentum 
 $p_T=0$, longitudinal momentum $p_L=x p$, energy $E'=xE$, mass $m=(x^2 E^2-x^2 p^2)^{1/2}=xM$ , where $p,E$ and $M$ are the 
 momentum, energy and mass of the nucleon.
 \item[iv)] One defines the parton momentum distribution~(represented on the RHS of Fig.~\ref{reacrate}) as $f_{i}(x) = 
 \frac{dP_{i}}{dx}$, where $f_{i}(x)$ is the probability that the struck charged parton $i$ carries a fraction $x$ of the 
 nucleon's four momentum $p$.
 
 \item[v)] These partons carry a fraction $x$ of the nucleon's momentum and energy. All the fractions $x$ add up to 1 such 
 that
 \begin{equation}
 \sum_{i^\prime} \int dx~x~f_{i^\prime}(x)=1 , \nonumber
 \end{equation}
 where $i^\prime$ is sum over  the charged~(quarks) as well as the neutral~(gluons) partons in the nucleon.

 \item[vi)] The cross sections and the structure functions $\nu W_2(\nu,Q^2)$ and $M W_1(\nu,Q^2)$ are then calculated as an 
 incoherent sum of the cross sections from all the partons with momentum $xp$, energy $xE$ and mass $m=xM$ integrated over 
 $x$ and weighted with the momentum distribution $f_i(x)$ for each parton $i$. Consequently we write
\begin{eqnarray}
 \nu W_2^{ep}(\nu,Q^2)&=&\sum_i e_i^2 \int dx \;f_i(x)\;\nu W_2^{e p_i \rightarrow e p_i}(\nu,Q^2)=\sum_i e_i^2 \int dx\;f_i(x)
 \;\delta\left(1-\frac{Q^2}{2Mx\nu}\right),\nonumber
\end{eqnarray}
where $e_i$ is the charge of the parton $q_i$ in units of $|e|$.
\begin{eqnarray}
 \nu W_2^{ep}(\nu,Q^2)&=&\sum_i e_i^2 \int dx \;f_i(x)\;x\;\delta\left(x-\frac{Q^2}{2M\nu} \right)\nonumber\\
 \Rightarrow \nu W_2^{ep}(\nu,Q^2) \rightarrow F_2(x) &=&\sum_i e_i^2 x f_i(x),\;\;x=\frac{Q^2}{2M\nu},\nonumber
\end{eqnarray}
where expression for $\nu W_2$ is used from Eq.~(\ref{eqc}), in case of electron scattering from point particles. Similarly,
\begin{eqnarray}
 W_1^{ep}(\nu,Q^2)&=& \sum_i e_i^2 \int dx\;f_i(x)\;\frac{Q^2}{2 M \nu}\;\frac{1}{2 x}\;\delta\left(x-\frac{Q^2}{2M\nu}\right),\nonumber\\
 M W_1^{ep}(\nu,Q^2)\rightarrow F_1(x) &=& \sum_i e_i^2 \;f_i(x)\;x \;\frac{1}{2x}=\frac{1}{2x}\nu W_2^{ep}(\nu,Q^2),
 \nonumber\\
 \label{cgra}
 \textrm{i.e.}\; F_2(x)&=&\sum_i e_i^2 x f_i(x),\;\;\;F_1(x)=\frac{1}{2x}\;F_2(x),
\end{eqnarray}
which is known as the Callan-Gross relation~($F_{2} (x) = 2xF_{1}(x)$). The application of the parton model to DIS of electrons from proton leads to the 
following conclusions:
\begin{itemize}
 \item [(a)] The parton model reproduces the phenomenon of Bjorken scaling and the Bjorken variable $x=\frac{Q^2}{2M\nu}$ is 
 identified as the fraction of the proton momentum carried by the partons.
 
 \item [(b)] It is well known that in case of the electron scattering from a spin zero point particle, there is no $W_1 
 (\nu,Q^2)$ term implying $F_1(x)=0$ which is not true experimentally. Therefore, partons have nonzero spin.
 
 \item [(c)] The separation of electron scattering from nucleons into the longitudinal~($\sigma_L$) and 
 transverse~($\sigma_T$) components of virtual photon scattering from nucleons shows that in the limit $\nu \rightarrow \infty$, 
 $q^2\rightarrow \infty$, with $x=\frac{Q^2}{2M\nu}$ fixed
 \begin{equation*}
  \sigma_L \rightarrow 0
 \end{equation*}
 in the case of the partons having spin 1/2. This is confirmed experimentally implying that the partons have spin 1/2, thus 
 identifying them with the quarks. 
 
 \item [(d)] Using Eq.~(\ref{cgra}), one writes
 \begin{eqnarray}
 \label{eqab1}
  \frac{1}{x}\;F_2^{ep}(x)&=&\frac{4}{9}\Big(u_p(x)+\bar u_p(x) \Big)+\frac{1}{9}\Big(d_p(x)+\bar d_p(x) \Big) +\frac{1}{9}
  \Big(s_p(x)+\bar s_p(x) \Big),\\
   \label{eqab2}
   \frac{1}{x}\;F_2^{en}(x)&=&\frac{4}{9}\Big(u_n(x)+\bar u_n(x) \Big)+\frac{1}{9}\Big(d_n(x)+\bar d_n(x) \Big) +\frac{1}{9}
   \Big(s_n(x)+\bar s_n(x) \Big),
 \end{eqnarray}
 where $u_p(x)$ and $\bar u_p(x)$ are the probability distributions of $u$ quarks and antiquarks within the proton. The 
 isospin invariance implies that $u_p(x)=d_n(x)=u(x)=u_v(x)+u_s(x)$, $d_p(x)=u_n(x)=d(x)=d_v(x)+d_s(x)$, where $q_{v,s}$ for 
 each quark are the valence and sea quarks. Assuming a symmetric sea i.e. all the sea quark constituents have similar 
 distribution i.e. $u_s(x)=\bar u_s(x)=d_s(x)=\bar d_s(x)=s_s(x)=\bar s_s(x)=S(x)$ (say), results
\begin{eqnarray}
 \label{eqab3}
 \frac{1}{x}F_2^{ep}(x)=\frac{1}{9}\;[4 u_v(x)+d_v(x)]+\frac{4}{3}S, \qquad \qquad
  \frac{1}{x}F_2^{en}(x)=\frac{1}{9}\;[ u_v(x)+4 d_v(x)]+\frac{4}{3}S
\end{eqnarray}
Eqs.~(\ref{eqab1})--(\ref{eqab3}) predict the following relations:
\begin{itemize}
 \item Neglecting sea quark contributions
 \begin{equation}
  \frac{F_2^{en}}{F_2^{ep}}=\frac{u_v+4 d_v+\frac{4}{3}S}{4 u_v+ d_v+\frac{4}{3}S} \;\;\Rightarrow\;\;\frac{1}{4}\;\le\; 
  \frac{F_2^{en}}{F_2^{ep}} \;\le4.
 \end{equation}
 lower (upper) limits due to the dominance of $u_v(d_v)$ independent of the value of $x$. If sea quarks dominates, then the 
 ratio would be 1. These predictions have been confirmed experimentally.
 
 \item Using Eq.~(\ref{eqab3}) for proton and neutron
\begin{eqnarray}
 \frac{1}{x}\;\Big(F_2^{ep}(x)- F_2^{en}(x)\Big) = \frac{1}{3} [u_v(x)-d_v(x)],
\end{eqnarray}
leading to 
\begin{equation}
 \int \frac{dx}{x}\;\Big( F_2^{ep}(x)- F_2^{en}(x) \Big)=\frac{1}{3},
\end{equation}
when there is contribution from the valence quarks only i.e., without their sea quark partners, then the peak should occur at 
$x=1/3$, if there are two valence $u$ quarks and one $d$ quark inside the proton and two valence $d$ quarks and one $u$ quark 
inside the neutron, which was found to be true in the experimental data from SLAC~\cite{Friedman:1972sy}.

\item Defining $\epsilon_q=\int_0^1 x\;(q+\bar q)\;dx,$ for $u$ and $d$ quarks and neglecting strangeness $\epsilon_s$ 
component, we get
\begin{eqnarray}
 \int dx F_2^{ep}(x)=\frac{4}{9} \epsilon_u +\frac{1}{9}\;\epsilon_d=0.18, \qquad \qquad
 \int dx F_2^{en}(x)=\frac{1}{9} \epsilon_u +\frac{4}{9}\;\epsilon_d=0.12,\nonumber
\end{eqnarray}
$0.18$ and $0.12$ are the experimentally observed values~\cite{Friedman:1972sy} resulting $\epsilon_u+\epsilon_d=0.54$, which 
implies that only $54\%$ of the momentum is carried by the valence quarks. The remaining fraction of the momentum is carried 
by the gluons as the momentum fraction carried by the strange quarks is small. This indicates significant participation of gluons 
and sea quarks in the DIS specially in the low $x$ region. This leads to the violation of Bjorken scaling. Moreover, with electron 
scattering from gluons, the QCD effects which describe the quark-gluon interactions also come into play to modify the 
predictions of QPM. These are discussed in the following sections in the context of neutrino-nucleon scattering.
\end{itemize}

\end{itemize}
\end{itemize}

\subsection{$\nu$--N scattering in the DIS region}
\begin{figure} 
\begin{center}
\includegraphics[height=4.5 cm, width=14 cm]{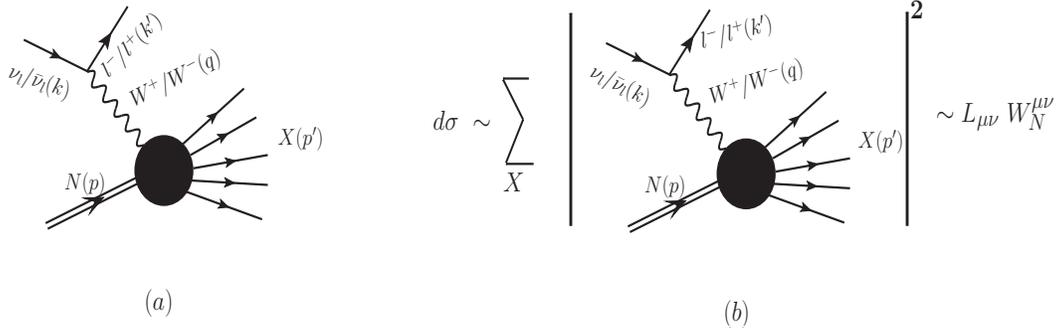}
\end{center}
\caption{ $\nu_l({\bar\nu}_l) - N$ inclusive scattering where the summation sign represents the sum over all the hadronic 
states such that the cross section~($d\sigma$) for the DIS  $\propto L_{\mu \nu} W_{N}^{\mu \nu}$. $X$ 
represents the jet of hadrons in the final state.}\label{fey_nun}
\end{figure}
The general expression of the double differential scattering cross section for CC induced $\nu_l 
({\bar\nu}_l)-N$ scattering in the laboratory frame corresponding to the reaction:
\begin{eqnarray}\label{reaction}
\nu_l(k) / \bar\nu_l(k) + N(p) \rightarrow l^-(k') / l^+(k') + X(p');~~~~~~~~~~~ l=e, \mu, \tau,
\end{eqnarray}
 shown in Fig.~\ref{fey_nun} is given by
 \begin{equation}
\label{eq:q2nu}
{ d^2\sigma_N^{WI} \over d\Omega' dE'} =  \frac{1}{2\pi^2 }~\frac{|{\vec{k}^\prime}|}{|{ \vec{k}}|}\;   \overline{\sum} \sum 
|{\cal M}|^2 \;.
\end{equation}
For CC induced process, the matrix element square i.e. $\overline{\sum} \sum |{\cal M}|^2$ in Eq.~(\ref{eq:q2nu}), 
averaged over the initial spin states and summed over the final spin states, is given in terms of the leptonic ($L_{\mu\nu}$) 
and hadronic ($ W^{\mu\nu}_N$) tensors as
\begin{equation}\label{amp_wk}
\overline{\sum} \sum |{\cal M}|^2 = \frac{G_F^2}{2}~\left(\frac{M_W^2}{Q^2+M_W^2}\right)^2 ~L_{\mu\nu} ~W^{\mu\nu}_N,
\end{equation}
where $G_F$ is the Fermi coupling constant and $M_W$ is the mass of the intermediate vector boson $W^\pm$. The leptonic tensor 
$L_{\mu \nu}$ is given in Eq.~(\ref{lep_tens}). In the case of DIS, the hadronic final state is unknown, 
therefore, the hadronic tensor $W^{\mu\nu}_N$ is written to parameterize our ignorance of the hadronic current. The most 
general form of the hadronic tensor is constructed by using the available four vectors at the disposal of hadronic vertex, 
i.e., the metric tensor $g^{\mu\nu}$, four momentum $p^{\mu}$ and the four momentum transfer $q^\mu$. 
Using the expression for the leptonic tensor $L_{\mu \nu}$ from Eq.~(\ref{lep_tens}) and the hadronic tensor $W^{\mu\nu}_N$ 
from Eq.~(\ref{hadtensor}), the expression of the differential cross section in terms of the nucleon structure functions 
$W_{iN}(\nu,Q^2);~(i=1-3)$ for the case of massless lepton ($m_l\to 0$) is obtained as~\cite{Athar:2020kqn}:
\begin{eqnarray}\label{xsec_q2nu}
 \frac{d^2\sigma}{d\Omega'\;dE'}&=&\frac{G_F^2 E'^2 \cos^2\left(\frac{\theta}{2}\right)}{2\pi^2\left(1+\frac{Q^2}{M_W^2}
 \right)^2}\;\Big[2 \tan^2\left( \frac{\theta}{2}\right) W_{1N}(\nu,Q^2)+ W_{2N}(\nu,Q^2)\pm \left(\frac{E+E'}{M}\right)
 \tan^2\left( \frac{\theta}{2}\right) W_{3N}(\nu,Q^2)\Big].
\end{eqnarray}
In the case of massive lepton, all the five structure functions (as mentioned in Eq.~(\ref{hadtensor})) would contribute 
$W_{iN}^{WI} (\nu,Q^2)~(i=1-5)$, while the contribution of the term with $W_{6N} (\nu, Q^2)$ vanishes when contracted with the 
leptonic tensor.

The scattering cross section in terms of the Bjorken scaling variable $x$ and inelasticity 
$y=\frac{\nu}{E_\nu}$ is expressed as
\begin{eqnarray}
 \frac{d^2\sigma}{dxdy}&=&\frac{G_F^2E_\nu Q^2}{2\pi x(1+\frac{Q^2}{M_W^2})^2}
 \Big\{\Big[y^2x+\frac{m_l^2 y}{2E_\nu M}\Big]\frac{M}{\nu}W_{1N}(x,Q^2)+
 \Big[\Big(1-\frac{m_l^2}{4E_\nu^2}\Big)-\Big(1+\frac{Mx}{2E_\nu}\Big)y\Big]W_{2N}(x,Q^2)\nonumber\\
 &\pm& 2 \Big[xy\Big(1-\frac{y}{2}\Big)-
 \frac{m_l^2 y}{4E_\nu M}\Big]W_{3N}(x,Q^2)
 +\frac{m_l^2(m_l^2+Q^2)}{4E_\nu^2M^2 }W_{4N}(x,Q^2)-\frac{m_l^2}{E_\nu M}W_{5N}(x,Q^2)\Big\}.\;\;\;~~~~~
\end{eqnarray}
Following the same analogy as discussed in Section~\ref{forep}, the weak nucleon structure functions $W_{iN}^{WI} (\nu,Q^2)~
(i=1-5)$ are written in terms of the dimensionless nucleon structure functions $F_{iN}^{WI}(x,Q^2)~(i=1-5)$ as: 
 \begin{eqnarray}\label{ch2:relation}
M W_{1N}(\nu,Q^2) &=& F_{1N}(x,Q^2)  ,\;\; \frac{Q^2}{2xM}W_{2N}(\nu,Q^2)=F_{2N}(x,Q^2),\;\;
 \frac{Q^2}{xM}W_{3N}(\nu,Q^2)  = F_{3N}(x,Q^2),\nonumber\\
\frac{Q^2}{2M}W_{4N}(\nu,Q^2) &=& F_{4N}(x,Q^2),\;\; 
 \frac{Q^2}{2xM}W_{5N}(\nu,Q^2)= F_{5N}(x,Q^2),\;\;\;
\end{eqnarray}
which leads to the following expression of the differential scattering cross section~\cite{Ansari:2020xne}:
\begin{eqnarray}\label{xsec:dis}
 \frac{d^2\sigma}{dxdy}&=&\frac{G_F^2ME_\nu}{\pi(1+\frac{Q^2}{M_W^2})^2}
 \Big\{\Big[y^2x+\frac{m_l^2 y}{2E_\nu M}\Big]F_{1N}(x,Q^2)+
 \Big[\Big(1-\frac{m_l^2}{4E_\nu^2}\Big)-\Big(1+\frac{Mx}{2E_\nu}\Big)y\Big]F_{2N}(x,Q^2)\nonumber\\
 &\pm& \Big[xy\Big(1-\frac{y}{2}\Big)-
 \frac{m_l^2 y}{4E_\nu M}\Big]F_{3N}(x,Q^2)
 +\frac{m_l^2(m_l^2+Q^2)}{4E_\nu^2M^2 x}F_{4N}(x,Q^2)-\frac{m_l^2}{E_\nu M}F_{5N}(x,Q^2)\Big\},
\end{eqnarray}
where $M$ is the mass of the target nucleon and $m_l$ is the mass of the final state charged lepton. $x$ and $y$ are the 
scaling variables which lie in the following ranges:
\begin{eqnarray}
\frac{m_l^2}{2M (E_\nu - m_l)} &\le& x \le 1;~~~a-b \le y\le a+b, \qquad \qquad \text{with} \\
 a&=&\frac{1-m_l^2\Big(\frac{1}{2ME_\nu x}+\frac{1}{2E_\nu^2} \Big)}{2\Big(1+\frac{M x}{2E_\nu}\Big)},\;\;
 b=\frac{\sqrt{\left(1-\frac{m_l^2}{2 M E_\nu x}\right)^2-\frac{m_l^2}{E_\nu^2}}}{2\Big(1+\frac{M x}{2E_\nu}\Big)}.
\end{eqnarray}
In general, the dimensionless nucleon structure functions are derived in the quark-parton model assuming Bjorken scaling and 
are functions of only one variable $x$. In this model, these structure functions obey Callan-Gross~\cite{Callan:1969uq} and 
Albright-Jarlskog~\cite{Albright:1974ts} relations, respectively, given by
\begin{eqnarray}
 F_{1}(x)&=&\frac{F_2(x)}{2 x}\;;\;\;
 F_{5}(x)=\frac{F_2(x)}{2 x}. \nonumber
\end{eqnarray}
At the leading order of perturbative QCD, the structure functions are derived in terms of the parton distribution functions 
$q_i(x)$ and $\bar q_i(x)$ as:
\begin{eqnarray}\label{parton_wk}
F_{2}(x)  &=& \sum_{i} x [q_i(x) +\bar q_i(x)] \;;\;~~~~~~~~
x F_3(x) =  \sum_i x [q_i(x) -\bar q_i(x)]\;;\;\;
F_4(x)=0.
\end{eqnarray} 
Generally, the proton ($F_{2,3}^{p}(x)$) and the neutron ($F_{2,3}^{n}(x)$) structure functions are obtained in the four 
flavor quark scheme, assuming that the heavy quark flavors ($b$ and $t$) do not contribute as they are massive in 
comparison to the nucleon's mass ($M << ~m_b ~\text{or}~ m_t$), as:
\begin{eqnarray}
\label{f2ep}
F_2^{\nu_l p}(x)&=& 2 x [d(x) + s(x) + \bar u(x) + \bar c(x) ]\;,\;
F_2^{\bar\nu_l p}(x)= 2 x [ u(x) + c(x) + \bar d(x) + \bar s(x) ],\nonumber\\
F_2^{\nu_l n}(x)&=& 2 x [ u(x) + s(x) + \bar d(x) + \bar c(x) ]\;,\;
F_2^{\bar\nu_l n}(x)= 2 x [ d(x) + c(x) +\bar u(x) + \bar s(x) ],\nonumber\\
x F_3^{\nu_l p}(x)&=& 2 x[d(x)+s(x)-\bar u(x)-\bar c(x)]\;,\;
xF_3^{\bar\nu_l p}(x)= 2 x [u(x)+c(x)-\bar d(x)-\bar s(x)],\nonumber\\
x F_3^{\nu_l n}(x)&=& 2 x[u(x)+s(x)-\bar d(x)-\bar c(x)]\;,\;
xF_3^{\bar\nu_l n}(x)= 2 x [d(x)+c(x)-\bar u(x)-\bar s(x)].\nonumber
\end{eqnarray}
In the above expressions, $x u(x)$ represents the probability of finding an up quark with the target nucleon's momentum 
fraction $x$ and similarly for other quark flavors. These probability distributions are also known as the parton distribution 
functions (PDFs). These PDFs for the nucleon have phenomenologically been determined by various groups 
and they are known in the literature by the acronyms MRST~\cite{Martin:1998sq}, GRV~\cite{Gluck:1998xa}, 
GJR~\cite{Gluck:2007ck}, MSTW~\cite{Martin:2009iq}, ABMP~\cite{Alekhin:2016uxn}, ZEUS~\cite{ZEUS:2002xjx},
HERAPDF~\cite{Zhang:2015tuh}, NNPDF~\cite{DelDebbio:2007ee}, CTEQ~\cite{Nadolsky:2008zw}, CTEQ-Jefferson 
Lab~(CJ)~\cite{Accardi:2009br}, MMHT~\cite{Harland-Lang:2014zoa}, etc. 

For an isoscalar nucleon target, the structure functions are defined for nucleon  $F_{iN}$ as:
\begin{equation}
 F_{iN}=\frac{F_{ip}+F_{in}}{2},~~~(i=1-5).
\end{equation}
The neutrino scattering experiments performed at CERN using heavy liquid bubble chamber Gargamelle also provided some 
conclusive evidence in support of the quark-parton model~\cite{Morfin:2021ujm}:
\begin{itemize}
 \item For an isoscalar nucleon target, the ratio of the $F_{2}(x)$ structure functions in electron and neutrino scattering in the 
 parton model depends only on the quark charges, i.e.,
\begin{equation}\label{emwc}
 \frac{{1 \over 2}\;\int\;[F_{2}^{\nu p}(x)+ F_{2}^{\nu n}(x)]dx}{{1 \over 2}\;\int\;[F_{2}^{e p}(x)+ F_{2}^{e n}(x)]dx}=
 \frac{2}{e_u^2+e_d^2}=\frac{18}{5},
\end{equation}
where the strange quark contribution has been neglected, and $e_u$ and $e_d$ are respectively the electric charges of the $u$ 
and $d$ quarks. The observed value for this ratio was $3.4\pm 0.7$~\cite{GargamelleNeutrino:1974exc}. This test provided the 
most convincing evidence that the nucleons are made up of the quarks which have fractional electric charge as real dynamical 
entities.

\item  In the quark-parton model for the point-like constituents, it was observed that the total (anti)neutrino scattering 
cross section is proportional to the energy in CM frame which was verified by the Gargamelle 
collaboration~\cite{GargamelleNeutrino:1974exc}.

\item The Gross-Llewellyn Smith sum rule~\cite{Milton:1998ct} in the quark-parton model states that:
\begin{equation}
 \int_0^1 F_{3N}(x) dx=\int_0^1[u(x)-\bar u(x)+d(x)-\bar d(x)] dx=3.
\end{equation}
which was reported by the Gargamelle collaboration to be $3.2\pm 0.6$~\cite{GargamelleNeutrino:1974exc}. 
\end{itemize}

\begin{figure}
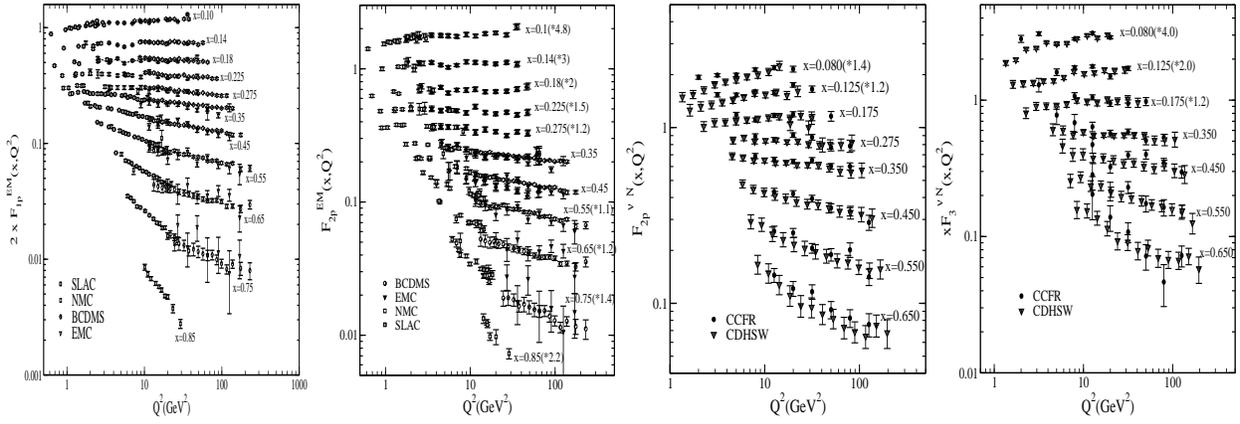
 
\begin{center}
 \includegraphics[height=5.5 cm, width=4 cm]{xf1.eps}
 \includegraphics[height=5.5 cm, width=4 cm]{f2_em_bodek.eps}
 \includegraphics[height=5.5 cm, width=4 cm]{f2.eps}
 \includegraphics[height=5.5 cm, width=4 cm]{xf3.eps}
 \end{center}
\caption{Experimental results of nucleon structure functions for electromagnetic~\cite{Whitlow:1991uw, Benvenuti:1989rh, 
Arneodo:1996rv, Aubert:1985fx, NewMuon:1995zeb} and weak~\cite{Oltman:1992pq, Berge:1989hr} interaction induced DIS
processes.} \label{f1f2}
 \end{figure}

In the lowest order of perturbative QCD, the partons are treated as free, noninteracting constituents of nucleon, but the 
partons present inside the nucleon may interact among themselves via the gluon exchange which can be described using QCD. The 
incorporation of contribution from the gluon emission induces the $Q^2$ dependence of the nucleon structure functions, leading to 
the violation of Bjorken scaling. The $Q^2$ evolution of the structure functions is determined by the 
Dokshitzer-Gribov-Lipatov-Altarelli-Parisi~(DGLAP) evolution 
equation~\cite{Altarelli:1977zs}. In Fig.~\ref{f1f2}, the experimental results for the nucleon structure functions from 
several electron scattering experiments~\cite{Whitlow:1991uw, Benvenuti:1989rh, 
Arneodo:1996rv, Aubert:1985fx, NewMuon:1995zeb}, and neutrino scattering experiments like CCFR~\cite{Oltman:1992pq}, CDHSW~\cite{Berge:1989hr}, 
etc. are presented in a wide range of $x$ and $Q^2$. One may notice from the figure that with the increase in $x$ and $Q^2$,
the structure functions decreases, while for lower $x$ and $Q^2$ there is a rise. This behavior of structure functions show 
scaling breakdown. In the next section, we discuss some of the experimental 
results for the total cross section from various neutrino scattering experiments. 

\subsection{Experimental results}
\begin{figure} 
\begin{center}
 \includegraphics[height=6 cm, width=11 cm]{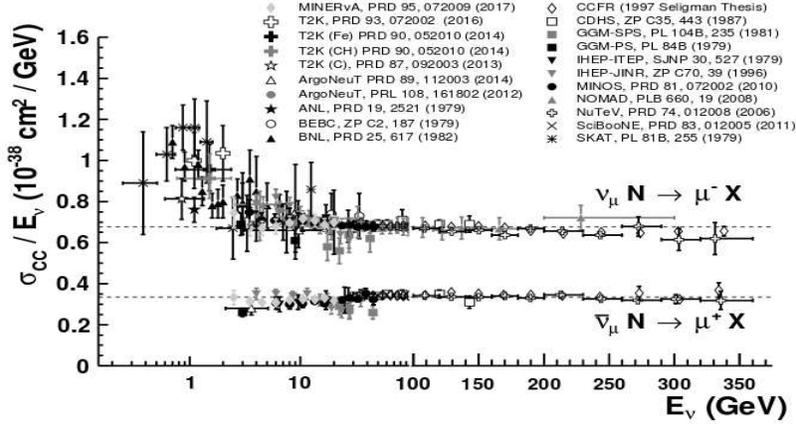}
 \end{center}
  \caption{Charged current total cross section for $\nu_\mu-N$ and $\bar\nu_\mu-N$ processes~\cite{ParticleDataGroup:2020ssz}.}
  \label{cctot_xsec}
\end{figure}

The total scattering cross section for CC DIS process in (anti)neutrino scattering has been experimentally 
measured by several experiments such as CCFRR~\cite{MacFarlane:1983ax}, CCFR90~\cite{Auchincloss:1990tu}, 
CCFR96~\cite{Seligman:1997fe}, CDHS~\cite{Berge:1987zw}, NuTeV~\cite{NuTeV:2005wsg}, BEBC-WBB~\cite{Colley:1979rt}, 
ANL~\cite{Barish:1978pj}, CHARM~\cite{Allaby:1987bb}, etc., and some of them have been 
shown here in Fig.~\ref{cctot_xsec}. These experiments have been performed on the various targets like hydrogen, deuterium, marble, 
iron, freon, freon-propane, etc. The world average values of the total scattering cross section for the neutrino and 
antineutrino interactions with nucleon/nuclear targets are~\cite{Nakamura:2009iq}:
\begin{eqnarray}
 \sigma^{\nu N}/E_{\nu} = 0.677 \pm 0.014\times 10^{-38} \rm{\text{cm}}^2\text{GeV}^{-1},\qquad \qquad
  \sigma^{\bar\nu N}/E_{\bar\nu} = 0.334 \pm 0.008 \times 10^{-38} \rm{\text{cm}}^2\text{GeV}^{-1}.\nonumber
\end{eqnarray}
By integrating $\frac{d^2\sigma^{\nu N}}{dx dy}$ over $x$ and $y$ between the limits 0 and 1, the expressions of the total 
scattering cross section for an isoscalar nucleon target for neutrino and antineutrino induced processes are obtained as
\begin{eqnarray}
 \sigma^{\nu N} &=& \frac{G_F^2 \;s}{2 \pi}\sum_{q}\int x\left(q(x)+\frac{\bar q(x)}{3}\right) dx \,,\hspace{3 mm}
 \sigma^{\bar\nu N} = \frac{G_F^2 \;s}{2 \pi}\sum_{q}\int x\left(\frac{q(x)}{3}+\bar q(x)\right) dx,
\end{eqnarray}
in 3q (viz. $u,d,s$) or 4q (viz. $u,d,s,c$) model. 
Through the total scattering cross section, one may directly determine the total momentum carried by all the quarks and 
antiquarks i.e. 
\begin{equation}
 \int x(q(x)+\bar q(x))\;dx = \frac{3\pi}{2\;G^2_F\;s}(\sigma^{\nu N}+\sigma^{\bar\nu N})
\end{equation}
and the fraction carried by the antiquarks as:
\begin{equation}
\frac{\int x \bar q(x)\;dx}{\int x(q(x)+\bar q(x))\;dx }= \frac{1}{2}\;\left(\frac{3\sigma^{\bar\nu N}-\sigma^{\nu N}}
{\sigma^{\nu N}+\sigma^{\bar\nu N}}\right),
\end{equation}
which were experimentally found to be~\cite{Allaby:1987bb}:
\begin{eqnarray}
 \int x(q(x)+\bar q(x))\;dx = 0.492 \pm 0.006 \pm 0.019, \qquad \qquad
  \frac{\int x \bar q(x)\;dx}{\int x(q(x)+\bar q(x))\;dx } = 0.154 \pm 0.005 \pm 0.011.\nonumber
\end{eqnarray}
From the above equations, it may be noticed that in the limits of high $Q^2$ and $\nu$, charged partons carry only $50\%$ of 
the nucleon's momentum and among them  antiquarks carry $15\%$ of the charged partons momentum, and the rest of 50\% of the 
momentum is carried by the gluons. It is therefore very important to understand the momentum distribution of gluons and the 
role of gluons in DIS processes.
\subsection{QCD corrections}\label{qcd}
 \begin{figure}
 \begin{center}
\includegraphics[height=3 cm, width=13 cm]{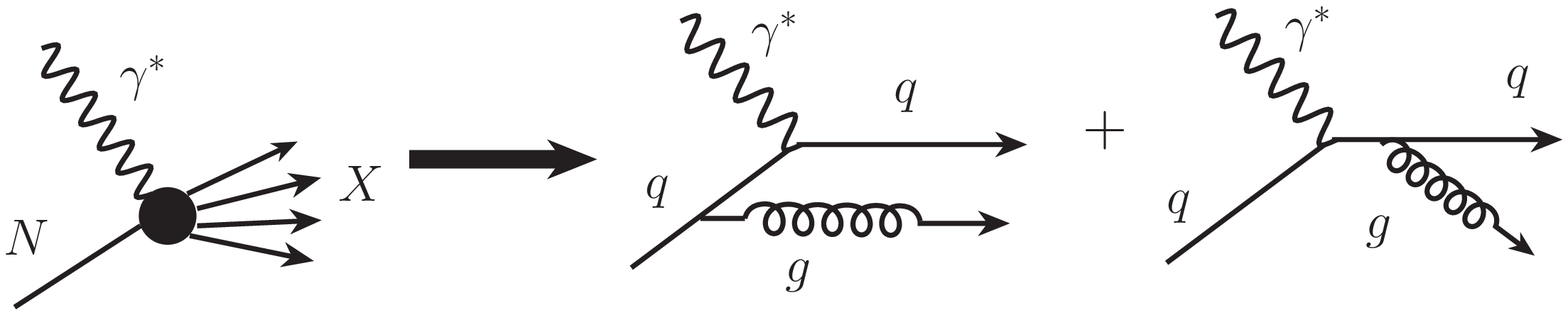}\\
\includegraphics[height=3 cm, width=13 cm]{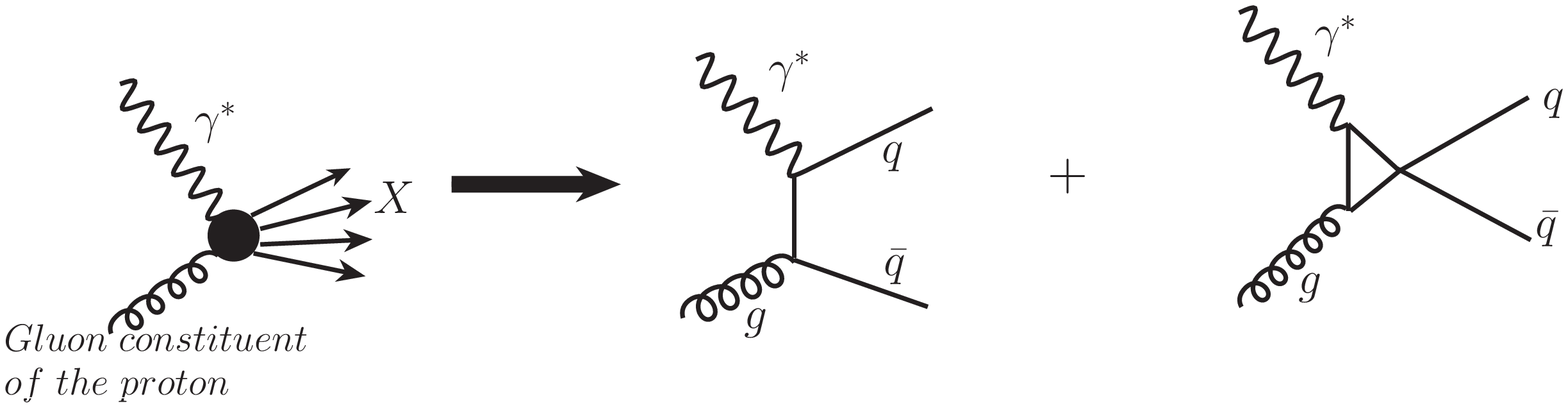}
\end{center}
\caption{Diagrammatic representation of the processes $ \gamma^\ast q \to q g$~(upper panel) and 
$\gamma^\ast g \to q \bar q$~(lower panel)~\cite{Halzen:1984mc}.}\label{fig_nlo}
\end{figure}
In the naive quark-parton model, a quark is treated as free fermion, while QCD tells us that quarks carry color and the color 
is exchanged by eight bicolored gluons. These gluons also interact among themselves. Therefore, the parton model of Feynman, 
Bjorken, Paschos, and others should be extended to envisage the dynamical role of gluons as the carrier of the strong force 
associated with the colored quarks. In the higher orders of perturbative QCD (pQCD), partons present inside the nucleon 
interact among themselves via the gluon exchange and the contribution from the gluons is responsible for the $Q^2$ dependence 
of the nucleon structure functions. For example, in the case of electromagnetic interactions, $ \gamma^\ast q \to q g$ and 
$\gamma^\ast g \to q \bar q$ are the possible channels, which are depicted in Fig.~\ref{fig_nlo}~\cite{Halzen:1984mc}. 
Generally, the $Q^2$ dependence of the structure functions is determined by evolving the $Q^2$ dependent parton densities using 
the DGLAP evolution equation~\cite{Altarelli:1977zs}. If we know the PDFs at 
some initial value of four momentum transfer square say, $Q_0^2$, then with the help of DGLAP evolution equation it is possible 
to know the value of PDFs at any other $Q^2\;(>\;Q_0^2)$ at all Bjorken $x$. 

Since the structure functions are expressed in terms of the parton density distribution functions, therefore, it is important 
to understand their behavior in the entire kinematic region of $x$ and $Q^2$. For lower values of $Q^2$, a few GeV$^2$ or 
less, the methods of the pQCD are not applicable due to the large value of the strong coupling $\alpha_s(Q^2)$ and the 
nonperturbative phenomena become important. In this region of $Q^2$, the quark-quark and the quark-gluon interaction effects 
play important role reflecting the dynamics of the internal constituents of the nucleon. Therefore, these dynamical effects 
deal with the interaction of struck quark with surrounding quarks via the gluon exchange. In DIS, by using a factorization 
theorem the nonperturbative physics is kept into a set of well defined, gauge invariant and universal quantities, which may be 
expressed by the matrix element of the parton operators between the hadron states, and the matrix element is expressed in terms 
of an expansion in inverse power of the momentum transfer. Basically, the quantity used to expand these matrix element is 
the parameter $\lambda/Q$, where $\lambda$ is a nonperturbative~(hadronic) scale. The perturbative scale $Q$ makes the 
coupling running and has to be large~(at least a few GeV) in order to make perturbation theory applicable, keeping $\lambda$ to 
be of the order of $\lambda_{QCD} \approx$ 200 MeV such that $\lambda/Q<<1$ and this expansion is called the {  twist} 
expansion. In the following, we discuss in brief the perturbative and nonperturbative corrections considered in this work:
\begin{itemize}
\item [(i)]{\bf NLO evolution}\\
In the naive parton model in the limit of $Q^2\to \infty,~\nu\to \infty$ with $x=\frac{Q^2}{2 M \nu}\to\textrm{``a finite 
value''}$, nucleon structure functions are the function of $x$. The probability of the gluon emission due 
to the interaction involves the strong coupling constant $\alpha_s(Q^2)$, which depends on $Q^2$. In the limit of $Q^2 \to 
\infty$, $\alpha_s(Q^2)$ becomes very small and, therefore, the higher order terms in a perturbative 
approach in which structure functions are expanded in orders of $\frac{\alpha_s(Q^2)}{2\pi}$ can be neglected. While for a 
moderate value of $Q^2$, $\alpha_s(Q^2)$ is large and higher order terms such as next-to-leading order~(NLO) give a 
significant contribution. The $Q^2$ evolution of structure functions is determined by the DGLAP evolution 
equation~\cite{Altarelli:1977zs}. In this approach, the nucleon structure functions are expressed in terms of the convolution 
of coefficient function ($C_{f,i}\;;\;(f=q,g; i=1-5)$) with the density distribution of partons ($f(x)$) inside the nucleon 
as~\cite{Athar:2020kqn}
\begin{equation}\label{f2_conv}
x^{-1} F_{i} (x) = \sum_{f=q,g} C_{f,i}^{(n)}(x) \otimes f(x)\; ,
\end{equation}
where the superscript $n=0,1,2,...$ for N$^{(n)}$LO evolution and the symbol $\otimes$ is the Mellin convolution 
\begin{equation}
C_{f,i}(x)\otimes f(x) =\int_x^1 C_{f,i}(y)\; f\left(\frac{x}{y} \right) {dy \over y}.
\end{equation}

The parton coefficient functions are generally expressed in power of $\frac{\alpha_s(Q^2)}{2\pi}$ 
as~\cite{vanNeerven:2000uj}:
\begin{eqnarray}
C_{f,i}(x,Q^2)=\sum_n\left(\frac{\alpha_s(Q^2)}{2\pi}\right)^n\;c_{f,i}^{(n)}(x).
\end{eqnarray}
By using the above expression in Eq.~(\ref{f2_conv}), one obtains
\begin{eqnarray}
\sum_{f=q,g} C_{f,i}(x,Q^2)\otimes f(x)&=&\sum_{f=q,g} \sum_n\left(\frac{\alpha_s(Q^2)}{2\pi}\right)^n\;c_{f,i}^{(n)}(x)
\otimes f(x)\nonumber\\
&=&\sum_n\left(\frac{\alpha_s(Q^2)}{2\pi}\right)^n\;\Big[c_{ns,i}^{(n)}(x) \otimes q_{ns}(x)+\langle e^2 \rangle \Big
\{c_{ns,i}^{(n)}(x) \otimes q_s(x)+c_{ps,i}^{(n)}(x) \otimes q_{s}(x)\nonumber\\
&+&c_{g,i}^{(n)}(x) \otimes g(x)\Big\}\Big],\nonumber
\end{eqnarray}
where $\langle e^2 \rangle$ is the average charge of partons which is $\langle e^2 \rangle=\frac{5}{18}$ for the 
electromagnetic interaction and $\langle e^2 \rangle=1$ for the weak interaction, for the four quark flavors ($u,d,s,c$). 
$q_s(x),~q_{ns}(x)$ are the singlet and the nonsinglet quark distributions and $g(x)$ is the gluon distribution. $c_{ns,i} 
(x)$ is the coefficient function for the nonsinglet and $c_{ps,i}^{(n)}(x)$ and $c_{g,i}^{(n)}(x)$ are the coefficient 
functions for the pure-singlet quark and gluon, respectively. For example, in the case of $F_2(x)$, one obtains the 
following expression~\cite{Vermaseren:2005qc}:
\begin{eqnarray}\label{f2or}
\sum_{f=q,g} C_{f,2}(x,Q^2)\otimes f(x)&=&x^{-1} F_{2N}^{EM,WI}(x) =\sum_n\left(\frac{\alpha_s(Q^2)}{2\pi}\right)^n\;
\Big[c_{ns,2}^{(n)}(x) \otimes q_{ns}(x)+\langle e^2 \rangle 
\Big\{c_{ns,2}^{(n)}(x) \otimes q_s(x)\nonumber\\
&+& c_{ps,2}^{(n)}(x) \otimes q_{s}(x)+c_{g,2}^{(n)}(x) \otimes g(x)\Big\}\Big],
\end{eqnarray}
where the singlet and nonsinglet quark distributions in 4-flavor scheme are given by
\begin{eqnarray}
 q_s(x) = u(x)+\bar u(x)+d(x)+\bar d(x)+s(x)+\bar s(x)+c(x)+\bar c(x)\;,\qquad \qquad
 q_{ns}(x) = F_{2N}^{EM,WI}(x) - \langle e^2 \rangle q_s .\nonumber
\end{eqnarray}
At the leading order ($n=0$), the coefficient functions for the quarks and gluons are, 
respectively $c_{ns,2}^{(0)}(x)=\delta(1-x)$, $c_{ps,2}^{(0)}=0$ and $c_{g,2}^{(0)}(x)=0$, which lead to
\begin{eqnarray}
x^{-1} F_{2N}^{EM,WI}(x) = c_{ns,2}^{(0)}(x)\otimes \Big\{q_{ns}(x) + \langle e^2 \rangle q_s(x)\Big\}, \nonumber
\end{eqnarray}
while at NLO ($n=1$), $c_{ps,2}^{(1)}=0$ using which in Eq.~(\ref{f2or}), the following expression is obtained: 
\begin{eqnarray}
x^{-1} F_{2N}^{EM,WI}(x) &=&\sum_n\left(\frac{\alpha_s(Q^2)}{2\pi}\right)\;\Big[c_{ns,2}^{(1)}(x) \otimes \Big\{q_{ns}(x)+
\langle e^2 \rangle q_s(x)\Big\}+ \langle e^2 \rangle c_{g,2}^{(1)}(x) \otimes g(x)\Big].\nonumber
\end{eqnarray}
Similarly, one may obtain the expression for $n=2$, which corresponds to NNLO term and so on. For the expressions of quark 
and gluon coefficient functions and other details, see Refs.~\cite{vanNeerven:2000uj, Vermaseren:2005qc, Moch:1999eb,
Moch:2004pa, Moch:2004xu}.

Moreover, the expression for the weak structure function $F_{3N}^{WI}(x)$ in terms of the coefficient function and the parton
density distribution function ($f(x)=q_v(x)$ as mainly valence quarks contribute in $F_{3N}(x)$) is given 
by~\cite{Moch:2008fj}: 
\begin{eqnarray}
 F_{3N}^{WI}(x) &=& \sum_n \left(\frac{\alpha_s(Q^2)}{2\pi}\right)^{(n)} c_{ns,3}^{(n)}(x) \otimes q_v(x),\nonumber
\end{eqnarray}
where $q_v(x)$ is the valence quark distribution and $c_{ns,3}^{(n)}(x)$ is the nonsinglet coefficient function corresponding 
to the different perturbative terms including leading order and the other higher order terms. For detailed discussion, 
see Refs.~\cite{Moch:2004pa, Moch:2008fj}.

\item [(ii)] {\bf Target mass corrections effect}\\
The target mass correction~(TMC) is a nonperturbative effect which comes into the picture at low $Q^2$, where perturbation theory 
fails. The TMC effect is significant at low $Q^2$ and high $x$, and has been found to be important in the determination of 
the distribution of valence quarks. Unfortunately, this kinematic region has not been explored much, unlike the region of 
high $Q^2$ and low $x$. The TMC effect is also known as ``kinematic higher twist effect''. In 1976, Georgi and Politzer 
determined the TMC to electroweak structure functions, using operator product expansion~(OPE), at the 
leading order of QCD~\cite{Georgi:1976ve}. Thus, these are subleading $\frac{1}{Q^2}$ corrections to the leading twist structure 
functions.

In the region of low $Q^2$, and in the presence of heavier quarks like charm, bottom, etc., $x$ 
is modified to $\xi$, which is known as the Nachtmann variable and is related to the Bjorken variable $x$ 
as~\cite{Schienbein:2007gr}: 
\begin{equation}\label{nach}
\xi=\frac{2x}{1+\rho}\;;\;\;\rho=\sqrt{1+4 \mu x^2} \;,\;\mu=\frac{M^2}{Q^2}\;,\;x=\frac{Q^2}{2M \nu}\;.
\end{equation}
$\xi$ depends only on the hadronic mass and does not have corrections due to the masses of final state quarks. However, for 
the massive partons, the Nachtmann variable $\xi$ gets modified to the slow rescaling variable $\bar\xi$. The variables 
$\xi$ and $\bar\xi$ are related as:
\begin{equation}\label{snach}
\bar\xi=\xi \left( 1+\frac{m_c^2}{Q^2}\right)=\frac{\xi}{\lambda}.
\end{equation}
It may be noticed from Eqs.~(\ref{nach}) and (\ref{snach}) that the Nachtmann variable corrects the Bjorken variable for the 
effect of hadronic mass while the generalized variable $\bar\xi$ further corrects $\xi$ for the effect of the partonic 
masses. The simplified expressions of target mass corrected structure functions for massless quarks ($u,~d,$ and $s$) are 
given in Ref.~\cite{Zaidi:2019mfd, Zaidi:2019asc}. TMC effect has been discussed by several authors such as Scheinbein et 
al.~\cite{Schienbein:2007gr}, Ellis et al.~\cite{Ellis:1982cd}, Aivazis et al.~\cite{Aivazis:1993kh}, Brady et 
al.~\cite{Brady:2011uy}, etc. by taking into account different approaches like OPE, collinear 
factorization, $\xi$-scaling, and the approach discussed by Ellis-Furmanski-Petronzio~\cite{Ellis:1982cd}. In our numerical 
calculations, the nucleon structure functions ($F_{iN}(x,Q^2);~(i=1-3)$) are evaluated by incorporating the TMC effect 
following the works of Scheinbein et al.~\cite{Schienbein:2007gr}.

\item [(iii)] {\bf Inclusion of charm quark mass effect:}\\
When the mass of the charm quark is included, the dimensionless structure functions at NLO are given by~\cite{Kretzer:2002fr, 
Kretzer:2001tc}
\begin{eqnarray}\label{eq_chm}
F_i^c(x,Q^2)=(1-\delta_{i4})\cdot s^\prime(\bar{\xi}, \mu^2)+\frac{\alpha_s(\mu^2)}{2\pi}\Big\{\int_{\bar{\xi}}^1 
\frac{dy^\prime}{y^\prime}
\Big[C^i_q\Big( y^\prime, \frac{Q^2}{\mu^2}, \lambda \Big)s^\prime\Big(\frac{\bar{\xi}}{y^\prime}, \mu^2\Big) + 
C^i_g\Big( y^\prime, \frac{Q^2}{\mu^2}, \lambda \Big)g^\prime\Big(\frac{\bar{\xi}}{y^\prime}, \mu^2\Big)   \Big]\Big\},
\end{eqnarray}
for scattering off the CKM rotated weak eigenstate~\cite{Ansari:2020xne}
\begin{eqnarray}
 s^\prime&=&s\cdot \cos^2\theta_C + d\cdot \sin^2\theta_C,
\end{eqnarray}
and its QCD evolution partner $g^\prime$
\begin{eqnarray}
g^\prime&=&g\cdot \cos^2\theta_C + g\cdot \sin^2\theta_C,
\end{eqnarray}
where $s$, $d$, and $g$ are the quarks and gluon density distributions and $\theta_C$ is the Cabibbo angle.
In Eq.~(\ref{eq_chm}), $C^i_q$ and $C^i_g$; ($i=1-5$) 
are respectively the fermionic and gluonic coefficient functions at NLO which are taken from Ref.~\cite{Kretzer:2003iu}. 
 $\bar\xi$ is the slow rescaling variable and the variables $\lambda$ and $y^\prime$ are defined as
\begin{equation}
\lambda=\frac{Q^2}{(Q^2+m_c^2)}\;,\;\;y^\prime=\frac{\bar\xi}{y},
\end{equation}
where $m_c$ is the charm quark mass.
The terms at NLO ($C^i_q$ and $C^{i}_g$ ; ($i=1-5$)) with strong coupling constant $\frac{\alpha_s 
(\mu^2)}{2\pi}$ give finite contribution. From the above expression, it may be noticed that though at the leading order 
$F_{4}(x)=0$, however, when the NLO terms are taken into account, a nonzero contribution for $F_{4}(x)$ is obtained~\cite{Ansari:2020xne}.

\item [(iv)] {\bf Dynamical higher twist effect:}\\
Higher twist~(HT: twist-4) is a dynamical effect arising due to the multiparton correlations~\cite{Wilson:1969zs, Shuryak:1981kj, Dasgupta:1996hh}. 
This effect involves the interaction of the 
struck quark with other quarks via the exchange of gluon and it suppresses by the power of $\left({1 \over Q^2} \right)^n$, 
where $n=1,2,..$. This effect is also pronounced in the region of low $Q^2$ and high $x$, like the TMC effect, but is 
negligible for high $Q^2$ and low $x$. In the formalism of OPE~\cite{Wilson:1969zs, Shuryak:1981kj}, the structure 
functions $F_{i}(x,Q^2)$ are expressed in terms of powers 
of $1/Q^2$ (power corrections):
\begin{equation}
F_{i}(x,Q^2) = F_{i}^{j = 2}(x,Q^2)
+ {{\cal H}_{i}^{j = 4}(x) \over Q^2}  + ..... ,   \;\;\; i=1,2,3, 
\label{eqn:ht}
\end{equation}
where the first term ($j=2$) is known as the twist-two or leading twist (LT) term, and it corresponds to the scattering off 
a free quark. This term obeys the Altarelli-Parisi equation and is expressed in terms of PDFs. It is responsible for the 
evolution of structure functions via perturbative QCD $\alpha_s(Q^2)$ corrections. The term corresponding to $j=4$ is known 
as the twist-4 or higher twist term and it reflects the multiparton correlations. It has been observed by Zaidi et 
al.~\cite{Zaidi:2019asc} that the scattering cross section obtained with TMC and HT corrections at NLO have negligible 
difference from the results obtained at NNLO with the TMC effect only.
\end{itemize}

\subsection{Results and discussion}\label{results}
\begin{figure} 
\centering
	\includegraphics[height=7cm, width=0.95\textwidth]{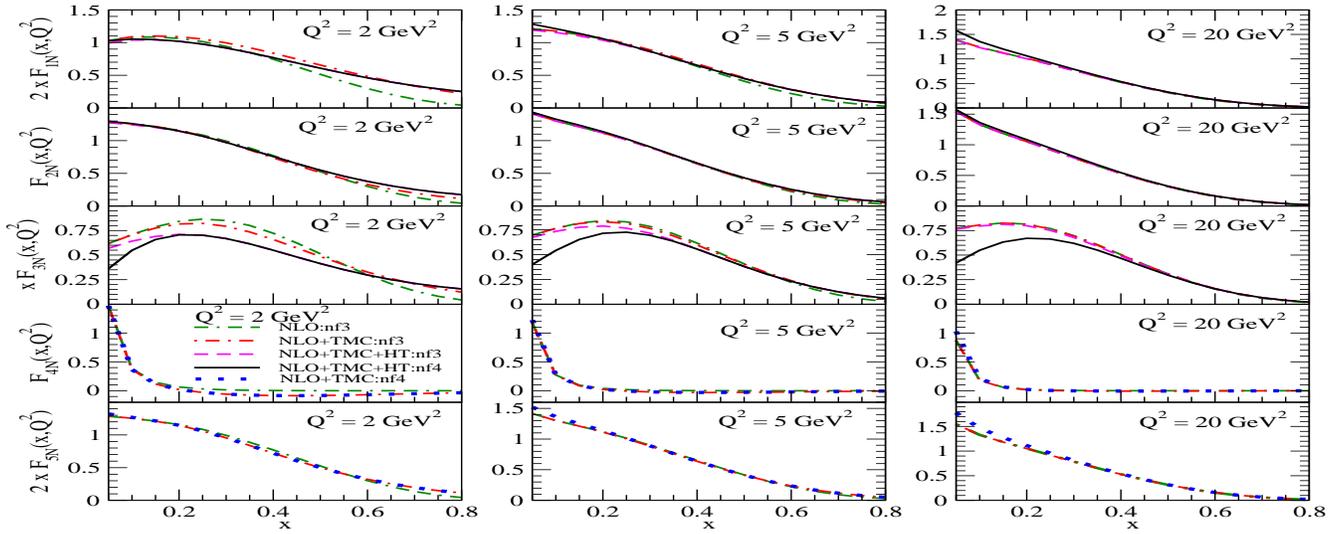}
\caption{ Results for the free nucleon structure functions $F_{iN}(x,Q^2)$;$(i=1-5)$~(top to bottom) at the different values of 
$Q^2$ viz. 2, 5 and 10 GeV$^2$~(left to right) are shown. These results are obtained at NLO using MMHT nucleon PDFs 
parameterization~\cite{Harland-Lang:2014zoa}. The results are shown without the TMC effect (double dashed-dotted line), with 
the TMC effect in the 3-flavor(nf3) scheme (dashed-dotted line) as well as four flavor(nf4) scheme(dotted line), with TMC 
and HT effects in the 3-flavor(nf3) scheme (dashed line) as well as four flavor(nf4) scheme(solid line).}\label{fig:sf_free}
\end{figure}
\begin{figure} 
 \centering
\includegraphics[height=9cm, width=0.8\textwidth]{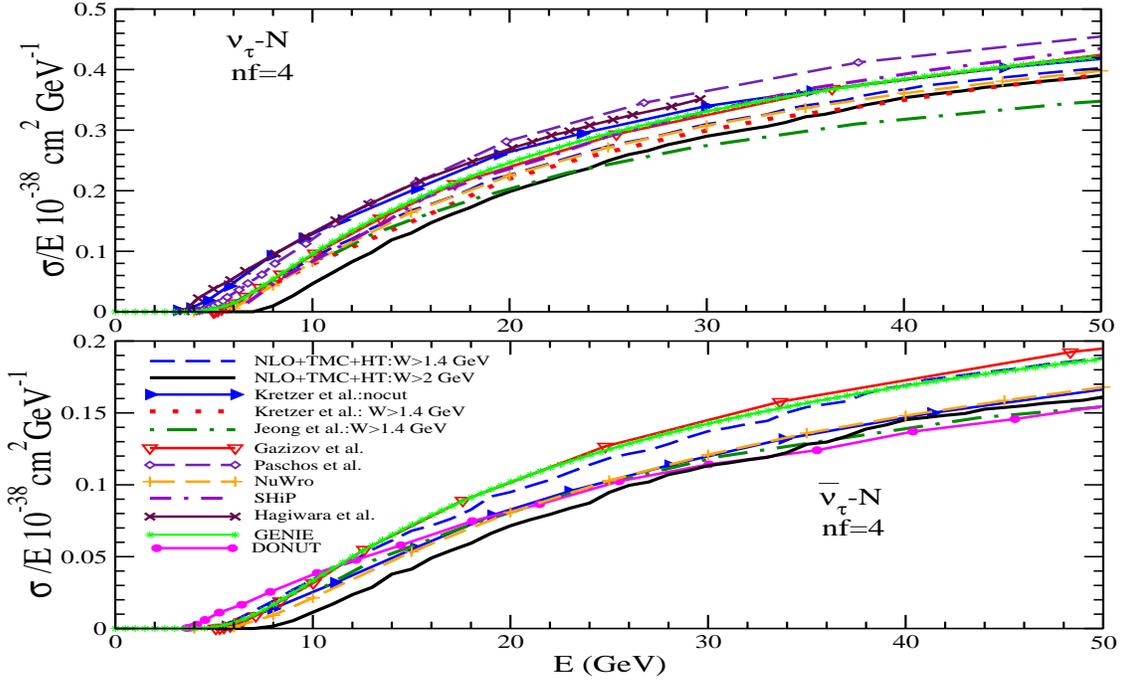}
\caption{$\frac{\sigma}{E}$ vs $E$ with a $W$ cut of 1.4 GeV(dashed line) and 2 GeV(solid line), for 
tau type neutrinos~(top panel) and antineutrinos~(bottom panel) with the TMC~\cite{Kretzer:2003iu} and 
HT~\cite{Dasgupta:1996hh} effects. These results are  compared with the results of different models available in the 
literature~\cite{Super-Kamiokande:2017edb, Kretzer:2002fr, Jeong:2010nt, Hagiwara:2003di, Paschos:2001np, Gazizov:2016dhn, 
Anelli:2015pba} as well as with the Monte Carlo generators GENIE~\cite{GENIE:2021npt} and 
NuWro~\cite{Juszczak:2005zs}.}\label{fig:sig_comp}
\end{figure}
\begin{figure} 
  \centering
 \includegraphics[height=6 cm, width=0.7\textwidth]{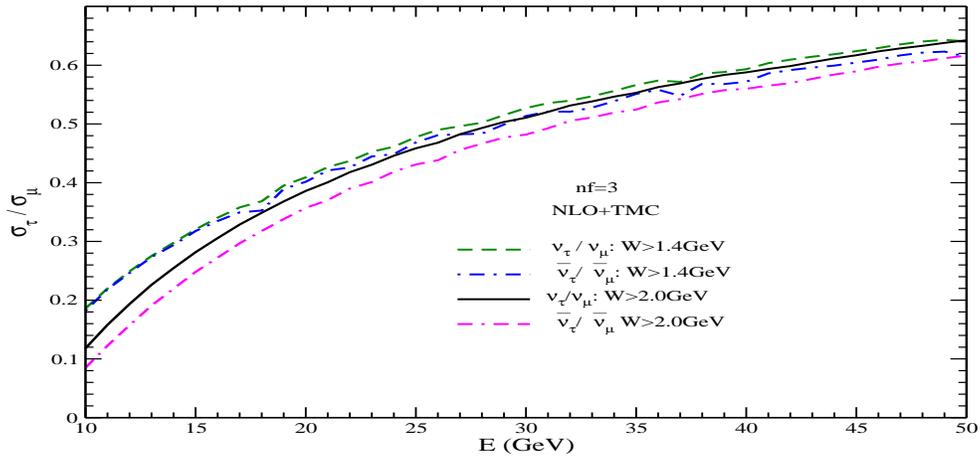}
 \caption{Ratio of the total scattering cross section $\frac{\sigma_{\nu_\tau}}{\sigma_{\nu_\mu}}$ vs $E$ are shown with 
 $W > 1.4$ GeV and $W > 2$ GeV for $\nu_\tau - N$ and ${\bar\nu}_\tau - N$ DIS. Dashed and dashed-dotted lines represent the 
 results with $W>1.4$ GeV while the solid and double dashed-dotted lines represent the results with $W>2$ GeV for neutrinos 
 and antineutrinos, respectively.  The effect of TMC~\cite{Kretzer:2003iu} is also included.}\label{fig:sig_ratio}
\end{figure}

In this section, we present the results for the nucleon structure functions (Fig.~\ref{fig:sf_free}) and the total 
scattering cross sections (Figs.~\ref{fig:sig_comp} and \ref{fig:sig_ratio}) obtained using the formalism discussed in 
the previous section. All the results are presented at NLO by incorporating the following considerations:
\begin{itemize}
\item The target mass correction effect for the massive as well as massless quarks.

\item The higher twist effect in terms of function ${\cal H}_{i}^{j = 4}(x)$ using Eq.~(\ref{eqn:ht}) in 
the evaluation of $F_{iN}(x,Q^2);~~i=1-3$. 

\item The effect of massive charm quark.
\end{itemize}
The results for the free $\nu_l(\bar\nu_l)-N$ scattering have been obtained in the three~(nf3)- and 
four~(nf4)- flavor schemes~\cite{Zaidi:2019asc}. All the results are presented using MMHT PDFs parameterization of Harland-Lang et 
al.~\cite{Harland-Lang:2014zoa}. A cut in $Q^2$ of $Q^2 \ge 1$~GeV$^2$ has been used to obtain the numerical results. 
The tau lepton mass has been considered in the numerical calculations for $\nu_\tau({\bar\nu}_\tau)-N$ scattering~\cite{Ansari:2020xne}.

In Fig.~\ref{fig:sf_free}, the results for the free nucleon structure functions $2xF_{1N}(x,Q^2)$, $F_{2N}(x,Q^2)$, $xF_{3N} 
(x,Q^2)$, $F_{4N}(x,Q^2)$ and $2xF_{5N}(x,Q^2)$ are shown at the three different values of $Q^2$ viz. 
2~GeV$^2$, 5~GeV$^2$ and 20~GeV$^2$. These results are presented at NLO without the 
TMC effect, with the TMC effect in 3-flavor and 4-flavor
schemes, with TMC and HT effects in 3-flavor and 4-flavor schemes. It may be observed that 
the TMC effect is dominant in the region of high $x$ and low $Q^2$ and it becomes small at low $x$ and high $Q^2$. 
Quantitatively, the TMC effect is found to be different in $F_{2N}(x,Q^2)$ from $F_{1N}(x,Q^2)$ while the TMC effect in 
$F_{5N}(x,Q^2)$ is similar to the effect in $F_{2N}(x,Q^2)$, whereas, in the case of $F_{4N}(x,Q^2)$ the whole contribution 
arises in the leading order due to the TMC effect. The contribution of $F_{4N}(x,Q^2)$ to the cross sections being dependent 
on the lepton mass (Eq.~(\ref{xsec:dis})), is important for the $\nu_\tau$ scattering and in the region of $x\le 0.2$. This 
contribution to the cross section becomes almost negligible for $x>0.2$ when TMC effect is not incorporated but 
with the inclusion of TMC effect a nonzero though small contribution in the region of high $x$ and low $Q^2$ has been found. 
The difference in the results of nucleon structure functions $F_{iN} (x, Q^2);~~ (i = 1-5)$ evaluated at NLO with and without 
the TMC effect at $x=0.3$ is $5\%(3\%)$ in $F_{1N} (x, Q^2)$, $2\%(<1\%)$ in $F_{2N} (x, Q^2)$, $7\%(\sim 3\%)$ in $F_{3N} 
(x, Q^2)$ and $4\%(\sim 2\%)$ in $F_{5N} (x, Q^2)$ for $Q^2=2(5)$ GeV$^2$.

In Fig.~\ref{fig:sig_comp}, we compare the results for $\sigma/E$ vs $E$ with the results of Li et al.~\cite{Super-Kamiokande:2017edb} 
(solid line with circles), Kretzer 
et al.~\cite{Kretzer:2002fr} (solid line with right triangle without a cut on $W$; dotted line with a cut of $W > 1.4$ GeV), 
Jeong et al.~\cite{Jeong:2010nt} (dash-dotted line), 
Hagiwara et al.~\cite{Hagiwara:2003di} (solid line with cross symbol), Paschos et al.~\cite{Paschos:2001np} (dashed line with diamond), 
Gazizov et al.~\cite{Gazizov:2016dhn}(solid line with down triangle),  Anelli et al.~\cite{Anelli:2015pba} (double dash-dotted 
line), as well as with the Monte Carlo generators 
GENIE~\cite{GENIE:2021npt} and NuWro~\cite{Juszczak:2005zs}. These results are presented for both cases of cuts on the CM energy taken 
to be 1.4 GeV~(dashed line) and 2 GeV~(solid line) by incorporating TMC and 
HT effects at NLO in the four flavor scheme. Our results with a cut of $W>1.4~ GeV$ is in 
good agreement with the result of Kretzer et al.~\cite{Kretzer:2002fr} while there are significant 
differences from the result of Jeong et al.~\cite{Jeong:2010nt}. Notice that the results of 
the total scattering cross section with the same CM energy cut reported by Kretzer and Reno~\cite{Kretzer:2002fr} and Jeong 
and Reno~\cite{Jeong:2010nt} are also different from each other. The difference is mainly due to the choice of lower cuts on 
$Q^2$ in the evaluation of PDFs. It is important to point out that the results given by the different 
models~\cite{Super-Kamiokande:2017edb, Kretzer:2002fr, Jeong:2010nt, Hagiwara:2003di, Paschos:2001np, Gazizov:2016dhn, 
Anelli:2015pba, Conrad:2010mh} have significant differences due to their choice of different kinematic regions. 
Furthermore, we have observed that the effect of CM energy cut is more pronounced in the case of $\bar\nu_\tau-N$ DIS than in 
$\nu_\tau-N$ DIS process. Moreover, one may also notice that the total scattering cross section gets suppressed with the 
increase in the kinematic cut on the  CM energy. It implies that a suitable choice of $W$ and $Q^2$ to define the DIS region 
and using them to calculate the nucleon structure 
functions, differential and total scattering cross sections are quite important. The constrains in the kinematic variables 
$Q^2$ and $W$ should be kept in mind while comparing the predictions of the cross sections in various theoretical models.

To understand the effect of lepton mass on the cross section, in Fig.~\ref{fig:sig_ratio}, the ratio of the total scattering 
cross sections $\frac{\sigma_{\nu_\tau - N}}{\sigma_{\nu_\mu - N}}$ vs $E$ (dashed and solid lines) and 
$\frac{\sigma_{\bar{\nu}_\tau - N}}{\sigma_{\bar{\nu}_\mu - N}}$ vs $E$ (dash-dotted and double dash-dotted lines) with cuts 
of $W>1.4$ GeV and $W>2$ GeV, are shown. These results are evaluated at NLO with TMC effect in the three 
flavor scheme. Notice that the lepton mass effect is important through out the energy region shown here. However, this effect 
becomes small with the increase in energy and therefore the ratio increases but does not reach unity even at 100 GeV. It is 
important to point out that for the ratio with CM energy cut of 2 GeV, the lepton mass effect is more pronounced than in the 
case of $W>1.4$ GeV. One may also notice that the lepton mass effect is quantitatively different for 
neutrino and antineutrino induced processes, though qualitatively it shows similar behavior. For example, the ratio obtained 
with a cut of $W>2$ GeV deviates from unity by $89\%(36\%)$ for neutrino and $91\%(38\%)$ for antineutrino at $E=10(50)$ GeV. 


\section{Neutrino scattering from nuclei}\label{nu:nuclei}
Most of the (anti)neutrino experiments are done with the nuclear targets in the entire energy region of $\nu_l 
({\bar\nu}_l)$ starting from a few MeV to several hundreds of GeV. A theoretical description of these processes requires a 
knowledge of nuclear structure of the initial and final nuclei in addition to the knowledge of the neutrino interactions 
with nucleons within the nucleus. While the theory of the basic neutrino interactions with nucleon is described by the 
SM of electroweak interactions, given in Section~\ref{nu_QE}, an appropriate knowledge of the nuclear structure of 
the initial and final nuclei depends upon the various energy scales used to study the (anti)neutrino-nucleus reactions. In 
the low energy region, neutrinos scatter elastically~(quasielastically) by the weak NC~(CC) interactions 
where the target nucleus can be in the ground state or can be excited to the higher excited states which then decays by 
emitting photons, electrons with neutrinos, or nucleons. As the energy increases, the IE processes occur in which 
leptons are produced along with the new particles like pions, kaons, or other mesons which are emitted along with the 
residual nucleus in the final state. In both cases, a realistic description of the nuclear wave functions in the initial 
and the final states corresponding to the various nuclear excitations is needed. However, in the case of inclusive reactions 
where only leptons are observed in the final state and a sum over all the nuclear states are performed, the nuclear 
wave functions of a large number of excited states in the final nucleus are needed. To obtain a reliable description of all 
the excited states is quite a difficult task. Alternatively, some approximation methods are used in the case of inclusive 
scattering like the closure approximation or the Fermi gas models, where only a reliable description of the initial state 
nuclear wave functions or nuclear density is needed~\cite{Athar:2020kqn}. With further increase in energy, jet of hadrons 
are produced in the final state along with a charged lepton, and the process is known as DIS. 
 
In general, NME depends upon the behavior of bound nucleons in the nucleus and their response to the 
external probes depending upon their energy and the type of reactions induced by them. In the case of QE and IE 
processes, the general consideration of NME include the Fermi motion as the nucleon inside the 
nucleus is not at rest, Pauli blocking effect arising due to the exclusion principle when more than one nucleon is involved 
in the reaction, and the multinucleon correlation effects due to the strong interaction of the nucleons, and the meson 
exchange current effect. In most of the calculations using the impulse approximation~(IA), some of these effects due to the 
Fermi motion, binding energy and Pauli principle on the nucleon are simulated by using a spectral function $S({\vec p}, E)$ { which is related to the imaginary part of the nucleon Green's function, 
and describes the energy and momentum distribution of the nucleons in the nuclear ground state.  
In 
the simplest model like the nonrelativistic Fermi gas model, it is given by theta function in the initial states i.e. $\Theta(p_F - p)$, where $p_F$ is the Fermi momentum and has been used most frequently in 
early calculations. 
In the nuclear models using the mean field approximation, the nucleon spectral function $S(\vec{p}, E)$ has been calculated without taking into account the multinucleon correlation effects. 
While some recent models include the multinucleon correlation effects. For example, see the discussion in Ref.~\cite{Benhar:2015wva}.
The effect of nucleon correlation helps in explaining the experimental data as observed earlier in experiments done at Saclay and NIKHEF~\cite{Kramer:1989uiu, Frullani:1984nn} and recently at JLab~\cite{JeffersonLabHallA:2022cit, JeffersonLabHallA:2020rcp} as seen in the case of $(e,e^\prime p)$ processes in nuclei. 
In view of this, the phenomenologically determined values of $S(\vec{p},E)$ from $(e,e^\prime p)$ experiments is used in the calculations of the quasielastic and inelastic scattering of (anti)neutrinos from the nuclear targets.}
In the case of IE reactions, where mesons like $\pi$, $K$, $\eta$, etc. are produced, the most 
studied process is the single pion production, which is dominated by the resonance production. One considers modification of 
the properties of the various excited resonances especially their masses and widths in the nuclear medium. However, these 
modifications are well studied only in the case of $\Delta$ resonance. In addition, the pion produced in the decay of these 
resonances undergo final state interaction with the residual nucleus, where elastic scattering, charge exchange 
process~(like $\pi^- p \rightarrow \pi^0 n$) or pion absorption~($\pi NN \rightarrow NN$) may take place. If the produced 
pion is absorbed in the nucleus, it mimics a QE-like event. With further increase in energy, the process of DIS takes place 
in which the (anti)neutrinos interact with the subnucleonic degrees of freedom like 
the mesons and quarks in the nucleons. In the case of DIS, shadowing and antishadowing corrections become important in the 
region of low Bjorken variable $x$~(see Section~\ref{dis:nucleus}). In the intermediate region of $x$, the mesonic 
contributions  become important where the interaction of an intermediate vector boson~($W$, $Z$) takes place with the 
virtual mesons in the nucleus and in the region of high $x$, Fermi motion effects are important. 
 
In the following, we take up the $\nu(\bar\nu)$-nucleus scattering in the low, intermediate, and high energy regions and 
discuss NME.

\subsection{Coherent elastic neutrino-nucleus scattering in the low energy region}\label{CEvNS}
The weak NC in the $\Delta S=0$ sector predicted in the SM allow the existence of the elastic scattering of 
(anti)neutrino from nucleons and nucleus without any threshold constrains and can take place even at very low energies. In 
the case of nuclear targets, it was pointed out by Freedman~\cite{Freedman:1973yd} and later by Kopeliovich and 
Frakfurt~\cite{Kopeliovich:1974mv} that if the (anti)neutrino energy and the momentum transfer are too low to induce any 
excitation or particle emission in the nucleus and the nucleus remains in the ground state, then it is possible that 
scattering from individual nucleons can be in phase leading to coherent scattering. This coherent scattering would lead to a 
considerable enhancement in the cross section, which grows with the increase in the number of nucleons. The necessary 
condition to observe the phenomenon of coherence in $\nu$-nucleus scattering is that at these (anti)neutrino energies, the 
momentum transfer $Q$ is low enough to satisfy the condition $Q \ll  \frac{1}{R}$, where $R$ is the radius of the nucleus. 
 
While the condition of coherence favors the use of the medium and heavy nuclear targets with larger $A$, it also presents 
formidable problems in its detection. In coherent reactions induced by NC interactions, the only observable 
is the recoiling nucleus with a very small kinetic energy i.e. in the energy region of keV for (anti)neutrinos of a few MeV 
energy, which is very difficult to measure experimentally. However, the latest developments in the detector technology have 
enabled the measurement of very low energy recoils of nuclei, resulting in the observation of the coherent reactions for 
example by the COHERENT collaboration at ORNL at its spallation neutron source facility using the intense muon neutrino 
beam obtained from the pions decaying at rest~\cite{Barbeau:2021exu}. The first observation of coherent elastic 
neutrino-nucleus scattering~(CEvNS) was reported by this collaboration in CsI(Na), using the scintillating crystal detector 
followed by its observation in Ar with a single-phase liquid Ar detector~\cite{COHERENT:2020iec} and later with the larger 
exposure of CsI(Na)~\cite{Collar:2014lya}. Many new experiments are planned to be done with other nuclear targets and further 
substantiate the observation of CEvNS in near future~\cite{Barbeau:2021exu, Akimov:2022oyb}.

The elastic $\nu(\bar \nu)$-nucleus scattering process represented by the reaction
\begin{equation}
 \nu(\bar \nu) \;\;+\;\; ^{A}_ZX_{N} \longrightarrow \nu(\bar \nu)  \;\;+\;\; ^{A}_ZX_{N}
\end{equation}
takes place when $\nu(\bar \nu)$ scatters elastically from the nucleus $ ^{A}_ZX_{N}$ which is a composite system of $A$ nucleons. 
Using the quantum mechanical principle of superposition, the scattering amplitude $F({\vec k}^{\prime},{\vec k})$ for a 
$\nu(\bar \nu)$ with initial and final momentum $\vec k$ and ${\vec k}^\prime$ respectively, is written as the sum of the 
scattering amplitude from each nucleon in the nucleus i.e. $f_j({\vec k}^{\prime},{\vec k})$ and is given in the lowest 
order as:
\begin{equation}\label{eqn82}
 F({\vec k}^{\prime},{\vec k})= \sum_{j=1}^A f_j({\vec k}^{\prime},{\vec k}) e^{i(\vec k - {\vec k}^\prime).{\vec x}_j}.
\end{equation}
In the low energy region of the $\nu(\bar \nu)$, if $Q$ is very small as compared to the inverse of $R$, i.e. $QR << 1$, 
then the scattering amplitude $F({\vec k}^{\prime},{\vec k})$ becomes the coherent sum of the individual amplitudes from 
nucleons $f_j({\vec k}^{\prime},{\vec k})$ i.e. 
\begin{equation}\label{eq:CE}
  F({\vec k}^{\prime},{\vec k})= \sum_{j=1}^Z f_j^p({\vec k}^{\prime},{\vec k}) + \sum_{j=1}^N f_j^n({\vec k}^{\prime},
  {\vec k})
\end{equation}
and the scattering is called the coherent elastic $\nu(\bar \nu)$-nucleus scattering. 

For the scattering processes induced by the electroweak probes like the electrons and (anti)neutrinos, where the 
contribution from the multiple scattering amplitudes  are small, Eq.~(\ref{eqn82}) gives a satisfactory description of the 
scattering in the lowest order. In general, the individual scattering amplitudes are different for the proton and neutron 
targets in nuclei, and the summations over nucleons in Eq.~(\ref{eq:CE}) runs separately over protons and neutrons. In 
order to calculate the scattering amplitudes $f_j({\vec k}^{\prime},{\vec k})$ from the individual nucleons in the low 
energy region, the interaction Lagrangian of the neutrino and the electron for their interaction with nucleon given by the 
SM of the electroweak interactions is used. 

In the SM, the coherent scattering is driven by NC interaction Lagrangian ${  L}^{NC}(x)$ given in 
Section~\ref{SM}, where $l_\mu^{NC}$ and $J_\mu^{NC}$ are the leptonic and hadronic currents whose matrix elements between 
the neutrino and the nucleon states is explicitly given in Section~\ref{QE:NC}. In the region of low energy of the scattering 
processes, the nucleons are treated as nonrelativistic particles and a nonrelativistic reduction of the matrix 
elements~\cite{Athar:2020kqn} can be used to calculate the cross sections. The nonrelativistic reduction of the matrix 
elements~\cite{Athar:2020kqn} shows that in the leading order of $\mathcal{O} (\frac{\vec q}{M})$ only the time component of 
the vector current with the weak charge operator and the space component of the axial-vector current with the spin operator 
contribute to the matrix element, which gives rise to the spin independent and the spin dependent parts of the coherent 
scattering. Further simplification is achieved if simple scalar and isoscalar nuclear targets with spin zero and $N=Z$ are 
used as considered in the early works of Freedman~\cite{Freedman:1973yd} and others, where the axial-vector currents do not 
contribute. In general, the contribution of the spin dependent amplitudes to the coherent process is quite small as compared 
to the contribution of the spin independent amplitude as the number of unpaired protons~(neutrons) in the spin space is quite 
small as compared to the total number of protons~(neutrons) in the nucleus. This has been explicitly shown recently by 
Hoferichter et al.~\cite{Hoferichter:2020osn} and many others quoted in this work. A simple calculation shows that the 
differential cross section $\frac{d\sigma}{dT}$, where the recoil energy of the nucleus $T\sim\frac{Q^2}{2M_A}$, with $M_{A}$ 
being mass of the target nucleus, is given by~\cite{Abdullah:2022zue}:
\begin{equation}
\left(\frac{d\sigma}{dT}\right)_{\nu A}= \frac{G_F^2 M_A }{2\pi}F^2(Q^2) \left[(f_{1}+g_{1})^2+(f_1 - g_1)^2\left(1-\frac{T}
{E_\nu}\right)^2 - (f_1^2 - g_1^2)\frac{M_AT}{{E_\nu}^2}\right],
\end{equation}
and $F(Q^2)$ is the nuclear form factor. $f_1$ and $g_1$ are the vector and axial-vector couplings of the neutral vector 
boson in the SM to the nucleus $A(Z,N)$ given by:
\begin{equation}
  f_{1} = {\tilde f}_{1}^p Z + {\tilde f}_{1}^n N \qquad \quad \text{and} \quad\qquad g_{1} = {\tilde g}_{1}^p Z + 
  {\tilde g}_{1}^n N,
\end{equation}
where ${\tilde f}_{1}^{p(n)}$ and ${\tilde g}_{1}^{p(n)}$, respectively, are the vector and axial-vector NC couplings of the 
(anti)neutrino to the proton~(neutron). Neglecting the axial-vector contribution as compared to the contribution of the 
vector current, we obtain in the case of isoscalar nuclear targets with spin zero, the expression for the kinetic energy 
distribution as:
\begin{equation}
\left(\frac{d\sigma}{dT}\right)_{\nu A}\simeq \frac{G_F^2 M_A }{2\pi}F^2(Q^2) f^2_1\left[1+\left(1-\frac{T}{E_\nu}\right)^2 
- \frac{M_AT}{{E_\nu}^2}\right],
\end{equation}
where $f_1$ is calculated using the SM values of ${\tilde f}_1^p$ and ${\tilde f}_1^n$ given by:
\begin{equation}
 {\tilde f}_1^p = \frac{1}{2}- 2\sin^2\theta_W, ~~~~~~~~~~~~~~~~~~~{\tilde f}_1^n = -\frac{1}{2}
\end{equation}
leading to
\begin{equation}
\left(\frac{d\sigma}{dT}\right)_{\nu A}\simeq \frac{G_F^2 M_A }{2\pi}F^2(Q^2)\frac{Q_W^2}{4} \left[1+\left(1-\frac{T}{E_\nu}
\right)^2 - 
\frac{M_AT}{{E_\nu}^2} \right],
\end{equation}
where $Q_W = Q_W^n N + Q_W^p Z$ is called the weak charge of the nucleus with $Q_W^p = 1-4\sin^2\theta_W$ and $Q_W^n = -1$ 
being the weak charges of proton and neutron, respectively. The nuclear form factor $F(Q^2)$ is the Fourier transform 
of the nucleon density distribution in the nucleus and is given by
\begin{equation}
 F(Q^2)=4\pi \int dr r^2 \frac{\sin qr}{qr}\rho_N(r),
\end{equation}
where $\rho_N(r)$ is the common density distribution for the protons and neutrons. In the case of $N \ne Z$ nuclei and 
neglecting the axial-vector contribution, $F(Q^2)Q_W$ is replaced by
\begin{equation}
 F(Q^2)Q_W \to Z(1-4\sin^2\theta_W) F_p(Q^2) - N F_n(Q^2)
\end{equation}
with
\begin{equation}
 F_{p}(Q^2)= \frac{4\pi}{Z} \int dr r^2 \frac{\sin qr}{qr}\rho_{p}(r), \qquad \qquad F_{n}(Q^2)= \frac{4\pi}{N} \int dr 
 r^2 \frac{\sin qr}{qr}\rho_{n}(r).
\end{equation}
The form factor $F_{p}(Q^2)$ obtained from the analysis of electron scattering experiments by Helm~\cite{Helm:1956zz} and 
Klein-Nystrand~\cite{Klein:1999qj} using the Gaussian or the surface diffuse density distribution are generally used. A 
similar density distribution for neutrons $F_{n}(Q^2)$ with a radius parameter larger than the proton is used. In recent 
years, theoretical calculations for these form factors  based on the relativistic mean field, energy density functional, shell 
model, and coupled-clusters theory, to describe the nuclear structure have been made~\cite{Abdullah:2022zue}. However it has 
been shown that~\cite{Hoferichter:2020osn} in the low energy region of relevance to the coherent scattering, the nuclear model 
dependence of the form factors $F_{p,n}(Q^2)$ is quite small.

The numerical values of the weak charges $Q_W^{p,n}$ of protons and neutrons including the radiative corrections are given 
by~\cite{Abdullah:2022zue, Erler:2013xha}; $Q_W^{\nu_e, p} = 0.0766$, and $Q_W^{\nu_e, n} = -1.0233$ making the contributions 
from the protons to the CEvNS very small. Further assuming $T<<E_\nu$ and neglecting the proton contribution, i.e., $ F^2(Q^2) 
Q_W^2 \simeq F^2_{n}(Q^2) N^2$, we can write the recoil energy distribution as 
\begin{equation}\label{recoil-eq}
\frac{d\sigma}{dT}\simeq \frac{G_F^2 M_A }{4 \pi}N^2 F_n^2(Q^2)\left(1- \frac{M_AT}{2E^2_\nu}\right).
\end{equation}
In the above expression, radiative corrections have not been taken into account. Since the CEvNS is an elastic neutral 
current reaction in which the recoil nucleus is the only observable, therefore, the energy and direction of the recoiling 
nucleus need to be measured. While the recoil energy distribution is given in Eq.~(\ref{recoil-eq}), the directional 
distribution is given by
\begin{equation}\label{direc-dist}
\frac{d\sigma}{d\Omega_R}= \frac{G_F^2 }{16 \pi^2} Q^2_W |F(Q^2)|^2 E_\nu (1+\cos\theta_R),
\end{equation}
showing that the CEvNS is peaked in the forward direction $\theta_R\sim 0$, where $\theta_R$ is the angle between the 
incoming and the outgoing neutrinos~\cite{Abdullah:2022zue}.

The first experimental program to measure the CEvNS cross section was started by the COHERENT collaboration at ORNL
using its SNS facility using prompt monoenergetic neutrinos from $\pi^+ 
\to \mu^+ \nu_\mu$ decays and the delayed neutrinos from the subsequent  decays of muons i.e. 
$\mu^+ \to e^+ \nu_e \bar \nu_\mu$ with continuous energy spectra. These spectra are described by $\phi_{\nu}(E_\nu)$ written 
as~\cite{Akimov:2022oyb}:
\begin{eqnarray}
 \phi_{\nu_\mu}(E_{\nu}) &=& \frac{2 m_\pi}{m_\pi^2 m_\mu^2}~\delta\left(1-\frac{2E_\nu m_\pi}{m_\pi^2 -m_\mu^2}\right),\\
 \phi_{\nu_e}(E_{\nu}) &=& \frac{192}{ m_\mu}\left(\frac{E_\nu }{m_\mu}\right)^2~\delta\left(\frac{1}{2}-\frac{E_\nu }{m_\mu}
 \right),\\
 \phi_{\bar \nu_\mu}(E_{\nu}) &=& \frac{64}{ m_\mu}\left(\frac{E_\nu }{m_\mu}\right)^2~\delta\left(\frac{3}{4}-\frac{E_\nu }
 {m_\mu}\right).
\end{eqnarray}
The COHERENT collaboration reported the CEvNS cross sections on CsI(Na) using a crystal detector with a 14.6 kg mass target 
exposed for 308 days with observation of 134$\pm$22 events while the SM prediction is 173$\pm$48 
events~\cite{COHERENT:2017ipa}. A later analysis of the full CsI(Na) data with higher statistics reported the number of events 
to be 306$\pm$20 while the SM prediction is $341 \pm 11$~(theory) $\pm 42$~(expt.) events~\cite{Akimov:2021dab}. The dominant 
contribution to the systematic uncertainty is due to the uncertainty in the simulated flux of neutrinos which is around 
10$\%$. This corresponds to a flux averaged cross section $<\sigma > = 165^{+30}_{-25}\times 10^{-40}$~cm$^2$ and $\sin^2 
\theta_W = 0.22\pm 0.028 \pm 0.026$ at $Q^2 = 50$~MeV$^2$ using Helm model for the nuclear form factor. 

The second result reported from the COHERENT collaboration is from COH Ar-10 experiment in which a single phase liquid argon
detector was used with 24kg LAr as target and reported the observation of 159$\pm$43 CEvNS events~\cite{COHERENT:2020iec}. The 
uncertainties are due to the uncertainties in the neutrino flux as well as in the neutrino-nucleus cross sections and are 
about 13$\%$~\cite{COHERENT:2020iec, COHERENT:2020ybo}. This corresponds to the flux averaged cross section $<\sigma>=2.2 
\pm 0.7\times 10^{-39}$~cm$^2$ consistent with the prediction of SM $1.8\times 10^{-39}$~cm$^2$. Both of these cross 
sections measured in Cs and Ar are within $1\sigma$ of the prediction of SM.  

Encouraged by these measurements, the COHERENT collaboration has already planned many CEvNS experiments to be done in future 
with LAr~(24kg), LAr~(612kg), Ge~(18kg) and NaI(Tl)~(3388kg), $D_2O$~(600kg), Fe, Pb~(1000kg), etc. detectors. The 
scintillator experiments at other laboratories around the world have been planned proposing other nuclear targets like Xe, 
Pb, Si, Ge, Cu, etc. However, various experiments proposed for the search of dark matter are developing detectors which can 
also be used to observe CEvNS using astrophysical neutrinos from supernovae, solar or atmospheric neutrinos. An excellent 
compilation of such future experiments is given in Ref.~\cite{Abdullah:2022zue}.

The physics reach of the CEvNS is very rich. A high precision determination of the CEvNS observables like the energy and 
angular distributions of the recoiling nucleus provides opportunity to explore various physics topics because the 
theoretical uncertainties in calculating these observables are quite small. This is because the value of the mixing angle 
$\theta_W$ in SM and the nuclear form factors in the region of very low momentum transfer are quite well known, the 
uncertainty arises only due to the determination of the neutrino flux, which can be improved in future. In the following, we list 
some physics topics in the weak interaction physics where the study of CEvNS is likely to make important contributions:
\begin{itemize}
 \item Establishes the occurrence of coherence phenomenon in (anti)neutrino-nucleus scattering cross sections by 
 confirming the $N^2$~(Eq.~(\ref{recoil-eq})) dependence of CEvNS.
 
 \item With high precision data and a knowledge of the nuclear form factors directly determine the weak mixing angle 
 $\theta_W$ and complement its determination from the polarized electron scattering measurements, both of which directly 
 measure the weak charge $Q_W$ in terms of $\theta_W$.
 
 \item CEvNS observables can help to determine the electromagnetic properties of (anti)neutrinos. The electromagnetic 
 interactions being the charge conserving interactions can also contribute to the CEvNS observable with different type of 
 energy and angular distributions of the recoil nucleus. A high precision determination of these observables will determine 
 the EM properties of (anti)neutrinos like the
 \begin{itemize}
  \item charge radius of (anti)neutrino and its flavor dependence if any using the $\nu_e(\bar \nu_e)$ and $\nu_\mu(\bar 
  \nu_\mu)$ beams with $\pi$DAR and $\mu$DAR neutrinos at the accelerators.
  
  \item magnetic moment of the neutrinos by observing the recoil energy and angular distributions of the target nucleus in 
  $\nu(\bar \nu)$-nucleus scattering and its flavor dependence. The differential cross section in the presence of magnetic 
  moment of neutrino gives an additional contribution~\cite{Giunti:2015gga, Singh:1995hv}: 
  \begin{equation}
  \left(\frac{d\sigma}{dQ^2}\right)_{\text{MagMom}}=\left(\frac{\mu_{\nu}}{\mu_{B}}\right)^{2} \frac{\pi \alpha^2 Z^2}{m_e^2 
  Q^2} \left(1 - \frac{p \cdot q}{2M_{A} E_{\nu}} \right)^{2} F_{\text{ch}}^2(Q^2),
 \end{equation}
 where $p \cdot q = \frac{Q^2}{2} = -M_{A}T$, which may be written as
 \begin{equation}
  \left(\frac{d\sigma}{dT}\right)_{\text{MagMom}}=\left(\frac{\mu_{\nu}}{\mu_{B}}\right)^{2} \frac{\pi \alpha^2 Z^2}{m_e^2} 
  \left(\frac{1 - T/{E_\nu}}{T} + \frac{T}{4E_\nu^2}\right)F_{\text{ch}}^2(Q^2),
 \end{equation}
 where $\alpha$ is the fine structure constant and $F_{\text{ch}}(Q^2)$ is the charge form factor of the nucleus, with 
 $F_{\text{ch}} (0) = 1$, $\mu_{\nu}$ is the magnetic moment of the neutrino and $\mu_{B}$ is the Bohr magneton.
   
 \item observation of a nonvanishing magnetic moment could also help to distinguish between Dirac and Majorana neutrinos.
 \end{itemize}
 
\item CEvNS can be used as strong probe to study NSI and BSM physics 
specially in the vector sector where the CEvNS gets the dominant contribution. Any significant deviation from the SM 
predictions would indicate the presence of NSI interactions. The CEvNS with $N\ne Z$ nuclei and nonzero spin can be used to 
explore the NSI in the axial-vector sector.

\item The CEvNS with relatively higher energy beams can be used to determine the neutron distribution of nuclei as has been 
done using the PV observables in electron scattering. For example, the COHERENT result implies a neutron radius $R_n = 
5.0 \pm 0.7 $ fm and a neutron skin i.e. $R_n-R_p = 0.2 \pm 0.7$~fm for Cs, which is consistent with the theoretical 
calculations~\cite{Coloma:2020nhf}. Such measurements can be extended to other nuclei and would compliment the studies made 
using PV electron scattering.
\end{itemize}

\subsection{Neutrino trident production}
The neutrino trident production is a process in which an (anti)neutrino scattering from nucleus ``A'' produces a pair of 
charged leptons of opposite charges along with the (anti)neutrinos in the final state. The charge lepton pair produced in
the final state may be of the same flavor i.e. $l^+l^- (l=e,~\mu,~\tau)$ or of the mixed flavor like $\mu^\pm e^\mp,~\mu^\pm 
\tau^\pm,~\tau^\pm e^\mp$, consistent with the LFN conservation for each flavor. The reactions are 
represented as
\begin{eqnarray}
\nu_l({\bar\nu}_l) + A &\rightarrow& \nu_l({\bar\nu}_l) + l^{\prime -}(l^{\prime +}) + l^{\prime +}(l^{\prime -}) + 
A~~~~~~~~~
({\textrm{NC interaction}}), \nonumber\\
\nu_l({\bar\nu}_l) + A &\rightarrow& l^-(l^+) + l^{\prime -}(l^{\prime +}) + \bar{\nu}_{l^\prime}({\nu}_{l^\prime}) + 
A~~~~~~~~~
({\textrm{CC interaction}}).
\end{eqnarray}
Some of the reactions are induced by both the CC as well as NC, while the others by either CC or NC. The various possible reactions 
induced by CC and NC are shown in Table~\ref{tab:trident}.
\begin{table}[htb]
\begin{center}
 \begin{tabular}{|c|c|}\hline
  Scattering process & SM contribution \\\hline
  $\nu_\mu(\bar \nu_{ \mu}) A \to \nu_\mu(\bar \nu_{\mu}) \mu^-\mu^+ A$& CC + NC\\
  $\nu_\mu(\bar \nu_{ \mu}) A \to \nu_e(\bar \nu_{e}) e^\pm \mu^\mp A$ & CC \\
  $\nu_\mu(\bar \nu_{ \mu}) A \to \nu_\mu(\bar \nu_{\mu}) e^-e^+ A$& NC\\
  $\nu_e(\bar \nu_{e}) A \to \nu_e(\bar \nu_{e}) e^-e^+ A$& CC+NC\\
  $\nu_e(\bar \nu_{e}) A \to \nu_\mu(\bar \nu_{\mu}) \mu^\pm e^\mp A$& CC\\
  $\nu_e(\bar \nu_{e}) A \to \nu_e(\bar \nu_{e}) \mu^-\mu^+ A$& NC\\\hline
 \end{tabular}
\caption{Various (anti)neutrino induced trident processes from a nucleus $A$.}\label{tab:trident}
\end{center}
\end{table}

The neutrino induced trident production from the nuclear targets has been theoretically studied for more than 60 years, 
but the first detailed calculations were made by Czyz et al.~\cite{Czyz:1964zz} followed by many others using the effective 
V-A theory of weak interactions with and without using the charged intermediate vector bosons $(W^\pm)$.
\begin{figure}[htbp]
\centering
 \includegraphics[height=7cm, width=15 cm]{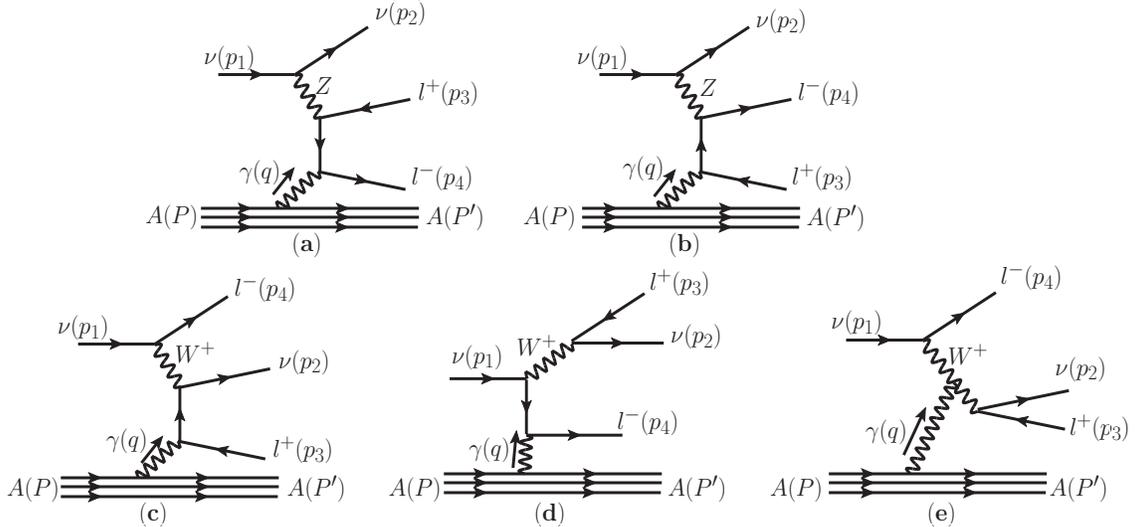}
 \caption{Leading hadronic contribution to the (anti)neutrino trident production in the SM.}
 \label{trident-production}
\end{figure}
A summary of the earlier calculations is given by Llewellyn Smith~\cite{LlewellynSmith:1971uhs}. After the prediction of 
NC in the SM and its contribution, it was found that NC mediated by $Z$ 
boson interferes destructively with the charged $W$ boson contribution in the $l^+l^- (l=e,~\mu,~\tau)$ channels, reducing 
the event rate of trident production by 40$\%$. Experimentally the first observation of (anti)neutrino induced trident 
production was reported by  the CHARM-II~\cite{CHARM-II:1990dvf} and CCFR~\cite{CCFR:1991lpl} collaborations followed by 
the NuTeV collaboration~\cite{NuTeV:1999wlw} in the $\nu_\mu(\bar \nu_{ \mu}) + Z\to \nu_\mu(\bar \nu_{ \mu}) + \mu^- + 
\mu^+ + Z$ channel. The ratios of these observed cross section to the SM predicted values 
are~\cite{Altmannshofer:2019zhy}:
\begin{equation}
\frac{\sigma(\nu_\mu\rightarrow \nu_\mu \mu^+ \mu^-)_{\textrm{experiment}}}{\sigma(\nu_\mu\rightarrow \nu_\mu \mu^+
\mu^-)_{\textrm{SM}}}=
\left\{
 \begin{array}{ll}
1.58 \pm 0.64 & \qquad \textrm{(CHARM-II)}\\
0.82 \pm 0.28 &  \qquad \textrm{(CCFR)}\\
0.72^{+1.73}_{-0.72} &  \qquad \textrm{(NuTeV)}
 \end{array}
\right.
\end{equation}
The (anti)neutrino trident production takes place in the Coulomb field of the nucleus and the charge leptons connect the 
nucleus through the photon exchange as shown in Fig.~\ref{trident-production}. Therefore, the hadronic part of the matrix 
element is described by the electromagnetic form factors. The dominant contribution to the trident production is given by 
the diagrams \ref{trident-production}a - \ref{trident-production}d, there are other additional contributions which are quite 
small i.e.
\begin{itemize}
 \item An additional contribution to all the diagrams in Fig.~\ref{trident-production}a - \ref{trident-production}d, due to 
 the $Z$ exchange of the hadronic vertex.
 
 \item At the leptonic vertex in the case of NC contributions, there could also be an electromagnetic production through the 
 photon exchange, if the neutrinos have magnetic moment $\mu_\nu$. The contribution of this term is very small, unless 
 $\mu_\nu$ is quite large. If sufficient number of trident events are observed, it could be used to put a limit on 
 $\mu_\nu$. With current experimental information on the trident events, the limits obtained are $10^{-8} 
 \mu_B$~\cite{Abdullah:2022zue} i.e two order of magnitude smaller than those obtained from the other 
 experiments~\cite{ParticleDataGroup:2020ssz}.
 
 \item There is an additional contribution from the diagram Fig.~\ref{trident-production}e, which is quite small due to two 
 $W$ boson propagators.
 
 \item In principle, the photon at the hadronic vertex can interact with the whole nucleus or the individual nucleons or 
 with the quarks in the nucleus. Depending upon the energy and the momentum carried by the virtual photon which is 
 transferred to the hadronic system leading to the coherent, diffractive or DIS production of 
 trileptons. It has been shown that the contribution to the total trident production is dominated by the coherent production 
 in the few GeV region of neutrino energies~\cite{Ballett:2018uuc}. 
\end{itemize}
In view of this, we focus on the coherent production of 
 tridents, which are contributed by the leading diagrams of Fig.~\ref{trident-production}a- \ref{trident-production}d with the 
 photon exchange at the hadronic vertex and the $W$ or $Z$ exchange at the leptonic vertex of the tridents and mention the 
 other processes only briefly. The reader is referred to the literature for the diffractive and the DIS production of 
 trident~\cite{Ballett:2018uuc}.
 
The earlier calculations of (anti)neutrinos induced trident production are done for the coherent production using the 
effective photon approximation~(EPA). In this approximation, the cross section for the full scattering process is 
calculated in two parts. In the first part, the cross section $\sigma(s)$ is obtained for the photo-trident production 
process i.e. $\gamma \nu \to \nu l^\pm l^{\prime \mp}$ with real photons with the CM energy~($\sqrt{s}$) of $\gamma \nu$ system 
using the effective $V-A$ theory or the SM. Quantitatively, the radiative contribution of these three processes to the trident 
production depends upon the specific pair of leptons produced in the final state and the energy region of the neutrinos. 
Typically the contribution of the diffractive scattering varies between $10\%-40\%$ being largest in the case of 
$\mu^-\mu^+$ channels but Magill~\cite{Magill:2016hgc} find a larger contribution. The contribution of DIS is 
smaller than $10\%$.

In the second part, this cross section is multiplied by the probability $P(s,q^2)$ of the nucleus producing a virtual photon 
with virtuality $q^2$ given by
\begin{equation}
 P(s,q^2)=\frac{Z^2e^2}{4\pi^2} \frac{ds}{s} \frac{dq^2}{q^2} |F(q^2)|^2,
\end{equation}
where $\sqrt{s}$ is the CM energy of the incoming neutrino and a real photon system, $Ze$ is the charge and $F(q^2)$ is the 
electromagnetic form factor of nucleus determined from the electron scattering experiments. In recent years, the applicability 
of EPA in the various kinematic regions of $s$ and $q^2$ has been described by some authors and a full calculation in the case 
of spin zero nuclei has been done as outlined below~\cite{Ballett:2018uuc, Magill:2016hgc}.

The full matrix element $\mathcal{M}$ for the trident production by the (anti)neutrinos corresponding to the 
diagrams \ref{trident-production}a-\ref{trident-production}d is given using the SM as
\begin{equation}
i \mathcal{M} = \mathrm{L}^\mu (\{p_j\},q) \, \frac{-ig_{\mu \nu}}{q^2} \, \mathrm{H}_{\rm X}^{\nu}(P,P^{\prime})\,; ~~~~
j=2-4 .
\end{equation}
The total leptonic amplitude $\mathrm{L}^\mu (\{p_j\},q)$ is given by
\begin{align}
\mathrm{L}^\mu & \equiv - \frac{ie G_F}{\sqrt{2}}[\bar{u}(p_2)\gamma^\tau (1-\gamma_5)u(p_1)] \times \bar{u}(p_4)
\left[\gamma_\tau (V_{\alpha\beta\kappa}-A_{\alpha\beta\kappa}\gamma_5)
\frac{1}{(\slashed{q}-\slashed{p}_3-m_3)}\gamma^\mu\right.\nonumber\\
& \left. + \gamma^\mu \frac{1}{(\slashed{p}_4-\slashed{q}-m_4)} \gamma_\tau (V_{\alpha\beta\kappa}-A_{\alpha\beta\kappa}
\gamma_5)\right] v(p_3)\, ,
\label{eq:Lmu}
\end{align}
and the total hadronic $\mathrm{H}_{\rm X}^{\nu}(P,P^{\prime})$ amplitude is given by
\begin{align}
H_{\rm X}^\nu  &\equiv \langle {H}(P) \vert J_\mathrm{EM}^\nu (q^2)\vert {H}(P^\prime)\rangle\, ,
\label{eq:Hmu}
\end{align}
where $q \equiv P - P^\prime$ is the four momentum transfer, $m_3$~($m_4$) the positively~(negatively) charged lepton mass,
$V_{\alpha\beta\kappa}~(A_{\alpha\beta\kappa})\equiv g_{V}^{\beta}(g_A^{\beta})\delta_{\beta\kappa}+\delta_{\alpha\beta}
\,(\beta=\alpha \, \mathrm{or} \; \kappa)$ the vector~(axial-vector) couplings, depending on the channel. ${J}^\nu_{\rm{EM}}
(q^2)$ is the electromagnetic current for the hadronic system $A$ (a nucleus or a nucleon). $V_{\alpha\beta k}$ and 
$A_{\alpha\beta k}$ for various processes are given in the Table~\ref{tab:neutrinoME}. 
\begin{table} 
\centering
\begin{tabular}{ cccccc }
\hline \hline
 $\nu$ Process                          ~~~&~~~ $\overline{\nu}$ Process
 ~~~&~~~ $V_{ijk}$                      ~~~&~~~ ~$A_{ijk}~$          ~~~&~~~ Mediator\\\hline\hline
 $\nu_e\rightarrow \nu_e e^+e^-$        & $\overline{\nu}_e\rightarrow \overline{\nu}_e e^+e^-$
 & $\frac{1}{2}+2\sin^2\theta_W$  & ~~$\frac{1}{2}$      & W,Z \\
 $\nu_\mu\rightarrow \nu_\mu \mu^+\mu^-$& $\overline{\nu}_\mu\rightarrow \overline{\nu}_\mu \mu^+\mu^-$
 & $\frac{1}{2}+2\sin^2\theta_W$  & ~~$\frac{1}{2}$      & W,Z \\
 $\nu_e\rightarrow \nu_\mu \mu^+e^-$    & $\overline{\nu}_e\rightarrow \overline{\nu}_\mu e^+\mu^-$
 & $1$                            & ~~$1$                        & W \\
 $\nu_\mu\rightarrow \nu_e e^+\mu^-$    & $\overline{\nu}_\mu\rightarrow \overline{\nu}_e \mu^+e^-$
 & $1$                            & ~~$1$                        & W \\
 $\nu_e\rightarrow \nu_e \mu^+\mu^-$    & $\overline{\nu}_e\rightarrow \overline{\nu}_e \mu^+\mu^-$
 & $-\frac{1}{2}+2\sin^2\theta_W$ & $-\frac{1}{2}$       & Z \\
 $\nu_\mu\rightarrow \nu_\mu e^+e^-$    & $\overline{\nu}_\mu\rightarrow \overline{\nu}_\mu e^+e^-$
 & $-\frac{1}{2}+2\sin^2\theta_W$ & $-\frac{1}{2}$       & Z \\
  $\nu_\mu \rightarrow \nu_\mu \tau^+ \tau^-$ & $\overline{\nu}_\mu \rightarrow  \overline{\nu}_\mu \tau^- \tau^+$
  & $-\frac{1}{2}+2\sin^2\theta_W$ &$-\frac{1}{2}$ & Z \\
 $\nu_\mu \rightarrow   \nu_\tau\mu^- \tau^+$ & $\overline{\nu}_\mu \rightarrow   \overline{\nu}_\tau\mu^+ \tau^-$ & 1
 & ~~1 & W\\
 $\nu_\tau \rightarrow  \nu_\mu \tau^- \mu^+ $ & $\overline{\nu}_\tau \rightarrow  \overline{\nu}_\mu \tau^+ \mu^- $ & 1
 & ~~1 & W\\
 $\nu_\tau \rightarrow  \nu_\tau\mu^+ \mu^-$ & $\overline{\nu}_\tau \rightarrow  \overline{\nu}_\tau\mu^- \mu^+$
 & $-\frac{1}{2}+2\sin^2\theta_W$ & $-\frac{1}{2}$ & Z\\
  $\nu_\tau \rightarrow  \nu_\tau e^+ e^-$ & $\overline{\nu}_\tau \rightarrow  \overline{\nu}_\tau e^- e ^+$
  & $-\frac{1}{2}+2\sin^2\theta_W$ & $-\frac{1}{2}$ & Z \\\hline\hline
\end{tabular}
\caption{Modified vector and axial coupling constants for the different combinations of incident neutrino flavors and final
states}\label{tab:neutrinoME}
\end{table}
Using this matrix element, the differential  cross section is given by
\begin{align}
\frac{\textrm{d}^2 \sigma_{\nu  {\rm X}}}{\textrm{d} Q^2 \textrm{d} \hat{s}}&= \frac{1}{32  \pi^2(s-M_{A}^2)^2}
\frac{\mathrm{H}_{\rm X}^{\mu\nu}\mathrm{L}_{\mu\nu}}{Q^4}\, ,
\end{align}
where
\begin{eqnarray}
 L_{\mu\nu}=\frac{1}{2s+1}\overline{\sum}\sum_{spins}L_\mu^\dag L_{\nu},\qquad \qquad
 H^{\mu\nu}=\frac{1}{2J+1}\overline{\sum}\sum_{spins}{H^{\mu}}^{\dagger} H^{\nu}.
\end{eqnarray}
In the case of coherent scattering, the hadronic matrix element is taken as the matrix element of electromagnetic current between nuclear 
states of initial and final momentum $P^\mu$ and $P^{\prime \mu}$, respectively, and for a spin zero nucleus ``$A$'', it is given by
\begin{equation}
 H=\langle A(P^\prime)|J_{EM}^\mu|A(P)\rangle F(Q^2)(P+P^\prime)^\mu,
\end{equation}
where $F(Q^2)$ is the nuclear form factor.

In the case of nucleons, the hadronic matrix element for the protons and neutrons is given as~\cite{Athar:2020kqn}:
\begin{equation}\label{eq:trident}
 \langle N(P)|J_{EM}^\mu|N(P^\prime)\rangle= e\bar u(P')\left[F_1^N(Q^2)\gamma^\mu+i\sigma^{\mu\nu}F_2^N(Q^2)\frac{q_\nu}
 {2M}\right]
 u(p)
\end{equation}
for $N=p,n$ and
\begin{equation}\label{eq:trident1}
 H^{\mu\nu}_N=ZH_p^{\mu\nu}(P,P^\prime)+(A-Z)H_n^{\mu\nu}(P,P^\prime).
\end{equation}
However, the nucleons are not free but bound with {a momentum and energy distribution} which is 
described by the spectral function $S(\vec{p},E)$. Such spectral functions for the nucleon have been determined experimentally 
from the electron-nucleus scattering from many nuclei and have been used extensively in the calculations of QE $\nu-A$ 
scattering. The simplest form of $S(\vec{p},E)$ is given by a theta function $\Theta(P_F-P)$ in the case of Fermi gas model 
in which the free nucleon cross section is multiplied by a quenching factor $R(\vec{q})$ to obtain the cross section from the 
bound nucleons~\cite{LlewellynSmith:1971uhs, Bodek:2021trq}, i.e.
\begin{equation}\label{eq:trident2}
 \frac{\textrm{d}^2 \sigma_{\nu - {\rm N}}}{\textrm{d} Q^2 \textrm{d} s}\to R(|{\vec q}|) \frac{\textrm{d}^2 \sigma_{\nu - 
 {\rm N}}}{\textrm{d} Q^2 \textrm{d} s},
\end{equation}
where
\begin{eqnarray}
R(|{\vec q}|)&=&\frac{3}{2}\frac{|{\vec q}|}{2p_F}-\frac{1}{2} \left(\frac{|{\vec q}|}{2p_F} \right)^3~~~~~~~~~~~\text{if} 
~~~|{\vec q}| < 2 p_F\\\nonumber 
&=&1,~~~~~~~~~~~~~~~~~~~~~~~~~~~~~~~~~~~\text{if} ~~~~|{\vec q}| > 2 p_F,
\end{eqnarray}
and $p_F$ is the Fermi momentum of the nucleons in the nucleus, generally taken to be different for proton and neutron in 
case of $N\ne Z$ nuclei. Taking the values of the hadronic current given in Eqs.~(\ref{eq:trident}) and (\ref{eq:trident1}), 
the cross section is calculated with help of Eq.~(\ref{eq:trident2}). Generally, the cross section for the trident production 
is calculated in terms of the photon flux for the longitudinal and transverse photons following the standard procedure. 
In the region of very low $Q^2,~Q^2\sim 0$ the contribution from the transverse photons dominates corresponding to the real 
photon, justifying the use of EPA in the coherent production. 

\begin{figure}[htbp]
\centering
 \includegraphics[height=7 cm, width=16 cm]{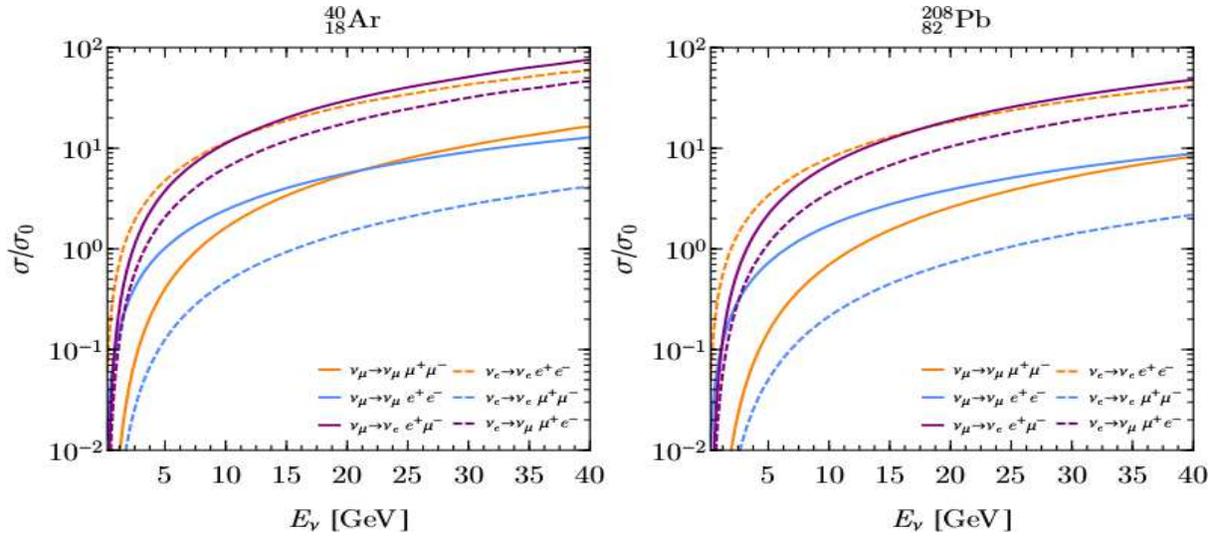}
 \caption{Coherent trident cross section for (anti)neutrino scattering on  $^{40}Ar$ and  $^{208}Pb$. The figures are taken 
 from Ref.~\cite{Ballett:2018uuc}.}\label{trident-production1}
\end{figure}

In the case of DIS, the cross section for the (anti)neutrino nucleon scattering $\frac{d\sigma_{\nu N}}{dQ}$ is 
obtained by convoluting the (anti)neutrino parton~(q) cross section $\frac{d\sigma_{\nu q}}{dQ}$ with the nucleon~($N=p,n$) 
parton distribution functions $f_q^N(\xi,Q)$, with $Q$ being the four momentum transfer and is given by~\cite{Magill:2016hgc}:
\begin{equation}\label{signu}
 \sigma_{\nu N}= \sum_q \int_{\xi_{min}}^1 d\xi \int_{Q_{min}}^{Q_{max}} dQ \frac{d\sigma_{\nu q}(\xi,q)}{dQ}f_q^N(\xi,Q),
\end{equation}
where $\xi$ is the fractional momentum of the partons~($q$) with $\xi_{min}\geq 0$ to enable the creation of lepton pair. 
With $\sigma_{\nu N}$ calculated from Eq.~(\ref{signu}), the DIS contribution to the nuclear cross section is obtained 
using
\begin{equation}
\sigma_{\nu A} = Z\sigma_{\nu p}+ (A-Z)\sigma_{\nu n}.
\end{equation}
The details for calculating $\frac{d\sigma_{\nu q}}{dQ}$ can be found in Ref.~\cite{Magill:2016hgc}. 
Fig.~\ref{trident-production1} shows some representative results~(reactions given in Table~\ref{tab:trident}) for coherent 
trident cross section of (anti)neutrino scattering on $^{40}$Ar and $^{208}$Pb~\cite{Ballett:2018uuc}.

In recent years, the rare process of the neutrino trident production has been studied in some detail as a probe of new 
physics proposed in many theoretical models of BSM physics, which propose the existence of new 
vector bosons~($Z^\prime$) and new scalar boson~($S^\prime$), which couple to the leptons and the 
quarks~\cite{Magill:2016hgc, Ge:2017poy}. The effect of coupling of these new bosons to the leptons at the leptonic vertex has 
been studied for the different values of mass and coupling of these bosons and varying them in the reasonable limits, which are 
constrained by the limits obtained mainly from the recent experiment in muon~($g-2$) measurements and the limits obtained 
from the search of such bosons in the BaBar and Belle experiments~\cite{Altmannshofer:2014pba}. In view of the importance of 
such theoretical studies, many experiments to measure the neutrino trident production with accelerator and atmospheric 
neutrinos have been proposed. 

A summary of the feasibility of seeing neutrino trident production events in these experiments using accelerator neutrinos 
has been given by Ballett et al.~\cite{Ballett:2018uuc}. On the other hand, the use of atmospheric neutrinos to see the 
neutrino trident production in IceCube, ARCA and DeepCore has been discussed by Ge et al.~\cite{Ge:2017poy}.


\subsection{Exclusive reactions in $\nu-$nucleus scattering in the low energy region}
With the increase in (anti)neutrino energies, corresponding to the de Broglie wavelength of the (anti)neutrinos being 
smaller than the nuclear radius, the interaction takes place with the individual nucleons in the nucleus, leading the 
target nucleus to be in any one of its excited states defined by a definite spin~($\vec{S}$), parity~($P$), and angular 
momentum~($\vec{J}$) or its breaking leading to a residual nucleus which can be in its ground state or in an excited state. 
These are called exclusive reactions. In both cases particles like photons, electrons, nucleons or alpha particles can be 
emitted as a decay product of the excited nuclear states through $\alpha, \beta$ and $\gamma$ decays or as the direct emission 
of these particles in the knock out reactions in case of the emission of the nucleons and alpha particles. With further 
increase in energy, new particles like mesons~($\pi,~2\pi, K, ~\eta,~\rho$, etc.) or associated production of strange 
particles~($K \Lambda, ~K\Sigma$, etc.) can take place. These reactions can be induced by the CC as well as 
NC in the $\nu-$nucleus reactions using $\nu_e, \nu_\mu$ and $\nu_\tau$ beams. However, in the low energy 
region~(below $\tau$ production threshold), the CC reactions are induced only by the $\nu_e (\bar\nu_{e})$ and $\nu_\mu 
(\bar\nu_{\mu})$, while NC reactions can be induced by (anti)neutrinos of all flavors. In this section, we discuss the 
low energy exclusive neutrino-nucleus reactions in which the final nucleus is either in the ground state or in an excited 
state, for example
\begin{eqnarray}
 \nu_{e(\mu)} + \;^{A}_ZX_{N} &\rightarrow& e^-(\mu^-) +\; ^{A}_{Z+1}Y_{N-1} \left(^{A}_{Z+1}Y_{N-1}^*\right) \nonumber \\
 \bar \nu_{e(\mu)} + \;^{A}_ZX_{N} &\rightarrow& e^+(\mu^+) +\; ^{A}_{Z-1}Y_{N+1} \left(^{A}_{Z-1}Y_{N+1}^*\right)\nonumber\\
 \nu_{e,\mu,\tau} + \;^{A}_ZX_{N}&\rightarrow& \nu_{e,\mu,\tau} + {^{A}_ZX_{N}} \left({^{A}_ZX_{N}^*}\right)
\end{eqnarray}
Historically, the first study of the exclusive $\bar\nu$-nucleus scattering was done by Cowan et al.~\cite{Cowan:1956rrn} with 
the reactor antineutrinos with $E_{\bar\nu_e}\leq 10$~MeV on Cl target leading to the detection of antineutrino. In the very 
low energy region, the ``superallowed'' Fermi and Gamow-Teller transitions to the ground state of final nucleus take place. 
The theoretical calculations of the neutrino-nucleus cross section of these reactions have minimal uncertainties arising due 
to the theoretical inputs. This is because most of the nuclei are theoretically well described in their ground state and the 
$Q^2$ dependence of the nuclear form factor is almost constant in this low energy region. Moreover, the parameters describing 
the nuclear transitions involved in these reactions, are also well determined from either the $\beta$ decays of these nuclei 
or the low energy ($p,n$) reactions. As the (anti)neutrino energy increases relevant to the $\nu_e(\bar\nu_e)$ and $\nu_\mu 
(\bar\nu_\mu)$ beams from the pion and muon decays at rest (as well as decays in flight) available at the particle accelerators, 
many nuclear states are excited needing information about the nuclear wave functions of various nuclei in their ground 
state as well as in the excited states, which are reachable by both the allowed and forbidden Fermi and Gamow-Teller 
transitions. In the simplest description of neutrino-nucleus reactions, the impulse approximation with the shell model wave 
functions calculated using different types of nucleon-nucleon potentials has been used in literature to study the exclusive 
neutrino reactions from nuclear targets. The effect of nucleon-nucleon correlations in the nuclear wave functions have also 
been included in a few calculations using various theoretical tools like the RPA, CRPA, QRPA etc. Moreover, the contribution 
of the meson exchange current~(MEC) and other subnuclear degrees of freedom have also been included in some 
calculations~\cite{Ruso:2022qes}.

We present in the following, the basics of the general formalism for describing the exclusive (anti)neutrino-nucleus 
reactions applicable at low and intermediate energies. The basic CC and NC reactions on nuclear targets take place on 
nucleon $N=n,p$ and are written as:
\begin{eqnarray}
 \nu_l + n &\longrightarrow& l^- + p, \qquad \qquad {\bar\nu}_l + p \longrightarrow l^+ + n, \qquad \quad 
 l=e,\mu,\tau \qquad ~\Delta S=0 \text{ (CC)} 
 \nonumber\\
 \nu_l ({\bar\nu}_l)+n(p) &\longrightarrow& \nu_l({\bar\nu}_l)+n(p), \qquad \qquad \qquad \qquad \qquad\qquad 
 \qquad~~\qquad \qquad \Delta S=0\text{ (NC)} \nonumber\\
 \label{lepton:reaction}
 {\bar\nu}_l +n(p) &\longrightarrow &l^{+} + \Sigma^{-}~(\Lambda,\Sigma^{0}), \qquad \qquad \qquad \qquad \qquad 
 ~~\qquad\qquad \quad\quad |\Delta S|=1 \text{ (CC)} 
\end{eqnarray}
for which the matrix elements of the weak leptonic $l^\mu$ and hadronic $J^\mu$ 
currents are given in Eqs.~(\ref{lep_curr}) and (\ref{had_curr}). 

In case of the nucleons bound in the nucleus, these matrix elements are to be taken between the initial~($\ket{i}$) and 
final~($\ket{f}$) nuclear states i.e.
\begin{eqnarray}
{\cal M}_{fi}^{CC(NC)}&=&\langle f|{\cal{H}}_W^{CC(NC)}|i\rangle,\nonumber\\
\text{where}~~~~~~~~{\cal{H}}&=& -\frac{G_F {\it a}}{\sqrt{2}}\int d\vec{x}~ l^\mu_{CC(NC)}~ J_{\mu}^{CC(NC)}(x),
\end{eqnarray}
$l^{\mu}_{CC(NC)}$ is the leptonic current, $J_{\mu}^{CC(NC)}$ is the hadronic current operator in the nucleus, 
which are given 
in Eqs.~(\ref{had_curr}) and (\ref{jmu:NC}), and
${\it a}=\cos\theta_C (1)$ for CC~(NC) induced processes. 

Since the leptons in Eq.~(\ref{lepton:reaction}) are free point particles, therefore, we can describe them by plane 
waves~(neglecting the Coulomb effect of the charged lepton in final state) to write 
\begin{eqnarray}
l_{\mu}(x)= l_{\mu}e^{-i\vec{q}\cdot \vec{x}},~~~~~~ q=k'-k=p-p',
\end{eqnarray}
such that the matrix element ${\cal M}_{fi}$ between the initial~($|i\rangle$) and final~($|f\rangle$) states, using the 
notation $l^\mu$ for $l^{\mu}_{CC(NC)}$ and $J^\mu$ for $J_{\mu}^{CC(NC)}$ for simplicity, is written as 
\begin{eqnarray}\label{Ch14:eq:had}
{\cal M}_{fi}&=& - \frac{G_F {\it a}}{\sqrt{2}}\langle f| \int e^{-i\vec{q}\cdot \vec{x}} l^\mu J_\mu(x)d\vec{x}|i\rangle=
-\frac{G_F {\it a}}{\sqrt{2}}\langle f|\int e^{-i\vec{q}\cdot \vec{x}} (l^0J_0-\vec{l}\cdot \vec{J})d\vec{x} |i\rangle.
\end{eqnarray}
The matrix element  ${\cal M}_{fi}$ is calculated using the multipole expansion of the $e^{-i\vec{q}\cdot \vec{x}}$ and 
$\vec{l}e^{-i\vec{q}\cdot \vec{x}}$. For that, we write
\begin{eqnarray}\label{Ch14:MPEl}
\vec{l}=\sum_{\lambda=0,\pm 1}l_{\lambda}\hat{e}_{\lambda}^\dagger,
\end{eqnarray}
where $\hat{e}_{\lambda}(\lambda=\pm 1,0)$ are the components of the unit vector $(\hat{e}_{x},\hat{e}_{y},\hat{e}_{z})$ in 
the spherical basis defined as 
\begin{eqnarray}
e_{\pm1}=\pm \frac{\hat{e}_{x}\pm i \hat{e}_{y}}{\sqrt{2}},~~~~~ \hat{e}_0=\hat{e}_z,~\vec{e}_\lambda \cdot 
\vec{e}_{\lambda^\prime}=\delta_{\lambda\lambda^\prime}, \qquad {\text{such that}} \qquad l_{\lambda}=\sum \vec{l}\cdot 
\hat{e}_{\lambda}
\end{eqnarray}
 and write
\begin{eqnarray}
e^{i\vec{q}\cdot \vec{x}}=\sum_{J=0}^{\infty}\sqrt{4\pi(2J+1})~ i^J j_J(qx)Y_{J0}(\Omega_{x}),
\end{eqnarray}
where $Y_{J0}(\Omega_x)$ are the spherical harmonics and $j_J(qx)$ are the spherical Bessel's function.

Using the definition of the vector spherical harmonics $\vec{\cal Y}_{Jl1}^M$ defined as
\begin{eqnarray}\label{ch14:cal_Y_VSH}
\vec{\cal Y}_{Jl1}^M= \sum_{m\lambda}\langle lm1\lambda|l1 JM\rangle Y_{lm}(\theta,\phi)\vec{e}_{\lambda},
\end{eqnarray}
where $\langle lm1|\lambda|l1J M\rangle$ is the Clebsch-Gordan~(CG) coefficients and choosing $\hat{q}\parallel \hat{e}_Z$ 
i.e. unit vector along the $Z$ axis, we can write
\begin{equation}\label{Ch14:MPE1}
 \vec{e}_\lambda e^{i \vec{q}\cdot \vec{x}}= \sum_{l}\sum_{J=0}^{\infty} \sqrt{4\pi(2l+1)}~ i^{J}j_{l}(qx)\langle l01
 \lambda|l1J\lambda\rangle\vec{\cal Y}_{JL1}^\lambda, ~~~~~x=|\vec{x}|,~~~~~
q=|\vec{q}|.
\end{equation}

In Eq.~(\ref{Ch14:MPE1}), we perform the expansion over $l$, using the values of the CG coefficients for $\lambda=\pm1$ and 
$\l=0$ explicitly. There would be, in general, three terms for each $\lambda(= \pm 1,0)$, corresponding to $l=J+1, J,J-1$. 
Using the nonvanishing values of CG coefficients in each case and the following properties of the vector spherical harmonics 
$\vec{\cal Y}_{Jl1}^M$ i.e.~\cite{Walecka:1975}:
\begin{eqnarray}
\vec{\nabla}_{r}\times j_{J}(r)\vec{\cal Y}_{JJ1}^M&=&-i\Big(\frac{J}{2J+1}\Big)^{\frac{1}{2}}j_{J+1}(r)\vec{\cal 
Y}_{J,J+1,1}^M+i\Big(\frac{J+1}{2J+1}\Big)^{\frac{1}{2}}j_{J-1}(r)\vec{\cal Y}_{J,J-1,1}^M,~~~~\\
\vec{\nabla}_{r} j_{J}(r) Y_{JM}&=&\Big(\frac{J+1}{2J+1}\Big)^{\frac{1}{2}}j_{J+1}(r)\vec{\cal Y}_{J,J+1,1}^M 
\Big(\frac{J}{2J+1}\Big)^{\frac{1}{2}}j_{J-1}(r)\vec{\cal Y}_{J,J-1,1}^M, ~~~~
\end{eqnarray}
the expression for $\vec{e}_{\lambda} e^{i \vec{q} \cdot \vec{x}}$ given in Eq.~(\ref{Ch14:MPE1}) is evaluated.

After performing some basic algebraic manipulations, the following expressions are obtained~\cite{Athar:2020kqn}:
\begin{eqnarray}\label{eq:lambda}
\vec{e}_{\vec{q}\lambda}e^{i\vec{q}\cdot\vec{x}}&=&-\frac{i}{q}\sum_{J=0}^\infty[4\pi(2J+1)]^{\frac{1}{2}}i^J\vec{\nabla}
(j_J(qx)Y_{J0}(\Omega_x)),~~\qquad \qquad \qquad\qquad ~~~~\text{for}~\lambda=0\nonumber\\
&=&-\sum_{J\geq 1}^\infty[2\pi(2J+1)]^{\frac{1}{2}}i^J\Big[\lambda_{j_J}(qx)\vec{\cal Y}_{JJ1}^\lambda+\frac{1}{q}
\vec{\nabla}\times(j_J(qx))\vec{\cal Y}_{JJ1}^\lambda\Big],~~~~~~~\text{for}~\lambda=\pm1.
\end{eqnarray}
Therefore, the matrix element in Eq.~(\ref{Ch14:eq:had}) is written using Eq.~(\ref{eq:lambda}) as~\cite{Athar:2020kqn}
\begin{eqnarray}
 \langle f |\hat{H}_W|i\rangle &=&+\frac{G_F {\it a}}{\sqrt{2}}\langle f|\Big(\sum_{J=0}^\infty[4\pi(2J+1)]^{\frac{1}{2}}(-i)^J 
 [l_3\hat{{ L}}_{J0}(q)-l_0\hat{{ C}}_{J0}(q)] \nonumber\\
 &&-\sum_{\lambda=\pm1}l_\lambda\sum_{J\geq 1}^\infty
 [2\pi(2J+1)]^{\frac{1}{2}}(-i)^J\times[\lambda \hat{{{T}}}_{J-\lambda}^{mag}(q)+\hat{T}_{J-\lambda}^{el}(q)]\Big)|i\rangle,
\end{eqnarray}
where
\begin{eqnarray}\label{Ch14-MPEC}
\hat{ C}_{JM}(q)&\equiv& \int d\vec{x}[j_J(qx)Y_{JM}(\Omega_x)]{ \hat{J}}_0(\vec{x}); \qquad \quad 
\hat{ L}_{JM}(q)\equiv  \frac{i}{q}\int d\vec{x}[\vec{\nabla}(j_J(qx)Y_{JM}(\Omega_x))]\cdot\hat{\vec{J}}(\vec{x}),\\
\label{Ch14-MPEmag}
\hat{T}_{JM}^{el}(q)&\equiv&  \frac{i}{q}\int d\vec{x}[\vec{\nabla}{\times}{\vec j}_J(qx)\vec{\cal Y}_{JJ1}^M]\cdot\hat{\vec{J}}
(\vec{x}); \qquad \quad
\hat{T}_{JM}^{mag}(q)\equiv \int d\vec{x}[j_J(qx)\vec{\cal Y}_{JJ1}^M]\cdot \hat{\vec{J}}(\vec{x}),
\end{eqnarray}
and are called multipoles.

In the following, we enumerate some features of the above multipoles: 
\begin{itemize}
 \item[(i)] The $C_{JM}(q)$, $L_{JM}(q)$, $T_{JM}^{el}(q)$ and $T_{JM}^{mag}(q)$ are called, respectively, the Coulomb, 
 longitudinal, transverse electric, and transverse magnetic multipoles of the current.
 
 \item[(ii)] The weak current operator $J^\mu(J^0,\vec{J})$ appearing in the definition of multipoles contains 
 vector~($V^\mu$) and axial-vector ($A^\mu$) currents in both cases of CC and NC reactions. Therefore, each 
 multipole~($M_{JM}=C_{JM},L_{JM},T^{el}_{JM},T^{mag}_{JM}$) consists of the vector and axial-vector multipoles and is 
 generally written as:
 \[M_{JM}\longrightarrow M_{JM}^V(q) +M_{JM}^{A}(q),\]
 where $M_{JM}^V(q)$ and $ M_{JM}^{A}(q)$ are the multipoles corresponding to vector and axial-vector currents.
 
 \item[(iii)]  The parity of $M_{JM}^V$ is defined in the conventional way with reference to the electromagnetic vector 
 current. The parity of the vector~($M_{JM}^V(q)$) and axial-vector~($M_{JM}^{A}(q)$) multipoles is opposite to each 
 other, which are shown in Table-\ref{multipole_parity}.
   \begin{table}[h]
\begin{center}
 \begin{tabular}{|c|c|c|c|c|c|c|c|c|}
 \hline
 Multipole&$C_{JM}^V$&$L_{JM}^{V}$&$T_{JM}^{el,V}$&$T_{JM}^{mag,V}$&$C_{JM}^A$& $L_{JM}^{A}$&$T_{JM}^{el,A}$&
 $T_{JM}^{mag,A}$\\\hline
 Parity&$(-1)^J$&$(-1)^J$&$(-1)^J$&$(-1)^{J+1}$&$(-1)^{J+1}$ & $(-1)^{J+1}$ & $(-1)^{J+1}$ & $(-1)^{J}$
\\\hline
 \end{tabular}
 \end{center}
  \caption{Parity of vector and axial-vector multipoles.}
 \label{multipole_parity}
\end{table}

 \item[(iv)] In general, there are 8 multipoles to be considered, four corresponding to the vector currents and four 
 corresponding to the axial-vector currents. However, since the vector current is conserved i.e. 
 \begin{equation}
  q_\mu J^\mu=0 \qquad \Rightarrow \qquad
 q_0J^0=\vec q\cdot \vec J.
 \end{equation}
Taking $\vec{q}\parallel \hat{e}_z$, we get a relation between the Coulomb and longitudinal multipoles i.e. 
\begin{equation}\label{ch14-MPE_jmu}
 q_0 \langle J_f |C_{JM}^V|J_i\rangle - {q_Z}\langle J_f |L_{JM}^V|J_i\rangle=0.
\end{equation}
Therefore, the $\nu(\bar\nu)$ cross sections are given in terms of seven multipoles while the electron scattering is 
described in terms of three multipoles.

\item[(v)] The single nucleon current operators~($J_{0}$, $\vec{J}$) to be used with the nuclear wave functions in the 
impulse approximation are derived from the definition of the vector and axial-vector current operators for the free nucleon 
given in Section~\ref{nu_QE}. In the case of a nucleus, the nucleons are treated nonrelativistically, therefore, the 
nonrelativistic reduction of the current operators can be used. In the case of CC reactions, we obtain $J^\mu_{CC}$ in the 
lowest order of momenta, neglecting the term $O\left(\frac{\vec{q\;}^2}{M^2}\right)$, $O\left(\frac{\vec{p\;}^2}{M^2} 
\right)$ as~\cite{Athar:2020kqn}
\begin{eqnarray}\label{Ch14-MPE0}
 J^0_{CC}&=&\left(f_1(q^2)+g_1(q^2)\vec \sigma \cdot \frac{2\vec p-\vec q}{2M}\right)\tau^{\pm},\\
 \label{Ch14-MPEvec}
 \vec{J}_{CC}&=&\left(g_1(q^2)\vec \sigma -i(f_1(q^2) + f_2(q^2))\frac{\vec \sigma\times\vec q}{2M}\right)\tau^{\pm}+f_1(q^2)
 \frac{2\vec{p}-\vec{q}}{2M}\tau^{\pm}.
\end{eqnarray}
The operator $\tau^{+(-)}$ corresponds to the $\nu(\bar \nu)$ scattering processes. Similar expressions are obtained for the 
NC interactions with $f_1(q^2)$, $f_2(q^2)$ and $g_1(q^2)$ replaced by NC form factors ${\tilde f}_1(q^2)$, 
${\tilde f}_2(q^2)$ and ${\tilde g}_1(q^2)$ and $\tau^{+(-)}$ replaced by the isoscalar~($\mathbb{I}_4$) and isovector 
operators~($\tau_3$) depending upon the isospin structure of $J_{NC}^{\mu}$.

It should be noted that the terms involving $\frac{q_0}{2M}$ are of the order of $O\big(\frac{{q}^{~2}}{4M^2}\big) $ as
$q_0=-\frac{q^2}{2M}$ for the elastic scattering and are, therefore, neglected in the case of nuclear transitions at low 
energies. The nuclear operators corresponding to the nucleon operators given in Eqs.~(\ref{Ch14-MPE0}) and 
(\ref{Ch14-MPEvec}) are, therefore, written in the impulse approximation as
\begin{eqnarray}\label{Ch14-MPE_j0}
 J^{0s}(\vec{x})&=&\sum_{j=1}^A\left[f_1(q^2) +g_1(q^2)\left( \frac{p(j)}{M}\delta(x-x_0)\right)_{sym}\right] \tau^{\pm} 
 \delta(\vec{x} - \vec{x}_j),\\
 \label{Ch14-MPE_jvec}
 \vec{J}^{s}(\vec{x})&=&\sum_{j=1}^A\left[ g_1(q^2)\vec\sigma + f_1(q^2)\frac{2\vec p\cdot \vec q}{2M} -i \frac{f_1(q^2) + 
 2M(q^2)}{2M}\vec\sigma(j)\times \vec q\right]\tau^{\pm} \delta(\vec{x} - \vec{x}_j),
\end{eqnarray}
where $\vec x_{j}$ is the position coordinate of the interacting nucleon. This shows that various operators in the nuclear 
space which enter in the current $J^0(x)$ and $\vec J(x)$ operators are the type $\tau^\pm(j), \tau^\pm(j)\sigma(j)$ and 
$\tau^\pm(j)\vec p(j)(=-i\tau^{\pm}(j) \vec{\nabla}(j))$ multiplied by the spherical harmonics~($Y_{JM}$), vector spherical 
harmonics~($\vec {Y}_{JMl}$), the gradient~($\vec{\nabla}\cdot \vec{Y}$), and the curl~($\vec{\nabla}\times 
\vec{Y}$) operators of the vector spherical harmonics as shown in the definition of the multipoles in 
Eqs.~(\ref{Ch14-MPEC}) and (\ref{Ch14-MPEmag}).
\end{itemize}
Using the matrix element in Eq.~(\ref{Ch14:eq:had}), the cross section is calculated for the transition between the initial 
state $\ket{i}$ and the final state $\ket{f}$ of the nucleus, which are defined by the definite angular momenta and parity as 
$\ket{J_i M_i}$ and $\ket{J_f M_f}$ and is given by
\begin{eqnarray}\label{MM:dsig}
\frac{d\sigma}{d\Omega}= \frac{k^\prime E^\prime}{4\pi^2} \sum_{\text{lepton spins}} \frac{1}{2J_i+1}\sum_{{\cal M}_i}
\sum_{{\cal M}_f} |\bra{J_f {\cal M}_f} {\cal H} \ket{J_i {\cal M}_i}|^2.
\end{eqnarray}
Since the matrix element of ${\cal H}$ is written in terms of the various multipoles with definite angular momentum and 
parity, $\ket{J \lambda}$ (see Eqs.~(\ref{Ch14-MPEC}) and (\ref{Ch14-MPEmag})), the standard angular momentum algebra can be 
used to calculate the cross section given in Eq.~(\ref{MM:dsig}). The general expression for the cross section is obtained as~\cite{Athar:2020kqn}: 
\begin{eqnarray}\label{dsig:Multi}
 \left(\frac{d\sigma}{d\Omega} \right)_{\nu\bar{\nu}}&=&\Big(\frac{q}{\epsilon}\Big)\frac{G^2\epsilon^2}{4\pi^2}\frac{4\pi}
 {2J_i+1}\Big[\Big[\sum_{J=0}^\infty\lbrace(1+\hat{\nu}\cdot\vec{\beta})|\langle J_f||\hat{M}_J||J_i\rangle|^2\nonumber\\
 &+&[1-\hat{\nu}\cdot\vec{\beta}+2(\hat{\nu}\cdot\vec{\beta})(\hat{q}\cdot\vec{\beta})]|\langle J_f||\hat{L}_J||J_i\rangle|^2\nonumber\\
 &-&[\hat{q}\cdot(\hat{\nu}+\vec{\beta})]~2~Re\langle J_f||\hat{L}_J||J_i\rangle\langle J_f||\hat{M}_J||J_i\rangle^*\rbrace\nonumber\\
 &+&\sum_{J\geq 1}^\infty\lbrace [1-(\hat{\nu}\cdot\hat{q})(\hat{q}\cdot\vec{\beta})][|\langle J_f||\hat{T}_J^{mag}||J_i\rangle|^2+|\langle J_f||\hat{T}_J^{el}||J_i|^2]\nonumber\\
&\pm &[\hat{q}\cdot(\hat{\nu}-\vec{\beta})]~2~Re\langle J_f||\hat{T}_J^{mag}||J_i\rangle\langle J_f||\hat{T}_{J}^{el}||J_i
\rangle^*\rbrace\Big]\Big],
\end{eqnarray}
where $\hat{\nu}\equiv \frac{\vec{\nu}}{|\vec{\nu}|}$, $\hat{q}\equiv \frac{\vec{q}}{|\vec{q}|}$,
$\bra{J_{f}} | \vec{J} | \ket{J_{i}}$ is the reduced matrix element of the multipole $M_{J}^{M}$ with angular momentum 
$\vec{J}$. In the relativistic limit, $\beta$ which depends upon the lepton velocity becomes $\beta \rightarrow 1$.
 
In order to calculate the nuclear matrix elements, we need the nuclear wave functions for $\ket{i}$ and $\ket{f}$ states 
which are essentially the nonrelativistic wave functions of the nucleons bound in a nucleus by a nucleon-nucleon potential. 
 
The nuclear states $\ket{i}$ and $\ket{f}$ characterized by the angular momentum~(and parity) $\ket{J_i M_i}$ and 
$\ket{J_f M_f}$ are generally expressed as:
\begin{eqnarray}
\ket{i}\equiv \ket{J_i M_i}\equiv \psi_{n l\frac{1}{2}J_i M_i}(\vec{x})&=& NR_{nl}(r)[Y_{lm_l}(\theta,\phi)\otimes 
Y_{\frac{1}{2}m_s}]_{J_iM_i} ,\\
\text{and} \qquad \qquad \ket{f}=\ket{J_f M_f}=\psi_{n^\prime l^\prime\frac{1}{2}J_fM_f}(\vec{x}) &=& N^\prime R_{n^\prime 
l^\prime}(r)[Y_{l^\prime m^\prime}(\theta,\phi)\otimes Y_{\frac{1}{2}m_{s^\prime}}]_{J_f M_f} ,
\end{eqnarray}
where $R_{nl}(r)$ and $R_{n^\prime l^\prime}(r)$ are the radial wave functions of the initial and final nucleus obtained by 
solving the Schr\"{o}dinger equation for nucleons moving in a central potential like the harmonic oscillator potential. In 
a more sophisticated description of the nuclear wave functions, various forms of the nucleon-nucleon potentials are included to 
describe the residual interactions to take into account the effect of pairing and other nucleon-nucleon correlation 
effects~\cite{Athar:2020kqn}. Using these wave functions and the transition current operators written in 
Eqs.~(\ref{Ch14-MPE_j0}) and (\ref{Ch14-MPE_jvec}), and the multipoles defined in 
Eqs.~(\ref{Ch14-MPEC}) and (\ref{Ch14-MPEmag}), the matrix element defined in Eq.~(\ref{Ch14:eq:had}) is calculated. The 
expressions for the matrix elements for various multipoles using the harmonic oscillator wave functions have been given by 
Haxton~\cite{Haxton:2008zza}.

Eq.~(\ref{dsig:Multi}) is the general result, which is used to calculate the nuclear cross section for QE $\nu$ and 
$\bar{\nu}$ reactions leading to discrete nuclear states in various nuclei from the deuteron to medium and heavy nuclei. Once 
the initial and final states are fixed, only the multipoles which are compatible with the change in angular momentum and 
parity would contribute. It should be noted that for an exclusive reaction leading to a specific transition, very few 
multipoles carrying spin and parity corresponding to $\Delta J$ and parity change in the transitions would contribute. 

Moreover, in case of light nuclei like H, D, $^3$He, $^4$He and $^{12}$C, these cross sections have also been calculated 
treating these nuclei as elementary particles. This method known as the elementary particle treatment~(EPT) was introduced 
by Fuji and Yamaguchi~\cite{10.1143/PTP.31.107}, and Kim and Primakoff~\cite{Kim:1965zzc} almost 60 years ago, and has been used 
later by Mintz et al.~\cite{Mintz:1980ez}, specially for $^2$D, $^3$He and $^{12}$C nuclei. In the case of deuteron target, 
many calculations have been made using EPT as well as the relativistic wave functions of the deuteron, with the parameters of 
these wave functions determined from extensive experimental efforts in the study of electrodisintegration of the deuteron 
i.e. $e+d\rightarrow e+n+p$. Very recently, an alternative approach based on the effective field theory~(EFT) has been used to 
calculate the low energy CC and NC weak processes on the deuteron like the $\nu_e+d \rightarrow e^- +p +p$ and $\nu_\mu +d 
\rightarrow \nu_\mu +n +p$ in the context of solar neutrino experiments at SNO and applied to antineutrino induced reactions 
relevant for reactor experiments. The deuteron target is also very useful for studying the effect of meson exchange currents 
in weak interactions, which were shown to play very important role in case of the electrodisintegration of deuteron. In the case of 
the low energy weak reactions from deuteron for energies up to 20~MeV relevant for the solar neutrinos and lower energies 
relevant for the reactor antineutrinos, all the above methods predict cross sections in agreement with each other within 
1--2 $\%$~\cite{Nozawa:1986cu, Nakamura:2000vp}.

Most of the exclusive (anti)neutrino-nucleus reactions in the low energy region have been done in hydrogen and deuteron 
targets while some experiments have been done with nuclear targets like $^{12}$C, $^{37}$Cl, $^{56}$Fe, $^{71}$Ga and 
$^{127}$I. The experimental and theoretical results have been summarized by Formaggio and Zeller~\cite{Formaggio:2012cpf}, 
and by Fukugita and Yamagida~\cite{Fukugita:2003en}. However, we show some low energy (anti)neutrino-nucleus cross sections 
for $^2$D, $^{12}$C, $^{37}$Cl and $^{71}$Ga calculated with the above formalism.

\begin{figure}[h]
 \centering
 \vspace{0.2cm}
\includegraphics[height=6cm,width=8cm]{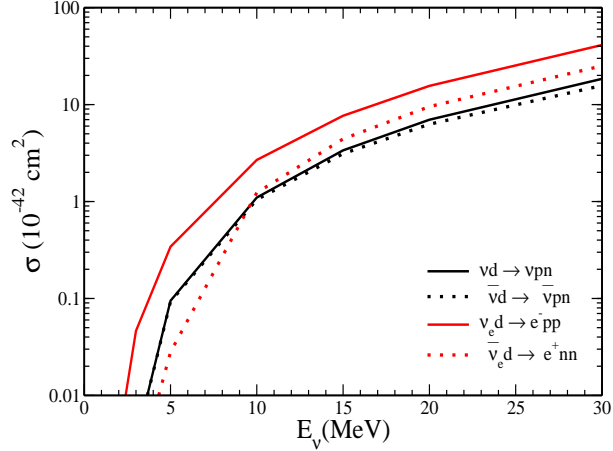}
 \caption{The total scattering cross section $\sigma$(in $10^{-42}{\text{cm}}^2$) vs $E_\nu$ for the (anti)neutrino NC and 
 CC reactions on deuteron target $\nu + D \rightarrow \nu + n + p$, $\bar\nu + D \rightarrow \bar\nu + n + p$, $\nu_e + D 
 \rightarrow e^- + p + p$ and $\bar\nu_e + D \rightarrow e^+ + n + n$~\cite{Nozawa:1986cu, Nakamura:2000vp}.}
 \label{Nakamura:D}
\end{figure}

Fig.~\ref{Nakamura:D} shows the total scattering cross section $\sigma$ vs $E_\nu$ for the (anti)neutrino NC and CC 
reactions on deuteron. These results are taken from Nakamura et al.~\cite{Nakamura:2000vp}. Table-\ref{table-deut} shows the 
total cross section folded over the reactor spectrum for the CC and NC induced processes on the deuteron target~\cite{Butler:2000zp}. The 
theoretical results are from Kubodera et al.~\cite{Kubodera:1993rk} while the experimental results are from 
Riley et al.~\cite{Riley:1998ca} and Willis et al.~\cite{Willis:1980pj}.

\begin{figure} 
 \centering
\includegraphics[height=5cm,width=8cm]{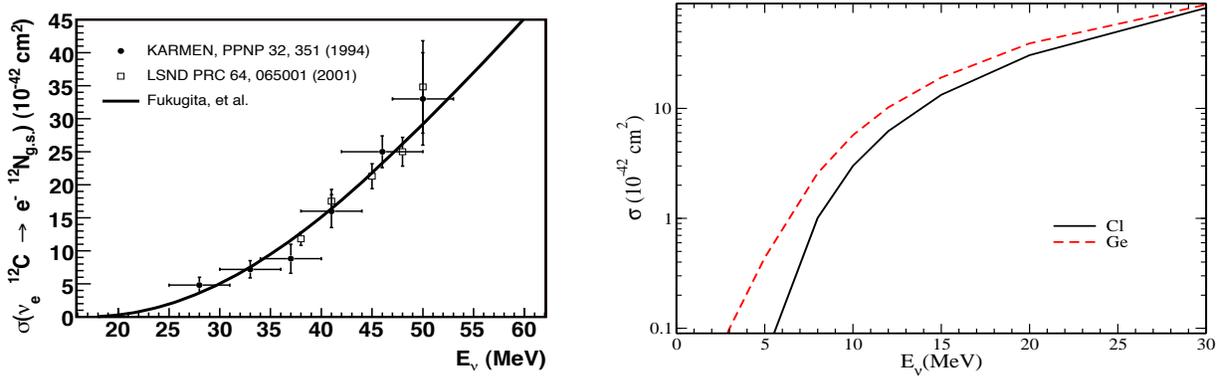}
\includegraphics[height=5cm,width=8cm]{XS.eps}
 \caption{(Left panel)~The total scattering cross section $\sigma$ vs $E_\nu$ for the exclusive reaction $\nu_e + ^{12}C \rightarrow e^- 
 + ^{12}N_{gs}$ from the muon decay at rest. The figure has been taken from Ref~\cite{Formaggio:2012cpf}. The data points are 
 from the KARMEN~\cite{Zeitnitz:1994kz} and LSND~\cite{LSND:2001fbw} experimental measurements. The theoretical curve is from 
 the works of Fukugita et al.~\cite{Fukugita:1988hg}. 
 (Right panel)~The total scattering cross section $\sigma$(in $10^{-42}{\text{cm}}^2$) vs $E_{\nu_e}$ for the reactions $\nu_e + 
 ^{37}Cl \rightarrow e^- + ^{37}Ar$ and $\nu_e + ^{71}Ga \rightarrow e^- + ^{71}Ge$~\cite{Ortiz:2000nf, Bahcall:1997eg}.}\label{Formaggio-Zeller:2012}
\end{figure}

Fig.~\ref{Formaggio-Zeller:2012} shows the results of $\sigma$ vs $E_\nu$ for the exclusive reaction $\nu_e + ^{12}C 
\rightarrow e^- + ^{12}N_{gs}$ from the muon decay at rest neutrinos. The figure has been taken from 
Ref.~\cite{Formaggio:2012cpf}. The experimental results are from the KARMEN~\cite{Zeitnitz:1994kz} and 
LSND~\cite{LSND:2001fbw} measurements. The theoretical curve is from the works of Fukugita et al.~\cite{Fukugita:1988hg} 
obtained in a model independent way with a direct evaluation of the nuclear matrix element from $\beta$ decays. In 
Fig.~\ref{Formaggio-Zeller:2012}, we also present the results of the cross section for CC $\nu_e$ induced process on heavier 
nuclei like $^{37}$Cl and $^{71}$Ga from Ortiz et al.~\cite{Ortiz:2000nf} and by Bahcall et al.~\cite{Bahcall:1997eg}.
\subsection{Inclusive quasielastic scattering in the low and intermediate energy regions}
Experimentally, the inclusive QE scattering has been studied by the accelerator as well as the atmospheric 
neutrinos. In the case of accelerator neutrinos, the first experiments were done following the suggestions from  Markov, Pontecorvo, and 
Schwartz, using the particle accelerators with energy of neutrinos in the few GeV region at BNL, ANL, and CERN 
using spark chambers with aluminum and iron nuclear targets and with the bubble chamber filled with freon and propane as 
targets. The theoretical interpretation of these experiments were initially done using the Fermi gas model to account for NME. 
Subsequently various versions of the shell model for describing the nuclear structure were used to study the 
NME in (anti)neutrino scattering from the nuclear targets following the theoretical techniques for 
describing the electron-nucleus scattering in this energy region. This approach is useful in calculating the inclusive as well 
as the exclusive reactions to specific nuclear states in the final nucleus in studying the low energy (anti)neutrino reactions 
induced by the solar and reactor (anti)neutrinos.

The second generation of the accelerator (anti)neutrino experiments were done with the hydrogen and deuteron filled bubble chambers at 
ANL and BNL followed by the experiments at FNAL, CERN, BNL and Serpukhov bubble chambers filled with heavier nuclear 
targets. These experiments with reasonably good statistics played very important role in determining the weak form factors 
of the nucleon and led to the study of the axial-vector response of the nucleons in the region of large energy~($\nu = E_\nu - E_l$) transfer 
and the four momentum transfer squared~($Q^2$) to the nuclear systems. 

In recent times, most of the experiments in QE inclusive scattering were done with the accelerator neutrinos 
obtained by the pion decay at rest in the case of $\nu_{e}$ with $E_{\nu_e} < 52.8$~MeV and pion decay in flight in the 
case of $\nu_{\mu}$ with $E_{\nu_\mu} < 286$~MeV. Although the first experiments were done at BNL in $^{12}$C which had very 
low statistics but later experiments done by the LSND and KARMEN collaborations at LANL and RAL in $^{12}$C produced results 
with better statistics. The LSND results reported in the ${\bar\nu}_e \longrightarrow e^+$ QE inclusive reactions 
supported the existence of neutrino oscillations proposed in the context of explaining the solar neutrino anomaly. 
Furthermore, the evidence of the existence of neutrino oscillations reported by the IMB and Kamiokande experiments with the 
atmospheric neutrinos in $\nu_\mu \longrightarrow \mu^-$ QE reactions, motivated the neutrino physics community to 
study the QE inclusive reactions like $\nu_e \longrightarrow e^-$ and $\nu_\mu \longrightarrow \mu^-$ 
with accelerator neutrinos in the intermediate energy region of around $E_{\nu(\bar\nu)} \sim 1$~GeV. Consequently, 
many experiments like the K2K and T2K at JPARC, MiniBooNE, MINOS, NOvA, MicroBooNE, ArgoNEUT, 
and MINERvA at the Fermilab in the few GeV energy region  and NOMAD at CERN at relatively higher energies have been done. For a general discussion on the 
historical development of the accelerator neutrino beams and detectors, see Refs.~\cite{Dore:2018ldz, Mahn:2018mai}.
\begin{table}
\begin{center}
 \begin{tabular}{|c|c|c|}\hline
 &$\sigma(\bar \nu+d\to e^+ +n +n )$&$\sigma(\bar \nu+d\to \bar \nu +p +n )$\\
 &($\times 10^{-45} cm^2$)&($\times 10^{-45} cm^2$)\\\hline
  Theory~\cite{Kubodera:1993rk}&10.02&6.02\\\hline
 Experiment~\cite{Riley:1998ca}&9.83$\pm$2.04&6.08$\pm$0.77\\\hline
 \end{tabular}
 \caption{CC and NC reactor averaged cross section $<\sigma> \left(10^{-45}{\text{cm}}^2\right)$ in deuteron.}
 \label{table-deut}
\end{center}
\end{table}

In the case of atmospheric neutrinos, the experiments on inclusive QE (anti)neutrino-nucleus scattering were done 
first by the underground experiments in the deep mines of India and South Africa in the context of cosmic ray 
studies~\cite{Athar:2020kqn}. In 1970's when the grand unified theory~(GUT) predicted proton decays, many experiments were 
designed to search for the proton decay events in which the atmospheric (anti)neutrino interactions producing the charged 
leptons are serious background. Though no proton decay event was observed, but the efforts started a comprehensive study of the 
atmospheric neutrinos with an energy spectrum, which is theoretically predicted to have peak around 500~MeV and have large 
tail extending up to few GeV.

Theoretically, the SM is used to describe the basic reactions of (anti)neutrinos on nucleons moving in a nucleus, with the 
NME taken into consideration using an appropriate nuclear model for describing the nuclear structure. In 
general, the appropriate nuclear model to describe NME depends upon the energy region and the type 
of reaction under consideration.

Generally, the following nuclear effects play important role in the case of inclusive QE reactions:
\begin{itemize}
\item [(i)] {\bf Fermi motion and nuclear binding}
  \begin{figure}
  \begin{center}
\includegraphics[height=3.5cm,width=5.5cm]{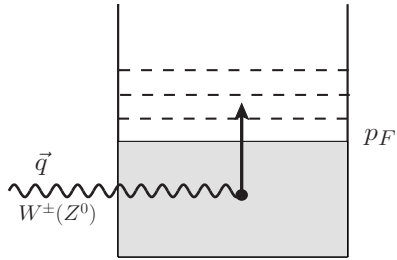}
 \end{center}
 \caption{Diagrammatic representation of Pauli blocking.}\label{Ch14-fig:2}
 \end{figure}
 
The nucleons are bound in a nucleus due to the strong nucleon-nucleon forces represented by a potential $V({\vec{r}})$, in 
which they are moving with a momentum ${\vec{p}}$. The momentum distribution of the nucleons bound in a nucleus is described 
by the wave function $\Psi({\vec{r}})$, which is obtained by solving the Schr\"{o}dinger equation in the nuclear 
potential $V(\vec{r})$ with a Hamiltonian $H$ 
given by
\begin{equation}\label{ch14:hamiltonian}
H=-\frac{\vec{\nabla}^2}{2M}+V(\vec{r}).
\end{equation}
Depending upon the potential $V(\vec{r})$, there are various approaches to obtain the wave function $\Psi(\vec{r})$. In the 
simplest approach of the shell model, $V(\vec{r})$ is taken to be a central potential but more sophisticated approaches 
also include the residual interactions in addition to the central potential $V(\vec{r})$ for describing the short and long 
range nucleon-nucleon correlations and pairing of the nucleons in nuclei. The parameters of the central potential and the residual 
interactions are fitted to reproduce the static properties of the nuclei like the binding energy, nuclear moments, nuclear 
deformations, etc. The effect of the Fermi motion, $\vec{p}_{F}$ and the binding energy is taken into account through these wave functions and 
through the kinematics of the reaction. 

In other approach of the Fermi gas model, the nucleon momentum $\vec{p}$ is constrained such that $|\vec{p}\;| \le p_F$,
the Fermi momentum which is given by
\begin{equation}
p_F=[3\pi^2\rho]^{\frac{1}{3}},
\end{equation}
where $\rho$ is the density of the nucleon in the nucleus. In such models, the momentum distribution of the initial nucleon 
is essentially given by a step function $\Theta(p_F - p)$ and the energy of the nucleon $E \ne \sqrt{{|\vec{p}\;|}^2 + M^2}$ 
but is modified by the binding energy. This { momentum and energy distribution} is called the spectral function of the nucleon $S(\vec{p},
E)$ and is given, in the Fermi gas model, by
\begin{equation}
 S(\vec{p},E)~\propto~ \Theta(p_F-p)\;\delta(E-\sqrt{{|\vec{p}\;|}^2 +M^2}+\epsilon),
\end{equation}
where $\epsilon$ is the separation energy, which depends upon the binding energy~(B.E.). In the modern Fermi gas models, 
the spectral function $S(\vec{p}, E)$, obtained phenomenologically from the electron-nucleus scattering experiments is used.

\item [(ii)] {\bf Pauli's principle} 
 
The Pauli principle guides the occupancy of the nucleons in various shell model states, which are predicted for a given 
central potential $V(\vec{r})$, and thus describes the nuclear states occupied by the valence nucleons and the structure of the core 
consisting of the closed shells. This is very important in theoretical calculations of various reactions in the shell model, 
where the interaction with the valence nucleons and the effect of core polarizations are considered. In the 
context of Fermi gas model, all the nuclear states in the Fermi sea up to the momentum $p_F$ are filled, thus constraining 
the final nucleon to have momentum $p^\prime > p_F$. Due to the interaction with a $W^{\pm}/Z^{0}$ boson, a nucleon may occupy 
a state above the Fermi sea with the momentum $p^\prime > p_F$ and creates a particle-hole (1p - 1h) pair in the nucleus as 
shown in Fig.~\ref{Ch14-fig:2}.

In the simplest versions of the Fermi gas model, the results of the free nucleon cross section are modified due to the 
above considerations on the momentum distribution of the initial and final nucleons by multiplying them with the spectral 
function $S(\vec{p}, E) \;
\Theta(p^\prime - p_F)$ with the corresponding modification on the energy of the final nucleon.

\item [(iii)] {\bf Meson exchange currents}

The QE reactions are generally calculated in the impulse approximation where the (anti)neutrino interacts with 
a single nucleon. The nuclear cross section is calculated as an incoherent or coherent sum of the transition amplitudes 
depending upon the kinematics and dynamics of the reactions (Fig.~\ref{Ch14:fig4}(a)). However, it has been shown in the 
electromagnetic reactions with photons and electrons from nuclei that the interaction of the external probes like $\gamma$ or 
$e$ can also take place with the nonnucleonic degrees of freedom of the nucleus like the meson or $\Delta$ degrees 
 of freedom.

These are called meson exchange currents~(MEC). The effect of MEC are important in some kinematic regions as shown in 
the case of photon and electron scattering from the nuclei as well as in the $\beta$ decays of nuclei leading to the 
quenching of axial-vector coupling. It could also play important role in the QE (anti)neutrino-nucleus scattering in which 
$W^{\pm}$ and $Z^{0}$ bosons interact with nonnucleonic degrees of freedom in nuclei like mesons and $\Delta$ resonances as 
shown in Fig.~\ref{Ch14:fig4}~(b) and (c) {or via the contact interaction term as shown in Fig.~\ref{Ch14:fig4}~(d)}. 

\item [(iv)] {\bf Nucleon-nucleon correlations} 
 \begin{figure}
    \begin{center}
  \includegraphics[height=4.5 cm, width= 16 cm]{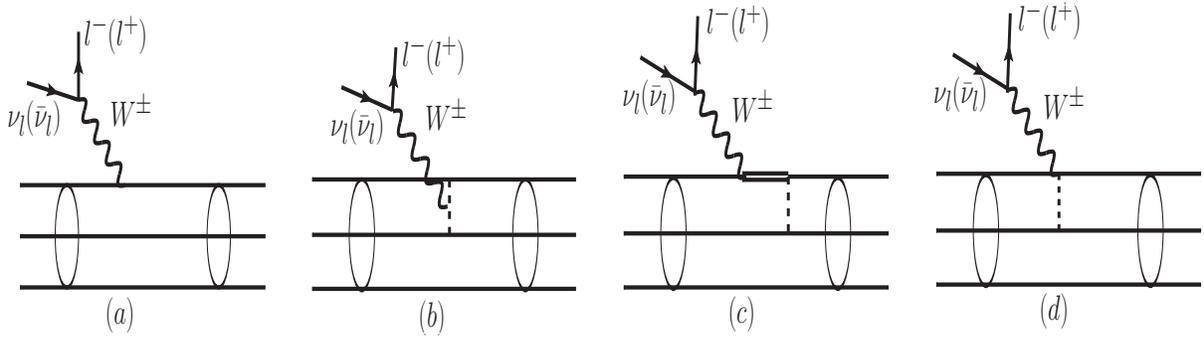}
   \end{center}
   \caption{Neutrino-nucleus scattering with MEC. {Diagrams (b), (c) and (d) also have exchange diagrams.}}\label{Ch14:fig4}
 \end{figure}
 
The nucleons which interact with (anti)neutrinos are highly correlated due to the pion exchange as well as the rho and 
omega exchanges leading to the long range and the short range correlations. Most of the calculations in the shell model 
are done using the nuclear wave functions with a central potential $V(\vec{r})$. These wave functions need to be calculated 
with potentials describing the short range as well as the long range correlations. Various attempts have been made to include 
them to calculate the QE inclusive reactions but most of them have been confined to the low energy region. For a 
general discussion, see Ref.~\cite{Athar:2020kqn}. In the case of Fermi gas model calculations, the long range correlation 
has been taken into account using the method of random phase approximation~(RPA) along with a realistic spectral function 
$S(\vec{p}, E)$ to describe { the energy and momentum distribution of the nucleons in the initial nucleus}, which is obtained phenomenologically from the 
electron-nucleus scattering experiments~\cite{JeffersonLabHallA:2022cit, JeffersonLabHallA:2020rcp}.

\item [(v)] {\bf Final state interaction~(FSI)}

This is one of the most important effects of nuclear medium in the IE reactions but is also important in the 
QE reactions, when a meson produced in the IE reaction is reabsorbed in the nucleus mimicking a 
QE-like reaction. In the case when only nucleons are produced in the final hadronic state inside the nucleus, the 
FSI of the nucleon is taken into account by calculating the final nuclear wave function with a 
nucleon-nucleon potential, which includes the nucleon-nucleon correlations as described in (iv) above. The QE-like process in which a pion 
produced in an IE reaction  is reabsorbed in the nucleus leads to an enhancement in the genuine QE 
cross section, which needs to be understood and corrected. Moreover, in the case of (anti)neutrino reactions, this 
phenomenon of QE-like events also affects the neutrino energy reconstruction where the initial neutrino energy is 
constructed using the QE kinematics of the reactions. These QE-like events contributing to the 
QE cross section do not come from the genuine QE reactions but from the reactions where an additional 
pion is produced corresponding to the IE pion production, which do not obey the QE kinematics. Therefore, 
a knowledge of FSI is very important both for the determination of the genuine QE cross sections as well as for 
the energy reconstruction of (anti)neutrinos.
 \end{itemize}
These nuclear effects have been taken into account with various degrees of sophistication in several calculations done 
within the shell model as well as in the Fermi gas model, and other models. In 
the low energy region corresponding to the supernova neutrinos and the neutrinos from 
the pion and muon decays at rest~(and also in flight in some cases), a small number of states are excited in the final 
nucleus, one sums over the cross section from each excited state. The theoretical calculation of the total cross section 
and the other observables in such cases requires a knowledge of the ground state of the initial nucleus as well as the wave function of 
the ground state and all the other excited states of the final nucleus. The NME described in (i)--(iv) 
above are taken into account in the shell model approach by calculating the nuclear wave functions in the initial and final 
states with various versions of the residual interaction describing the nucleon-nucleon potentials, using different 
approaches like the RPA~\cite{Auerbach:1997ay, Singh:1993rg, Kosmas:1996fh, Singh:1998md, 
Volpe:2000zn, Volpe:2001gy}, continuous RPA~(CRPA)~\cite{Kolbe:1992xu, Kolbe:1995af, 
Jachowicz:1998fn, Jachowicz:2002rr, Botrugno:2005kn}, quasi particle RPA~(QRPA)~\cite{Lazauskas:2007bs, Cheoun:2010zzc, 
Chasioti:2009fby, Tsakstara:2011zzd, Tsakstara:2012yd}, 
projected QRPA~\cite{Samana:2010up}, relativistic RPA~\cite{Paar:2007fi} and relativistic nuclear energy density 
functional~(RNEDF)~\cite{Paar:2012dj}, RPA with Hartree-Fock~(HF-RPA) wave functions~\cite{Auerbach:1997ay}, etc. The results obtained in these 
approaches have been reviewed by Kolbe et al.~\cite{Kolbe:2003ys}, 
and recently by Balasi et al.~\cite{Balasi:2015dba}, and Jachowicz and Nikolakopoulos~\cite{Jachowicz:2021ieb}. Alternate 
approaches, using the relativistic distorted wave impulse approximation methods using relativistic mean fields as 
well as the Green's function approaches using the nucleon-nucleon optical model to describe the final state interactions of 
the nucleons have been used by the Spanish-Italian groups~\cite{PhysRevD.97.116006, PhysRevD.85.093002, PhysRevC.88.025502, 
Gonzalez-Jimenez:2019qhq}. In recent years, the methods of SuperScaling Approach~(SuSA) based on the scaling behavior of the 
nuclear response functions observed in the electron-nucleus scattering have been applied to study the QE 
(anti)neutrino-nucleus scattering by the MIT-Spanish-Italian groups~\cite{Gonzalez-Jimenez:2014eqa, Megias:2014qva}.

In the case of light nuclei like $^{4}$He, $^{6}$Li, and $^{12}$C, the ab initio calculations of the nuclear response functions 
based on the Green's function Monte Carlo methods have been used by the Argonne-Rome group~\cite{Lovato:2020kba, 
Lovato:2015qka}. In these approaches, some authors have also calculated the effect of MEC. These microscopic methods were used to 
calculate the cross sections for transitions to all the accessible states~(ground and excited states) in the final nucleus and 
sum over them to obtain the total cross section. These methods become quite intractable when (anti)neutrino energies approach 
the GeV region in which case some approximation methods are used.

Historically, the approximation methods based on the Fermi gas model~\cite{Gatto:1953, Gatto:1955, Berman} and the closure 
approximation with the shell model wave functions~\cite{LlewellynSmith:1971uhs, Bell:1971ayw} have been used to analyze the 
early (anti)neutrino-nucleus experiments from CERN, ANL, and BNL laboratories. In view of the recent experiments in the few GeV region 
done at Fermilab, JPARC, and CERN, various improved versions of the Fermi gas model have been used. It is not possible to 
summarize in this report, all the microscopic approaches, which consider NME 
and we focus in the following only on the latest work done using the Fermi gas model to describe NME 
in (anti)neutrino-nucleus QE scattering. For details, see Ref.~\cite{Athar:2020kqn}.

\subsubsection{(Anti)neutrino-nucleus quasielastic scattering in Fermi gas models}\label{LFG}
The first application of the Fermi gas model to the (anti)neutrino-nucleus scattering results from CERN was done by Berman et 
al.~\cite{Berman} using the works of Gatto~\cite{Gatto:1953, Gatto:1955}, where the free nucleon differential  cross section $\frac{d\sigma}{dQ^2}$ is 
multiplied by  a factor $\left(1- \frac{D}{N}\right)$ and $D$ is given by~\cite{LlewellynSmith:1971uhs, Bodek:2021trq, 
Berman}:
\begin{eqnarray}\label{llewellyn_factor}
D&=&Z ~~for~~ x < u-v\nonumber\\
&=&\frac{1}{2}A\left\{1-\frac{3x}{4}(u^{2}+v^{2})+\frac{x^3}{2}-
\frac{3}{32x}(u^{2}-v^{2})^{2}\right\}~~~~~~~~for~~ u-v < x < u+v\nonumber\\
&=& 0 ~~~ for ~~x > u+v
\end{eqnarray}
with $x=\frac{|\vec q|}{2 p_{F}}$, $u=(\frac{2N}{A})^{1/3}$, $v=(\frac{2Z}{A})^{1/3}$ and $N(=A-Z),~ Z,~ A$ are neutron, 
proton and mass numbers of the initial nucleus, respectively. $p_{F}$ is the Fermi momentum and the three momentum transfer 
$|\vec q|=\sqrt{q_0^2+Q^2}$, $Q^2 = -q^2 \ge 0$. 

Smith and Moniz~\cite{Smith:1972xh} improved the Fermi gas model calculations and used the following expression for the double differential cross section:
\begin{eqnarray}\label{diff_xsect_smithmoniz}
\frac{d^{2}\sigma}{dk^ \prime d\Omega_l} &=&\frac{G_F^{2} {k^ \prime}^{2} \cos^{2}
(\frac{1}{2}\chi)}{2\pi^{2}M}\left\{W_{2}+[2W_{1}+\frac{m_{l}^{2}}{M^{2}}W_{\alpha}] \tan^{2}(\frac{1}{2}\chi)+
(W_{\beta}+W_{8})m_{l}^{2}/(ME_l\; \cos^{2}(\frac{1}{2}\chi)) \right. \nonumber\\
&-& \left. 2W_{8}/M\; \tan(\frac{1}{2}\chi) \sec(\frac{1}{2}\chi)[-Q^2 \; 
\cos^{2}(\frac{1}{2}\chi)+|{\vec q}|^{2}\; \sin^{2}(\frac{1}{2}\chi)+m_{l}^{2}]^{\frac{1}{2}}\right\},
\end{eqnarray}
where $\cos\chi = \frac{k^\prime}{E_l}  \cos\theta$. The form of $W_i$'s and other  details are given in 
Ref.~\cite{Smith:1972xh}.  

Gaisser and O'Connell~\cite{Gaisser:1986bv} have used the  relativistic response function $R(\vec{q},q_0)$, in a Fermi gas model to take 
into account NME. The expression for the double differential scattering cross section is given by
\begin{eqnarray}
 \frac{d^2\sigma}{d\Omega_ldE_l}&=&C~\frac{d\sigma_{free}}{d\Omega_l}~R(\vec{q},q_0),\nonumber\\
 R(\vec{q},q_0)&=&\frac{1}{\frac{4}{3}\pi p_{F_{N}}^3} \int \frac{d^3p_N~M^2}{E_NE_{N^\prime}}
 \delta(E_N + q_0 - E_B - E_{N^\prime}) \theta(p_{F_{N}} - |{\vec p_N}|)\theta(|{\vec p_N}+{\vec q}| - p_{F_{N^\prime}}),~~~~~~
\end{eqnarray}
where $p_{F_{N}}~(p_{F_{N^{\prime}}})$ is the Fermi momentum for the initial~(final) nucleon, $N,N^\prime$=n or p and $C=A-Z$ for neutrino induced 
process and $C=Z$ for the antineutrino induced process. $\frac{d \sigma_{free}}{d \Omega_l}$ is the differential scattering 
cross section for the (anti)neutrino reaction on free (proton)~neutron target.  

In 1990's, the local Fermi gas model~(LFGM) was used to study the (anti)neutrino-nucleus scattering in the low 
as well as in the intermediate energy regions to study NME on the cross section and possible 
modification of $M_{A}$ due to NME~\cite{Singh:1993rg, Singh:1998md, 
Athar:1999fu}. In recent years, LFGM 
has been improved by taking into account the relativistic effects, the effects of long range nucleon-nucleon correlations and 
the use of a realistic spectral function $S(\vec{p},E)$ to describe the { nucleon energy and momentum distribution} instead of a step 
function in momentum space used in the earlier calculations~\cite{Sobczyk:2020dkn, Vagnoni:2017hll}. The effect of $2p-2h$ 
excitations as well 
as MEC have also been taken into account in this approach by using different formalisms given 
by Martini et al.~\cite{Martini:2009uj, Martini:2010ex, Martini:2011wp, Martini:2013sha}, and Nieves et 
al.~\cite{Nieves:2011yp, Nieves:2013fr}, which have demonstrated the importance of NME in the determination 
of the cross section and the effective $M_{A}$ in the nuclear medium. In the following we describe very 
briefly, the (anti)neutrino-nucleus QE scattering in the LFGM. 

In the local density approximation, the cross section is evaluated as a function of local Fermi momentum $p_{F}(r)$ and 
integrated over the whole nucleus. In this approach, the incoming neutrino scatters from a neutron moving in a finite nucleus 
of neutron density $\rho_{n}(r)$, such that the differential cross section is given by 
 \begin{equation}\label{sig_4}
\left(\frac{d^2 \sigma}{ d E_l \; d \Omega_l }\right)_{\nu A}= 
2\int d{\vec r} \rho_n(r) \left(\frac{d^2 \sigma}{ d E_l \; d \Omega_l }\right)_{\nu n},
\end{equation}
where $r$ is the radius of the nucleus and the factor of 2 is to take into account the spin degrees of freedom.
 
It is assured that the nucleons in a nucleus~(or nuclear matter) occupy one nucleon per unit cell in 
phase space so that the total number of nucleons $N$ is given by 
\begin{equation}\label{delta2}
 N=2V\int_0^{p_F}\frac{d^3 p}{(2\pi)^3}~~\Longrightarrow \rho = \frac{N}{V} = 2 \int_0^{p_F}\frac{d^3 p}{(2\pi)^3},
\end{equation}
 where the factor of 2 is to take into account isospin degrees of freedom of the nucleon. 
 All the states up to a maximum momentum $p_F~(p<p_F)$ are filled. The momentum states higher than $|\vec p |> |\vec p_F|$ 
 are unoccupied such that the occupation number $n(\vec p, \vec r)$ is defined as:
\begin{equation}
 n(\vec p, \vec r)=\left\{
 \begin{array}{l}
  1,\qquad  p <  p_F \\
      0,\qquad p > p_F 
 \end{array}
\right..
\end{equation}
In  this model, the Fermi momentum is a function of $r$ and is not a constant, protons and neutrons are supposed to have 
different Fermi sphere such that
$$
{p_F}_p(r) = \left(  3 \pi^2 \rho_p(r) \right)^\frac13;  \qquad \qquad {p_F}_n(\vec{r}) = \left(  3 \pi^2 \rho_n(r) 
\right)^\frac13 ,
$$
where, $\rho_p(r)$ and $\rho_n(r)$ are, respectively, the proton and the neutron densities inside the nucleus and are, 
in turn, expressed in terms of the nuclear density $\rho(r)$ as
\begin{eqnarray}\label{eq:rho}
  \rho_{p}(r) &\rightarrow& \frac{Z}{A} \rho(r);  \qquad \qquad
  \rho_{n}(r) \rightarrow \frac{A-Z}{A} \rho(r),
 \end{eqnarray}
where $\rho(r)$ is generally parameterized in terms of harmonic oscillator density, 
two parameter Fermi density, Gaussian density, etc. and the density parameters are determined in the electron scattering 
experiments. For our numerical calculations, we have taken the density parameters from Refs.~\cite{DeVries:1987atn, 
DeJager:1974liz, Garcia-Recio:1991ocp}, which are summarized in Table-\ref{tab:nuc_para}. For the antineutrino induced reaction 
on the free nucleon or nucleons bound in a nucleus, the role of neutron and proton gets interchanged.
\begin{table}
 \begin{center}
\begin{tabular}{lccccccc} \\  \hline \hline 
   & \multicolumn{2}{c}{ $c_1$ }  &   \multicolumn{2}{c}{ $c_2$ }  &   \multicolumn{2}{c}{ $Q-$value } \\
  \text{Nucleus} &    $c_1^n$ & $c_1^p$ & $c_2^n$ & $c_2^p$  &  $\nu$ & $\bar \nu$   & \\	\hline		 
$^{12}C$  &  1.692   & 1.692    & 1.082$^{\ast}$   &   1.082$^{\ast}$   &   16.8  &   13.9   \\
$^{16}O$  &  1.833   & 1.833    & 1.544$^{\ast}$   &   1.544$^{\ast}$   &   14.9  &   10.9   \\
$^{40}Ar$ &  3.64   & 3.47    &  0.569     &   0.569    &    2.5    &   8.0   \\
$^{56}Fe$ &  4.05  & 3.971   &  0.5935    &   0.5935   &    6.8   &   4.8    \\
$^{208}Pb$ &  6.89   & 6.624    &  0.549   &  0.549    &    2.4  &   5.5   \\ \hline \hline 
\end{tabular}
\end{center}
\caption{Different parameters used for the numerical calculations of the nuclear density for the various nuclei. $c_1$ and $c_2$ are the density 
parameters~(in Fermi units)  defined for modified harmonic oscillator as $\rho(r)=\rho_0 (1 + c_2 (\frac{r}{c_1} )^2) 
exp(-(\frac{r}{c_1})^2 )$ and for 2-parameter Fermi density as $\rho(r)=\rho_0 /(1+exp(\frac{r-c_1}{c_2}))$. For $^{12}C$ 
and $^{16}O$ we have used modified harmonic oscillator density($^{\ast}$ $c_2$ is dimensionless) and for $^{40}Ar$,$^{56}Fe$ 
and $^{208}Pb$ nuclei, 2-parameter Fermi density have been used, where superscript $n$ and $p$ in density 
parameters~($c_{i}^{n,p}$; $i$=1,2) stand for neutron and proton, respectively. The $Q-$value of the reaction for different 
nuclei are given in MeV.}\label{tab:nuc_para}
\end{table}

The expression of the differential scattering cross section for say neutrino-nucleus scattering is written as 
\begin{equation}\label{sig_4}
\sigma_{\nu A}(E_l,\Omega_l)= \left(\frac{d^2 \sigma}{ d E_l \; d \Omega_l }\right)_{\nu A}= 
2\int d{\vec r}d{\vec p}\frac{1}{(2\pi)^3}n_n({\vec p},{\vec r})\left(\frac{d^2 \sigma}{ d E_l \; d \Omega_l }
\right)_{\nu n},
\end{equation}
where the expression of $\left(\frac{d^2 \sigma}{ d E_l \; d \Omega_l }\right)_{\nu n}$ is given in Eq.~(\ref{sig_zero}).

Now in Eq.~(\ref{sig_zero}), the neutron energy $E_n$ and the proton energy $E_p$ are replaced by $E_n(|\vec p|)$ and 
$E_p(|\vec{p}+\vec{q}|)$, where $\vec{p}$ is the momentum of the target nucleon inside the nucleus. This is because inside 
the nucleus the nucleons are not free and their momenta are constrained to satisfy the Pauli principle as discussed above, 
i.e., ${p}<{p_{F_{n}}}$ and ${p^\prime}(=|{\vec p}+{\vec q}|) > p_{F_{p}}$.

Moreover, in the finite nucleus, there is a threshold energy for the reaction to proceed known also as the $Q-$value of 
the reaction, which we have taken to be the value corresponding to the lowest 
allowed Fermi or Gamow-Teller transition. In Table-\ref{tab:nuc_para}, we have also tabulated $Q-$value of 
the reaction for the nuclear targets 
for which the numerical results have been presented.

\begin{figure} 
\centering
\includegraphics[height=6 cm, width=12 cm]{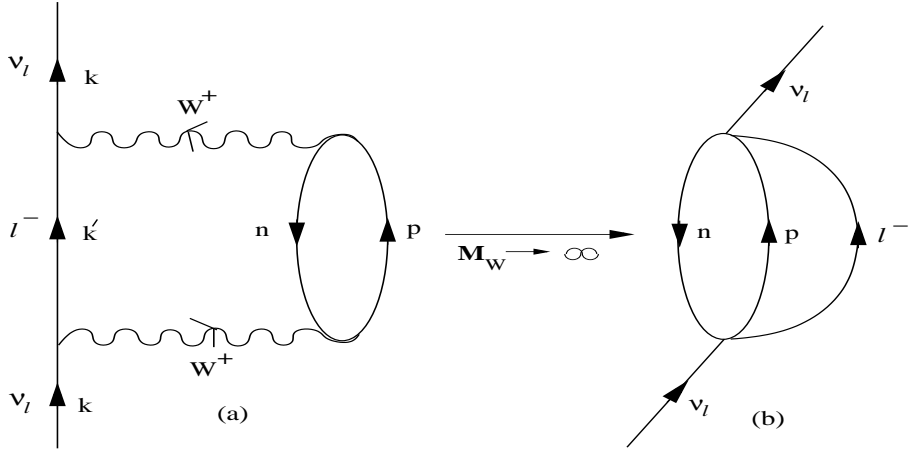}
\caption{Diagrammatic representation of the neutrino self-energy corresponding to the ph-excitation leading to $\nu_l +n 
\rightarrow l^- + p$ in nuclei. In the large mass limit of the intermediate vector boson~(i.e. $M_W\rightarrow \infty$) the 
diagram (a) is reduced to (b) which is used to calculate ${|{\cal M}|^2}$ in Eq.~(\ref{mat_quasi}).}
\label{fg:fig1}
\end{figure}

These considerations lead to a modification in the $\delta$ function used in Eq.~(\ref{sig_zero}) i.e.  $\delta[q_0 + E_n - 
E_p]$ is modified to $\delta[q_0+E_n(\vec{p})-E_p(\vec{p}+\vec{q})-Q]$ and the factor
\begin{equation}\label{delta_modi}
\int \frac{d\vec{p}}{(2\pi)^3}{n_n(\vec{p},\vec{r})}\frac{M_nM_p}{E_n E_p}\delta[q_0+E_n-E_p]
\end{equation}
occurring in Eq.~(\ref{sig_4}) is replaced by $-(1/{\pi})$Im${{U_N}(q_0,\vec{q})}$, where ${{U_N}(q_0,\vec{q})}$ is the 
Lindhard function corresponding to the 1particle-1hole~(ph) excitation 
shown in Fig.~\ref{fg:fig1}, and is given by~\cite{Singh:1993rg}:
\begin{equation}\label{lindhard}
{U_N}(q_0,\vec{q}) = {\int \frac{d\vec{p}}{(2\pi)^3}\frac{M_nM_p}{E_nE_p}\frac{n_n(p)
\left[1-n_p(\vec p + \vec q) \right]}{q_0+{E_n(p)}-{E_p(\vec p+\vec q)}+i\epsilon}}
\end{equation}
where $q_0$=$E_{\nu}-E_l-Q$. For the antineutrino reaction the suffix n and p will get interchanged.

The imaginary part of the Lindhard function~(Eq.~(\ref{lindhard})) corresponds to the intermediate particles 
in Fig.~\ref{fg:fig1} to be on shell, thereby describing the process $\nu_l + n \rightarrow l^- + p$. If we consider the initial 
nucleon at rest~(the static limit) i.e. $E_n=M_n$ and neglect any Pauli blocking for the proton, 
then the expression for the free neutrino-neutron cross section will be obtained. Therefore, the role of the Lindhard 
function is to take into account Pauli blocking as well as the Fermi motion of the nucleon in the nucleus when the neutrino 
interaction takes place.
 
The imaginary part of the Lindhard function is obtained to be~\cite{Singh:1993rg}: 
\begin{equation}\label{lindhard_imag}
Im{U_N}(q_0, \vec{q}) = -\frac{1}{2\pi}\frac{M_nM_p}{|\vec{q}|}\left[E_{F_1}-A\right]
\end{equation}
with $Q^2\geq0$, $E_{F_2}-q_0<E_{F_1}$ and $\frac{-q_0+|\vec{q}|{\sqrt{1+\frac{4{M^2}}{Q^2}}}}{2}<{E_{F_1}}$, 
where $E_{F_1}=\sqrt{p{_{F_n}}^2+{M_n}^2}$, $E_{F_2}=\sqrt{{p_{F_p}}^2+{M_p}^2}$ and \\
$A$ = $Max\left[M_n,\;\; E_{F_2}-q_0,\;\; \dfrac{-q_0+|\vec{q}| \sqrt{1+\frac{4{M^2}}{Q^2}}}{2}  \right]$.

With the inclusion of these nuclear effects, the total cross section $\sigma(E_\nu)$ is written as
\begin{eqnarray}\label{xsection_medeffects}
\sigma(E_\nu)&=&-2{G_F}^2\cos^2{\theta_C}\int^{r_{max}}_{r_{min}} r^2 dr \int^{{k^\prime}_{max}}_{{k^\prime}_{min}}k^\prime 
dk^\prime \int_{Q^{2}_{min}}^{Q^{2}_{max}}dQ^{2}\frac{1}{E_\nu^{2} E_l} L_{\mu\nu}J^{\mu\nu} Im{U_N}[E_\nu - E_l - Q, 
\vec{q}].
\end{eqnarray}
In the above expression $r_{min}$ and $r_{max}$ are the minimum and maximum limits of nuclear size. $k^\prime_{min}$ and 
$k^\prime_{max}$ are minimum and maximum values of outgoing lepton momenta. The energy and momentum of the outgoing lepton 
get modified due to the Coulomb interaction, which is taken into account in a modified effective momentum 
approximation~(MEMA)~\cite{Singh:1993rg}.

\begin{figure} 
\begin{center}
\includegraphics[height=.2\textheight,width=0.70\textwidth]{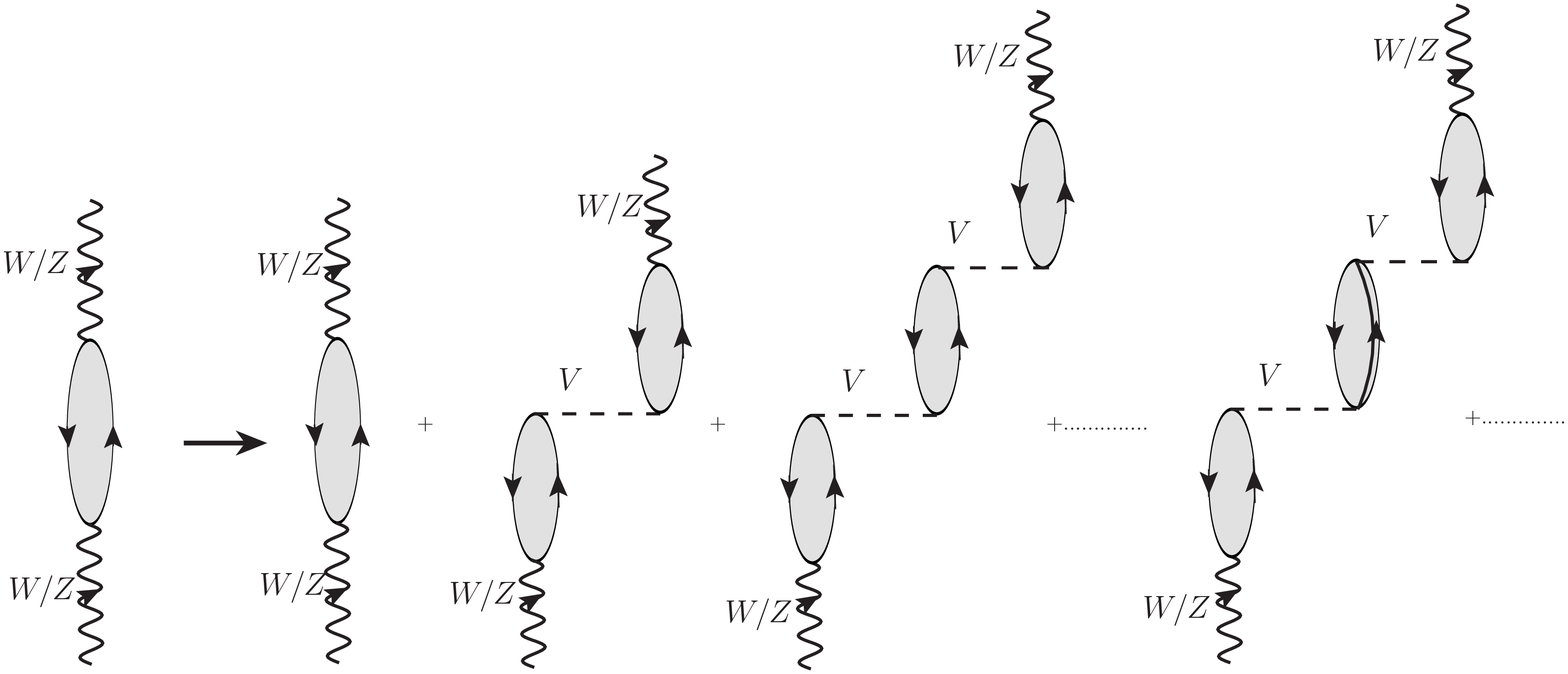}
\caption{RPA effects in the ph contribution to the W/Z self-energy, where particle-hole, $\Delta$-hole, $\Delta$-$\Delta$, 
etc. excitations contribute.}\label{fg:fig2}
\end{center}
\end{figure}

In the local density approximation, the effective energy of the lepton in the Coulomb field of the final nucleus is given 
by~\cite{Singh:1993rg, Engel:1997fy}:
\begin{equation}\label{effective_coulomb}
E_{eff} = E_l + V_c(r), \qquad \text{where} \qquad V_c(r)=4\pi\alpha\;Z_f\left(\frac{1}{r}\int_0^r\frac{\rho_p(r^\prime)}{Z_f}{r^\prime}^2dr^\prime +
\int_r^\infty\frac{\rho_p(r^\prime)}{Z_f}{r^\prime}dr^\prime\right)
\end{equation}
with $\alpha$ as fine structure constant and $Z_f$ as the charge of outgoing lepton, taken as $-1$ for neutrino and $+1$ for 
antineutrino. This leads to a change in the imaginary part of the Lindhard function occurring in 
Eq.~(\ref{xsection_medeffects})
\begin{equation*}\label{changed_lindhard}
Im{U_N}[E_\nu - E_l - Q, \vec{q}] \rightarrow Im{U_N}(E_\nu - E_l - Q - V_c(r), {\vec q}).
\end{equation*}
When the electroweak interactions take place in nuclei, the strength of the electroweak couplings may change from their free 
nucleon values due to the presence of strongly interacting nucleons. Though CVC forbids 
any change in the charge coupling, other couplings like  the magnetic, axial charge and pseudoscalar are likely to 
change from their free nucleon values. There exists considerable work in understanding the quenching of the  magnetic moment and the 
axial charge in nuclei due to the nucleon-nucleon correlations. In our approach, the nucleon-nucleon correlation effects are 
reflected in the modification  of nuclear response in the longitudinal and transverse channels. Due to PCAC, the axial 
current is strongly coupled to the pion field in the nuclear medium and therefore axial couplings are more likely to change 
due to the pionic effects modifying the nuclear response functions. To demonstrate an idea of these effects, we perform nonrelativistic 
reduction of the hadronic current~($J_\mu$ in Eq.~(\ref{had_curr}))~\cite{Athar:2020kqn}, and see the occurrence of $g_1 
{\vec\sigma} {\vec\tau}$, $f_2{{\vec\sigma}\times{\vec q}}{\vec\tau}$ and $g_3{\vec\sigma}\cdot{\vec q}{\vec\tau}$ terms in 
the weak current, which are linked to the spin-isospin excitations in nuclei, while $f_2$ and $g_3$ are coupled to the transverse 
and longitudinal channels, respectively, $g_1$ is coupled to both~\cite{Athar:2020kqn}. In a nuclear target, the coupling of these terms to 
the mesonic channels can be described through the diagram shown in Fig.~\ref{fg:fig2}.

The $ph-ph$ interaction is shown by the dashed line in Fig.~\ref{fg:fig2} and is described by the $\pi$ and $\rho$ exchanges 
modulated by the effect of short range correlations. For the $ph-ph$ potential, we use $V_{N}(q)= V_{\pi}(q) + V_{\rho}(q)$ 
in terms of the longitudinal and transverse components expressed as
\begin{equation}
V_{N}(q)=\frac{f^2}{m_{\pi}^2}\left[V_t(q)(\delta_{ij}-\hat{q_i}\hat{q_j})+V_l(q)\hat{q_i}\hat{q_j}\right](\sigma_i\sigma_j)
({\vec \tau}\cdot{\vec\tau})
\end{equation}
for the $ph$ case, ${\vec \sigma}$ and $\vec \tau$ are Pauli matrices acting on the nucleon spin and isospin spaces, 
respectively. A similar potential $V_{\Delta}$ in the case of $ph-\Delta $h interaction is obtained by substituting 
${\vec\sigma} \rightarrow {\vec S}$, ${\vec\tau} \rightarrow {\vec T}$ and $f \rightarrow f^*=2.15f$. $V_l$ is the strength 
of the potential in the longitudinal channel and $V_t$ is the strength of the potential in the transverse channel. Thus, we 
calculate this reduction in the vector-axial-vector~(VA) and the  axial-vector-axial-vector~($AA$) response functions due to the long range 
nucleon-nucleon correlations treated in the RPA. The representation into the longitudinal and the transverse channels is useful when 
one tries to sum the geometric series in Fig.~\ref{fg:fig2}, where the longitudinal and the  transverse channels decouple and can 
be summed independently.
  
The potential $V(q) $ is explicitly written as:
\begin{eqnarray}\label{longi_part}
V_l(q) = \frac{f^2}{m_\pi^2}\left[\frac{q^2}{-q^2+m_\pi^2}{\left(\frac{\Lambda_{\pi}^2-m_\pi^2}{\Lambda_\pi^2-q^2
}\right)^2}+g^\prime\right],\qquad \qquad
V_t(q) = \frac{f^2}{m_\pi^2}\left[\frac{q^2}{-q^2+m^2_\rho}{C_\rho}{\left(\frac{{\Lambda_\rho}^2-m^2_\rho}
{{\Lambda_\rho}^2-q^2}\right)^2}+g^\prime
\right],
\end{eqnarray}
$\frac{f^{2}}{4\pi}$ = 0.8, $\Lambda_\pi$ = 1.3 GeV, $C_\rho$ = 2, $\Lambda_\rho$ = 2.5 GeV, $m_\pi$ and $m_\rho$ are the 
pion and rho meson masses, and $g^\prime$ is the Landau-Migdal spin-isospin parameter taken to be $0.7$, which has been used 
quite successfully to explain many electromagnetic and weak processes in nuclei \cite{Singh:1998md, Gil:1997bm, 
Carrasco:1989vq}.

Using the matrix elements at the weak $WNN$ vertex and the $ph-ph$ potential, the contribution of Fig.~\ref{fg:fig2} is 
written as
\begin{eqnarray}\label{Ch14-GP:eq}
 U(q) = U(q) + U(q)V_{N}(q)U(q) + U(q) V_{N}(q) U(q)V_{N} (q)U(q) + ...
\end{eqnarray}
Writing the potential $V_{N}(q)$ in terms of $V_{l}$ and $V_{t}$, the above series can be separated in the 
longitudinal and the transverse components. The longitudinal component is then written as~\cite{Athar:2020kqn}: 
\begin{eqnarray}\label{Ch14-RPA_L}
U_L(q)&=&\left[\frac{U(q)}{1-U(q)V_l}\right]~\hat{q_i}\hat{q_j}~\sigma_i\sigma_j~{\vec{\tau}_1}\cdot{\vec{\tau}_2}.
\end{eqnarray}
Similarly, the transverse component is given by~\cite{Athar:2020kqn}:
\begin{equation}
U_T(q)=\left[\frac{U(q)}{1-U(q)V_t}\right]~(\delta_{ij}-\hat{q_i}\hat{q_j})~\sigma_i\sigma_j~{\vec{\tau}_1}\cdot{\vec{\tau}_2}.
\end{equation}
Therefore, we can write Eq.~(\ref{Ch14-GP:eq}) as:
\begin{equation}
{U}(q) \rightarrow \bar{U}(q)=\left[\left(\frac{U(q)}{1-U(q)V_t}\right)~(\delta_{ij}-\hat{q_i}\hat{q_j})+\left(\frac{U(q)}{1-U(q)V_l}\right)~
\hat{q_i}\hat{q_j}\right]~{\sigma_i}~{\sigma_j}{\vec{\tau}_1}\cdot{\vec{\tau}_2},
\end{equation}
where $U=U_N+U_\Delta$, with $U_N$ and $U_\Delta$ as the Lindhard functions for $ph$ and 
$\Delta h$ excitations, respectively, in the medium and the expressions for $U_N$ and $U_\Delta$ are taken from 
Ref.~\cite{Oset:1993sm, Oset:1989ey}. The different couplings of $N$ and $\Delta$ are incorporated in $U_N$ and $U_\Delta$ and 
then the same interaction strengths $V_l(q)$ and $V_t(q)$ are used to calculate the RPA response. These effects have been discussed by Nieves et 
al.~\cite{Nieves:2004wx} as well as by Athar et al.~\cite{SajjadAthar:2005ke}.

By using the above method of renormalization, we consider the different components of the hadronic tensor $J_{\mu\nu}$~(Eq.~(\ref{had_tens})) and 
sum up the RPA series shown in Fig.~\ref{fg:fig2}. For convenience we take ${\vec q}$ to be along the z direction and neglect all the corrections 
of order 
$O\left(\frac{p_F {\vec p}}{M^2}, \frac{p_F {\vec p^\prime}}{M^2} , \frac{p_F {q_0}}{M^{2}}  \right)$, and the different components of 
$J_{\mu\nu}$ like $J_{00}$, $J_{0z}$, $J_{zz}$, etc. with renormalization effect are obtained as~\cite{Nieves:2004wx, SajjadAthar:2005ke}:
 \begin{eqnarray}\label{Ch14-RPA1}
\frac{J_{00}^{RPA}}{M^2}&=&\left(f_1(Q^2)\right)^2\left[\left(\frac{E(\vec{p})}{M}\right)^2+\left(\frac{q_0E(\vec{p})-
{Q^2}/{4}}{M^2}\right)\right]
+\frac{Q^2}{M^2}\left(\frac{f_2(Q^2)}{2}\right)^2\left[\frac{{\vec p}^2+q_0E(\vec{p})+{q_0^2}/4}{M^2}-\frac{q_0^2}{Q^2}\right]\nonumber\\
&-&\frac{1}{2}\left(f_1(Q^2)f_2(Q^2)\right)
\left(\frac{|\vec q|}{M}\right)^2+g_1^2(Q^2)\left[\frac{{\vec p}^2+q_0E(\vec{p})-{Q^2}/4}{M^2}+{U_L}\left(\frac{q_0^2}{m_\pi^2+Q^2}\right)\left(\frac{Q^2}
{m_\pi^2+Q^2}\right)\right]\\
\frac{J_{0z}^{RPA}}{M^2}&=&\frac{1}{2}\left(f_1(Q^2)\right)^2\left[\frac{E({\vec p})}{M}\left(\frac{2p_z+|\vec q|}{M}\right)+
\frac{q_o p_z}{M^2}\right]+\frac{1}{2}\frac{Q^2}{M^2}\left(\frac{f_2(Q^2)}{2}\right)^2\left[\frac{E({\vec p})}{M}
\left(\frac{2p_z+|\vec q|}{M}\right) \right. \nonumber\\
&-& \left.\frac{2q_0|\vec q|}{Q^2}+\frac{q_0\left(2p_z+|\vec q|
\right)}{2M^2}\right]-\frac{1}{2}\left(f_1(Q^2)f_2(Q^2)\right)\left[\frac{q_0|\vec q|}{M^2}\right]+g_1^2(Q^2)\left[{U_L}\frac{E({\vec p})}{M}
\left(\frac{2p_z+|\vec q|}{2M}\right) \right. \nonumber \\
&+&\left. \frac{q_op_z}{2M^2}+{U_L}\left(\frac{q_0|\vec q|}{m_\pi^2+Q^2}\right)\left(\frac{q^2}{m_\pi^2+Q^2}\right)\right]\\
\frac{J_{zz}^{RPA}}{M^2}&=&\left(f_1(Q^2)\right)^2\left[\frac{p_z^2+|\vec q|p_z+Q^2/4}{M^2}\right]+\frac{1}{4}\frac{Q^2}{M^2}
\left(\frac{f_2(Q^2)}{2}\right)^2\left[\left(\frac{2p_z+|\vec q|}{M}\right)^2-\frac{q_0^2}{Q^2}\right] \nonumber \\
&-&\frac{1}{2}\left(f_1(Q^2)f_2(Q^2)\right)\left(\frac{q_0}{M}\right)^2+g_1^2(Q^2)\left[{U_L}+
\frac{p_z^2+|\vec q|p_z+Q^2/4}{M^2}+{U_L}\left(\frac{|\vec q|}{m_\pi^2+Q^2}\right)\left(\frac{Q^2}{m_\pi^2+Q^2}\right)\right]\\
\frac{J_{RPA}^{xx}}{M^2}&=&\left(f_1(Q^2)\right)^2\left[\frac{p_x^2+Q^2/4}{M^2}\right]+\frac{Q^2}{M^2}
\left(\frac{f_2(Q^2)}{2}\right)^2\left[{U_T}+\frac{p_x^2}{M^2}\right]\nonumber\\
&+&\frac{1}{2}\left(f_1(Q^2)f_2(Q^2)\right){U_T}\left(\frac{Q^2}{M^2}\right)+g_1^2(Q^2)\left[{U_L}+\frac{p_x^2+Q^2/4}{M^2}
\right]\\
\label{Ch14-RPA11}
\frac{J_{xy}^{RPA}}{M^2}&=&ig_1(Q^2)\left[f_1(Q^2)+f_2(Q^2)\right]\left[\frac{q_0p_z}{M^2}-{U_T}\frac{|\vec q|E({\vec p})}
{M^2}\right].
\end{eqnarray}
Thus, in a local density approximation in the presence of NME including the RPA effect,
the total cross section $\sigma(E_\nu)$, is written as
 \begin{eqnarray}\label{cross_section_quasi}
\sigma(E_\nu)=-2{G_F}^2\cos^2{\theta_C}\int^{r_{max}}_{r_{min}} r^2 dr 
\int^{{k^\prime}_{max}}_{{k^\prime}_{min}}k^\prime dk^\prime 
\int_{Q_{min}^{2}}^{Q_{max}^{2}}dQ^2\frac{1}{E_{\nu}^2 E_l}L_{\mu\nu}{J^{\mu\nu}_{RPA}}  Im{U_N}[E_{\nu} - E_l - Q - 
V_c(r), \vec{q}]\;\;
\end{eqnarray}
where $J^{\mu\nu}_{RPA}$ is the modified hadronic tensor when RPA effect is incorporated and the energy transferred to the 
hadronic tensor also gets modified from $q_0=E_\nu-E_l$ to $q_0=E_\nu-E_l-Q-V_c$.

\subsubsection{Inclusive quasielastic scattering at low energy}
\begin{itemize}
 \item [{\bf 5.4.2.1}] {\bf Theoretical results and comparison with experimental data}\\
We summarize in this section, the experimental and theoretical results~\cite{Auerbach:1997ay, Singh:1993rg, 
Kosmas:1996fh, Singh:1998md, Volpe:2000zn, Kolbe:1995af, Kolbe:2003ys, 
Nieves:2004wx, 
SajjadAthar:2005ke, Paar:2008zza, LSND:1997tqo, 
Nieves:2017lij, LSND:2002oco, Kolbe:1999au, Auerbach:2002tw, Umino:1995bql, Umino:1994wu, Hayes:1999ew, 
Kolbe:1994xb, Donnelly:1978tz, Donnelly:1973enn, LSND:1997lta, Suzuki:1987jf, Krmpotic:2004gx} for the low energy inclusive $(\nu_e,e^-),
(\nu_\mu,\mu^-)$ scattering cross sections from the KARMEN and LSND collaborations in $^{12}$C, presented in 
Table-\ref{KDAR2} along with the 
theoretical predictions for these processes in various nuclear models. Most of the theoretical methods used to obtain these 
results were earlier developed to study the related process of the inclusive muon capture for which the results are also 
included in Table~\ref{KDAR2}. We have included theoretical results only from those calculations, which quote results for the 
relevant physical observable for all the three processes i.e. the total cross section for the  $(\nu_e,e^-)$ and $(\nu_\mu,
\mu^-)$ inclusive scattering as well as the total rate for muon capture in $^{12}$C using the same nuclear model.

The various microscopic nuclear models referred in the column 3 of Table-\ref{KDAR2}, in general use the shell model  with varying model 
spaces including states from 1 $\hbar \omega$ to 4 $\hbar \omega$   excitations to calculate the ground state of the initial 
as well as the ground state and excited states of the  final nuclei and transitions between them.  The various models use different 
forms of the phenomenological 
nucleon-nucleon potentials like the Bonn potential, Landau-Migdal potential or the Skymre model potential to calculate the 
nuclear states. Moreover, the residual interaction between the nucleons which leads to the pairing and also to the 
quasi-particle  excitations of nucleons in the nuclei are also included in some nuclear models to calculate the nuclear wave 
functions of the higher excited states and the continuum state in the final nucleus. Depending upon the various assumptions 
and approximations, many nuclear models have been used to calculate the total cross sections for ($\nu_{e} ~~e^{-}$) 
and ($\nu_{\mu} ~~\mu^{-}$) inclusive reactions as well as the muon capture rate. The results for the inclusive cross 
sections, and muon capture rates from these calculations are 
shown in Table-\ref{KDAR2}, column-3. In column-4, we show the results for these observables obtained in various 
versions of the Fermi gas model.
It should be noted that the different entries for the total cross sections or the capture rate quoted under the same 
reference in column-3 and 4 of Table-\ref{KDAR2} corresponds to the various versions of the nuclear models and the 
parameters used in that model.
 
The following general observations about the theoretical results and their comparisons with the experimental results shown 
here in Table~\ref{KDAR2}, can be made:
\begin{itemize}
\item [(i)] There is no microscopic nuclear model, which  is able to explain all the three weak processes involving 
$\nu_{e}$, $\nu_{\mu}$ and muon capture. Even in a given model by changing the parameters of the model i.e. the various 
versions of the model, it is possible to explain either the cross section for the inclusive QE reactions or the inclusive 
muon capture but not both of them simultaneously~\cite{Auerbach:1997ay, Volpe:2000zn, Kolbe:1995af, Kolbe:2003ys, Paar:2008zza, 
Kolbe:1999au, Auerbach:2002tw, Hayes:1999ew, 
Kolbe:1994xb, Donnelly:1978tz, Donnelly:1973enn, 
Krmpotic:2004gx}.

\item [(ii)]  In the case of LFGM also the earlier nonrelativistic calculations by Singh and 
Oset~\cite{Singh:1993rg, Singh:1998md}, 
Umino et al.~\cite{Umino:1995bql, Umino:1994wu}, Kosmas and Oset~\cite{Kosmas:1996fh}, 
Nieves et al.~\cite{Nieves:2004wx}, and Athar et al.~\cite{SajjadAthar:2005ke} including RPA, reproduce ($\nu_e, e^-)$ 
but overestimate the ($\nu_\mu, \mu^-$) reaction cross section and underestimate $(\mu^-, \nu_\mu)$ capture rate. The LFGM 
has been considerably improved in the latest calculations by Nieves and Sobczyk~\cite{Nieves:2017lij}, where relativistic 
transition operators and 
the spectral function of nucleons  are used to calculate all the three processes. These models reproduce ($\nu_e, e^-$) and 
($\nu_\mu, \mu^-$) inclusive cross sections quite well but underestimate the inclusive ($\mu^-, \nu_\mu$) capture rate.
\end{itemize}
In summary, a satisfactory understanding of the nuclear reactions with $\nu_\mu$, $\nu_e$ scattering, and $\mu$ capture 
in nuclear targets in the energy region of $E < 230$~MeV is desirable.
\begin{table} 
\centering
  \begin{tabular}{|c|c|c|c|}\hline
   Process & Experiments& Microscopic theories & Fermi gas model \\ \hline
   $\sigma(\nu_\mu)$&& 28.8, 22.4, 14.5, 15.2~\cite{Auerbach:1997ay} & 16.65$\pm$1.37~\cite{Singh:1998md}\\
   $(\times 10^{-40})$ && 27.0, 21.1, 13.5, 14.3~\cite{Auerbach:1997ay} & 19~\cite{Kosmas:1996fh}\\
   $cm^2$ & $11.2 \pm 0.3 \pm 1.8$~\cite{LSND:1997tqo} & 19.25~\cite{Kolbe:1995af}, 
   19.59~\cite{Paar:2008zza} &
   13.2 $\pm$ 0.7, 9.7$\pm$0.3, 12.2~\cite{Nieves:2017lij}\\
   $\nu_{\mu} (^{12}C,X)\mu^{-}$ & $10.6 \pm 0.3 \pm 1.8$~\cite{LSND:2002oco} & 18.18, 17.80~\cite{Kolbe:1999au}, 30.0, 19.2~\cite{Auerbach:2002tw}
   &22.7-24.1~\cite{Umino:1995bql, Umino:1994wu}\\
   &&15.6, 13.2, 17.0, 31.3, 19.1~\cite{Hayes:1999ew} &25~\cite{Singh:1993rg}, 11.9~\cite{Nieves:2004wx}\\
   &&15.18, 19.23, 20.29, 21.08~\cite{Volpe:2000zn} & \\
\hline
   $\sigma(\nu_e)$ && 15.2, 15.6~\cite{Kolbe:1994xb}, 14.6~\cite{Donnelly:1978tz, Donnelly:1973enn} &14~\cite{Nieves:2004wx}, 15.48 $\pm$ 
   1.13~\cite{Singh:1998md}\\
   $(\times 10^{-42})$&& 16.42, 16.70, 55.1, 52.0~\cite{Volpe:2000zn} & 14~\cite{Kosmas:1996fh} \\
   $cm^2$ && 19.28, 18.15~\cite{Kolbe:1999au} & 13.8$\pm$0.4, 14.3$\pm$0.6, 8.6~\cite{Nieves:2017lij} \\ 
   && 6.9, 3.5, 4.1, 5.4, 3.1~\cite{Hayes:1999ew} & \\
   $\nu_{e} (^{12}C,X)e^{-}$ & 14.8$\pm$0.7$\pm$1.4~\cite{LSND:1997lta} & 23.7, 15.1~\cite{Auerbach:2002tw},
   15~\cite{Kolbe:2003ys},    12.14~\cite{Paar:2008zza}& 15.3~\cite{Singh:1993rg} \\
   &&114.4, 76.3, 16.5, 22.7~\cite{Auerbach:1997ay} & 13.6~\cite{SajjadAthar:2005ke}\\
   &&90.6, 63.2, 12.9, 17.6~\cite{Auerbach:1997ay}&\\
  \hline
   $\Gamma(\mu^-)$  &3.88 $\pm$ 0.05~\cite{Suzuki:1987jf} & 4.82, 4.26, 4.07, 4.47~\cite{Krmpotic:2004gx} & 
   3.3~\cite{Singh:1993rg}\\
   $(\times 10^4) $ && 5.24, 3.35~\cite{Auerbach:2002tw}, 3.56, 4.53~\cite{Hayes:1999ew} &3.37$\pm$0.16, 3.22, 3.19$\pm$0.06~\cite{Nieves:2017lij}\\
  $sec^{-1}$ && 2.98, 2.99, 3.17, 3.40~\cite{Hayes:1999ew} & 3.60$\pm$0.22~\cite{Singh:1998md}, 3.21~\cite{Nieves:2004wx}\\
   $\mu^{-} (^{12}C,X)\nu_{\mu}$ && 8.0, 6.87, 3.09, 3.48~\cite{Auerbach:1997ay}&\\
   &&8.4, 7.22, 3.23, 3.64~\cite{Auerbach:1997ay}&\\
  &&3.32, 4.06, 5.12~\cite{Volpe:2000zn}, 5.79~\cite{Kolbe:1999au}& \\
  \hline
   
  \end{tabular}
\caption{Latest experimental results and various theoretical results in different nuclear models for inclusive cross section 
for $\nu_e$ and $\nu_\mu $ scattering and muon capture rates in $^{12}$C.}\label{KDAR2}
 \end{table}

\item [{\bf 5.4.2.2}] {\bf Inclusive cross sections with monoenergetic KDAR neutrinos with $E_{\nu_\mu}=236$~MeV}\\
The monoenergetic muon neutrinos from KDAR  are identified as an ideal probe to study the neutrino-nucleus cross sections in 
the low energy region in order to benchmark NME in the exclusive as well as in the 
inclusive reactions in this energy region. An experiment for measuring neutrino cross section $\sigma(E_{\nu_\mu})$ with 
monoenergetic neutrinos would be free from the uncertainties arising from the reconstruction procedure of the initial 
neutrino energy present in most of the experiments using beams of continuous energy of the muon neutrinos from 
accelerators. Some new experiments have been planned to measure inclusive cross sections in $^{12}$C and $^{40}$Ar using the 
monoenergetic neutrinos from kaons decaying at rest. The monoenergetic neutrinos from kaons decaying at rest $K^+\rightarrow 
\mu^+ \nu_\mu$ are copiously produced with an energy $E_{\nu_{\mu}}=236$ MeV, along with a continuous 
energy spectrum of $\nu_e$ and $\nu_\mu$ from $Kl_3$ decays like $K^+ \rightarrow \pi^0 e^+ \nu_e$ and $K^+\rightarrow 
\pi^0 \mu^+ \nu_\mu$.

The first measurement of the inclusive cross section in $^{12}$C nucleus with the monoenergetic KDAR muon neutrinos has 
been recently reported by the MiniBooNE collaboration to be $\sigma =(2.7\pm1.2)\times 10^{-39} 
\text{ cm}^2$~\cite{MiniBooNE:2018dus}. Theoretically, this reaction in $^{12}$C in the energy region of few hundreds of 
MeV, has been studied by many authors, but specific calculations and discussions of the inclusive cross section for 
$E_{\nu_{\mu}}=236$ MeV have been done recently by Akbar et al.~\cite{Akbar:2017dih} in the relativistic Fermi gas model with 
RPA to include the effect of correlations and Nikolakopoulos et al.~\cite{Nikolakopoulos:2020alk} in a microscopic model using 
CRPA to include the effect of nucleon-nucleon correlations. Nikolakopoulos et al.~\cite{Nikolakopoulos:2020alk} have also 
extrapolated the results of some earlier calculations to predict the inclusive neutrino cross sections at $E_{\nu_\mu} = 
236$~MeV and presented a comparative study of the theoretical and experimental results. In Table~\ref{KDAR3}, we present a 
list of the theoretical results for the inclusive cross sections at $E_{\nu_\mu}$=236MeV in the process $\nu_\mu +^{12}C 
\rightarrow \mu^{-}+X$ obtained in various theoretical calculations along with the Monte Carlo predictions~\cite{Andreopoulos:2009rq, 
Juszczak:2009qa, Golan:2012wx} and 
the experimental result from the MiniBooNE 
experiment~\cite{MiniBooNE:2018dus} for comparison. 
 \begin{table}
\centering
 \begin{tabular}{|c|c|}\hline
  Experimental and Theoretical Models~~~~~~~~~~~~~~~~& Cross section\\\hline
  MiniBooNE Exp.~\cite{MiniBooNE:2018dus}  & 2.7$\pm$1.2\\\hline
  Akbar et al.~\cite{Akbar:2017dih} & 0.91\\\hline
  Martini et al. ~\cite{Martini:2011wp, LSND:2001fbw}& 1.3+0.2(np-nh)\\\hline
  
  GENIE~\cite{Andreopoulos:2009rq} & 1.75\\\hline
  NuWro~\cite{Juszczak:2009qa, Golan:2012wx} & 1.3+0.4(np-nh)\\\hline
  NUANCE~\cite{Casper:2002sd} & 1.4\\
  \hline
  CRPA~\cite{Gonzalez-Jimenez:2019qhq} & 1.58\\\hline
  RMF~\cite{Gonzalez-Jimenez:2019qhq} & 1.56\\\hline
  RFG~\cite{Nikolakopoulos:2020alk} &1.66\\\hline
  RFG 34~\cite{Nikolakopoulos:2020alk} &1.38\\\hline
 \end{tabular}
\caption{Experimental and theoretical results for the inclusive cross section for KDAR neutrinos. The cross sections are in 
units of $10^{-39}cm^2$.}\label{KDAR3}
\end{table}

We see that all the theoretical predictions for the cross section lie in the wide range of $(0.91~ \text{to} 
~1.66) \times 10^{-39}cm^2/neutron$ and underestimate the experimental results for the inclusive cross sections at 
$E_{\nu_\mu} = 236$~MeV. A comparison of the theoretical results of the inclusive neutrino cross sections in $^{12}$C in 
the case of monoenergetic neutrinos at $E_{\nu_\mu}$=236MeV and the earlier theoretical results in the energy region of the 
LSND experiment i.e. $E_{\nu_\mu} <$ 280MeV with the experimental data show that:
\begin{itemize}
\item[(i)] The theoretical predictions using various nuclear models for the inclusive cross sections for the reaction 
$\nu_\mu +^{12}$C $\rightarrow \mu^- +X$ with monoenergetic neutrinos have a large range of variation. This is surprising 
in the case of $^{12}$C which is 
one of the theoretically better studied nucleus.
 
\item[(ii)] Most of the theoretical predictions for the inclusive cross section for this reaction overestimate the 
experimental results for the LSND experiment with neutrino energy $E_{\nu_{\mu}}$ in the range of $120 MeV < E_{\nu_{\mu}} < 
280$ MeV, while the theoretical predictions in the case of the KDAR neutrino with $E_{\nu_{\mu}}=236$ MeV underestimate the 
experimental result. 

\item[(iii)] The latest experimental as well as theoretical results for the inclusive cross sections using the monoenergetic 
KDAR muon neutrinos together with the results obtained in the case of LSND and KARMEN experiments with electron and muon 
neutrinos along with the capture rate of $(\mu^-, \nu_\mu)$ process in $^{12}$C show that the theoretical calculations done 
in the impulse approximation for all the weak nuclear processes of $(\mu^- , \nu_\mu), ~(\nu_e , e^-)$ and $(\nu_\mu , 
\mu^- )$ in the low energy region are not theoretically understood satisfactorily, with a given nuclear model used to 
describe the structure of $^{12}$C nucleus.
\end{itemize}
\begin{figure}[h]
\centering
 \includegraphics[height=14cm,width=15 cm]{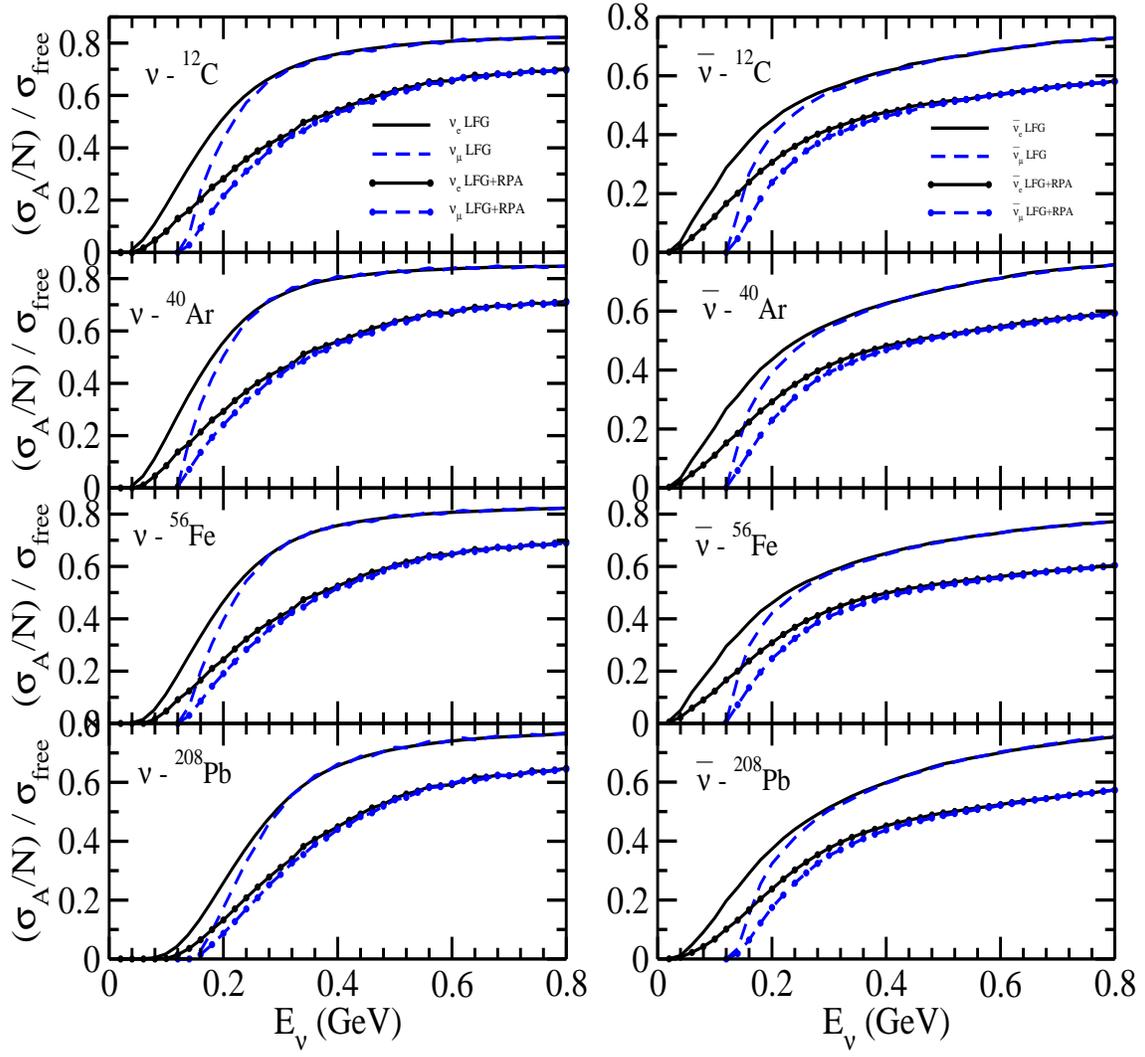}
\caption{ Ratio $\frac{\sigma_A/N}{\sigma_{free}}$  vs $E_{\nu}$, for neutrino~(left panel) and antineutrino~(right panel) 
induced processes in $^{12}$C, $^{40}$Ar, $^{56}$Fe, and $^{208}$Pb. The solid~(dashed) line represents cross section obtained 
from electron~(muon) type neutrino and antineutrino. For neutrino induced process $N=A-Z$, is neutron number and for 
antineutrino induced process $N=Z$, is proton number. $\sigma_A$ is cross section in nuclear target and has been evaluated 
using LFGM and LFG with RPA effect~(LFG+RPA) and $\sigma_{free}$ is the cross section for the free 
nucleon case.}\label{fig3c12}
\end{figure}
\end{itemize}

\subsubsection{Quasielastic (anti)neutrino scattering at intermediate energy with $\nu_{e} (\bar{\nu}_{e})$ and $\nu_{\mu} 
(\bar{\nu}_{\mu})$}\label{QE_nucleus}
\begin{itemize}
 \item [(i)] {\bf Nuclear model dependence}\\
In this section, the results obtained using Eq.~(\ref{cross_section_quasi}) with and without RPA effects are presented and 
the findings are discussed. In Fig.~\ref{fig3c12}, the results are presented for the ratio of scattering cross section per 
interacting nucleon obtained using LFG model~(Eq.~(\ref{xsection_medeffects})) and LFG model with RPA effect~(LFG + 
RPA)~(Eq.~(\ref{cross_section_quasi})) for (anti)neutrino induced processes in $^{12}$C, 
$^{40}$Ar, $^{56}$Fe and $^{208}$Pb to the scattering cross section on free nucleon target in the energy region from threshold to $0.8$~GeV. 
Performing calculations using LFG, we find that in $^{12}C$ NME like Fermi motion, Pauli blocking, 
binding energy, result in the reduction of cross section by $\sim 30(42)\%$ at $E_{\nu} = 0.3~GeV$ and around $20(30)\%$ at 
$E_{\nu} = 0.6~GeV$ from free nucleon case for $\nu_e(\bar\nu_e)$ induced processes. Inclusion of RPA correlation in LFG, 
reduces the cross section for $\nu_e(\bar\nu_e)$ scattering from free nucleon by $\sim 55(56)\%$ at $E_{\nu} = 0.3~GeV$ and 
$35(45)\%$ at $E_{\nu} = 0.6~GeV$. Similar results may be observed for $^{40}Ar$, $^{56}Fe$ and $^{208}Pb$ nuclear targets. 
In general, the reduction in the cross section increases with the increase in mass number. For $\nu_{\mu}$ and 
$\bar{\nu}_{\mu}$ induced processes at lower energies the reduction is larger and for $E_{\nu} > 0.4~GeV$, the reduction in 
$\nu_{e}$~($\bar \nu_{e}$) and $\nu_{\mu}$~($\bar \nu_{\mu}$) cross sections is almost the same. 

To compare our results with other variants of the Fermi gas model, we have obtained total scattering cross section in $^{40}Ar$ 
using Fermi gas model of Smith and Moniz~\cite{Smith:1972xh}, Llewellyn Smith~\cite{LlewellynSmith:1971uhs} and Gaisser and 
O'Connell~\cite{Gaisser:1986bv} and calculated fractional  difference $\delta{\sigma_{Model}}(=\frac{\sigma_{free} - 
\sigma_{Model}}{\sigma_{free}})$, the results for which are shown in Fig.~\ref{fig:del_mod}. Here $\sigma_{free}$ stands for 
the (anti)neutrino induced interaction cross section on free nucleon target and $\sigma_{Model}$  stands for the 
(anti)neutrino induced  interaction cross section for the nucleons bound inside the nucleus. The results for neutrino~($\nu_{e}$, 
$\nu_{\mu}$) is 
different from antineutrino~($\bar{\nu}_{e}$, 
$\bar{\nu}_{\mu}$) and is mainly due to the interference terms with $g_1$ which come with an opposite sign. In the 
case of LFG with RPA effects, the effect of renormalization is large and this suppresses the terms with $f_2$ and $g_1$, 
which results in a large change in neutrino vs antineutrino results. We find appreciable difference in the results when 
various nuclear models are used, and that may be observed from Fig.~\ref{fig:del_mod}. 

\begin{figure}[h]
\centering
\includegraphics[height=10 cm,width=12.5 cm]{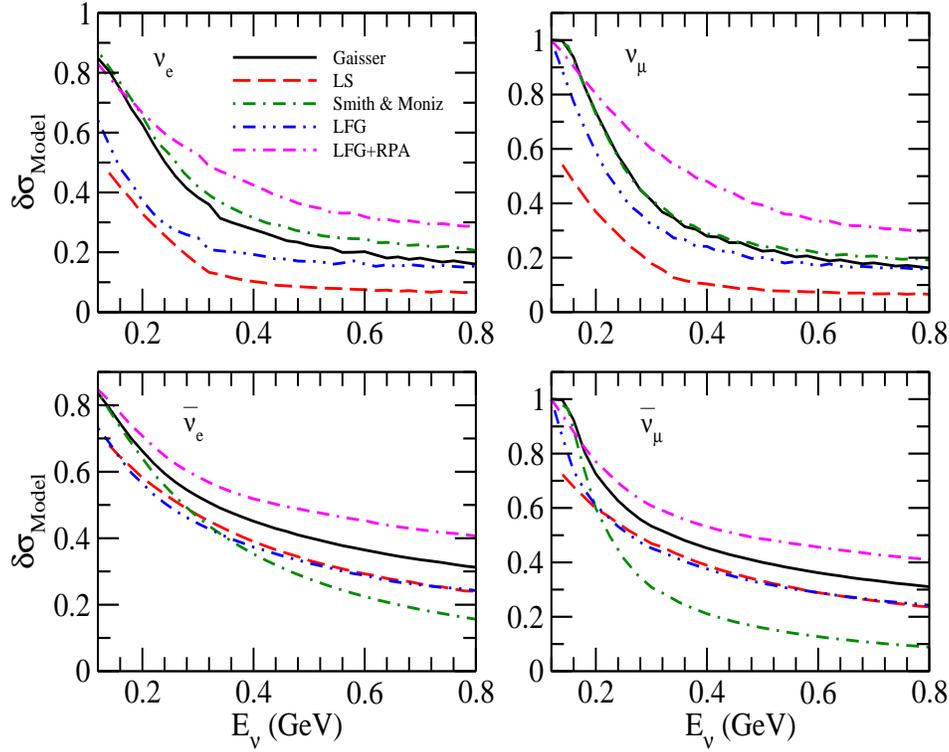}
\caption{The fractional suppression in cross section $\delta\sigma_{\rm Model}\left( = \dfrac{\sigma_{free}-\sigma_{Model}}
{\sigma_{free}}\right)$ vs $E_{\nu}$, where $\sigma_{free}$ is the cross section obtained for the free nucleon and $\sigma_{Model}$ is 
per interacting nucleon cross section in $^{40}$Ar obtained using the different nuclear models. The results are presented for 
the cross sections obtained from the different models of Fermi gas~($\sigma_{Model}$) viz. Smith and 
Moniz~\cite{Smith:1972xh}~(dashed dotted line), Llewellyn Smith~\cite{LlewellynSmith:1971uhs}~(dashed line), Gaisser O' 
Connell~\cite{Gaisser:1986bv}~(solid line), and with~(double dashed dotted line) \& without RPA~(dashed double dotted line) 
effect using LFGM. The top panel is for neutrino and bottom panel is for antineutrino induced processes.}
\label{fig:del_mod}
\end{figure}
 \begin{figure} 
 \centering
\includegraphics[height=10 cm,width=12.5 cm]{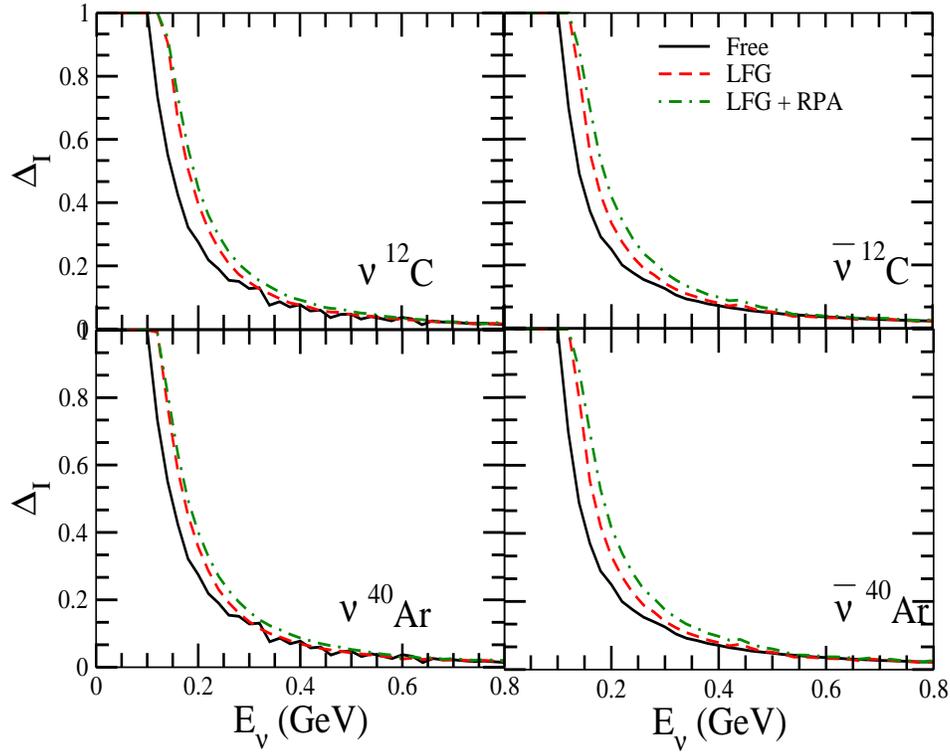}
\caption{$\Delta_{I}= \frac{\sigma_{\nu_e(\bar\nu_e)}-\sigma_{\nu_\mu(\bar\nu_\mu)}}{\sigma_{\nu_e(\bar\nu_e)}} $ for 
neutrino~(left panel) and antineutrino~(right panel) induced processes in $^{12}C$  and $^{40}Ar$ targets. 
Here $I$ stands for the results of the cross sections obtained (i) for the free nucleon case(solid line)~ (ii) 
in the LFGM~(dashed line) and (iii) for LFG with RPA effect(dashed dotted line).}
\label{fig:delta_sigma}
\end{figure}

\item [(ii)] {\bf Effect of lepton mass and its kinematic implications}\\
There are two types of corrections which appear when lepton mass $m_l~(l=e,\mu)$ is taken into account in the cross section 
calculations for the reaction $\nu_l(\bar\nu_l) + N \rightarrow l^-(l^+) + N^\prime$, ($N,N^\prime=n,p$) which can be 
classified as kinematical and dynamical in origin. The kinematical effects arise due to $E_l \ne |\vec k^{\prime}|$ in 
presence of $m_l$ and the minimum and maximum values of four momentum transfer square ($Q^2=-q^2 \ge 0$) i.e. $Q^2_{min}$ 
and $Q^2_{max}$ are modified, affecting the calculations of total cross sections. These effects are negligible for highly 
relativistic leptons but could become important at low energies near threshold specially for muons. On the other hand, the 
dynamical corrections arise as additional term proportional to $\frac{m_l^2}{M^2}$ in the existing contribution of vector 
and axial-vector form factors as well as new contributions due to induced pseudoscalar and other form factors associated 
with the SCC come into play. In fact all the contributions from the pseudoscalar form factor $g_3(Q^2)$ 
are proportional to $\frac{m_l^2}{M^2}$ while the contribution from the second class axial-vector form factor $g_2(Q^2)$ is 
proportional either to $\frac{m_l^2}{M^2}$ or $\frac{Q^2}{M^2}$ or both.

To study the lepton mass dependence on  $\nu_e(\bar\nu_e)$ and $\nu_\mu(\bar\nu_\mu)$ induced scattering cross sections in the 
free nucleon as well as in the nuclear targets, we define $\Delta_{I}= \frac{\sigma_{\nu_e(\bar\nu_e)}-\sigma_{\nu_\mu
(\bar\nu_\mu)}}{\sigma_{\nu_e(\bar\nu_e)}}$ for (anti)neutrino induced reaction in $^{12}C$ and $^{40}Ar$ nuclear 
targets, where $I =  i, ~ ii, ~ iii$, which respectively stands for the cross sections obtained in ($i$)~free 
(anti)neutrino case, ($ii$)~ the LFG model and ($iii$)~the LFG + RPA model.    
The results are presented in Fig.~\ref{fig:delta_sigma}, which show that the differences in the electron and muon production
cross sections for $\nu_l(\bar\nu_l)$ induced reactions on $^{12}C$ target are appreciable at low energies 
$E_{\nu}<0.4~GeV$. 
 
\item [(iii)] {\bf Vector form factors}\\
The various parameterizations of the vector form factors have been discussed in Section~\ref{para_FF}, and we find the  
dependence on the choice of the different parameterizations of the vector form factors on the (anti)neutrino-nucleus cross 
sections to be negligible. 

\item [(iv)] {\bf Axial vector form factor}\\
It is believed that NME due to 2p-2h excitations, MEC and multinucleon 
correlations are taken into account then the recent experimental results can also be considered to be consistent with a 
smaller value of $M_A$~\cite{Martini:2011wp, Nieves:2011yp, Martini:2012uc, Ankowski:2014yfa, Lalakulich:2012hs}. However, it 
may be observed from Table-\ref{tab:axial_mass:MA} that even with the same nuclear target, different values of $M_A$ have been 
obtained from the neutrino experiments done in different energy regions highlighting the energy region in which NME play significant role.

To study the explicit dependence of the cross section on the value of $M_{A}$, we define $\delta_{M_A}$ as
\begin{equation}\label{sdelta}
\delta_{M_A}=\frac{\sigma_{\nu_l}(M_A^{modified}) - \sigma_{\nu_l}(M_A= 
WA)}{\sigma_{\nu_l}(M_A= WA)} ,\qquad \qquad \text{ WA = world average = 1.026~GeV}
\end{equation}
where $l=e$ or $\mu$. 
We observe from Fig.~\ref{fig:partaldel_ma} that for the free nucleon, when a modified value of $M_A$ i.e. 
$M_A^{modified}= 0.9(1.2)~GeV$ is used instead of world average value of $1.026~GeV$ then a decrease(increase) of $5-15\%$ 
is obtained for $\nu_e / \nu_\mu$ reactions in the energy range of $0.2-0.8$~GeV. In the case of $\bar\nu_e / 
\bar\nu_\mu$-nucleon reactions this decrease(increase) is about $5-10\%$ in the same energy range. When NME are 
taken into account, for example, in the case of $^{40}Ar$ nucleus this decrease(increase) remains almost the same. 
Therefore, the uncertainty in the (anti)neutrino-nucleus cross sections is the same as in the case of free 
(anti)neutrino-nucleon scattering processes. 

\begin{figure} 
\centering
\includegraphics[height=12 cm, width=15 cm]{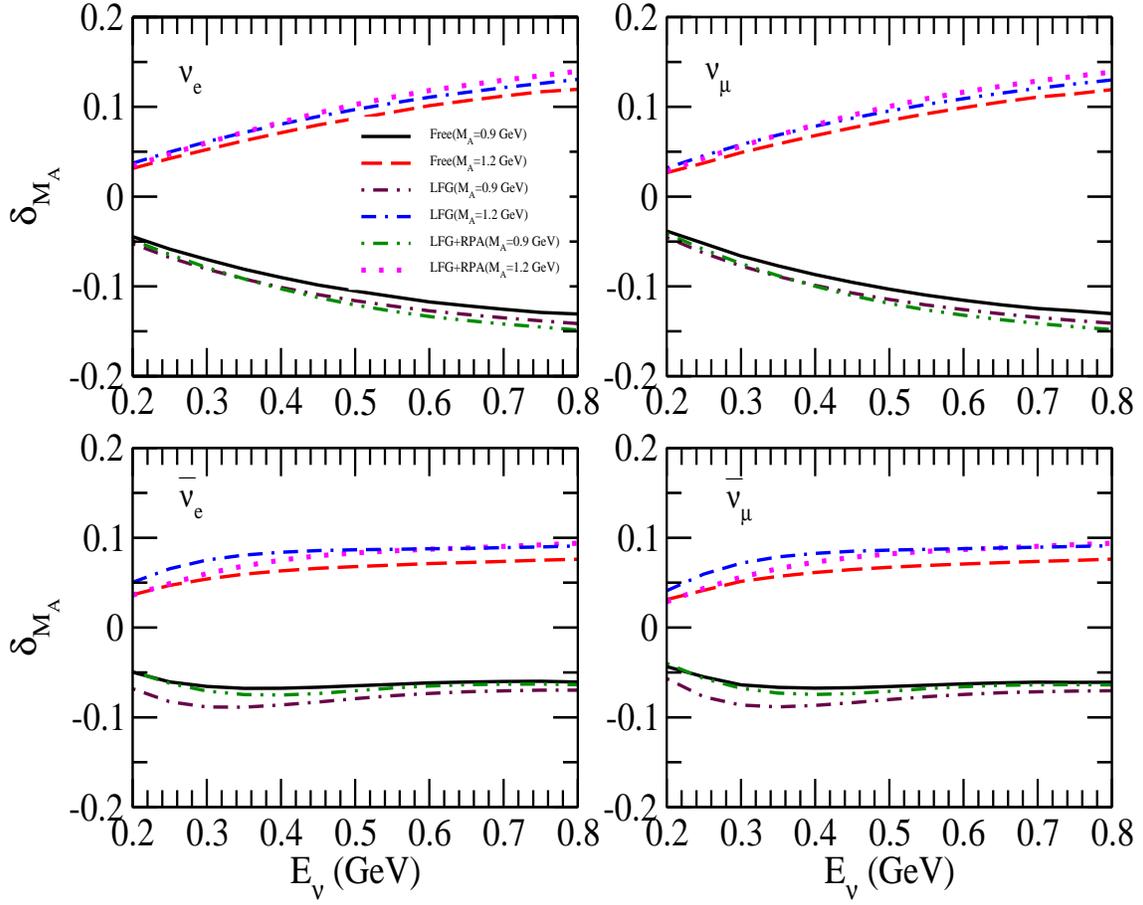}
\caption{The dependence of cross section on $M_A$ obtained using Eq.~(\ref{sdelta}). The results are shown for $\nu_e 
(\bar\nu_e)$ and $\nu_\mu(\bar\nu_\mu)$ induced processes on free nucleon as well as on $^{40}Ar$ target using LFG with and 
without RPA effect. Solid~(dashed) line denotes results for the free nucleon case with $M_A=0.9~GeV$~($1.2~GeV$), results 
obtained using LFG are shown by dash-dotted~(double dash-dotted) line with $M_A=0.9~GeV$~($1.2~GeV$) and results for LFG 
with RPA effect are shown by dash-double dotted~(dotted) line with $M_A=0.9~GeV$~($1.2~GeV$).}\label{fig:partaldel_ma}
\end{figure}
\begin{figure}
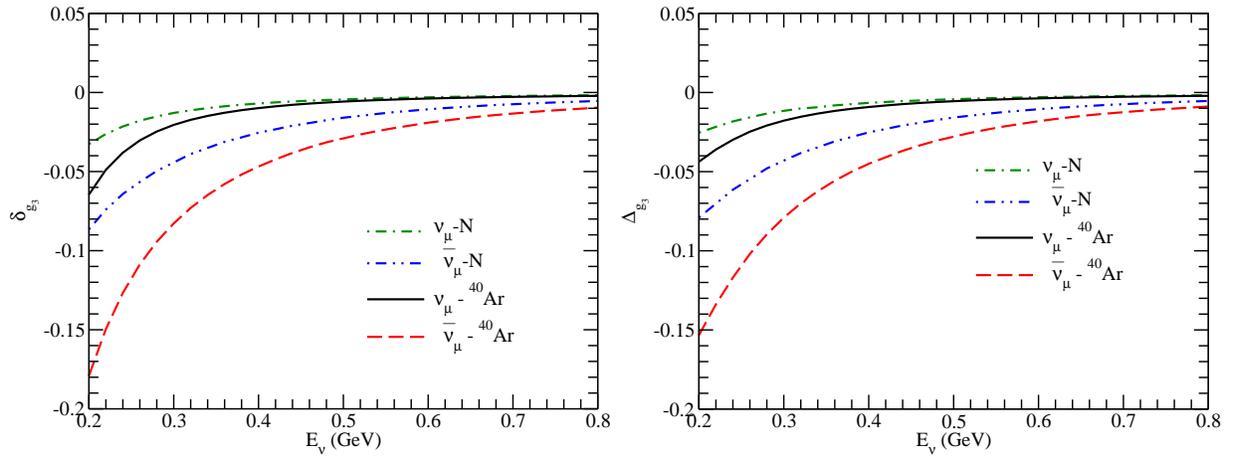
 
\centering
 \includegraphics[height=6 cm, width=8 cm]{sdelta_fp.eps}
 \includegraphics[height=6 cm, width=8 cm]{argon_fp0.eps}
\caption{(Left panel)~Results of the fractional change $\delta_{g_3}$ defined in Eq.~(\ref{eq:deltaFpnew}) as a function of 
(anti)neutrino energy. (Right panel)~Results of the fractional change $\Delta_{g_3}$ defined in 
Eq.~(\ref{eq:deltaFpnew}) as a function of (anti)neutrino energy. The results are shown for the $\nu_\mu$ induced cross 
section for the free nucleon 
case~(dashed dotted line), as well as for the nucleons bound in $^{40}Ar$~(solid line) nuclear target obtained using 
LFG with RPA effect. The results corresponding to $\bar\nu_\mu$ induced CCQE process are shown by dashed double dotted 
line~(free nucleon case) and dashed line~($^{40}Ar$ target). }\label{fig:del_FpAr}
\end{figure}
\item [(v)] {\bf Pseudoscalar form factor}\\
To study the effect of the pseudoscalar form factor $g_3(Q^2)$ on muon production cross section, we define
\begin{equation}\label{eq:deltaFpnew}
   \delta_{g_3}{(E_\nu}) = \frac{{\sigma_{\nu_\mu}}(g_3\neq0) - {\sigma_{\nu_\mu}}(g_3=0)}{{\sigma_{\nu_\mu}}(g_3=0)},
  \end{equation}
and similar expression for antineutrino is used. For the numerical calculations, the expression of $g_3(Q^2)$ given in 
Eq.~(\ref{fp_NN}) has been used. The results are presented in the left panel of  Fig.~\ref{fig:del_FpAr}. We find that 
$\delta_{g_3}$ is more sensitive in the case of $\bar\nu_\mu$ induced CCQE process than $\nu_\mu$ induced process for the 
free nucleon case as well as for $^{40}Ar$ nuclear target. This sensitivity decreases with the increase in $\nu_\mu/
\bar\nu_\mu$ energy and almost vanishes beyond $0.6$~GeV. 

We also study the sensitivity of pseudoscalar form factor $g_3(Q^2)$ to find out the difference in the electron vs muon 
production cross sections that are 
obtained using Eq.~(\ref{fp_NN}). For this purpose we define
\begin{eqnarray}\label{eq:deltaFp1}
   \Delta_1{(E_\nu}) = \frac{{\sigma_{\nu_\mu}}(g_3\neq0) - {\sigma_{\nu_e}}(g_3\neq0)}{{\sigma_{\nu_e}}(g_3\neq0)};~
  \Delta_2{(E_\nu}) = \frac{{\sigma_{\nu_\mu}}(g_3=0) - {\sigma_{\nu_e}}(g_3=0)}{{\sigma_{\nu_e}}(g_3=0)};~
  \Delta_{g_3} = \Delta_1(E_\nu) - \Delta_2(E_\nu).
  \end{eqnarray}
and the results for $\Delta_{g_3}$ are shown in the right panel of Fig.~\ref{fig:del_FpAr}. 

We have calculated the fractional difference $\Delta_{g_3}$ as given in Eq.~(\ref{eq:deltaFp1}) for the free nucleon case as well 
as for the nucleon bound in $^{40}Ar$ nuclear target using the LFG with RPA effect. We observe that the inclusion of 
pseudoscalar form factor decreases the  fractional change~($\Delta_{g_3}$) by about $3(8)\%$ at $E_{\nu(\bar\nu)}$ $\sim 
0.2GeV$ and becomes smaller with the increase in energy. When NME~(LFG+RPA) are taken into account 
in the evaluation of cross sections in $^{40}Ar$ then this difference increases to $4(15)\%$ at the same energy for 
neutrino(antineutrino) induced processes. 

\item [(vi)] {\bf Second class axial-vector form factor}\\
In the case of QE (anti)neutrino scattering from nuclear targets, the effect of the axial-vector form factor associated with SCC 
on the total scattering cross section is similar to that observed in the case of (anti)neutrino-nucleon QE scattering, and 
the effect is found to be less than a percent.  

\item [(viii)] {\bf Radiative corrections}\\
Radiative corrections are potential source of difference between electron and muon production cross sections in 
(anti)neutrino reactions due to their logarithmic dependence on the lepton mass through terms like log($\frac{E_l^\ast}
{m_l}$), where $E_l^\ast$ is some energy scale in the reaction. The radiative corrections in CC 
QE neutrino-nucleon reactions relevant for the present oscillation experiments in the energy region of few GeV 
have been recently calculated by Bodek~\cite{Bodek:2007wb}, Day and McFarland~\cite{Day:2012gb}, 
Graczyk~\cite{Graczyk:2013fha}, and Tomalak et al.~\cite{Tomalak:2022xup}. Bodek~\cite{Bodek:2007wb}, and Day and 
McFarland~\cite{Day:2012gb} make use of leading log approximation given by De Rujula et al.~\cite{DeRujula:1979grv} to 
calculate the contribution of soft  photon emission by the lepton leg bremsstrahlung diagram which gives major contribution 
to the radiative corrections depending on the lepton mass $m_l$. On the other hand, Graczyk~\cite{Graczyk:2013fha} includes 
the contribution of other diagrams like two boson exchange involving W and $\gamma$, propagator correction in addition to the 
soft photon bremsstrahlung. These effects have also been discussed by us in Ref.~\cite{Akbar:2015yda}.
\end{itemize}

\subsubsection{MiniBooNE axial dipole mass anomaly and nuclear medium effects}\label{sec3}
In the axial vector sector, the dipole mass $M_A$ is generally taken to be the world average value, determined from the 
QE scattering 
or $M_A=1.014\pm0.016$ GeV determined from the threshold pion electroproduction from proton/deuteron \cite{Bernard:2001rs, 
Bodek:2007ym}. However, using these values of $M_A$, the inclusive total and differential cross sections, 
obtained from the high statistics experiment in $^{12}$C, performed by the MiniBooNE collaboration, were 
underestimated~\cite{MiniBooNE:2010bsu, Katori:2009du, Sanchez:2009zz}. The MiniBooNE results were analyzed using the 
relativistic Fermi gas model and the microscopic nuclear models, which failed to explain the observed cross sections using 
the world average value of $M_A$~\cite{Singh:2011zzp}. It was also reported that a higher value of $M_A=1.35\pm 
0.17$~\cite{Benhar:2006nr, Benhar:2013bwa} can explain both the total and differential cross sections. 
This value of $M_A$ is considerably larger than the world average
value of $M_A$ determined from earlier experiments. The higher 
value of $M_A$ is also in disagreement with the results of another high statistics experiment performed by the 
NOMAD collaboration in $^{56}Fe$, which reported a value of $M_A=1.05\pm0.02\pm0.06$~GeV~\cite{Lyubushkin:2008pe}, 
consistent with the world average value. This is known as the MiniBooNE axial dipole mass anomaly. 

Assuming that the uncertainties in the neutrino flux at the MiniBooNE detector were well estimated and are reflected in the 
uncertainties quoted in the cross section measurements, there was a general consensus that NME are 
not adequately taken into account. This may be because:
\begin{itemize}
\item[1.] The effects of nuclear medium beyond the impulse approximation like MEC, 
$np-nh$, and $ph-\Delta h$ excitations are not included adequately in the impulse approximation, despite indications that 
they are 
quite important in the region of the low and intermediate energy neutrino-nucleus reactions.

\item[2.] In the intermediate energy region of the MiniBooNE experiment, where the $\nu_\mu$ spectrum peaks around 750 MeV, 
the real pions would be produced which could be reabsorbed in the nuclear medium mimicking the genuine QE 
inclusive events leading to an enhancement in the observed inclusive cross section for QE reactions. The effect 
of these events called the QE-like events were not included adequately in the theoretical calculations.

\item[3.] In most of the neutrino reactions, the energy of the initial neutrinos is reconstructed using free particle 
QE kinematics of neutrino-nucleon reactions in the nuclear medium. This kinematics is affected by the entanglement 
of the kinematics of the QE-like events due to the IE processes i.e. $\nu_\mu N N\rightarrow \mu \Delta N 
\rightarrow \mu N N$ or scattering from the correlated pair $\nu_\mu N N \rightarrow  \mu N N$ in the nucleus with the 
genuine lepton events produced in the real QE $\nu_\mu N\rightarrow \mu N$ scattering. The effect of this 
entanglement was not included in reconstructing the neutrino energy leading to underestimate the flux averaged cross 
sections.
\end{itemize}
A careful investigation of the above NME beyond impulse approximation was undertaken in view of the 
MiniBooNE axial dipole mass anomaly. The earlier calculations of Singh and 
Oset~\cite{Singh:1993rg}, Marteau et al.~\cite{Marteau:1999kt}, and Nieves et al.~\cite{Nieves:2004wx, Valverde:2006zn} were, 
respectively, improved by Martini et 
al.~\cite{Martini:2010ex,Martini:2011wp} and Nieves et al.~\cite{Nieves:2011pp} in which the $2p-2h$, $ph-\Delta h$, MEC 
effects as well as the pion reabsorption effects were taken into account. It was shown that the contribution of these 
effects is quite substantial in the energy region of MiniBooNE experiment and the observed cross section is reproduced 
quite well when the above mentioned NME are taken into account using the world average value of $M_A$ 
and can explain the axial dipole mass anomaly. These results were further improved by the calculations of Rocco et 
al.~\cite{Rocco:2018mwt} and Ivanov et al.~\cite{Ivanov:2018nlm} using the nucleon spectral functions to describe the { energy and 
momentum distribution} of the nucleons in the nucleus. 

\subsubsection{Nuclear medium effects due to two particle-two hole excitations}\label{2p2h-miniboone}
\begin{figure}
 \begin{center}
 \includegraphics[height=2cm,width=8cm]{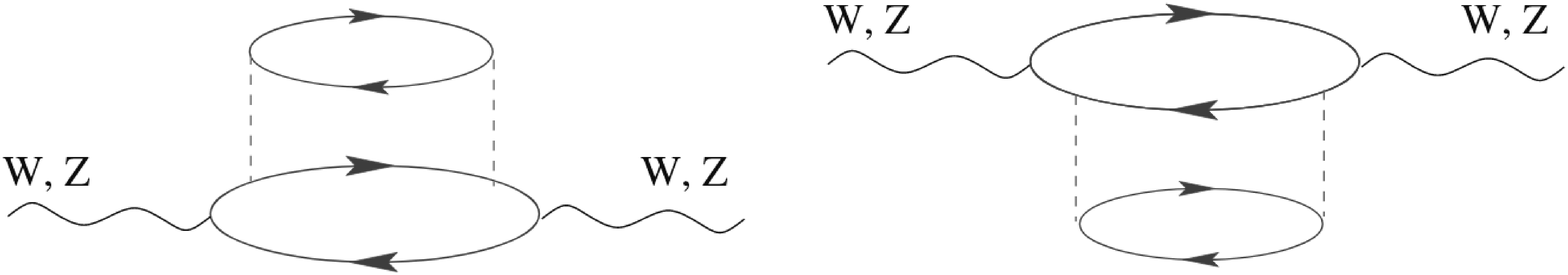}
 \includegraphics[height=2cm,width=8cm]{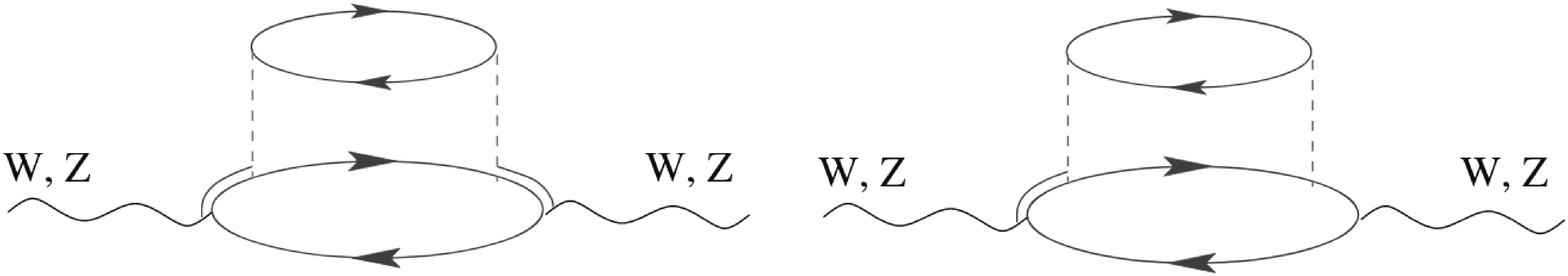}
 \end{center}
 \caption{Diagrams showing some typical 2p-2h contributions arising due to the $N-N$ and $N-\Delta$ correlations. 
 Solid~(dashed) lines denote nucleon~(pion) propagators. Double lines represent $\Delta(1232)$ propagators. Arrows pointing 
 to the right~(left) denote particle~(hole) states.}\label{Ch14-2p2h1}
 \end{figure}

\begin{figure}
 \centering
 \includegraphics[height=2cm,width=8cm]{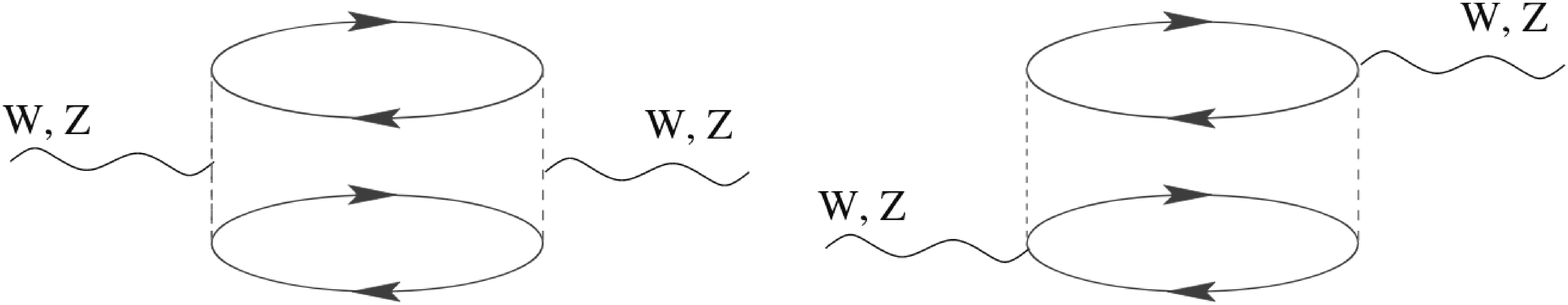}
 \includegraphics[height=2cm,width=8cm]{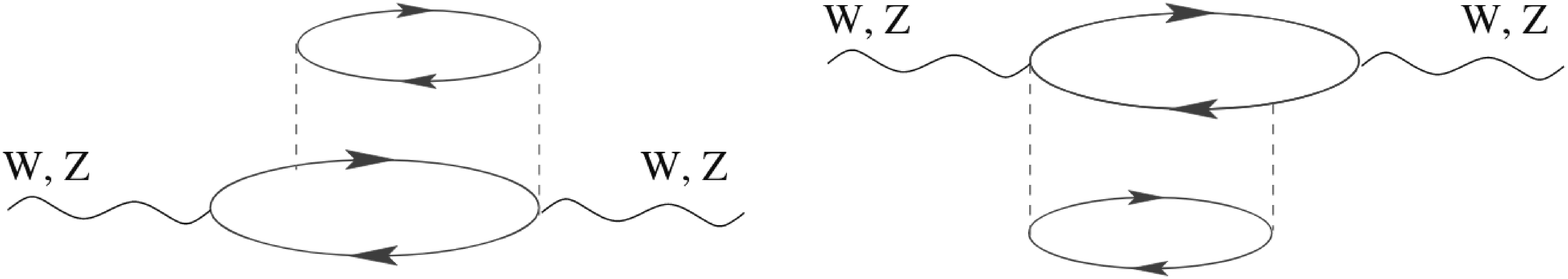}
 \caption{Diagrams showing some typical 2p-2h contributions arising due to the meson exchange. Solid~(dashed) lines denote 
 nucleon~(pion) propagators. Double lines represent $\Delta(1232)$ propagators. Arrows pointing to the right~(left) denote 
 particle~(hole) states.}\label{Ch14-2p2h2}
 \end{figure}

The MiniBooNE puzzle initiated an extensive debate on NME in (anti)neutrino-nucleus scattering. Since 
the $^{12}$C nucleus used in many neutrino oscillation experiments is one of the better understood nucleus theoretically as well as 
experimentally from the study of electron nucleus scattering regarding its wave function, the discrepancy in the measurement 
of neutrino cross section at MiniBooNE was very difficult to be explained. It was attributed to other nuclear effects 
beyond the impulse approximation due to MEC and nucleon-nucleon correlations or FSI effects leading to QE like events. 
In earlier treatments of including such effects, the diagrams corresponding to Fig.~\ref{Ch14-2p2h1} were taken into 
consideration while the diagrams corresponding to Fig.~\ref{Ch14-2p2h2}, where $W$ and $Z$ bosons interacting 
directly with the nonnucleonic degrees of freedom in the nucleus were not fully incorporated.  

\begin{figure}
 \centering
 \includegraphics[height=5cm,,width=8cm]{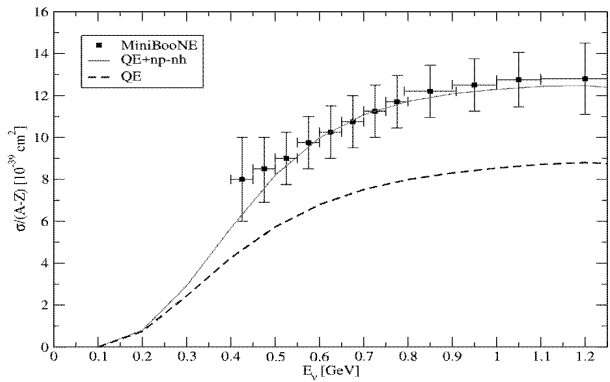}
 \caption{\label{fig_sig_tot_miniboone}``QE-like'' $\nu_\mu$-$^{12}$C cross sections measured by the 
 MiniBooNE collaboration~\cite{MiniBooNE:2010bsu, MiniBooNE:2008yuf} compared to Martini \textit{et 
 al.}~\cite{Martini:2009uj} calculations.}\label{np-nh-miniboone}
\end{figure}

It was suggested by the Lyon group~\cite{Martini:2009uj, Martini:2010ex, Martini:2011wp} for the first time that the processes 
like 2p-2h, or in general n particle-n hole (np-nh), which are multi-nucleon correlation effects could be 
important. Fig.~\ref{np-nh-miniboone} shows the results of QE-like'' $\nu_\mu$-$^{12}$C cross sections measured by the 
MiniBooNE collaboration~\cite{MiniBooNE:2010bsu, MiniBooNE:2008yuf} and the theoretical curves are the results with and without 2p-2h excitations 
using the model of Martini {et al.}~\cite{Martini:2009uj}. This was followed by the works of Valencia group~\cite{Nieves:2011yp, Nieves:2011pp}, 
which were in agreement with the theoretical observations made by the  Lyon group~\cite{Martini:2009uj}. This was in addition 
to the long range nuclear correlations discussed in Section~\ref{LFG} using RPA. Both these groups use  microscopic approach. This 
led to lots of interest among the scientific community to understand the multinucleon correlation effects. Presently these 
studies may be broadly divided into three categories:
\begin{itemize}
\item In the first approach, one starts from an independent particle model (IPM). For example, LFGM has been 
used by the Lyon~\cite{Martini:2009uj, Martini:2010ex, Martini:2011wp, Martini:2013sha, Martini:2014dqa, Ericson:2015cva, 
Martini:2016eec} and the Valencia~\cite{Sobczyk:2020dkn, Nieves:2011yp, Nieves:2013fr, Nieves:2011pp, Gran:2013kda, 
Bourguille:2020bvw} groups and their umbrella collaborations. The Ghent~\cite{Pandey:2014tza, VanCuyck:2016fab, VanCuyck:2017wfn} and the La 
Plata~\cite{BerruetaMartinez:2021rpf} groups use nonrelativistic and relativistic mean field approaches. In addition to that 
they took into account 2p-2h contributions to the neutrino-nucleus cross section. 

\item In the second approach, one starts from a correlated wave function and the 2p-2h excitations self evolve due to the short 
range correlations. In addition to that, the contribution from MEC in some of the works have also been 
included. For example, the works of group 
using Green's function Monte Carlo method~\cite{Lovato:2020kba, Lovato:2015qka, Lovato:2014eva, Lovato:2017cux} or the group 
using spectral function  approach~\cite{Rocco:2018mwt, Benhar:2005dj, Benhar:2015ula, 
Rocco:2015cil, Barbieri:2019ual} are based on this approach.

\item The third approach is more phenomenological as these methods are constrained by the electron scattering phenomenology, for 
example, the SuSA~\cite{Gonzalez-Jimenez:2014eqa, Amaro:2004bs}, and the model developed by Geissen 
group~(GiBUU)~\cite{Gallmeister:2016dnq}. Amaro et al.~\cite{Amaro:2011aa, Martinez-Consentino:2021avr} and 
Megias et al.~\cite{Megias:2014qva} calculated the MEC effect in the SuSA model. In the works of 
Refs.~\cite{RuizSimo:2014qgg, RuizSimo:2016rtu, Megias:2016fjk, 
RuizSimo:2016ikw, RuizSimo:2017onb, Amaro:2017eah, RuizSimo:2017hlc, Megias:2017cuh}, the multinucleon excitations 
are included via a microscopic fully relativistic calculation of the 2p-2h excitations. While the  
GiBUU includes 2p-2h excitations via an empirical spin-isospin response deduced from the electron scattering 
data~\cite{Gallmeister:2016dnq}.
\end{itemize} 
\begin{figure}[h]
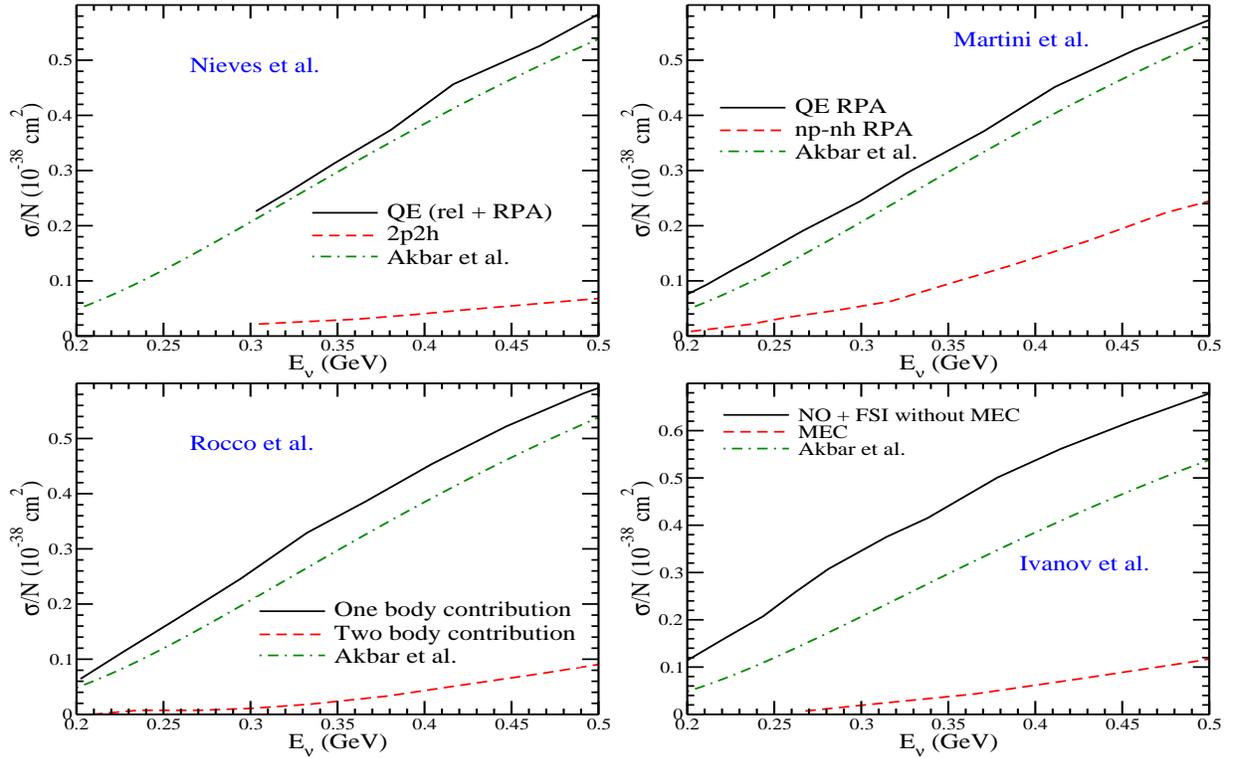

\centering
\includegraphics[height=5cm,width=8cm]{Nieves-1.eps}
\includegraphics[height=5cm,width=8cm]{Martini-1.eps}
\includegraphics[height=5cm,width=8cm]{Rocco-1.eps}
\includegraphics[height=5cm,width=8cm]{Ivanov-1.eps}
\caption{$\nu_\mu$ scattering cross section in $^{12}C$ per neutron target. Clockwise from the top left: (1) Nieves et 
al.~\cite{Nieves:2011pp} in the relativistic Fermi gas model with RPA effect and without 2p-2h contribution~(solid line) and 
only the 2p-2h contribution~(dashed line); (2) Martini et al.~\cite{Martini:2010ex, Martini:2011wp} relativistic Fermi gas 
model with RPA effect and without np-nh contribution~(solid line) and only the np-nh contribution~(dashed line); (3) Ivanov 
et al.~\cite{Ivanov:2018nlm} using realistic spectral function with nucleon-nucleon correlations and without MEC 
contribution~(solid line) and only the MEC contribution~(dashed line), and (4)  Rocco et al.~\cite{Rocco:2018mwt} using 
impulse approximation with spectral function and without 2p-2h contribution~(solid line), and only the 2p-2h contribution~(dashed line). 
The results of Akbar et al.~\cite{Akbar:2017dih} has been shown by dash-dotted line using the relativistic Fermi gas model 
with RPA effect.}\label{fig:KDAR}
\end{figure}

\subsubsection{Nuclear medium effects beyond the impulse approximation}
We have discussed the importance of NME like the 2p-2h, ph-$\Delta$h and MEC 
beyond the impulse approximation in the case of inclusive neutrino scattering in $^{12}$C in the intermediate energy region 
of several hundreds of MeV. These effects were also shown earlier to be important in the very low energy region of the 
nuclear beta decays~\cite{Nieves:2011yp, Dautry:1975xq, Ohta:1974fa, Ericson:1973vj, Riska:1970jxh, Delorme:1985ps, 
Muller:1981dc}, solar neutrino reaction and muon capture~\cite{Suzuki:1987jf, Mukhopadhyay:1998me}. 

In the case of $\pi$-DIF neutrinos corresponding to the low energy neutrinos of $E_{\nu}<236$~MeV, the calculations of 
Hayes and Towner~\cite{Hayes:1999ew} were performed in a microscopic 
nuclear model using a multiparticle shell model with large basis space, while the calculations of Umino et 
al.~\cite{Umino:1995bql, Umino:1994wu} were done using the relativistic Fermi gas model. Both the calculations find a 
reduction of about 20$\%$ in the inclusive cross sections while in the case of KDAR neutrinos, the work of Nikolakopoulos et 
al.~\cite{Nikolakopoulos:2020alk} finds an increase of about 20-25$\%$ obtained from extrapolating the works of 
NuWro~\cite{Juszczak:2009qa, Golan:2012wx} and Martini et al.~\cite{Martini:2010ex, Martini:2011wp}. 
 
Recently an ab initio calculation of the inclusive $\nu_\mu$ cross section in $^{12}$C has been done by Rocco et 
al.~\cite{Rocco:2018mwt} including the contribution of some two body effects. Moreover, Ivanov et al.~\cite{Ivanov:2018nlm} 
have made an improvement over the calculations of Nieves et al.~\cite{Nieves:2011pp} by using an spectral function 
$S(\vec{p},E)$ for the nucleon { energy and  momentum } distribution to calculate the inclusive cross sections in the relativistic Fermi gas 
model. We show in Fig.~\ref{fig:KDAR}, the results for the inclusive cross section $\sigma(E_{\nu_\mu})$ as a function of the 
neutrino energy E$_{\nu_\mu}$ in the energy range $0<E_{\nu_\mu}<500$~MeV in various models. It is clear from 
Fig.~\ref{fig:KDAR} that the calculations by Rocco et al.~\cite{Rocco:2018mwt} and Ivanov et al.~\cite{Ivanov:2018nlm} show 
an enhancement in the inclusive cross section at $E_{\nu_\mu}=236$~MeV which are quantitatively small as compared to the 
results quoted by Nikolakopoulos et al.~\cite{Nikolakopoulos:2020alk}.
  
We observe from the results shown in Table~\ref{KDAR3}, that:
\begin{itemize}
\item[(i)] The contribution of NME beyond the impulse approximation  is to increase the inclusive 
cross sections but the increase is not sufficient enough to explain the results of the KDAR neutrinos.
 
\item[(ii)] Such an increase in the inclusive cross section in the theoretical predictions due to NME, in the similar 
energy region of the $\pi$-DIF neutrinos would further enhance the disagreement between the 
theoretical and the experimental results in the case of LSND experiment. 

\item[(iii)] Moreover, this would also be in contradiction with earlier results of such effects calculated in the work of 
Hayes and Towner~\cite{Hayes:1999ew} in a microscopic model and Umino et al.~\cite{Umino:1995bql, Umino:1994wu} in the case of 
Fermi gas model.
\end{itemize}

It is clear that present status of the theoretical calculations for the inclusive cross section in the process $\nu_\mu 
+^{12}C\rightarrow\mu^- +X$ in the low energy region of few hundreds of MeV is not satisfactory even with the inclusion of 
NME beyond the impulse approximation calculated in various models available in the literature.

\subsubsection{$|\Delta S|=1$ quasielastic scattering in nuclei}\label{hyperon:nucleus}
We have discussed in Section~\ref{qe_hyperon} single hyperon production in the antineutrino induced CC 
interaction from the free nucleon target. 
When the reactions shown in Eqs.~(\ref{process1}) and (\ref{process3}) take place on nucleons which are bound in the nucleus, 
Fermi motion and Pauli blocking effects of initial nucleons play important role, which have been recently discussed in the 
literature~\cite{Fatima:2018wsy, Fatima:2021ctt, Singh:2006xp}. In the work of Ref.~\cite{Singh:2006xp}, the Fermi motion 
effects are calculated in LFGM which has been discussed in detail in Section~\ref{LFG}. For example, for 
$\Lambda$ or $\Sigma^0$ production in an antineutrino interaction with the nucleus, the differential scattering cross section 
is expressed in terms of the differential scattering cross section for an antineutrino scattering from a free nucleon i.e. 
$\frac{d\sigma}{dQ^{2}}|_{\bar\nu N}$~(Eq.~(\ref{dsig})) and integrated over the whole nucleus, which for example in the case 
of an antineutrino interaction on a proton target is given by 
\begin{equation}\label{diff-sec-hyperon-1}
\frac{d\sigma}{dQ^{2}}|_{\bar\nu A}= 2{\int d^3r} \rho_p(r) \frac{d\sigma}{dQ^{2}} |_{\bar\nu N},
\end{equation}
where a factor of 2 is to account for the spin degrees of freedom and the expression for $\rho_p(r)$ is given in 
Eq.~(\ref{eq:rho}). Similarly for a $\Sigma^-$ production from an antineutrino interaction from a neutron target $\rho_p(r)$ 
is replaced by $\rho_n(r)$.
 
Following Eqs.~(\ref{delta2}) and (\ref{sig_4}), we may write Eq.~(\ref{diff-sec-hyperon-1}) as
\begin{equation}\label{hyp-nucl}
\frac{d\sigma}{dQ^{2}}|_{\bar\nu A}=2{\int d^3r \int 
\frac{d^3p}{{(2\pi)}^3}n_p(p,r)\left[\frac{d\sigma}{dQ^{2}}\right]_{\bar\nu N}},
\end{equation}
where $n_p(p,r)$ is the occupation number of the nucleon. $n_p(p,r)=1$ for $p\le p_{F_p}$ and is equal to zero for $p>
p_{F_p}$, where $p_{F_p}$ is the Fermi momentum of the proton.
\begin{figure}[h]
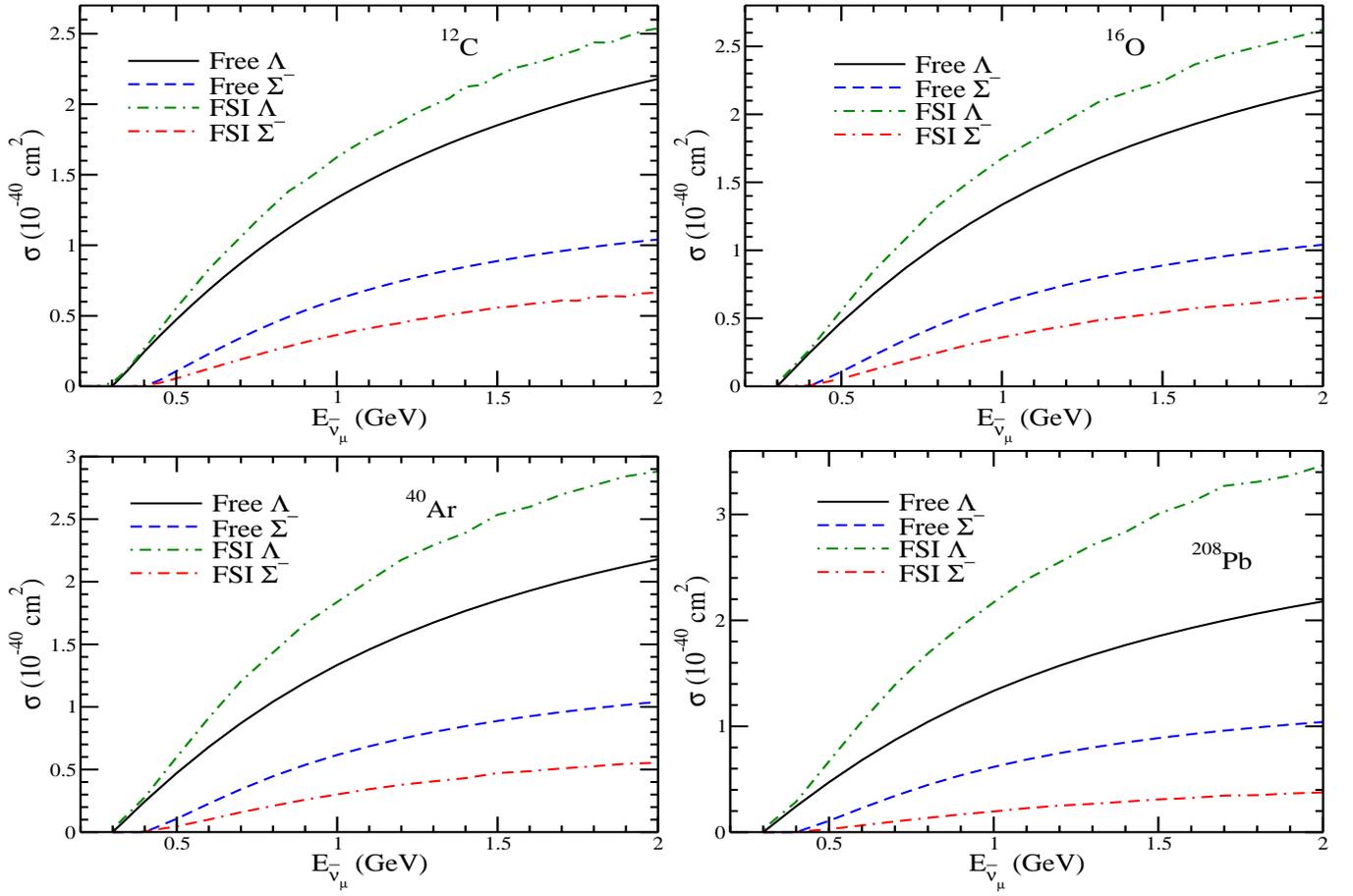

 \centering
\includegraphics[height=0.24\textheight,width=0.48\textwidth]{hyperon_nucleus_carbon.eps}
\includegraphics[height=0.24\textheight,width=0.48\textwidth]{hyperon_nucleus_oxygen.eps}\\
\includegraphics[height=0.24\textheight,width=0.48\textwidth]{hyperon_nucleus_argon.eps}
\includegraphics[height=0.24\textheight,width=0.48\textwidth]{hyperon_nucleus_lead.eps}
\caption{$\sigma$~(per active nucleon) vs $E_{\bar\nu_\mu}$ in $^{12}$C~(top-left panel),$^{16}$O~(top-right 
panel), $^{40}$Ar~(bottom-left panel) and $^{208}$~Pb(bottom-right panel) nuclear targets, for the QE hyperon production. The results for 
$\Lambda$ production~(per proton) are shown without~(solid line) and with FSI~(dash-dotted line). Corresponding results are shown for 
$\Sigma^-$ production~(per neutron) without~(dashed line) and with FSI~(double dash-dotted 
line)~\cite{Fatima:2018wsy, Fatima:2021ctt}.}\label{fg:xsec_c12}
\end{figure}

\subsubsection{Final state interaction}\label{FSI:hyperon}
The produced hyperons are affected by the final state interaction~(FSI) within the nucleus through the hyperon-nucleon 
elastic processes like $\Lambda N \rightarrow  \Lambda N$, $\Sigma N \rightarrow  \Sigma N$, etc. and the charge exchange 
scattering processes like $\Lambda n \rightarrow \Sigma^- p$, $\Lambda n \rightarrow \Sigma^0 n$, $\Sigma^- p 
\rightarrow \Lambda n$, $\Sigma^- p \rightarrow \Sigma^0 n$, etc. Because of these interactions in the nucleus, the 
probability of $\Lambda$ or $\Sigma$ production changes. This has been taken into account by using the prescription given 
in Ref.~\cite{Singh:2006xp}. In this prescription, an initial hyperon produced at a position $r$ within the nucleus, 
interacts with a nucleon to produce a new hyperon state within a short distance $dl$ with a probability $P = P_{Y}dl$, 
where $P_Y$ is the probability per unit length given by~\cite{Singh:2006xp}:
\[P_Y=\sigma_{Y+n \rightarrow f}(E)~\rho_{n}(r)~+~\sigma_{Y+p \rightarrow f}(E)~\rho_{p}(r),\] 
$f$ denotes a possible final hyperon-nucleon [$Y_f(\Sigma ~\text{ or }~ \Lambda) + N(n~\text{ or }~p)$] state with energy $E$ 
in the hyperon-nucleon CM system, $\rho_{n}(r)~(\rho_{p}(r))$ is the local density of the neutron~(proton) in the 
nucleus, and $\sigma$ is the total scattering cross section for CC channel like $Y(\Sigma ~\text{ or }~ 
\Lambda)~+~N(n~ \text{ or }~p) \rightarrow f$~\cite{Singh:2006xp}. Now a particular channel is selected, which gives rise to 
a hyperon $Y_f$ in the final state with probability $P$. For the selected channel, the Pauli blocking effect is taken into 
account by first randomly selecting a nucleon in the local Fermi sea, then a random scattering angle is generated in the 
hyperon-nucleon CM system assuming the cross sections to be isotropic. By using this information, hyperon and 
nucleon momenta are calculated and Lorentz boosted to the lab frame. If the nucleon in the final state has momentum above the 
Fermi momentum, we have a new hyperon type~($Y_f$) and/or a new direction and energy of the initial hyperon~($Y_i$). This 
process is continued until the hyperon gets out of the nucleus.
 
The results for the total cross section $\sigma(E_{\bar \nu_\mu})$ vs $E_{\bar \nu_\mu}$ for $\Lambda$ and $\Sigma^-$ 
production are obtained by integrating Eq.~(\ref{hyp-nucl}) over $Q^2$, for various nuclei of interest like $^{12}$C, 
$^{16}$O, $^{40}$Ar and $^{208}$Pb relevant to ongoing or proposed antineutrino experiments. It is found that NME due 
to Pauli blocking are very small~(not shown here)~\cite{Fatima:2018wsy, Fatima:2021ctt, Singh:2006xp}. 
However, the final state interactions due to $\Sigma-N$ and $\Lambda - N$ interactions in various channels tend to increase 
the $\Lambda$ production and decrease the $\Sigma^-$ production. The quantitative increase~(decrease) in $\Lambda~(\Sigma)$ 
yield due to FSI increases with the increase in nucleon number. Due to the FSI effect of the hyperon with the nucleon in the 
nucleus, the enhancement in the $\Lambda$ production cross section is 22--25$\%$ in $^{12}$C and $^{16}$O for 
$E_{{\bar\nu}_\mu}=0.6-1$~GeV, which increases to 34--38$\%$ in $^{40}$Ar and 52--62$\%$ in $^{208}$Pb. While the decrease 
in $\Sigma^-$ production cross section is about 40--46$\%$ in $^{12}$C and $^{16}$O for $E_{{\bar\nu}_\mu}=0.6-1$~GeV, which 
becomes 50--56$\%$ in $^{40}$Ar and 68--70$\%$ in $^{208}$Pb. $\Sigma^0$ production cross section is separately affected and 
the relation $ \sigma \left(\bar \nu_\mu + p \rightarrow \mu^+ + \Sigma^0 \right) = \frac12  \sigma \left(\bar \nu_\mu + n 
\rightarrow \mu^+ + \Sigma^- \right)$ is modified in the nucleus due to the presence of other nucleons. For example, the 
decrease~(not shown here) in $\Sigma^0$ production cross section is 28--32$\%$ in $^{12}$C and $^{16}$O for 
$E_{{\bar\nu}_\mu}=0.6-1$~GeV, which becomes 45--52$\%$ in $^{40}$Ar and 64--70$\%$ in $^{208}$Pb. 

\begin{figure} 
\centering
\includegraphics[height=0.24\textheight,width=0.55\textwidth]{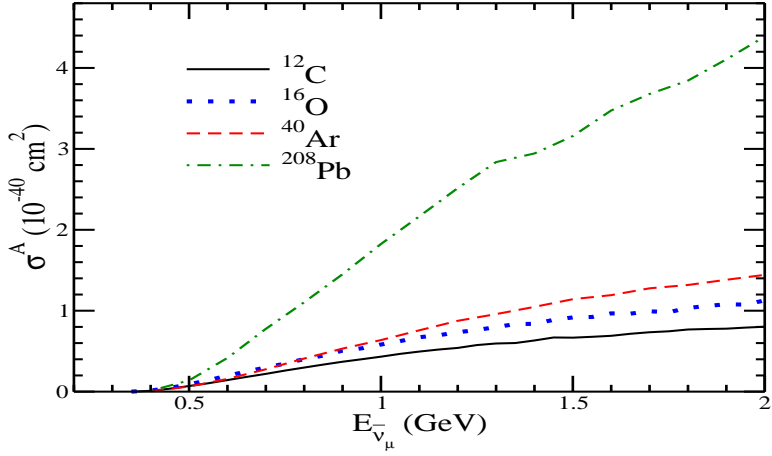}
\caption{$\sigma$ vs $E_{\bar\nu_\mu}$, for $\Sigma^+$ production arising due to final state interaction 
effect of $\Lambda$ and $\Sigma^0$ hyperons in nuclei. Solid line is the result in $^{12}C$, dotted is the result in 
$^{16}O$, dashed line is the result in $^{40}Ar$ and dash-dotted line is the result in 
$^{208}Pb$~\cite{Fatima:2018wsy, Fatima:2021ctt}.}\label{fg:xsec_sig_plus}
\end{figure}

We also see that though $\Sigma^{+}$ is not produced in the basic reactions, it can appear due to the final state interaction 
processes like $\Lambda p \rightarrow \Sigma^+ n$ and $\Sigma^0 p \rightarrow \Sigma^+ n$. In Fig.~\ref{fg:xsec_sig_plus}, 
the results for the $\Sigma^+$ production cross section are presented as a function of antineutrino energy in various 
nuclei. The total scattering cross section for $\Sigma^+$ production increases with the increase in nucleon number though per 
nucleon target it decreases, for example a suppression is observed in $^{208}$Pb than $^{40}$Ar. This may be due to 
considerably higher Fermi energy of neutrons than protons in $^{208}$Pb which inhibits the production of $\Sigma^+$ through 
$\Lambda p \rightarrow \Sigma^+ n$ and $\Sigma^0 p \rightarrow \Sigma^+ n$ reactions in $^{208}$Pb due to threshold 
considerations. It will be interesting to test these predictions whenever the experimental results are available in future.

\subsection{Inelastic scattering and pion production in the $\Delta$ dominance model}\label{SPP:nucleus}
We have discussed in Section~\ref{sec:1pion}, single pion production in (anti)neutrino induced reactions on the nucleon targets. 
In the case of free nucleon, it is observed that the single pion production in the sub-GeV region is dominated by the $\Delta$ 
resonance excitation which decays subsequently to a pion and a nucleon. The NR terms and the higher resonances also 
play important role especially in the $p\pi^0$ and $n\pi^+$ channels in the neutrino modes~(Fig.~\ref{fig:sigma_Nu_CC1pion}) 
and $n\pi^{0}$ and $p\pi^{-}$ channels in the antineutrino modes~(Fig.~\ref{fig:sigma_Nubar_CC1pion}). These results for 
$\sigma$ {  vs.} $E_{{\nu}_{\mu}}$ are obtained with $C_{3}^{A}=0$, $C_5^A(0)=1$, $C_{4}^{A} (Q^2)= -\frac{1}{4} C_{5}^{A} 
(Q^2)$ and $M_A=1.03$~GeV for the $N-\Delta$ transition form factor in the axial-vector sector. When the processes shown in 
Eqs.~(\ref{eq:CC_all}) and (\ref{eq:NC_all}) take place inside the nucleus, there are two possibilities of the pion production 
i.e. the target nucleus remains in the ground state leading to coherent production of pions or is excited and/or broken up 
leading to incoherent production of pions. As these processes take place inside the nucleus, NME come 
into play, which modifies the resonance properties like its mass and width. In literature, the nuclear medium modifications 
have been studied in the weak sector only in the case of $\Delta$ excitation, therefore, for the pion production in 
(anti)neutrino scattering from the nucleus, the calculations have been done in the $\Delta$ dominance model. Both the 
production processes in the $\Delta$ dominance model using the local density approximation have been considered to calculate 
single pion production from the several nuclear targets like $^{12}$C, $^{16}$O, $^{40}$Ar and $^{208}$Pb. In the delta dominance 
model, if one take $C_5^A(0)=1.2$ and $M_A =1.05$~GeV, then the results obtained with these parameters for the single pion 
production from free nucleons almost resemble the pion production from the free nucleons when the contribution from the higher 
resonances and NRB as well as their interferences are taken with $C_5^A(0)=1$ and $M_A = 
1.03$~GeV~(Fig.~\ref{fig:sigma_Nu_CC1pion}). In this section, we are discussing incoherent pion production. The effect of 
nuclear medium on the production of $\Delta$ is treated by including the modification of $\Delta$ properties in the medium. 
Once the pions are produced, they undergo final state interactions with the residual nucleus, which has been taken into 
account.

In the local density approximation~(Section~\ref{LFG}), the cross section for the reaction say $\nu_l(k) + N(p) \rightarrow 
l^-(k^\prime) + N^\prime(p^\prime) + \pi^i(p_\pi)$, where $i=\pm, 0 $ and $N, N^\prime=p \text{ or } n$ inside a nuclear 
target is evaluated as a function of local Fermi momentum~($p_{F}(r)$) and integrated over the size of whole nucleus i.e.    
  \begin{eqnarray}\label{single-pion-diff-xsec}
 \left(\frac{d\sigma}{dE_{\pi} d\Omega_{\pi}}\right)_{\nu A} &=& \int d\vec r ~\rho_N(r)~  \left(\frac{d\sigma}{dE_{\pi} 
 d\Omega_{\pi}}\right)_{\nu N}, \nonumber
 \end{eqnarray}
 where the expression for $\left(\frac{d\sigma}{dE_{\pi} 
 d\Omega_{\pi}}\right)_{\nu N}$ is given in Eq.~(\ref{dsigma:pion}) for the free nucleon target.
 While for the scattering of (anti)neutrino with a nucleon inside the nucleus, the interacting nucleon is not as rest 
 {  i.e.} $E_{N}=\sqrt{|{\vec p_N}|^{2} + M^{2}}$ and
 $ E_{N}^{\prime}=\sqrt{|{\vec p_N}^{\;\prime}|^{2} + M^{2}} = \sqrt{|\vec q - \vec p_{\pi} + \vec p_{N}|^{2} + M^{2}} $, and 
 therefore
 \begin{equation}
  \left(\frac{d\sigma}{dE_{\pi} d\Omega_{\pi}}\right)_{\nu N}= \frac{1}{32\left(2\pi\right)^5} \int d\Omega^\prime dE^\prime
  \delta^0\left(E_N({\vec p}) + q_o - E_{N^\prime}({\vec p}^\prime) - E_{\pi}(\vec{p}_{\pi}) \right) \frac{|{\vec k}^\prime| 
  |{\vec p}_\pi|}{E E_N E_{N^\prime}}\left(\frac{G_F^2}{2}cos^2\theta_c L_{\mu\nu} J^{\mu\nu} \right),
 \end{equation}
 where $L_{\mu\nu}$ is the leptonic tensor, the expression for which is given in Eq.~(\ref{lep_tens}) and the hadronic tensor 
 $J^{\mu\nu}= \frac{1}{2} \sum j^\mu {j^\nu}^\dagger$ where the hadronic current 
 \begin{equation}
  j^\mu= {\mathcal I}~ \bar u({\vec{p}}^{~\prime})\frac{f_{\pi N \Delta}}{m_{\pi}}  p^{\sigma}_{\pi}
  {\mathcal P}_{\sigma \lambda} \mathcal O^{\lambda \mu} u({\vec{p}}).
 \end{equation}
 ${\mathcal I} = \sqrt{3}$ for $\Delta^{++}$ and $\Delta^{-}$ excitations, otherwise ${\mathcal I} = 1$. In the above 
expression $\mathcal O^{\lambda \mu}$ is the $N-\Delta$ transition operator, the expression for which is given in 
Eq.~(\ref{eq:gamma_3half_pos}) and ${\mathcal P}_{\sigma \lambda}$ is the $\Delta$ propagator in the momentum space, is given in Eq.~(\ref{propagator:32}).

Following Eqs.~(\ref{delta2}) and (\ref{sig_4}), we may write Eq.~(\ref{single-pion-diff-xsec}) as 
 \begin{eqnarray}
 \left( \frac{d\sigma}{dE_{\pi} d\Omega_{\pi}} \right)_{\nu A} &=& 2 \int d\vec r \sum_{N=n,p}\frac{d\vec p_N}{(2 
 \pi)^3}~\Theta_{1}(E_{F}^{N}(r)-E_N) \Theta_{2}(E_{N} + q_{0} - E_{\pi} - E_{F}^{N^\prime}(r))
 \times\left( \frac{d\sigma}{dE_{\pi} d\Omega_{\pi}} \right)_{\nu N}. \nonumber
 \end{eqnarray}
Thus, in the local density approximation the expression for the total cross section for the neutrino induced CC 1$\pi^+$ 
production from the nuclear target is written as
\begin{eqnarray}\label{sigma_inelas}
\sigma_A(E)&=& \frac{1}{(4\pi)^5}\int_{r_{min}}^{r_{max}}\left(\rho_{p}(r) + \frac{1}{9}\rho_{n} (r)\right)  d\vec r 
\int_{Q^{2}_{min}}^{Q^{2}_{max}}dQ^{2}
\times \int^{k^\prime_{max}}_{k^\prime_{min}} d{k^\prime} \int_{-1}^{+1}dcos\theta_{\pi } \nonumber\\
&&\times \int_{0}^{2\pi}d\phi_{\pi} 
\frac{\pi|\vec  k^\prime||\vec k_{\pi}|}{M E_{\nu}^2 E_{l}}\frac{1}{E_{p}^{\prime}+E_{\pi}\left(1-\frac{|\vec q|}
{|\vec k_{\pi}|}cos\theta_{\pi }\right)}\left(\frac{G_F^2}{4}cos^2\theta_c L_{\mu\nu} J^{\mu\nu} \right).
\end{eqnarray}
For $1\pi^-$ production $\left(\rho_{p}(r) + \frac{1}{9}\rho_{n} (r)\right)$ in the above expression, is replaced by $\left(\rho_{n}(r) + 
\frac{1}{9}\rho_{p} (r)\right)$ and for the $\pi^0$ production, it is replaced by $\frac{2}{3}\left(\rho_{p}(r) + \rho_{n} (r)
\right)$, where $\rho_{p} (r)$ and $\rho_{n} (r)$ are already defined in Section~\ref{LFG}.

The nuclear medium modification on the $\Delta$ properties like the modification in its mass and width arises from the 
following sources:
\begin{itemize}
 \item [(a)] The intermediate nucleon state is partly blocked for the $\Delta$ decay because some of these states are 
 occupied~(Pauli blocking). The decayed nucleon must be in an unoccupied state. The Pauli correction is taken into account 
 by assuming a local Fermi sea at each point of the nucleus of density $\rho({r})$, and forcing the nucleon to be 
 above the Fermi sea. This leads to an energy dependent modification in the $\Delta$ decay width given as~\cite{Oset:1987re}:
\begin{equation}
{\Gamma} \rightarrow{\tilde {\Gamma}}-2\mbox{Im}{\Sigma}_{\Delta},
\end{equation}
where ${\tilde\Gamma}$ is the Pauli blocked width of $\Delta$ in the nuclear medium and its relativistic form 
is~\cite{Oset:1987re, Garcia-Recio:1989hyf}:
\begin{equation}\label{chapter15-eq3}
{\tilde \Gamma}=\frac{1}{6\pi}\left (\frac{f_{\pi N \Delta}}{m_{\pi}}\right )^{2}~\frac{M}{\sqrt{s}}~|{\vec p^{\,
\prime}}_{cm}|^{3}~F(p_{F}, E_{\Delta}, p_{\Delta}) , \qquad \text{where} \qquad 
|{\vec p^{\,\prime}}_{cm}|=\frac{\sqrt{(s-M^2-m_\pi^2)^2-4M^2m_\pi^2}}{2\sqrt{s}},
\end{equation}
and $F(p_{F}, E_{\Delta}, p_{\Delta})$, the Pauli correction factor is written as~\cite{Oset:1987re, Garcia-Recio:1989hyf}:
\begin{equation}
F(p_{F}, E_{\Delta}, p_{\Delta})= \frac{p_{\Delta}|{\vec p^{\,\prime}_{cm}}|+E_{\Delta}{E^\prime_p}_{cm}-E_{F}{\sqrt s}}
{2p_{\Delta}|{\vec p^{\,\prime}}_{cm}|}, 
\end{equation}
with $p_F$ as the Fermi momentum, $E_F=\sqrt{M^2+p_F^2}$ and ${\vec p^{\,\prime}}_{cm}$, ${E^\prime_p}_{cm}$ the nucleon 
momentum and the relativistic nucleon energy in the final $\pi$N CM frame.
If $F(p_{F}, E_{\Delta}, p_{\Delta})>1$ it is replaced by 1 in Eq.~(\ref{chapter15-eq3}), and similarly, if 
$F(p_{F}, E_{\Delta},p_{\Delta})<0$ then it is 
replaced by 0 in Eq.~(\ref{chapter15-eq3}).
 
In the above expression $\sqrt{s}$ is CM energy in the $\Delta$ rest frame averaged over the Fermi sea, 
$\bar{s}$ and is given as~\cite{Oset:1987re, Garcia-Recio:1989hyf}
\begin{equation}
\bar s=M^2+m_\pi^2+2E_\pi\left(M+\frac{3}{5}\frac{p_F^2}{2M}\right).
\end{equation}
\item [(b)]The produced nucleon in the $\Delta$ decay inside the nuclear medium feels a single particle potential due to 
all the other nucleons in the nucleus, known as the binding effect, which is taken care by the real part of the $\Delta$ self 
energy. This effect modifies the mass of $\Delta$ in the medium as~\cite{Oset:1987re, Garcia-Recio:1989hyf}:
\begin{equation}
M_{\Delta} \rightarrow {\tilde M_{\Delta}}=M_{\Delta}+\mbox{Re}\Sigma_{\Delta}.
\end{equation}
The $\Delta$ self energy plays a very important role in the different pion nuclear reactions. For a thorough study of the 
$\Delta$ self energy, readers are referred to the model developed by Oset and Salcedo~\cite{Oset:1987re}. For the scalar part 
of the $\Delta$ self energy, the 
numerical results are parameterized in the approximate analytical form (excluding the Pauli corrected width), and are given 
as~\cite{Oset:1987re, Garcia-Recio:1989hyf}:
\begin{equation}\label{chap15-eq6}
-\mbox{Im}{{\Sigma}_{\Delta}}=C_{Q}\left (\frac{\rho}{{\rho}_{0}}\right )^{\alpha}+C_{A2}\left (\frac{\rho}{{\rho}_{0}}
\right)^{\beta}+C_{A3}\left (\frac{\rho}{{\rho}_{0}}\right )^{\gamma},
\end{equation}
which is determined mainly by the one pion interaction in the nuclear medium. This includes the two body, three body and 
the QE absorption contributions for the produced pions in the nucleus. The coefficients $C_{Q}$ accounts for the 
QE part, the term with $C_{A2}$ for two body absorption and the one with $C_{A3}$ for three body absorption, and 
are parameterized in the range of energies 80~MeV$<T_{\pi}<320$~MeV, where $T_{\pi}$ is the pion kinetic energy, 
as~\cite{Oset:1987re, Garcia-Recio:1989hyf}:
\begin{equation}\label{chap15-eq7}
 C(x)=ax^{2}+bx+c,~~~~~~~x=\frac{T_{\pi}}{m_{\pi}},
\end{equation}
where C stands for all the coefficients i.e. $C_{Q},~C_{A2},~C_{A3},~\alpha$ and $\beta$($\gamma=2\beta$). The different 
coefficients used have been taken from Ref.~\cite{Oset:1987re, Garcia-Recio:1989hyf}.

The real part of the $\Delta$ self energy has been approximately taken as~\cite{Oset:1987re, Garcia-Recio:1989hyf}
\begin{equation}
\mbox{Re}{\Sigma}_{\Delta}\simeq40.0~\left(\frac{\rho}{\rho_{0}}\right)\mbox{MeV} .
\end{equation}
\item [(c)] It should be noted that $\tilde\Gamma$ describes the $\Delta$ decaying into nucleon and pion. The various terms 
in the Im$\Sigma_\Delta$ correspond to the different responses of $\Delta$ in the nuclear medium as explained earlier. 
$C_Q$ term in Im$\Sigma_\Delta$ gives additional contribution to the pion production which arises solely due to NME. Some 
of the $\Delta$s are absorbed through two body and three body absorption processes and do not lead to 
pion production. These are described by $C_{A2}$ and $C_{A3}$ terms in the expression for $Im{{\Sigma}_{\Delta}}$ given in 
Eq.~(\ref{chap15-eq6}) and do not contribute to the lepton production accompanied by pions. These constitute QE-like 
events besides the pions which are physically produced but reabsorbed in the nucleus due to FSI, which shall be discussed 
later in the text. Only the $C_Q$ term in the expression for Im$\Sigma_\Delta$~(Eq.~(\ref{chap15-eq6})) contributes to the 
lepton production accompanied by a pion. These have been discussed by us in Refs.~\cite{SajjadAthar:2009rc, SajjadAthar:2009rd}.
\end{itemize}
\begin{figure} 
 \centering
     \includegraphics[height=4.5cm,width=.45\linewidth]{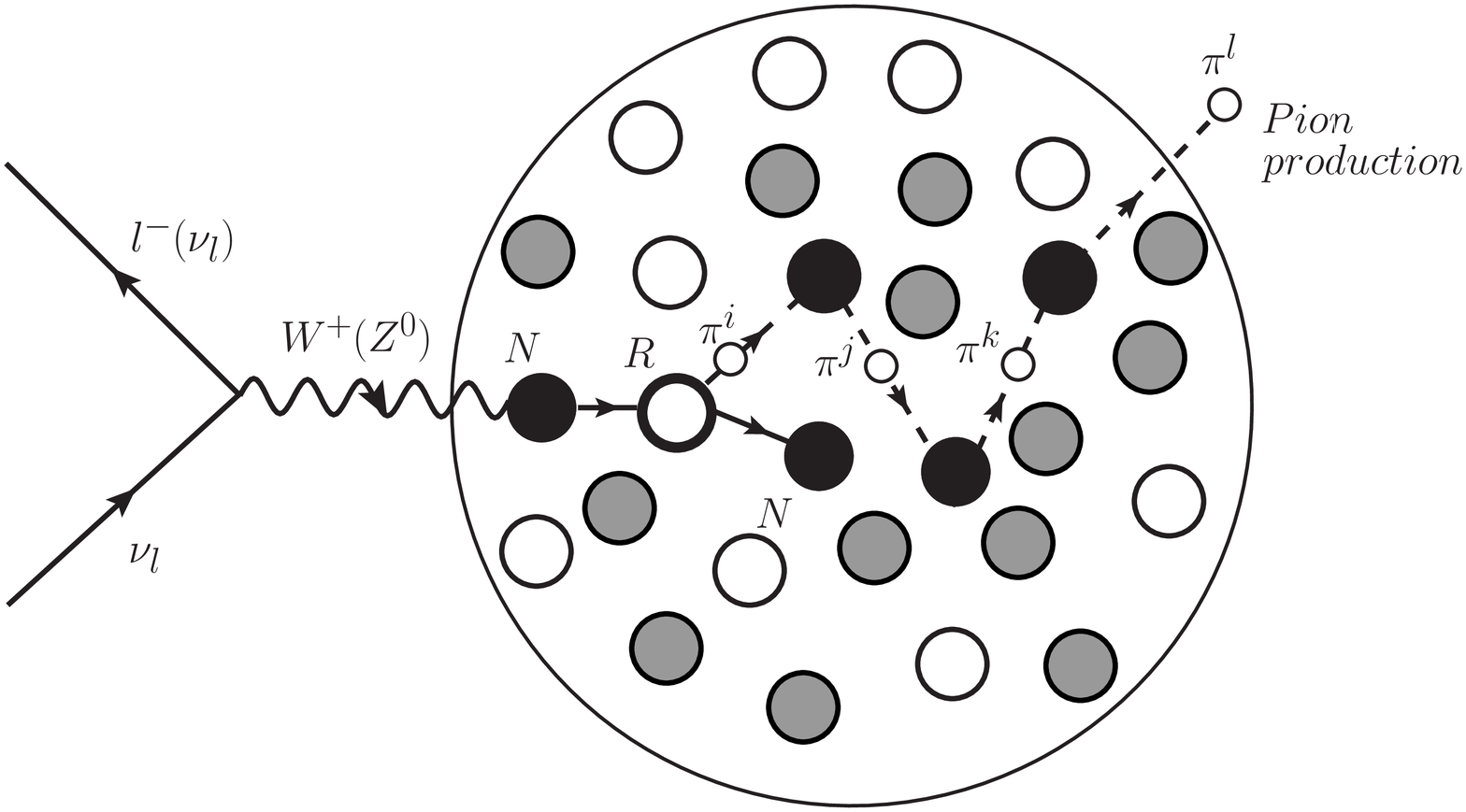}
     \includegraphics[height=4.5cm,width=.45\linewidth]{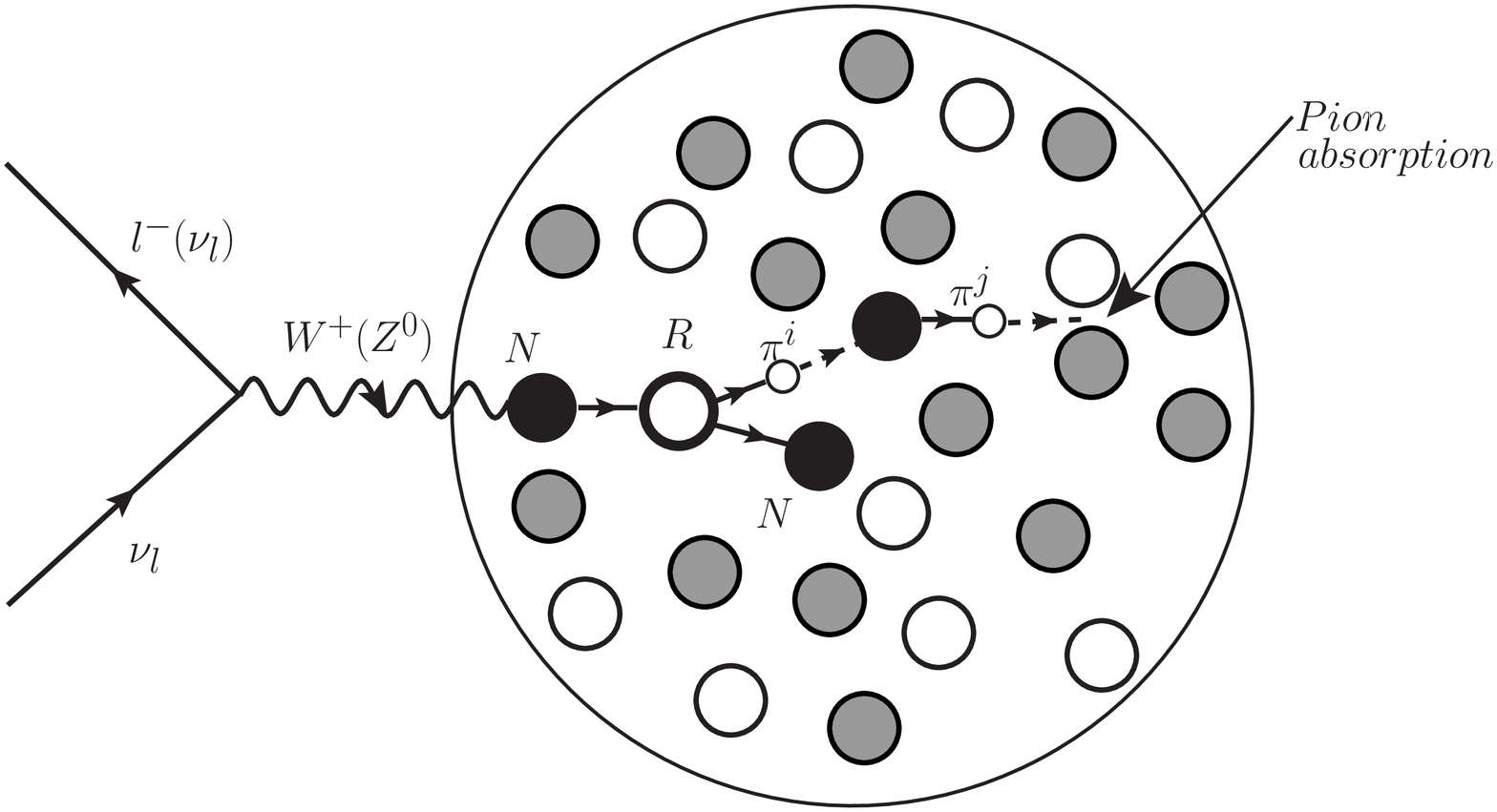}
\caption{Pion production inside a nuclear target and its interaction with the nucleons in the nucleus while coming out. The 
pions may undergo elastic scattering, charge exchange~(for example $\pi^+ + n \rightarrow \pi^0 + p$) reaction, etc. and 
therefore the charge of pion may change on way before it comes out of the nucleus~(left panel). In a neutrino induced 
reaction on a nucleon target when a pion is absorbed in the nuclear medium while coming out and constitutes the QE like process~(right panel).
The open~(shaded) circles represent 
protons~(neutrons) inside a nucleus and the dark shaded circle represents a nucleon~($p$ or $n$) with which a neutrino 
interacts through a CC~(NC) reaction and a nonresonant~($NR$) or a resonant~($R$) state is formed, which gives rise to a proton or 
a neutron and a pion~($\pi^{i}$), where $i$ represents the charge state.}\label{chapter15-CC_pi_FSI}
\end{figure}

\subsubsection{Final state interaction effect}\label{FSI:pion}
{ The effective pions which are produced after considering the $\Delta$ renormalization effect 
undergo final state interaction with the residual nucleus. Due to strong interactions with the nucleons, some of these pions can change direction, 
energy, charge, or even produce more pions through $\pi N \rightarrow \pi\pi N$ like reactions or may be lost through the $\pi N N \rightarrow N N$ reaction. 
These effects need to be taken into account~\cite{VicenteVacas:1994sm}.} Therefore, the production cross sections for the pions from 
the nuclear targets are affected by the presence of strong interactions of final state pions in the nuclear 
medium~(Fig.~\ref{chapter15-CC_pi_FSI}). For example, a pion produced in the nuclear medium may get absorbed by the nucleons 
and thus mimicking a QE-like event~(see Fig.~\ref{chapter15-CC_pi_FSI}) or can suffer elastic, and charge 
exchange scattering with the nucleons. There are generally two approaches to treat these final state interactions of pions. 
In one approach, the distortion of pion wave functions is calculated in an optical potential by solving the Schrodinger or 
Klein-Gordon equation of motion or approximation methods using Glauber model is used. In another approach, a microscopic 
method is used in which the motion of the pion inside the nucleus is followed step by step in which the pion suffers 
interaction with the nucleon. This approach has been discussed by Vicente Vacas et al.~\cite{VicenteVacas:1994sm} and is used by 
us in many calculations of treating the FSI of pions~\cite{Ahmad:2006cy, SajjadAthar:2009rd, Singh:1998ha, Athar:2007wd}. 

{ In the second approach, the final state interaction of pions is treated} using Monte Carlo simulations by generating pion
of given momentum and charge at point $\vec{r}$ in the nucleus. Assuming 
the real part of the pion nuclear potential to be weak as compared to their kinetic energies they are propagated in straight lines till they are out 
of the nucleus. At the beginning they are placed at $({\vec{b}}, z_{in})$, with a random 
impact parameter ${\vec{b}}$, with $|{\vec{b}}|<R$, where R is the nuclear radius which is taken to be a point where 
nuclear density $\rho(R)$ falls to ${10}^{-3}\rho_0$, where $\rho_0$ is the central density, and $z_{in}=-\sqrt{R^2-|{\vec{b}}|^2}$.
 Then it is moved in small steps 
$\delta l$ along the z-direction until it comes out of the nucleus or interact. If $P(p_\pi,r,\lambda)$ is the 
probability per unit length at the point r of a pion of momentum ${\vec{p}}_\pi$ and charge $\lambda$, then $P\delta l 
<<1$. A random number x is generated such that $x\in [0,1]$ and if $x > P\delta l$, then it is assumed that pion has 
not interacted while traveling a distance $\delta l$, however, if $x < P\delta l$ then the pion has interacted and 
depending upon the weight factor of each channel given by its cross section it is decided whether the interaction 
was QE, charge exchange reaction, pion production or pion absorption~\cite{VicenteVacas:1994sm}. For example, for 
the QE scattering
\begin{equation*}
P_{N(\pi^\lambda,\pi^{\lambda^\prime})N^\prime}=\sigma_{N(\pi^\lambda,\pi^{\lambda^\prime})N^\prime}\times
\rho_N,
\end{equation*}
where N is a nucleon, $\rho_N$ is its density and $\sigma$ is the elementary cross section for the reaction 
$\pi^\lambda +N \rightarrow \pi^{\lambda^\prime} + N^\prime$ obtained from the phase shift analysis. 
For a pion to be absorbed, $P$ is expressed in terms of the imaginary part of the pion self energy $\Pi$ i.e. 
$P_{abs}=-\frac{Im\Pi_{abs}(p_\pi)}{p_\pi}$, where the self energy $\Pi$ is related to the pion optical potential 
$V(\vec{r})$~\cite{VicenteVacas:1994sm}.

\begin{figure}[h]
\begin{center}
     \includegraphics[height=12cm,width=.95\linewidth]{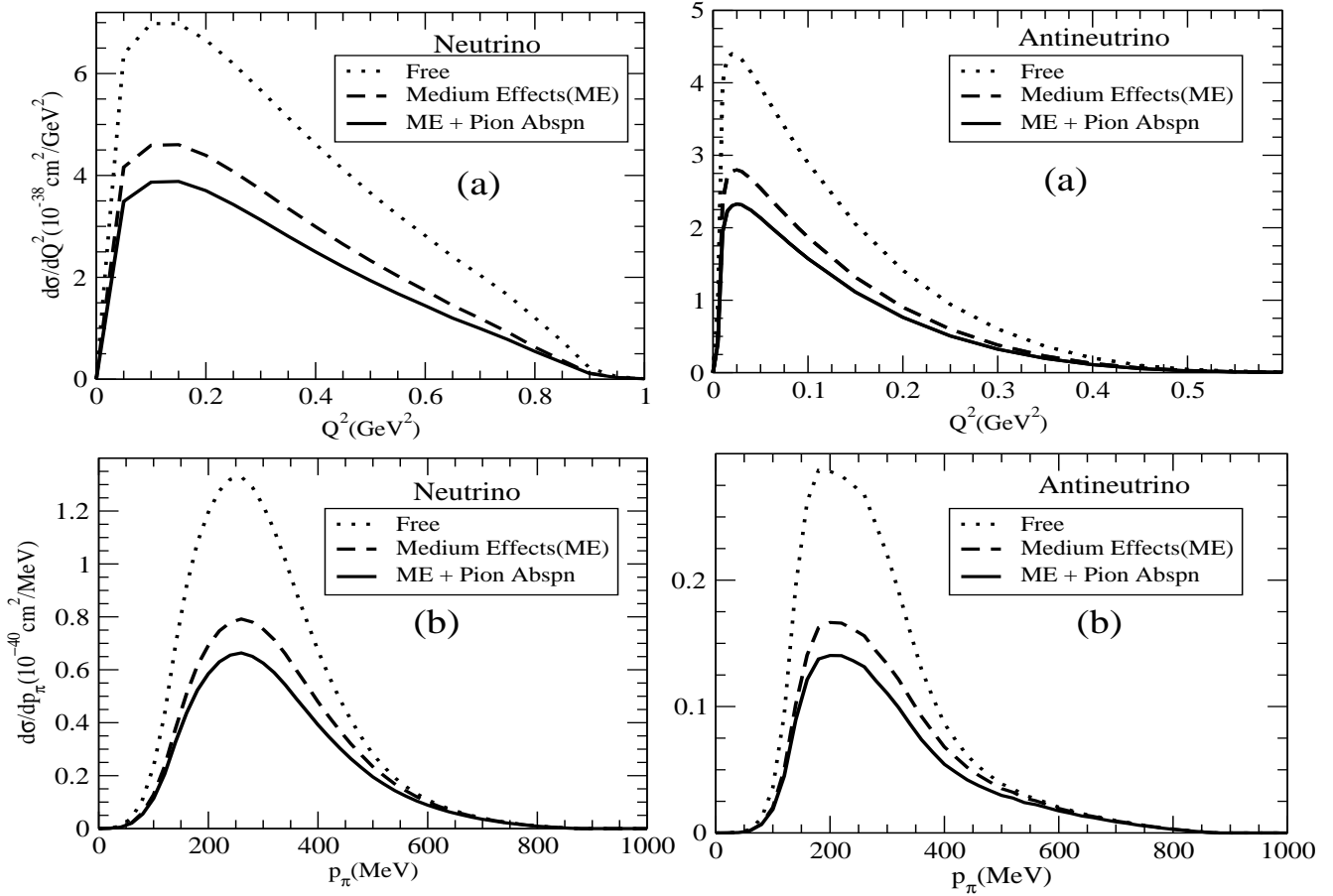}
\end{center}
     \caption{$\frac{d\sigma}{dQ^2}$ and $\frac{d\sigma}{dp_\pi}$ for the $\nu_\mu$(${\bar\nu}_\mu$) induced CC one 
     $\pi^+(\pi^-)$ process on $^{12}C$ target at $E_{\nu}=1GeV$~\cite{Ahmad:2006cy, SajjadAthar:2009rd, Athar:2007wd}.}
\label{chapter19-Incoh}
\end{figure}
\begin{figure}[h]
\begin{center}
 \includegraphics[height=6cm,width=.85\linewidth]{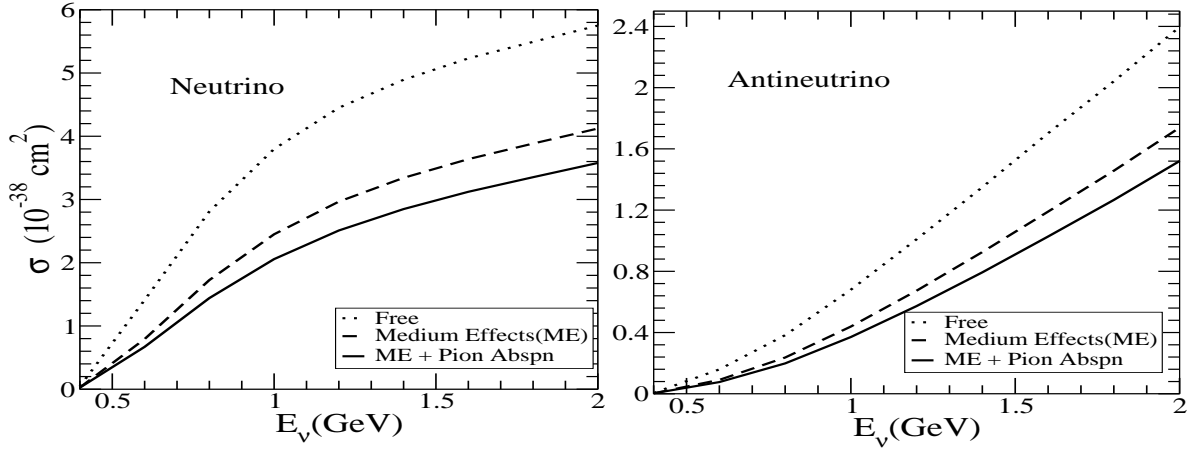}
\end{center}
\caption{$\sigma$ for $\nu_\mu$ (${\bar\nu}_\mu$) induced CC incoherent $\pi^+(\pi^-)$ production on $^{12}C$ 
target~\cite{Ahmad:2006cy, SajjadAthar:2009rd, Athar:2007wd}.}\label{chapter19-Incoh-Total}
\end{figure}

In Fig.~\ref{chapter19-Incoh}, results for the $Q^2$ distribution $\frac{d\sigma}{dQ^2}$ and the pion momentum distribution 
$\frac{d\sigma}{dp_{\pi}}$ are shown for CC $\nu_\mu$(${\bar\nu}_\mu$) induced incoherent one $\pi^{+}$
($\pi^{-}$) production cross section. These results are presented for the differential scattering cross section calculated 
with and without NME, and with NME including the pion absorption effects. For the 
$Q^2$-distribution shown in Fig.~\ref{chapter19-Incoh}, it may be seen that the reduction in the cross section as compared 
to the cross section calculated without NME is around $35\%$ in the peak region. When pion absorption 
effects are also taken into account there is a further reduction of around $15\%$. The results for the antineutrino induced 
one $\pi^{-}$ production cross section are qualitatively similar in nature but quantitatively we find that the peak shifts 
towards a slightly lower value of $Q^2$. Also in this figure, the results for the pion momentum distribution have 
been shown. In this case, the reduction in the cross section in the peak region is around $40\%$ when NME 
are taken into account, which further reduces by about $15\%$ when pion absorption effects are also taken into account.

In Fig.~\ref{chapter19-Incoh-Total}, the results for the total scattering cross section $\sigma$ for CC  
$\nu_\mu$~(${\bar\nu}_\mu$) induced one $\pi^{+}$~($\pi^{-}$) production cross section are shown. With the inclusion of 
NME the reduction in the cross section from the cross section calculated without NME 
for the neutrino energies between 1-2 GeV is 30-35$\%$ which further reduces by about 15$\%$ when pion absorption effects are 
also taken into account. The results with antineutrinos are qualitatively similar to the results obtained in the case of 
neutrino scattering. For more details, see Refs.~\cite{Ahmad:2006cy, Athar:2007wd}.

\subsection{A comparative discussion of results for quasielastic hyperon and delta production from nuclei leading to 
pions}\label{hyperon:Delta:nucleus}
We have seen in Section~\ref{hyperon:nucleus} that in the case of antineutrinos, the QE hyperon~($\Lambda$, 
$\Sigma^0$ and $\Sigma^-$) production also contributes to the total scattering cross section. The produced hyperons then 
decay into a nucleon and a pion, thus, giving additional contribution to the single pion production i.e.
\begin{eqnarray}\label{channels_numubar_pi-_lam}
\bar \nu_l + p &\rightarrow& l^{+} + \Lambda; \qquad\quad \Lambda \rightarrow n + \pi^0~~~\left[35.8\%\right], \qquad \quad
p + \pi^-~
\left[63.9\%\right]\nonumber\\
\bar \nu_l + p &\rightarrow& l^{+} + \Sigma^0; \qquad \Sigma^0 \rightarrow \gamma + \Lambda \qquad \left[100\%\right], 
\qquad \Lambda \rightarrow n + \pi^0,~ p + \pi^-\nonumber\\
\bar\nu_l + n &\rightarrow& l^{+} + \Sigma^-; \qquad \Sigma^- \rightarrow n + \pi^-,~\left[99.85\%\right]
\end{eqnarray}
where the quantities in the square brackets represent the branching ratios of the respective decay modes.
The pions are produced as a result of hyperon decays as shown in Eq.~(\ref{channels_numubar_pi-_lam}). However, when the 
hyperons are produced in a nuclear medium, some of them disappear through the hyperon-nucleon interaction processes like $YN 
\rightarrow NN$, though it is suppressed due to the nuclear effects. The pionic modes of hyperons are Pauli blocked as the 
momentum of the nucleons available in these decays is considerably below the Fermi level of energy for most nuclei leading to 
a long lifetime for the hyperons in the nuclear medium. Therefore, the hyperons which survive the $YN \rightarrow NN$ decay in 
the nuclear medium live long enough to travel and decay outside the nucleus. In view of this, no final state interaction of the 
produced pions with the nucleons inside the nuclear medium in the case of these pion producing reactions is considered. 
 
This mode of pion production is important in the low energy region~($E_{\bar{\nu}_{\mu}} < 1$~GeV), even though, the dominant contribution to
the single pion production comes from the $\Delta$ excitations in the few GeV energy region as discussed in 
Section~\ref{SPP:nucleus}. The cumulative effects of the lower threshold energy for the hyperon production compared to the 
delta production, and the near absence of the FSI for the pions coming from the hyperon decay compensate for the Cabibbo 
suppression as compared to the pions coming from the $\Delta$ excitations in the low energy region. This makes the study of 
hyperon production processes important in the context of oscillation experiments with antineutrinos in the sub-GeV energy 
region.
 
\begin{figure}[h]
\begin{center}
    \includegraphics[height=6cm,width=9cm]{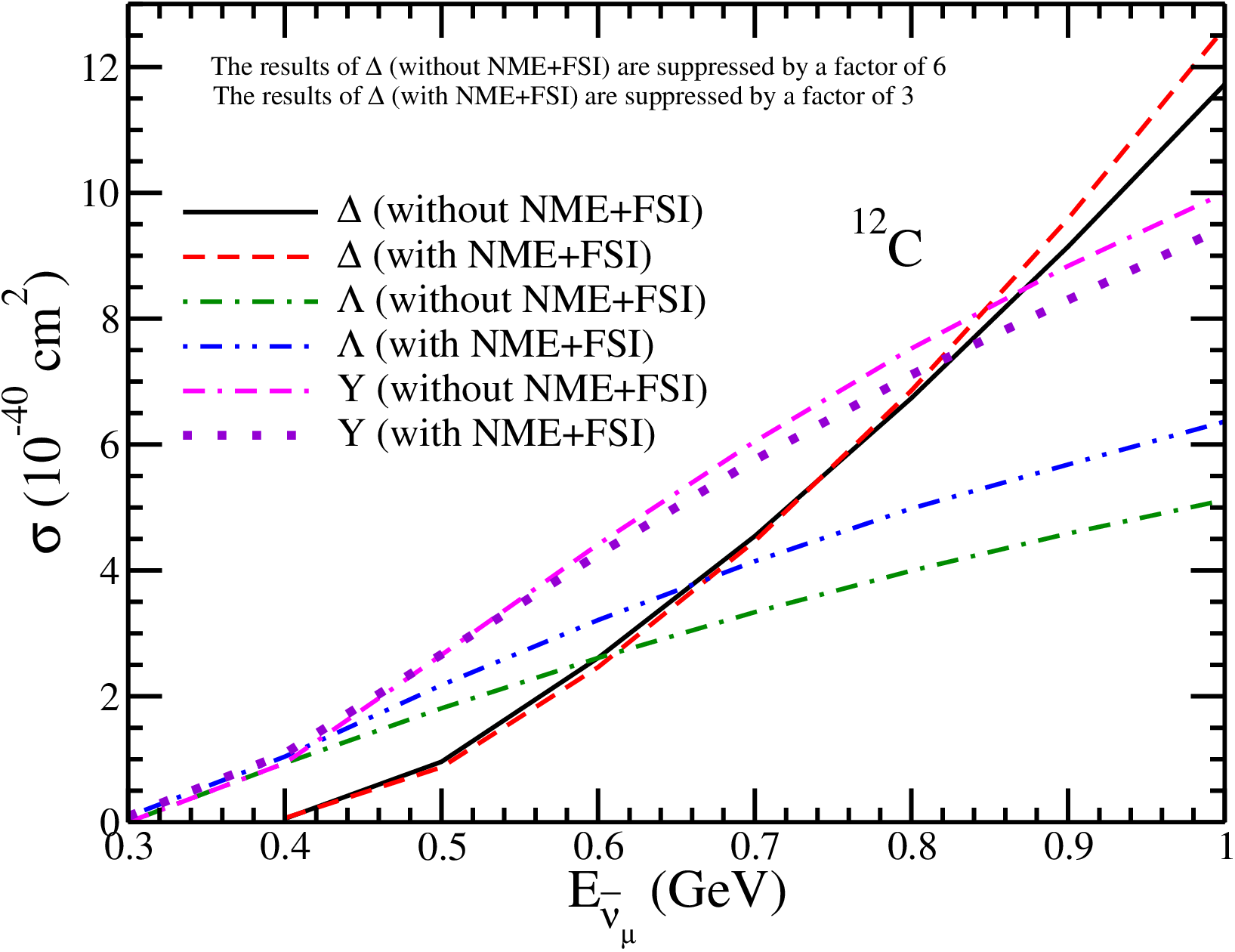} 
      \includegraphics[height=6cm,width=9cm]{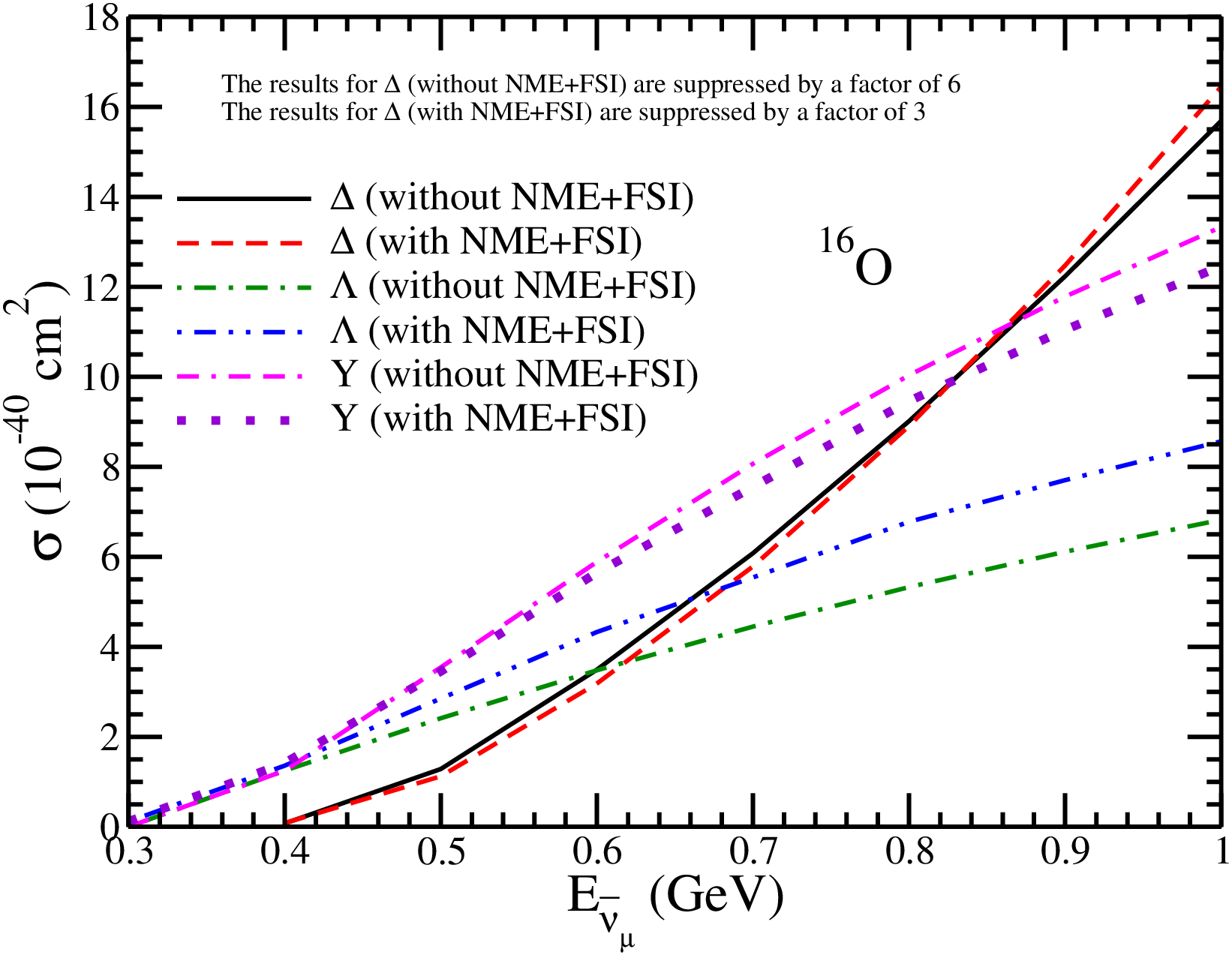}
       \includegraphics[height=6cm,width=9cm]{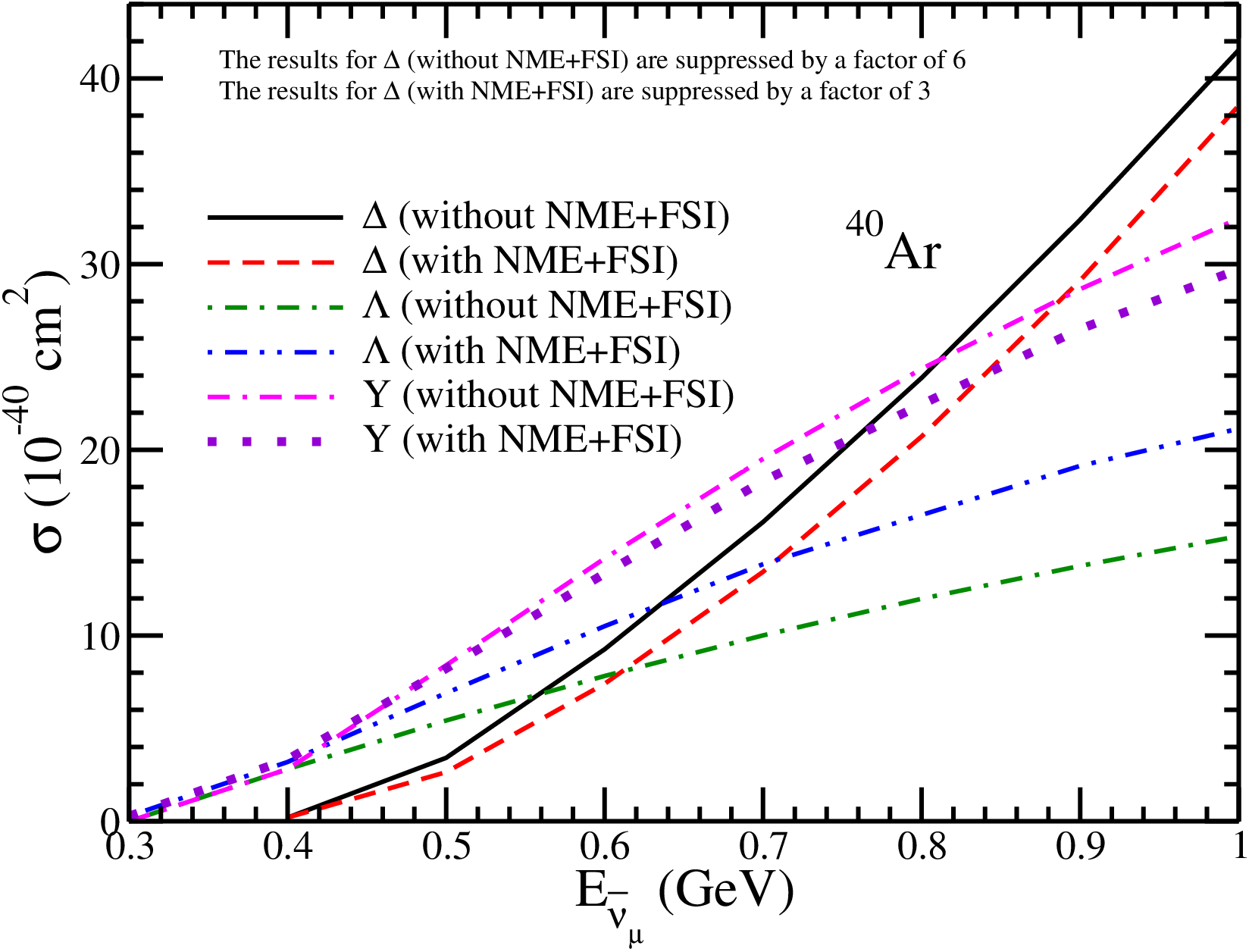}
           \includegraphics[height=6cm,width=9cm]{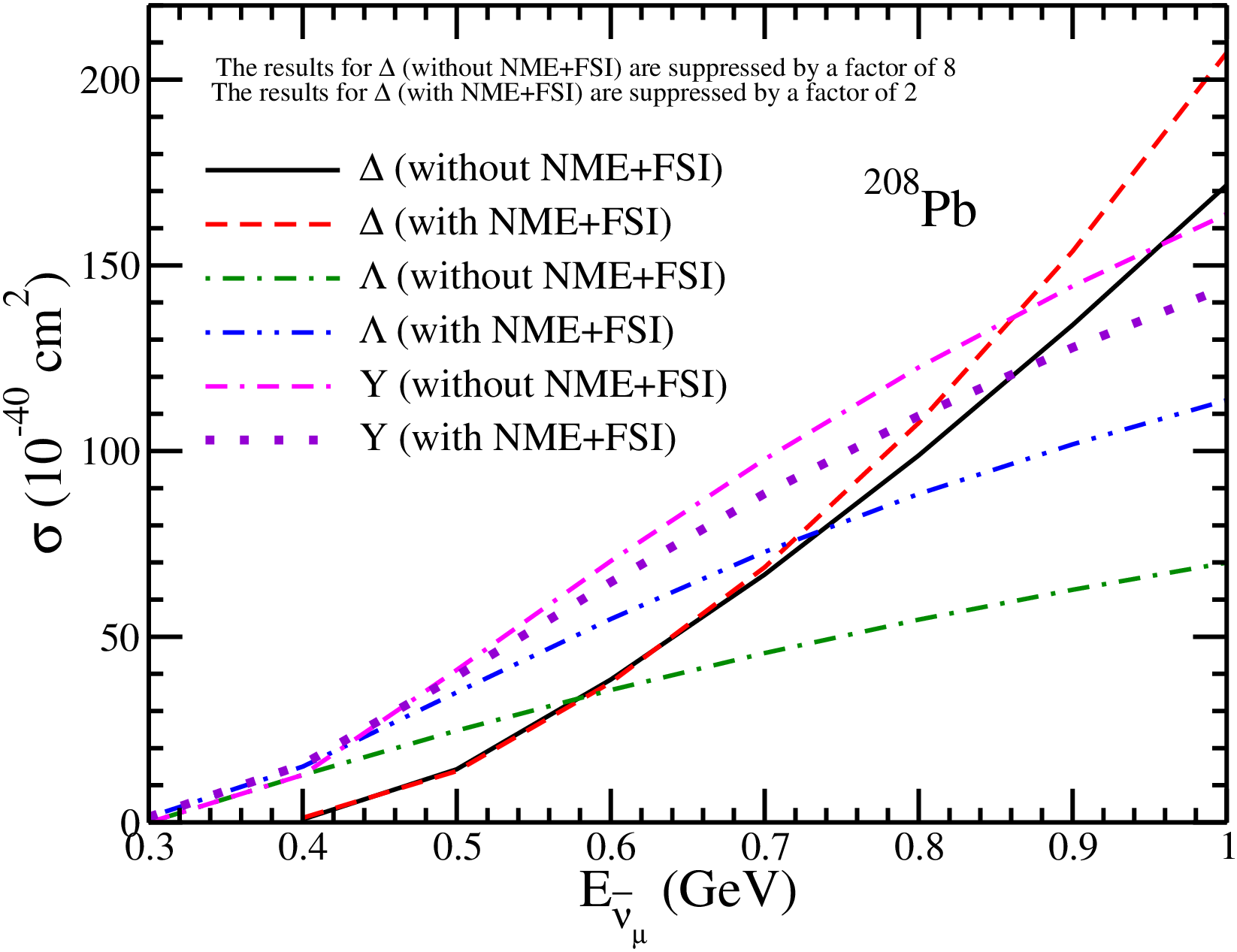}
\end{center}
\caption{Results for CC $\pi^-$ production in $^{12}$C~(upper left panel), $^{16}$O~(upper right panel), 
$^{40}$Ar~(lower left panel) and $^{208}$Pb~(lower right panel) with and without NME+FSI. The results are presented for the 
pion production from the $\Delta$, $\Lambda$ and total hyperon $Y(=\Lambda + \Sigma)$. Notice that 
in the case of $^{12}$C, $^{16}$O and $^{40}$Ar, the results of $\Delta$ without NME+FSI are suppressed by a factor of 6 and 
the results with NME+FSI are suppressed by a factor of 3, while in the case of $^{208}Pb$, the results of $\Delta$ without 
NME+FSI are suppressed by a factor of 8 and the results with NME+FSI are suppressed by a factor of 2 to bring them on the 
same scale~\cite{Fatima:2018wsy}.}\label{c12-ME}
 \end{figure}

\begin{figure}[h]
\begin{center}
    \includegraphics[height=6cm,width=9cm]{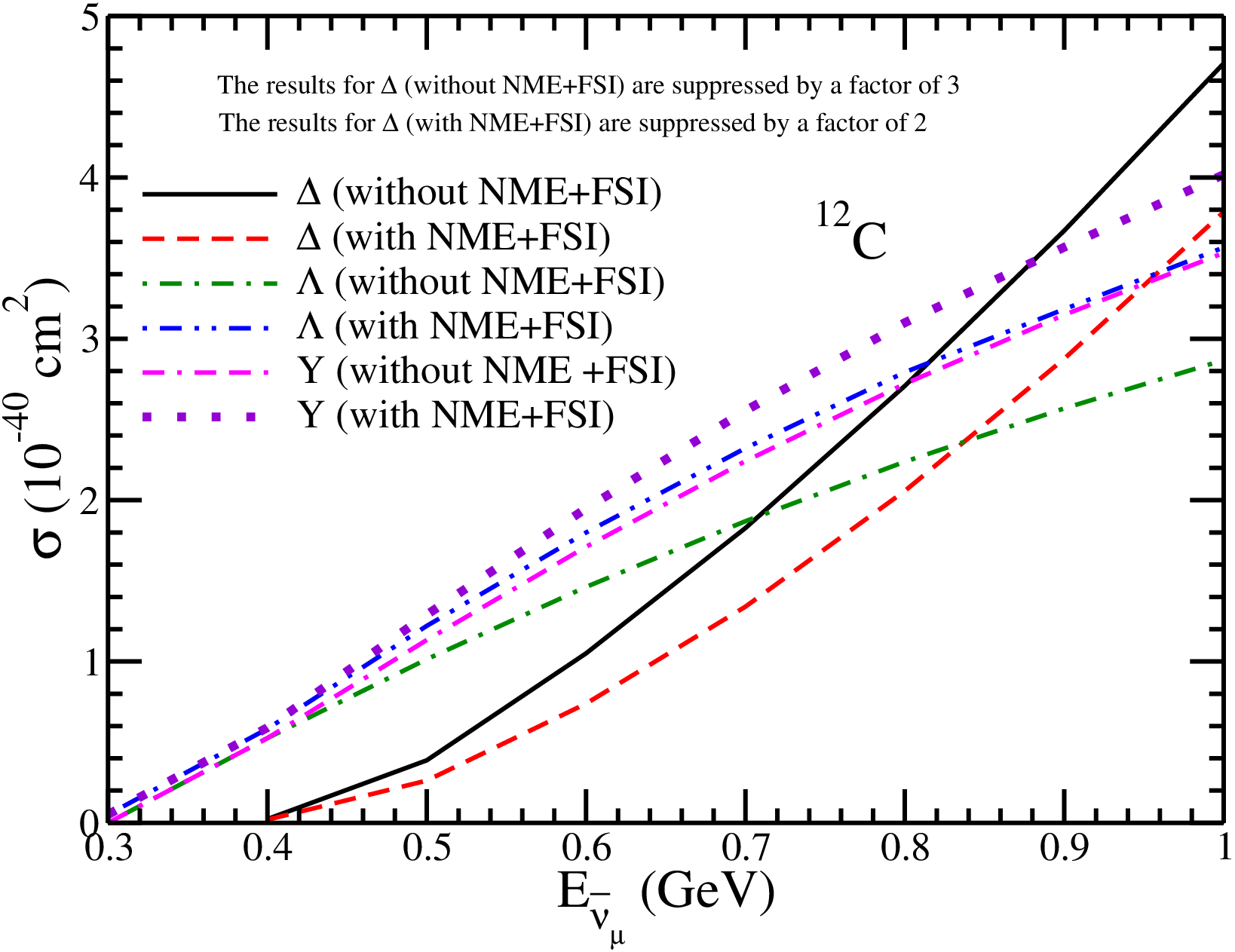}
    \includegraphics[height=6cm,width=9cm]{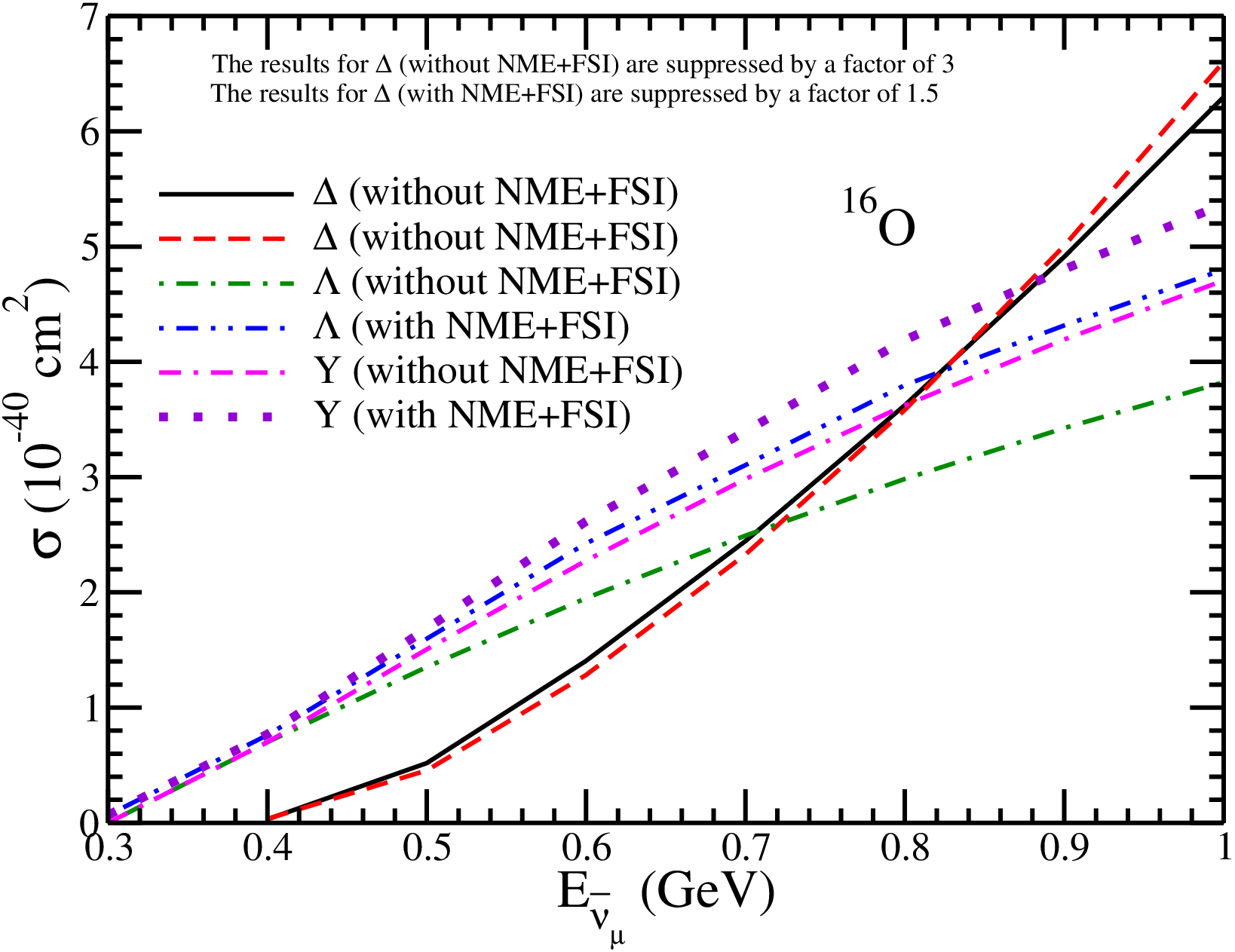}
    \includegraphics[height=6cm,width=9cm]{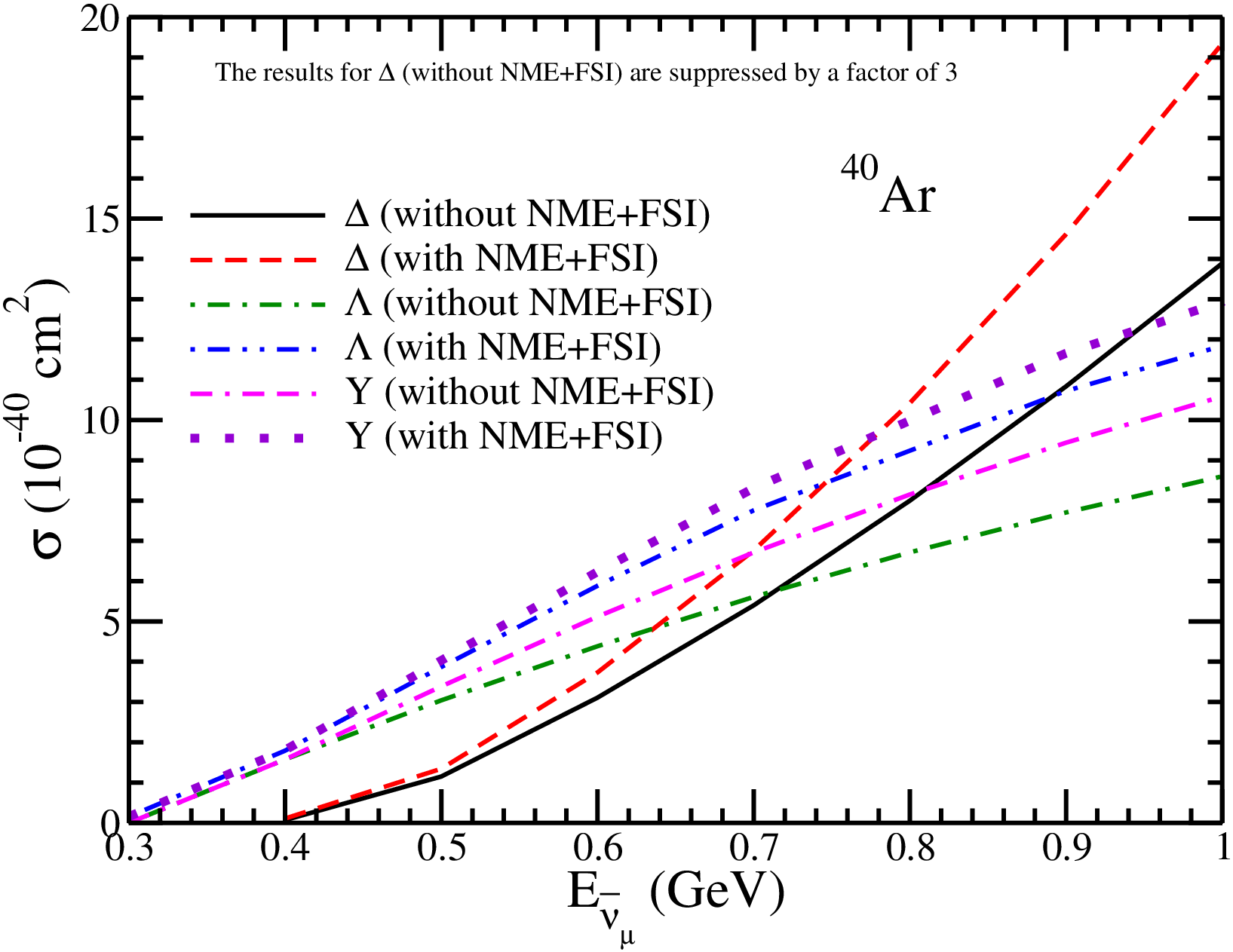}
    \includegraphics[height=6cm,width=9cm]{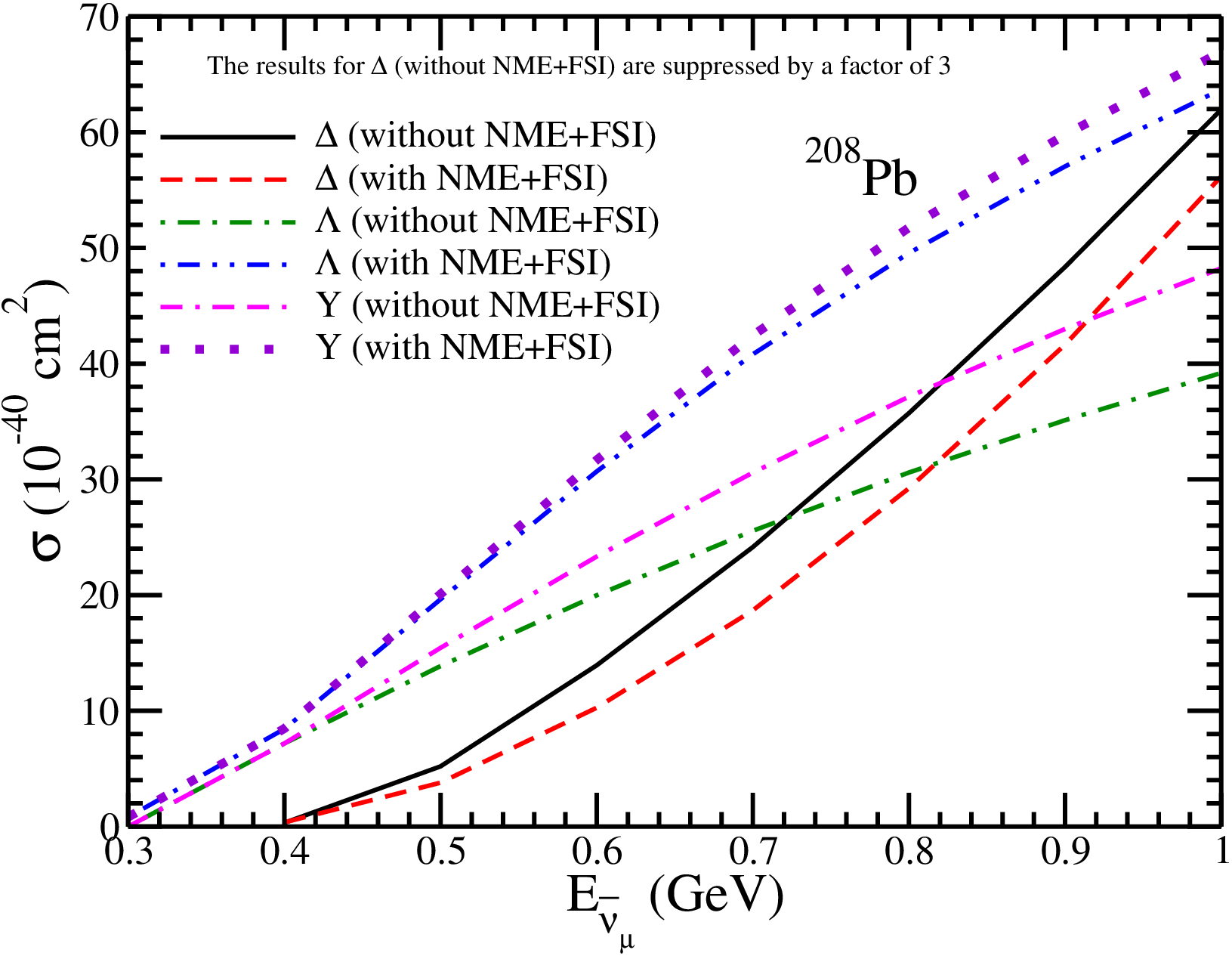}
\end{center}
\caption{Results for CC $\pi^0$ production in $^{12}$C~(upper left panel), $^{16}$O~(upper right panel), 
$^{40}$Ar~(lower left panel) and $^{208}$Pb~(lower right panel) with and without NME+FSI. The results are presented for 
the pion production from $\Delta$, $\Lambda$ and total hyperon $Y=\Lambda + \Sigma$. Notice that 
in the case of $^{12}$C, the results of $\Delta$ without NME+FSI are suppressed by a factor of 3 and the results with 
NME+FSI are suppressed by a factor of 2 while in the case of $^{16}$O, the results of $\Delta$ without NME+FSI are 
suppressed by a factor of 3 and the results with NME+FSI are suppressed by a factor of 1.5. Moreover, in the case of 
$^{40}$Ar and $^{208}$Pb, the results of $\Delta$ without NME+FSI are suppressed by a factor of 
3~\cite{Fatima:2018wsy}.}\label{c12-ME-NC}
 \end{figure}

To quantify our statement, in Fig.~\ref{c12-ME} and Figs.~\ref{c12-ME-NC} the results for $\pi^-$ and $\pi^0$ productions in 
nuclei like $^{12}$C, $^{16}$O, $^{40}$Ar and $^{208}$Pb are presented. These results are shown for the cross sections 
obtained without and with NME+FSI effect, for the pion production arising due to the $\Lambda$ production, total hyperon~($Y$) 
production and the $\Delta$ production. For the hyperon production, NMEs in the production process as well as the FSI due to 
hyperon-nucleon interactions have been taken into account. Similarly, for the pions arising from $\Delta$, the nuclear medium 
modification on $\Delta$ properties and the pion FSI effect have been considered, which results in large reduction in the 
pion production cross section.
 
Using the results of $\sigma$, the results for the ratio of hyperon to delta production cross sections are obtained, with and 
without NME+FSI, for $\pi^-$ as well as $\pi^0$ productions for all the nuclear targets considered here by defining 
  \begin{equation}\label{r_without}
 \left. {\it R}_N={\frac{\sigma(Y \rightarrow N\pi)}{\sigma(\Delta \rightarrow N\pi)}}\right 
 \rvert_{\text{without NME+FSI effects}}, \qquad \quad
  \left. {\it R}_A=\frac{\sigma(Y \rightarrow N\pi)}{\sigma(\Delta \rightarrow N\pi)}\right 
  \rvert_{\text{with NME+FSI effects}}.
 \end{equation}
This ratio directly tells us the enhancement of the ratio $R_A$ due to NME+FSI with the increase in the mass number of the 
nuclear targets as the pions getting produced through the $\Delta$-resonant channel undergo a suppression due to NME+FSI 
effect, while the pions getting produced from the hyperons~(all the interactions taken together i.e. $\Lambda$ as well as 
$\Sigma$ contributions) have comparatively smaller NME+FSI effect.
 
In Fig.~\ref{c12-ME}, the results for the total scattering cross section $\sigma$ vs $E_{{\bar\nu}_\mu}$, for ${\bar\nu}_\mu$ 
scattering off the nucleon in $^{12}$C, $^{16}$O, $^{40}$Ar and $^{208}$Pb nuclear targets giving rise to $\pi^-$ 
through the $\Delta$, $\Lambda$ and $Y$ productions with and without NME and FSI are presented. In the case of hyperon production 
for $^{12}$C, the effect of FSI due to $Y-N$ interaction leads to increase in the cross section of $\Lambda$ production from 
the free case, which is about $23-24\%$ for $E_{{\bar\nu}_\mu}=0.6 - 1$ GeV, while the change in the total hyperon production 
cross section results in a decrease in the cross section due to the FSI effect which is about $3 - 5\%$ at these energies. In 
the case of pions produced through $\Delta$ excitations, NME+FSI lead to an overall reduction of around 50$\%$ in the $\pi^-$ 
production for the antineutrino energies 0.6 $< E_{\bar\nu_\mu} <$ 1GeV. This results in the change in the ratio of 
${\it R}_N$~(Eq.~(\ref{r_without})) from 0.28 and 0.14 respectively at $ E_{\bar\nu_\mu}$=0.6 and 1~GeV to 
${\it R}_A$~(Eq.~(\ref{r_without})) $\rightarrow$ 0.58 and 0.25 at these energies. In the case of $^{16}$O nuclear target, the 
observations are similar to what has been discussed in the case of $^{12}$C nuclear target. For ${\bar\nu}_\mu$ scattering 
off $^{40}$Ar, in the 
case of $\Lambda$ production, the effect of FSI leads to increase in the cross section which is about $34-38\%$ for 
$E_{{\bar\nu}_\mu}=0.6 - 1$~GeV, however, the overall change in the $\pi^-$ production from the hyperons results in a net 
reduction in the cross section from the free case, which is about $6 - 8\%$ at these energies. In the case of pions produced 
through $\Delta$ excitations, NME+FSI leads to a reduction of around $55 - 60\%$ in the $\pi^-$ production for the 
antineutrino energies 0.6 $\le E_{\bar\nu_\mu} \le$ 1 GeV, and the reduction is less at higher energies. This results in the 
change in the ratio of ${\it R}_N$ from 0.25 and 0.13 respectively at $ E_{\bar\nu_\mu} =0.6$ and 1 GeV to ${\it R}_A$, 0.6 
and 0.26 at the corresponding energies. For a heavy nuclear target like $^{208}$Pb, the change in the cross section 
due to NME+FSI is quite large. For example, the reduction in the cross section due to 
NME+FSI when a $\Delta$ is produced as the resonant state, is about 75$\%$ at $ E_{\bar\nu_\mu}=$ 0.6 GeV and 70$\%$ at 
$E_{\bar\nu_\mu}=$ 1 GeV from the cross sections calculated without the medium effect. The enhancement in the $\Lambda$ 
production cross section is about $55 - 60\%$ at these energies. While the overall change in the $\pi^-$ production from the 
hyperons results in a net reduction which is about $8 - 12\%$. This results in the change in the ratio of ${\it R}_N$ from 
0.23 and 0.12 respectively at $ E_{\bar\nu_\mu}=0.6$ and 1 GeV to ${\it R}_A~\rightarrow$ 0.86 and 0.35.    
 
In Fig.~\ref{c12-ME-NC}, the results for the total scattering cross section $\sigma$ vs $E_{{\bar\nu}_\mu}$, for 
${\bar\nu}_\mu$ scattering off nucleon in $^{12}$C, $^{16}$O, $^{40}$Ar and $^{208}$Pb nuclear targets giving rise to $\pi^0$ 
 through the $\Delta$, $\Lambda$ and $Y$ productions with and without NME+FSI are presented. In the 
case of $\pi^0$ arising due to hyperon decay, the contribution comes from the $\Lambda$ and $\Sigma^0$ decay, while there is 
no contribution from $\Sigma^-$. Due to the FSI effect in $Y-N$~($Y=\Lambda, \Sigma^{-,0}$), there is substantial increase 
in the $\Lambda$ production cross section and reduction in the $\Sigma^0$ production cross section from the free case, which 
leads to an overall increase in the $\pi^0$ production. Therefore, unlike the $\pi^-$ production where there is an overall 
reduction, in the case of $\pi^0$ production, there is a net increase in the cross section which is about $13 - 14\%$ in 
$^{12}$C and $^{16}$O, $22 - 23\%$ in $^{40}$Ar and $26 - 38\%$ in $^{208}$Pb for $E_{\bar\nu_\mu}$ = 0.6 to 1 GeV. The 
different Clebsch-Gordan coefficients for $\Delta$ and the branching ratios for the hyperons give a different value of 
${\it R}_N$ and ${\it R}_A$. The ratio ${\it R}_N$ from 0.58 and 0.26 at 
$E_{\bar\nu_\mu} = $ 0.6 and 1 GeV,  respectively, changes in nucleus to ${\it R}_A~\rightarrow$ $\sim$1.3 and $\sim$0.5, 
in $^{12}$C and $^{16}$O, from 0.55 and 0.25 
respectively at $E_{\bar\nu_\mu}=0.6$ and 1 GeV to 1.68 and 0.66 in $^{40}$Ar, and from 0.56 and 0.26 respectively at 
$E_{\bar\nu_\mu}=0.6$ and 1 GeV to 3 and 1.2 in $^{208}$Pb. Thus, in the case of $\pi^0$ production, there is significant 
increase in $Y \rightarrow N \pi$ to $\Delta \rightarrow N \pi$ ratio when NME+FSI are taken into account specially in the 
case of heavier nuclear targets. Therefore, these pions arising due to hyperon production and its subsequent decay must be 
taken into account while doing the analysis for CC antineutrino induced single pion production from nuclear 
targets. For more details, see Refs.~\cite{Fatima:2021ctt, Fatima:2018wsy}.
 
\subsection{Coherent production of mesons}\label{coherent:meson}
Coherent meson production is the process in which (anti)neutrino scatters from the nucleus producing mesons but the nucleus 
stays in the ground state and such processes can take 
place via CC as well as NC induced reactions like
\begin{eqnarray}\label{coh-kine}
    \nu_l (\bar{\nu}_l) (k) + A(p_A) &\rightarrow& l^- (l^+) (k^\prime) +  m^+ (m^-) (k_m)  + A ({p_A}^\prime), \;\;(\text{CC})\\
    \nu_l (\bar{\nu}_l) (k) + A(p_A) &\rightarrow& \nu_l (\bar{\nu}_l) (k^\prime) +  m^0 (k_m)  + A ({p_A}^\prime), \;\;
    (\text{NC})
\end{eqnarray}
where $m = \pi, ~K$, etc., $q=k - k^\prime$, and the quantities in the bracket represent the respective momenta of the 
particles. The momentum transferred to the nuclear target is very small.

\begin{figure} 
 \centering
 \includegraphics[height=6cm,width=0.45\textwidth]{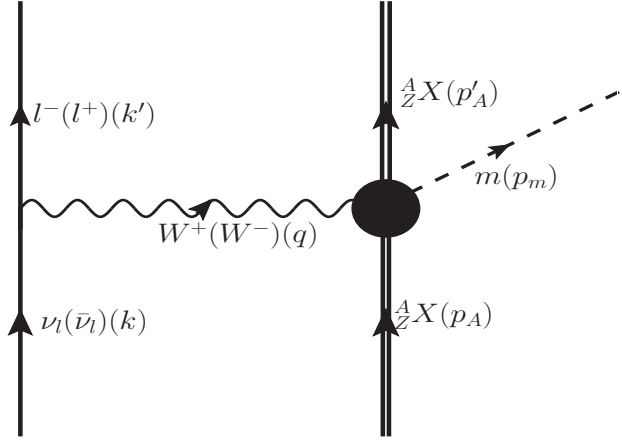}
  \caption{Feynman diagram depicting coherent production of mesons in CC (anti)neutrino scattering from nuclear 
  target.}\label{coherent-feyn}
 \end{figure}

In a coherent meson production, almost all the energy transferred $(q_0)$ from the leptonic 
vertex~(Fig.~\ref{coherent-feyn}) is taken by the outgoing meson $(E_m)$ i.e. $q_0 \approx E_m$. The momentum transfer 
squared to the nucleus is given by $t=(q - k_m)^2=({p_A}^\prime - p_A)^2\simeq -2 M_A T_A$, for $M_A \gg T_A$, with $M_A$ 
and $T_A$ being, respectively, the mass of the target nucleus and kinetic energy of the recoiling nucleus in the laboratory 
frame. 
In these reactions, the momentum transfer to the nucleus is so small that the coherence condition $Q << \frac{1}
{R}$~(Section~\ref{CEvNS}) is satisfied and the individual amplitudes for the pion production from each nucleon in the nucleus 
add coherently. In view of the smallness of the momentum and energy transfer, most of the energy-momentum from the lepton 
system is transferred to the meson making $t$ very small. Since the nuclear form factor falls very rapidly with $t$, the 
coherent meson production is dominant in the forward direction i.e. $t \approx 0$.

The importance of studying coherent pion production has been realized in the context of neutrino oscillation experiments. 
In the analysis of neutrino oscillation experiments, it is important to reconstruct the initial neutrino energy spectrum 
from the observed charged lepton energy spectrum using the kinematics of the QE reactions from the nuclear 
targets in which most of the leptons~($e$ and $\mu$) are produced in the forward direction. Most of the $\pi^{\pm}$, 
$\pi^0$~(through the electromagnetic shower as their decay products) produced in the forward direction through the coherent 
reactions mimic the real $\mu^+$, $\mu^-$ and $e^-$, which introduce uncertainty in the QE cross section for lepton 
production and the reconstruction of the neutrino energy. For example, around 1GeV this uncertainty in the (anti)neutrino estimated energy 
could be in the range 
150-200~MeV. A knowledge of these uncertainties is crucial in interpreting the results for $\nu_\mu({\bar\nu}_\mu)$ 
disappearance or $\nu_e({\bar\nu}_e)$ appearance in the context of neutrino oscillation studies making the study of coherent 
pion production very important in the neutrino-nucleus reactions. 

\subsubsection{Coherent pion production}\label{coherent:pion}
Experimentally, the coherent pion production in the high energy region was first reported by the Aachen-Padova 
collaboration~\cite{Faissner:1983ng} in 1983 while studying isolated $\pi^0$s produced in the $\nu_\mu$ and ${\bar\nu}_\mu$ 
induced processes. This was followed by a study performed by the Aachen-Gargamelle group~\cite{Isiksal:1984vh} where 
coherent NC $\pi^0$ events in the Gargamelle heavy freon exposure were isolated. Later there were several $\nu_\mu$ and 
${\bar\nu}_\mu$ experiments like CHARM~\cite{CHARM:1985bva, CHARM-II:1993xmz}, SKAT~\cite{SKAT:1985uch, Nahnhauer:1986xh}, 
where NC induced coherent pions were observed over a wide range of neutrino energies using different nuclear targets. In 
the recent accelerator experiments being performed to study neutrino oscillations like the K2K~\cite{K2K:2005uiu}, 
SciBooNE~\cite{SciBooNE:2008bzb}, MiniBooNE~\cite{MiniBooNE:2008mmr}, NOvA~\cite{NOvA:2019bdw}, T2K~\cite{T2K:2016soz}, etc., 
as well as the dedicated $\nu_{l}-A$ cross section experiments MINERvA~\cite{MINERvA:2014ani} and 
NOMAD~\cite{NOMAD:2009idt} have put either a limit on coherent pion cross section or have provided cross sections 
at some energies. For example:
\begin{itemize}
 \item The K2K experiment~\cite{K2K:2005uiu} at a neutrino energy of 1.3 GeV, has put an upper limit of $0.6 \times 10^{-2}$ 
 on the cross section ratio of coherent pion production to the total CC interaction cross section at $90\%$CL.
 
 \item The SciBooNE experiment~\cite{SciBooNE:2008bzb} at neutrino energy of 1.1 GeV, has put an upper limit of $0.67 \times 
 10^{-2}$ on the cross section ratio of coherent pion production to the total CC interaction cross section at 
 $90\%$CL.
 
 \item The MiniBooNE collaboration~\cite{MiniBooNE:2008mmr} has studied the coherent fraction  $\frac{\sigma_{\text{(coh)}}}
 {\sigma_{(\text{coh} + \text{incoh})}}=19.5 \pm 1.1({\text{stat}}) \pm 2.5({\text{sys}})\%$ for the $\pi^0$ production at 
 $E_{\nu} < 2$~GeV.
 
 \item  The NOMAD experiment~\cite{NOMAD:2009idt} at the CERN SPS has reported NC induced coherent $\pi^0$ 
 production cross section to be $72.6 \pm 8.1({\text{stat}}) \pm 6.9({\text{sys}})\times 10^{-40}\frac{\text{cm}^2}
 {\text{nucleus}}$ for $<E_\nu> \simeq 25$~GeV.
 
 \item The T2K experiment~\cite{T2K:2016soz} has reported a value of $1.03 \pm 0.25({\text{stat}}) \pm 0.70({\text{sys}})
 \times 10^{-39}\frac{\text{cm}^2}{\text{nucleus}}$ for $<E_\nu> \simeq 1.5$~GeV.
 
 \item MINERvA collaboration~\cite{MINERvA:2014ani} has reported the flux-averaged cross sections for the coherent $\pi^{+} (\pi^{-})$ production 
 induced by $\nu_{\mu} (\bar{\nu}_{\mu})$ on the carbon target to be
 $[3.49 \pm 0.11({\text{stat}}) \pm 0.37({\text{flux}}) \pm 0.20({\text{other-sys}})] 
 ([2.65 \pm 0.15({\text{stat}}) \pm 0.31({\text{flux}}) \pm 0.30({\text{other-sys}})])\times 10^{-39}\frac{\text{cm}^2}
 {^{12}C}$, respectively, for $E_\nu$ in the range of 1.5 to 20 GeV.
 
 \item  Recently the NOvA collaboration~\cite{NOvA:2019bdw} has reported the result of the flux averaged cross section for 
 the neutrino induced NC coherent $\pi^0$ production corresponding to an average energy $<E_{\nu_\mu}> = 
 2.7$~GeV to be $13.8 \pm 0.9({\text{stat}}) \pm 2.3({\text{syst}}) \times 10^{-40} {\text{cm}}^2/{\text{nucleus}}$.
\end{itemize}

Theoretically there are two different approaches which have been used to study the coherent pion production. The first 
approach is based on the Adler's PCAC model~\cite{Adler:1968tw} which relates the coherent pion scattering cross section at 
$Q^2=0$ with the pion-nucleus elastic scattering cross section. This approach takes NME into account only 
through the final state interaction of the outgoing pions with the nucleus. The first calculation based on this approach 
was done by Rein and Sehgal~\cite{Rein:1980wg} followed by many others in later years~\cite{Rein:1982pf, Rein:2006di, 
Belkov:1986hn, Berger:2008xs, Paschos:2009ag}. 
 
In this approach, the triple differential cross section for $\pi^0$ production is given by \cite{Rein:1980wg}:
\begin{align}\label{RS_eq4}
\frac{d\sigma^{\pi^0}}{dx\ dy\ d|t|}= \frac{G^{2}_{F}Mf^{2}_{\pi}A^{2}}{2\pi^{2}}E(1-y)\frac{1}{16\pi}(\sigma^{\pi 
N}_{tot})^{2} (1+r^{2}) \left(\frac{m^{2}_{A}}{m^{2}_{A}+Q^{2}}\right)^{2}e^{-b|t|}F_{abs},
\end{align}
where $x = Q^{2}/2M\nu$, $y= \nu/E$, $\sigma^{\pi N} $ is the pion-nucleon cross section, $f_\pi$ is the pion decay 
constant, $m_A$ is 1GeV, $M$ denotes the nucleon mass, $A$ is the number of nucleons within the nucleus, $r\left(=Re 
f_{\pi N}(0)/Im f_{\pi N}(0)\right)$ is defined as the ratio of the real to imaginary part of the forward pion-nucleon 
scattering amplitude, $b = 1/3R^{2}$, with $R(R= R_{0}A^{1/3})$ as the nuclear radius, and $F_{abs}$ describes the effects 
of pion absorption in the nucleus, and is given by
\begin{equation}\label{RS_eq5}
F_{abs} = exp\left\{-\frac{9A^{1/3}} {16\pi R^{2}_{0}}\sigma^{\pi N}_{inel} \right\},
\end{equation}
where the experimental results for the average pion-nucleon cross section were used.

The above expression was obtained for the massless leptons even for the CC process. In a later work, Rein and 
Sehgal~\cite{Rein:2006di} took into account the lepton mass by introducing a multiplicative correction factor.

Later, Berger and Sehgal \cite{Berger:2008xs} used experimental data for pion-carbon scattering in the low energy region 
relevant for the contemporary cross section measurements of the coherent pion production in this energy region to describe 
the pion absorption effect in the nucleus and obtained the results for the coherent pion production. In an another model, 
Kartavtsev et al.~\cite{Paschos:2005km}, Paschos and Schalla~\cite{Paschos:2009ag}, and Higuera and 
Paschos~\cite{Higuera:2013pra} have included the lepton mass in all their kinematical calculations for the CC coherent pion 
production and used pion-nucleus cross section in neutrino scattering. These authors have compared their results for the 
charged and NC neutrino induced coherent pion production cross section with the results from the experimental 
collaborations of MiniBooNE~\cite{Raaf:2004ty, Raaf:2005up}, K2K~\cite{K2K:2005uiu}, Aachen-Padova~\cite{Faissner:1983ng} and 
Gargamelle~\cite{Isiksal:1984vh} experiments.

\begin{figure} 
 \centering
 \includegraphics[height=6cm,width=0.75\textwidth]{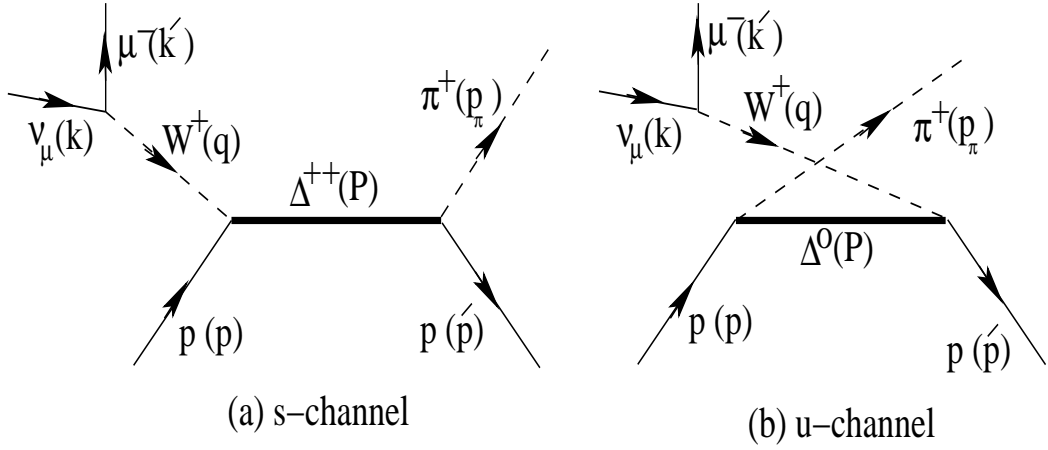}
  \caption{Feynman diagram depicting coherent production of mesons in CC (anti)neutrino scattering from nuclear 
  target.}\label{coherent-prl}
 \end{figure}
 
The second approach is the use of microscopic models for pion production that have been developed by various 
groups~\cite{Kelkar:1996iv, Singh:2006br, Alvarez-Ruso:2007rcg, Alvarez-Ruso:2007kwp, Amaro:2008hd, Leitner:2009ph, 
Hernandez:2009vm, Zhang:2012xi, Nakamura:2009iq}, which are based on the single nucleon process $\nu_l({\bar\nu}_l) + N 
\rightarrow l^-(l^+) + N + \pi^+(\pi^-)$, dominated by $\Delta$ production in the nucleus as shown in 
Fig.~\ref{coherent-prl}. The total cross section is then obtained by coherently summing the contribution of the pion 
production amplitudes from all the nucleons in the nucleus. The nuclear medium modification of the $\Delta$ properties in 
the nucleus~(Section~\ref{SPP:nucleus}) and the FSI of the outgoing pion with the nuclear target are taken into account. 
Different treatments for the FSI of pion with the nucleus have been taken. For example the work of~\cite{Kelkar:1996iv, 
Amaro:2008hd,Alvarez-Ruso:2007rcg} uses Klein-Gordon equation, while Refs.~\cite{Singh:2006br, Zhang:2012xi} use semiclassical 
eikonal approximation and Nakamura et al~\cite{Nakamura:2009iq} uses Lippmann-Schwinger equation.
  
In this approach, the first calculation was done by the Aligarh group~\cite{Singh:2006br} using 
relativistic formalism in the $\Delta$-dominance model. They performed the calculations for CC and NC 
(anti)neutrino induced coherent pion productions for several nuclear targets in the intermediate neutrino energy region of a 
few GeV. The NME are taken into account in the weak production process as well as in the final state 
interaction of the outgoing pions with the nucleus. The calculation uses the local density approximation to the delta hole 
model which was initially developed for photo and electro production of pions from nuclei~\cite{Carrasco:1989vq}. The final 
state interaction of pions has been treated in eikonal approximation with the pion optical potential described in terms of the 
self energy of a pion in a nuclear medium calculated in this model~\cite{Oset:1987re, Garcia-Recio:1989hyf}. 
The amplitude for CC 1$\pi^+$ production from the proton is written using the Feynman diagram shown in 
Fig.~\ref{coherent-prl} and is given by~\cite{Singh:2006br}:
\begin{equation}\label{coh-mat}
{\cal M} = {\cal I}\frac{G_F}{\sqrt{2}}~\cos\theta_C~l^\mu~{\cal J}_\mu~{\cal F}_{p}({\vec{q}}-{\vec{k}_\pi}),
\end{equation}
where ${\cal I}=\sqrt{3}~(1)$ for s~(u) channel, $l^\mu$ is the leptonic current given in Eq.~(\ref{lep_curr}) and the 
hadronic current ${\cal J}_\mu$ is given by
\begin{eqnarray}\label{jmu:coherent}
{\cal J}_\mu=\frac{f_{{\pi} N {\Delta}}}{m_{\pi}}\sum_s{\bar {\Psi}}^s(p^\prime)\left[k_{\pi\sigma}\Lambda^{\sigma \lambda}
\Gamma_{\lambda \mu}^{\frac32}\right]\Psi^s(p),
\end{eqnarray}
where $\Lambda^{\sigma \lambda}$ is the relativistic $\Delta$ propagator given by 
\begin{eqnarray}\label{prop:coherent}
{\Lambda}^{\sigma\lambda}={\frac{{\slashed{P}}+M_\Delta}{P^2-M^{2}_{\Delta}+i\Gamma M_\Delta}}\left[g^{\sigma \lambda}-
\frac{1}{3}\gamma^\sigma\gamma^\lambda-\frac{2}{3M^2_\Delta}P^\sigma P^\lambda+\frac{(P^\sigma \gamma^\lambda-\gamma^\sigma 
P^\lambda)}{3M_\Delta}\right],
\end{eqnarray}
and $\Gamma_{\lambda \mu}^{\frac32}$ is the weak N-$\Delta$ transition vertex given as the sum of vector and axial part using 
Eq.~(\ref{eq:gamma_3half_pos})~(Section~\ref{res:inelastic}). The nuclear form factor ${\cal F}_{p}({\vec{q}}-{\vec{k}_\pi})$ 
in Eq.~(\ref{coh-mat}) is given as~\cite{Singh:2006br}:
\begin{equation}
{\cal F}_{p}({\vec{q}}-{\vec{k}}_\pi)=\int d{\vec{ r}}~\rho_p({r})~e^{-i({\vec q}-{\vec k}_\pi).{\vec r}},
\end{equation}
with $\rho_p({r})$ is the proton density in the nucleus. For production from nuclear 
targets, the contributions from the protons and neutrons are considered. Incorporating the isospin factors for charged pion 
production from proton and neutron targets corresponding to $W^+$ exchange diagram, the nuclear form factor is obtained as
\begin{equation}
{\cal F}({\vec q}-{\vec k}_\pi)=\int d{\vec r} \left[{\rho_p (r)}+\frac{1}{3}{\rho_n (r)}\right]e^{-i({\vec q}-
{\vec k}_\pi).{\vec r}}.
\end{equation} 
Using these expressions, the differential cross section is given as:
\begin{eqnarray}
\frac{d^5\sigma}{d\Omega_\pi d\Omega_{{\bar\nu}\mu}dE_\mu}&=&\frac{1}{8}\frac{1}{(2\pi)^5}\frac{|~{\vec{k}^\prime}|~|~{\vec{k}_\pi}|}
{E_{\bar\nu}}~{\cal R}~{\bar{\sum}}\sum|{\cal M}|^{2}, \qquad \quad \text{where} \\
{\cal R}&=&\left[\frac{M~|{\vec{k}_\pi}|}{E_{p^\prime}|{\vec{k}_\pi}|+E_\pi(|{\vec{k}_\pi}|-|{\vec{q}}|\cos\theta_\pi)}\right],
\end{eqnarray}
is a kinematical factor incorporating the recoil effects, which is very close to unity for low $Q^2$, relevant for the 
coherent reactions. In this model, the 
results of the cross sections for the antineutrino processes are obtained by replacing $\left[{\rho_p ({r})} 
+\frac{1}{3}{\rho_n (r)}\right]$ by $\left[{\rho_n (r)}+ 
\frac{1}{3}{\rho_p ({r})}\right]$. For NC coherent pion production, it is replaced by 
$\frac{2}{3}\left[\rho_p ({r}) + \rho_n ({r})\right]$. 
\begin{figure}
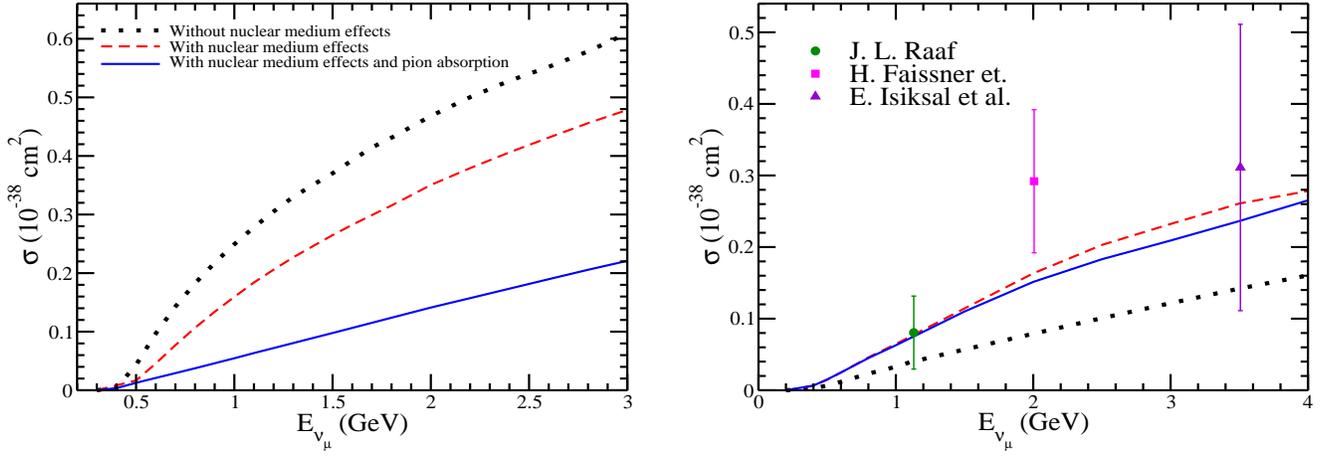
 
 \centering
 \includegraphics[height=6cm,width=0.45\textwidth]{coherent_sigma_CC.eps}
 \hspace{5mm}
 \includegraphics[height=6cm,width=0.45\textwidth]{coherent_sigma_NC.eps}
\caption{$\sigma$ vs $E_{\nu_\mu}$  for CC coherent 1$\pi^+$ production in $^{12}$C~(left panel). The results 
are shown without NME~(dotted line) and with NME~(dashed line). When both the pion 
absorption and NME are taken into account, the results are shown by the solid line. In the right 
panel, the results are shown for the total cross section $\sigma$ vs $\text{E}_{\nu}$ for the coherent $\pi^0$ production in 
$^{12}\text{C}$~(solid), $^{27}\text{Al}$~(dashed) and freon~(dotted) with nuclear medium and pion absorption effects, 
along with the experimental results of Refs.~\cite{Faissner:1983ng, Isiksal:1984vh, Raaf:2004ty, Raaf:2005up}.}\label{coherent:total_cc}
 \end{figure}
The NME due to renormalization of $\Delta$ properties in the nuclear medium have been treated in the same 
manner as discussed in Section~\ref{SPP:nucleus}. Accordingly the $\Delta$ propagator $\Lambda^{\sigma \lambda}$ in ${\cal 
J}_\mu$ given by Eq.~(\ref{prop:coherent}) is modified due to the modifications in mass and width of the $\Delta$ in the 
nuclear medium discussed in Section~\ref{SPP:nucleus}. However, the final state interaction of the pions with the residual 
nucleus has been treated in a different way. The final state interaction in coherent production of pions is taken into account 
by replacing the plane wave pion by a distorted wave pion. The distortion of the pion is calculated in the eikonal 
approximation in which the distorted pion wave function is written as~\cite{Singh:2006br}:
\begin{equation}
e^{i({\vec{q}}-{\vec{k}}_\pi)\cdot{\vec{r}}}\rightarrow \mbox{exp}\left[i({{\vec{q}-\vec{k}_\pi}})\cdot{\vec{r}}-\frac{i}{v}\int_{-\infty}^z 
V_{opt}({\vec{b}}, z^\prime)dz^\prime\right],
\end{equation}
where $\vec{r}=(\vec{b}, z)$, $\vec{q}$ and $\vec{k}_\pi$ are the momentum transfer and the pion momentum, respectively. The 
pion optical potential $V_{opt}$ is related with the pion self-energy $\Pi$ as $\Pi=2\omega~V_{opt}$, with $\omega$ as the 
energy of the pion and $|{\vec{v}}|={|{\vec{k}_\pi}|}/{\omega}$. The pion self-energy is calculated in local density 
approximation of the $\Delta$-hole model and is given as~\cite{VicenteVacas:1994sm}:
\begin{equation}\label{pion-self}
\Pi(\rho({\vec{b}}, z^\prime))=\frac{4}{9}\left(\frac{f_{\pi N\Delta}}{m_\pi}\right)^2\frac{M^2}{\bar s}|{\vec{k}_\pi}|^2~ 
\rho({\vec{b}}, 
z^\prime)~G_{\Delta h}({\bar s}, \rho),
\end{equation}
where $\bar{s}$ is the CM energy in the $\Delta$ decay averaged over the Fermi sea and $G_{\Delta h}({\bar s},
\rho)$ is the $\Delta$-hole propagator given by:
\begin{eqnarray}
G_{\Delta h}(s,\rho(\vec{b}, z^\prime))=\frac{1}{\sqrt{{\bar s}}-M_\Delta+\frac{1}{2}i\tilde\Gamma({\bar s},\rho)-i\mbox{Im}
\Sigma_\Delta({\bar s}, \rho)-\mbox{Re}\Sigma_\Delta({\bar s},\rho)}.
\end{eqnarray}
When the pion absorption effect is taken into account the nuclear form factor ${\cal F}(\vec{q}-\vec{k}_\pi)$ modifies to 
$\tilde{\cal F}(\vec{q}-\vec{k}_\pi)$ and is given as~\cite{Singh:2006br}:
\begin{eqnarray}
\tilde{\cal F}(\vec{q}-\vec{k}_\pi)=2\pi\int_0^\infty b~db\int_{-\infty}^\infty dz~\rho(\vec{b}, z)\left[J_0(k_\pi^tb)~
e^{i(|\vec{q}|-k_\pi^l)z}~e^{-if(\vec{b}, z)}\right],
\end{eqnarray}
where 
\begin{equation}
f(\vec{b}, z)=\int_z^{\infty} \frac{1}{2|{\vec{k}_\pi}|}{\Pi(\rho(\vec{b}, z^\prime))}dz^\prime,
\end{equation}
and the pion self-energy $\Pi$ is defined in Eq.~(\ref{pion-self}).
\begin{figure}
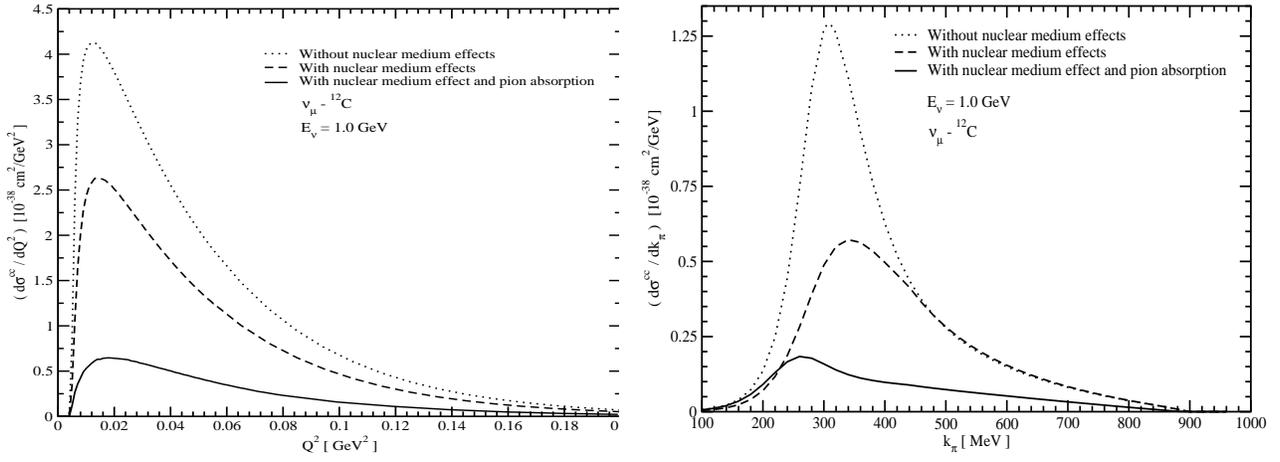
 
 \centering
 \includegraphics[height=6cm,width=0.45\textwidth]{dsdq_c12.eps}
 \includegraphics[height=6cm,width=0.45\textwidth]{figc_car.eps}
 \caption{$\frac{d\sigma}{dQ^2}$ vs $Q^2$ for CC coherent 1$\pi^+$ production in $^{12}$C (left panel).
 $\frac{d\sigma}{dk_{\pi}}$ vs $k_{\pi}$ for CC coherent 1$\pi^+$ production in $^{12}$C (right panel).
 Line and points have the same meaning as in Fig.~\ref{coherent:total_cc}~\cite{Athar:2007wd, Singh:2006br}.}\label{coherent:dsigma_dq2}
 \end{figure}
 
These modifications lead to the following expression for the total scattering cross section 
\begin{equation}
\sigma=\frac{1}{8}\frac{1}{(2\pi)^5}\int d\Omega_\pi\int d\Omega_{{\bar\nu}\mu}\int dE_\mu\frac{|~\vec{k}^\prime|~|~\vec{k}_\pi|}
{E_{\bar\nu}}~{\cal R}~{\overline{\sum}}\sum|{\tilde {\cal M}}|^{2},
\end{equation}
where
\begin{eqnarray}
{\tilde {\cal M}}={\cal I} \frac{G_F}{\sqrt{2}}~\cos\theta_C~l^\mu~\tilde{\cal J}_\mu~\tilde{\cal F}(\vec{q}-\vec{k}_\pi), \qquad 
\tilde{\cal J}_\mu=\frac{f_{{\pi} N {\Delta}}}{m_{\pi}}\sum_s{\bar {\Psi}}^s(p)\left[k_{\pi\sigma}\tilde\Lambda^{\sigma 
\lambda}{\cal{O}}_{\lambda \mu}\right]\Psi^s(p),
\end{eqnarray}
where $\tilde\Lambda^{\sigma \lambda}$ is the modified $\Delta$ propagator inside the nuclear medium.

In Fig.~\ref{coherent:total_cc}, the results are presented for the total scattering cross section $\sigma$ for the 
coherent CC reaction induced by $\nu_\mu$ in $^{12}$C. These results are shown without NME, with NME and when the pion 
absorption effect is taken into account. It is found that the 
NME lead to a reduction of around 45$\%$ for E$_\nu$=0.7 GeV, 25-35$\%$ around E$_\nu$=1.0 - 2.0 GeV and 
it is about 20$\%$ at E$_\nu$=3.0 GeV while the reduction due to final state interaction is quite large. This suppression in 
the cross section due to the nuclear medium and the pion absorption effects is about 80$\%$ for E$_\nu$ $\sim$ 1.0 GeV, 
70$\%$ for E$_\nu$ $\sim$ 2.0 GeV and 65$\%$ for E$_\nu$ $\sim$ 3.0 GeV.  We show in the right panel of 
Fig.~\ref{coherent:total_cc}, the total cross section $\sigma^\text{NC}(\text E_\nu)$ for NC induced 
$\pi^0$ production from $^{12}\text{C}$, $^{27}\text{Al}$ and $\text{CF}_3\text{Br}$(freon), along with the experimental 
results from the MiniBooNE collaboration for $^{12}\text{C}$~\cite{Raaf:2004ty, Raaf:2005up}, from the Aachen collaboration 
for $^{27}\text{Al}$~\cite{Faissner:1983ng} and from the Gargamelle collaboration for freon~\cite{Isiksal:1984vh}. It may be 
seen that the theoretical results for NC induced coherent $\pi^0$ production are in reasonable agreement with 
presently available experimental results in the intermediate energy region. The recently reported result by the NOvA 
collaboration for NC induced coherent $1\pi^0$ production cross section with mass number $A=13.8$ at 
$<E_{\nu_\mu}> = 2.7$~GeV is $13.8 \pm 0.9 ({\text{stat}}) \pm 2.3({\text{syst}}) \times 10^{-40} {\text{cm}}^2/
{\text{nucleus}}$~\cite{NOvA:2019bdw}, and is found to be in very good agreement with the results of Ref.~\cite{Singh:2006br}.

In Fig.~\ref{coherent:dsigma_dq2}, the results are presented for the $Q^2$-distribution i.e. $\left(\frac{d\sigma^{CC}}
{dQ^2} \right)$ for the coherent CC reaction induced by $\nu_\mu$ in $^{12}$C~ (left panel) at neutrino energy 
E$_{\nu_\mu}$=1.0 GeV where NME, and NME+FSI  effects are shown explicitly. 
It may be observed that the reduction in the cross section in the peak region is around 35$\%$, and decreases further 
uniformly. The total reduction in the cross section is around 85$\%$ in the peak region when pion absorption effect is also 
taken into account, and decreases further uniformly. In the right panel of this figure, we present the results for 
the momentum distribution of pion $\left(\frac{d\sigma^{CC}}{dk_\pi} \right)$ for the coherent CC reaction 
induced by $\nu_\mu$ in $^{12}$C at neutrino energy E$_{\nu_\mu}$=1 GeV where NME and NME+FSI effects are shown explicitly. 
We find that the reduction in the cross section due to the NME increases with the pion momentum ${k_\pi}$ and just before the 
peak region it starts decreasing, in the peak region 
it is about 60$\%$ and decreases further, for example, it is about 45$\%$, 20$\%$ and 5$\%$ at ${k_\pi}$=350 MeV, 400 MeV and 
450 MeV, respectively, after which both are approximately the same. The effect of the pion absorption show the further 
strong reduction in the cross section, and in the peak region (${k_\pi}$=320 - 360 MeV of nuclear effects) it is about 
75-80$\%$, accompanied by the shift in the peak towards the lower value of the pion momentum $\vec{k}_\pi$, and then 
decreases further. Similar trend is observed in case of $^{16}$O. For a detailed discussion, see Refs.~\cite{Singh:2006br, 
Athar:2007wd, Ahmad:2006cy}.
  
Alvarez-Ruso et al.~\cite{Alvarez-Ruso:2007kwp} and Amaro et al.~\cite{Amaro:2008hd} have also included NR 
contributions~(Section~\ref{NRB}) besides the delta resonance and found the contribution of NR terms to be very 
small. For the FSI of pion with the nucleus, they solved the Klein-Gordon equation. To see the difference in the results 
obtained by our group~\cite{Singh:2006br, Athar:2007wd, Ahmad:2006cy} and by solving Klein-Gordon equation for the treatment 
of pion FSI~\cite{Alvarez-Ruso:2007kwp}, a comparison was done by Alvarez-Ruso et al.~\cite{Alvarez-Ruso:2011hvm} and found 
the difference to be very small. Therefore, while comparing the experimental data for the coherent pion production, Monte Carlo 
generators generally  use the prescription of Ref.~\cite{Singh:2006br} for its simplicity.
 
Nakamura et al.~\cite{Nakamura:2009iq} have used dynamical model in coupled channels using the prescription of 
Ref.~\cite{Sato:2003rq} where the bare $N-\Delta$ transition from a constituent quark model is renormalized by meson clouds. 
Then the medium modification of the $\Delta$ properties and pion FSI have been taken into account. They have fitted the free 
parameters of the scattering potential and pion-nucleus optical potential to the pion-nucleus elastic scattering data. For 
CC process the flux averaged cross section corresponding to K2K experiment was found to be 
$\sigma_{CC}^{avg} = 6.3 \times 10^{-40}{\text{cm}}^2$ corresponding to the K2K observed result of $\sigma_{K2K}  < 7.7 
\times 10^{-40}{\text{cm}}^2$~\cite{K2K:2005uiu}. For NC reaction, the flux averaged cross section for the 
$\pi^0$ production calculated in this model gives $\sigma_{\text{NC}}^{\text{avg}}=2.8 \times 10^{-40}{\text{cm}}^2$ while 
the experimentally observed number from the MiniBooNE collaboration is $\sigma_{\text{MiniBooNE}}=7.7 \pm 1.6 \pm 3.6 \times  
10^{-40}{\text{cm}}^2$~\cite{Raaf:2004ty, Raaf:2005up}.

\subsubsection{Coherent kaon production}
The coherent kaon production has been studied by Alvarez-Ruso et al.~\cite{Alvarez-Ruso:2012kmi} using the formalism 
discussed here in Sections~\ref{kaon} and \ref{sec:1antiKaon} for the kaon and antikaon productions off the nucleon target, respectively. The differential 
cross section for reaction~(\ref{Ch12_reaction}) in the laboratory frame has been taken as:
\begin{equation}
\label{eq:sec}
\frac{d^{\,5}\sigma}{d\Omega_l dk_0' d\Omega_K } = \frac{1}{4 (2 \pi)^5}  
\frac{|\vec{k}^\prime||\vec{p}_K|}{|\vec{k}| M^2}\frac{G^2}{2}
L_{\mu\nu}\, {\cal M}^\mu_{K^+}(q,p_K) \left({\cal M}^\nu_{K^+}(q,p_K)\right)^*,
\end{equation}
where the nuclear current ${\cal M}^{\mu}_{K^+}$ is obtained as the coherent sum over all nucleons, 
leading to the nuclear densities   
\begin{equation}
{\cal M}^\mu_{K^+}(q,p_K) = 
\int d^3\vec{r}\ e^{{\rm i} \vec{q}\cdot\vec{r}}
\left\{\rho_p(\vec{r}\,) {\cal J}^\mu_{pK^+}(q,\hat{p}_K) + 
       \rho_n(\vec{r}\,) {\cal J}^\mu_{nK^+}(q,\hat{p}_K) \right \} \phi^*_{>}(\vec{p}_K,\vec{r}).
\label{eq:A}
\end{equation}
In the above expression
\begin{equation}
{\cal{J}}^\mu_{NK^+}(q,\hat{p}_K) = \frac12 \sum_i \mathrm{Tr} \left[ (\slashed{p} + M)\gamma^0 \Gamma_{i; NK^+}^\mu(q,
\hat{p}_K) \right] \frac{M}{p_0}, 
\label{eq:J}
\end{equation}
where the index $i$ refers to all the possible mechanisms in Figs.~\ref{Ch12_fg:terms} and \ref{Ch12_fg:terms_antiKaon}; 
$\Gamma_{i; NK^+}^\mu$ is given in Eq.~(\ref{Ch12_NRB:kaon}) with $j^\mu_i = \bar{N}(p') \Gamma_{i; NK^+}^\mu  N(p)$. The 
initial and final nucleons in the nucleus, are assumed to be on shell with $\vec{p} = (\vec{p}_K - \vec{q})/2$ and 
$\vec{p}\,' = - \vec{p}$. 

$\phi^*_{>}(\vec{p}_K,\vec{r})$ in Eq.~(\ref{eq:sec}) denotes the outgoing kaon wave function obtained by solving the 
Klein-Gordon equation for the kaons moving in an optical potential $V_{\mathrm{opt}}$:
\begin{equation}
\left( - \vec{\nabla}^2 - \vec{p}_{K}^{\,2} + 2 p_K^0 V_{\mathrm{opt}} \right) \phi^*_{>}(\vec{p}_K,\vec{r}) =0.
\label{eq:kg}
\end{equation}
\begin{figure} 
\begin{center}
\includegraphics[width=0.45\textwidth,height=6cm]{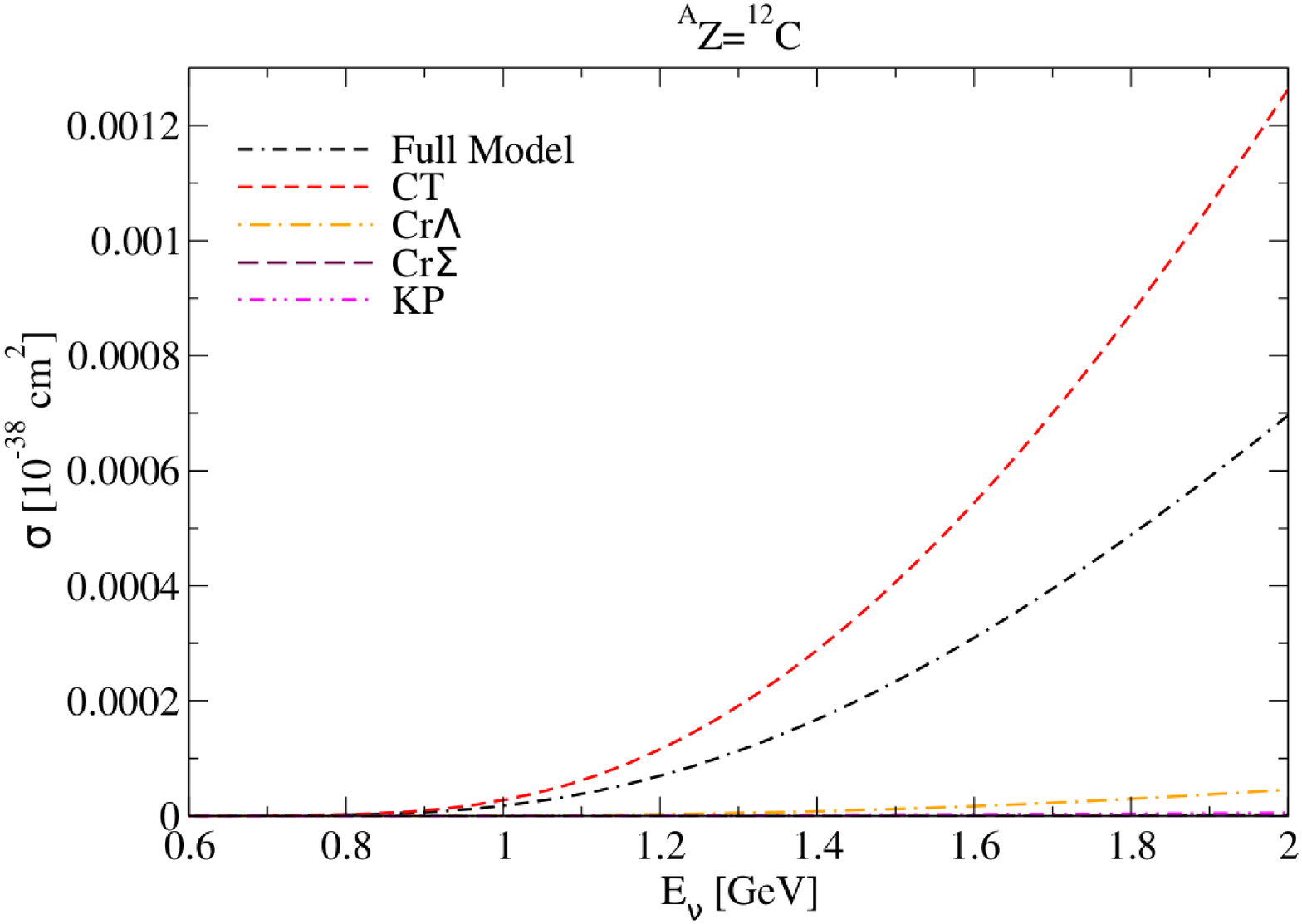}
\includegraphics[width=0.45\textwidth,height=6cm]{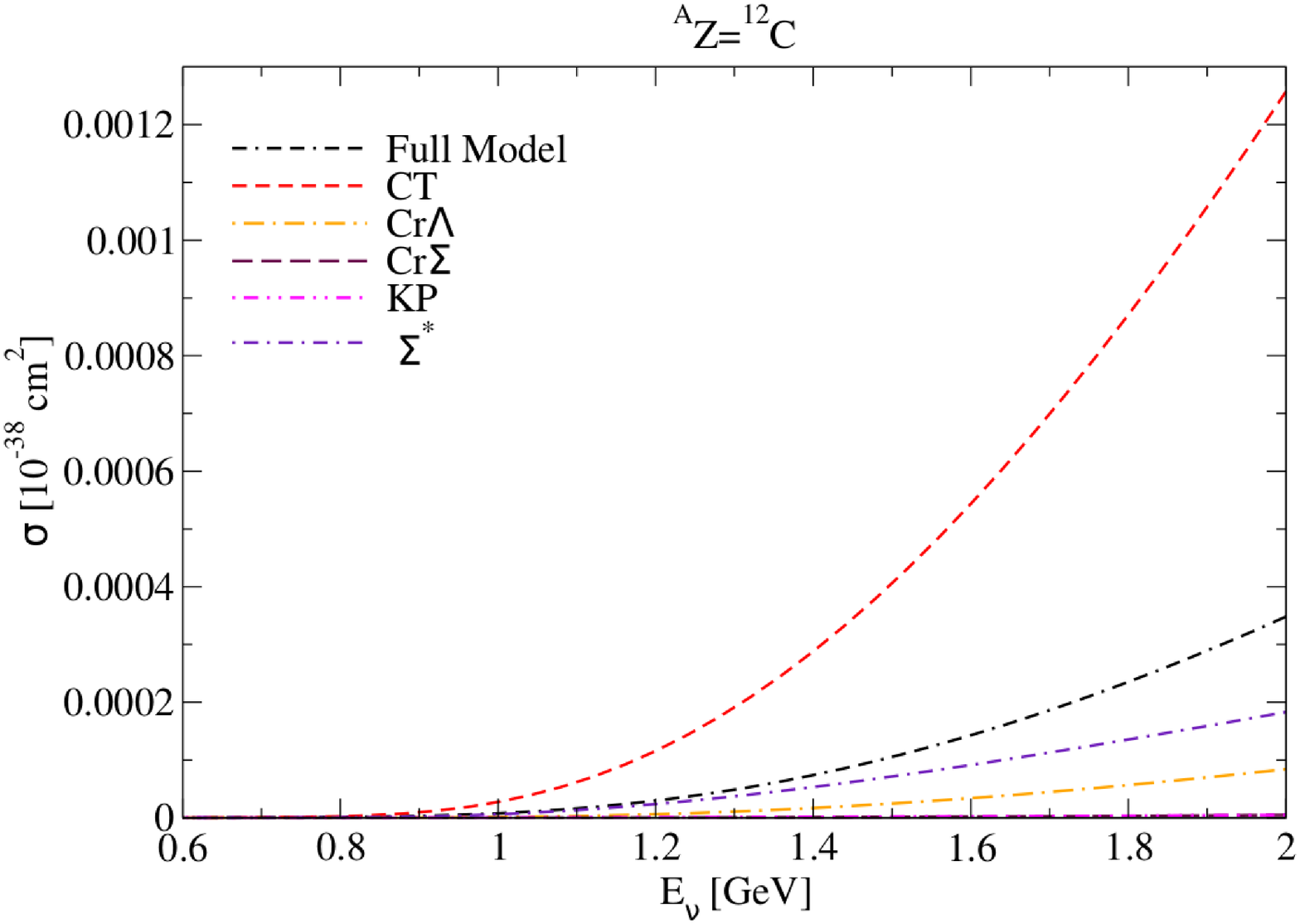}
\end{center}
\caption{\label{fig:mechs}Total cross section as a function of the neutrino energy to the coherent $K^{+}$~(left panel) and 
$K^{-}$~(right panel) reaction on $^{12}$C, when kaon distortion effect is taken into account. Figure has been taken from Ref.~\cite{Alvarez-Ruso:2012kmi}.}
\end{figure}
Fig.~\ref{fig:mechs} shows the results for the coherent $K^+$ and $K^-$ production cross sections vs neutrino energy, 
respectively for $\nu_\mu$(left panel) and  ${\bar\nu}_\mu$(right panel) reaction on $^{12}$C nuclear target. Like the 
$\nu_\mu$ induced single kaon production off nucleon target discussed here in Section~\ref{kaon}, here also the dominant 
contribution is from the contact term and due to the destructive interference there is reduction in the total cross sections 
when all the contributions~(Figs~\ref{Ch12_fg:terms} and \ref{Ch12_fg:terms_antiKaon}) are taken into account. It has been observed by these 
authors~\cite{Alvarez-Ruso:2012kmi} that at $E_{\nu_\mu}$=2GeV, the cross section per nucleon for carbon nucleus is a 
factor of about forty smaller than the one obtained for the free nucleon case. Recently MINERvA 
collaboration~\cite{MINERvA:2016cun} has reported at 3$\sigma$ C.L., the evidence for coherent kaon production in the neutrino 
induced scattering on carbon nuclear target but no real events were reported to be observed.

%
\subsection{Deep inelastic $\nu_l/\bar\nu_l-A$ scattering}\label{dis:nucleus}  
\subsubsection{Introduction}
\begin{figure} 
        \begin{center}
 \includegraphics[height=4. cm, width=6.8 cm]{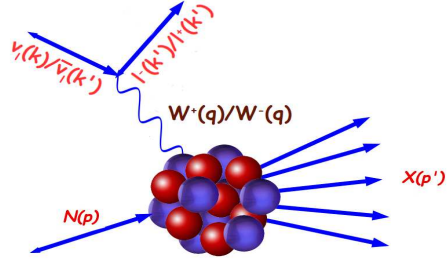}
 \end{center}
 \caption{The Feynman diagram showing the charged (anti)neutrino induced DIS process with bound 
 nucleons.}\label{nuclear_dis}
\end{figure}

When a (anti)neutrino interacts with a bound nucleon inside a nucleus, the scattering cross sections and the nucleon 
structure functions get modified due to NME. The reaction for this interaction process via CC DIS channel is represented as
\begin{equation}
 \nu_l/\bar\nu_l(k)+A(p)\longrightarrow l^-/l^+(k')+X(p')\;,\;\;\;
\end{equation}
where $A$ is the target nucleus, $X$ is jet of hadrons in the final state, $l=e, \mu, \tau$, and the quantities in the 
parentheses are the four momenta of the corresponding particles, and is shown in Fig.~\ref{nuclear_dis}.

The experimental evidence of NME in DIS channel was first reported 
by the European Muon Collaboration (EMC) in 1983 when it measured the scattering cross sections by using the very high 
energy muon beams off $^{56}Fe$ and $^{2}D$ targets and found that the ratio of the cross sections per nucleon in $^{56}Fe$ 
 and $^{2}D$
to be considerably different from unity~\cite{EuropeanMuon:1983wih}. As the DIS cross sections are 
generally expressed in terms of the nucleon structure functions, the EMC observation implied that the structure functions for 
a nucleon bound inside a nucleus are different from the structure functions of a free nucleon. This effect is famously known 
as the EMC effect. The spectacular discovery of EMC effect motivated physicists to perform similar DIS experiments with 
(anti)neutrino beams using different nuclear targets. The first experiment using neutrino beam were done at CERN on 
$^{20}$Ne and $^{56}$Fe targets by the BEBC and CDHS collaborations followed by the CDHSW and NOMAD collaborations using 
$^{208}$Pb and $^{12}$C targets, respectively. Similar experiments at FNAL were done first by the CCFR, NuTeV, followed by MINOS and 
MINERvA, collaborations using $^{56}$Fe. Several other experiments using charged lepton beam were performed by the different 
collaborations like SLAC~\cite{Gomez:1993ri}, HERMES~\cite{HERMES:1999bwb}, BCDMS~\cite{BCDMS:1985dor, BCDMS:1987upi}, 
NMC~\cite{NewMuon:1995tgs, NewMuon:1995cua}, JLab~\cite{Seely:2009gt}, etc. using nuclear targets, both moderate and heavy, 
for a wide range of Bjorken variable $x$($0 < x < 1$) and $Q^2$. From the experimental 
observations, some general features of the ratio $R(x,Q^2)=\frac{F_{2A}(x,Q^2)}{F_{2D}(x,Q^2)}$ may be inferred: 
\begin{itemize}
 \item The $x$ dependence of $R(x,Q^2)$ has considerable structure, i.e., it is different in the different regions of $x$.
 \item The shape of the effect is almost independent of $A$.
 \item The strength of the NME increases with the increase in mass number $A$.
 \item The functional form of $R(x,Q^2)$ has a very weak dependence on $Q^2$.
\end{itemize}
Generally, NME manifested through the ratio $R(x,Q^2)$ are broadly divided into four regions of $x$ 
as shown in Fig.~\ref{nuclear_effect}~\cite{Gomez:1993ri, HERMES:1999bwb, BCDMS:1987upi, 
 NewMuon:1995cua, EuropeanMuon:1992pyr} in which the $x$ dependence is attributed to different physical effects. These are:
 \begin{figure} 
\begin{center}
 \includegraphics[height=6 cm, width=12 cm]{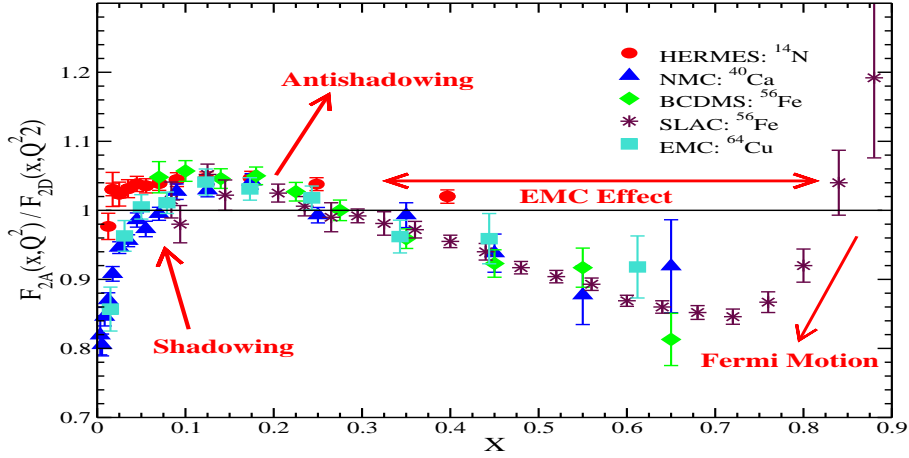}
  \end{center}
 \caption{Ratio $R(x,Q^2)=\frac{F_{2A}(x,Q^2)}{F_{2D}(x,Q^2)}$; ($A=$target nucleus) vs $x$ shows NME in structure function. 
 Experimental data are taken from Refs.~\cite{Gomez:1993ri, HERMES:1999bwb, BCDMS:1987upi, 
 NewMuon:1995cua, EuropeanMuon:1992pyr}. }\label{nuclear_effect}
\end{figure}
\begin{enumerate}
 \item {\bf Shadowing Effect:} In the region of low $x(< 0.1)$, a suppression is
 found in the ratio $R(x,Q^2)$ which is known as the shadowing effect.
 This suppression becomes more pronounced with the increase in the mass number $A$.
 
 \item {\bf Antishadowing Effect:} This is the region of $0.1 \le x \le 0.2$, where there is an enhancement in the ratio of 
 structure functions ($R(x,Q^2)$) and has been found to have almost no nuclear mass dependence.
 
 \item {\bf EMC Effect:} The ratio $R(x,Q^2)$ shows a dip in the region of $0.2<x<0.7$ and this is known as EMC effect, 
 named after the first experimental observation by the EMC collaboration~\cite{EuropeanMuon:1983wih}. 
 \item {\bf Fermi Motion:} The nucleons bound inside the nuclear target are moving with some Fermi momentum which increases 
 with the increase in the mass number. This is responsible for the abrupt rise in the ratio of structure functions in the 
 region of $x\ge 0.7$.
\end{enumerate}
It was observed that the results for NME on mass dependence $A$ were consistent with $log(A)$ and 
average nuclear density $\bar\rho_A=\frac{\rho_A}{A}$, where $\rho_A$ is the nuclear density~\cite{Malace:2014uea}. In 
recent years the MINERvA collaboration has measured the (anti)neutrino-nucleus cross sections using several targets like 
carbon, water, iron and lead in a wide energy spectrum, where the contribution to the cross section comes from different 
reaction channels. MINERvA has explicitly studied DIS channel~\cite{MINERvA:2020zzv, 
MINERvA:2021owq, Mousseau:2016snl, Tice:2014pgu} as it aims to perform EMC kind of measurements in the weak sector covering 
a wide range of $x$ and $Q^2$. For this purpose, MINERvA performed the analysis of (anti)neutrino-nucleus DIS data and 
reported the results for the ratio of scattering cross sections $\frac{d\sigma_A/dx}{d\sigma_{CH}/dx};~(A=^{12}C,~^{56}Fe,
~^{208}Pb)$ vs $x$. It has been observed that MINERvA's experimental results on $R(x,Q^2)$ are not satisfactorily explained by 
either the phenomenological models or theoretical models in the entire range of Bjorken $x$~\cite{Mousseau:2016snl, 
Tice:2014pgu}. Hence, it is crucial to develop a better understanding of NME both theoretically, 
where dynamics of the nucleons in the nuclear medium is taken into consideration, as well as phenomenologically, which 
involve the determination of the effective parton distribution of nucleons within a nucleus. In the following, we first discuss the theoretical approach.
 
Theoretically several attempts have been made to understand these effects and many models have been proposed. These models 
are based on the basis of nuclear binding, nuclear medium modification including short range correlations in 
nuclei~\cite{Malace:2014uea, Akulinichev:1985cfw}, pion excess in nuclei~\cite{Bickerstaff:1989ch, Kulagin:1989mu, 
Marco:1995vb, Ericson:1983um, Bickerstaff:1985mp, Berger:1987er}, multi-quark clusters~\cite{Jaffe:1982rr, Mineo:2003vc, 
Cloet:2005rt}, dynamical rescaling~\cite{Nachtmann:1983py, Close:1983tn}, nuclear shadowing~\cite{Frankfurt:1988nt, 
Armesto:2006ph}, etc. However, no comprehensive theoretical or phenomenological understanding of the nuclear modifications 
of the  bound nucleon properties across the complete range of $x$ and $Q^2$ consistent with the presently available 
experimental data exists~\cite{Arneodo:1992wf, Geesaman:1995yd, Hen:2013oha, Piller:1999wx}. Furthermore, initially it was 
thought that NME in electromagnetic and weak structure functions should be similar in $F_{1A}(x,Q^2)$ 
and $F_{2A}(x,Q^2)$ despite the additional contribution from the axial current in the weak sector. Recently in a 
phenomenological study Kalantarians et al.~\cite{Kalantarians:2017mkj} have made a comparison of electromagnetic vs weak 
nuclear structure functions ($F_{2A}^{EM}(x,Q^2)$ vs $F_{2A}^{WI}(x,Q^2)$) and found out that at low $x$ these two structure 
functions are different. Theoretically there have been very few calculations to study NME in the weak 
structure functions, where explicitly a comparative study of electromagnetic and weak nuclear structure functions have been 
made~\cite{Zaidi:2019asc, Haider:2016tev}. Our group~\cite{Zaidi:2019asc, Haider:2016tev} has explicitly studied the 
difference in the electromagnetic and weak nuclear structure functions, both for $F_{2A}^{EM}(x,Q^2)$ and $F_{2A}^{WI} 
(x,Q^2)$,  $F_{1A}^{EM}(x,Q^2)$ and $F_{1A}^{WI}(x,Q^2)$. More theoretical as well experimental studies are needed in the 
weak sector to understand NME for a wide range of $x$ and $Q^2$ for moderate as well as heavy nuclear 
targets. 
 
In the weak sector, there are only two groups who have theoretically studied NME in the weak nuclear 
structure functions, one is Kulagin and Petti~\cite{Kulagin:2007ju} and the other is Athar et al.~(Aligarh-Valencia 
group)~\cite{Ansari:2020xne, Zaidi:2019asc, SajjadAthar:2007bz, SajjadAthar:2009cr, Haider:2011qs, Haider:2012ic, Haider:2012nf, Haider:2016zrk, 
Ansari:2021cao, Zaidi:2021iam, AtharSajjad:2022ipr}. Kulagin and Petti in their model of 
nuclear DIS took into account nuclear effects like the nuclear shadowing, Fermi motion, binding energy, nuclear pion excess 
and off-shell corrections to bound structure functions. While the Aligarh-Valencia group~\cite{Ansari:2020xne, Zaidi:2019asc, 
SajjadAthar:2007bz, SajjadAthar:2009cr, Haider:2011qs, Haider:2012ic, Haider:2012nf, Haider:2016zrk, 
Ansari:2021cao, Zaidi:2021iam, AtharSajjad:2022ipr} have used a microscopic model which uses relativistic nucleon spectral 
function to describe target nucleon momentum distribution incorporating the effects of Fermi motion, binding energy and 
nucleon correlations in a field theoretical model. The spectral function that describes the energy and momentum distribution 
of the nucleons in nuclei is obtained by using the Lehmann's representation for the relativistic nucleon propagator and 
nuclear many body theory is used to calculate it for an interacting Fermi sea in the nuclear 
matter~\cite{FernandezdeCordoba:1991wf}. A local density approximation is then applied to translate these results to a 
finite nucleus. Furthermore, the contributions of the pion and rho meson clouds in a many body field theoretical approach 
have also been considered. In Section~\ref{formalism-dis-a}, we discuss in brief the theoretical approach of the 
Aligarh-Valencia group to understand NME in $\nu_l(\bar\nu_l)-$nucleus scattering.
\subsubsection{Formalism}\label{formalism-dis-a}
The general expression of the differential scattering cross section for (anti)neutrino-nucleus 
DIS process 
\begin{eqnarray}\label{reac}
\nu_l(k)+A(p_A) \rightarrow l^-(k^\prime) +X(p^\prime_A); ~~l=e~\textrm{or}~\mu~\textrm{or}~\tau,
\end{eqnarray}
is written in analogy
with CC $\nu_l(\bar\nu_l)-N$ scattering discussed in Section~\ref{dis:nucleon} 
by replacing the nucleon hadronic tensor $W^{\mu\nu}_N$ with the 
 nuclear hadronic tensor $W^{\mu\nu}_A$ and is given by:
 \begin{equation}
\label{eq:q2nua}
\frac{ d^2\sigma_A }{ dQ^2 d\nu } =  \frac{G_F^2}{4\pi E_\nu E_l}~\left(\frac{M_W^2}{Q^2+M_W^2}\right)^2~\frac{|
\vec{k}^\prime|}{|\vec{k}|}\; L_{\mu\nu} ~W^{\mu\nu}_A,
\end{equation}
or it may be expressed in terms of the scaling variables as 
\begin{eqnarray}\label{xecA}
{ d^2\sigma_A \over dx dy }&=& \left({G_F^2 y M E_l  \over 2\pi E_{\nu}}\right) {\left(M_W^2\over M_W^2+Q^2 \right)^2} 
\; {|\vec{k}^\prime| \over|\vec{k}|}\;L_{\mu\nu}\; W^{\mu\nu}_A,
\end{eqnarray}
 $W^{\mu\nu}_A$ is written in terms of the weak nuclear structure functions $W_{iA}^{WI}(\nu,Q^2)$ ($i=1,2,3$) as
\begin{eqnarray}
 \label{nuc_had_weak}
W_{A}^{\mu \nu} &=&
\left( \frac{q^{\mu} q^{\nu}}{q^2} - g^{\mu \nu} \right) 
W_{1A} (\nu_A, Q^2)
+ \frac{W_{2A} (\nu_A, Q^2)}{M_A^2}\left( p^{\mu}_A - \frac{p_A . q}{q^2}  q^{\mu} \right)\left( p^{\nu}_A - 
\frac{p_A . q}{q^2} q^{\nu} \right)\pm \frac{i}{2M_A^2} \epsilon^{\mu \nu \rho \sigma} p_{A \rho} q_{\sigma}
W_{3A} (\nu_A, Q^2) \nonumber \\
&+& \frac{W_{4A} (\nu_A, Q^2)}{M_A^2} q^{\mu} q^{\nu}+\frac{W_{5A} (\nu_A, Q^2)}{M_A^2} (p^{\mu}_A 
q^{\nu} + q^{\mu} p^{\nu}_A)+ \frac{i}{M_A^2} (p^{\mu}_A q^{\nu} - q^{\mu} p^{\nu}_A)
W_{6A} (\nu_A, Q^2)\,.
\end{eqnarray}
In the above expression $M_A$ is the mass and $p_A$ is the four momentum of the initial nuclear target and the 
positive/negative sign is for the $\nu_l/\bar\nu_l$. The leptonic tensor in Eq.~(\ref{xecA}) has the same form as given in 
Eq.~(\ref{lep_tens}). $W_{6A}(\nu_A,Q^2)$ does not contribute to the cross section as it vanishes when contracted with the 
leptonic tensor $L_{\mu \nu}$. The nuclear structure functions $W_{iA}(\nu_A,Q^2)~(i=1-5)$ are written in terms of the 
dimensionless nuclear structure functions $F_{iA}(x_A);~(i=1-5)$ as~\cite{Zaidi:2019asc, Kretzer:2003iu}:
\begin{eqnarray}\label{relation1}
 &&F_{1A}(x_A) =W_{1A}(\nu_A,Q^2);\;\; F_{2A}(x_A) = \frac{Q^2}{2xM_A^2}W_{2A}(\nu_A,Q^2);\;\; F_{3A}(x_A) = 
 \frac{Q^2}{xM_A^2}W_{3A}(\nu_A,Q^2);\\
 &&F_{4A}(x_A) = \frac{Q^2}{2M_A^2}W_{4A}(\nu_A,Q^2);\;\; F_{5A}(x_A) = \frac{Q^2}{2xM_A^2}W_{5A}(\nu_A,Q^2), 
 \end{eqnarray}
where $\nu_A$(=$\frac{p_{_A}\cdot q}{M_{_A}}=q^{0}$) is the energy transferred to the nuclear target in the rest frame 
of the nucleus i.e. $p_A=(p_A^0,~\vec{p}_A= 0)$ and $x_A(=\frac{Q^2}{2 p_A \cdot q}=\frac{Q^2}{2 p_{A}^0  q^0 } = 
\frac{Q^2}{2 A~M q^0}=\frac{x}{A})$ is the Bjorken scaling variable corresponding to the nucleus.

The expression for the differential cross section for the $\nu_l/{\bar\nu}_l - A$ scattering 
is then written as~\cite{Zaidi:2021iam}:
\begin{eqnarray}\label{xsecsf}
 \frac{d^2\sigma_A}{dxdy}&=&\frac{G_F^2ME_\nu}{\pi(1+\frac{Q^2}{M_W^2})^2}
 \Big\{\Big[y^2x+\frac{m_l^2 y}{2E_\nu M}\Big]F_{1A}(x,Q^2)+
 \Big[\Big(1-\frac{m_l^2}{4E_\nu^2}\Big)-\Big(1+\frac{Mx}{2E_\nu}\Big)y\Big]F_{2A}(x,Q^2)\nonumber\\
 &\pm& \Big[xy\Big(1-\frac{y}{2}\Big)-
 \frac{m_l^2 y}{4E_\nu M}\Big]F_{3A}(x,Q^2)
 +\frac{m_l^2(m_l^2+Q^2)}{4E_\nu^2M^2 x}F_{4A}(x,Q^2)-\frac{m_l^2}{E_\nu M}F_{5A}(x,Q^2)\Big\},\;\;\;
\end{eqnarray}
where the kinematic variables have the same meaning as defined in Section~\ref{dis:nucleon}. For $\nu_e/\bar{\nu}_e$ and 
$\nu_\mu/\bar{\nu}_\mu$ interactions with a nuclear target (i.e. in the limit $m_l\to 0$), only the first three terms of 
Eq.~(\ref{xsecsf}), i.e. the terms with $F_{1A}(x,Q^2)$, $F_{2A}(x,Q^2)$ and $F_{3A}(x,Q^2)$ would contribute. However, 
for the $\nu_\tau/\bar{\nu}_\tau$ induced processes, all the five structure functions ($F_{iA}(x,Q^2);~(i=1-5)$) contribute and this has 
been discussed recently by Zaidi et al.~\cite{Zaidi:2021iam}. Here the discussions are made only for the massless 
lepton case.

The nucleons bound inside the nucleus are moving continuously with a finite momentum, i.e. $\vec{p}_N$ is nonzero and the 
motion of such nucleons corresponds to the Fermi motion. Therefore, these nucleons are off shell. If the momentum transfer 
is taken to be along the $Z$-axis then $q^\mu=(q^0,0,0,q^z)$ and the Bjorken variable $x_N$ corresponding to the nucleon 
bound inside a nucleus is written as:
 \begin{equation}
 x_N = \frac{Q^2}{2 p_N \cdot q} = \frac{Q^2}{2 (p_N^0 q^0 - p_N^z q^z)}.
\end{equation}
The momentum~($p_N\ne 0$) of the initial nucleon is constrained by the Fermi momentum ($p_{F_N}$) of the nucleon in the 
nucleus, i.e., $p_N \le p_{F_N}$. These bound nucleons interact among themselves via the strong interaction and thus various 
NME come into play which are effective in the different regions of the Bjorken variable $x$.

\subsubsection{Fermi motion, binding and nucleon correlation effects}\label{spec}
In the local density approximation, using many body field theoretical approach, the scattering cross section for a 
(anti)neutrino interacting with a bound nucleon ($\nu_l + N \rightarrow l^- + X$) is obtained by considering a flux of 
neutrinos hitting a collection of target nucleons over a given length of time. Now a majority will simply pass through the 
target without interacting while a certain fraction will interact with the target nucleons leaving the pass-through 
fraction and entering the fraction of neutrinos yielding final state leptons and hadrons. Then the concept of "neutrino 
self energy" is used which has a real and imaginary part. The real part modifies the lepton mass while the imaginary part 
is related to this fraction of interacting neutrinos and gives the total number of neutrinos that have participated in the 
interactions that give rise to the charged leptons and hadrons.
 
 \begin{figure} 
\begin{center}
 \includegraphics[height=4.0 cm, width=11 cm]{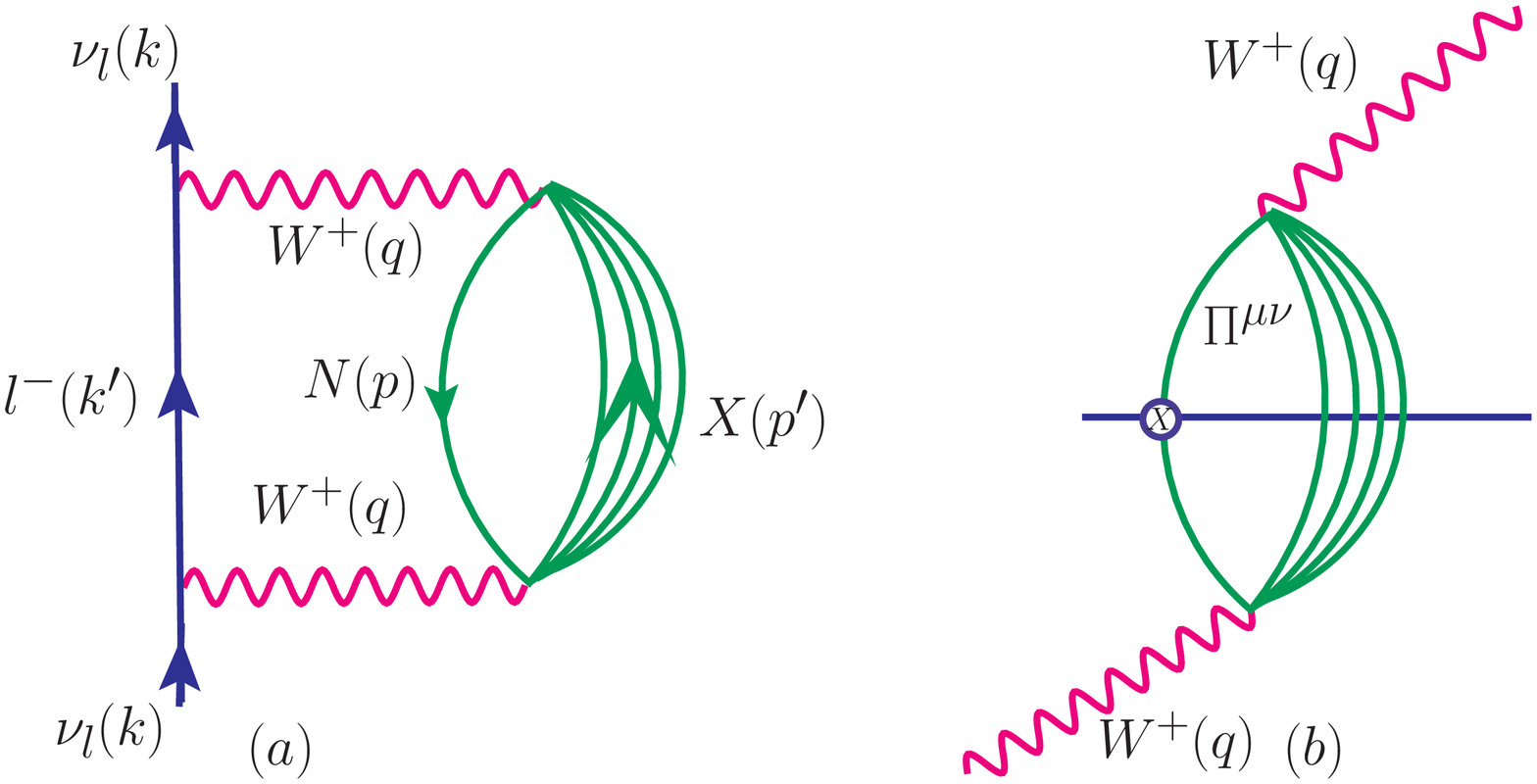}
 \end{center}
 \caption{Diagrammatic representation of {\bf (a)} the neutrino self-energy and {\bf (b)} the intermediate vector boson 
 $W^+$ self-energy.}\label{wself_energy}
\end{figure}
\begin{figure} 
\begin{center}
 \includegraphics[height=2.5 cm, width=12 cm]{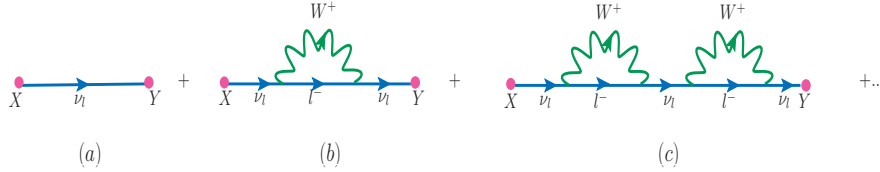}
 \end{center}
 \caption{Fermion two point function and its modification.}
 \label{self_fig}
\end{figure}
The neutrino self energy is evaluated corresponding to the diagram shown in Fig.~\ref{wself_energy}~(left panel). The cross 
section for an element of volume $dV$ in the rest frame of the nucleus is related to the probability per unit time 
($\Gamma$) of the $\nu_l$ interacting with a nucleon bound inside a nucleus. $\Gamma dt dS$ provides probability times a 
differential of area ($dS$) which is nothing but the cross section ($d\sigma$)~\cite{Marco:1995vb}, i.e.
\begin{equation}\label{defxsec}
d\sigma=\Gamma dt ds=\Gamma\frac{dt}{dl}ds dl=\Gamma \frac{1}{v}dV = \Gamma \frac {E_l}{\mid \vec{k} \mid}d^3 r,
\end{equation}
where $v\left(=\frac{\mid \vec{k} \mid}{E_l}\right)$ is the velocity of the incoming $\nu_l$. The probability per unit time 
of the interaction of $\nu_l$ with the nucleons in the nuclear medium to give the final state is related to the imaginary 
part of the $\nu_l$ self energy $\Sigma(k)$, as~\cite{Marco:1995vb}:
\begin{equation}\label{eqr}
-\frac{\Gamma}{2} = \frac{m_\nu}{E_\nu(\vec{k})}\; Im \Sigma(k),
\end{equation}
where $Im(\Sigma(k))$ is the imaginary part of the neutrino self energy (shown in Fig.~\ref{wself_energy} (left panel)).
By using Eq.~(\ref{eqr}) in Eq.~(\ref{defxsec}), we obtain
\begin{equation}\label{eqq}
 d\sigma= \frac{-2m_\nu}{\mid \vec{k} \mid} Im \Sigma (k) d^3 r.
\end{equation}
 In  many body field theory the interaction of neutrino with a potential provided by a nucleus
can be explained as the modification to the fermion two point function as depicted  in Fig.~\ref{self_fig}.
Figure \ref{self_fig}(a) corresponds to the free field fermion propagator while Figure \ref{self_fig}(b,c) constitutes to 
the neutrino self-energy. Thus to get $d\sigma$, we are required to evaluate the imaginary part of the neutrino self energy 
$Im \Sigma (k)$ which is obtained by following the Feynman rules~\cite{Athar:2020kqn}:
\begin{equation}\label{nu_imslf1}
Im \Sigma(k)=\frac{ G_F}{\sqrt{2}} {4 \over m_\nu} \int \frac{d^3 k^\prime}{(2 \pi)^4} {\pi \over E(\vec{k}^\prime)} 
\theta(q^0) \left(\frac{M_W}{Q^2+M_W^2}\right)^2\;Im[L_{\mu\nu}^{WI} \Pi^{\mu\nu}(q)].
\end{equation}
In the above expression, $\Pi^{\mu\nu}(q)$ is the $W$ boson self-energy, which is written in terms of the nucleon ($G_l$) 
and meson ($D_j$) propagators (depicted in Fig.~\ref{wself_energy} (right panel)) following the Feynman rules and is given 
by
\begin{eqnarray}\label{wboson}
 \Pi^{\mu \nu} (q)&=& \left(\frac{G_F M_W^2}{\sqrt{2}}\right) \times \int \frac{d^4 p}{(2 \pi)^4} G (p) 
\sum_X \; \sum_{s_p, s_l} \prod_{i = 1}^{N} \int \frac{d^4 p'_i}{(2 \pi)^4} \; \prod_{_l} G_l (p'_l)\; \nonumber \\  
&\times&  \prod_{_j} \; D_j (p'_j)<X | J^{\mu} | N >  <X | J^{\nu} | N >^* (2 \pi)^4 \times\delta^4 (k + p - k^\prime - 
\sum^N_{i = 1} p'_i),\;\;\;
\end{eqnarray}
where $s_p$ is the spin of the nucleon, $s_l$ is the spin of the fermions in $X$, $<X | J^{\mu} | N >$ is the hadronic 
current for the initial state nucleon to the final state hadrons, index $l,~j$ are respectively, stands for the fermions 
and bosons in the final hadronic state $X$, and $\delta^4 (k + p - k^\prime - \sum^N_{i = 1} p'_i)$ ensures the 
conservation of four momentum at the vertex. $G(p^0,\vec{p})$ is the nucleon propagator which inside the nuclear medium 
provides information about the propagation of the nucleon from the initial state to the final state or vice versa. 

\begin{figure} 
\begin{center}
 \includegraphics[height=5 cm, width=12.5 cm]{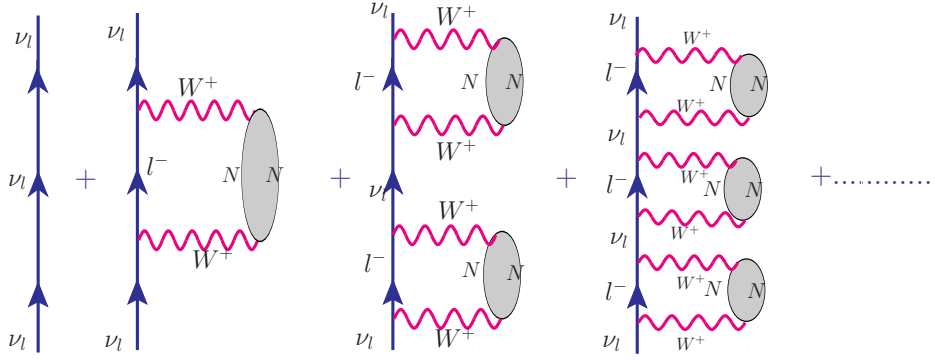}
 \end{center}
 \caption{Diagrammatic representation of neutrino self-energy in the nuclear medium.}
 \label{n_self}
\end{figure}

The relativistic nucleon propagator $G(p^0,\vec{p})$ in a nuclear medium is obtained by starting with the relativistic free 
nucleon Dirac propagator $G^{0}(p^{0},{\vec{p}})$, which is written in terms of the contribution from the positive and 
negative energy components of the nucleon described by the Dirac spinors $u(\vec{p})$ and $v(\vec{p})$~\cite{Marco:1995vb, 
FernandezdeCordoba:1991wf}. Only the positive energy contributions are retained as the negative energy contributions are 
suppressed. In the interacting Fermi sea, the relativistic nucleon propagator is then written in terms of the nucleon self 
energy $\Sigma^N(p^0,\vec{p})$ which is shown in Fig.~\ref{n_self}. In nuclear many body technique, the quantity that contains 
all the information on the single nucleon properties is the nucleon self energy $\Sigma^N(p^0,\vec{p})$. For an interacting Fermi 
sea the relativistic nucleon propagator is written in terms of the nucleon self energy and in nuclear matter the interaction 
is taken into account through Dyson series expansion which is the quantum field theoretical analogue of the 
Lippmann-Schwinger equation for the dressed nucleons, and in principle an infinite series in perturbation theory. This 
perturbative expansion is summed in a ladder approximation as 
\begin{eqnarray}\label{gp1}
G(p^0,\vec{p})&=&\frac{M}{E(\vec{p})}\frac{\sum_{r}u_{r}(\vec{p})\bar u_{r}(\vec{p})}{p^{0}-E(\vec{p})}+\frac{M}
{E(\vec{p})}\frac{\sum_{r}u_{r}(\vec{p})\bar
u_{r}(\vec{p})}{p^{0}-E(\vec{p})}\Sigma^N(p^{0},\vec{p})\times \frac{M}{E(\vec{p})} \frac{\sum_{s}u_{s}(\vec{p})\bar 
u_{s}(\vec{p})}{p^{0}-E(P)}+..... \nonumber \\
&=&\frac{M}{E(\vec{p})}\frac{\sum_{r} u_{r}(\vec{p})\bar u_{r}(\vec{p})}{p^{0}-E(\vec{p})-\sum_{r}\bar u_{r}(\vec{p})
\Sigma^N (p^{0},\vec{p})u_{r}(\vec{p})
\frac{M}{{E(\vec{p})}}},\;\;\;
\end{eqnarray}
where $\Sigma^N(p^0,\vec{p})$ is the nucleon self energy which is obtained following the techniques of many body theory. 
This has been taken from Ref.~\cite{FernandezdeCordoba:1991wf, Oset:1981mk} which uses the nucleon-nucleon scattering cross 
section and the spin-isospin effective interaction with random phase approximation(RPA) correlation as inputs. In this 
approach, the real part of the self energy of nucleon is obtained by means of the dispersion relations using the expressions 
for the imaginary part. The Fock term, which does not have imaginary part, does not contribute either to $Im 
\Sigma^N(p^{0},\vec{p})$ or to $Re \Sigma^N(p^{0},\vec{p})$ through the dispersion relation and its contribution to 
$\Sigma^N(p^{0},\vec{p})$ is explicitly calculated and added to $Re \Sigma^N(p^{0},
\vec{p})$~\cite{FernandezdeCordoba:1991wf}. The model however misses some contributions from similar terms of Hartree type which 
are independent of nucleon momentum $p$. This semi-phenomenological model of nucleon self energy is found to be in 
reasonable agreement with those obtained in sophisticated  many body calculations and has been successfully used in the 
past to study NME in many processes induced by photons, pions and leptons~\cite{Gil:1997bm, Gil:1997jg}.
 The expression for the nucleon self energy in the nuclear matter i.e. $\Sigma^N(p^0,{\vec{p}})$ is taken from 
 Ref.~\cite{FernandezdeCordoba:1991wf}, and the dressed nucleon propagator is expressed as~\cite{Marco:1995vb}:
 \begin{small}
 \begin{eqnarray}\label{Gp}
G(p^0,\vec{p}) &=& \frac{M}{E(\vec{p})} 
\sum_r u_r (\vec{p}) \bar{u}_r(\vec{p})
\left[\int^{\mu}_{- \infty} d \omega 
\frac{S_h (\omega, \vec{p})}{p^0 - \omega - i \eta}
+ \int^{\infty}_{\mu} d  \omega 
\frac{S_p (\omega, {\vec{p}})}{p^0 - \omega + i \eta}\right],\;\;~~~
\end{eqnarray}
\end{small}
where $S_h (\omega, {\vec{p}})$ and $S_p (\omega, {\vec{p}})$ are the hole
and particle spectral functions, respectively. $\mu=\epsilon_F+M$ is the chemical potential, $\omega=p^0-M$ is the 
removal energy and $\eta$ is the infinitesimal small quantity, i.e. $\eta \to 0$. 
The hole and particle spectral functions are given by~\cite{Marco:1995vb, FernandezdeCordoba:1991wf}:
\begin{eqnarray}\label{sh}
 S_h(p^0,\vec{p})&=&\frac{1}{\pi}
 \frac{\frac{M}{E_N(\vec{p})}\textrm{Im}\Sigma^N(p^0,\vec{p})}{\left(p^0-
 E_N(\vec{p})-\frac{M}{E_N(\vec{p})}\textrm{Re}\Sigma^N(p^0,\vec{p})\right)^2+
 \left(\frac{M}{E_N(\vec{p})}\textrm{Im}\Sigma^N(p^0,\vec{p})\right)^2};\;\;\text{for} ~~~~p^0 \le \mu \\
 \label{sp}
 S_p(p_0,\vec{p})&=&-\frac{1}{\pi}
 \frac{\frac{M}{E(\vec{p})}\textrm{Im}\Sigma^N(p_0,\vec{p})}{\left(p_0-
 E(\vec{p})-\frac{M}{E(\vec{p})}\textrm{Re}\Sigma^N(p_0,\vec{p})\right)^2+
 \left(\frac{M}{E(\vec{p})}\textrm{Im}\Sigma^N(p_0,\vec{p})\right)^2};\;\;\text{for} ~~~~p_0 > \mu
\end{eqnarray}
which obey the following relations
 \begin{equation}
 \int_{-\infty}^\mu\,dp_0\;S_h(p_0,\vec{p})=n(\vec{p})\;;\;\;\;\; \int^{\infty}_\mu\,dp_0\;S_p(p_0,\vec{p})=1-n
 (\vec{p}) ,
\end{equation}
where $n(\vec{p})$ is the Fermi occupation number.

Hence, one may obtain the spectral function sum rule which is given by
\begin{equation}
 \int_{-\infty}^{\mu}~S_h (\omega, {\vec{p}})~d\omega~+~ \int_{\mu}^{+\infty}~S_p (\omega, {\vec{p}})~d\omega~=~1
\end{equation}
with  the removal energy $\omega(=p^0-M)$. For the numerical calculations, the expression of chemical potential, i.e.
\begin{equation}\label{chem}
 \mu=\frac{p_{F_N}^2}{2M}+Re\Sigma^N\Big[\frac{p_{F_N}^2}{2M},p_{F_N} \Big]
\end{equation}
has been taken from Ref.~\cite{FernandezdeCordoba:1991wf} which is defined in terms of Fermi momentum ($p_{F_N}$) and the 
nucleon self energy ($\Sigma^N$)~\cite{FernandezdeCordoba:1991wf}.

For an inclusive process, only the hole spectral function contributes. 
In Fig.~\ref{spec}, following Ref.~\cite{Marco:1995vb}, we have plotted $S_h(\omega,\vec{p})$ vs $\omega$ (where 
$\omega=p_0-M$), for $p < p_F$ and $p > p_F$ in $^{12}C$ and $^{56}Fe$ nuclei. It may be observed that for $p < p_F$ the hole 
spectral function $S_h$ almost mimics a delta function as it corresponds to a Lorentzian distribution with a very narrow 
width. While for $p > p_F$, $S_h$ is not exactly zero, although very small in magnitude but has a longer range. This behavior 
is different from independent particle model, where it is exactly zero and this difference arises due to nucleon 
correlation~\cite{Mahaux:1985zz}. The details are given in Ref.~\cite{Zaidi:2019asc, Haider:2015vea}. 

\begin{figure} 
\begin{center} 
 \includegraphics[height=0.25\textheight,width=0.8\textwidth]{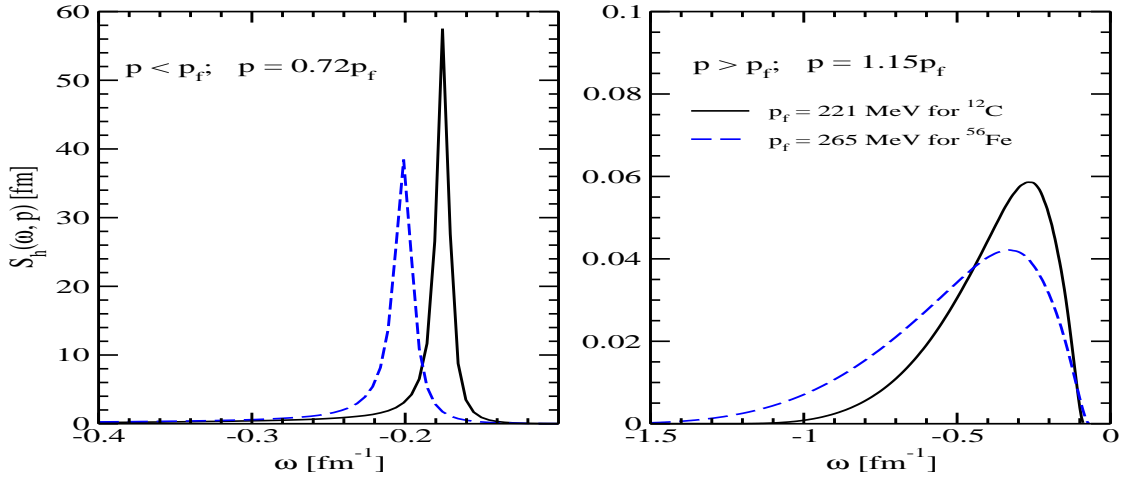}
  \caption{$S_h(\omega,\vec{p})$ vs $\omega$ for $p < p_F$(Left panel) and $p > p_F$(Right panel) in $^{12}C$(solid line) 
  and $^{56}Fe$(dashed line).}\label{spec}
 \end{center}
\end{figure}
The cross section is then obtained by using Eqs.~(\ref{eqq}) and (\ref{nu_imslf1}) as: 
\begin{equation}\label{dsigma_3}
\frac {d^2\sigma_A}{dx dy}=-\frac{G_F^2\;M\;y}{2\pi}\;\frac{E_l}{E_\nu}\;\frac{|\vec{k}^\prime|}{|\vec{k}|}
\left(\frac{M_W^2}{Q^2+M_W^2}\right)^2 L_{\mu\nu} \int  Im \Pi^{\mu\nu}(q) d^{3}r.
\end{equation}
Comparing Eq.~(\ref{dsigma_3}) with Eqs.~(\ref{xecA}), and using Eqs.~(\ref{wboson}) and (\ref{Gp}), the expression of the 
nuclear hadronic tensor for an isospin symmetric nucleus in terms of the nucleonic hadronic tensor and spectral function, 
is obtained as~\cite{Haider:2015vea}
\begin{equation}\label{conv_WAa}
W^{\mu \nu}_{A} = 4 \int \, d^3 r \, \int \frac{d^3 p}{(2 \pi)^3} \, 
\frac{M}{E (\vec{p})} \, \int^{\mu}_{- \infty} d p^0 S_h (p^0, \vec{p}, \rho(r))
W^{\mu \nu}_{N} (p, q), \,
\end{equation}
where the factor of 4 is for the spin-isospin of nucleon and $\rho(r)$ is the charge density of the nucleon in the nucleus 
which is given in Appendix~\ref{app_lda}. $S_h$ incorporates the effects like Fermi 
motion, binding energy and nucleon correlations and is physically interpreted as equal to the joint probability of (i) 
removing a nucleon with momentum $\vec{p}$ from the correlated ground state, and (ii) finding the resulting system of 
(A-1) nucleons with energy in the interval $p_0$ and $p_0 + dp_0$. 

Moreover, we have ensured that the spectral function is 
properly normalized and checked it by obtaining the correct baryon number ($A$) for a given nucleus~\cite{Haider:2015vea}:
\begin{eqnarray} \label{norm3}
4\int\frac{d^{3}p}{(2\pi)^{3}}\int_{-\infty}^{\mu}S_{h}(\omega,p,p_{F_N}(\vec{{r}})) d\omega= \rho({\vec{r}}),
\end{eqnarray}
that leads to the following normalization condition 
\begin{equation}\label{norm5}
4 \int d^3 r \;  \int \frac{d^3 p}{(2 \pi)^3} 
\int^{\mu}_{- \infty} \; S_h (\omega, {\vec{p}}, \rho(r)) 
\; d \omega = A\,.
\end{equation}

 The binding energy per nucleon for a nucleus is given by~\cite{Haider:2015vea}:
\begin{equation}\label{benergy}
 |E_A|=-\frac{1}{2}\;\Big(<E_N-M>+\frac{A-2}{A-1}\;<T> \Big),
\end{equation}
with $<T>$ as the average kinetic energy, $<E_N>$ as the total nucleon energy and have been tabulated in 
Appendix~\ref{app_lda} for the nuclei for which numerical calculations have been made. Thus, by normalizing the spectral function to 
a given number of nucleons in the nucleus and getting a binding energy very close to the experimental value, no free 
parameter is left. Details are given in Ref.~\cite{FernandezdeCordoba:1991wf, Haider:2015vea}.

For a nonisoscalar nuclear target $W^{\mu \nu}_{A}$ is written in terms of the proton/neutron hole spectral 
function~($S_h^j;~j=p,n$) and the corresponding hadronic tensor~($W^{\mu \nu}_{j};~j=p,n$) is expressed as
\begin{equation}\label{conv_WAa}
W^{\mu \nu}_{A} = 2\sum_{j=p,n} \int \, d^3 r \, \int \frac{d^3 p}{(2 \pi)^3} \, 
\frac{M}{E_N ({\vec{p}})} \, \int^{\mu_j}_{- \infty} d p^0 S_h^{j} (p^0, \vec{p}, \rho(r))
W^{\mu \nu}_{j} (p, q), \,
\end{equation}
where the factor of 2 is due to the two possible projections of nucleon spin and $\mu_j;~(j=p,n)$ is the chemical potential 
for the proton/neutron. In the local density approximation, the hole spectral
functions of protons and neutrons are the function of local Fermi momentum and the equivalent normalization is written as
\begin{eqnarray} \label{norm2}
2\int\frac{d^{3}p}{(2\pi)^{3}}\int_{-\infty}^{\mu}S_{h}^{p,n}(\omega,p,p_{F_{p,n}}({\vec{r}})) d\omega= \rho_{p,n}({\vec{r}}),
\end{eqnarray}
$p_{F_{p(n)}}$ is the Fermi momentum of proton/neutron inside the nucleus which is expressed in terms of the proton/neutron 
density as discussed above. The hole spectral functions are normalized separately to the respective proton and neutron numbers 
in a nuclear target as~\cite{Zaidi:2019asc, Haider:2015vea}:
\begin{eqnarray}
\label{specp}
  2 \int d^3r\;\int \frac{d^3 p}{(2\pi)^3} \;\int_{-\infty}^{\mu_p}\;S_h^p(\omega,\vec{p},\rho_p(r))\;d\omega &=& Z\;, \\
  \label{specn}
    2 \int d^3r\;\int \frac{d^3 p}{(2\pi)^3} \;\int_{-\infty}^{\mu_n}\;S_h^n(\omega,\vec{p},\rho_n(r))\;d\omega &=& N\;.
 \end{eqnarray}
 
 The hadronic tensor ($W^{\mu \nu}_{j}$) is then written in terms of the dimensionless proton and neutron structure 
functions~($F_{ij}(x,Q^2);$ $i=1-5;~j=p,n$). By using Eq.~(\ref{conv_WAa}) and the general form of hadronic tensor 
with an appropriate choice of $x,y,z$ components, we obtain the following expressions of the dimensionless nuclear structure 
functions for a nonisoscalar nuclear target~\cite{Zaidi:2019mfd, Zaidi:2019asc, Haider:2015vea}:
  \begin{eqnarray}\label{spect_funct}
F_{iA,N}  (x_A, Q^2) &=& 4\int \, d^3 r \, \int \frac{d^3 p}{(2 \pi)^3} \, 
\frac{M}{E_N (\vec{p})} \, \int^{\mu}_{- \infty} d p^0~ S_h(p^0, \vec{p}, \rho(r))~
\times f_{iN} (x,Q^2) ,
\end{eqnarray}
where $i=1-5$ and
\begin{eqnarray}
 f_{1N}(x,Q^2)&=&AM\left[\frac{F_{1N} (x_N, Q^2)}{M} + \left(\frac{p^x}{M}\right)^2 \frac{F_{2N} (x_N, Q^2)}{\nu_N}
 \right],\\
f_{2N}(x,Q^2)&=&\left( \frac{F_{2N}(x_N,Q^2)}{M^2 \nu_N}\right)\left[ \frac{Q^4}{q^0 {(q^z)}^2}\left(p^z+\frac{q^0 
(p^0-\gamma p^z) }{Q^2}{ q^z} \right)^2+\frac{q^0 Q^2 (p^x)^2}{{(q^z)}^2}\right],
 \end{eqnarray}
 \begin{eqnarray}
f_{3N}(x,Q^2)&=&A\frac{q^0}{q^z} \;\times\left({p^0 q^z - p^z q^0  \over p \cdot q} \right)F_{3N} (x_N,Q^2),\\
f_{4N}(x,Q^2)&=&A\; \left[F_{4N}(x_N,Q^2) +\frac{p^z Q^2}{{ q^z}} \frac{F_{5N}(x,Q^2)}{M \nu_N}\right],\\
f_{5N}(x,Q^2)&=&A\;\;\frac{F_{5N}(x_N,Q^2)}{M \nu_N}\times\left[q^0(p^0-\gamma p^z)+Q^2 \frac{p^z}{{ q^z}} \right],
\end{eqnarray}
where $\nu_N=\frac{p\cdot q}{M}=\frac{p^0 q^0 - p^z q^z}{M}$, $\gamma=\frac{q^0}{q^z}$. 
Using these expressions, the effect of Fermi motion, binding energy and nucleon correlations have been included through the 
use of hole spectral function. Furthermore, bound nucleons may interact with each other via meson exchange such as 
$\pi,~\rho,$ etc., and the interaction of the intermediate vector boson 
with the mesons play an important role in the evaluation of nuclear structure functions~\cite{Zaidi:2019asc, Haider:2012nf}. 
Hence, we have incorporated these effects in the numerical calculations and discussed the mesonic contributions in the 
following Section~\ref{meson_form}.

\subsubsection{Mesonic contribution}\label{meson_form}
There are virtual mesons (mainly $\pi$ and $\rho$ mesons) associated with each nucleon bound inside the nucleus. This 
mesonic cloud gets strengthened by the strong attractive nature of the nucleon-nucleon interaction, which leads to a 
reasonably good probability of interaction of virtual bosons(IVB) with a meson instead of a nucleon~\cite{Marco:1995vb, 
Ericson:1983um, Kulagin:2004ie, LlewellynSmith:1983vzz}. It has been observed by us that the mesonic contribution, which is 
mainly dominated by the pion cloud, becomes more pronounced in the heavier nuclear targets and significant in the intermediate 
region of $x~ (0.2 < x < 0.6)$. It may be pointed out that calculations performed with only the spectral function, result 
in a reduction in the nuclear structure function from the free nucleon structure function, while the inclusion of mesonic 
cloud contribution leads to an enhancement of the nuclear structure function.
 \begin{figure} 
\begin{center}
\includegraphics[height=4. cm, width=4. cm]{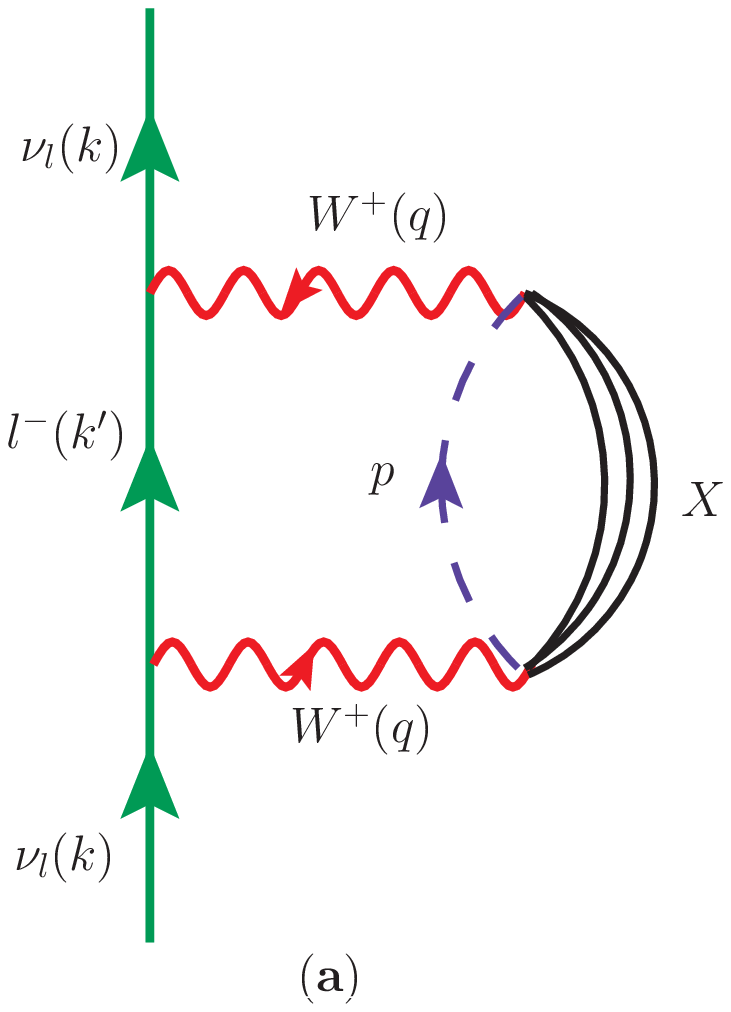}
 \includegraphics[height=4. cm, width=12 cm]{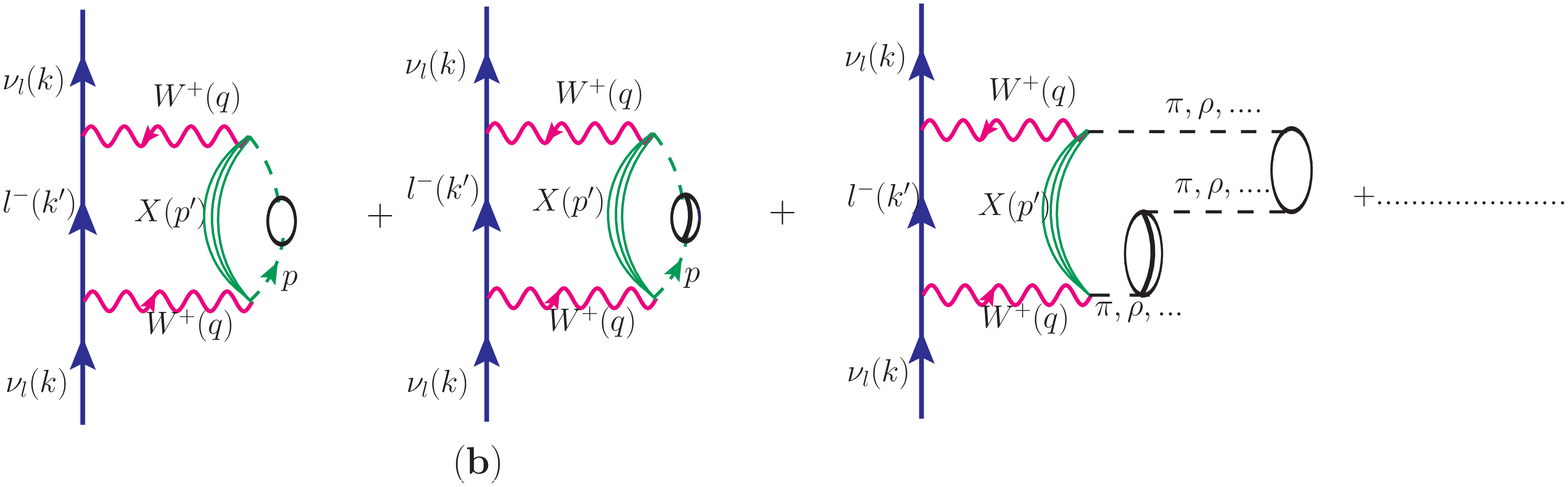}
 \end{center}
 \caption{Neutrino self-energy diagram accounting for lepton-meson DIS (a) the bound nucleon propagator is substituted 
 with a meson($\pi$ or $\rho$) propagator with momentum $p$ represented here by dashed line
 (b) by including particle-hole $(1p–1h)$, delta-hole $(1\Delta–1h)$,
 $1p1h-1\Delta1h$, etc. interactions.}
 \label{n_self-1}
\end{figure}
 
To obtain the contribution from the virtual mesons, the neutrino self energy is again evaluated using many body 
techniques~\cite{Marco:1995vb}, and to take into account mesonic effects a diagram similar to the one shown in 
Fig.~\ref{wself_energy} is drawn, except that instead of a nucleon now there is a meson which results in the change of a 
nucleon propagator by a meson propagator. 
 
The meson propagator does not corresponds to the free mesons but it corresponds to the 
mesons which arise due to NME~\cite{FernandezdeCordoba:1991wf}. In the nuclear medium these 
mesons arise through $ph$, $\Delta h$, $ph-\Delta h$, $2p-2h$, etc. interactions 
as shown in Fig.~\ref{n_self-1}. The mesonic structure functions $F_{{i A,a}}(x_a,Q^2), ~~(i=1,2,5;a=\pi,\rho)$ are obtained 
as~\cite{Marco:1995vb, Zaidi:2021iam}:
 \begin{eqnarray} 
\label{pion_f21}
F_{{i A,a}} (x_a,Q^2)  &=&  -6 \kappa \int \, d^3 r  \int \frac{d^4 p}{(2 \pi)^4} 
        \theta (p^0) ~\delta I m D_a (p) \;2m_a~\;f_{ia} (x_a), \qquad \quad \text{where} \qquad
        \end{eqnarray}
        \begin{eqnarray}
\label{F1rho_wk}
f_{1a} (x_a) &=&  A m_a \left[\frac{F_{1a} (x_a)}{m_a}~+~\frac{{|\vec{p}|^2~-~(p^{z})^2}}{2(p^z~q^z~-~p^0~q^0)}
\frac{F_{2a} (x_a)}{m_a}\right],\\
 \label{F2rho_wk}
 f_{2a} (x_a) &=&\left( \frac{F_{2a} (x_a)}{m_a^2 \nu}\right)\left[ \frac{Q^4}{q^0 {( q^z)}^2}\left(p^z+
 \frac{q^0(\gamma p^z-p^0) }{Q^2}{ q^z} \right)^2+\frac{q^0 Q^2 {(|\vec{p}|^2~-~(p^{z})^2)}}{2{( q^z)}^2}\right],\\
\label{F5rho_wk}
f_{5a} (x_a)&=&A\;\;\frac{F_{5a} (x_a)}{m_a \nu}\times\left[q^0(\gamma p^z-p^0)+Q^2 \frac{p^z}{{ q^z}} \right].
\end{eqnarray}
In Eqs.~(\ref{pion_f21}), (\ref{F1rho_wk}) and (\ref{F2rho_wk}) $\kappa=1(2)$ for pion~(rho meson), 
$\nu=\frac{q_0(\gamma p^z-p^0)}{m_a}$, $x_a=-\frac{Q^2}{2p \cdot q}$, $m_a$ is the mass of the meson~($\pi$ or $\rho$). 
$D_a(p)$ is the meson($\pi$ or $\rho$) propagator in the nuclear medium and is written as 
 \begin{equation}\label{dpi}
D_a (p) = [ p_0^2 - {\vec{p}}\,^{2} - m_a^2 - \Pi_{a} (p_0, \vec{p}) ]^{- 1}\,, \qquad \quad \text{with} \qquad 
\Pi_a(p_0, \vec{p})=\frac{f^2}{m_\pi^2}\;\frac{C_\rho\;F^2_a(p){\vec{p}}\,^{2}\Pi^*}{1-{f^2\over m_\pi^2} V'_j\Pi^*}\,.
\end{equation}
In the above expression, $f=1.01$, $C_\rho=1$ for the pion and $C_\rho=3.94$ for the rho meson. $F_a(p)={(\Lambda_a^2-M_a^2) \over 
(\Lambda_a^2 - p^2)}$ is the $\pi NN$ or $\rho NN$ form factor, $p^2=p_0^2~-~\vec{p}^2$, $\Lambda_a$=1~$GeV$(this was fixed by 
Aligarh-Valencia group to describe NME in electromagnetic nuclear structure functions to explain 
experimental data from JLab and other experiments for a wide range of nuclear targets~\cite{SajjadAthar:2009cr}). For 
pion~(rho meson), $V_j'$ is the longitudinal (transverse) part of the spin-isospin interaction and $\Pi^*$ is the irreducible 
meson self energy that contains the contribution of particle-hole and delta-hole excitations. Various quark and antiquark PDFs 
parameterizations for pions are available in the literature such as given by Conway et al.~\cite{Conway:1989fs}, Martin et 
al.~\cite{Martin:1998sq}, Sutton et al.~\cite{Sutton:1991ay}, Wijesooriya et al.~\cite{Wijesooriya:2005ir}, Gluck et 
al.~\cite{Gluck:1991ey}, etc. Aligarh-Valencia group has observed~\cite{Zaidi:2019mfd} that the choice of pionic PDF 
parameterizations would not make any significant difference in the event rates. In this work, the parameterization given by 
Gluck et al.~\cite{Gluck:1991ey} has been taken into account for pions and for the rho mesons same PDFs as for the pions have 
been used. It is important to mention that mesonic contribution does not play any role to $F_{3A}(x,Q^2)$. The reason is that 
$F_{3A}(x,Q^2)$ depends mainly on the valence quark distribution and not on the sea quarks distribution. In the evaluation of 
$F_{4A}(x,Q^2)$, the mesonic contribution has not been incorporated because the mesonic PDFs for $F_{4A}(x,Q^2)$ are not 
available in the literature and for $F_{5A}(x,Q^2)$ mesonic effect is included by using the Albright-Jarlskog 
relation~\cite{Albright:1974ts} at the leading order as the parameterization for mesonic PDFs for $F_{2A}(x,Q^2)$ is available 
in the literature.

\subsubsection{Shadowing and antishadowing effects}\label{shad_form}
The (anti)shadowing effect in the nuclear structure functions is a leading twist effect which arises due to the (constructive)~destructive 
interference of amplitudes in the multiple parton scattering processes. It is a coherent effect as it results from coherent 
scattering of hadronic fluctuations from at least two nucleons in the target nucleus. There are two broad approaches to 
understand it, one is known as Glauber-Gribov formalism~\cite{Glauber:1959, Gribov:1968jf, Gribov:1968gs}, and the other is 
known as Regge-Gribov framework~\cite{Gribov:1967vfb, Abramovsky:1973fm, Karmanov:1973va}. For shadowing the initial works 
used generalized vector dominance(GVD) model~\cite{Gribov:1968jf, Gribov:1968gs, Sakurai:1972wk, Bjorken:1971, Bramon:1972vv} 
in the Glauber-Gribov formalism, while recently color dipole model has also been used~\cite{Nikolaev:1990ja, Mueller:1993rr}. 
These coherent corrections are found to be different in the electromagnetic and weak interaction channels because the 
hadronization process of the corresponding intermediate vector bosons are different. In the literature, different approaches 
are available to incorporate these coherent corrections and discussed by Nikoleav and Zakharov~\cite{Nikolaev:1990ja}, Armesto 
et al.~\cite{Armesto:2002ny, Brodsky:2004hh}, Kopeliovich et al.~\cite{Kopeliovich:2012kw}, Kulagin and 
Petti~\cite{Kulagin:2007ju, Kulagin:2004ie}, etc. We have followed the prescription of 
Kulagin and Petti~\cite{Kulagin:2007ju, Kulagin:2004ie} who have used the formalism developed by Glauber and Gribov, and 
considers the multiple scattering of the hadronic components of the virtual photon(in electromagnetic interaction induced 
processes) or W/Z (in weak interaction induced processes) with the target nucleus. Then it considers the structure 
functions at small $x$ as a superposition of contributions from different hadronic states. In the case of (anti)neutrino 
induced DIS processes, they have treated (anti)shadowing differently from the prescription applied in the case of 
electromagnetic structure functions~\cite{Kulagin:2007ju, Kulagin:2004ie}, due to the presence of the axial-vector current in 
the neutrino interactions. The interference between the vector and the axial-vector currents introduces C-odd  terms  in  the 
neutrino   cross  sections,  which  are  described by structure function $F_{3A} (x,Q^2)$. In their calculation of 
nuclear corrections, separate contributions to different structure functions according to their C-parity have been taken 
into account. This results in a different dependence of medium effects on the nuclear structure functions depending upon 
their C-parity specially in  the  nuclear  (anti)shadowing  region~\cite{Haider:2011qs}. For example, to determine the 
nuclear structure function $F_{iA} (x,Q^2); ~(i=1-3,5)$ with the shadowing effect:
\begin{equation}
 F_{iA}^{S}(x,Q^2) = \delta R(x,Q^2) \times F_{iN} (x,Q^2)\; ,
 \label{shdw11}
\end{equation}
where $F_{iA}^{S}(x,Q^2);~(i=1-3,5)$ is the nuclear structure function with shadowing effect and the factor 
$\delta R(x,Q^2)$ is given in Ref.~\cite{Kulagin:2004ie}.
 
The expression for $F_{iA} (x,Q^2),~(i=1,2,5)$ in the full theoretical model is given by
\begin{equation}\label{sf_full}
  F_{iA} (x,Q^2)= F_{iA,N} (x,Q^2) + \;F_{iA, \pi} (x,Q^2)  + F_{iA, \rho} (x,Q^2) \;+ F_{iA}^{S}(x,Q^2)\;,
\end{equation}
where $F_{iA,N} (x,Q^2)$ is the structure function with only the hole spectral function which takes care of Fermi motion, 
binding energy and nucleon correlations. Through $F_{iA,\pi(\rho)} (x,Q^2)$ pion(rho) meson cloud contributions have been 
included and the shadowing effect is incorporated by using $F_{iA}^{WI,S}(x,Q^2)$. 
The final expression for $F_{3A}(x,Q^2)$ is given by
\begin{eqnarray}\label{f3_tot}
 F_{3A}(x,Q^2)= F_{3A,N}(x,Q^2) + F_{3A,shd}(x,Q^2).
\end{eqnarray}
In view of $F_{4N}(x,Q^2)$ being very small as it vanishes in the leading order and contributes only due to higher order
corrections we have evaluated $F_{4A}(x,Q^2)$ using only the spectral function and therefore write
\begin{equation}\label{f4_tot}
 F_{4A}(x,Q^2)=F_{4A,N}(x,Q^2).
\end{equation}
For $F_{5A}(x,Q^2)$, the Albright-Jarlskog relation is used.
  \begin{table}[h]
\begin{center}
\begin{tabular}{llllll}
\hline
Phenomenological group & data type used\\
 EKS98~\cite{Eskola:1998df,Eskola:1998iy}  &$l$+$A$ DIS, p+$A$ DY \\
 HKM~\cite{Hirai:2001np}                 & $l$+$A$ DIS \\
 HKN04~\cite{Hirai:2004wq}	& $l$+$A$ DIS, p+$A$ DY \\ 
 nDS~\cite{deFlorian:2003qf} 	& $l$+$A$ DIS, $p$+$A$ DY\\
 EKPS~\cite{Eskola:2007my}	 & $l$+$A$ DIS, p+$A$ DY \\
 HKN07~\cite{Hirai:2007sx} 	& $l$+$A$ DIS, p+$A$ DY \\
 EPS08~\cite{Eskola:2007my}	& $l$+$A$ DIS, p+$A$ DY, $h^{\pm},\pi^0,\pi^{\pm}$ in d+Au \\
EPS09~\cite{Eskola:2009uj} & $l$+$A$ DIS, p+$A$ DY, $\pi^0$ in d+Au \\
EPPS16~\cite{Eskola:2016oht}&$l$+$A$ and $\nu$+$A$ DIS, p+$A$ and $\pi-A$ DY, $d-A$, and LHC proton-lead collisions data\\
nCTEQ~\cite{Schienbein:2009kk, Stavreva:2010mw} & $l$+$A$ DIS, p+$A$ DY\\
nCTEQ~\cite{Kovarik:2010uv}	& $l$+$A$ and $\nu$+$A$ DIS, p+$A$ DY \\
DSSZ~\cite{deFlorian:2011fp}& $l$+$A$ and $\nu$+$A$ DIS, p+$A$ DY,$\pi^0,\pi^{\pm}$ in d+Au, computed with nFFs \\
TUJU21~\cite{Helenius:2021tof}&$l$+$A$ and $\nu$+$A$ DIS, data for $W ^\pm$ and $Z^0$ boson production in $p+Pb$\\
KSASG20~\cite{Khanpour:2020zyu}& $l$+$A$ and $\nu$+$A$ DIS, p+$A$ DY\\
nNNPDF2.0~\cite{AbdulKhalek:2020yuc}& NC DIS and CC $\nu_l-A$ DIS, data for $W ^\pm$ and $Z^0$ boson production in $p+Pb$\\
\hline
\end{tabular}
\end{center}
\vspace{-0.5cm}
\caption{The developments in the global DGLAP analysis of nPDFs since 1998.  DY = Drell-Yan 
dilepton production; nFFS = nuclear fragmentation functions~\cite{Eskola:2012rg}.}
\label{tab2:nPDFs}
\end{table}
\subsubsection{Phenomenological approach to understand NME in DIS}
The phenomenological studies of nuclear parton distribution functions~(nPDFs)
are broadly based on the analysis of experimental data on charged lepton-nucleus DIS, 
Drell-Yan processes with $\pi$ and $p$ and neutrino-nucleus DIS, etc. Several studies have been made to understand
nPDFs~\cite{Eskola:1998df, Eskola:1998iy, Hirai:2001np, Hirai:2004wq, 
deFlorian:2003qf, Eskola:2007my, Hirai:2007sx, Eskola:2009uj, Eskola:2016oht, Schienbein:2009kk, Stavreva:2010mw, 
Kovarik:2010uv, deFlorian:2011fp, Helenius:2021tof, Khanpour:2020zyu, AbdulKhalek:2020yuc, Eskola:2012rg, Kovarik:2015cma, 
Owens:2007kp, Pumplin:2002vw, Stump:2003yu}. In these studies, the approaches which have been used are the following:
\begin{itemize}
 \item In the first approach used by Eskola et al.~\cite{Eskola:1998df, Eskola:1998iy, Eskola:2009uj}, Hirai et 
 al.~\cite{Hirai:2004wq} and de Florian and Sassot~\cite{deFlorian:2003qf} mainly the charged lepton-nucleus and Drell-Yan 
 proton-nucleus scattering data (for detail see Table~\ref{tab2:nPDFs}) have been used. In this approach a set of free nucleon 
 PDFs given by any standard parameterization available in literature is taken to calculate the free proton ($f_i^p(x, Q_0)$) 
 and free neutron ($f_i^n(x, Q_0)$) structure functions. Then using some global fitting techniques the nuclear correction 
 factors are found and that is multiplied with the free nucleon PDFs to provide agreement with the nuclear experimental data. 
 The free nucleon PDFs multiplied with this nuclear correction factor $R_i^A(x_i,Q_0)$ give nuclear PDFs $F_2^A$ and $F_3^A$, 
 i.e.,
 \begin{equation}
 F_{2,3}^A(x,Q) = R_i(x,Q,A) F_{2,3}^N (x,Q).\nonumber 
 \end{equation}
 In an analysis de Florian et al.~\cite{deFlorian:2011fp} also included $\nu$-$A$ DIS data in their analysis along with 
 $l$-$A$, $p$-$A$, $d$-$A$ data and reported that there is no conflict between the nuclear modification of the $l^{\pm}$-$A$ 
 DIS and $\nu$-$A$ DIS data.
 
 \item  In the second approach the nCTEQ group~\cite{Kovarik:2010uv} have obtained $F_2^A$  and $F_3^A$ by analyzing charged 
 lepton-A DIS data and DY $p$-$A$ data sets, and $\nu(\bar \nu)$-A DIS data sets separately. In the nCTEQ 
 framework~\cite{Kovarik:2015cma}, the parton distributions of the nucleus are constructed as:
 \begin{equation}
 F_{i}^{(A,Z)}(x,Q) = \frac{Z}{A}F_{i}^{p/A}(x,Q) + \frac{A-Z}{A}F_{i}^{n/A}(x,Q),
 \label{eq:nucleusPDF}
 \end{equation} 
 Isospin symmetry is used to construct the PDFs of a neutron in the nucleus, $F_{i}^{n/A}(x,Q)$, by exchanging up- and 
 down-quark distributions from those of the proton. The observation of nCTEQ group from this analysis is that $F_2^A$ in 
 electromagnetic interaction is different in nature than $F_2^A$ in weak interaction in some regions of Bjorken $x$. Thus the 
 results in these two approaches are not in complete agreement with each other~\cite{deFlorian:2011fp, Kovarik:2010uv}.
\end{itemize}

  \begin{figure} 
\begin{center}
\includegraphics[height=4.1 cm, width=6 cm]{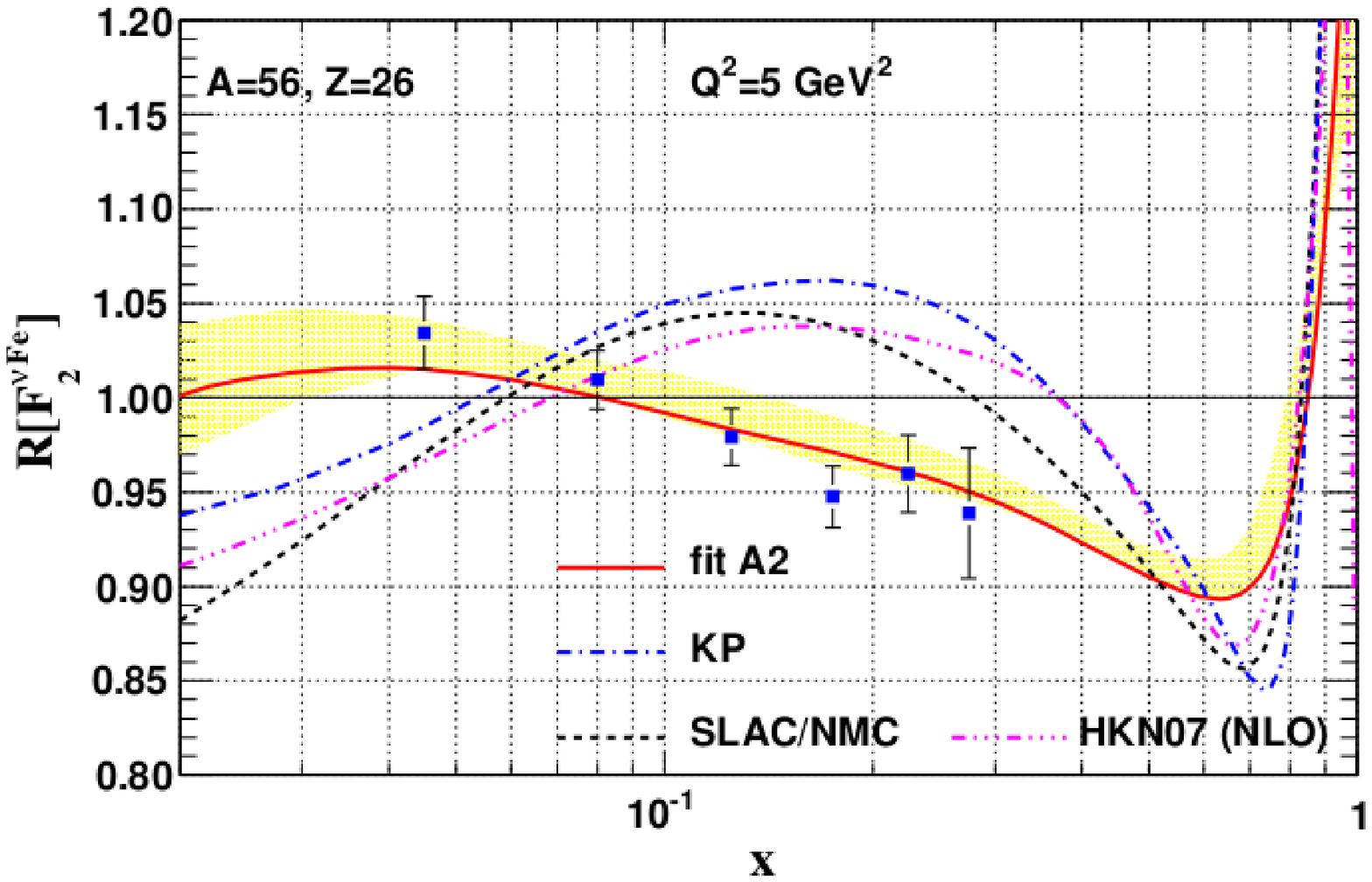}
\includegraphics[height=4.1 cm, width=6 cm]{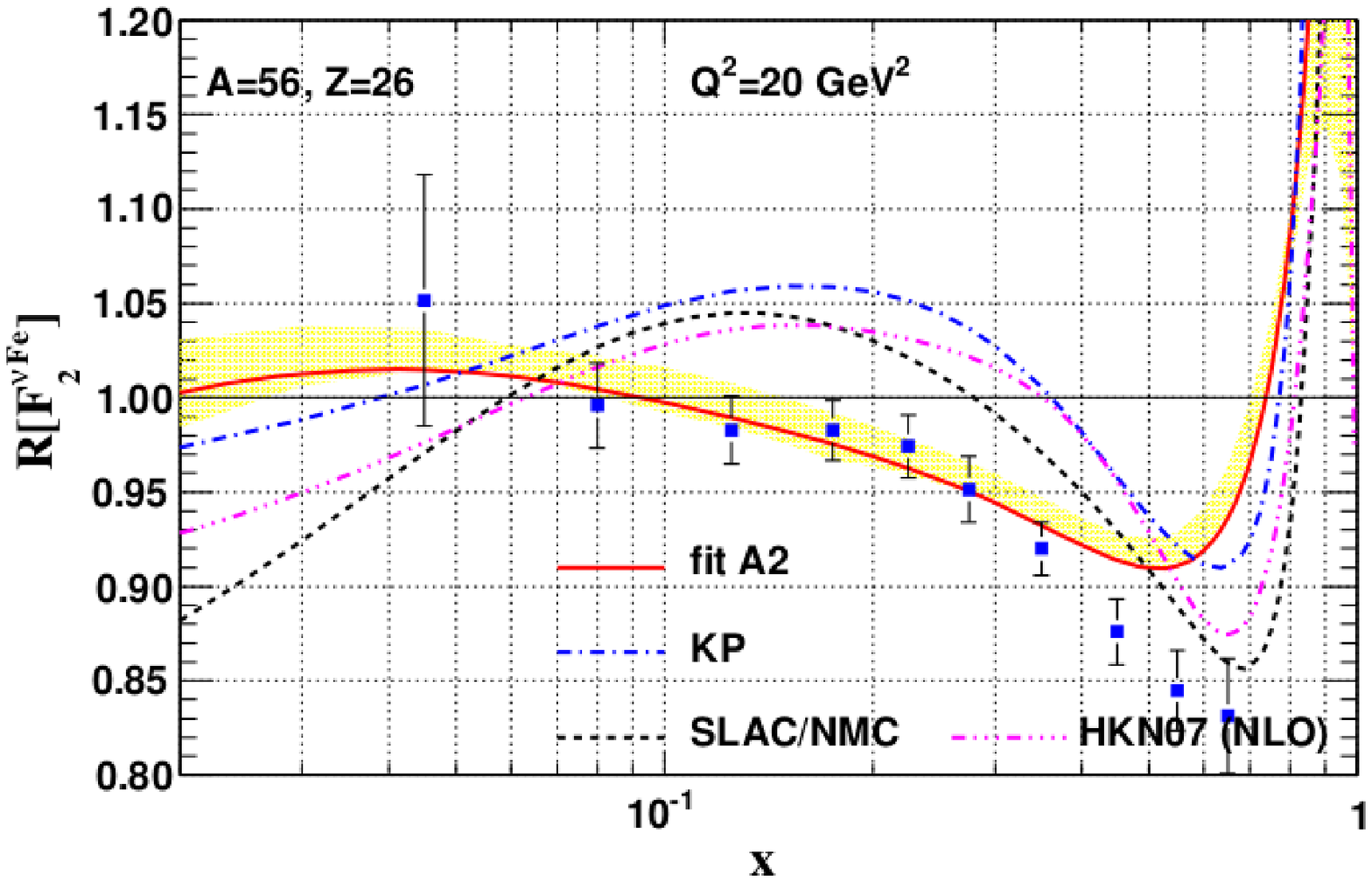}\\
\includegraphics[height=4.1 cm, width=6 cm]{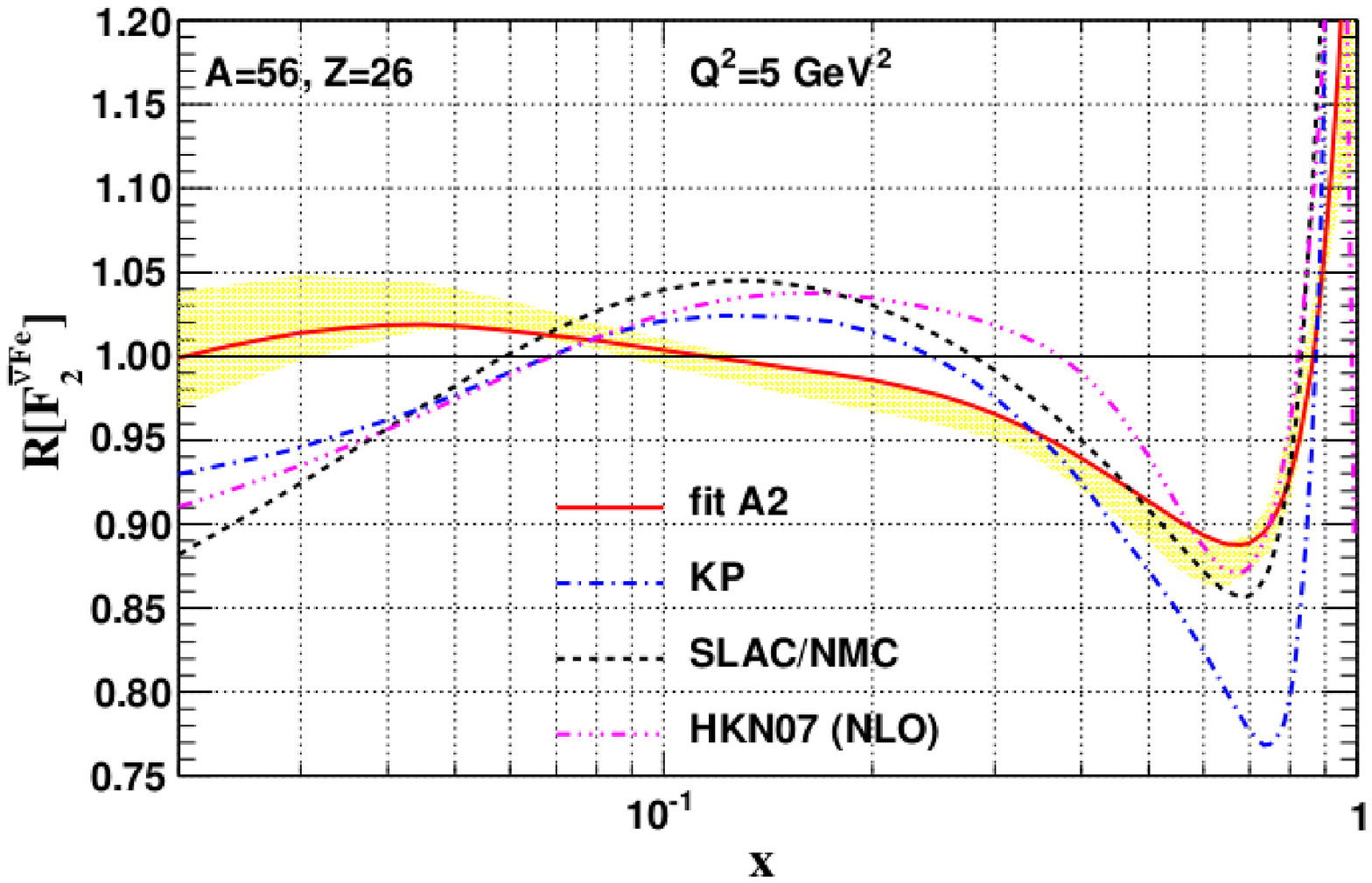}
\includegraphics[height=4.1 cm, width=6 cm]{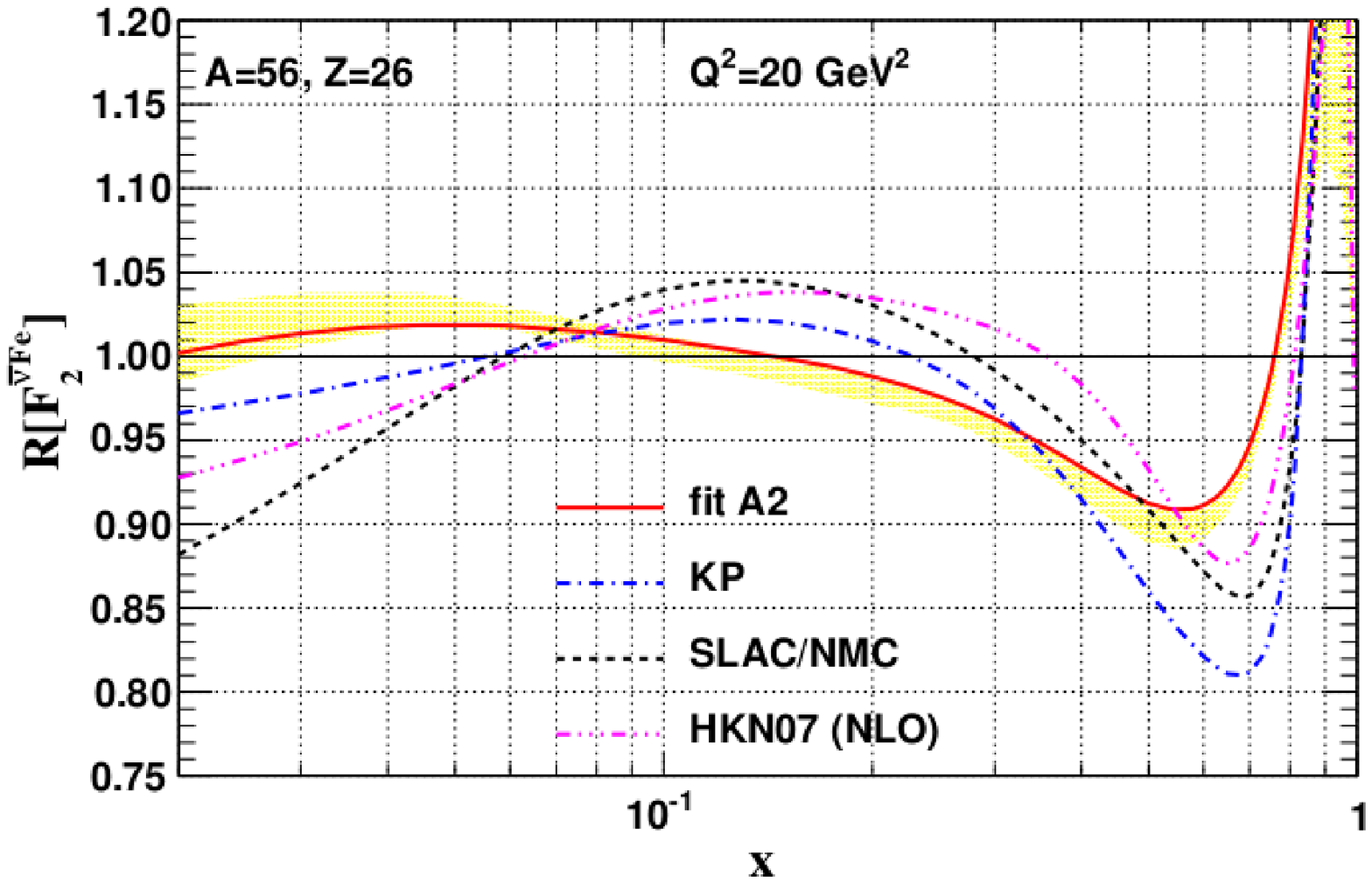}
\end{center}
\caption{ 
Nuclear correction factor $R$ for the structure function $F_2$ in CC $\nu Fe$ scattering at  a)~$Q^2=5~GeV^2$ and 
b)~$Q^2=20~GeV^2$.  The  solid curve shows the result of the nCTEQ analysis of NuTeV differential cross sections (labeled fit 
A2), divided by the results obtained with the reference fit (free-proton) PDFs;  the uncertainty from the A2 fit is 
represented by the yellow band. Plotted also are NuTeV data points of the average $F_2$ to illustrate the consistency of the 
fit with the input points. For comparison the correction factor from the Kulagin--Petti (KP) model 
\protect\cite{Kulagin:2006dg}~(dashed-dotted line), Hirai et al. fit~\cite{Hirai:2007sx}~(double-dashed-dotted line), and the 
SLAC/NMC parameterization (dashed line) of the charged-lepton nuclear 
correction factor are also shown. We compute this for $\{A=56,Z=26\}$. Figure has been taken from~\cite{Schienbein:2009kk}.}
 \label{fig:fig6a}
\end{figure}

In Fig.~\ref{fig:fig6a}, taken from Ref.~\cite{Schienbein:2009kk}, the results for the nuclear correction factors for 
$F_2^{\nu A}$ and $F_2^{\bar\nu A}$ are shown at $Q^2=5$ and 20 GeV$^2$ obtained from the various groups, such as 
Hirai et al.~\cite{Hirai:2004wq, Hirai:2007sx} who have used phenomenological analysis of experimental data from 
lepton-nucleus and Drell-Yan experiments, the results of Kulagin and Petti~\cite{Kulagin:2004ie, Kulagin:2007ju, 
Kulagin:2006dg} obtained in a theoretical model using spectral function, the 
SLAC/NMC~\cite{Kopeliovich:2012kw} curve obtained from an $A$ and $Q^2$ independent parameterization of calcium and iron 
charged lepton DIS data~\cite{Kopeliovich:2012kw}, and the phenomenological analysis of CTEQ group~\cite{Kovarik:2010uv}. 
From the figure, it may be noticed that the nuclear correction factor has large variation and the present phenomenological 
results do differ among themselves and particularly from the analysis of CTEQ group~\cite{Kovarik:2010uv}. It may be noticed 
that the results of CTEQ analysis do not show shadowing at low $Q^2$, while the correction factor shows the antishadowing 
shifts towards lower values of $x$. Furthermore, CTEQ results of the nuclear correction factor has also been found to be 
smaller than those obtained from charged lepton nucleus scattering data as well as obtained in the theoretical study of 
Kulagin and Petti~\cite{Kulagin:2007ju, Kulagin:2004ie}. 
\begin{figure}[ht]
  \begin{center}
  \includegraphics[height=5.cm, width=0.95\textwidth]{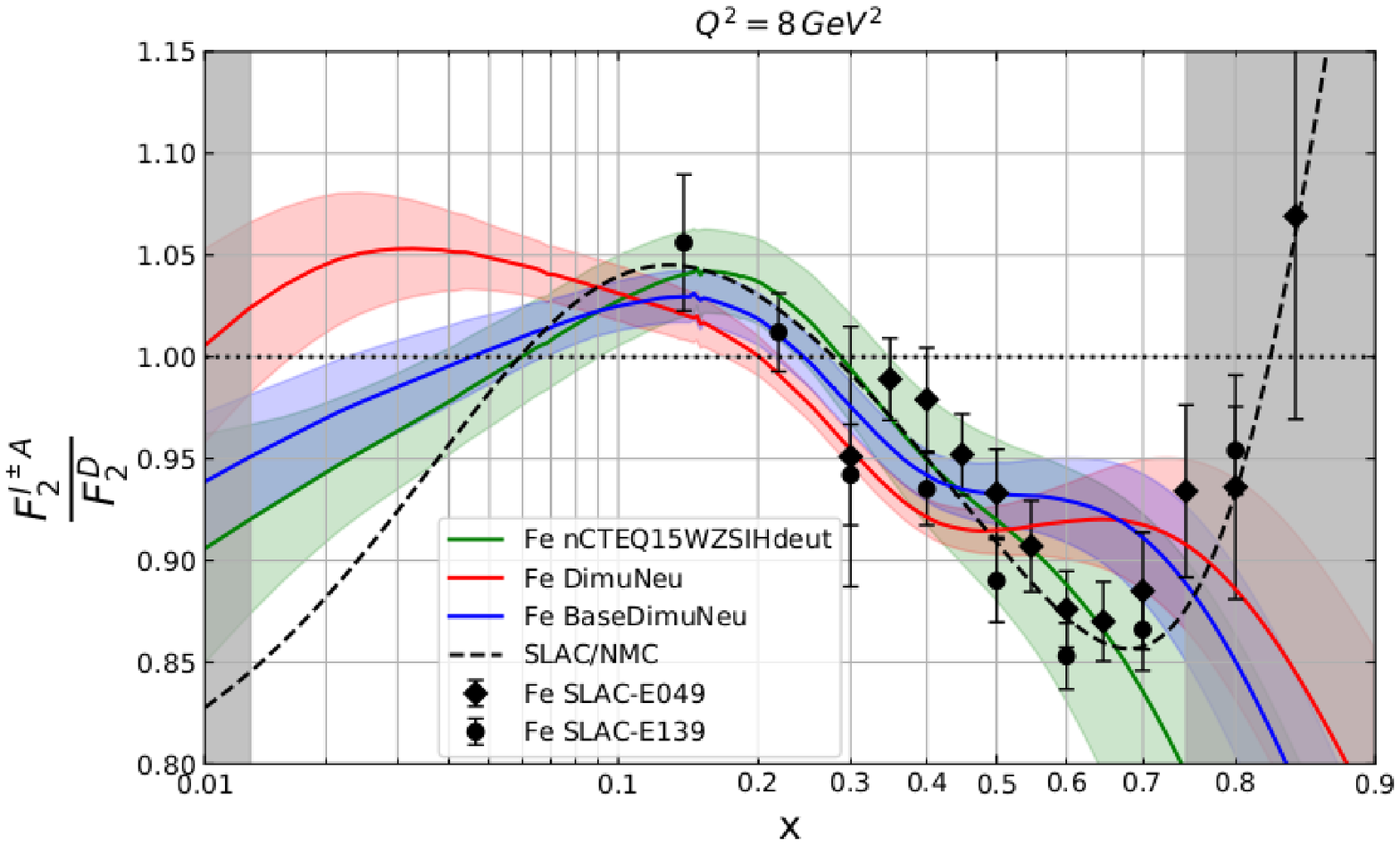}\\
  \includegraphics[height=5.cm, width=0.9\textwidth]{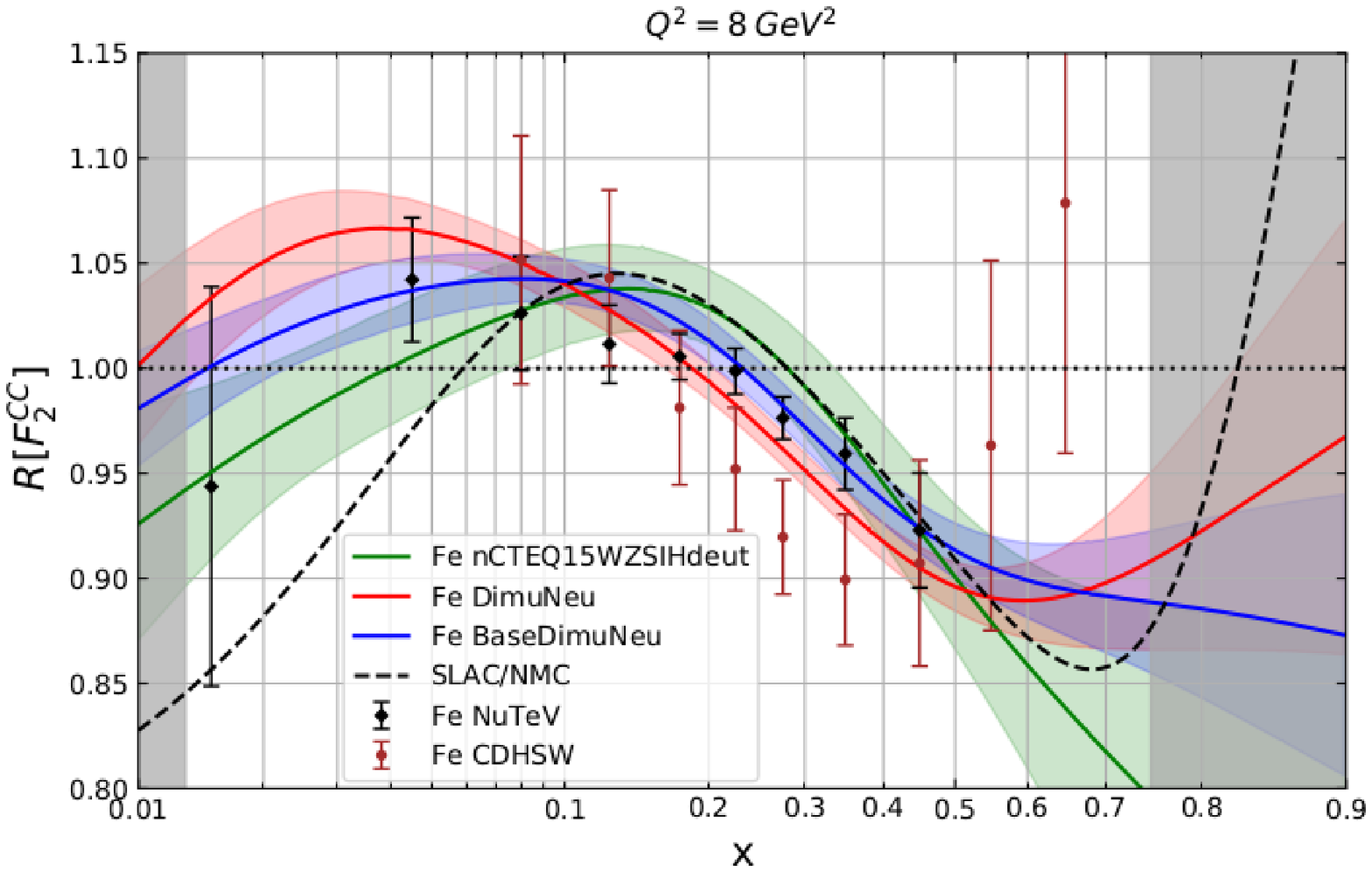}\\
  \includegraphics[height=5.cm, width=0.9\textwidth]{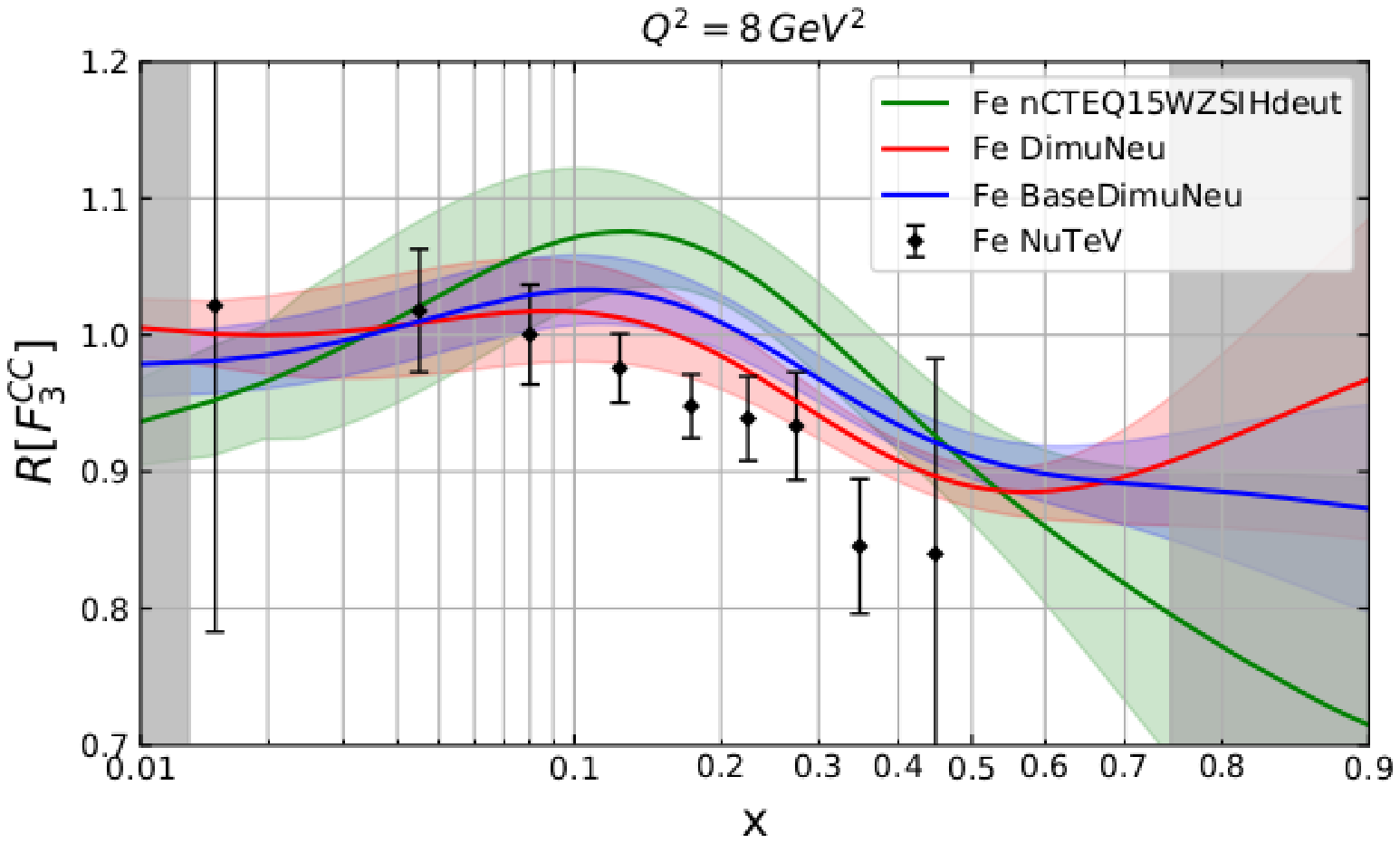}
  \end{center}
\caption{$R^{l^\pm A}=\frac{F_2^{l^\pm A}}{F_2^D}$ and $R^{(\nu A+\bar\nu A)}=\frac{F_i^{(\nu A+\bar\nu A) }}{f_i^p+f_i^n};~
(i=2,3)$  using the fitted nuclear PDFs, i.e., nCTEQ15WZSIHdeut~\cite{Muzakka:2022wey}. $F_2^D$ includes the deuteron 
correction factor and $f_i^p$ and $f_i^n$ are the free proton and neutron structure functions. This figure is taken from  
Ref.~\cite{Muzakka:2022wey}~(courtesy Jorge G. Morfin).}\label{fig:mz}
\end{figure}
Recently, using the nCTEQ framework, this group has included more neutrino data and made a comparative analysis of $l^{\pm}-A$ 
and $\nu_l/\bar\nu_l-A$ DIS cross sections and obtained the nuclear correction factor~\cite{Muzakka:2022wey}. In their 
work~\cite{Muzakka:2022wey}, the global analysis used to extract the nuclear PDFs is based on the nCTEQ15WZSIH analysis which 
incorporates the charged lepton DIS data, LHC $W$ and $Z$ boson production data and single inclusive hadron production data 
from RHIC and LHC. Furthermore, nCTEQ collaboration~\cite{Muzakka:2022wey} has also incorporated the deuteron corrections by 
using the CJ15 analysis while determining the fitted nPDFs which are labeled as nCTEQ15WZSIHdeut. In order to study the 
compatibility between the $l^{\pm}-A$ and $\nu_l/\bar\nu_l-A$ data they have compared the results of nuclear structure 
function ratios obtained by using the nCTEQ15WZSIHdeut analysis to the results obtained by using the
\begin{itemize}
 \item DimuNeu analysis, based on inclusive and semi-inclusive neutrino data only from CDHSW, CCFR, NuTeV and CHORUS 
 experiments.
 
 \item BaseDimuNeu analysis, based on inclusive neutrino and charged lepton data along with the other data
sets incorporated in nCTEQ15WZSIHdeut analysis.
\end{itemize}
In Fig.~\ref{fig:mz}, the ratios are presented for $\frac{F_2^{l^\pm A}}{F_2^D}$ (top panel), $R[F_2]=\frac{F_2^{(\nu A + 
\bar\nu A) }}{f_2^p + f_2^n}$ (middle panel) and $R[F_3]=\frac{F_3^{(\nu A+\bar\nu A) }}{f_3^p + f_3^n}$ (bottom panel) at 
$Q^2=8$ GeV$^2$, where $A=^{56}$Fe is the nuclear target and $f_i^{p/n};~(i=2,3)$ are the free proton/neutron structure 
functions. These authors~\cite{Muzakka:2022wey} included the deuteron corrections in $F_2^D$, while obtaining the ratio 
for $\frac{F_2^{l^\pm A}}{F_2^D}$, however, these corrections are absent in the case of $R[F_2]$ and $R[F_3]$. It may be observed from 
the figure that the predictions of the nCTEQ15WZSIHdeut and the DimuNeu analyses are incompatible with each other. The 
predictions from BaseDimuNeu for $l^{\pm}-A$ are compatible up to $1\sigma$ level with the results from nCTEQ15WZSIHdeut, 
however it is incompatible for $\nu_l/\bar\nu_l-A$ at $x=0.025$. It is important to notice that the tension in the case of 
(anti)neutrino observables is larger as compared to the charged lepton case. One may notice similar observations 
while comparing the predictions from the BaseDimuNeu and DimuNeu analyses. These results are also compared with the 
corresponding experimental data from SLAC, NuTeV and CDHSW. The tension between $l^{\pm}-A$ and $\nu_l/\bar\nu_l-A$ data sets 
observed from Fig.~\ref{fig:mz} requires more study to understand NME in neutrino and 
antineutrino reactions on nuclear targets.

\subsubsection{Isoscalarity correction}\label{isoscalarity_form}
For heavy nuclear targets, where the number of neutrons is greater than the number of protons $N>Z$, the isoscalarity 
corrections become important. Hence, it is required to study the effect of the corrections arising due to neutron excess on 
nuclear structure functions for a given nuclear target by treating it to be isoscalar ($N=Z$) as well as nonisoscalar 
($N\ne Z$). These corrections are phenomenologically taken into account by multiplying the results by a correction factor 
$R_{A}^{Iso}$ defined as
\begin{eqnarray}\label{isocorr_em}
R_{A}^{Iso} &=& \frac{[{\sigma^{\nu/\bar\nu p}+\sigma^{\nu/\bar\nu n}] /2}}{{[Z \sigma^{\nu/\bar\nu p} + (A-Z) \sigma^{\nu/
\bar\nu n}]/ A}} =  \frac{[{F_2^{\nu/\bar\nu p}+F_2^{\nu/\bar\nu n}] /2}}{{[Z F_2^{\nu/\bar\nu p} + (A-Z) F_2^{\nu/\bar\nu n}]/ 
A}},
\end{eqnarray}
where $F_2^{\nu/\bar\nu \;\;{p, n}}$ and $\sigma^{{\nu/\bar\nu} \;\;{p, n}}$ are, respectively, the weak structure functions and scattering 
cross sections for proton and neutron.

While in the Aligarh-Valencia model isoscalarity corrections have been taken in an entirely different way by separately 
normalizing the hole spectral function for the proton (Eq.~(\ref{specp})) and neutron (Eq.~(\ref{specn})) numbers for a 
nonisoscalar nuclear target, and to the nucleon numbers for an isoscalar nuclear target (Eq.~(\ref{norm5})) as discussed in 
Sec.~\ref{formalism-dis-a} and it has been observed that these two different prescriptions (Eq.~(\ref{isocorr_em}), and 
using Eqs.~(\ref{specp}) and (\ref{specn})) give different isoscalarity correction effect, which has been discussed in Ref~\cite{Athar:2020kqn}. 
 \begin{figure} 
\begin{center}
 \includegraphics[height= 7 cm , width= 0.95\textwidth]{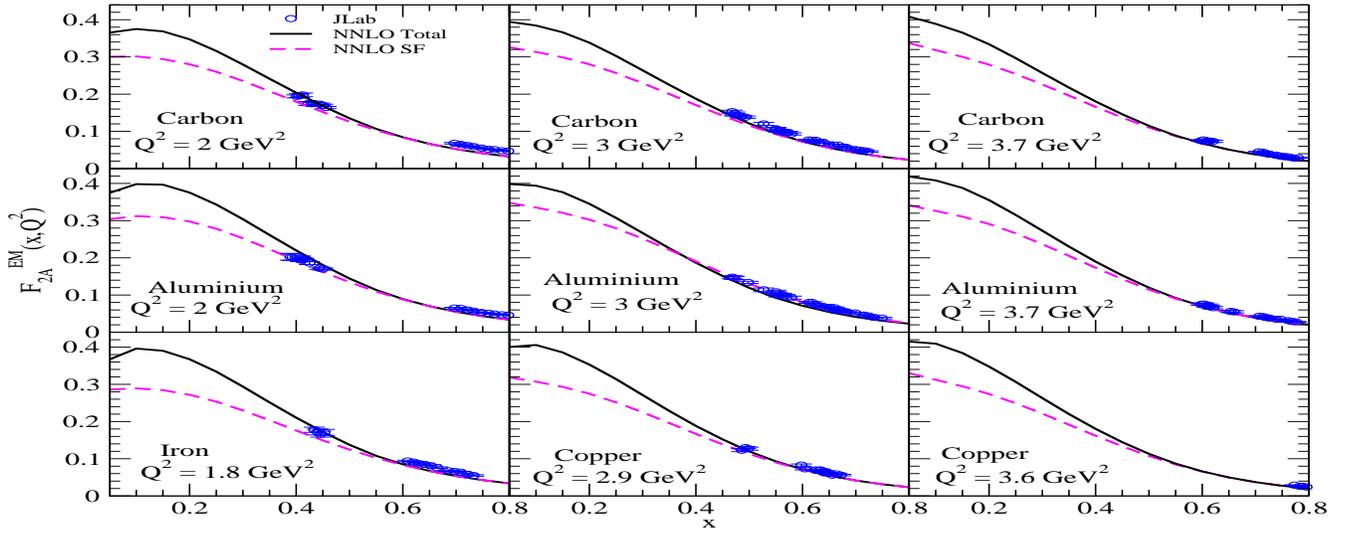}
\end{center}
\caption{$ F_{2A}^{EM} (x,Q^2)$ ($A = ^{12}C$, $^{27}Al$, $^{56}Fe$ and $^{63}Cu$) vs $x$ are shown at different $Q^2$. 
 The results are obtained for the spectral function only (dashed line) and for the full model (solid line) with TMC effect 
 using MMHT nucleonic PDFs at NNLO. 
 The results are compared with the experimental data of JLab~\cite{Mamyan:2012th} (empty circles). For the numerical 
 calculations MMHT PDFs parameterization~\cite{Harland-Lang:2014zoa} has been used.}
\label{prc:fig2}
\end{figure}

\subsubsection{Results and discussion}\label{res}
Before applying the formalism to understand NME in weak interaction induced processes, Aligarh-Valencia 
group applied their formalism to the electromagnetic interaction induced processes~\cite{Zaidi:2019mfd, Haider:2016zrk, 
Haider:2015vea}.

The numerical results are presented for the nuclear structure functions $F_{2A}^{EM}(x,Q^2)$, $F_{2A}^{WI}(x,Q^2)$ and 
$F_{3A}^{WI}(x,Q^2)$ calculated in the Aligarh-Valencia model and compared with the experimental results. In case of 
$F_{2A}^{EM}(x,Q^2)$ where ample data are available in several nuclear targets for a wide range of $x$ and $Q^2$, the results 
are presented in Fig.~\ref{prc:fig2} for some of the nuclear targets like $^{12}$C, $^{27}$Al, $^{56}$Fe and $^{63}$Cu which 
are obtained using the MMHT nucleonic PDFs at NNLO by incorporating the target mass corrections as discussed in 
Section~\ref{dis:nucleon}. These results for $F_{2A}^{EM}(x,Q^2)$ vs $x$, at the different values of $Q^2$~($\approx 
2-4~GeV^2$) are compared with the experimental observations of JLab~\cite{Mamyan:2012th}. The theoretical results are 
presented for the spectral function only (SF) and for the full model (total) which includes the shadowing and mesonic cloud 
contributions. The results obtained in the full model get enhanced due to mesonic effect which is significant in the region of 
low and intermediate $x$. For example, at $Q^2\simeq3$ GeV$^2$ this enhancement in carbon is found to be $23\%$ at $x=0.1$, 
$21\%$ at $x=0.2$ and $10\%$ at $x=0.4$ while in copper it becomes $32\%$ at $x=0.1$, $\sim 29\%$ at $x=0.2$ and $13\%$ at 
$x=0.4$. It may be noticed that the mesonic cloud contributions decreases with the increase in $x$. However, it becomes more 
pronounced for the heavier nuclear targets. It may be noticed from the figure that our theoretical results show a good 
agreement with the experimental data~\cite{Mamyan:2012th} in the region of intermediate $x$, however, for $x>0.6$ and $Q^2 
\approx 2~GeV^2$ they slightly underestimate the experimental results. One should remember that the region of high $x$ and 
low $Q^2$ is the transition region of nucleon resonance excitations and DIS, therefore, the theoretical results are expected 
to underestimate the experimental data as the theoretical results do not include the resonance contribution. However, with the 
increase in $Q^2$, theoretical results show better agreement with the experimental observations of JLab~\cite{Mamyan:2012th} 
in the entire range of $x$.
\begin{figure}[h] 
\begin{center} 
     \includegraphics[height=0.3\textheight,width=0.95\textwidth]{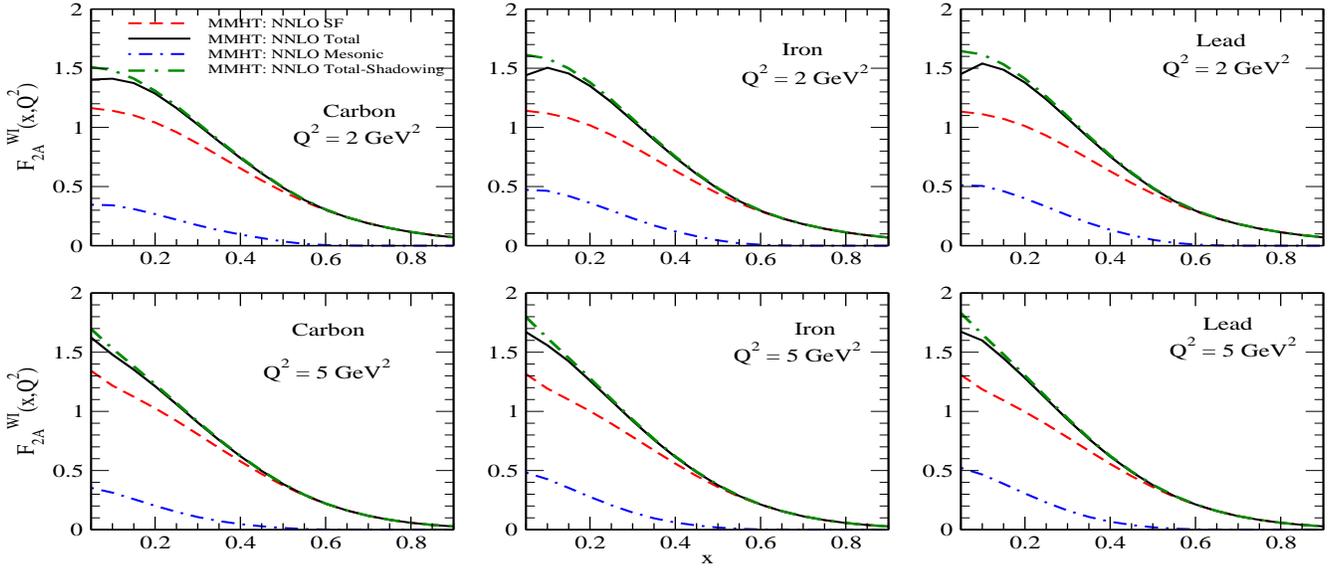}
 \caption{Results are shown for the weak nuclear structure function $F_{2A}^{WI}(x,Q^2)$ vs $x$ at $Q^2=2,~5~GeV^2$, in 
 $^{12}$C, $^{56}$Fe and $^{208}$Pb for {\bf (i)} only the spectral function (dashed line), {\bf (ii)} only the mesonic 
 contribution (dash-dotted line) using Eq.~(\ref{pion_f21}), {\bf (iii)} the full calculation (solid line) using 
 Eq.~(\ref{sf_full}) and {\bf (iv)} the double dash-dotted line is the result without the shadowing and antishadowing 
 effects. The numerical calculations have been performed at NNLO using the MMHT~\cite{Harland-Lang:2014zoa} nucleon PDFs 
 parameterizations.}\label{figurewk}
 \end{center}
\end{figure} 
\begin{figure}[h]
 \begin{center}
\includegraphics[height=0.25\textheight,width=0.75\textwidth]{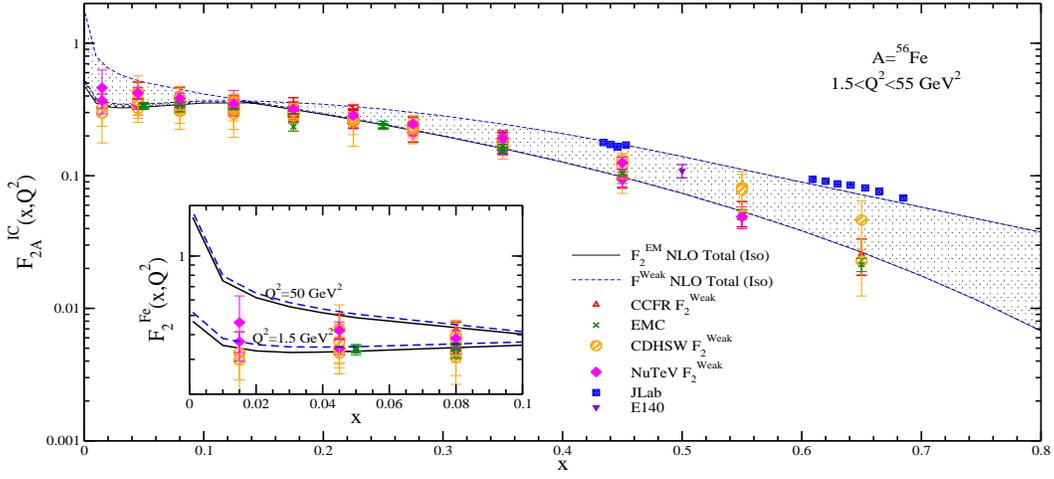}
\end{center}
\caption{Results of EM and Weak nuclear structure functions in $^{56}Fe$ obtained with the full theoretical model 
at NLO are shown. The results of $F_2^{Weak}$ are scaled by a factor of $\frac{5}{18}$.
The results are also compared with the experimental data of Refs.~\cite{Oltman:1992pq, Berge:1989hr, NuTeV:2005wsg, 
Mamyan:2012th, EuropeanMuon:1986xsr}. For the numerical calculations CTEQ6.6 PDFs 
 parameterization~\cite{Nadolsky:2008zw} has been used.}
\label{fig3}
\end{figure}

In Fig.~\ref{figurewk}, the results are presented for $F_{2A}^{WI}(x,Q^2)$ vs $x$ for $^{12}$C, $^{56}$Fe and $^{208}$Pb, for the 
isoscalar nuclear targets,  at the different values of $Q^2$ relevant for the current neutrino experiments. From the figure, 
it may be observed that the mesonic contributions result in an enhancement in the nuclear structure functions and is 
significant in the low and intermediate region of $x$. Moreover, the effect is more pronounced at low $Q^2$, which becomes 
larger with the increase in mass number $A$. For example, in comparison to the total contributions (solid line) in  carbon, 
the mesonic contribution at $x=0.1$  is found to be $24\%$ in iron which increases to $33\%$ in lead. With the increase in 
$x$~(say $x=0.4$), the enhancement due to the mesonic contributions reduces to $13\%$ and $18\%$ respectively and becomes almost negligible for $x\ge 0.6$ at 
$Q^2=2~GeV^2$. To depict the coherent nuclear effects(shadowing) which results in suppression of the structure functions at 
low $x$, the results without shadowing are shown with the double-dash-dotted line, and with the 
increase in mass number of the nuclear target($^{56}Fe$ vs $^{208}Pb$), the strength of suppression becomes larger in the 
region of low $x$.

In Fig.~\ref{fig3}, we present a comparison of the results for the electromagnetic ($F_{2A}^{EM}(x,Q^2)$) and weak 
(($F_{2A}^{Weak}(x,Q^2)$)) nuclear structure functions vs $x$ in iron nucleus for a wide range of $Q^2$ viz. $1.5 < Q^2 < 
55$~GeV$^2$ using the full model at NLO. The numerical calculations are performed using the CTEQ6.6 nucleonic PDFs 
parameterization~\cite{Nadolsky:2008zw}. The theoretical results for weak nuclear structure functions (dashed line) are scaled 
by a factor of $\frac{5}{18}$~(see Eq.~(\ref{emwc})) in order to make it comparable with the results from the electromagnetic 
interaction channel (solid line). These results are also compared with the available charged lepton-nucleus scattering data 
from the JLab~\cite{Mamyan:2012th}, EMC~\cite{EuropeanMuon:1986xsr} and SLAC~\cite{Gomez:1993ri, Hen:2013oha, Whitlow:1990dr, 
Dasu:1993vk} experiments as well as with the (anti)neutrino-nucleus data from CDHSW~\cite{Berge:1989hr}, 
CCFR~\cite{Oltman:1992pq} and NuTeV~\cite{NuTeV:2005wsg} collaborations. The theoretical results of nuclear structure 
functions in Fig.~\ref{fig3} are shown by the band for the above mentioned range of $Q^2$. We have observed that the present 
results are consistent with the CCFR~\cite{Oltman:1992pq}, JLab~\cite{Mamyan:2012th}, NuTeV~\cite{NuTeV:2005wsg} data at 
medium and high values of $Q^2$ but not in good agreement at low $Q^2$ with the older experiments like 
CDHSW~\cite{Berge:1989hr} and EMC~\cite{EuropeanMuon:1986xsr}. In the inset of the figure, we have shown the curves for the 
two different values of $Q^2$ viz. $Q^2=1.5$ GeV$^2$ and $Q^2=50$ GeV$^2$ up to $x\le0.1$. One may observe that at low $x$, EM 
nuclear structure function is slightly lower than the weak nuclear structure function which is about $\sim 4\%$ in iron at 
$x=0.1$, and the difference decreases with the increase in $x$ and almost becomes negligible for $x~>~0.3$. We find that~(not 
shown here) with the increase in mass number this difference increases. For example, in $^{208}Pb$ it becomes $\sim 7\%$ while 
for low mass nuclei like carbon this difference decreases to $\sim 1-2\%$ at $x=0.1$.

 \begin{figure}
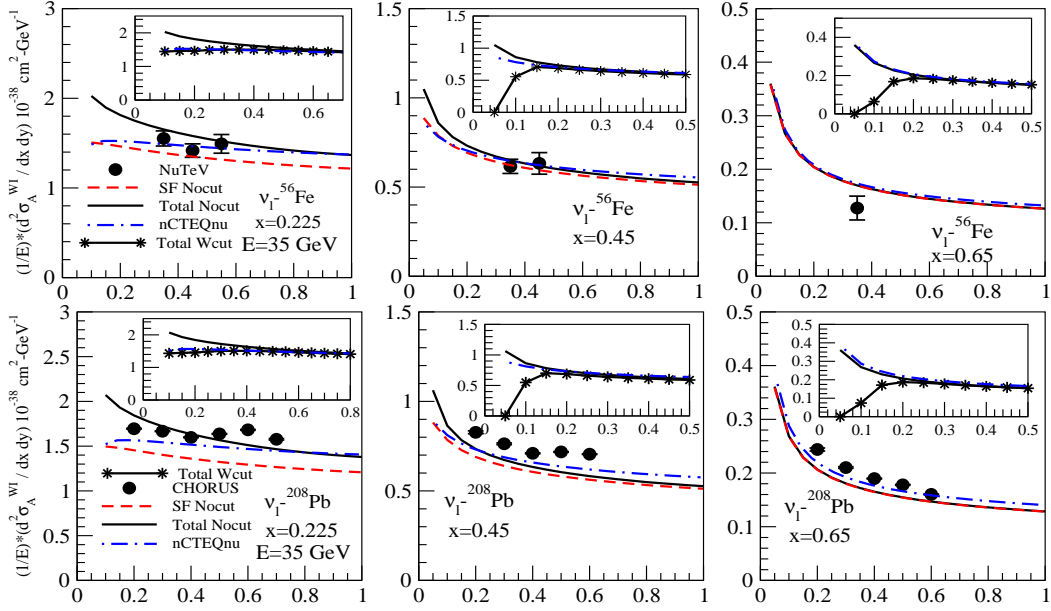
 
\begin{center}
\includegraphics[width=0.75\textwidth,height=4cm]{d2sigma_35gev_iron_iso_v3.eps}\\
\includegraphics[width=0.75\textwidth,height=4cm]{d2sigma_35gev_lead_iso_v3.eps}
\caption{Differential cross section vs $y$, for the different values of $x$ for $\nu_\mu-Fe$~(top row)
and $\nu_\mu-Pb$~(bottom row) DIS.  
Theoretical predictions are shown with the spectral function only (dashed line) and with the full model (solid line) at NNLO. 
In the inset the effects of an additional kinematical cut of W $\ge$ 2 GeV (solid line with star) for the full theoretical 
model are shown. The blue dash-dotted line in the top row is the result from nCTEQnu nPDFs with $Q^2 \ge 1.0 ~GeV^2$. Solid 
circles with error bars are the limited experimental data points of NuTeV at this lower energy.}
\label{fig:35GeVFe-nu&nub}
\end{center}
\end{figure}

In Figs.~\ref{fig:35GeVFe-nu&nub}, the theoretical predictions for $\nu-$Fe and $\nu-$Pb deep inelastic differential 
scattering cross sections of the Aligarh-Valencia group as well as the predictions using the phenomenological approach of 
nCTEQnu nuclear PDFs are presented at E$_\nu$ = 35 GeV relevant for the MINERvA experiment. The results of Aligarh-Valencia
group are shown using only the spectral function as well as using the full model (Eqs.~(\ref{sf_full}) and (\ref{f3_tot})). 
It can be observed that the mesonic contributions play important role in the region of $x \le 0.5$. Comparing the two 
approaches for $\nu_l$ scattering, the nCTEQnu-based results are somewhat lower than the theoretical prediction from 
Aligarh-Valencia group at the low $x$, while the results of the two approaches are in reasonable agreement with each 
other in the region of higher $x$. Similar observations have been made for antineutrino induced scattering on these nuclear 
targets which have been discussed in Ref.~\cite{SajjadAthar:2020nvy}.

\begin{figure}
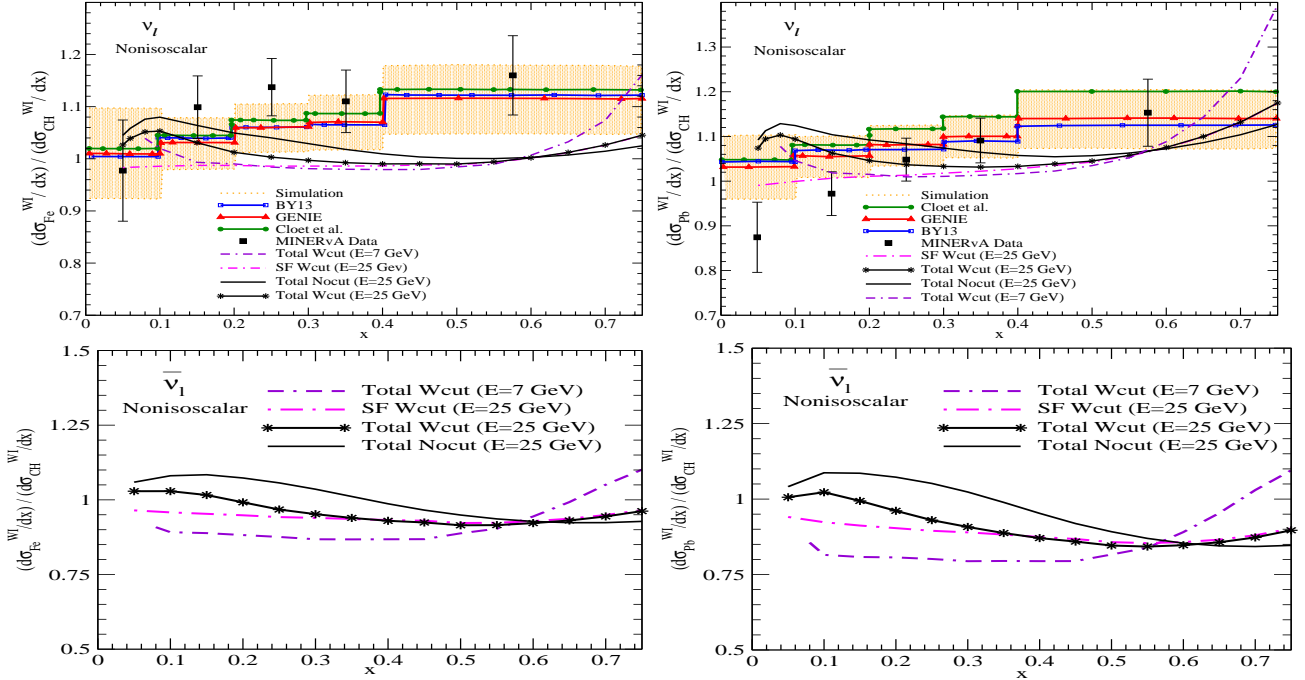
 
\begin{center}
  \includegraphics[height= 4.5 cm , width= 0.45\textwidth]{dsigma_fetoch.eps}
  \includegraphics[height= 4.5 cm , width= 0.45\textwidth]{dsigma_pbtoch.eps}\\
  \includegraphics[height= 4.5 cm , width= 0.93\textwidth]{dsigma_nubar_25gev_forppnp.eps}
\end{center}
 \caption{$\frac{d\sigma^{WI}_{A}/dx}{d\sigma^{WI}_{CH}/dx}~(A=$ $^{56}$Fe, $^{208}$Pb) vs $x$ for incoming neutrino (top 
 panel) and antineutrino (bottom panel) beam of energies $E=7$ GeV and $25$ GeV. The numerical results are obtained with the 
 spectral function only (dash-dotted line: $E=25$ GeV) as well as with the full model (solid line: $E=25$ GeV, solid line with 
 star: $E=25$ GeV and double dash-dotted line: $E=7$ GeV) at NNLO and are compared with the phenomenological results of Cloet 
 et al.~\cite{Cloet:2006bq}, Bodek-Yang~\cite{Bodek:2010km}, GENIE Monte Carlo~\cite{Andreopoulos:2009rq} and with the 
 simulated results~\cite{Mousseau:2016snl}. The solid squares are the experimental points of MINERvA~\cite{Mousseau:2016snl}.
 The results are shown for the nonisoscalar nuclear targets.}\label{fig6}
\end{figure}

The MINERvA experiment at the Fermilab is studying NME in several nuclear targets like carbon, hydrocarbon, 
water, iron and lead for a wide range of Bjorken $x$ and $Q^2$, and have recently presented the results for the ratio of 
differential cross section in several nuclear targets~\cite{Mousseau:2016snl}. In Fig.~\ref{fig6}, we show the 
results for the ratio ($\frac{d\sigma^{WI}_{A}/dx}{d\sigma^{WI}_{CH}/dx}$) vs $x$ in the case of $\nu_l(\bar\nu_l)-A$ DIS 
scattering for $^{56}Fe$ and $^{208}Pb$ which are summarized below:
\begin{itemize}
 \item The deviation of the ratio from unity is significant in iron as well as in lead although it is comparatively smaller 
 in $ \frac{d\sigma^{WI}_{Fe}/dx}{d\sigma^{WI}_{CH}/dx}$ than in $\frac{d\sigma^{WI}_{Pb}/dx}{d\sigma^{WI}_{CH}/dx}$. This 
 reflects the $A$ dependence of NME in which the contribution of mesons increases with $A$ at low and 
 intermediate $x$. For example, at $E=25$ GeV the contribution of mesons is found to be $9\%(\sim 7\%)$ at $x=0.1$ and 
 $1\%(1\%)$ at $x=0.3$ in lead(iron). It is important to notice that even for high energy neutrino beams the effect of nuclear 
 medium on the differential scattering cross section are significant.
  
 \item To quantify the mass dependence, the difference between the results of $\frac{d\sigma^{WI}_{Fe}/dx}{d\sigma^{WI}_{CH}/
 dx}$ and $\frac{d\sigma^{WI}_{Pb}/dx}{d\sigma^{WI}_{CH}/dx}$ obtained by using the full model at $E=25$ GeV (solid line) has 
 been found to be $\simeq 5\%$ at $x=0.05$, $4\%$ at $x=0.1$ and $\sim 7\%$ at $x=0.6$. 
 
 \item  The isoscalarity correction is found to be significant. For example, it has been found that at $E= 25$ GeV, this 
 effect is $2\%(5\%)$, and $5\%(13\%)$ at $x=0.3$ and $0.7$, respectively, in iron(lead) when no kinematical cut is applied on 
 $W$.
  \begin{figure}[h]
    \centering
    \includegraphics[width=0.95\textwidth,height=0.3\textheight]{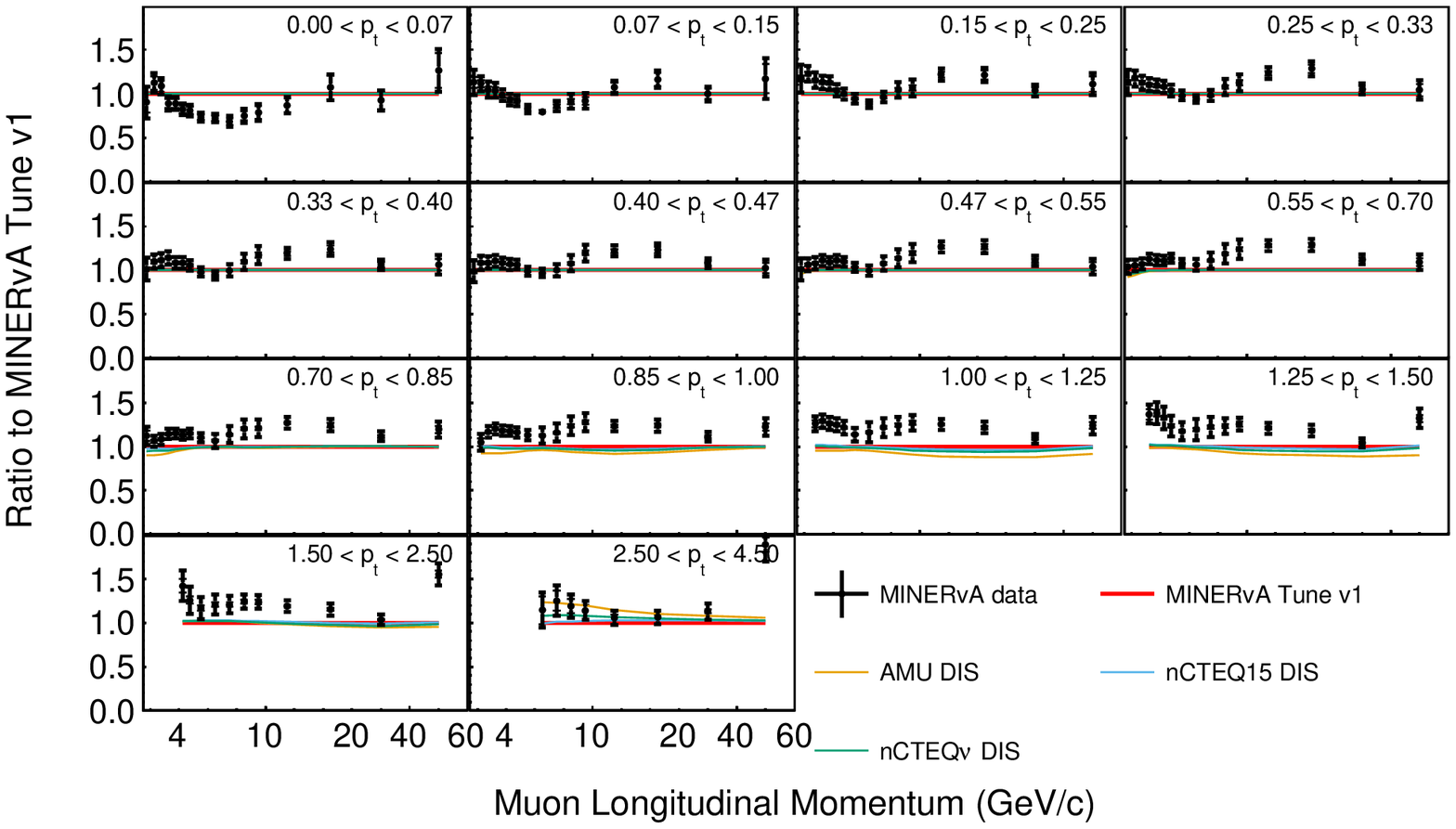}
    \includegraphics[width=0.95\textwidth,height=0.3\textheight]{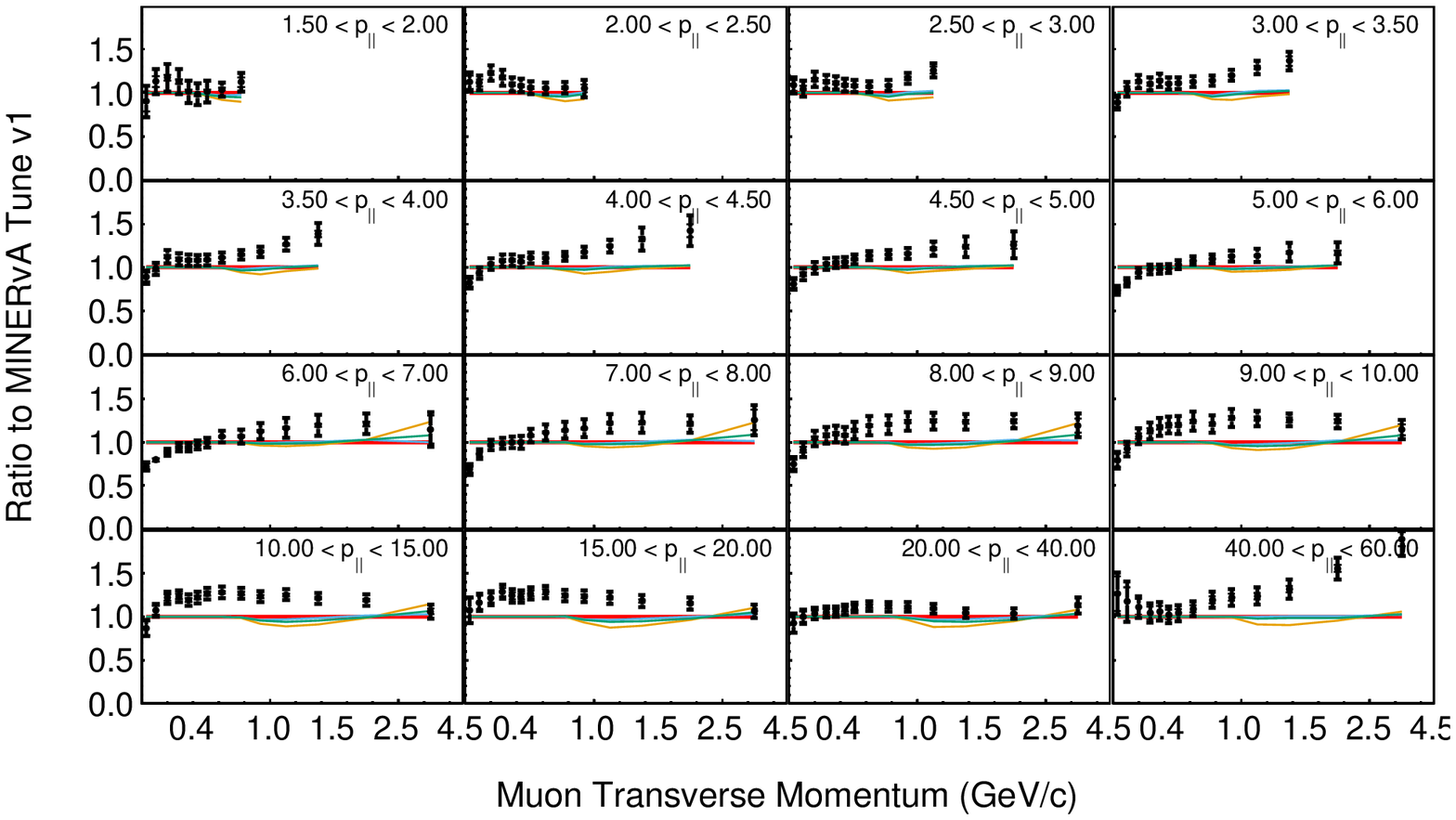}
    \caption{The scattering cross section ratio from different DIS models to MINERvA's simulated 
    results~\cite{MINERvA:2021owq}.}\label{fig:set1comp}
\end{figure}

In a recent MINERvA analyses for $\nu_\mu$ induced CC inclusive scattering process off hydrocarbon target,
the results for the differential cross sections as a function of lepton kinematics like the longitudinal and transverse 
momenta of muons have been reported by Filkins et al.~\cite{MINERvA:2020zzv} at $\langle E_\nu \rangle=3.5$ GeV and by 
Ruterbories et al.~\cite{MINERvA:2021owq} at $\langle E_\nu \rangle\sim 6$ GeV, where they have compared the results from 
MINERvA experiment with the results from the theoretical model developed by Aligarh-Valencia group (labeled as AMU 
DIS)~\cite{Zaidi:2021iam, Zaidi:2019asc, Haider:2016zrk} and also with the phenomenological results of 
nCTEQ15~\cite{Kovarik:2015cma}, Cloet 
 et al.~\cite{Cloet:2006bq} and nCTEQnu~\cite{Schienbein:2007fs}. Here we will only discuss the results of 
Ref.~\cite{MINERvA:2021owq} for the ratio of DIS cross sections from different DIS models employing the kinematic 
constrain $W\ge$~2.0 GeV/$c^2$ and $Q^{2}\ge$~1.0 GeV$^{2}$/c$^{4}$ to the MINERvA's simulated results (labeled as MINERvA 
tune v1)~\cite{MINERvA:2021owq} which are shown in Fig.~\ref{fig:set1comp}. From the figure, it may be noticed that although 
the theoretical predictions~\cite{Haider:2016zrk} and the phenomenological results~\cite{Kovarik:2015cma, Schienbein:2007fs} 
show reasonable agreement among themselves but none of them are able to completely explain the MINERvA's experimental data in 
the entire range of the charged lepton momentum.

 \item The energy dependence of the scattering cross section has also been shown by comparing the numerical results which are 
 obtained by using the full model with $Q^2>1$ GeV$^2$ and $W>2$ GeV at $E=25$ GeV (solid line with star) with the 
 corresponding results obtained at $E=7$ GeV (double dash-dotted line). It may be observed that in the region of low and 
 intermediate $x$ the results for the ratio of $ \frac{d\sigma^{WI}_{A}/dx}{d\sigma^{WI}_{CH}/dx}$ at $E=7$ GeV are smaller in 
 magnitude from the results at $E=25$ GeV, however, with the increase in $x$ the ratio $ \frac{d\sigma^{WI}_{A}/dx}{d 
 \sigma^{WI}_{CH}/dx}$ obtained for $E=7$ GeV increases. While moving towards $E=25$ GeV from $E=7$ GeV, we have observed that 
 there is a difference of about $3\%(\sim 5\%)$, $\sim 2\%(2 \%)$ and $10\%(16 \%)$ at $x=0.1$, $x=0.3$ and $x=0.75$, 
 respectively in iron(lead).
 
 \item It may be noticed that MINERvA's experimental data have large error bars which is due to the statistical uncertainties
 and the wide band around the simulation results is due to the systematic errors which shows an uncertainty up to $\sim 
 20\%$~\cite{Mousseau:2016snl}. The present numerical results have been compared with the MINERvA's experimental data, 
 simulated results as well as with the results of Cloet et al.~\cite{Cloet:2006bq} (solid line with circle), Bodek et 
 al.~\cite{Bodek:2010km} (solid line with square) and the GENIE MC generator~\cite{Andreopoulos:2009rq} (solid line with 
 triangle). It may be observed that neither the theoretical predictions nor the phenomenological results are able to 
 satisfactorily describe the observed ratios of the differential cross sections in the entire region of Bjorken $x$. 
 
 \item The NME in $\frac{d\sigma^{WI}_{A}}{dx}$ for $\bar\nu_l-A$ scattering (Fig.~\ref{fig6}, bottom 
 panel) are found to be qualitatively similar to $\nu_l-A$ scattering when no cut on CM energy is applied, however, 
 quantitatively they are different specially at low and mid values of $x$. For example, at $E=7$ GeV the enhancement in the 
 cross section when full calculation is applied vs. using the cross section results with spectral function only is about 24$\%$ at 
 $x=0.25$ in $\nu_l-^{208}Pb$ scattering, while it is 65$\%$ in ${\bar\nu}_l-^{208}Pb$ scattering, and the difference in the 
 two results~(full calculation vs. SF only) decreases with the increase in $x$. At $E=25$ GeV the enhancement in the cross section is about 20$\%$ at 
 $x=0.25$ in $\nu_l-^{208}Pb$ scattering, while it is $\sim$45$\%$ in ${\bar\nu}_l-^{208}Pb$ scattering.
 
 \item When a cut of 2 GeV is applied on the CM energy $W$, then a suppression in the region of low and mid $x$ is observed 
 in the $\bar\nu_l-A$ differential cross section, resulting in a lesser enhancement due to mesonic effects. For example, at 
 $E=25$ GeV, the enhancement due to the mesonic contributions becomes $\sim$18$\%$ (vs 20\% without cut) in ${\nu}_l-
 ^{208}$Pb scattering while $\sim 28\%$ (vs 45\% without cut) in ${\bar\nu}_l-^{208}Pb$ scattering at $x=0.25$. At $E=7$ GeV, 
 with a cut of 2 GeV on $W$, the enhancement is about 2$\%$ at $x=0.25$ in $\nu_l-^{208}Pb$ scattering, while there is 
 reduction in $\bar\nu_l-A$  scattering, implying small contribution from the mesonic part. This reduction in $\frac{d
 \sigma^{WI}_{A}}{dx}$ for $\bar\nu_l-A$ scattering is about 15$\%$ in a wide region of $x$($\le$ 0.6).
\end{itemize} 
  
\section{Quark-hadron duality in $\nu_l$ scattering}\label{QH:duality}
QCD is the theory which describes strong interactions in terms of quarks and gluons with remarkable 
features of asymptotic freedom at high energies~($E$) and $Q^2$ and confinement at low energies and 
$Q^2$. At low $E$ and $Q^2$, the effective degrees of freedom to describe strong interactions are the mesons and 
nucleons using effective Lagrangian motivated by the symmetry properties of QCD while at high $E$ and $Q^2$, the quark and gluon 
degrees of freedom are used to describe the strong interactions using perturbative QCD. In the case of lepton scattering 
processes induced by charged leptons and (anti)neutrinos on nucleons and nuclei, the inclusive cross sections at low energy 
are dominated by the QE (CC induced) and elastic (NC induced) scattering processes but as the energy increases 
beyond 1 GeV, the inclusive cross sections are expressed in terms of the structure functions corresponding to the excitation of 
various resonances like $\Delta,~N^\ast$, etc., lying in the first or higher resonance regions depending on $W$ of the final hadrons. 
On the other hand, at high energy and $Q^2$,  the inclusive cross sections are 
expressed in terms of the structure functions corresponding to DIS processes. In the 
intermediate energy region corresponding to the transition between resonance excitations and DIS, we are yet to find a method 
best suited to describe the inclusive lepton scattering. 

The phenomenon of Quark-Hadron~(QH) duality in electron scattering from proton was first observed at SLAC  by Bloom and 
Gilman~\cite{Bloom:1970xb, Bloom:1971ye} more than fifty years ago, and provides a connection between the low energy and the 
high energy description of electron-proton scattering. They found a connection between the structure function $\nu\;W_2(\nu, 
Q^2)$ in the nucleon resonance region and that in the deep inelastic continuum. QH duality states that the structure functions 
in low $Q^2$ region of resonance excitation suitably averaged over an energy interval is the same as the structure function at 
high $Q^2$ region corresponding to the DIS in the same energy interval. At the basic level it may be understood as the 
equality between two integrals
\begin{equation}
 \int_{\xi_{min}}^{\xi_{max}} d\xi {\it F}^{\text{Res}} (\xi,Q^2_{\text{Res}}) \approx \int_{\xi_{min}}^{\xi_{max}} d\xi 
 {\it F}^{\text{DIS}}(\xi,Q^2_{\text{DIS}}), 
\end{equation}
where ${\it F}^{\text{Res}}$ and ${\it F}^{\text{DIS}}$ represent the structure functions in the resonance and the DIS regions, respectively,  
and $\xi$ is the Nachtmann variable and its minimum and maximum values depend upon the choice of $W$. Generally the minimum 
value of $W$ is taken to be pion production threshold i.e. $W_{min}=M + m_\pi$, while the maximum value of $W$ may vary from 
1.6GeV to 2.2GeV.  

\begin{figure}[h]
\begin{center}
\includegraphics[height=6cm,width=10cm]{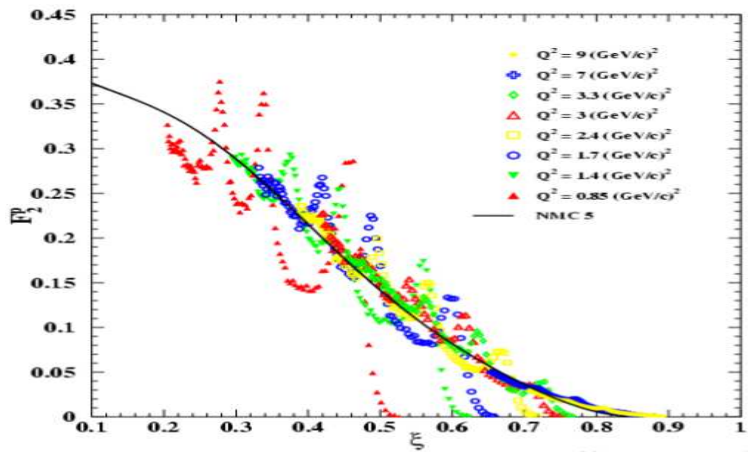}
\caption{Comparison of $F_2^p$ from the series of resonances measured by E94-110~\cite{JeffersonLabE00-115:2009jll} vs the 
Nachtmann variable $\xi$ at the indicated $Q^2$ compared to the extrapolated DIS measurement from the NMC 
collaboration at 5 $GeV^2$. Figure has been taken from Ref.~\cite{Niculescu:2000tk}.}\label{eN_duality94110}
\end{center}
\end{figure}

Therefore, the QH-duality implies that the average over resonance structure functions produced in the inclusive eN scattering has 
a close resemblance with the scaling structure functions measured in the DIS region and with the increase in $Q^2$ 
the average over resonance structure functions approaches the asymptotic scaling structure functions. In other words, it 
establishes a connection between the quark-gluon description of $e-p$ scattering at high $Q^2$ in the region of DIS with the 
hadronic description of the same phenomenon at low $Q^2$ in the region of resonance excitations. This seems to be valid in 
each resonance region individually as well as in the entire resonance region when the structure functions are summed over 
higher resonances. This is termed as local duality. When the phenomenon of the local QH duality is also observed in the case 
of higher moments of structure functions in electron-nucleus scattering, it is termed as the global duality, which was 
observed in early Jefferson Lab measurement~(E94-110)~\cite{JeffersonLabE00-115:2009jll} for $Q^2 \ge 0.5 ~GeV^2$, as can be 
seen in Fig.~\ref{eN_duality94110}, with resonances following the extrapolated DIS curve i.e. the DIS scaling curve 
extrapolated down into the resonance region passes through the average of the peaks and valleys of the resonance structure. 
This implies a connection between the behavior of the resonance excitations and DIS which ultimately signals that there is 
perhaps a common origin in terms of a point-like structure for both resonance and DIS interactions.

It is also observed that the ratio of resonance peak to background remains almost constant as $Q^2$ is varied from low to high 
$Q^2$. It is conjectured that there may exist two component duality where the resonance contribution and the background 
contribution to the structure functions in the resonance excitation region corresponds respectively to the valence quarks and 
sea quarks contribution in structure functions in the DIS region. This has been tested at JLab~\cite{Niculescu:2000tk} for 
$e-N$ interaction where it has been found that $F_2$ structure function averaged over resonances at low values of $\xi$ 
behaves like the valence quark contribution obtained from the DIS with scaling. However, these observations are to be verified 
by model calculations as well as by the further experimental data when they become available with higher precision. The 
phenomenon of local duality and global duality has been observed in JLab, NMC and other experiments on electron nucleon and 
electron nucleus scattering~\cite{Melnitchouk:2005zr}.

\begin{figure}[h]
\begin{center}
\includegraphics[height=6.5cm,width=10cm]{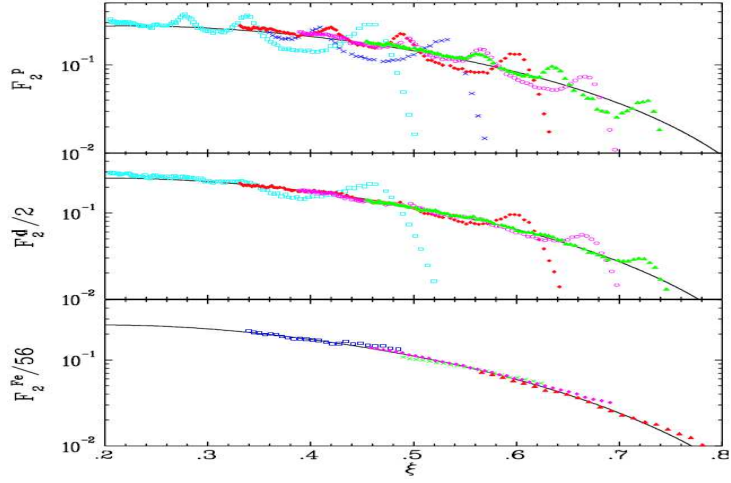}
\caption{ Comparison of $F_2$ structure function per nucleon for proton, deuteron and iron targets. The curves are GRV 
parameterization~\cite{Gluck:1998xa} at $Q^2=1$ GeV$^2$, corrected for the nuclear EMC effect. Figure has been taken  
from Ref.~\cite{Melnitchouk:2005zr}.}\label{fig:duality}
\end{center}
\end{figure}

Melnitchouk et al.~\cite{Melnitchouk:2005zr} have complied and analyzed the experimental data of the structure function $F_2(x)$ 
above $W^2=1.2$ GeV$^2$ for hydrogen, deuterium and iron targets for $0.5\le Q^2\le 7$ GeV$^2$ which are shown in 
Fig.~\ref{fig:duality}. The solid line is the curves obtained by them, which represents $F_2(x)$ scaling curve for the 
nucleon that is corrected for the known nuclear medium modifications to the structure function and have been obtained using 
the GRV PDFs parameterization~\cite{Gluck:1998xa}. They observed that in the case of proton, the resonance structure is 
clearly visible and $F_2(x)$ is seen to oscillate around the scaling curve, however, for the deuterium, the resonances become less 
pronounced, and in iron further diminishing. In the middle panel of the figure, the prominent peak for the deuterium data is 
identified as the contribution from the $\Delta$ resonance and follows the scaling curve similar to those observed for the proton. 
However, the other resonance peaks are smeared (deuteron vs proton) so much as to be indistinguishable from the scaling 
structure function. Moreover, in iron it has been observed that even the QE peak is washed out by the smearing at 
higher $Q^2$, and scaling is seen at all values of $\xi$ and the resonance region becomes indistinguishable from the scaling
regime.

Presently the different experimental observations obtained from the charged-lepton scattering results in the following features of 
the QH duality~\cite{Melnitchouk:2005zr, Lalakulich:2009zza}:
 \begin{itemize}
 \item the resonance excitation structure functions data oscillate around the scaling DIS structure functions data  
 \item the resonance excitation structure functions data are on an average equivalent to the DIS structure functions data  
 \item the resonance excitation structure functions data move towards the DIS structure functions data with the increase in 
 $Q^2$.
 \end{itemize}

More experimental data with high precision are needed even for the inclusive electron-nucleon and electron-nucleus scattering, for 
a wide range of $\xi$ and $Q^2$ and using the different nuclear targets, before the validity of QH-duality can be established 
conclusively in $e-N$ scattering. For sure with its verification, the description of lepton-nucleon and lepton-nucleus 
scattering over the entire SIS region will become much easier.

The (anti)neutrino-nucleus scattering in the energy region of SIS is currently an important topic to be explored both 
theoretically and experimentally as the accelerator and atmospheric neutrino experiments have significant contribution from 
the few GeV energy region corresponding to the kinematic region of resonance excitations and DIS. The 
validity of QH duality in CC and NC sectors of weak interaction may provide a way to obtain (anti)neutrino-nucleon 
and (anti)neutrino-nucleus scattering cross sections in the transition region where either the use of effective Lagrangian or 
the quark-parton description is not adequate. More importantly, if duality does hold for neutrino-nucleon interactions then it 
would be possible to extrapolate the better-known neutrino DIS structure into the SIS region and give an indication of how 
well current event simulators are modeling the (anti)neutrino-nucleus cross sections in the SIS region.  

Since the experimental data available from the hydrogen and deuterium bubble chamber experiments from the 70's and 80's for the
resonance production in $\nu$-N scattering lack the level of precision of the electron-nucleus scattering, the experimental 
study of duality in neutrino-nucleon scattering is not conclusive. Moreover, due to dearth of experimental data on $\nu$-A 
scattering above the $\Delta$ resonance in the SIS region, the study of QH duality in the transition region is also very 
limited. Recently, the phenomenon of QH duality has been experimentally observed in the NC sector through the
observation of parity violating asymmetry in the scattering of polarized electron from proton and deuteron targets at 
JLab~\cite{JeffersonLabHallA:2013yxz}. Theoretically not much progress has been made except the early works of Lalakulich et 
al.~\cite{Lalakulich:2009zza, Lalakulich:2006yn}, Sato et al.~\cite{Sato:2003rq} and Gross et al.~\cite{Gross:1991pi}. This 
has also limited the development of the neutrino event generators as the modern $\nu$ interaction simulation efforts are not 
able to compare their results with the duality predictions for $\nu-N$ and $\nu-A$ interaction events in the transition region.
\begin{figure} 
\begin{center}
\includegraphics[height=6cm,width=0.3\textwidth]{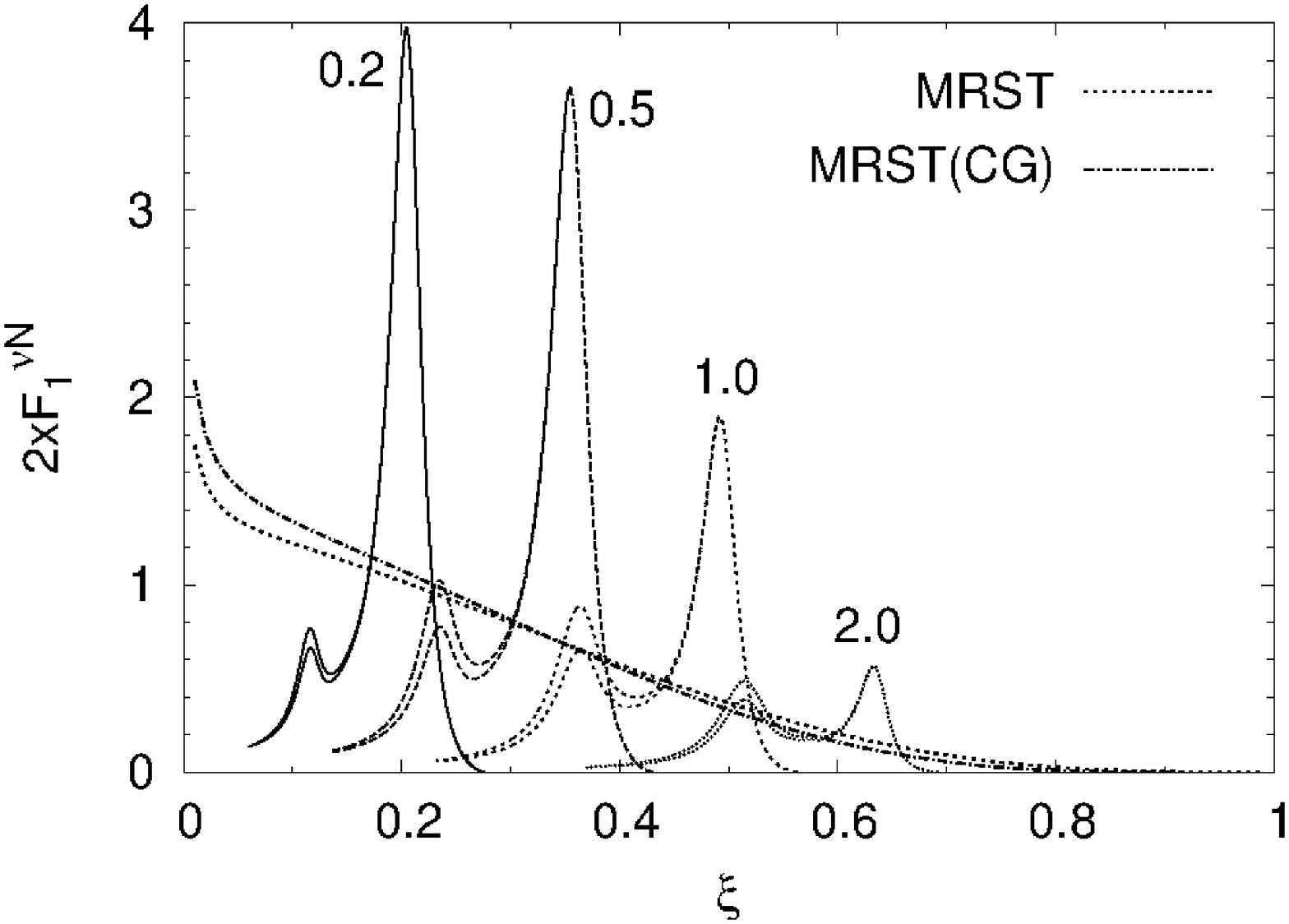}
\includegraphics[height=6cm,width=0.3\textwidth]{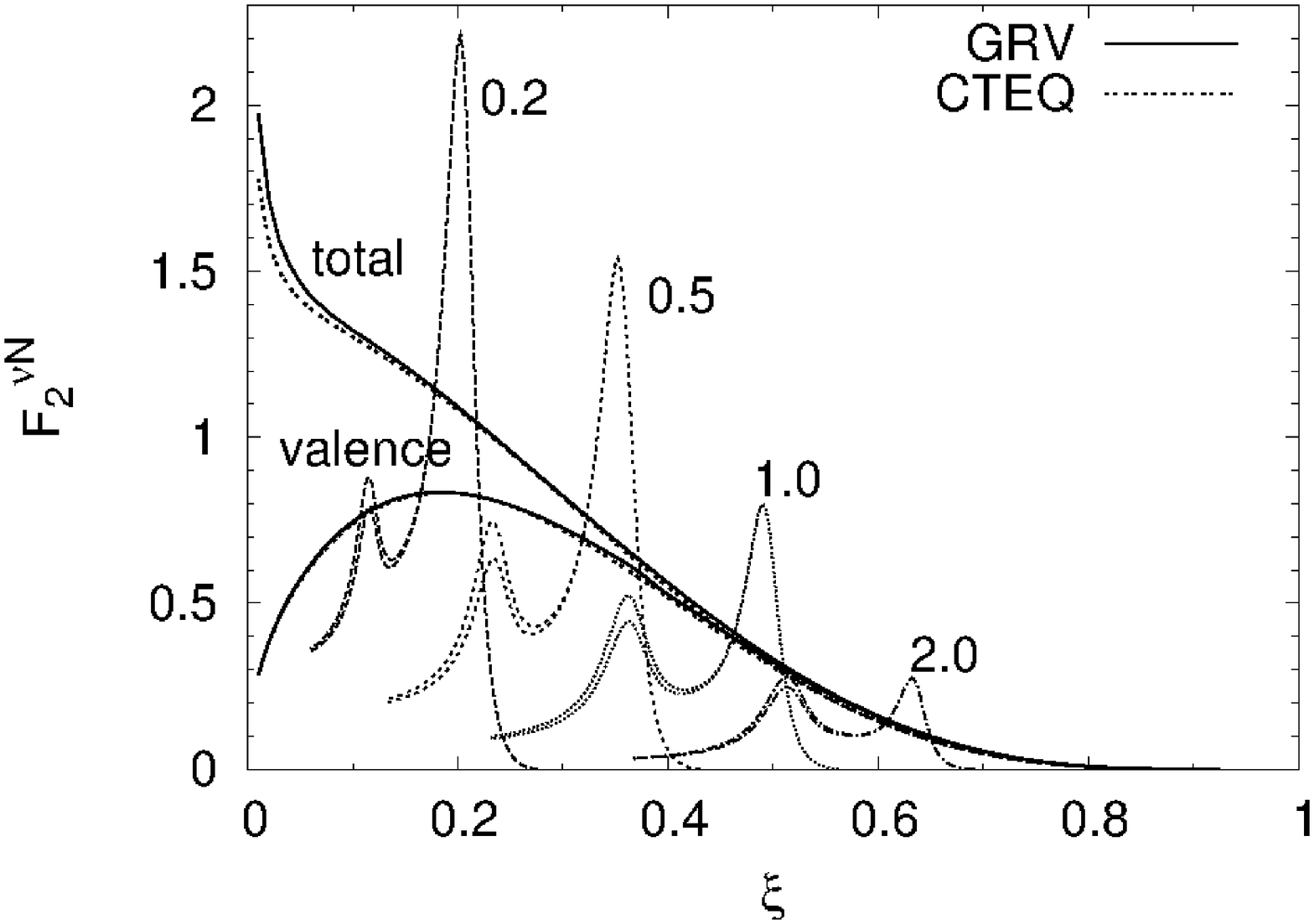}
\includegraphics[height=6cm,width=0.3\textwidth]{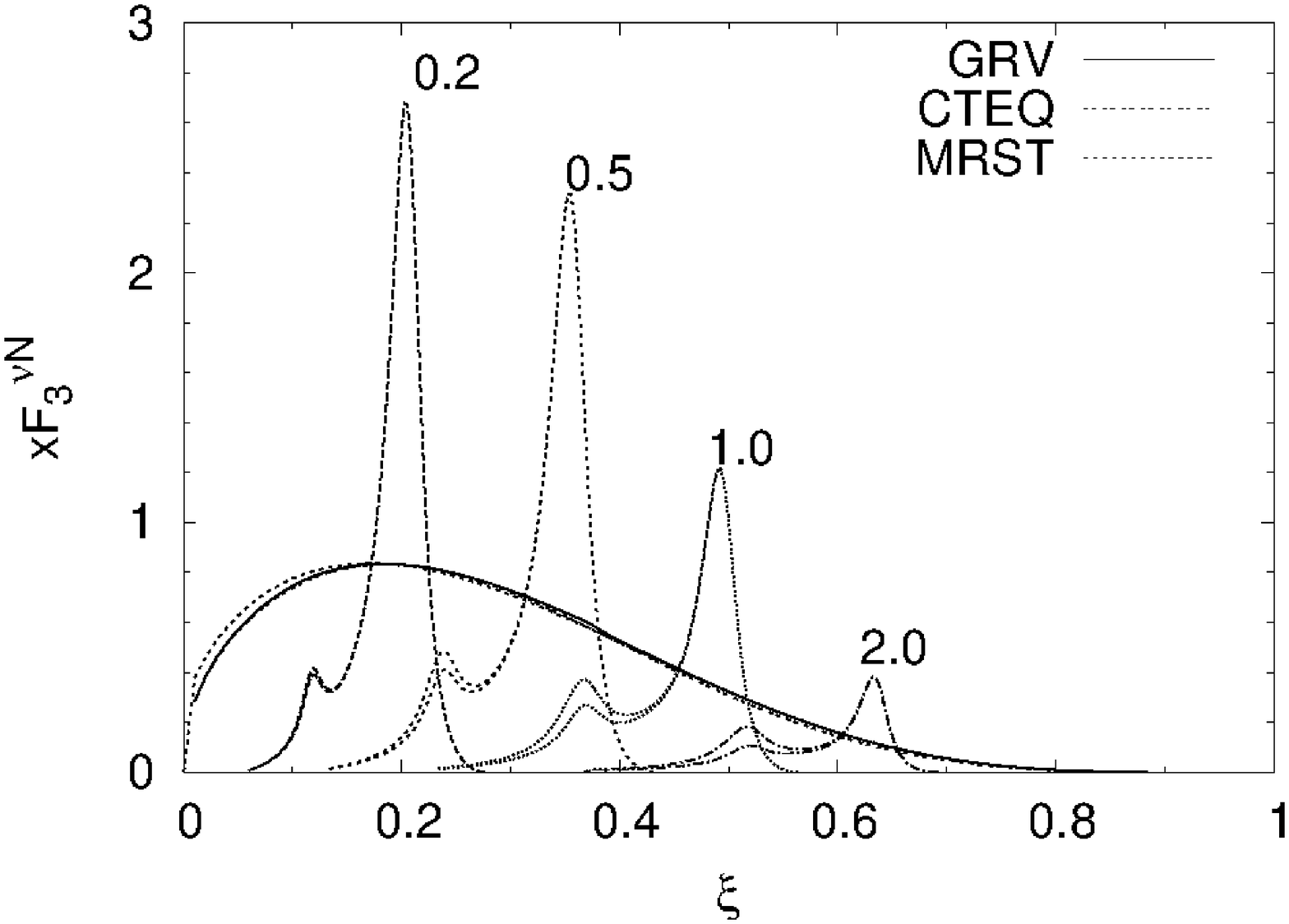}
\caption{Duality for the neutrino-nucleon ${F_1}^{\nu-N}$, $2x{F_2}^{\nu-N}$ and $x{F_3}^{\nu-N}$ structure functions as a 
function of $\xi$ at different $Q^2$. Figure has been taken  
from Ref.~\cite{Lalakulich:2006yn}.}\label{Fig-f1f2f3}
\vspace{-0.5cm}
\end{center}
\end{figure}

It has been argued by Close and Melnitchouk~\cite{Close:2003wz} that the isospin symmetry constraints the QH duality 
not to hold locally for protons and neutrons separately even if one includes several resonances with both even and odd 
parities as the neutrino interaction on the proton target and the antineutrino interaction on the neutron target in the few 
GeV energy region is dominated by the different charged states of $\Delta(1232)$ resonance which has been discussed here in 
detail in Section~\ref{sec:1pion}. For example, neutrino-proton structure functions are three times larger than the 
neutrino-neutron structure functions and therefore resonance structure functions are significantly larger than the leading 
twist functions i.e. $F_i^{\nu p({\text{Res}})} > F_i^{\nu p({\text{LT}})}$. This clearly indicates the violation of 
QH-duality for neutrino interaction on a proton target. Similarly for a neutrino-neutron scattering besides the contribution 
from the isospin $\frac{3}{2}$ resonances, there is also significant contribution from the isospin $\frac{1}{2}$ resonances, 
but the total contribution from isospin $\frac{1}{2}$ resonances is smaller than that from isospin $\frac{3}{2}$ resonances 
due to $\Delta(1232)$ dominance which results $F_i^{\nu n({\text{Res}})} ~<~ F_i^{\nu n({\text{LT}})}$. Thus QH-duality is 
also not valid for $\nu-n$ scattering.
 
Therefore, one considers duality for the average of proton and neutron structure functions. This has been studied by 
Lalakulich et al.~\cite{Lalakulich:2006yn} in neutrino-nucleon scattering as well as in neutrino-nucleus 
interactions~\cite{Lalakulich:2009zza} and they find that for an isoscalar nucleon target duality holds, which has been shown 
 in Fig.~\ref{Fig-f1f2f3} for the neutrino-nucleon ${F_1}^{\nu N}$, $2x{F_2}^{\nu N}$ and $x{F_3}^{\nu N}$ structure 
functions at several values of $Q^2$. It may be observed that for $\nu N$ scattering duality holds good for $F_2$ and $F_3$ 
and not so well for $F_1$.

Geissen-Ghent collaboration~\cite{Lalakulich:2009zza} have studied Bloom-Gilman duality, in electron and neutrino scattering 
on nuclei, and found that the ratio
\begin{equation}\label{eqg}
 I(Q^2_{Res},Q^2_{DIS})=\frac{\int_{\xi_{min}}^{\xi_{max}}d\xi\; F_j^{Res}(\xi,Q^2_{Res})}{\int_{\xi_{min}}^{\xi_{max}}d\xi\; 
 F_j^{DIS}(\xi,Q^2_{DIS})}\;<\;1,
\end{equation}
where $F_j$ represents $2 x F_{1},$ $F_2,$ and $x F_3$, and $Res$~($DIS$) represent resonance~(DIS) structure functions at the 
same $\xi$. This collaboration~\cite{Lalakulich:2009zza} has emphasized the importance of including NR as well as resonance 
contributions while evaluating the numerator in Eq.~(\ref{eqg}).
  
To conclude, the study of QH duality needs serious attention especially by the neutrino physics community as a substantial 
contribution to the events is expected to come from the transition region for all the next generation planned accelerator and 
atmospheric neutrino experiments.

\section{Monte Carlo event generators and some of the recent results from the accelerator experiments}\label{MC}
Monte Carlo event generators are scientific programs/libraries to simulate events for the neutrino interactions with 
matter~(electrons, nucleons and nuclei). In the neutrino sector, the early event generators were 
NEUGEN~\cite{Gallagher:2002sf}, NUANCE~\cite{Casper:2002sd}, NUX~\cite{NUX-Rubbia}, NEUT~\cite{Hayato:2021heg, Hayato:2009zz} 
and Geneve~\cite{Cavanna:2002se}. These were initially developed by the experimenters to simulate events for a particular 
experiment. For example, the earliest version of the NEUGEN event generator was written for the 
Soudan 2 experiment, in the mid-1980's, to simulate the neutrino backgrounds in the proton decay searches. In the earlier 
version of NUANCE and NEUGEN, to simulate neutrino interactions with the nuclear targets, Smith and Moniz model was used for 
the QE scattering from nuclei, Rein and Sehgal model was used for the resonance excitation, and for the DIS, the 
NME like the shadowing, anti-shadowing, Fermi motion and EMC were not considered. Later with the need of 
more sophisticated and robust event generators, several collaborative projects were started which led to the amalgamation of 
theorists, phenomenologists as well as experimenters in the development of the neutrino event generators. The recent Monte 
Carlo generators widely used in the accelerator and atmospheric neutrino experiments are GENIE~\cite{Andreopoulos:2009rq, 
GENIE:2021npt}, NEUT~\cite{Hayato:2021heg, Hayato:2009zz}, 
NuWro~\cite{Golan:2012wx}, GiBUU~\cite{Gallmeister:2016dnq}, FLUKA~\cite{Battistoni:2009jen}, etc., which are 
updated regularly by the respective developers. 

There are now provisions of alternative nuclear models for the QE scattering like the Smith and Moniz Fermi gas 
model, LFGM of the Valencia group, Superscaling model of the Donnelly group, more sophisticated models to 
take into account many body nucleon correlation effects like the inclusion of 2p-2h effect using either the formalism of 
Martini et al.~\cite{Martini:2009uj, Martini:2010ex, Martini:2011wp, Martini:2013sha} or Nieves et al.~\cite{Nieves:2011yp, 
Nieves:2012yz}, and the final state interaction effects, etc. 

Most of these modern generators~(GENIE, NEUT, NuWro) have common inputs. However, the differences in their implementation, the 
value of the parameters used, and the approaches to avoid double counting yield different predictions. For example, in the earlier 
version of GENIE, the QE scattering is modeled using the relativistic Fermi gas model of Llewellyn Smith, for the 
baryon resonance excitations in NC and CC channels Rein and Sehgal model is used in which 16 resonances were 
considered and DIS is calculated using Bodek and Yang prescription. Recent version of GENIE 3 uses different models of NME 
for the QE neutrino induced processes like LFGM  of the Valencia group with 
1p-1h and 2p-2h excitations~\cite{Nieves:2011yp, 
Nieves:2012yz}, Superscaling approach~($1p-1h + 2p-2h$) of Donnelly et 
al.~\cite{Amaro:2021sec}, etc., GENIE is widely being used by experimenters involved in the Fermilab neutrino program like 
MINERvA, NOvA, MicroBooNE collaborations.
These MC generators have now become essential in analyzing the neutrino events as they make use of the latest developments
in nuclear theory to the description of $\nu-A$ interaction cross sections.

NEUT was developed initially by the Kamiokande collaborators to simulate atmospheric neutrino events, it is now being used by 
the Super-Kamiokande as well as the T2K collaborations and is continuously being updated. One among the many revised versions, NEUT 
version 5.3.2 describes CCQE neutrino-nucleon interactions using the spectral function~(SF) approach of Benhar et 
al.~\cite{Benhar:1994hw} with $M_A$ =1.21GeV. The resonant pion production process is  described by the Rein-Sehgal 
model~\cite{Rein:1980wg} with updated  nucleon-resonance transition form factors~\cite{Graczyk:2007bc} and 
$M_A^{\text{Res}}$=0.95GeV. For the 2p-2h interactions, they have used the microscopic model developed by Nieves et 
al.~\cite{Nieves:2011yp}. DIS is modeled using GRV98 PDFs ~\cite{Gluck:1998xa} with the corrections by Bodek and 
Yang~\cite{Bodek:2003wc}. The final state interactions describing the transport of the hadrons produced in the elementary 
neutrino interaction through the nucleus, are simulated using a semi-classical intranuclear cascade model.

GiBUU~\cite{Buss:2011mx} has been developed by Mosel and his collaborators at Giessen, and uses local Thomas-Fermi gas in a 
mean field potential for the QE scattering, for the resonances it uses MAID analysis 
of electron-nucleon pion production as input for the vector part of both the resonant and NR amplitudes. It is a 
transport model where FSI is implemented by solving Giessen Boltzmann-Uehling-Uhlenbeck~(GiBUU) equation. It encompasses a 
unified framework for hadron, lepton and neutrino interactions with nuclei from a few hundreds of MeV to a few tens of GeV. 
In recent years, tuned generators like MINERvA tune~\cite{MINERvA:2015ydy}, MicroBooNE tune~\cite{MicroBooNE:2021sfa, 
MicroBooNE:2021ccs}, etc. have been used, where modifications are made in the GENIE or other versions of the generators. For 
example, it was found by the MINERvA collaboration~\cite{MINERvA:2015ydy}, that the model of Nieves et 
al.~\cite{Nieves:2011yp}, underestimates the 2p-2h strength in the dip region and therefore they increased the flux-folded 
strength by a significant amount. For a general discussion, see Ref.~\cite{Mosel:2019vhx}.

Recently the T2K collaboration~\cite{T2K:2020jav} plotted the ratio of the double differential cross section per nucleon for 
$\nu_\mu$ induced CC reaction for oxygen to carbon nuclear targets i.e. $R_{\frac{O}{C}}$ (which has been shown in 
Fig.~\ref{t2k-prd101} for the two muon angle bins), and compared their data with the simulated results from the MC generators 
like NEUT 5.4.1 LFG, GENIE v3- SuSAv2, NuWro SF, GiBUU, NEUT 5.4.0 SF, GENIE v3 LFG, NuWro LFG and RMF~(1p-1h) + 
SuSAv2~(2p-2h). They have also presented the results for the integrated cross sections per nucleon and compared them with the 
MC predictions~\cite{T2K:2020jav}.

ArgoNeuT collaboration~\cite{ArgoNeuT:2018und} has reported the CC $\nu_\mu({\bar\nu}_\mu)$ induced 1$\pi^+$($\pi^-$) 
differential cross sections on $^{40}$Ar and compared their results with the MC generators, which are presented in 
Fig.~\ref{argoneut-1pi}. The data for the pion angular distribution and the muon angular distribution obtained by the ArgoNeuT 
collaboration has been shown, where the comparisons with the different MCs like GENIE v$2\_12\_2$, NuWro $17.01.1$, GiBUU and 
NEUT$5.3.7$ have also been made~\cite{ArgoNeuT:2018und}. 

To highlight the difference among the predictions of the various MCs as well as their comparison with the data, in 
Fig.~\ref{minerva-pt-pl}, we present an analysis of the MINERvA collaboration where they have shown absolute normalized ratios of 
data, and the comparison with the GENIE 2.8.4, NuWro, and GiBUU to MnvGENIEv1 for muon transverse~($p_T$) and 
longitudinal~($p_L$) momenta. It may be observed from Fig.~\ref{minerva-pt-pl} that the transverse momentum projection shows 
tension among all models and data in the 0.55\;$<\;p_T\;<$\;1.5~{GeV} range. While in the case of longitudinal momentum, all 
models underpredict the cross section.

In Fig.~\ref{microboone-enu-emu}, we show the recent results of the MicroBooNE collaboration~\cite{MicroBooNE:2021sfa}, for 
the total cross section divided by the bin-center neutrino energy vs neutrino energy i.e. $\frac{\sigma(E_{\nu_\mu})}{\langle 
E_{\nu_\mu}\rangle}$ vs $E_{\nu_\mu}$. These results are compared with the MicroBooNE MC~\cite{MicroBooNE:2021ccs}, 
predictions from GENIE v3.0.6~\cite{Andreopoulos:2009rq, GENIE:2021npt}, NuWro 19.02.01~\cite{Golan:2012rfa}, NEUT 
5.4.0.1~\cite{Hayato:2009zz}, and GiBUU 2019.08~\cite{Buss:2011mx}.

It may be observed from the above results that there is agreement among the different MC generators with the data nevertheless 
more work is needed to understand medium effects in the nuclear targets. It needs more collaborative efforts. 
	\begin{figure} 
 \centering
\includegraphics[height=0.25\textheight,width=0.49\textwidth]{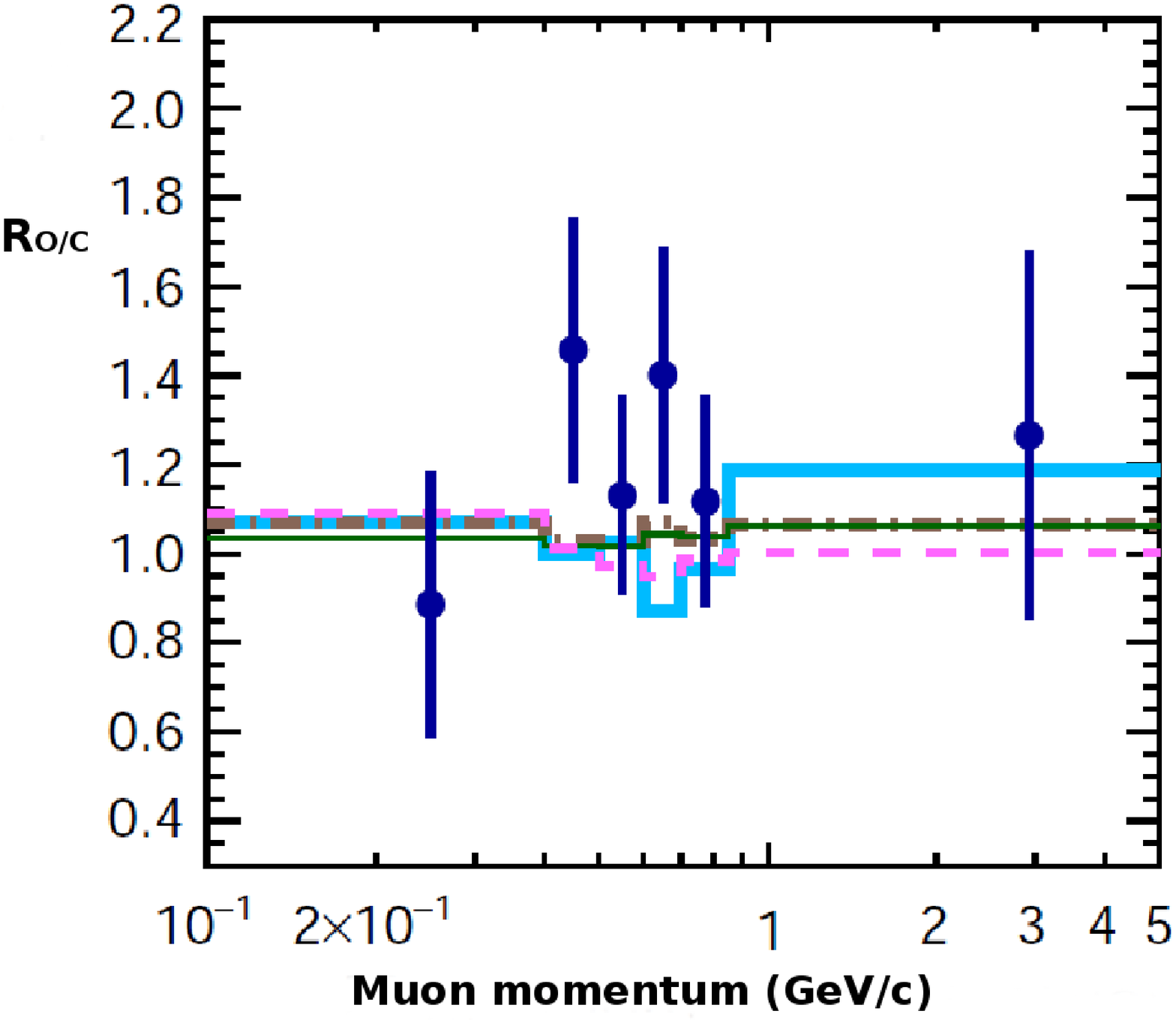}
\includegraphics[height=0.25\textheight,width=0.49\textwidth]{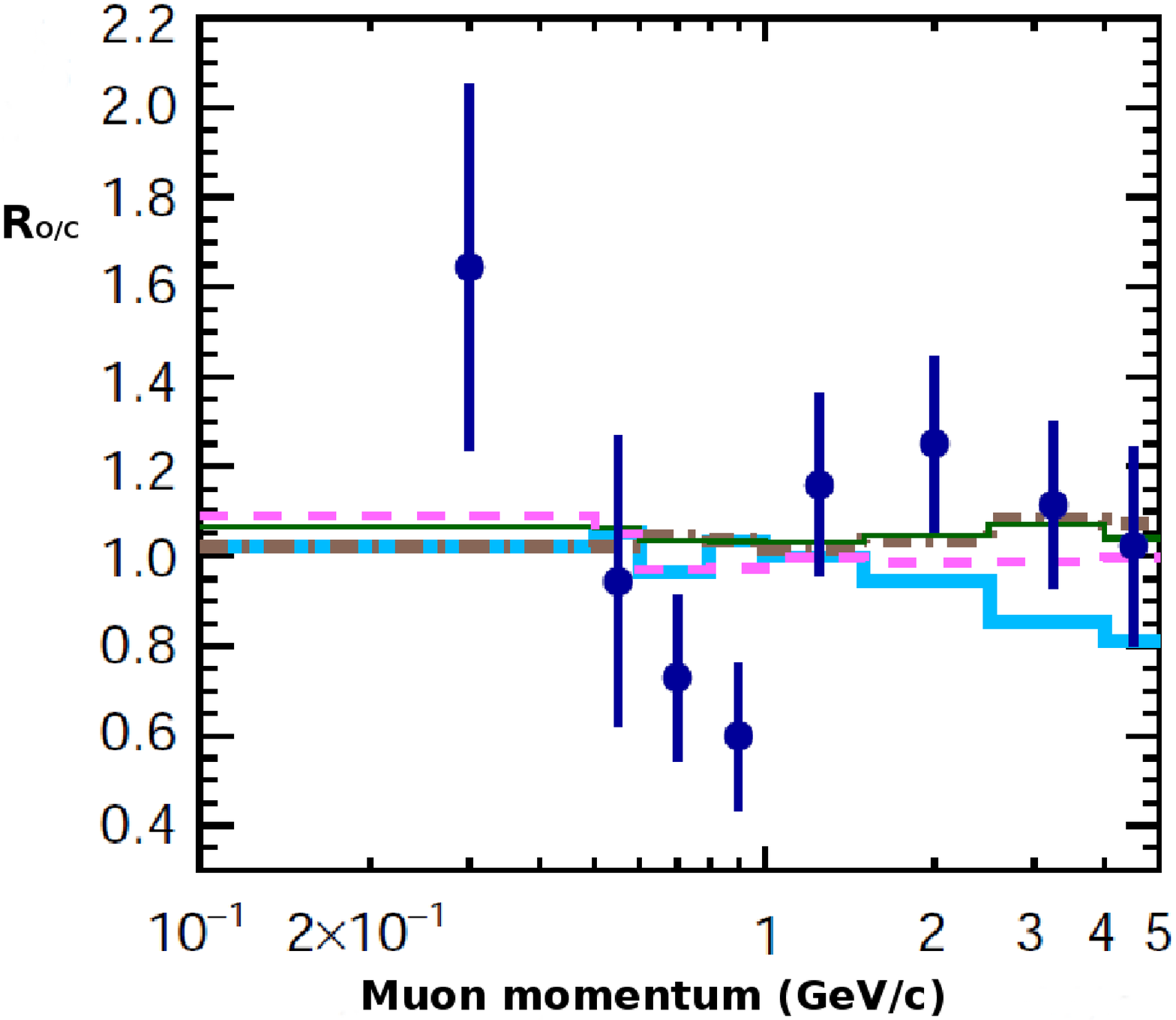}\\
\includegraphics[height=0.25\textheight,width=0.49\textwidth]{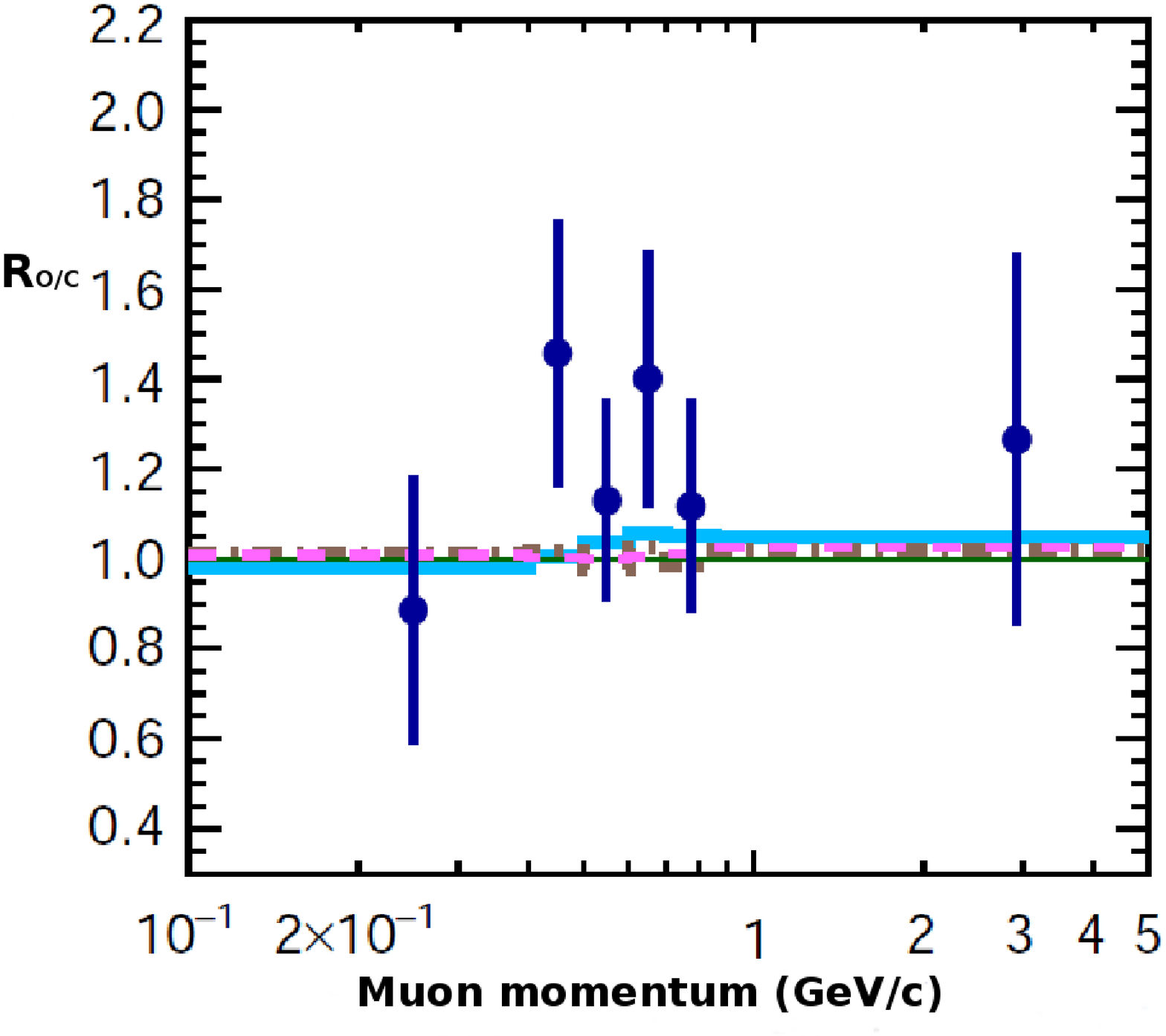}
\includegraphics[height=0.25\textheight,width=0.49\textwidth]{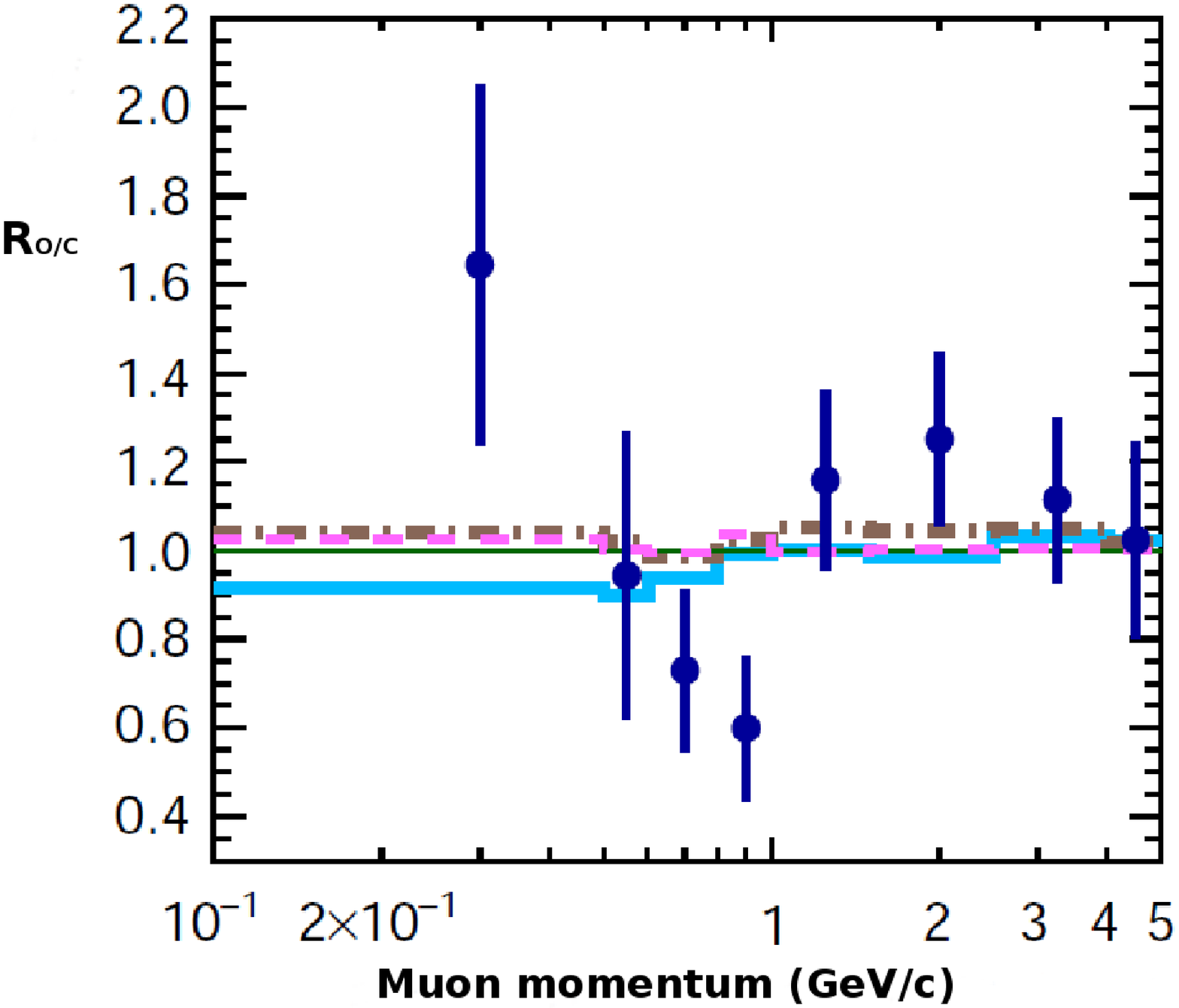}
\caption{Ratio of the double differential cross section per nucleon for $\nu_\mu$ induced CC reaction in oxygen and carbon i.e. 
$R_{\frac{O}{C}}$. T2K collaboration~\cite{T2K:2020jav} has compared their data with~(upper panel) NEUT 5.4.1 LFG~(brown), 
GENIE v3- SuSAv2~(green), NuWro SF~(magenta) and GiBUU~(light blue) MC predictions. The left panel is for the muon angle bin 
$0.75  <\cos\theta_{\mu}<0.86$ and the right panel is for $0.93 <\cos\theta_{\mu}<1$. The lower panel is the same comparison 
with other MC generators like NEUT 5.4.0 SF~(brown), GENIE v3 LFG~(green), NuWro LFG~(magenta) and RMF~(1p1h) + 
SuSAv2~(2p2h)~(light blue). This figure has been taken from Ref.~\cite{T2K:2020jav}.}\label{t2k-prd101}
\label{t2k-prd101}
\end{figure}
\begin{figure}[h]
 \centering
\includegraphics[height=0.22\textheight,width=0.49\textwidth]{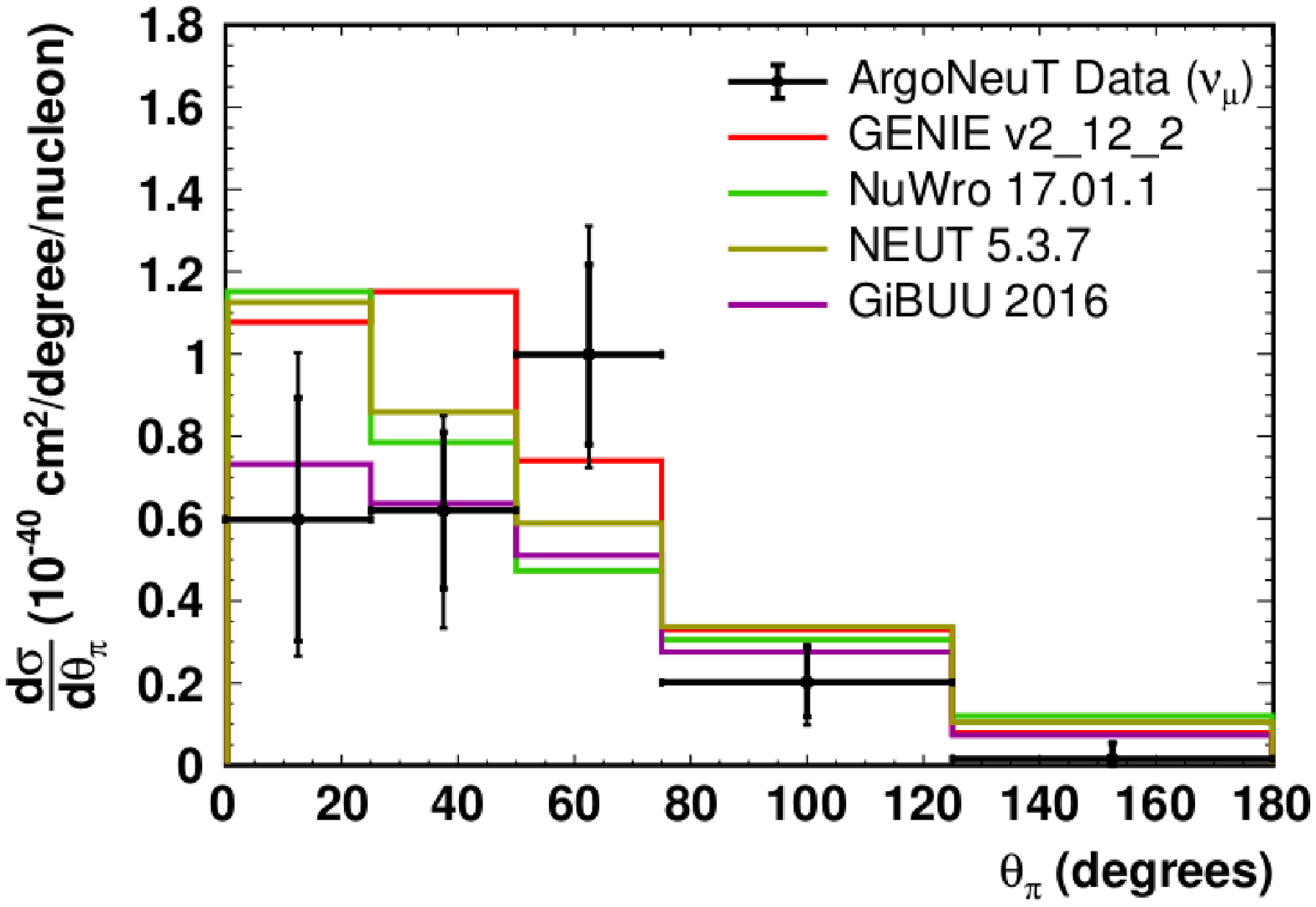}
\includegraphics[height=0.22\textheight,width=0.49\textwidth]{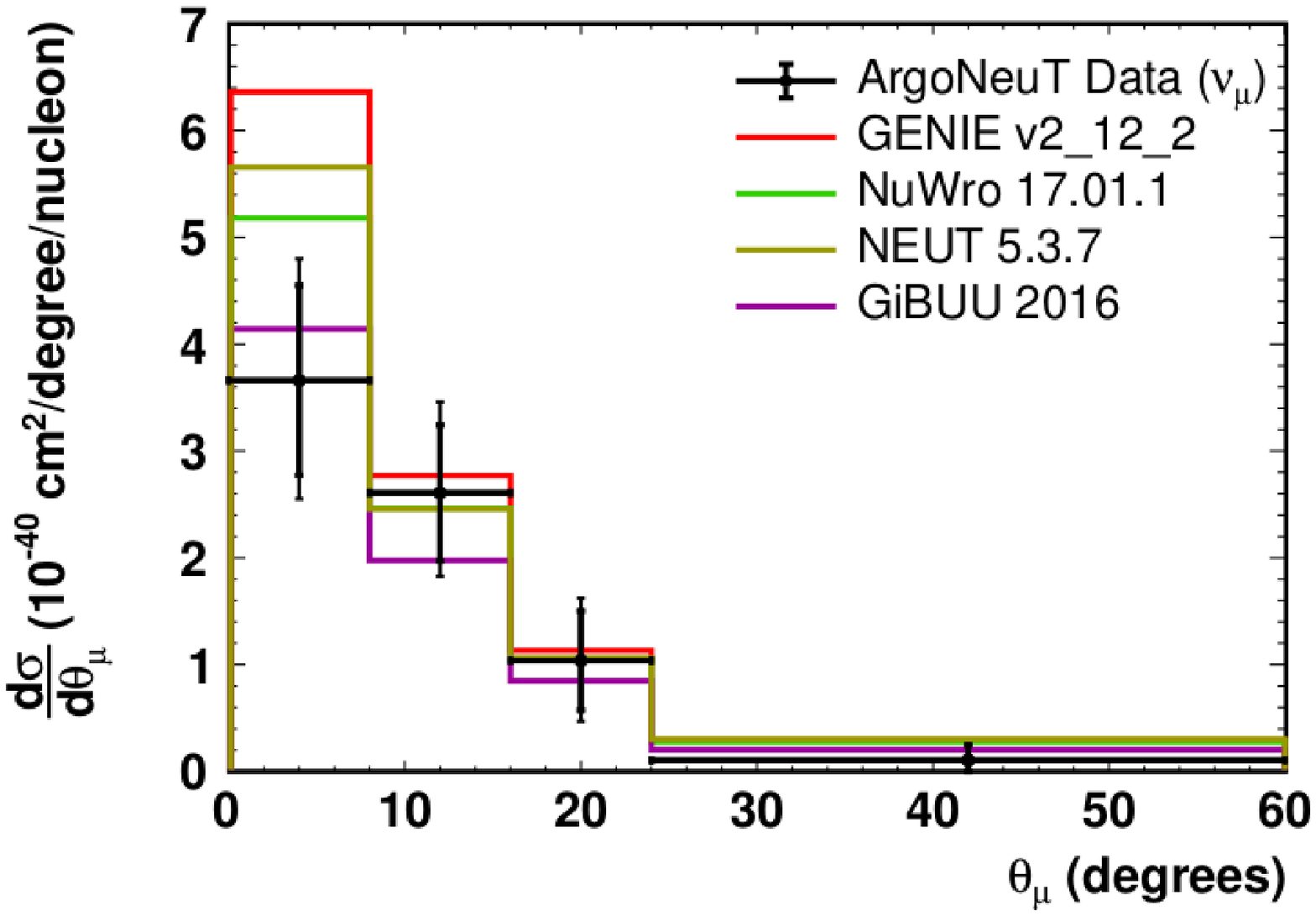}\\
\includegraphics[height=0.22\textheight,width=0.49\textwidth]{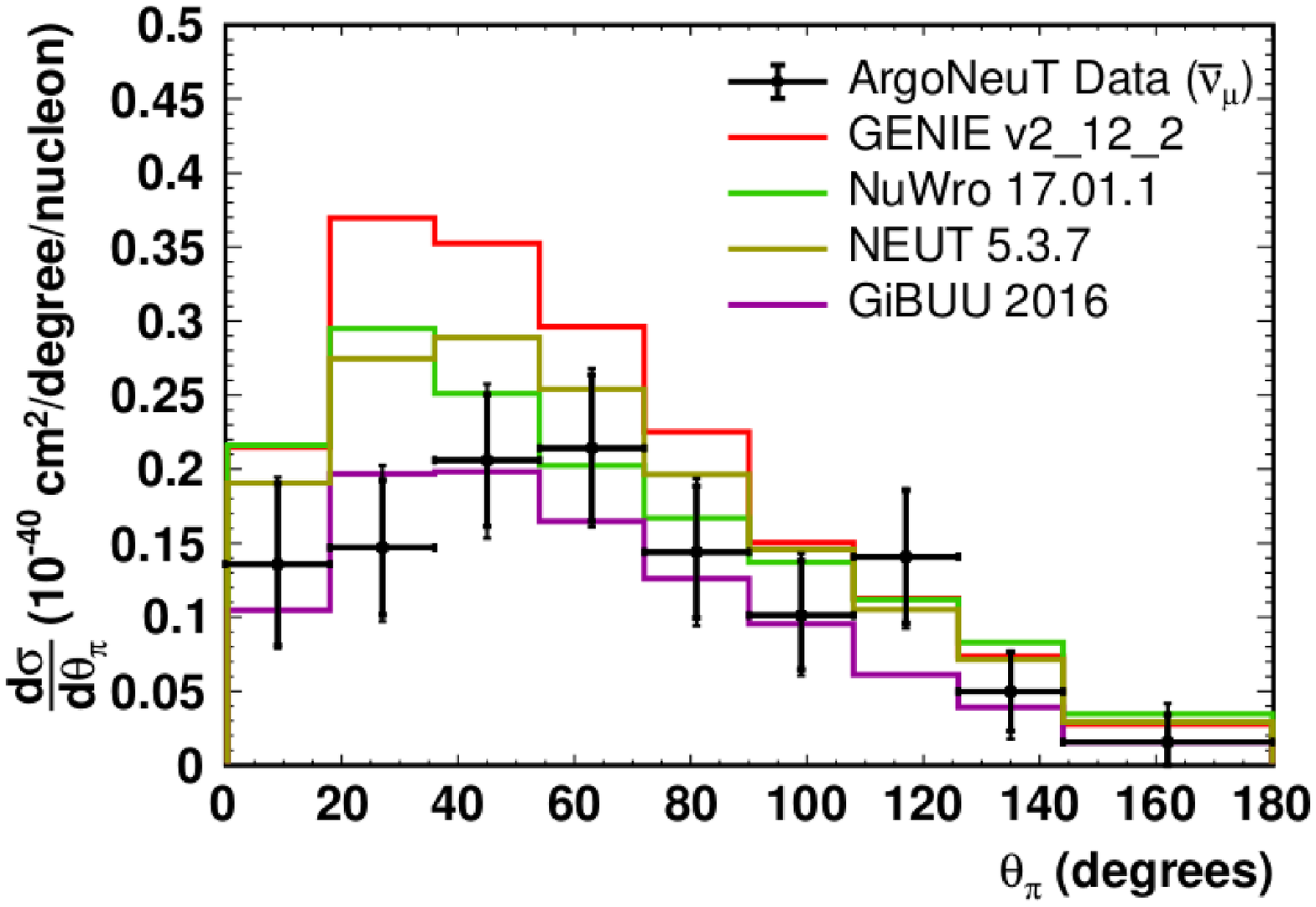}
\includegraphics[height=0.22\textheight,width=0.49\textwidth]{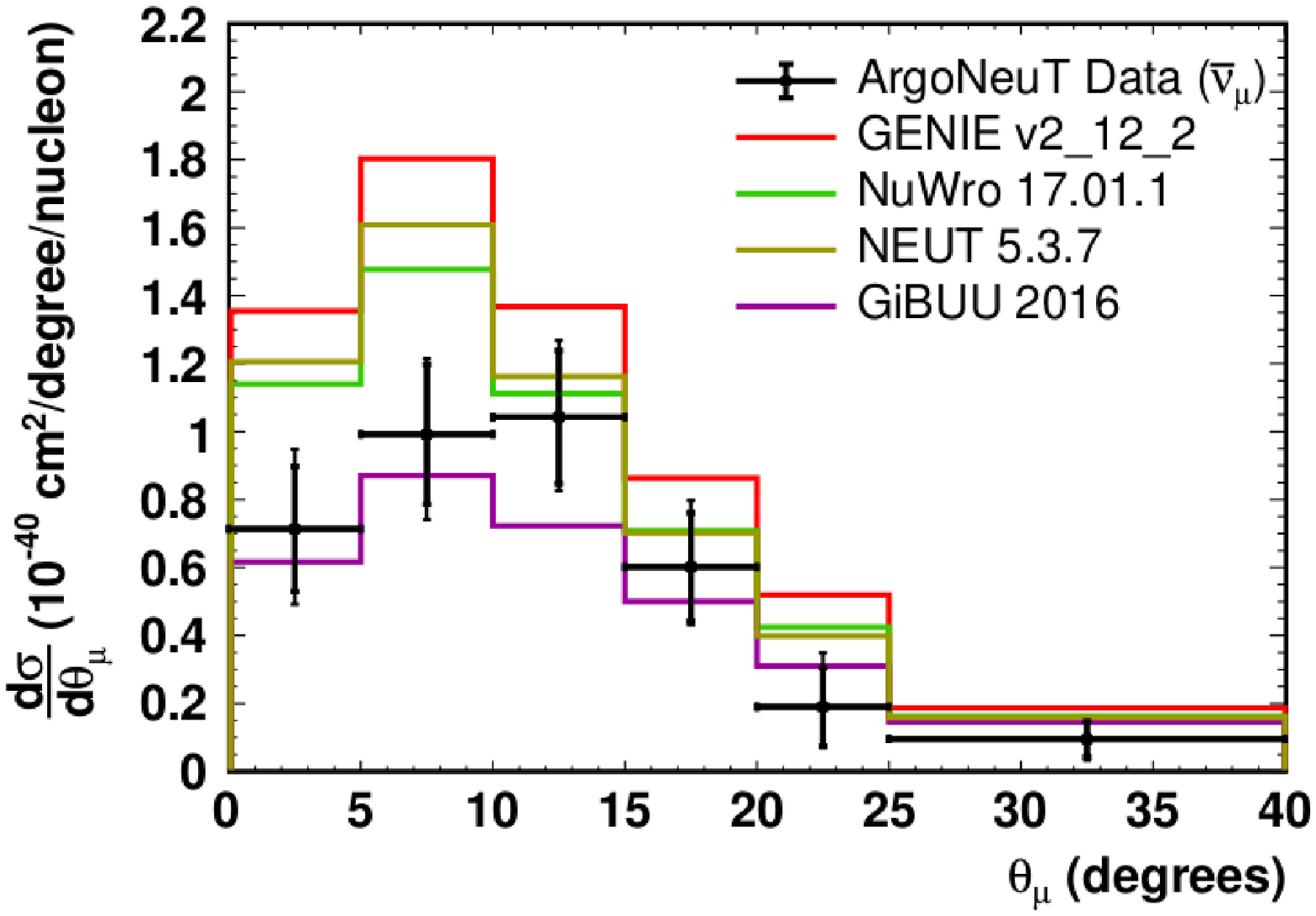}
 \caption{ArgoNeuT CC $\nu_\mu({\bar\nu}_\mu)$ induced 1$\pi^+$~(upper panel) and 1$\pi^-$~(lower panel) differential cross 
 sections on $^{40}{Ar}$. Left panel is for pion angular distribution i.e. the outgoing pion angle~($\theta_\pi$) w.r.t. the 
 initial neutrino direction and the right panel is for muon angular distribution i.e. the outgoing muon angle~($\theta_\mu$) 
 w.r.t. the initial neutrino direction. Comparisons are with the different MCs like GENIE 
 v$2\_12\_2$~\cite{Andreopoulos:2009rq}, NuWro $17.01.1$~\cite{Golan:2012wx}, GiBUU 2016~\cite{Gallmeister:2016dnq} and 
 NEUT~$5.3.7$~\cite{Hayato:2009zz}. This figure has been taken from Ref.~\cite{ArgoNeuT:2018und}.}\label{argoneut-1pi}
\end{figure}
 \begin{figure}[h]
 \centering
\includegraphics[height=0.22\textheight,width=0.49\textwidth]{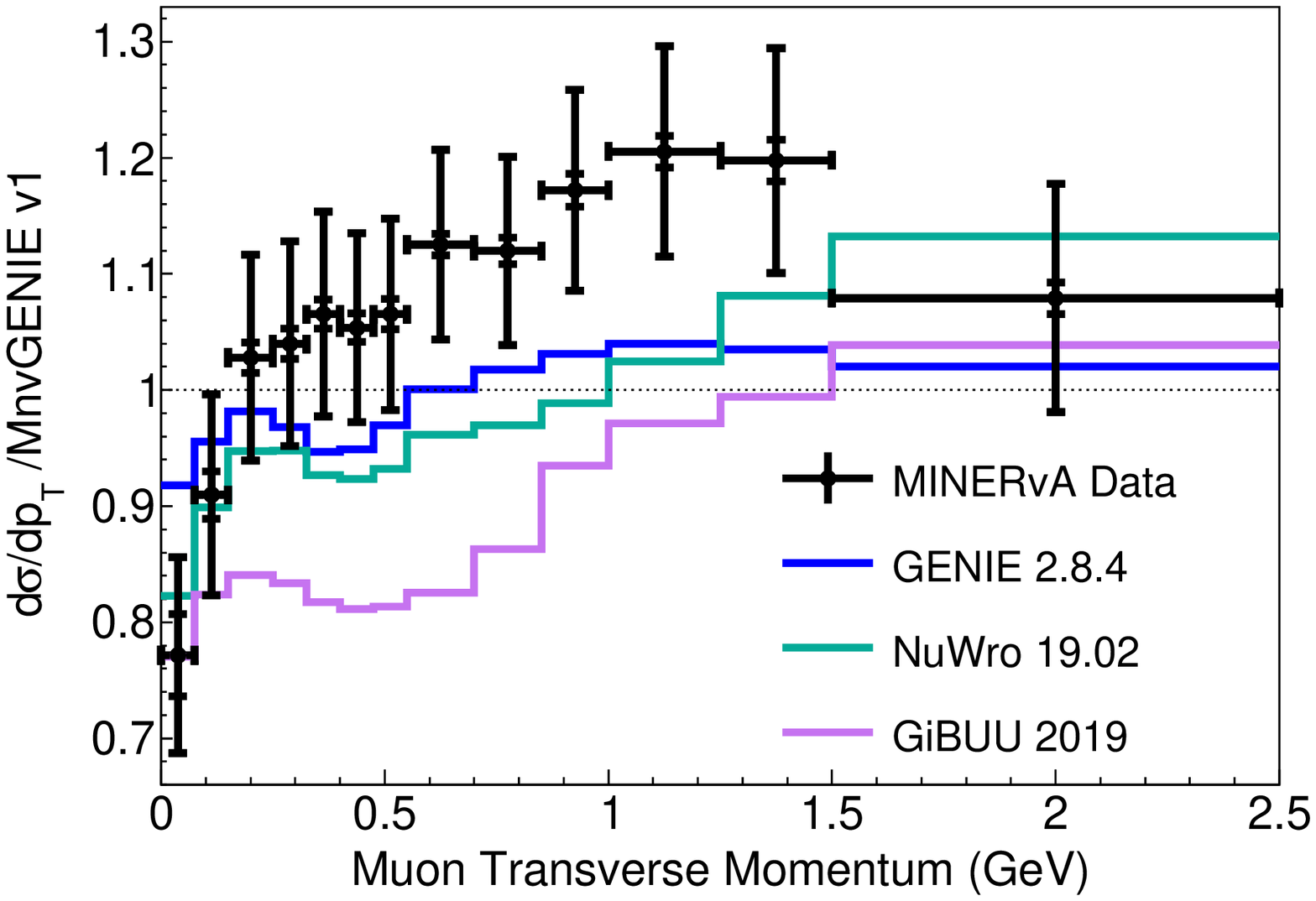}
\includegraphics[height=0.22\textheight,width=0.49\textwidth]{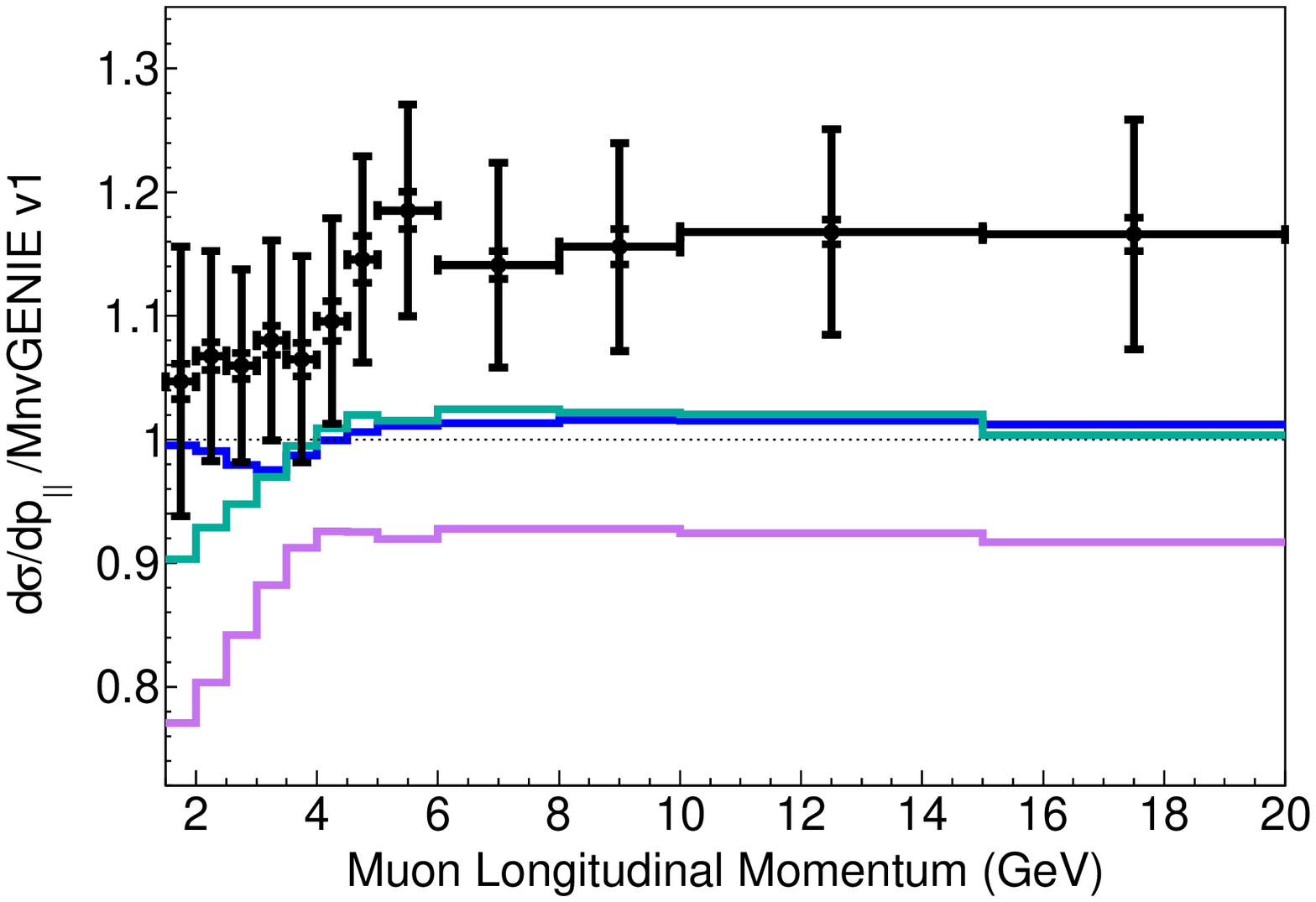}\\
\caption{MINERvA inclusive CC differential cross sections for muon neutrinos on hydrocarbon in terms of the 
transverse and longitudinal muon momentum distributions~\cite{MINERvA:2020zzv}. The ratio is for absolutely 
normalized ratios of data, GENIE 2.8.4~\cite{Andreopoulos:2009rq}, NuWro~\cite{Golan:2012wx}, and 
GiBUU~\cite{Buss:2011mx} to MnvGENIEv1~\cite{MINERvA:2020zzv} for $p_T$ and $p_L$.}
\label{minerva-pt-pl}
\end{figure}
\begin{figure}
 \centering
\includegraphics[height=0.22\textheight,width=0.49\textwidth]{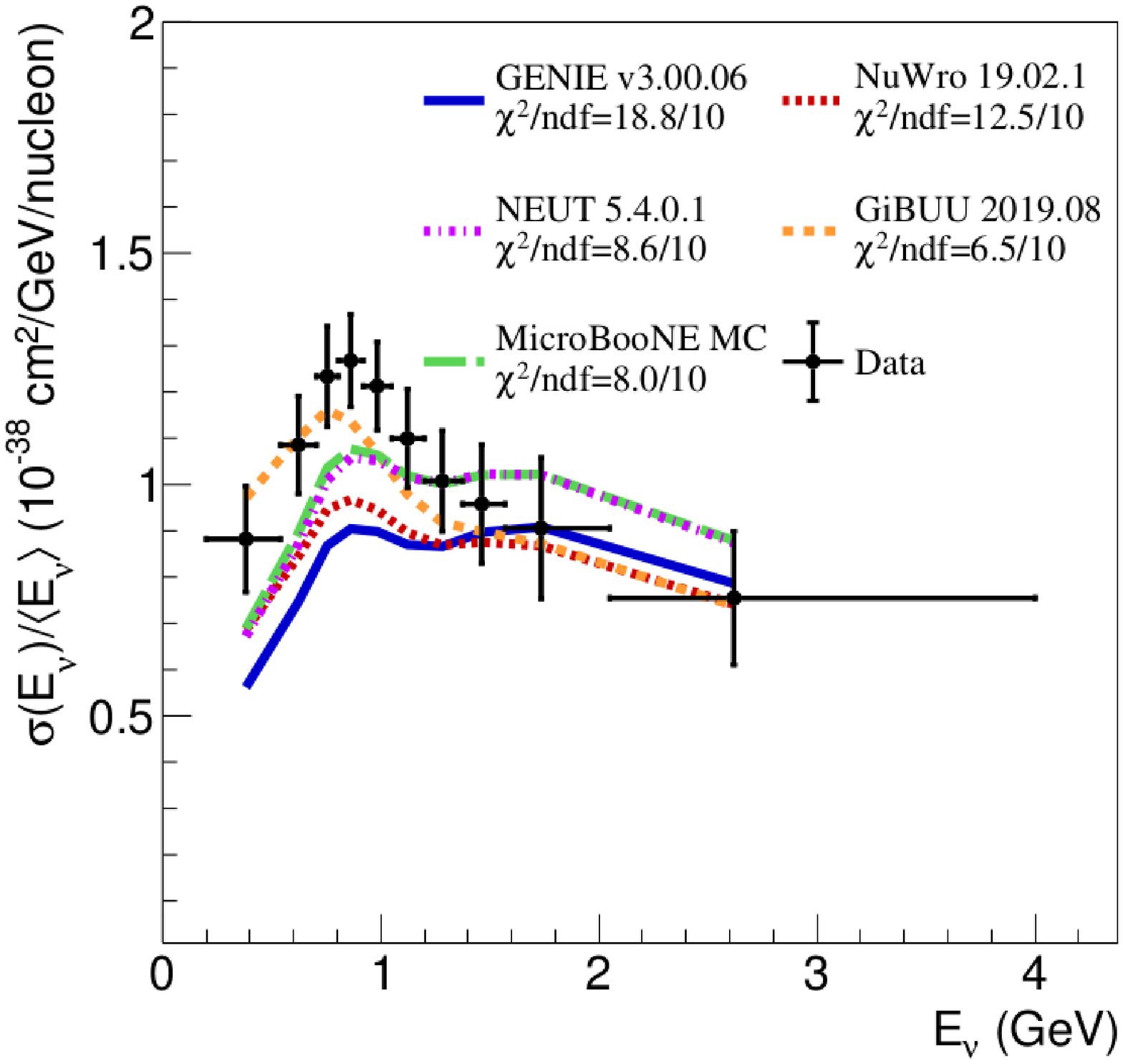}
\includegraphics[height=0.22\textheight,width=0.49\textwidth]{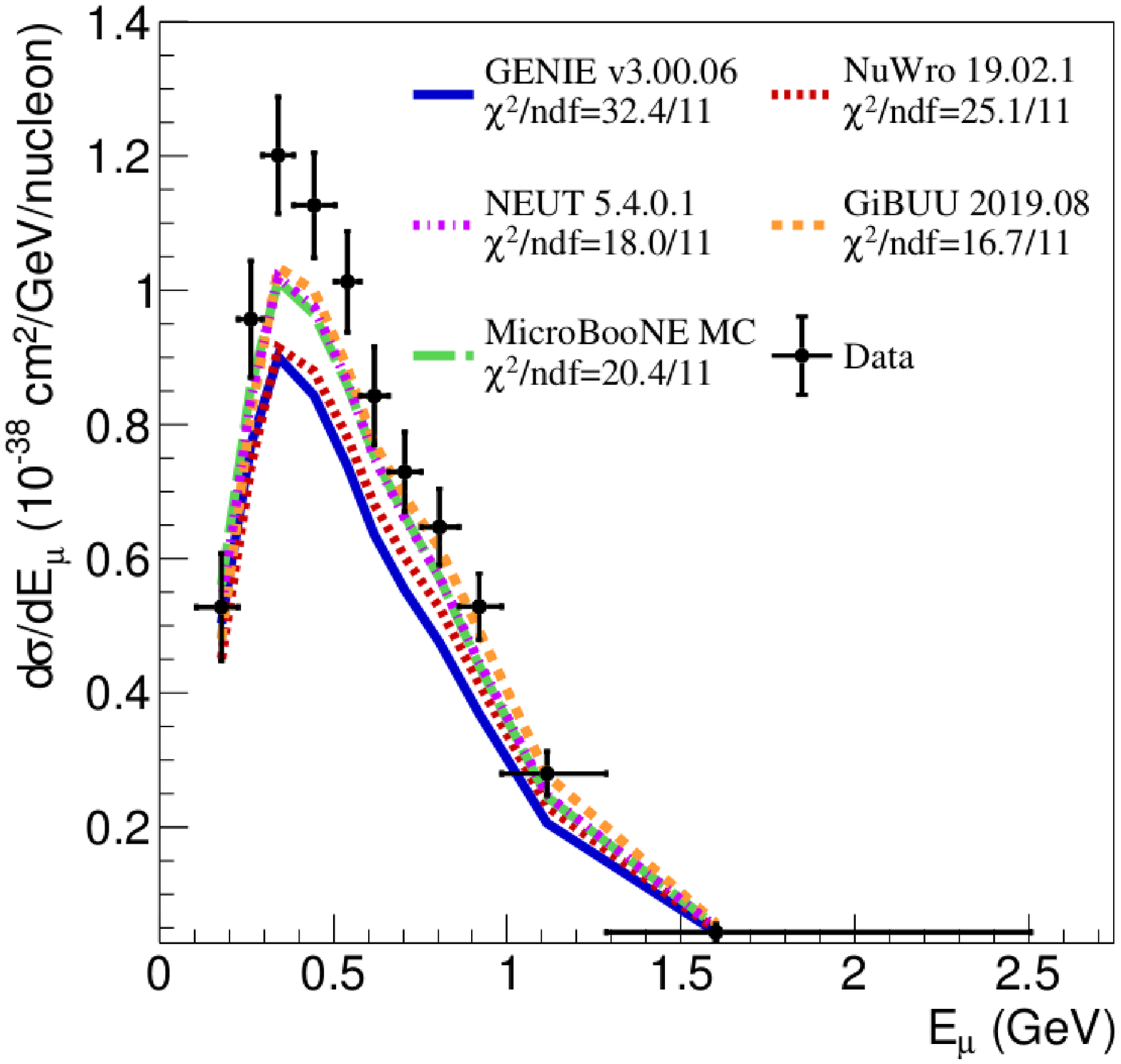}
\caption{Results obtained by the MicroBooNE collaboration~\cite{MicroBooNE:2021sfa} using 
LArTPC located in the Fermilab Booster Neutrino Beam with a mean neutrino energy of approximately 0.8~GeV. (Left panel) The 
extracted $\nu_{\mu}$CC inclusive scattering cross section per nucleon divided by the bin-center neutrino energy i.e. 
$\frac{\sigma(E_{\nu_\mu})}{\langle E_{\nu_\mu}\rangle}$ vs $E_{\nu_\mu}$. (Right panel) The measured $\nu_{\mu}$ CC 
differential cross section per nucleon as a function of muon energy i.e. $d\sigma/dE_\mu$ vs $E_\mu$. These results are 
compared with the MicroBooNE MC~\cite{MicroBooNE:2021ccs}, predictions from GENIE v3.0.6~\cite{Andreopoulos:2009rq, 
GENIE:2021npt}, NuWro 19.02.01~\cite{Golan:2012rfa}, NEUT 5.4.0.1~\cite{Hayato:2009zz}, and GiBUU 
2019.08~\cite{Buss:2011mx}.}\label{microboone-enu-emu}
\end{figure}

\section{Summary and outlook}\label{sec:sum}
The physics of neutrino interactions with matter has many aspects both in theory and experiment. This is because the neutrinos 
play very significant role in various areas of physics i.e. astrophysics, cosmology, nuclear physics, particle physics and 
geophysics. The neutrino physics originated with the attempts to understand the nuclear $\beta$ decays which led to the 
discovery of weak interactions and its role in astrophysical processes of energy generation in stars, synthesis of elements, 
supernova explosions, and the formation of neutron stars and white dwarfs, etc. With the discovery of new particles and their weak 
decays, the neutrino physics became an important component of particle physics both theoretically and experimentally. The 
advent of high energy particle accelerators  which produced unstable mesons like pions and kaons leading to the neutrino beams of 
$\nu_\mu(\bar{\nu}_{\mu})$ and $\nu_e(\bar{\nu}_{e})$ as their decay products, started the 
era of $\nu_{l}(\bar{\nu}_{l})$ scattering from the nucleons and nuclei in various energy regions starting from MeV to GeV 
which was earlier restricted to the very low energy region of (anti)neutrinos in the region of a few tens of MeV corresponding 
to the reactor antineutrinos. The processes of (anti)neutrino-nucleon scattering being a weak process has small cross section 
and therefore requires intense (anti)neutrino beams and large volume target-detectors to enhance the (anti)neutrino induced 
production of charged leptons, mesons and photon events to make them statistically significant for physical interpretations. 
The requirement of large volume target-detectors necessitated the use of medium or heavy nuclear material as targets to 
perform the (anti)neutrino scattering. This led to various neutrino-nucleus experiments being done at CERN, ANL, BNL, 
Fermilab, and SKAT using high energy neutrino beams. The confirmation of the phenomenon of neutrino flavor oscillations with 
the solar neutrinos and reactor antineutrinos in the low energy region and with the accelerator and atmospheric neutrinos in 
the intermediate and high energy regions has started great interest in studying the (anti)neutrino-nucleus scattering in the 
entire energy region of (anti)neutrino spectrum. Moreover, the study of these processes in the very low energy region is of 
immense interest in various astrophysical processes, while in the very high energy region, they are relevant for studying the 
origin of cosmic rays.

In view of this, we have presented a review of the (anti)neutrino reactions with nucleons and nuclei in this work. After 
presenting an updated summary of the neutrino properties and its sources from natural and man made origin in 
Section~\ref{intro}, we describe briefly the SM of neutrino interactions and apply it to study the (anti)neutrino 
scattering from point particles like leptons and quarks in this section. In this context, the resonance scattering of 
neutrinos from electrons and the observation of Glashow resonance in very high energy region is discussed. In 
Section~\ref{nu_interaction}, we apply the SM to study the various (anti)neutrino reactions on nucleons like 
QE~(elastic) scattering induced by CC~(NC) weak interactions, IE production of mesons 
like $\pi, K, \eta$ and hyperons like $\Lambda, \Sigma^0, \Sigma^-,\Xi^-, \Xi^0$ followed by the DIS in 
the region of very high energy and $Q^2$ corresponding to Bjorken scaling. In this region, QPM is 
used in the leading order of perturbative QCD to obtain results for the nucleon structure functions and the scattering cross 
sections. The effect of the evolution of structure functions to lower $Q^2$ in NLO and other 
corrections like TMC and HT are also discussed.
  
In the case of QE scattering induced by charged weak currents, the matrix element is described in terms of three 
vector $f_i (Q^2)$ and three axial-vector $g_i(Q^2)~ (i=1,2,3)$ form factors, which are all real due to T-invariance. Using 
the isotriplet and CVC hypotheses, the vector form factors $f_1(Q^2)$ and $f_2(Q^2)$ are related with the Sachs' 
electric~($G_{E}^{p,n}(Q^2)$) and magnetic~($G_{M}^{p,n}(Q^2)$) form factors of the nucleon, and $f_3(Q^2)=0$. The axial 
vector form factor $g_1 (Q^2)$ at $Q^2=0$ is derived in terms of $g_{NN\pi}$, the pion nucleon coupling constant and $f_\pi$, 
the pion decay constant using PCAC and GT relation. The pseudoscalar form factor $g_3(Q^2)$ is related to 
$g_1(Q^2)$ with the help of the PPDAC. The principle of 
$G (= C e^{i\pi I_2})$ invariance has been used to set $g_2(Q^2)=f_3(Q^2)=0$. The $Q^2$ dependence of both the 
vector~(axial-vector) form factors is parameterized by a dipole form using the dipole mass $M_V~(M_A)$ where the 
phenomenological values of $M_V$ and $M_A$ obtained from the analysis of electron and neutrino scattering experiments 
have been used. In the case of vector form factors, some recent parameterizations used in the analysis of electron scattering 
have also been used. With these inputs, the theoretical formulation of various QE reactions are studied in the strangeness 
conserving $\Delta S=0$ sector for $\nu_l$ and $\bar{\nu}_{l}$ scattering, and in the strangeness changing $\Delta S=1$ sector 
for $\bar{\nu}_{l}$ scattering. In the $\Delta S=1$ sector where the hyperons $Y = \Lambda, \Sigma^{0(-)}$ are produced, $SU(3)$ 
symmetry has been used to obtain the $N-Y$ transition form factors. The numerical results are presented for:
\begin{itemize}
 \item [(i)] The differential scattering cross section $\frac{d\sigma}{dQ^2}$, total scattering cross section $\sigma(E)$ and 
 various components of the hadron polarizations $P_L^h,~P_T^h,~P_P^h$ and the lepton polarizations $P_L^l,~P_T^l,~P_P^l$, 
 where the subscripts $L$, $T$ and $P$ refer to the longitudinal, transverse and perpendicular components of the polarization 
 vector and $h=N,Y(=\Lambda,\Sigma^{0(-)})$ and $l=\mu ~\text{and}~ \tau$.
 
 \item [(ii)] The sensitivity of these observables to the use of various parameterizations of the vector form factors $f_1 
 (Q^2)$ and $f_2(Q^2)$.
 
 \item [(iii)] The dependence of these observables on numerical values of $M_A$ by varying it within 
 10\% of the world average value.
 
 \item [(iv)] The dependence of these observables on the pseudoscalar form factor, which is important in the case of the final 
 state lepton becoming massive like the $\tau $ lepton. 
 
 \item [(v)] The effect of G-noninvariance by taking $g_2(Q^2)\neq 0$ and parameterizing it in a dipole form with some 
 representative values of $g_2(0)$ taken in the range of $-3$ to $3$.
 
 \item [(vi)] The effect of T-noninvariance by making $g_2(Q^2)$ imaginary with same representative numerical values taken 
 for $Re~g_2(0)$.
\end{itemize} 
From the results presented in Section~\ref{nu_interaction}, it may be concluded that 
\begin{itemize}
 \item [(i)] The total and differential cross sections as well as the polarization observables of the final hadrons in the 
 $\Delta S=0$ QE scattering are almost insensitive to the different parameterizations of the weak vector form 
 factors.
 
 \item [(ii)] There is a significant dependence of $M_{A}$ on the total and differential cross sections while the polarization 
 observables show a little effect on the variation in $M_{A}$, especially in the case of antineutrino induced QE 
 scattering.
 
 \item [(iii)] The presence of SCC shows a strong dependence on the total cross section as well as on the 
 polarization observables, irrespective of the nature of the form factor $g_{2}(Q^2)$~(real or imaginary) for both neutrino 
 and antineutrino induced reactions. 
 
 \item [(iv)] In the case of $\Delta S=0$ reactions, the effect of pseudoscalar form factor is almost negligible for the 
 $\nu_{\mu}~(\bar{\nu}_{\mu})$ induced processes due to the small mass of muon. However, in the case of $\nu_{\tau} ~ 
 (\bar{\nu}_{\tau})$ induced processes, there is some dependence of $g_{3} (Q^2)$ on the polarization observables, especially 
 in the threshold region.
 
 \item [(v)] In the case of polarization observables, it is possible to study T violation by taking the imaginary values of 
 the form factor associated with the SCC. The transverse component of polarization of the final lepton or 
 hadron, perpendicular to the reaction plane, arises due to the imaginary values of $g_{2} (0)$. Therefore, the finite 
 value of $P_{T} (Q^2)$ gives evidence of T violation in the QE reactions. We have found a strong dependence of 
 $g_{2}^{I}$ on the transverse polarization.
 \end{itemize}

In the case of QE induced $1Y$ production, with the increase in $M_A$, $\sigma$ for ${{\bar\nu}_\mu + p 
\longrightarrow \mu^+ + \Lambda}$ increase by about 10\% with $M_A = 1.1$~GeV and $\sim 20\%$ with $M_A = 1.2$~GeV at 
$E_{\bar{\nu}_\mu}=1$~GeV when compared with the cross section obtained using $M_A=1.026$~GeV. We observe that in the case of 
$Q^2$-distribution, the longitudinal $P_L (Q^2)$ and the perpendicular $P_P (Q^2)$ components of polarizations show large 
variations as we change $|g_2^{R} (0)|$ from 0 to 3, which is about 50$\%$ in the peak region of $Q^2$ distribution. With an 
imaginary $g_2(Q^2)$, there is contribution to the transverse polarization also. We find that for the 
process $\bar{\nu}_{\mu} + p \longrightarrow \mu^{+} + \Lambda$, $P_L(Q^2)$ the results for $P_L (Q^2)$ are less sensitive to $g_2^I (0)$ at 
low antineutrino energies, $P_P(Q^2)$ is sensitive to $g_2^{I}(0)$ at $E_{\bar{\nu}_{\mu}} =$ 1~GeV. Moreover, $P_T(Q^2)$ 
shows 40$\%$ variation at $Q^2 = $ 0.4 GeV$^2$, $E_{\bar{\nu}_{\mu}} =$ 1 GeV, when $g_2^{I} (0)$ is varied from 0 to 3.
 
In Sections~\ref{sec:inelastic:nucleon} and \ref{dis:nucleon}, IE and DIS processes on 
nucleons have been discussed. In the case of IE reactions, the production of mesons like $\pi,\eta$ and $K$, and the 
production of hyperons like $\Lambda, \Sigma$, etc. have been taken up. The production of hyperons along with pions like 
$\Sigma \pi, \Lambda \pi$ are also discussed briefly. These IE processes take place through the NR as well as 
the resonance excitation mechanisms. In the case of NR mechanism, an interaction Lagrangian based on the nonlinear 
realization of chiral $SU(3)$ symmetry has been used to describe the interaction of nucleons and hyperons with nonstrange and 
strange mesons like pion, eta and kaon. The meson decay constants $f_\pi, ~f_{K},~ f_\eta,$ and meson masses are treated as 
parameters and their experimental values have been used. The strong couplings like $g_{NN\pi}$, $g_{NKY}$, etc. are expressed 
in terms of the meson decay constants and the symmetric and antisymmetric axial-vector couplings $D$ and $F$, the values of 
these constants have been taken from the PDG~\cite{ParticleDataGroup:2020ssz} for numerical evaluations. 

In the case of resonance excitation mechanisms all the resonance $R$ with $J^{P}=\frac{1}{2}^\pm$ and $J=\frac{3}{2}^\pm$ with 
$I=\frac{1}{2}$ and $\frac{3}{2}$ up to $W=2$~GeV have been included. Specifically we have considered $P_{33}(1232)$, $P_{11}
(1440)$, $D_{13}(1520)$, $S_{11}(1535)$, $S_{31} (1620)$, $S_{11}(1650)$, $D_{33}(1700)$, $P_{11}(1710)$, $P_{13}(1720)$, 
$P_{11}(1880)$, $S_{11}(1895)$, and $P_{13} (1900)$ resonances. We have considered the contribution of individual resonances
and emphasize the role of each resonance in the case of pion, $\eta$ and associated particle productions. The $N\rightarrow R$ 
transition form factors in the vector sector have been calculated from the experimental values of the helicity amplitudes 
taken from the PDG~\cite{ParticleDataGroup:2020ssz}. The $Q^2$ dependence of the vector form factors has been taken from the 
earlier works on electroproduction. In the axial-vector sector, the leading axial-vector form factor is obtained in terms of 
the $R\rightarrow N\pi$ decay width and $\pi\rightarrow \mu\nu$ decay constant $f_\pi$. The  pseudoscalar form factor is 
obtained in terms of the leading axial-vector form factor using PCAC and PDDAC as obtained in the case of $N-N$ transition. 
In case of spin $\frac{3}{2}^{\pm}$ resonances, there are additional form factors whose contributions are neglected. The 
$Q^2$ dependence of the vector and axial-vector form factors for $N - R_{\frac{3}{2}^\pm}$ resonances are treated in analogy 
with $\Delta$ resonance which has been extensively discussed in the literature, while the $Q^2$ dependence of $J=\frac{1}
{2}^\pm$ resonances are treated in analogy with the nucleon case using a dipole parameterization. All the strong couplings 
like the $g_{RN\pi}$, $g_{RKY}$ etc. are taken from the experiments as reported in PDG~\cite{ParticleDataGroup:2020ssz}. The 
results have been presented for the total scattering cross sections for the following processes:
\begin{itemize}
 \item $\nu_{l}({\bar\nu}_{l}) N \rightarrow l^\mp N^\prime \pi^i$, \qquad \qquad 
 ${\bar\nu}_{l} N \rightarrow l^+ Y \pi^i$, $i=\pm, 0$ and $N, N^\prime=p$ or $n$.
 
 \item $\nu_{l}({\bar\nu}_{l}) N \rightarrow l^\mp N^\prime K^j$, \qquad \qquad 
 ${\bar\nu}_{l} N \rightarrow l^+ \Xi K^j$, $j=+, 0$.
 
 \item $\nu_{l}({\bar\nu}_{l}) N \rightarrow l^\mp N^\prime \eta$, 
 \qquad \qquad $\nu_{l}({\bar\nu}_\mu) N \rightarrow l^\mp \Lambda K^j$.
\end{itemize}
We find that in the case of $1\pi^+$ production in $\nu_\mu  p \longrightarrow \mu^-  p \pi^+$ reaction, $\Delta(1232)$ 
resonance has the most dominant contribution. In the case when no cut on the hadronic CM energy $W$ is applied, the presence 
of the NRB terms increase the cross section by about $14\%$ at $E_{\nu_\mu}=1$~GeV which decreases with the 
increase in energy and becomes $\sim 9\%$ at $E_{\nu_\mu}$=2GeV. However, when the cuts on $W$ are applied, then due to the 
presence of background contributions, this increase in the cross section further increases and becomes $\sim 13\%$ at 2~GeV 
for $W < 1.4$~GeV and $12\%$ for $W<1.6$~GeV. While for $1\pi^0$ production in the reaction $\nu_\mu  n \longrightarrow \mu^-  
p  \pi^0$ and $1\pi^+$ production in $\nu_\mu  n \longrightarrow \mu^-  n  \pi^+$ we observe significant contributions from 
the NRB terms as well as from other higher resonance excitations besides the $\Delta(1232)$ dominance. We 
find that the inclusion of the background terms, increases $\sigma$ by about $32\%$ at $E_{\nu_\mu}=1$~GeV 
which becomes $20\%$ at $E_{\nu_\mu}=2$~GeV. When higher resonances are included then there is a further increase of 
about $3\%$ at $E_{\nu_\mu} = 1$~GeV and $40\%$ at $E_{\nu_\mu}=2$~GeV. Similar observations have been made for the antineutrino 
induced processes leading to one pion production.

For $\eta$ production in $\nu_{\mu} n \longrightarrow \mu^{-} \eta p$ and $\bar{\nu}_{\mu} p \longrightarrow \mu^{+} \eta n$ reactions, we find that $S_{11}(1535)$ resonance excitation gives the most dominant contribution and the 
contribution from the NRB terms and from the $S_{11}(1650)$ and $P_{11} (1710)$ resonance excitations are 
almost negligible. For the single kaon production in $\nu_\mu p \longrightarrow \mu^{-} K^+ p $ and $ \nu_\mu n 
\longrightarrow \mu^{-} K^0 p$ reactions, the contact term has the dominant contribution to the total scattering cross section. 
Similar observation is made for the antineutrino induced process where an antikaon is produced.
  
For the associated particle production in the reactions $\nu_{\mu} n \longrightarrow \mu^{-} \Lambda K^{+}$ and 
$\bar{\nu}_{\mu} p \longrightarrow \mu^{+} \Lambda K^0$, the background terms give the largest contribution and the 
contributions from the resonances like $S_{11}(1650)$, $P_{11} (1710)$ and $P_{13} (1720)$ considered in this work are small. 
In the case of neutrino induced $K\Lambda$ production, $P_{11} (1710)$ has the largest contribution among the resonances. While 
in the case of antineutrino induced associated particle production process, a destructive interference between the 
background and the resonance terms occurs and the results obtained with background terms only are almost two times the results of 
full model in the entire energy range. Among the resonances, the most dominant contribution is from $P_{13} (1720)$ followed 
by $P_{11} (1710)$ and $S_{11}(1650)$ in the low energy region~($E_{\bar{\nu}_{\mu}} <1.5$~GeV). We would like to emphasize 
that the results of single hyperon production cross section are larger at low antineutrino energies in the region of 1.2 GeV 
and are comparable even at $E_{{\bar\nu}_\mu}=2$~GeV with the cross section for the associated particle production of 
hyperons obtained in the present model.

In the case of DIS, the cross sections are calculated in the leading order of the perturbative QCD 
using the QPM in terms of the nucleon structure functions $F_i^N(x,Q^2); i=1-5$, using the Callan-Gross relation between 
$F_2(x,Q^2)$ and $F_1(x,Q^2)$, and Albright-Jarlskog relation between $F_5(x,Q^2)$ and $F_2(x,Q^2)$. The evaluation of the 
structure functions $F_i(x,Q^2)~(i=1,2,3,5)$ at NLO and NNLO has been obtained using DGLAP equation for which results have 
been quoted. The corrections to the nucleon structure functions due to TMC and HT effects have been described qualitatively 
so that they can be applied to lower $Q^2$ connecting to SIS or the transition region. We find 
these corrections to be substantial in some regions of $x$ and $Q^2$. 

Finally in Section~\ref{nu:nuclei}, we study the NME in the QE, IE and DIS. In the case of QE scattering, NME are described in the low energy as 
well as in the intermediate energy regions. In the low energy region relevant to the solar, reactor and accelerator 
neutrinos~(for 
(anti)neutrinos obtained from pions and kaons decaying at rest), the exclusive reactions have been studied in some nuclei like 
$^{12}$C and $^{56}$Fe, using the method based on multipole expansion. The NME are included by using the 
nuclear wave functions calculated with realistic nucleon-nucleon potentials which include the effect of nucleon correlations 
and pairing, etc. and reproduce the binding energy and other static properties of the nuclei. 

In the region of intermediate and higher energies, the method based on LFGM has been used in which the 
long range correlations are included in a RPA approach to augment the results obtained by calculating the cross sections and 
angular as well as the energy distributions of the final state leptons based on the 1p-1h excitation. Further modifications of 
the cross sections and energy/angular distributions due to the 2p-2h and MEC have been discussed and 
the results are compared with the experimental results obtained by the MiniBooNE, T2K, MINERvA, etc. collaborations.
 
The effects due to Fermi motion, Pauli blocking, binding energy, etc. for CCQE processes reduce the cross section by $\sim 30(42)\%$ at 
$E_{\nu} = 0.3$~GeV which becomes $20(30)\%$ at $E_{\nu} = 0.6$~GeV, compared to the free nucleon case. When RPA correlation 
effects are included a further reduction of $\sim 55(56)\%$ at $E_{\nu} = 0.3$~GeV and $35(45)\%$ at $E_{\nu} = 0.6$~GeV is 
obtained. This reduction further gets enhanced when $A$ is increased in the case of heavier nuclei like $^{40}$Ar, $^{56}$Fe 
and $^{208}$Pb. Due to threshold effect, the reduction is larger for $\nu_{\mu}$ and $\bar{\nu}_{\mu}$ induced CCQE processes 
at lower energies, which has been discussed quantitatively in Section~\ref{QE_nucleus}.

For the single hyperon production in ${\bar\nu}_\mu$ induced CCQE process, the effects of Fermi motion and Pauli blocking are 
found to be negligible, while the effect of FSI is substantial, for example, the enhancement in the $\Lambda$ production 
cross section is 22--25$\%$ in $^{12}$C and $^{16}$O for $E_{{\bar\nu}_\mu}=0.6-1$~GeV, which increases to 34--38$\%$ in 
$^{40}$Ar and 52--62$\%$ in $^{208}$Pb. While the decrease in $\Sigma^-$ production cross section is about 40--46$\%$ in 
$^{12}$C and $^{16}$O for $E_{{\bar\nu}_\mu}=0.6-1$~GeV, which becomes 50--56$\%$ in $^{40}$Ar and 68--70$\%$ in $^{208}$Pb. 
The FSI effect also results in the production of $\Sigma^+$, which is forbidden in the free nucleon case due to the $\Delta S 
= \Delta Q$ rule. These hyperons decay to pions in the nucleus and these pions contribute significantly to the total 1$\pi$ 
production for $E_{{\bar\nu}_\mu} < 1.2$GeV, which is generally assumed to be dominated by the $\Delta$ resonance.

In the case of IE processes, NME are discussed only in the case of pion production which are 
dominated by the $\Delta$-excitation. The nuclear medium modifications of the properties of $\Delta$, specially in their mass 
and decay width have been included in calculating the $\Delta$ excitation and its decays. After pions are produced in the 
nuclear medium, the final state interaction of pions with nucleons in the medium is included in a Monte Carlo approach in 
which the pions go through the process of elastic scattering, charge exchange scattering and absorption as they travel through 
the medium. We find that NME and FSI effects together results in a net reduction of about 40$\%$ to the total pion production 
for neutrinos of 1~GeV energy range. 

In the case of DIS, NME are included by calculating the nucleon spectral function $S({\vec p}, E)$ for 
the initial nucleon in a relativistic field theoretical approach to describe the { energy and momentum distribution of the nucleons in the nucleus} which 
includes the effect of nuclear binding, Fermi motion and nucleon-nucleon correlations specially the long range correlation. 
The nucleon structure functions $F_i^N (x, Q^2)~(i=1-5)$ are then convoluted with the spectral function to include NME and 
the nuclear structure functions $F_i^A (x, Q^2)~(i=1-5)$ are calculated. The effect of TMC and HT are 
calculated at the nucleon level before convoluting it with the nucleon spectral function $S({\vec p}, E)$ in the nuclei. At 
the nuclear level, the effect of shadowing and antishadowing is also included in a multiple scattering model used by Kulagin 
and Petti. These nuclear structure functions are then used to calculate the differential and total scattering cross sections. 
We find that the inclusion of NME are also important in the DIS region.
 
In addition to these considerations of NME, some new processes also become relevant in the case of 
(anti)neutrino scattering with nuclear targets like
\begin{itemize}
 \item [(i)] Coherent elastic neutrino-nucleus scattering,
 \item [(ii)] Coherent production of $\pi$ and $K$ mesons,
 \item [(iii)] Trident production and its enhancement in the case of nuclear targets,
\end{itemize}
which are also discussed in view of their importance in contemporary studies of neutrino interactions with nuclei.
 
The review article attempts to present an overall picture of the properties of neutrinos and their interactions with nucleons 
and nuclear targets and provides experimenters, phenomenologists as well as theorists the current understanding of 
neutrino interactions with matter in the few GeV energy region.


\section*{Acknowledgements}

It is a great pleasure to thank Prof. Amanda Faessler, the then Editor-in-Chief of Progress in Particle and Nuclear
Physics, who invited MSA to write the review article. We want to thank the present Editor-in-Chief, Prof. Christian Fischer for
being considerate in granting us time.
We would like to thank our collaborators Eulogio Oset, Manuel J.
Vicente Vacas, and Luis Alvarez Ruso from the University of Valencia, Spain; Ignacio Ruiz Simo, University of Granada,
Spain, Takaaki Kajita and (Late)Morihiro Honda at the University of Tokyo, Japan, Zubair Ahmad Dar at
William and Mary, US, Faiza Akbar at the University of Rochester, US, and Farhana Zaidi, Huma Haider, M. Rafi Alam,
Shakeb Ahmad and Shikha Chauhan at the Aligarh Muslim University, India. Farhana Zaidi, Vaniya Ansari, Sayeed Akhter and
Prameet Gaur are acknowledged for their help in the preparation of the present manuscript. MSA and AF are thankful to the
Department of Science and Technology (DST), Government of India for providing financial assistance under Grant No.
SR/MF/PS-01/2016-AMU.
	


	
	\appendix
	\renewcommand*{\thesection}{\Alph{section}}
	

\section{Expression of the hadronic current $J_{\mu\nu}$}\label{appendix}
\begin{eqnarray}
J_{\mu\nu}&=&\frac{1}{2} a^2~\Bigg[4f_1^2(Q^2) \left(p_\mu^\prime p_\nu + p_\nu^\prime p_\mu - (p\cdot p^\prime- M M^\prime) 
g_{\mu\nu} \right)\nonumber\\ 
&+& \left. 4\frac{f_2^2 (Q^2)}{(M + M^\prime)^2} \left(- M M^\prime Q^2 g_{\mu \nu} + q_{\mu} \left(-q_{\nu} (M M^\prime 
+p \cdot p^\prime)+p^{\prime}_{\nu} \left(p \cdot q\right) + p_{\nu} \left(p^\prime \cdot q\right) \right)\right. \right. 
\nonumber \\
&-& \left. \left. 2 g_{\mu \nu} \left(p \cdot q\right) \left(p^\prime \cdot q\right) - Q^2 g_{\mu \nu} \left(p \cdot 
p^\prime \right) + Q^2 p_{\mu} p^{\prime}_{\nu} +p_{\mu} q_{\nu} \left(p^\prime \cdot q\right) + p^{\prime}_{\mu} 
\left(q_{\nu} \left(p \cdot q\right) + Q^2 p_{\nu} \right) \right) \right. \nonumber \\
&+& \frac{16 f_3^2 (Q^2)}{(M+M^\prime)^2} \left(q_{\mu} q_{\nu} (M M^\prime + p \cdot p^\prime) \right) + 4g_1^2 
(Q^2) \left((p_\mu^\prime p_\nu + p_\mu p_\nu^\prime) - (p\cdot p^\prime +M M^\prime) g_{\mu\nu} \right) \nonumber\\
&+& \frac{4 g_2^2 (Q^2)}{(M + M^\prime)^2} \left(M M^\prime Q^2 g_{\mu \nu} + q_{\mu} \left(q_{\nu} (M M^\prime- p\cdot 
p^\prime) + p^{\prime}_{\nu} \left(p\cdot q\right) + p_{\nu} \left(p^\prime \cdot q\right) \right)\right. \nonumber \\
&-& \left. 2 g_{\mu \nu} \left(p\cdot q\right) \left(p^\prime \cdot q\right) - Q^2 g_{\mu \nu} \left(p\cdot p^\prime\right) 
+ Q^2 p_{\mu} p^{\prime}_{\beta} + p_{\mu} q_{\nu} \left(p^\prime \cdot q\right) + p^{\prime}_{\mu} \left(q_{\nu} \left(p 
\cdot q\right) + Q^2 p_{\nu}\right) \right) \nonumber \\
&+& \frac{16 g_3^2 (Q^2)}{(M+M^\prime)^2} \left(q_\mu q_\nu(p^\prime\cdot p -M M^\prime) \right) \nonumber \\
&+&\frac{4f_1 (Q^2) f_2 (Q^2)}{(M+M^\prime)} \left(q_{\mu} \left(M^\prime p_{\nu}- M p^{\prime}_{\nu} \right) + 2 M g_{\mu 
\nu} \left(p^{\prime} \cdot q\right) - M p^{\prime}_{\mu} q_{\nu} - 2 M^\prime g_{\mu \nu} \left(p \cdot q \right) + M^\prime 
p_{\mu} q_{\nu} \right) \nonumber\\
&+&\frac{8f_1(Q^2) f_3 (Q^2)}{(M+M^\prime)} \left(q_{\mu} \left(M p^{\prime}_{\nu} + M^\prime p_{\nu}\right)+ q_{\nu} 
\left(M p^{\prime}_{\mu}+ M^\prime p_{\mu}\right) \right) +8 i f_1 (Q^2) g_1(Q^2) \left(\epsilon_{\mu \nu \alpha \beta} ~
{p}^\alpha {p^{\prime}}^\delta \right) \nonumber \\
&+& \frac{ 8i f_1(Q^2) g_2 (Q^2)}{(M+M^\prime)} \left(M^\prime \epsilon_{\mu \nu \alpha \beta} p^{\alpha} q^{\beta} - 
M \epsilon_{\mu \nu \alpha \beta} {p^{\prime}}^\alpha q^{\beta} \right) + \frac{8 f_2(Q^2) f_3 (Q^2)}{(M+M^\prime)^2} 
\left(q_{\nu} \left(p_{\mu} \left(p^\prime \cdot q\right)- p^{\prime}_{\mu} \left(p \cdot q\right)\right) \right. \nonumber \\
&+& \left. q_{\mu} \left(p_{\nu} \left(p^\prime \cdot q\right) - p^{\prime}_{\nu} \left(p \cdot q\right)\right) \right) + 8 i 
\left(\frac{f_2 (Q^2) g_1(Q^2)}{(M+M^\prime)}\right) \left(M \epsilon_{\mu \nu \alpha \beta} {p^{\prime}}^\alpha q^{\beta} +
M^\prime \epsilon_{\mu \nu \alpha \beta} p^{\alpha} q^{\beta} \right) \nonumber\\
&+& \frac{8 i f_2 (Q^2) g_2 (Q^2)}{(M+M^\prime)^2} \left(q_{\mu} \epsilon_{\nu \alpha \beta \delta} p^{\alpha} 
{p^{\prime}}^\beta q^{\delta} -q_{\nu} \epsilon_{\mu \alpha \beta \delta} p^{\alpha} {p^{\prime}}^\beta q^{\delta} - Q^2 
\epsilon^{\mu \nu \alpha \beta} p^{\alpha} {p^{\prime}}^\beta + 2 \left(p \cdot q\right) \epsilon^{\mu \nu \alpha \beta} 
p^{\prime \alpha} q^{\beta} \right) \nonumber \\
&+& \frac{8 i f_2 (Q^2) g_3 (Q^2)}{(M + M^\prime)^2} \left(q_{\mu} \epsilon_{\nu \alpha \beta \delta} p^{\alpha} 
{p^{\prime}}^\beta q^{\delta} - q_{\nu} \epsilon_{\mu \alpha \beta \delta} p^{\alpha} {p^{\prime}}^\beta q^{\delta} \right)
\nonumber\\
&+& \frac{8 i f_3 (Q^2) g_2 (Q^2)}{(M+M^\prime)^2} \left(q_{\mu} \epsilon_{\nu \alpha \beta \delta} p^{\alpha} 
{p^{\prime}}^\beta q^{\delta} - q_{\nu} \epsilon_{\mu \alpha \beta \delta} p^{\alpha} {p^{\prime}}^\beta q^{\delta} \right) 
\nonumber \\
&+& \frac{4 g_1(Q^2) g_2 (Q^2)}{(M+M^\prime)} \left(q_{\mu} \left(M p^{\prime}_{\nu} + M^\prime p_{\nu}\right)-2 M 
g_{\mu \nu} \left(p^{\prime} \cdot q\right) + M p^{\prime}_{\mu} q_{\nu} -2 M^\prime g_{\mu \nu} \left(p \cdot q\right) + 
M^\prime p_{\mu} q_{\nu} \right) \nonumber \\
&+&\frac{8 g_1(Q^2) g_3(Q^2)}{(M+M^\prime)} \left(q^{\mu } \left(M^\prime p_{\nu} - M p^{\prime}_{\nu}\right) + q_{\nu} 
\left(M^\prime p_{\mu}- M p^{\prime}_{\mu}\right) \right) \nonumber \\
&+& \frac{8 g_2(Q^2) g_3(Q^2)}{(M+M^\prime)^2} \left(q_{\nu} \left(p_{\mu} \left(p^{\prime} \cdot q\right) - p^{\prime}_{\mu} 
\left(p \cdot q\right)\right)+  q_{\mu} \left(p_\nu \left(p^{\prime} \cdot q\right) - p^{\prime}_{\nu} \left(p \cdot q\right) 
\right) \right) \Bigg],
\end{eqnarray}
where $a=\cos\theta_C(\sin\theta_C)$ for $\Delta S =0(1)$ CC induced processes and $a=1$ for NC induced processes.
\subsection{Expressions of $N(Q^2)$, $A^h(Q^2)$, $B^h(Q^2)$, and $C^h(Q^2)$}\label{appendix1}
The expressions $N(Q^2)$, $A^h(Q^2)$, $B^h(Q^2)$, and $C^h(Q^2)$ are expressed in terms of the Mandelstam variables and 
the form factors as:
 \begin{eqnarray}
N(Q^2) &=& a^2\left\{f_1^2(Q^2)\left(\frac{1}{2} \left(2 \left(M^2-s\right) \left({M^\prime}^2-s\right)-t \left(\Delta^2-2 s
\right) +t^2 +  m_{l}^2 \left(\Delta^2-2 s-t\right) \right) \right) \right. \nonumber \\
&+& \frac{f_2^2(Q^2)}{(M + M^\prime)^2} \left(\frac{1}{4} \left(-2 t \left(M^4-2 s \left(M^2 + {M^\prime}^2\right) + 
{M^\prime}^4+2 s^2\right)+2 t^2 \left((M + M^\prime)^2-2 s\right) \right. \right. \nonumber \\
&+& \left. \left. m_{l}^2 \left(2 \Delta (M + M^\prime) \left(M^2 + {M^\prime}^2-2 s\right)+t \left(
(M - 3 M^\prime) (M + M^\prime)+4 s)+t^2\right) \right. \right. \right. \nonumber
\end{eqnarray}
\begin{eqnarray}
&+& \left. \left. m_{l}^4 (-((3 M-M^\prime) (M+ M^\prime)+t)) \right)\right) \nonumber \\
&+& g_1^2 (Q^2) \left(\frac{1}{2} \left(2 \left(M^2-s\right) \left({M^\prime}^2-s\right)-t \left((M + M^\prime)^2-2 s 
\right)+t^2 + m_{l}^2 \left((M + M^\prime)^2-2 s-t\right) \right) \right)\nonumber \\
&+& \frac{|g_2(Q^2)|^2}{(M + M^\prime)^2} \left( \frac{1}{4} \left(4 \left(\Delta^2-t\right) \left(\left(M^2-s\right) 
\left({M^\prime}^2-s\right)+s t\right)+ m_{l}^2 \left(4 \Delta \left(M^3+ M^2 M^\prime - M (3 s+t) + M^\prime s
\right) \right. \right. \right. \nonumber\\
&+& \left. \left. \left.2 \Delta ^2 \left((M + M^\prime)^2-2 s-t\right)-(4 s+t) \left(\Delta^2-t\right)\right) 
+ 2 \Delta ^2 \left(- 2 \left(M^2-s\right) \left({M^\prime}^2-s\right)-t \left((M + M^\prime)^2+2 s\right)
+t^2\right) \right. \right. \nonumber \\
&+& \left.  m_{l}^4 \left(\Delta^2+4 M \Delta -t\right) \right) \Big) + \frac{g_3^2(Q^2)}{(M+M^\prime)^2} \left(m_{l}^2 
\left(m_{l}^2-t\right) \left(\Delta^2-t\right) \right) \nonumber \\
&+& \frac{f_1 (Q^2) f_2 (Q^2)}{(M+M^\prime)} \left(- \left(t (M + M^\prime) \left(\Delta^2-t\right) + m_{l}^2 
\left(- \Delta \left({M^\prime}^2-s \right)+M^\prime t\right) + m_{l}^4 M \right) \right) \nonumber \\
&\pm& f_1(Q^2) g_1 (Q^2) \left( - \left(t \left(M^2+{M^\prime}^2-2 s-t\right) + m_{l}^2 \left(M^2-{M^\prime}^2+t\right)
\right) \right) \nonumber \\
&\pm& \frac{Re[f_1 (Q^2) g_2(Q^2)]}{(M+M^\prime)} \left( - \Delta  \left(t \left(M^2+ {M^\prime}^2-2 s-t\right) + 
m_{l}^2 \left(M^2- {M^\prime}^2+t\right)\right) \right) \nonumber \\
&\pm& \frac{f_2 (Q^2) g_1 (Q^2)}{(M+M^\prime)} \left( -(M + M^\prime) \left(t \left(M^2 + {M^\prime}^2-2 s-t\right) + 
m_{l}^2 \left(M^2 - {M^\prime}^2+t\right)\right) \right) \nonumber \\
&\pm& \frac{Re[f_2 (Q^2) g_2(Q^2)]}{(M+M^\prime)^2} \left( \Delta  (-(M + M^\prime)) \left( t \left(M^2 + {M^\prime}^2 
-2 s-t\right) + m_{l}^2 \left(M^2 - {M^\prime}^2+t\right)\right) \right) \nonumber \\
&+& \frac{Re[g_1 (Q^2) g_2(Q^2)]}{(M+M^\prime)} \left( \left(\Delta  \left(-t (M + M^\prime)^2 +t^2\right)  + 
m_{l}^2 \left(M^3 + M^2 M^\prime + \Delta \left((M + M^\prime)^2-2 s-t\right) \right. \right. \right. \nonumber \\
&-& \left. 3 M s- M t+ M^\prime s\right) +m_{l}^4 M \left. \left. \right) \right) + \frac{g_1(Q^2) g_3 (Q^2)}{(M+M^\prime)} 
\left( -2  m_{l}^2 \left(m_{l}^2 M + M^3 - M^2 M^\prime - M (s+t)+M^\prime s\right) \right) \nonumber \\
&+&\left.\frac{Re[g_2 (Q^2) g_3 (Q^2)]}{(M+M^\prime)^2} \left( m_{l}^2 \left(- 2 \Delta \left(m_{l}^2 M + M^3 - 
M^2 M^\prime - M (s+t ) + M^\prime s \right) - \left(\Delta^2-t\right) \left(m_{l}^2+2 M^2-2 s-t\right)   \right) \right) \right\}
\end{eqnarray}
where $(+)-$ sign represents the (anti)neutrino induced scattering and the Mandelstam variables are defined as,
\begin{eqnarray}
 s = M^2 + 2 ME, \qquad \quad t = M^2 + {M^\prime}^2 - 2 M E^\prime, 
\end{eqnarray}
with $\Delta = M^\prime - M$.
\begin{eqnarray}\label{Ah}
 A^h(Q^2) &=& - 2a^2\left[ f_1^2 (Q^2) \left(\pm \frac{1}{2} (M + M^\prime) \left(\Delta^2-t\right) \right) 
 \pm \frac{f_2^2 (Q^2)}{(M+M^\prime)^2} \left( \frac{1}{2} t (M + M^\prime) \left(\Delta^2-t\right) \right) \right. 
 \nonumber \\
 &\pm & g_1^2 (Q^2) \left( \frac{1}{2} \Delta \left((M + M^\prime)^2-t\right) \right) 
 \pm \frac{|g_2 (Q^2)|^2}{(M+M^\prime)^2} \left(\frac{1}{2} t \Delta \left((M + M^\prime)^2-t\right) \right) 
 \nonumber \\
 &\pm& \frac{f_1 (Q^2) f_2 (Q^2)}{(M+M^\prime)} \left( \frac{1}{2} \left(4 M M^\prime t+t^2- \Delta^2 \left(M + 
 M^\prime \right)^2 \right) \right) + f_1 (Q^2) g_1 (Q^2) \left( - M^\prime \left(M^2 + {M^\prime}^2-2 s-t\right) 
 \right) \nonumber \\
 &+& \frac{Re[f_1 (Q^2) g_2 (Q^2)]}{(M+M^\prime)} \left( \frac{1}{2} \left(\left( M^2 + {M^\prime}^2-2 s-t\right) 
 \left(-t -2 M^\prime \Delta + \Delta^2 \right) + m_{l}^2 \left(\Delta^2-t\right)\right) \right) \nonumber\\
 &+& \frac{f_1 (Q^2) g_3 (Q^2)}{(M+M^\prime)} \left( -m_{l}^2 \left(\Delta^2-t\right) \right) + \frac{f_2 (Q^2) g_1 (Q^2)}
 {(M+M^\prime)} \left( \frac{1}{2} \left (\left(M^2 - {M^\prime}^2-t\right) \left(M^2 + 
 {M^\prime}^2-2 s-t\right) \right.\right. \nonumber\\
 &+& \left.\left. m_{l}^2 \left((M + M^\prime)^2-t\right)\right) \right) + \frac{Re[f_2 (Q^2) g_2 (Q^2)]}{(M+M^\prime)^2} 
 \left( \frac{1}{2} \left((M + M^\prime) \left(\Delta^2-t\right) \left(m_{l}^2 + M^2 + {M^\prime}^2-2 s-t\right) \right. 
 \right. \nonumber \\
 &+&  \left. \Delta  \left(m_{l}^2 \left((M + M^\prime)^2-t\right)+\left(M^2 - {M^\prime}^2-t\right) 
 \left(M^2 + {M^\prime}^2-2 s-t\right)\right)\right) \Bigg) \nonumber \\ 
 &+& \frac{f_2 (Q^2) g_3 (Q^2)}{(M+M^\prime)^2} \left( -m_{l}^2 (M + M^\prime) \left(\Delta^2-t\right) \right) \pm
 \frac{Re[g_1 (Q^2) g_2 (Q^2)]}{(M+M^\prime)} \left( \frac{1}{2} \left((M + M^\prime)^2-t\right) 
 (- \Delta^2 -t) \right) \Bigg]\\
 \label{Bh}
 B^h(Q^2) &=& a^2\frac{2}{M^\prime} \left[ f_1^2 (Q^2) \left(\pm \frac{1}{4} \left(t \left(\Delta^2-2 s\right)-t^2 - 2 
 M^\prime \Delta \left(M^2-s\right) +m_{l}^2 \left(M^2+2 M M^\prime - {M^\prime}^2+t\right) \right) \right) \right. 
 \nonumber \\
 &\pm& \frac{f_2^2 (Q^2)}{(M+M^\prime)^2} \left( \frac{1}{4} \left(t (M + M^\prime) \left(M^3 + M^2 M^\prime - M 
 \left({M^\prime}^2+2 s+t\right) + {M^\prime}^3 - M^\prime t \right) \right. \right. \nonumber
 \end{eqnarray}
 \begin{eqnarray}
 &+& \left.  m_{l}^2 \left(M^4+t (M + M^\prime)^2 - {M^\prime}^4 \right) \right)\Bigg) \pm g_1^2 (Q^2) \left( \frac{1}{4} ( 
 \left(-2 M^\prime (M +  M^\prime) \left(M^2-s\right) + t \left((M + M^\prime)^2
 -2 s\right)\right. \right. \nonumber\\
 &-& \left. \left. t^2 + m_{l}^2 \left(M^2-2 M M^\prime - {M^\prime}^2+t\right)\right) \right) 
 \pm \frac{|g_2 (Q^2)|^2}{(M+M^\prime)^2} \left( \frac{1}{4} \Delta  \left(2 M^\prime \left(-2 m_{l}^2 M^2 - M^4 + 
 M^2 \left({M^\prime}^2+s+t\right)\right. \right. \right. \nonumber\\
 &+& \left. \left.\left. s \left(t- {M^\prime}^2\right) \right) + \Delta \left(- 2 M^\prime (M + M^\prime) \left(M^2-s\right)
 +t \left((M + M^\prime)^2-2 s\right)-t^2\right) + m_{l}^2 \left(M^2 - 2 M M^\prime - {M^\prime}^2+t\right) \right) \right) 
 \nonumber \\
  &\pm& \frac{f_1(Q^2) f_2 (Q^2)}{(M+M^\prime)} \left( \frac{1}{2} \left(M^4 M^\prime + M^3 t- M^2 M^\prime 
 \left({M^\prime}^2+ s\right)-M t \left({M^\prime}^2+2 s+t\right)+ M^\prime \left({M^\prime}^2-t\right) (s+t) \right. 
 \right. \nonumber \\
&+& \left. \left. m_{l}^2 \left(M^3 + M^2 M^\prime + M \left({M^\prime}^2+t\right) - {M^\prime}^3 + M^\prime t\right) 
\right) \right) + f_1 (Q^2) g_1 (Q^2) \left( \frac{1}{2} \left(t \left(M^2 + {M^\prime}^2-2 s\right) - 2 s 
\left(s-M^2\right) - t^2 \right.\right. \nonumber\\
&-&  \left. \left. m_{l}^2 \left(M^2 + {M^\prime}^2-2 s-t\right) \right) \right) + \frac{Re[f_1 (Q^2) g_2 (Q^2)]}
{(M+M^\prime)} \left( -\frac{1}{2} \left(M^4 M^\prime -2 M^3 s + M^2 \left(
 {M^\prime}^3 - M^\prime (s+t)-\Delta  (2 s+t)\right) \right. \right. \nonumber \\
 &+& \left. \left. 2 M s (s+t) - {M^\prime}^3 s - {M^\prime}^2 \Delta  t + M^\prime s t + 2 \Delta  s^2+2 \Delta  s t+\Delta  
 t^2 + m_{l}^2 
 \left(M^3 + M^2 \Delta -M (3 s+t)+\Delta  \left({M^\prime}^2-2 s-t\right)\right) \right. \right. \nonumber \\
 &+&\left. \left. m_{l}^4 M \right) \right) + \frac{f_1(Q^2) g_3 (Q^2)}{(M+M^\prime)} \left( m_{l}^2 M \left(m_{l}^2 + M^2 - 
 s - t\right) \right) +\frac{f_2 (Q^2) g_1 (Q^2)}{(M+M^\prime)} \left( \frac{1}{2} \left( - M^\prime \left(M^2 - s\right) 
 \left(M^2 + {M^\prime}^2 - 2 s\right) \right. \right. \nonumber \\
 &+& \left. \left. t \left(M^3+2 M^2 M^\prime + M {M^\prime}^2 + {M^\prime}^3 - 3 M^\prime s\right) -
 t^2 (M + M^\prime) - m_{l}^2 \left(M^2 M^\prime + M \left({M^\prime}^2+s\right) + M^\prime 
 \left({M^\prime}^2 - 2 s - t\right)\right) \right. \right. \nonumber\\
 &+& \left. \left. m_{l}^4 M \right) \right) + \frac{Re[f_2 (Q^2) g_2 (Q^2)]}{(M+M^\prime)^2} \left( \frac{1}{4} \left(2 
 \left(M^3 \left(- {M^\prime}^3 + M^\prime (3 s+t) + \Delta  t\right) + M^2 \left({M^\prime}^4 - {M^\prime}^3 \Delta - 
 {M^\prime}^2 (3 s+t) \right. \right. \right. \right. \nonumber \\
 &+& \left. \left. \left. \left. M^\prime \Delta (3 s+2 t)+2 s t\right) + M M^\prime s \left( {M^\prime}^2 - 2 s-3 
 t\right) + s \left({M^\prime}^2-t\right) \left(2 (s+t)-{M^\prime}^2\right) + M \Delta  t \left({M^\prime}^2 
 -t\right) \right. \right. \right. \nonumber \\
 &+& \left. \left. \left. M^\prime \Delta  (s+t) \left( {M^\prime}^2 - 2 s-t\right) \right) +  m_{l}^2 \left(-2 M^4 - 2 M^3 
 M^\prime + M^2 (-2 M^\prime  \Delta +2 s+t)-2 M \Delta \left({M^\prime}^2 + s\right) + 2 M M^\prime (3 s+t) \right. \right. 
 \right.  \nonumber \\
 &+& \left. \left.  2 M^\prime \Delta  \left(- {M^\prime}^2 + 2 s+t\right) - \left({M^\prime}^2-t\right) 
 (4 s+t)\right) - m_{l}^4 \left(3 M^2 - {M^\prime}^2 + t\right) \right) \Bigg)   \nonumber \\
 &+& \frac{f_2 (Q^2) g_3 (Q^2)}{(M+M^\prime)^2} \left( \frac{1}{2} m_{l}^2 \left(m_{l}^2 \left(M^2+2 M M^\prime - 
 {M^\prime}^2 + t\right) - 2 M^\prime \Delta \left(M^2-s\right)+t \left(\Delta^2 - 2 s\right) - t^2\right) 
 \right) \nonumber \\
 &\pm& \frac{Re[g_1(Q^2) g_2 (Q^2)]}{(M+M^\prime)} \left( \frac{1}{2} \left(M^\prime \left(-2 m_{l}^2 M^2 - M^4 + 
 M^2 \left({M^\prime}^2 + s + t\right) + s \left(t - {M^\prime}^2\right)\right) \right. \right. \nonumber \\
 &+& \left. \left. \Delta  \left(m_{l}^2 \left(M^2 - 2 M M^\prime - {M^\prime}^2 + t\right) - 2 M^\prime (M + 
 M^\prime) \left(M^2 - s\right) + t \left((M + M^\prime)^2 - 2 s\right) - t^2\right) \right) \right)  \Big] 
 \end{eqnarray}
 \begin{eqnarray}
 C^h(Q^2) &=& 2a^2 \left[\pm \frac{Im[f_1(Q^2) g_2(Q^2)]}{(M+M^\prime)}\left(-t +2 M \Delta + \Delta^2  \right) \pm
 \frac{Im[f_2(Q^2) g_2(Q^2)]}{(M+M^\prime)}\left(-t +\frac{\Delta  \left(M^2-{M^\prime}^2+t\right)}{M+ M^\prime}+
 \Delta^2 \right) \right. \nonumber \\
 &+& \left. \frac{Im[g_1(Q^2) g_2(Q^2)]}{(M+M^\prime)}\left(M^2 + {M^\prime}^2-2 s-t +  m_{l}^2  \right) 
 + \frac{Im[g_3(Q^2) g_2(Q^2)]}{(M+M^\prime)^2}\left( 2 m_{l}^2 \Delta \right) \right] 
 \end{eqnarray}
 \subsection{Expressions of $A^l(Q^2)$, $B^l(Q^2)$ and $C^l(Q^2)$ }\label{appendix2}
The expressions $A^l(Q^2)$, $B^l(Q^2)$ and $C^l(Q^2)$ are expressed in terms of the Mandelstam variables and the form 
factors as:
\begin{eqnarray}\label{Al}
 A^l(Q^2)&=& 2a^2 \left[f_1^2(Q^2) \left(\frac{1}{2} m_{l} \left(M^2+2 M M^\prime-s\right) \right) \right. \nonumber \\
 &+&  \frac{f_2^2(Q^2)}{(M + M^\prime)^2} \left( \frac{1}{4} m_{l} \left(-m_{l}^2 ((3 M-M^\prime) 
 (M+M^\prime)+t)-2  M^4+2 M^2 \left({M^\prime}^2+s\right)+4 M M^\prime t \right. \right.  \nonumber \\
 &-& \left. \left. 2 \left({M^\prime}^2-t\right) (s+t)\right) \right) +g_1^2(Q^2) \left( \frac{1}{2} m_{l} \left(M^2-2 M 
 M^\prime-s\right)\right) \nonumber \\
 &+& \frac{|g_2 (Q^2)|^2}{(M+M^\prime)^2} \left(\frac{1}{4} m_{l} \left(\left(\Delta^2-t\right) \left(m_{l}^2
 +4 M^2+2 {M^\prime}^2-2 s-2 t\right)+2 \Delta  \left(2 m_{l}^2 M - 4M^2 \Delta \right. \right. \right. 
 \nonumber \\
 &+& \left. \left.  3 M \left({M^\prime}^2-s- t\right)+M^\prime \left(-{M^\prime}^2+s+t\right)\right)+2 
 \Delta ^2 \left(M^2-2 M M^\prime-s\right)\right) \Bigg) \nonumber 
 \end{eqnarray}
 \begin{eqnarray}
 &+& \frac{g_3^2 (Q^2)}{(M+M^\prime)^2} \left(m_{l}^3 \left(\Delta^2-t\right) \right) +\frac{f_1(Q^2) f_2(Q^2)}{(M+M^\prime)} 
 \left(-\frac{1}{2} m_{l} \left(2 m_{l}^2 M - \Delta \left(2 M^2-
 {M^\prime}^2-s\right)-t (3 M+M^\prime)\right) \right) \nonumber \\
 &+& f_1(Q^2) g_1(Q^2) \left(-m_{l} \left(M^2-s\right) \right) + \frac{Re[f_1(Q^2) g_2(Q^2)]}{(M+M^\prime)} 
 \left(-m_{l} \Delta  \left(M^2-s\right) \right) \nonumber \\
 &+& \frac{f_2(Q^2) g_1(Q^2)}{(M+M^\prime)} \left(-m_{l} (M+M^\prime) \left(M^2-s\right) \right) + \frac{Re[f_2(Q^2)
 g_2 (Q^2)]}{(M+M^\prime)^2} \left( -m_{l} \Delta  (M+M^\prime) \left(M^2-s\right) \right) \nonumber \\
  &+& \frac{Re[g_1(Q^2) g_2(Q^2)]}{(M+M^\prime)} \left(\frac{1}{2} m_{l}  \left(2 m_{l}^2 M +4 M^3+2 M^2 (\Delta -
 2 M^\prime)+M M^\prime (3 M^\prime-4 \Delta ) \right. \right. \nonumber \\
 &-& \left. \left. 3 M (s+t)-{M^\prime}^3+M^\prime (s+ t)-2 \Delta  s\right) \right) + \frac{g_1(Q^2) g_3(Q^2)}{(M+M^\prime)} 
 \left(- \left(2 m_{l}^3 M - m_{l} \Delta \left({M^\prime}^2-s-t\right)\right) \right) \nonumber \\
 &+& \frac{Re[g_3(Q^2) g_2(Q^2)]}{(M+M^\prime)^2} \left(-m_{l} \left(m_{l}^2 \left(\Delta^2+2 M \Delta - t\right) - t
 ({M^\prime}^2-s-t)\right) \right) \Big]\\
 B^l(Q^2)&=& a^2\frac{2}{M^\prime} \left[f_1^2 (Q^2) \left( \frac{1}{4 m_{l}} M^\prime \left(m_{l}^2 \left(-\left(M^2+
 2 M M^\prime-{M^\prime}^2+t\right)\right)+2 \left(M^2-s\right) \left({M^\prime}^2-s\right)-t \Delta^2+2 s t+t^2\right)
 \right) \right. \nonumber \\
 &+& \frac{f_2^2 (Q^2)}{(M+M^\prime)^2} \left(\frac{1}{8 m_{l}} M^\prime \left(m_{l}^4 ((3 M-M^\prime) (M+
 M^\prime)+t) + m_{l}^2 \left(\Delta^2-t\right) \left(2 (M+M^\prime)^2+t\right) \right. \right. \nonumber \\
 &-& \left. \left. 2 t \left(M^4-2 s \left(M^2+{M^\prime}^2\right)+{M^\prime}^4+2 s^2\right)+2 t^2 \left((M+
 M^\prime)^2-2 s\right)\right) \right) \nonumber \\
 &+& g_1^2 (Q^2) \left( \frac{1}{4 m_{l}}M^\prime \left(m_{l}^2 \left(-\left(M^2-2 M M^\prime-{M^\prime}^2+t\right)
 \right) +2 \left(M^2-s\right) \left({M^\prime}^2-s\right)-t (M+M^\prime)^2+2 s t+t^2\right) \right) \nonumber \\
 &+& \frac{|g_2(Q^2)|^2}{(M+M^\prime)^2} \left(\frac{1}{8 m_{l}} M^\prime \left(m_{l}^4 \left(-\left(\Delta^2
 +4 M \Delta -t\right)\right)+m_{l}^2 \left(-2 \Delta ^2 \left(M^2-2 M M^\prime-{M^\prime}^2+t\right) \right. \right. 
 \right. \nonumber \\
 &-& \left. \left. \left. \left(4 M^2 -t\right) \left(\Delta^2-t\right)+4 M \Delta (2 M \Delta + t)\right)+2 \Delta^2
 \left(2 \left(M^2-s\right) \left({M^\prime}^2-s\right)-t (M+M^\prime)^2+2 s t+t^2\right) \right. \right. \nonumber \\
 &-& \left. 8 \Delta^2 \left(\left(M^2 -s\right) \left({M^\prime}^2-s\right)+s t\right)+4 \left(\Delta^2-
 t\right) \left(\left(M^2-s\right) \left({M^\prime}^2-s\right)+s t\right)\right) \Big) \nonumber \\
 &+& \frac{g_3^2 (Q^2)}{(M+M^\prime)^2} \left(-\frac{1}{2} m_{l} M^\prime \left(m_{l}^2-t\right) \left(\Delta^2
 -t\right) \right) \nonumber \\
 &+& \frac{f_1(Q^2) f_2(Q^2)}{(M+M^\prime)} \left(\frac{1}{2 m_{l}} M^\prime \left(m_{l}^2-t\right) \left(m_{l}^2 M+
 (M+M^\prime) \left((M-M^\prime)^2-t\right)\right) \right) \nonumber\\
 &+& f_1(Q^2) g_1 (Q^2) \left(\frac{1}{2 m_{l}} M^\prime \left(m_{l}^2-t\right) \left(M^2+{M^\prime}^2-2 s-t\right)
 \right) \nonumber \\
 &+& \frac{Re[f_1(Q^2) g_2 (Q^2)]}{(M+M^\prime)} \left(\frac{1}{2 m_{l}} M^\prime \Delta  \left(m_{l}^2-t\right) 
 \left(M^2+{M^\prime}^2-2 s-t\right) \right) \nonumber \\
 &+& \frac{f_2 (Q^2) g_1(Q^2)}{(M+M^\prime)} \left(\frac{1}{2 m_{l}} M^\prime \left(m_{l}^2-t\right) (M+M^\prime) 
 \left(M^2+{M^\prime}^2-2 s-t\right) \right) \nonumber \\
 &+& \frac{Re[f_2 (Q^2) g_2 (Q^2)]}{(M+M^\prime)^2} \left(\frac{1}{2 m_{l}} M^\prime \Delta  \left(m_{l}^2-t\right) 
 (M+M^\prime)\left(M^2+{M^\prime}^2-2 s-t\right) \right) \nonumber \\
 &+& \frac{Re[g_1(Q^2) g_2(Q^2)]}{(M+M^\prime)} \left( \frac{1}{2 m_{l}} M^\prime \left(-m_{l}^4 M+m_{l}^2 
 \left(-2 M^3+M^2 (2 M^\prime-\Delta )+M (2 M^\prime \Delta +t)+\Delta  \left({M^\prime}^2-t\right)\right) \right. 
 \right. \nonumber \\
 &+&  \left. \Delta  \left(2 \left(M^2-s\right) \left({M^\prime}^2-s\right)-t \left((M+M^\prime)^2-2 s \right)+t^2
 \right) -2 \Delta \left(\left(M^2-s\right) \left({M^\prime}^2-s\right)+s t\right)\right) \Bigg) \nonumber \\
 &+& \frac{g_1(Q^2) g_3(Q^2)}{(M+M^\prime)} \left(m_{l} M M^\prime \left(m_{l}^2-t\right) \right) +\frac{Re[g_2 (Q^2) g_3 
 (Q^2)]}{(M+M^\prime)^2} \left(\frac{1}{2} m_{l} M^\prime \left(m_{l}^2-t\right) \left(\Delta^2+2 M \Delta -t\right) \right) 
 \Bigg] 
 \end{eqnarray}
 \begin{eqnarray}
 \label{Cl}
 C^l(Q^2) &=& a^22 \left[\frac{Im[g_1(Q^2) g_2(Q^2)]}{(M+M^\prime)} \left(m_{l} (M+M^\prime) \right) + 
 \frac{Im[g_3(Q^2) g_2(Q^2)]}{(M+M^\prime)^2} (2 m_{l} t) \right]
\end{eqnarray}

\section{Cabibbo theory, $SU(3)$ symmetry and weak $N-Y$ transition form factors}\label{Cabibbo}
In the Cabibbo theory, the weak vector and the axial-vector currents corresponding to the $\Delta S=0$ 
and $\Delta S = 1$ hadronic currents whose matrix elements are defined between the states $|N\rangle$ and $|N^{\prime}
\rangle$ are assumed to belong to the octet representation of $SU(3)$.  
Accordingly, they are defined as:
\begin{eqnarray}\label{APB:su3}
V^\mu_i=\bar{q}F_i\gamma^\mu q,\qquad \qquad 
A^\mu_i=\bar{q}F_i\gamma^\mu\gamma^5 q,
\end{eqnarray}
where $F_i=\frac{\lambda_i}{2}$($i=1-8$) are the generators of flavor $SU(3)$ and $\lambda_i$s are the well known Gell-Mann 
matrices. 
 
From the property of $SU(3)$ group, it follows that there are three corresponding $SU(2)$ subgroups of $SU(3)$ which must be 
invariant under the interchange of quark pairs $ud$, $ds$ and $us$ respectively, if the group is invariant under the 
interchange of $u$, $d$ and $s$ quarks. Each of these $SU(2)$ subgroups has raising and lowering operators. One of them is 
SU(2)$_I$, generated by the generators $(\lambda_{1}, \lambda_{2}, \lambda_{3})$ to be identified with the isospin operators 
$(I_{1}, I_{2}, I_{3})$ in the isospin space. For example, $I_\pm$ of isospin is given by
\begin{equation}
 I_\pm~=~I_1\pm i I_2~=~ F_1\pm i F_2~=~\frac{1}{2}(\lambda_1\pm i \lambda_2).
\end{equation}
The other two are defined as $SU(2)_U$ and $SU(2)_V$ generated by the generators $(\lambda_{6}, \lambda_{7}, \frac{1}{2} 
(\sqrt{3} \lambda_{8} - \lambda_{3}))$ and $(\lambda_{4}, \lambda_{5}, \frac{1}{2} (\sqrt{3} \lambda_{8} + \lambda_{3}))$, 
respectively, in U-spin and V-spin space with $(d~~s)$ and $(u~~s)$ forming the basic doublet representation of $SU(2)_U$ 
and $SU(2)_V$. For more details, see Ref.~\cite{Athar:2020kqn}.
 
From the Gell-Mann matrices $\lambda_i$, one may obtain the raising and lowering operators with U-spin and V-spin in 
analogy with I-spin as:
\begin{eqnarray}
 U_\pm=U_1\pm i U_2= F_6\pm i F_7,\qquad \qquad
 V_\pm=V_1\pm i V_2= F_4\pm i F_5.\nonumber
\end{eqnarray}
In neutron $\beta$-decay, 
the vector and the axial-vector currents for this transition can be written as 
\begin{eqnarray}
 \bar \psi_u\gamma_\mu\psi_d=\bar q\gamma_\mu\left(
 \frac{\lambda_1+i \lambda_2}{2}\right)q=V_\mu^{1+i 2} ,\qquad \qquad
 \bar \psi_u\gamma_\mu \gamma_{5} \psi_d=\bar q
 \gamma_\mu \gamma_{5}\left(\frac{\lambda_1+i \lambda_2}{2}\right)q=A_\mu^{1+i 2}. \nonumber
\end{eqnarray}
Similarly,  $s \rightarrow u$ and $u \rightarrow s$ transformations are written as
\begin{eqnarray}
 \label{APB:j4}
  \bar \psi_u\gamma_\mu\psi_s&=&\bar q\gamma_\mu\left(\frac{\lambda_4+i \lambda_5}{2}\right)q  = V_\mu^{4+i 5}, \qquad \quad
  \bar \psi_s\gamma_\mu\psi_u=\bar q\gamma_\mu\left(\frac{\lambda_4-i \lambda_5}{2}\right)q  = V_\mu^{4-i 5}, 
 \\
 \label{APB:j5}
  \bar \psi_u\gamma_\mu \gamma_{5} \psi_s&=&\bar q\gamma_\mu \gamma_{5}\left(\frac{\lambda_4+i \lambda_5}{2}\right)q  = 
 A_\mu^{4+i 5}, \qquad \quad \bar \psi_s\gamma_\mu \gamma_{5} \psi_u=\bar q\gamma_\mu \gamma_{5}\left(\frac{\lambda_4-i 
 \lambda_5}{2}\right)q  = A_\mu^{4-i 5}. 
\end{eqnarray}
The electromagnetic current which is vector current is written using the $SU(3)$ content
of the charge operator $e= I_3 + \frac{Y}{2}
= \lambda_3 + \frac{1}{2 \sqrt{3}} \lambda_8$ as: 
\begin{eqnarray}\label{APB:jmu}
 J_\mu^{em}&=&V_\mu^3+\frac{1}{\sqrt{3}}V_\mu^8.
\end{eqnarray}
Therefore, the charge changing weak vector and axial-vector currents are written, in Cabibbo theory, as:
\begin{eqnarray}\label{APB:jmu}
 V_\mu^{\pm}&=&\left[V_\mu^1~\pm~ i V_{\mu}^{2}\right]\cos\theta_c~+~\left[V_{\mu}^{4}~\pm~ i V_{\mu}^{5}\right]\sin
 \theta_c, \nonumber \\
 A_{\mu}^{\pm}&=&\left[A_{\mu}^{1}~\pm~ i A_{\mu}^{2}\right]\cos\theta_c~+~\left[A_{\mu}^{4}~\pm~ i A_{\mu}^{5}\right]
 \sin\theta_c.
 \end{eqnarray}
In the Cabibbo theory, the isovector part of the electromagnetic current $J_{em}^\mu$ i.e. $V_{\mu}^{3}$ along with the 
weak vector currents $V_{\pm}^\mu$ are assumed to transform as an octet of vector currents under $SU(3)$. Similarly, the axial 
vector currents are also assumed to transform as an octet under $SU(3)$. The weak transition form factors $f_{i} (q^2)$ and 
$g_{i} (q^2);~ i=1-3$ are determined using Cabibbo theory of $V-A$ interaction extended to the strange sector.
  
 In general, the expression for the matrix element of the transition 
between the two states of baryons (say $B_i$ and $B_k$), through the $SU(3)$ octet ($V_j$ or $A_j$) of currents can be written 
as:
 \begin{eqnarray}\label{APB:bb}
< B_i | V_j | B_k > = if_{ijk}F^V + d_{ijk} D^V, \qquad \qquad
< B_i | A_j | B_k > = if_{ijk}F^A + d_{ijk} D^A.
 \end{eqnarray}
$F^{V}$ and $D^{V}$ are determined from the experimental data on the electromagnetic form factors, and $F^{A}$ and $D^{A}$ 
are determined from the experimental data on the semileptonic decays of neutron and hyperons. The physical baryon octet 
states are written in terms of their octet state $B_{i}$ as discussed in Section~\ref{NRB}.
The matrix element of electromagnetic currents between proton, and neutron states are obtained as
\begin{eqnarray}
 \langle p|J^{em}|p\rangle  =\frac{D^V}{3}+F^V, \qquad \qquad \langle n|J_{em}|n\rangle=\frac{1}{2}\left[-\frac{2}{3}D^V-\frac{2}{3}D^V\right] = -\frac{2D^V}{3}.
 \end{eqnarray}
\begin{table}[htb]
\begin{center}
 \begin{tabular}{|c|c|c|c|}\hline
  ~{{Interaction}}~&~{{Transition}}~&~~~~{{$a$}}~~~~&~~~~{{$b$}}~~~~\\
  \hline
  Electromagnetic&$p\rightarrow p$&1&$\frac{1}{3}$\\
  interaction&$n\rightarrow n$&0&-$\frac{2}{3}$\\\hline
  Weak vector&$n\rightarrow p$&1&1\\
   and axial-vector&$ \Lambda \rightarrow p$&-$\sqrt{\frac{3}{2}}$&-$\frac{1}{\sqrt{6}}$\\
  &$ \Sigma^0 \rightarrow p$&-$\frac{1}{\sqrt{2}}$&$\frac{1}{\sqrt{2}}$\\
  &$ \Sigma^- \rightarrow n $&-1&1\\
  &$\Sigma^\pm \rightarrow \Lambda$&0&$\sqrt{\frac{2}{3}}$\\
  &$\Sigma^- \rightarrow \Sigma^0$&$\sqrt{2}$&0\\
  &$\Xi^- \rightarrow \Lambda$&$\sqrt{\frac{3}{2}}$&-$\frac{1}{\sqrt{6}}$\\
  &$\Xi^- \rightarrow \Sigma^0$&$\frac{1}{\sqrt{2}}$&$\frac{1}{\sqrt{2}}$\\
  &$\Xi^0 \rightarrow \Sigma^+$&1&1\\
  &$\Xi^- \rightarrow \Xi^0$&1&-1\\\hline
 \end{tabular}
 \caption{Values of the coefficients $a$ and $b$ given in Eq.~(\ref{APB:fi}).}\label{APB:table3}
 \end{center}
\end{table}
 The nucleon electromagnetic form factors  can be written 
 in terms of their $SU(3)$ coupling constants as 
\begin{eqnarray}\label{APB:fi}
f_i (Q^2) = a F_i^V (Q^2) + b D_i^V  (Q^2), \qquad \qquad
g_i (Q^2) = a F_i^A (Q^2) + b D_i^A  (Q^2), \qquad i =1,2,3.
\end{eqnarray}
We obtain the Clebsch-Gordan coefficients for 
the electromagnetic $p\rightarrow p$ and $n\rightarrow n$ transitions, which are tabulated in Table-\ref{APB:table3}, which leads to
\begin{eqnarray}\label{APB:FiV}
 F_i^V(Q^2)=F_i^p(Q^2)-F_i^n(Q^2),\qquad \qquad
  D_i^V(Q^2)=-\frac{3}{2}F_i^n(Q^2).
\end{eqnarray}
$F_{i}^{V} (Q^2)$ and $D_{i}^{V} (Q^2)$ determined in terms of the electromagnetic form factors of the nucleon, are used to determine all the form factors in the case of the matrix element of the weak vector current for the various 
$\Delta S=0,1$ transitions.
The coefficients $a$ and $b$ for the various transitions are listed in Table-\ref{APB:table3}.

\section{Density parameters}\label{app_lda}
The modified harmonic oscillator~(MHO) density for carbon while two-parameter Fermi (2pF) density for aluminium, argon, iron, 
lead, etc. are given by:
 \begin{eqnarray}
  \textrm{MHO ~density}:~~~\rho_N(r)&=&\rho_0\Big[1+c_2\left(\frac{r}{c_1} \right)^2 \Big]~,~~~\nonumber
 \textrm{2pF ~density}:~~~ \rho_N(r)=\frac{\rho_0}{1+e^{(r-c_1)/c_2}}\nonumber
 \end{eqnarray}
with $c_1$ and $c_2$ as the density parameters and $\rho_0$ as the central density~\cite{DeVries:1987atn, 
DeJager:1974liz, Garcia-Recio:1991ocp}. These 
parameters are individually tabulated in Table~\ref{table1} for the proton and neutron in the case of nonisoscalar nuclear target 
as well as for the nucleon in the case of isoscalar nuclear target.
  \begin{table}
 \centering
\begin{tabular}{|c|cc|cc|c|c|c|c|}
\hline
 \multirow{2}{*}{ Nucleus}  & \multicolumn{4}{c|}{Nonisoscalar }   & \multicolumn{2}{c|}{Isoscalar} &$ \multirow{2}{*}
 {B.E./A}$ &$ \multirow{2}{*}{T/A}$\\\cline{2-7}  &    $c_1^n$ & $c_1^p$ & $c_2^n$ & $c_2^p$  &  $c_1$   &  $c_2$& & \\\hline  
$^{12}$C  &  -  & -    &   - &    -   &  1.692& 1.082$^{\ast}$& 7.6   &  20.0 \\\hline 
$^{27}$Al&-&-&-&-&3.07&0.519&7.6&20.2\\\hline 
$^{40}$Ar  &  3.64  & 3.47    &   0.569 &    0.569   &  3.53& 0.542& 8.6   &  29.0 \\\hline 
$^{56}$Fe &  4.050  & 3.971   &  0.5935    &   0.5935     & 4.106&0.519&  8.8    &  30.0   \\\hline 
$^{63}$Cu&4.31&4.214&0.586&0.586&4.163&0.606&8.7&29.3\\\hline 
$^{118}$Sn&5.55&5.40&0.543&0.543&5.442&0.543&8.5&31.2\\\hline 
$^{197}$Au&6.79&6.55&0.522&0.522&6.38&0.535&7.9&33.8\\\hline 
$^{208}$Pb &  6.890   & 6.624    &  0.549   &  0.549     & 6.624&0.549&  7.8    & 32.6 \\\hline 
\end{tabular}

\caption{Different parameters used for the numerical calculations for various nuclei. For $^{12}$C
we have used modified harmonic oscillator density($^{\ast}$ $c_2$ is dimensionless) and for $^{40}$Ar, $^{56}$Fe and 
$^{208}$Pb nuclei, 2-parameter Fermi density have been used, where superscript $n$ and $p$ in density 
parameters($c_{i}^{n,p}$; $i=1,2$) stand for the neutron and proton, respectively. Density parameters for isoscalar and 
nonisoscalar nuclear targets are given separately in units of fm. The kinetic energy per nucleon($T/A$) and the binding 
energy per nucleon ($B.E./A$) obtained using Eq.~(\ref{benergy}) for the different nuclei are given in MeV. }
\label{table1}
\end{table}

\bibliography{athar_review}

\end{document}